**Università del Salento**

Dipartimento di Matematica e Fisica “Ennio De Giorgi”

Corso di Dottorato di Ricerca in Fisica e Nanoscienze

# Anomalies and Parity-Violating Interactions:
# From Conformal to Thermal Field Theory

*Supervisor*
Prof. Claudio Corianò

*Candidate*
Stefano Lionetti

*"Whenever I find myself growing grim about the mouth; whenever it is a damp, drizzly November in my soul; whenever I find myself involuntarily pausing before coffin warehouses, and bringing up the rear of every funeral I meet; and especially whenever my hypos get such an upper hand of me, that it requires a strong moral principle to prevent me from deliberately stepping into the street, and methodically knocking people's hats off - then, I account it high time to get to sea as soon as I can."*

---

Herman Melville, *Moby Dick*

# Contents

# List of Publications

The chapters of this thesis are based on the following research papers:

- C. Corianò, S. Lionetti and D. Melle, "Topological Sum Rules and Spectral Flows of Chiral and Gravitational Axion-like Interactions", [arXiv:2502.03182 [hep-ph]].
- C. Corianó and S. Lionetti, "CFT Constraints on Parity-odd Interactions with Axions and Dilatons", Phys. Rev. D **110** (2024) no.12, 125008 [arXiv:2408.02580 [hep-th]].
- C. Corianò, M. Cretì, S. Lionetti and R. Tommasi, "Gravitational chiral anomaly at finite temperature and density", Phys. Rev. D **110** (2024) no.2, 025008 [arXiv:2404.06272 [hep-th]].
- C. Corianò, M. Cretì, S. Lionetti and R. Tommasi, "Axionlike quasiparticles and topological states of matter: Finite density corrections of the chiral anomaly vertex", Phys. Rev. D **110** (2024) no.2, 025014 [arXiv:2402.03151 [hep-ph]].
- C. Corianò, S. Lionetti and M. M. Maglio, "Parity-violating CFT and the gravitational chiral anomaly", Phys. Rev. D **109** (2024) no.4, 045004 [arXiv:2309.05374 [hep-th]].
- C. Corianò, S. Lionetti and M. M. Maglio, "CFT correlators and CP-violating trace anomalies", Eur. Phys. J. C **83** (2023) no.9, 839 [arXiv:2307.03038 [hep-th]].
- C. Corianò, S. Lionetti and M. M. Maglio, "Parity-odd 3-point functions from CFT in momentum-space and the chiral anomaly", Eur. Phys. J. C **83** (2023) no.6, 502 [arXiv:2303.10710 [hep-th]].

The following papers are related to the topics presented in this thesis but are not discussed in detail:

- C. Corianò, S. Lionetti, D. Melle and L. Torcellini, "A Dilaton Sum Rule for the Conformal Anomaly Form Factor in QCD at Order $\alpha_s$", [arXiv:2504.01904 [hep-ph]].
- C. Corianò, S. Lionetti, D. Melle and R. Tommasi, "The Gravitational Form Factors of Hadrons from CFT in momentum-space and the Dilaton in Perturbative QCD", Eur. Phys. J. C **85** (2025) no.5, 498 [arXiv:2409.05609 [hep-ph]].
- C. Corianò, M. Cretì, S. Lionetti and M. M. Maglio, "Three-wave and four-wave interactions in the 4d Einstein Gauss-Bonnet (EGB) and Lovelock theories", Nucl. Phys. B **998** (2024), 116420 [arXiv:2302.02103 [hep-th]].

## Peer-reviewed proceedings

- C. Corianò, S. Lionetti, D. Melle, R. Tommasi and L. Torcellini, "Gravitational Form Factors and the QCD Dilaton at Large Momentum Transfer", PoS **CORFU2024**, [arXiv:2504.20884 [hep-ph]].

- C. Corianò, S. Lionetti, D. Melle and R. Tommasi, "The gravitational form factor of the pion and proton and the conformal anomaly", EPJ Web Conf. **314** (2024), 00030 [arXiv:2409.19586 [hep-ph]].
- C. Corianò, S. Lionetti, D. Melle, R. Tommasi and L. Torcellini, "Conformal Backreaction, Chiral and Conformal Anomalies in the Early Universe", Proceeding of 17th Marcel Grossmann Meeting, [arXiv:2409.18004 [astro-ph.CO]].
- C. Corianó and S. Lionetti, "Quantum anomalies and parity-odd CFT correlators for chiral states of matter," EPJ Web Conf. **314** (2024), 00026 [arXiv:2409.10480 [hep-th]].
- C. Corianò, S. Lionetti, M. M. Maglio and R. Tommasi, "Nonlocal Gravity, Dark Energy and Conformal Symmetry: Testing the Hierarchies of Anomaly-Induced Actions", PoS **CORFU2023** (2024), 165 [arXiv:2404.09225 [hep-th]].
- C. Corianò, M. Cretì, S. Lionetti, D. Melle and R. Tommasi, "Axion-like Interactions and CFT in Topological Matter, Anomaly Sum Rules and the Faraday Effect", [arXiv:2403.15641 [hep-ph]].
- C. Corianò, M. Cretì, S. Lionetti, M. M. Maglio and R. Tommasi, "4D Einstein Gauss-Bonnet Gravity without a Dilaton", PoS **CORFU2022** (2023), 099 [arXiv:2305.19554 [hep-th]].
- S. Lionetti, "Asymptotic symmetries and soft theorems in higher-dimensional gravity", EPJ Web Conf. **270** (2022), 00034 [arXiv:2209.10889 [hep-th]].
- C. Corianò, M. Cretì, S. Lionetti, M. M. Maglio and R. Tommasi, "Dimensional Regularization of Topological Terms in Dilaton Gravity", PoS **CORFU2021** (2022), 025 [arXiv:2205.03535 [hep-th]].

# Introduction

Symmetries play an important role in physics in general and in quantum field theory (QFT) in particular. A symmetry of the classical action is a transformation of the fields that leaves the action invariant. A critical question arises when we inquire whether this symmetry holds true in the quantum realm. In QFT, a quantum anomaly occurs when a symmetry of the classical theory fails to persist at the quantum level. In the path integral formulation, such phenomenon emerges if the action is invariant under the transformations but the measure of the path integral is not.
Of all the anomalies, the chiral and conformal ones are among the most discussed. The violation of the chiral symmetry at the quantum level manifests as a breakdown in the conservation of the axial current $J_A^\mu = -ig\,\bar{\psi}\gamma^\mu\gamma^5\psi$

$$\nabla_\mu\langle J_A^\mu\rangle = a_1\,\varepsilon^{\mu\nu\rho\sigma}F_{\mu\nu}F_{\rho\sigma} + a_2\,\varepsilon^{\mu\nu\rho\sigma}R^\alpha_{\ \beta\mu\nu}R^\beta_{\ \alpha\rho\sigma}$$

where $a_1$ and $a_2$ are numerical coefficients. This anomaly includes both a gauge and a gravitational contribution. Initially discovered in the context of high-energy physics, the chiral anomaly's significance extends far beyond. Its implications span from cosmology [1–8] to condensed matter physics [9–19], with phenomena such as the quantum Hall effect, the chiral magnetic effect and applications to topological insulators (see Fig. 1).
On the other hand, the violation of conformal symmetry at the quantum level is reflected by the energy-momentum tensor's failure to be traceless

$$g_{\mu\nu}\langle T^{\mu\nu}\rangle = b_1 E_4 + b_2 C^{\mu\nu\rho\sigma}C_{\mu\nu\rho\sigma} + b_3\nabla^2 R + b_4 F^{\mu\nu}F_{\mu\nu} + f_1\,\varepsilon^{\mu\nu\rho\sigma}R_{\alpha\beta\mu\nu}R^{\alpha\beta}_{\ \ \rho\sigma} + f_2\varepsilon^{\mu\nu\rho\sigma}F_{\mu\nu}F_{\rho\sigma}$$

where $C^{\mu\nu\rho\sigma}$ is the Weyl tensor, $E_4$ is the Gauss-Bonnet term and $b_i$ and $f_i$ are numerical coefficients. The last two terms on the right-hand side of the equation are parity-odd and have been the center of recent discussions and controversies (see for example [20–24]). Although they are in general allowed, it remains debated whether they exist in a theory with Weyl fermions.
The anomalous laws described above translate into Ward identities that dictate the behavior of various three-point correlation functions. In this thesis, we investigate parity-odd interactions which are affected by chiral and conformal anomalies in the context of conformal and thermal field theories.
In general, by imposing (anomalous) Ward identities such as those presented above, one can fully determine the longitudinal and the trace part of a correlator. Specifically, correlators affected by an anomaly exhibit a pole in the anomalous momentum $\left(1/q^2\right)$ within the longitudinal-trace part. On the other hand, the transverse-traceless component of a correlator is not directly constrained by these Ward identities. In order to fix such component, we can impose the conformal constraints following from the invariance under dilatations and special conformal transformations.

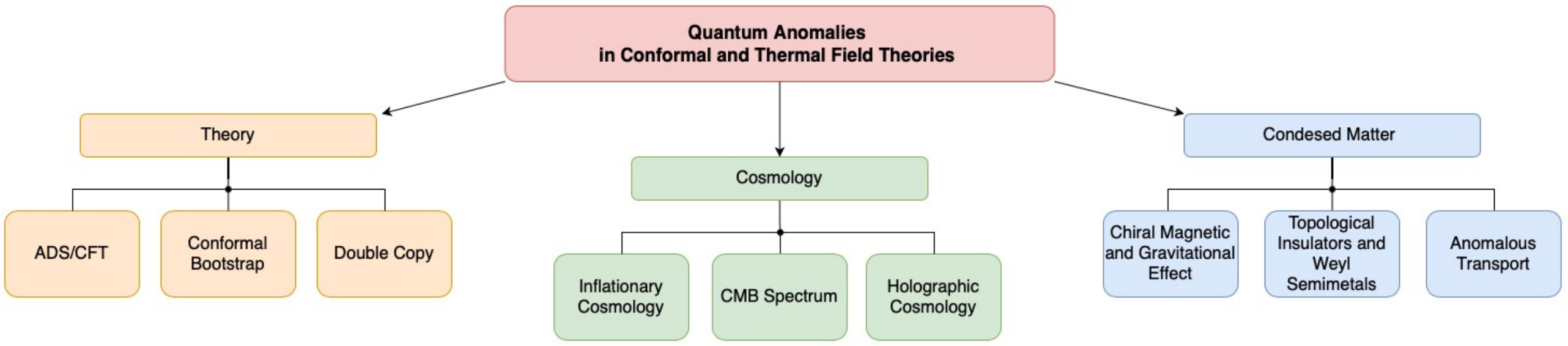

Figure 1: Examples of topics interconnected with quantum anomalies in conformal and thermal field theories

## Parity-odd correlators in CFT

The Poincaré group is a fundamental symmetry in physics, encompassing translations and Lorentz transformations. These transformations leave the metric invariant, making the Poincaré group essential in the study of special relativity and quantum field theory. However, certain physical theories, such as those involving massless particles (like photons) or in the study of critical phenomena, exhibit a broader set of symmetries. This broader symmetry group is known as the conformal group, which extends the Poincaré group to include additional transformations that preserve the angle between spacetime intervals, rather than their exact length. In addition to the Poincaré transformations, the conformal group includes dilations and special conformal transformations.

Conformal field theories (CFTs) have played a central role for over half a century in different areas of physics, from the theory of critical phenomena to string theory and the AdS/CFT correspondence. Conformal invariance imposes strong constraints on correlation functions. Indeed the general structure of two-point and three-point correlation functions is completely determined up to a few constants by such symmetry. Their expression can be easily obtained by methods in coordinate-space. However, such methods are generally valid only when the correlation functions are computed at separate points. The coordinate-space approach provides limited information on the origin of quantum anomalies which occur at short distances and thus with correlators in which some points coalesce. Therefore, one of the main reasons for studying CFT correlation functions directly in momentum-space is to see the effects of anomalies more directly [25–27]. Furthermore, from the perspective of perturbative field theory which is naturally formulated in momentum-space, it is of interest to study CFTs in the same setting.

The idea of using conformal Ward identities to determine the structure of three-point functions in momentum-space was presented for the first time independently in [28] and [29], the second of which outlines a method that includes the tensor case. Investigations into parity-odd CFT correlators have only recently emerged [30–37]. In this thesis, we focus in particular on parity-odd CFT correlators that are affected by chiral and conformal anomalies. The conformal constraints in momentum-space take the form of differential equations that a correlator must satisfy. By solving these equations, we can determine the most general expression of a correlator in CFT. The general solution can be expressed in terms of integrals involving a product of three Bessel functions ("3K integrals"), which can be mapped to ordinary Feynman integrals for free field theory realizations. The analysis of the parity-odd sector in the presence of chiral and conformal anomalies is discussed in [34–37].

CFTs in momentum-space, especially with broken parity, find applications in cosmology, condensed matter physics, study of anomalies, holography and the conformal bootstrap program. Indeed, the recent discovery of applications to topological insulators and Weyl semimetals has sparked new interest in the topic. Another significant application is in the cosmological setting where the CMB bispectrum, which is a measure of nongaussianity, is given by the 3-point correlators in momentum-space [38–41].

## Parity-odd correlators in thermal field theories

In recent years, it has become increasingly clear that quantum anomalies play a pivotal role in the dynamics of fundamental interactions. This influence extends not only to the vacuum case, but also to scenarios involving a finite temperature and chemical potential, as demonstrated for example in the context of the chiral magnetic effect. The potential for experimental verification of these interactions, which could reveal the nature of the pseudoparticles emerging from virtual corrections, as predicted by recent and past analyses, should be taken seriously at the experimental level.

In the second part of this thesis, we examine gauge and gravitational chiral anomaly interactions in a hot and dense medium. The correlators under scrutiny are $\langle J_V J_V J_A \rangle$ and $\langle T T J_A \rangle$, where $J_V$ is a vector current, $J_A$ is an axial-vector current and $T$ is the energy-momentum tensor. We compute the correlators perturbatively using the same Feynman diagrams of the vacuum case. However, in this case we need to replace the usual fermionic propagator with its generalization in a hot medium in the real time formulation. We use the expression

$$S_F = (\not{k} + m)\left\{ \frac{1}{k^2 - m^2} + 2\pi i \delta\left(k^2 - m^2\right)\left[ \frac{\theta(k_0)}{e^{\beta(E-\mu)} + 1} + \frac{\theta(-k_0)}{e^{\beta(E+\mu)} + 1} \right] \right\},$$

where $\mu$ is the chemical potential and $\beta = 1/(k_B T_{emp})$. Such expression can also be formulated covariantly by introducing the four-vector $\eta$, defining the velocity of the heat-bath. Thermal effects break both conformal invariance—and more significantly, Lorentz invariance—resulting in outcomes that differ from those discussed in the previous section.

Considerable attention has been devoted to examining the effects of finite temperature and density on the chiral gauge anomaly in several contexts [1, 42–47]. Despite the diverse approaches taken by the various authors to address this issue, a unanimous consensus emerges: the chiral gauge anomaly remains insensitive to corrections from finite temperature and density. In this thesis, we confirm such results with an independent perturbative computation of the $\langle J_V J_V J_A \rangle$ in a thermal medium. Furthermore, we determine the full form of the correlator, i.e., both its longitudinal and transverse sectors. We find that thermal corrections appear only in the transverse sector, while the longitudinal anomalous sector remains topologically protected from such corrections In addition, when the photons are on-shell, we show the entire correlators reduce to a longitudinal anomalous sector.

The gravitational axial anomaly has received relatively little attention in the context of finite temperature and density. In this thesis, we address this gap by investigating the $\langle T T J_A \rangle$ interaction. Our goal is to demonstrate that the gravitational axial anomaly remains unaffected, with no contributions from either density or temperature effects. This result underlines the robust protection of chiral anomalies against thermal corrections. The findings have significant implications for both theoretical and experimental physics. For instance, the protection of the gravitational chiral anomaly influences processes such as the decay of axions or axion-like particles into gravitational waves and the generation of chiral currents by gravitational waves in dense and hot environments. Additionally, this protection opens intriguing experimental avenues in condensed matter physics, where analogous phenomena might be observed and studied.

## Overview of the thesis

This thesis examines parity-odd anomalous interactions primarily in two distinct contexts: conformal field theory and thermal field theory.
In the first two chapters, we introduce the fundamental concepts necessary for our analysis. In particular, Chapter 1 explores Ward identities and quantum anomalies, which arise when a classical symmetry is not preserved in the quantum regime. Chapter 2 then provides an overview of conformal field theory, first focusing on its coordinate-space formulation before transitioning to a momentum-space perspective. Here, we introduce the methodology required to determine the structure of correlation functions by imposing conformal Ward identities. These techniques will be systematically applied to various correlators in the subsequent chapters.
Equipped with these foundational concepts, in Chapter 3, we consider as a first example the simple case of parity-odd correlators built with at least one scalar operator. This choice allows us to begin with relatively simple cases that are nonetheless highly relevant, as they are directly connected to chiral and conformal anomalies.
Chapter 4 examines chiral anomaly interactions, specifically the correlators $\langle J_V J_V J_A \rangle$ and $\langle J_A J_A J_A \rangle$. We demonstrate how conformal equations fully determine these correlators in terms of the chiral anomaly coefficient. Similarly, in Chapter 5, we analyze the chiral gravitational anomaly correlator $\langle TTJ_A \rangle$ and once again illustrate how conformal equations uniquely determine its structure based on the gravitational anomaly coefficient.
Chapter 6 shifts focus to a different type of anomaly: trace anomalies. We specifically study the parity-odd $\langle JJT \rangle$ correlator and investigate how it is constrained by conformal invariance, both in the presence and absence of parity-violating trace anomalies.
Subsequently, we transition to the study of quantum anomalies in a finite-temperature and finite-density background. Chapter 7 introduces key concepts of thermal field theory and applies them to the chiral anomaly interaction $\langle J_V J_V J_A \rangle$. We compute the correlator perturbatively, analyzing both its anomalous longitudinal and transverse components.
In Chapter 8, we extend our analysis to the gravitational anomaly at finite temperature and density. Specifically, we compute the anomaly of the $\langle TTJ_A \rangle$ correlator perturbatively and investigate possible thermal corrections.
Chapter 9 explores the perturbative structure of the standard $\langle J_V J_V J_A \rangle$ and $\langle TTJ_A \rangle$ correlators in the presence of massive fermions. These results are used to validate the presence of anomaly sum rules, which are essential area laws governing the anomaly vertices.
Finally, in the concluding chapter, we discuss and reflect on our findings, highlighting their implications and possible extensions. Additional technical details are provided in the Appendix.

# Chapter 1

# Ward identities and quantum anomalies

Symmetries play an important role in physics in general and in quantum field theory in particular. A symmetry of the classical action is a transformation of the fields that leaves the action invariant. Key examples include Poincaré transformations and gauge symmetries. An essential question is whether a symmetry persists at the quantum level.
In the functional integral formulation of quantum field theory, classical symmetries translate into Ward identities for correlation functions of the quantum effective action. A crucial assumption in this derivation is that the functional integral measure remains invariant under the symmetry. If this assumption fails, the Ward identities are violated by a so-called anomaly.
Alternatively, if one is computing (diverging) Feynman diagrams, one has to introduce some regularization and it may happen that no regularization preserves all of the symmetries. We are then faced with a breakdown of Ward identities, which again signals the presence of an anomaly.
Another perspective involves checking whether a classically conserved current remains conserved at the quantum level. If the current is not conserved, the quantum effective action is not invariant, indicating an anomaly.
The consequences of anomalies depend on the nature of the symmetry. If a global symmetry is anomalous, classical selection rules may no longer hold, allowing classically forbidden processes to occur. In contrast, anomalies in gauge symmetries pose a severe problem, as gauge invariance is essential for unitarity and renormalizability. A consistent gauge theory must ensure that such anomalies cancel when summing over all particle contributions.
This chapter provides an introduction to Ward identities and anomalies. We begin with the concepts of the partition function and effective action, followed by the derivation of Ward identities. We then examine anomalies, focusing in particular on the chiral anomaly, which we derive using both Feynman diagrams and the Fujikawa method. Finally, we explore broader topics, including gravitational and conformal anomalies.

## 1.1 The effective action

We begin our discussion by defining the partition function of a theory with action $S_0$ in Minkowski space as

$$Z = \mathcal{N} \int d\chi\, e^{iS_0} \tag{1.1.1}$$

where $\mathcal{N}$ is a normalization constant. Here, $\chi$ represents a collection of fields, which can include scalars and fermions. We now define the effective action as

$$e^{i\mathcal{S}} = Z \quad \leftrightarrow \quad \mathcal{S} = -i \log Z. \tag{1.1.2}$$

According to these definitions, in the Feynman diagrammatic expansion, $Z$ contains both connected and disconnected graphs, while $\mathcal{S}$ collects only connected graphs.

The effective action can depend on a set of external fields that are not integrated over in the path integral. In this thesis, we consider the gravitational field as an external field, thereby avoiding quantization issues. The axial field $A_\mu$ should also be regarded as external since, as we will see, it is affected by anomalies and therefore cannot serve as a proper gauge field. Regarding the $V_\mu$ field, it can be treated either as an external field or not, as it does not affect our main objectives.

One can now define correlation functions through functional differentiation with respect to the sources. For instance, the quantum averages of 1-point functions are defined as

$$\begin{aligned} \langle T^{\mu\nu}(x)\rangle &= \frac{2}{\sqrt{-g(x)}}\frac{\delta\mathcal{S}}{\delta g_{\mu\nu}(x)}\bigg|_0, & \langle O(x)\rangle &= \frac{1}{\sqrt{-g(x)}}\frac{\delta\mathcal{S}}{\delta\phi(x)}\bigg|_0, \\ \left\langle J_A^\mu(x)\right\rangle &= \frac{1}{\sqrt{-g(x)}}\frac{\delta\mathcal{S}}{\delta A_\mu(x)}\bigg|_0, & \left\langle J_V^\mu(x)\right\rangle &= \frac{1}{\sqrt{-g(x)}}\frac{\delta\mathcal{S}}{\delta V_\mu(x)}\bigg|_0, \end{aligned} \tag{1.1.3}$$

where the subscript zero indicates that the sources have been switched off and the metric is taken to be flat. Note that we have introduced two distinct fields that couple to currents. The vector field $V_\mu$ couples to the parity-even vector current $J_V$ which can be realized using fermions as

$$J_V^\mu = -ig\bar{\psi}\gamma^\mu\psi \tag{1.1.4}$$

where $g$ denotes the charge of the fermion field. The axial field $A_\mu$ couples to the parity-odd axial current $J_A$, which can be realized as follows

$$J_A^\mu = -ig\bar{\psi}\gamma^\mu\gamma^5\psi \tag{1.1.5}$$

where we have introduced the Dirac matrix

$$\gamma^5 = i\gamma^0\gamma^1\gamma^2\gamma^3 \tag{1.1.6}$$

In this thesis, we will use the notation $J$ without a subscript when the specific type of current is not relevant. As we will explore in this chapter, although both $J_V$ and $J_A$ are classically conserved, the conservation of the axial current is broken at the quantum level.

Higher-point correlation functions can be defined by applying multiple functional derivatives to the effective action. For example, a correlator involving $n$ energy-momentum tensors is given by

$$\langle T^{\mu_1\nu_1}(x_1)\dots T^{\mu_n\nu_n}(x_n)\rangle = \frac{2^n}{\sqrt{-g(x_1)}\dots\sqrt{-g(x_n)}}\,\frac{\delta^n\mathcal{S}}{\delta g_{\mu_1\nu_1}(x_1)\dots\delta g_{\mu_n\nu_n}(x_n)}\bigg|_0 \tag{1.1.7}$$

To preserve symmetry under permutations of operator insertions, we position all $\sqrt{g}$ factors outside the functional derivatives. With this convention, these $\sqrt{g}$ factors are effectively set to unity.

Focusing on the gravitational sector as an example, the effective action can be expanded in a Taylor series around the flat metric $\eta$ as follows

$$\begin{aligned} &\mathcal{S}[\eta+h] = \mathcal{S}[\eta] + \\ &\sum_{n=1}^{\infty}\frac{1}{2^n\,n!}\int d^4x_1\dots d^4x_n\,\sqrt{-g(x_1)}\dots\sqrt{-g(x_n)}\,\langle T^{\mu_1\nu_1}(x_1)\dots T^{\mu_n\nu_n}(x_n)\rangle\, h_{\mu_1\nu_1}(x_1)\dots h_{\mu_n\nu_n}(x_n), \end{aligned} \tag{1.1.8}$$

where the coefficients of the expansion are precisely the correlation functions of the energy-momentum tensor. A similar expansion can also be performed with vector fields and their associated currents.

## 1.2 Ward identities

In this section, we illustrate the procedure for obtaining the canonical Ward identities related to the three local symmetry that one can have in the calculation of correlation functions of scalar, conserved current and conserved and traceless energy-momentum tensor operators. The first step is to couple the system to some background fields as we have discussed in the previous section. Then, we require that the resulting generating functional is invariant under diffeomorphisms, gauge and Weyl transformations. Our derivation is general and we employ non-Abelian vector currents

$$\langle J^{a\mu}(x)\rangle = \frac{1}{\sqrt{-g(x)}}\frac{\delta Z}{\delta V^a_\mu(x)}\bigg|_0 \tag{1.2.1}$$

The Abelian case can be obtained by ignoring the color indices and setting the structure constants to zero.

### 1.2.1 Transverse Ward identities

We know that under a diffeomorphism $\xi^\mu$ the sources in the generating functional transform as

$$\delta g^{\mu\nu} = -(\nabla^\mu\xi^\nu + \nabla^\nu\xi^\mu), \tag{1.2.2}$$
$$\delta V^a_\mu = \xi^\nu\nabla_\nu V^a_\mu + \nabla_\mu\xi^\nu\, V^a_\nu, \tag{1.2.3}$$
$$\delta\phi^I = \xi^\nu\partial_\nu\phi^I, \tag{1.2.4}$$

Under a gauge symmetry transformation with parameter $\alpha^a$, the sources transform as

$$\delta g^{\mu\nu} = 0, \tag{1.2.5}$$
$$\delta V^a_\mu = -D^{ac}_\mu\alpha^c = -\partial_\mu\alpha^a - f^{abc}V^b_\mu\alpha^c, \tag{1.2.6}$$
$$\delta\phi^I = -i\alpha^a(T^a_R)^{IJ}\phi^J, \tag{1.2.7}$$

where $T^a_R$ are matrices of a representation $R$ and $f^{abc}$ are structure constants of the group $G$. The gauge field transforms in the adjoint representation, while $\phi^I$ may transform in any representation $R$. The covariant derivative is $D^{IJ}_\mu = \delta^{IJ}\partial_\mu - iV^a_\mu(T^a_R)^{IJ}$.
The Ward Identities follow from the requirement that the generating functional $Z$ is invariant under these transformations and in particular under the variation

$$\delta_\xi = \int d^dx\left[-(\nabla^\mu\xi^\nu + \nabla^\nu\xi^\mu)\frac{\delta}{\delta g^{\mu\nu}} + (\xi^\nu\nabla_\nu V^a_\mu + \nabla_\mu\xi^\nu\, V^a_\nu)\frac{\delta}{\delta V^a_\mu} + \xi^\nu\partial_\nu\phi^I\frac{\delta}{\delta\phi^I}\right], \tag{1.2.8}$$
$$\delta_\alpha = -\int d^dx\left[(\partial_\mu\alpha^a + f^{abc}V^b_\mu\alpha^c)\frac{\delta}{\delta V^a_\mu} + i\alpha^a(T^a_R)^{IJ}\phi^J\frac{\delta}{\delta\phi^I}\right], \tag{1.2.9}$$

so that the canonical Ward Identities for the diffeomorphism and gauge transformations are respectively given by

$$\delta_\xi Z = 0, \qquad \delta_\alpha Z = 0. \tag{1.2.10}$$

By integrating by parts and using the property $\frac{1}{\sqrt{-g}}\partial_\mu\sqrt{-g}=\Gamma^\lambda_{\lambda\mu}$, we find

$$\begin{aligned}\delta_\alpha Z = 0 &= -\int d^dx\left[(\partial_\mu\alpha^a + f^{abc}V^b_\mu\alpha^c)\frac{\delta}{\delta V^a_\mu} + i\alpha^a(T^a_R)^{IJ}\phi^J\frac{\delta}{\delta\phi^I}\right]Z\\ &= -\int d^dx\sqrt{-g}\left[(\partial_\mu\alpha^a + f^{abc}V^b_\mu\alpha^c)\frac{1}{\sqrt{-g}}\frac{\delta}{\delta V^a_\mu} + i\alpha^a(T^a_R)^{IJ}\phi^J\,\frac{1}{\sqrt{-g}}\frac{\delta}{\delta\phi^I}\right]Z\\ &= -\int d^dx\sqrt{-g}\,\alpha^a\left[-\Gamma^\lambda_{\mu\lambda}\langle J^{\mu a}(x)\rangle - (\partial_\mu\delta^{ab} + f^{acb}V^c_\mu)\langle J^{\mu b}(x)\rangle + i(T^a_R)^{IJ}\phi^J\,\langle\mathcal{O}_I(x)\rangle\right],\end{aligned}\tag{1.2.11}$$

Using the arbitrariness of the parameter $\alpha^a$, we obtain the first Ward Identity related to the gauge symmetry

$$\begin{aligned}0 &= D^{ab}_\mu\langle J^{\mu b}\rangle + \Gamma^\lambda_{\mu\lambda}\langle J^{\mu a}(x)\rangle - i(T^a_R)^{IJ}\phi^J\langle\mathcal{O}_I\rangle\\ &= \nabla_\mu\langle J^{\mu a}\rangle + f^{abc}V^b_\mu\langle J^{\mu c}\rangle - i(T^a_R)^{IJ}\phi^J\langle\mathcal{O}_I\rangle.\end{aligned}\tag{1.2.12}$$

The other Ward identities related to the diffeomorphism invariance will be

$$\begin{aligned}0 = \delta_\xi Z &= \int d^dx\left[-(\nabla^\mu\xi^\nu + \nabla^\nu\xi^\mu)\frac{\delta}{\delta g^{\mu\nu}} + (\xi^\nu\nabla_\nu V^a_\mu + \nabla_\mu\xi^\nu\,V^a_\nu)\frac{\delta}{\delta V^a_\mu} + \xi^\nu\partial_\nu\phi^I\frac{\delta}{\delta\phi^I}\right]Z\\ &= -\int d^dx\sqrt{-g}\left[\frac{1}{2}(\nabla^\mu\xi^\nu + \nabla^\nu\xi^\mu)\frac{2}{\sqrt{-g}}\frac{\delta}{\delta g^{\mu\nu}} - (\xi^\nu\nabla_\nu V^a_\mu + \nabla_\mu\xi^\nu\,V^a_\nu)\frac{1}{\sqrt{-g}}\frac{\delta}{\delta V^a_\mu} - \xi^\nu\partial_\nu\phi^I\frac{1}{\sqrt{-g}}\frac{\delta}{\delta\phi^I}\right]Z\\ &= -\int d^dx\sqrt{-g}\left[\frac{1}{2}(\nabla^\mu\xi^\nu + \nabla^\nu\xi^\mu)\langle T_{\mu\nu}(x)\rangle - (\xi^\nu\nabla_\nu V^a_\mu + \nabla_\mu\xi^\nu\,V^a_\nu)\langle J^{\mu a}(x)\rangle - \xi^\nu\partial_\nu\phi^I\langle\mathcal{O}_I(x)\rangle\right]\\ &= -\int d^dx\sqrt{-g}\,\xi^\nu\left[-\nabla^\mu\langle T_{\mu\nu}(x)\rangle - \nabla_\nu V^a_\mu\langle J^{\mu a}(x)\rangle + \nabla_\mu(V^a_\nu\langle J^{\mu a}(x)\rangle) - \partial_\nu\phi^I\,\langle\mathcal{O}_I(x)\rangle\right]\end{aligned}\tag{1.2.13}$$

that leads to

$$\begin{aligned}0 &= \nabla^\mu\langle T_{\mu\nu}(x)\rangle + \nabla_\nu V^a_\mu\langle J^{\mu a}(x)\rangle - \nabla_\mu V^a_\nu\langle J^{\mu a}(x)\rangle - V^a_\nu\nabla_\mu\langle J^{\mu a}(x)\rangle + \partial_\nu\phi^I\langle\mathcal{O}_I(x)\rangle\\ &= \nabla^\mu\langle T_{\mu\nu}(x)\rangle - F^a_{\mu\nu}\langle J^{\mu a}(x)\rangle - f^{abc}V^a_\nu V^b_\mu\langle J^{\mu c}\rangle - V^a_\nu\nabla_\mu\langle J^{\mu a}(x)\rangle + \partial_\nu\phi^I\langle\mathcal{O}_I(x)\rangle.\end{aligned}\tag{1.2.14}$$

Using the Ward Identity related to the gauge symmetry we obtain the final result

$$\begin{aligned}0 &= \nabla^\mu\langle T_{\mu\nu}(x)\rangle - F^a_{\mu\nu}\langle J^{\mu a}(x)\rangle - f^{abc}V^a_\nu V^b_\mu\langle J^{\mu c}\rangle - V^a_\nu\nabla_\mu\langle J^{\mu a}(x)\rangle + \partial_\nu\phi^I\langle\mathcal{O}_I(x)\rangle\\ &= \nabla^\mu\langle T_{\mu\nu}(x)\rangle - F^a_{\mu\nu}\langle J^{\mu a}(x)\rangle - f^{abc}V^a_\nu V^b_\mu\langle J^{\mu c}\rangle + V^a_\nu\left[f^{abc}V^b_\mu\langle J^{\mu c}\rangle - i(T^a_R)^{IJ}\phi^J\langle\mathcal{O}_I\rangle\right] + \partial_\nu\phi^I\langle\mathcal{O}_I(x)\rangle\\ &= \nabla^\mu\langle T_{\mu\nu}(x)\rangle - F^a_{\mu\nu}\langle J^{\mu a}(x)\rangle + D^{IJ}_\nu\,\phi^I\langle\mathcal{O}_J(x)\rangle\end{aligned}\tag{1.2.15}$$

The equations (1.2.12) and (1.2.14) are the Ward Identities for 1-point functions with sources turned on. These equations may then be differentiated with respect to the sources, with the aim of obtaining the corresponding Ward Identities for higher point functions.

### 1.2.2 Trace Ward identity

In this section, we discuss the Ward identities that arise from Weyl invariance. The meaning and relationship between Weyl and conformal invariance will be explored in greater detail in Chapter 2. For now, it suffices to note that Weyl invariance is defined for quantum field theories that can be coupled to a background metric $g_{\mu\nu}$ in a diffeomorphism-invariant manner. For such theories, Weyl transformations correspond to a local rescaling of the metric combined with a transformation of the local operators

$$\delta_\sigma g_{\mu\nu} = 2g_{\mu\nu}\sigma, \tag{1.2.16}$$

$$\delta_\sigma V^a_\mu = 0, \tag{1.2.17}$$

$$\delta_\sigma \phi = (d-\Delta)\phi\,\sigma. \tag{1.2.18}$$

Let's then consider the case in which the generating functional is invariant under Weyl transformation and free of the Weyl anomaly, for which

$$\delta_\sigma Z = 0. \tag{1.2.19}$$

The variation of the generating functional is realized by the following operator

$$\delta_\sigma = \int d^dx\;\sigma\left[2g_{\mu\nu}\frac{\delta}{\delta g_{\mu\nu}} + (d-\Delta)\phi^I\,\frac{\delta}{\delta\phi^I}\right], \tag{1.2.20}$$

Therefore, we can expand (1.2.19) as

$$\begin{aligned}0 = \delta_\sigma Z &= \int d^dx\;\sigma\left[2g_{\mu\nu}\frac{\delta}{\delta g_{\mu\nu}} + (d-\Delta)\phi^I\,\frac{\delta}{\delta\phi^I}\right]Z\\ &= \int d^dx\;\sqrt{-g}\,\sigma\left[g_{\mu\nu}\langle T^{\mu\nu}(x)\rangle + (d-\Delta)\phi^I\,\langle\mathcal{O}^I(x)\rangle\right].\end{aligned} \tag{1.2.21}$$

In this case we find the following trace, or Weyl, Ward identity in the presence of sources

$$\langle T^\mu_\mu(x)\rangle = (\Delta-d)\phi^I\langle\mathcal{O}^I(x)\rangle. \tag{1.2.22}$$

Also in this case, we can differentiate with respect to the sources in order to obtain the trace Ward identities for $n$-point functions.

## 1.3 Perturbative analysis of the chiral anomaly

In the previous section, we have derived the Ward identities corresponding to various symmetries. Here, we explore whether a symmetry of the classical action is always preserved in the quantum realm. As an example, we consider the well-known chiral symmetry. As we will see, quantum fluctuations break such symmetry. We will compute the chiral anomaly through a detailed examination of Feynman diagrams.
We start by considering the Lagrangian of a single massless fermion

$$\mathcal{L} = \bar\psi i\gamma^\mu\partial_\mu\psi \tag{1.3.1}$$

This theory is manifestly invariant under the separate transformations

$$\psi \to e^{i\theta}\psi, \qquad\qquad \psi \to e^{i\theta\gamma^5}\psi \tag{1.3.2}$$

corresponding to the conserved vector and axial-vector currents respectively

$$J^\mu_V = -ig\bar\psi\gamma^\mu\psi, \qquad\qquad J^\mu_A = -ig\bar\psi\gamma^\mu\gamma^5\psi \tag{1.3.3}$$

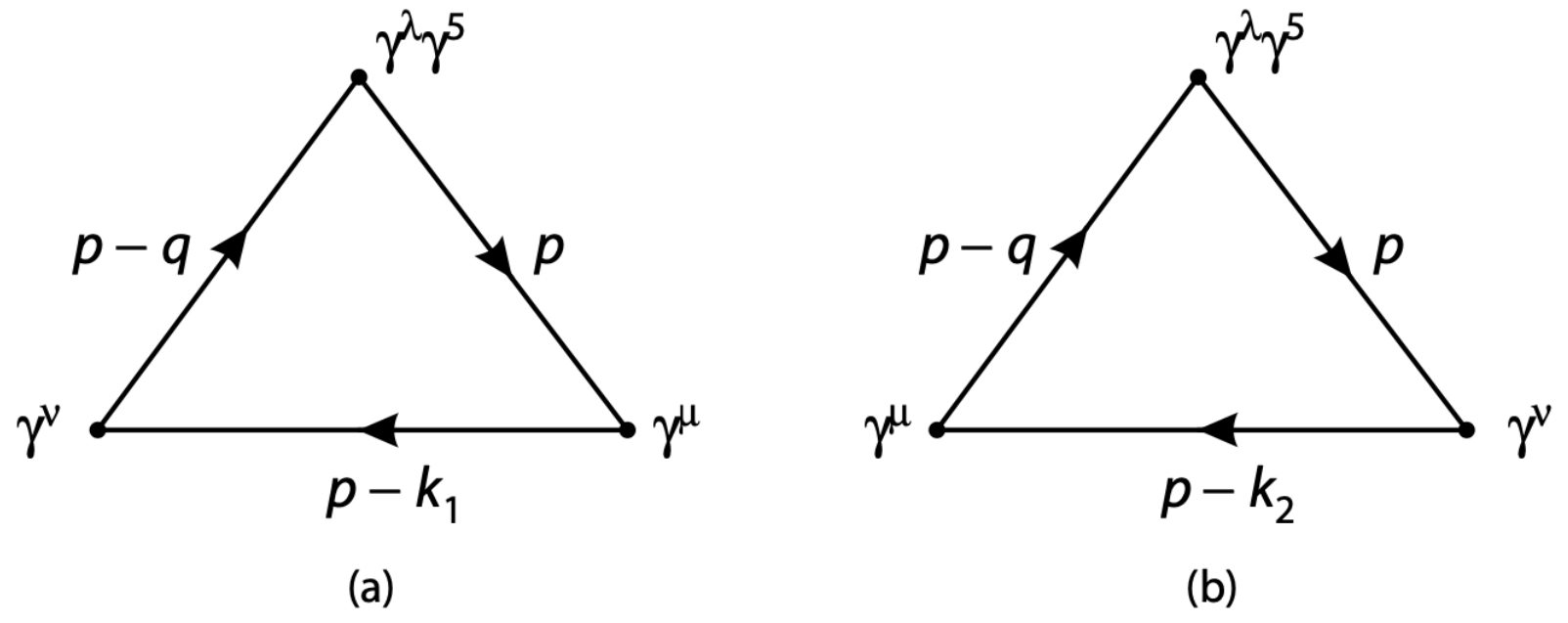

Figure 1.1: Feynman diagrams for the $\langle J_V J_V J_A\rangle$ interaction

These currents can be coupled to external source fields $V_\mu$ and $A_\mu$, as we have discussed previously. One can verify that $\partial_\mu J_V^\mu = 0$ and $\partial_\mu J_A^\mu = 0$ follow immediately from the classical equation of motion $i\gamma^\mu\partial_\mu\psi = 0$.
Let us now calculate the amplitude $\langle J_V^{\mu_1}(x_1)J_V^{\mu_2}(x_2)J_A^{\mu_3}(x_3)\rangle$. According to such interaction a fermion-antifermion pair is created at $x_1$ and another such pair is created at $x_2$ by the vector current. Then the fermion from one pair annihilates the antifermion from the other pair and the remaining fermion-antifermion pair is subsequently annihilated by the axial current. The Feynman diagrams of corresponding to such interaction are reported in Fig. 1.1 and can be expressed as

$$\Delta^{\lambda\mu\nu}(k_1,k_2) \equiv \langle J_V^\mu(k_1)J_V^\nu(k_2)J_A^\lambda(q)\rangle = \\ -g^3\int\frac{d^4p}{(2\pi)^4}\,\mathrm{tr}\left(\gamma^\lambda\gamma^5\frac{1}{\not p-\not q}\gamma^\nu\frac{1}{\not p-\not k_1}\gamma^\mu\frac{1}{\not p}+\gamma^\lambda\gamma^5\frac{1}{\not p-\not q}\gamma^\mu\frac{1}{\not p-\not k_2}\gamma^\nu\frac{1}{\not p}\right) \tag{1.3.4}$$

with $q = k_1 + k_2$. Note that the two terms are required by Bose statistics. The overall factor of $(-1)$ comes from the closed fermion loop.
Classically, we recognize two symmetries that lead to the conservation equations $\partial_\mu J_V^\mu = 0$ and $\partial_\mu J_A^\mu = 0$. However, when transitioning to the quantum theory, these conservation laws are not guaranteed to hold automatically. If $\partial_\mu J_V^\mu = 0$ remains valid in the quantum theory, then it must follow that

$$k_{1\mu}\Delta^{\lambda\mu\nu} = 0, \qquad k_{2\nu}\Delta^{\lambda\mu\nu} = 0. \tag{1.3.5}$$

Similarly, if the axial current remains conserved, we require

$$q_\lambda\Delta^{\lambda\mu\nu} = 0. \tag{1.3.6}$$

Having set up the formalism, the next logical step is to compute $\Delta^{\lambda\mu\nu}$ explicitly to determine whether these symmetries persist under quantum fluctuations. Before engaging in the explicit calculation, however, let us reflect on the implications of a violation of these conservation laws. The non-conservation of the vector current would be particularly concerning. The associated charge,

$$Q_V = \int d^3x J_V^0, \tag{1.3.7}$$

represents the number of fermions. If $\partial_\mu J_V^\mu \neq 0$, this would imply that fermions could spontaneously appear or disappear, violating a fundamental principle of particle physics. Moreover, the gauge invariance of electromagnetism relies crucially on this conservation law.

On the other hand, the violation of axial current conservation is a less alarming prospect. The axial charge

$$Q_A = \int d^3x J_A^0, \tag{1.3.8}$$

is associated with chiral symmetry, which is susceptible to quantum anomalies, as we are going to see. Unlike the vector charge, whose conservation is critical for fundamental principles like the gauge symmetry, the non-conservation of the axial charge does not lead to such drastic consequences. Ultimately, the calculation of $\Delta^{\lambda\mu\nu}$ will reveal whether these conservation laws hold or whether quantum effects introduce anomalies. While the breakdown of vector current conservation would pose serious theoretical and experimental issues, a violation of axial current conservation would be a far less troubling outcome.

### 1.3.1 Shifting integration variable

We start by contracting $k_{1\mu}$ with $\Delta^{\lambda\mu\nu}$ as written in (1.3.4). We then express $\not{k}_1$ in the first term as $\not{p} - (\not{p} - \not{k}_1)$ and in the second term as $(\not{p} - \not{k}_2) - (\not{p} - \not{q})$, thus obtaining

$$\begin{aligned} & k_{1\mu}\Delta^{\lambda\mu\nu}(k_1,k_2) \\ &= -g^3 \int \frac{d^4p}{(2\pi)^4} \operatorname{tr}\left(\gamma^\lambda\gamma^5 \frac{1}{\not{p}-\not{q}}\gamma^\nu \frac{1}{\not{p}-\not{k}_1} - \gamma^\lambda\gamma^5 \frac{1}{\not{p}-\not{k}_2}\gamma^\nu \frac{1}{\not{p}}\right) \end{aligned} \tag{1.3.9}$$

In the integrand, the first term is simply the second term with the shift of the integration variable $p \to p - k_1$. The two terms appear to cancel, suggesting that $k_{1\mu}\Delta^{\lambda\mu\nu} = 0$, as expected. However, we must carefully consider whether shifting the integration variable is legitimate.
When is $\int_{-\infty}^{+\infty} dp f(p+a)$ equal to $\int_{-\infty}^{+\infty} dp f(p)$? The difference between these two integrals is

$$\int_{-\infty}^{+\infty} dp \left(a\frac{d}{dp}f(p) + \cdots\right) = a(f(+\infty) - f(-\infty)) + \cdots \tag{1.3.10}$$

Clearly, if $f(+\infty)$ and $f(-\infty)$ are two different constants, the shift is not valid. However, if the integral $\int_{-\infty}^{+\infty} dp f(p)$ converges, or even if it is only logarithmically divergent, the shift is permissible. Since this condition is not met in our case, the shift is not allowed.
As is customary, we rotate the Feynman integrand to Euclidean space. Generalizing our observation above to $d$-dimensional Euclidean space, we obtain

$$\int d_E^d p[f(p+a) - f(p)] = \int d_E^d p\left[a^\mu \partial_\mu f(p) + \cdots\right] \tag{1.3.11}$$

Applying Gauss's theorem, this integral is transformed into a surface integral over an infinitely large sphere enclosing all of Euclidean spacetime

$$\lim_{P\to\infty} a^\mu \left(\frac{P_\mu}{P}\right) f(P) S_{d-1}(P) \tag{1.3.12}$$

where $S_{d-1}(P)$ is the area of a $(d-1)$-dimensional sphere. Averaging over the surface of the sphere, and recalling from Feynman diagram calculations that the average of $P^\mu P^\nu/P^2$ is $\frac{1}{4}\eta^{\mu\nu}$ by symmetry arguments, we obtain, after rotating back to Minkowskian space,

$$\int d^4p[f(p+a) - f(p)] = \lim_{P\to\infty} i a^\mu \left(\frac{P_\mu}{P}\right) f(P)\left(2\pi^2 P^3\right) \tag{1.3.13}$$

Note the $i$ from Wick rotating back. Using this result with

$$f(p) = \mathrm{tr}\left(\gamma^\lambda\gamma^5 \frac{1}{\not p - \not k_2}\gamma^\nu \frac{1}{\not p}\right) = \frac{\mathrm{tr}\left[\gamma^5(\not p - \not k_2)\gamma^\nu \not p \gamma^\lambda\right]}{(p-k_2)^2 p^2} = \frac{4i\varepsilon^{\tau\nu\sigma\lambda}k_{2\tau}p_\sigma}{(p-k_2)^2 p^2} \tag{1.3.14}$$

we find

$$k_{1\mu}\Delta^{\lambda\mu\nu} = -\frac{g^3}{(2\pi)^4}\lim_{P\to\infty} i(-k_1)^\mu \frac{P_\mu}{P}\frac{4i\varepsilon^{\tau\nu\sigma\lambda}k_{2\tau}P_\sigma}{P^4}2\pi^2P^3 = -\frac{g^3}{8\pi^2}\varepsilon^{\lambda\nu\tau\sigma}k_{1\tau}k_{2\sigma} \tag{1.3.15}$$

so that $k_{1\mu}\Delta^{\lambda\mu\nu} \neq 0$. As previously discussed, this would have serious consequences, as it would imply a violation of fermion number conservation. The question, then, is: what went wrong?
The root of the issue lies in the integral defining $\Delta^{\lambda\mu\nu}$, as presented in equation (1.3.4). This integral is linearly divergent, which renders it ill-defined.
Before addressing the specific calculations of $k_{1\mu}\Delta^{\lambda\mu\nu}$ and $k_{2\nu}\Delta^{\lambda\mu\nu}$, it is more pertinent to examine whether the expression for $\Delta^{\lambda\mu\nu}$ is invariant with respect to the choice of the physicist performing the calculation. To illustrate this concern, consider the case in which a different physicist chooses to shift the integration variable $p$ in the divergent integral of equation (1.3.4) by an arbitrary 4-vector $a$. This arbitrary shift raises the question of whether the value of $\Delta^{\lambda\mu\nu}$ can depend on such choices, highlighting the need for a more rigorous treatment of the integral. Therefore, we introduce the following quantity

$$\Delta^{\lambda\mu\nu}(a,k_1,k_2) = -g^3\int\frac{d^4p}{(2\pi)^4}\mathrm{tr}\left(\gamma^\lambda\gamma^5\frac{1}{\not p + \not a - \not q}\gamma^\nu\frac{1}{\not p + \not a - \not k_1}\gamma^\mu\frac{1}{\not p + \not a}\right) + \{\mu, k_1 \leftrightarrow \nu, k_2\} \tag{1.3.16}$$

It seems that there can be multiple results for the Feynman diagrams shown in Fig. 1.1. Which result should be considered correct? The only reasonable approach is to select the value of $a$ such that both $k_{1\mu}\Delta^{\lambda\mu\nu}(a,k_1,k_2)$ and $k_{2\nu}\Delta^{\lambda\mu\nu}(a,k_1,k_2)$ vanish.
To proceed, let us compute the difference $\Delta^{\lambda\mu\nu}(a,k_1,k_2) - \Delta^{\lambda\mu\nu}(k_1,k_2)$ by using eq. (1.3.13) with the function $f(p) = \mathrm{tr}\left(\gamma^\lambda\gamma^5\frac{1}{\not p - \not q}\gamma^\nu\frac{1}{\not p - \not k_1}\gamma^\mu\frac{1}{\not p}\right)$. Noting that

$$\begin{aligned} f(P) &= \lim_{P\to\infty}\frac{\mathrm{tr}\left(\gamma^\lambda\gamma^5 \not P\gamma^\nu \not P\gamma^\mu \not P\right)}{P^6} \\ &= \frac{2P^\mu\,\mathrm{tr}\left(\gamma^\lambda\gamma^5 \not P\gamma^\nu \not P\right) - P^2\,\mathrm{tr}\left(\gamma^\lambda\gamma^5 \not P\gamma^\nu\gamma^\mu\right)}{P^6} = \frac{+4iP^2P_\sigma\varepsilon^{\sigma\nu\mu\lambda}}{P^6} \end{aligned} \tag{1.3.17}$$

we find

$$\begin{aligned} \Delta^{\lambda\mu\nu}(a,k_1,k_2) - \Delta^{\lambda\mu\nu}(k_1,k_2) &= -\frac{4g^3}{8\pi^2}\lim_{P\to\infty} a^\omega\frac{P_\omega P_\sigma}{P^2}\varepsilon^{\sigma\nu\mu\lambda} + \{\mu,k_1\leftrightarrow\nu,k_2\} \\ &= -\frac{g^3}{8\pi^2}\varepsilon^{\sigma\nu\mu\lambda}a_\sigma + \{\mu,k_1\leftrightarrow\nu,k_2\} \end{aligned} \tag{1.3.18}$$

Since there are two independent momenta $k_1$ and $k_2$, we can take $a = \alpha(k_1+k_2) + \beta(k_1-k_2)$. With this choice, we obtain

$$\Delta^{\lambda\mu\nu}(a,k_1,k_2) = \Delta^{\lambda\mu\nu}(k_1,k_2) - \frac{\beta g^3}{4\pi^2}\varepsilon^{\lambda\mu\nu\sigma}(k_1-k_2)_\sigma \tag{1.3.19}$$

Note that $\alpha$ drops out of the expression. As expected, $\Delta^{\lambda\mu\nu}(a,k_1,k_2)$ depends on $\beta$, and hence on $a$. The requirement for the conservation of the vector current, i.e. $k_{1\mu}\Delta^{\lambda\mu\nu}(a,k_1,k_2) = 0$, now fixes the parameter $\beta$. Recalling

$$k_{1\mu}\Delta^{\lambda\mu\nu}(k_1,k_2) = -\frac{g^3}{8\pi^2}\varepsilon^{\lambda\nu\tau\sigma}k_{1\tau}k_{2\sigma} \tag{1.3.20}$$

we must choose to deal with $\Delta^{\lambda\mu\nu}(a,k_1,k_2)$ with $\beta = -\frac{1}{2}$. One way of viewing all this is to say that the Feynman rules do not suffice in determining $\langle J_V J_V J_A \rangle$. They have to be supplemented by vector current conservation.

### 1.3.2 Violation of the axial current conservation

We now arrive at the central conclusion of our discussion. While we have insisted on the conservation of the vector current, we must also ask: is the axial current conserved? To answer this, we compute the following expression

$$q_\lambda \Delta^{\lambda\mu\nu}(a,k_1,k_2) = q_\lambda \Delta^{\lambda\mu\nu}(k_1,k_2) - \frac{g^3}{4\pi^2}\varepsilon^{\mu\nu\lambda\sigma}k_{1\lambda}k_{2\sigma} \tag{1.3.21}$$

At this stage, we are familiar with the procedure and can proceed with the calculation

$$\begin{aligned} &q_\lambda \Delta^{\lambda\mu\nu}(k_1,k_2) \\ &= -g^3 \int \frac{d^4p}{(2\pi)^4}\,\mathrm{tr}\left(\gamma^5 \frac{1}{\not p - \not q}\gamma^\nu \frac{1}{\not p - \not k_1}\gamma^\mu - \gamma^5 \frac{1}{\not p - \not k_2}\gamma^\nu \frac{1}{\not p}\gamma^\mu\right) + \{\mu, k_1 \leftrightarrow \nu, k_2\} = -\frac{g^3}{4\pi^2}\varepsilon^{\mu\nu\lambda\sigma}k_{1\lambda}k_{2\sigma} \end{aligned} \tag{1.3.22}$$

Thus, we obtain the result

$$q_\lambda \Delta^{\lambda\mu\nu}(a,k_1,k_2) = -\frac{g^3}{2\pi^2}\varepsilon^{\mu\nu\lambda\sigma}k_{1\lambda}k_{2\sigma} \tag{1.3.23}$$

In summary, in the simple theory described by the Lagrangian $\mathcal{L} = \bar\psi i\gamma^\mu\partial_\mu\psi$ both the vector and axial currents are conserved classically. However, quantum fluctuations lead to the breakdown of axial current conservation. This phenomenon is known as the chiral anomaly, axial anomaly or Adler-Bell-Jackiw (ABJ) anomaly.
Anomalies represent a profoundly rich and complex topic in quantum field theory. Let us focus on a few key observations that highlight their significance

1. Suppose we gauge our simple theory $\mathcal{L} = \bar\psi i\gamma^\mu\left(\partial_\mu + igV_\mu\right)\psi$, where $V_\mu$ is the photon field. Then in Figure 1.1 we can think of two photon lines coming out of the vertices labeled $\mu$ and $\nu$. One can then write

$$\partial_\mu \langle J_A^\mu \rangle = \frac{ig^3}{(4\pi)^2}\varepsilon^{\mu\nu\lambda\sigma}F_{\mu\nu}F_{\lambda\sigma} \tag{1.3.24}$$

   where $F_{\mu\nu}$ the field stregth tensor of the vector field. The divergence of the axial current is not zero, but is an operator capable of producing two photons.

2. Writing the Lagrangian in terms of left and right handed fields $\psi_R$ and $\psi_L$ and introducing the left and right handed currents $J_R^\mu \equiv \bar\psi_R\gamma^\mu\psi_R$ and $J_L^\mu \equiv \bar\psi_L\gamma^\mu\psi_L$, we can repackage the anomaly as

$$\partial_\mu \langle J_R^\mu \rangle = \frac{1}{2}\frac{ig^3}{(4\pi)^2}\varepsilon^{\mu\nu\lambda\sigma}F_{\mu\nu}F_{\lambda\sigma} \tag{1.3.25}$$

   and

$$\partial_\mu \langle J_L^\mu \rangle = -\frac{1}{2}\frac{ig^3}{(4\pi)^2}\varepsilon^{\mu\nu\lambda\sigma}F_{\mu\nu}F_{\lambda\sigma} \tag{1.3.26}$$

   We can think of left handed and right handed fermions running around the loop in Fig. 1.1, contributing oppositely to the anomaly.

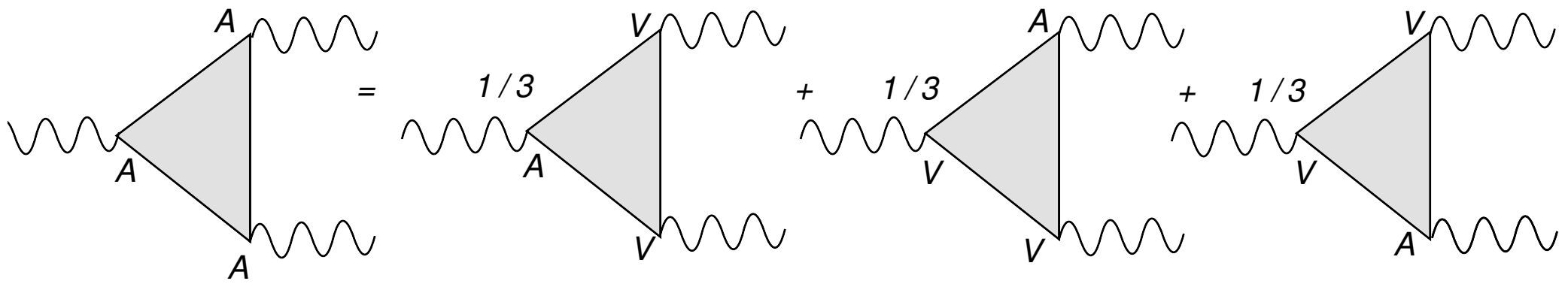

Figure 1.2: Distribution of the axial anomaly for the $\langle J_A J_A J_A \rangle$ diagram

3. We now consider the following Lagrangian

$$\mathcal{L} = \bar{\psi}\Big[i\gamma^\mu\Big(\partial_\mu + igV_\mu\Big) - m\Big]\psi. \tag{1.3.27}$$

The invariance under the chiral transformation $\psi \to e^{i\theta\gamma^5}\psi$ is now spoiled by the mass term. Classically, $\partial_\mu J_5^\mu = 2m\bar{\psi}i\gamma^5\psi$ and the axial current is explicitly not conserved. Then, taking into account the quantum fluctuations we have an additional term

$$\partial_\mu\langle J_A^\mu\rangle = 2m\langle J_A\rangle + \frac{ig^3}{(4\pi)^2}\varepsilon^{\mu\nu\lambda\sigma}F_{\mu\nu}F_{\lambda\sigma} \tag{1.3.28}$$

where

$$J_A = -ig\bar{\psi}\gamma^5\psi \tag{1.3.29}$$

4. The rate of the decay $\pi^0 \to \gamma + \gamma$ is strictly related to the chiral anomaly. Indeed, historically people used the erroneous result $\partial_\mu J_A^\mu = 0$ to deduce that this experimentally observed decay cannot occur![1] The resolution of this apparent paradox led to the correct result (1.3.24).

5. We have analyzed the $\langle J_V J_V J_A \rangle$ interaction. A natural extension of this analysis is to consider the correlator involving three axial currents, $\langle J_A J_A J_A \rangle$. This case will also be discussed in Section 4.4. By computing this interaction, one finds that the anomaly is distributed evenly among the three currents, as illustrated in Fig. 1.2

$$\langle J_A J_A J_A \rangle = \frac{1}{3}\Big[\langle J_A J_V J_V \rangle + \langle J_V J_A J_V \rangle + \langle J_V J_V J_A \rangle\Big] \tag{1.3.30}$$

To account for contributions from both axial and vector field strengths, it is preferable to use the more general expression

$$\partial_\mu\langle J_A^\mu\rangle = a_1\varepsilon^{\mu\nu\rho\sigma}F^V_{\mu\nu}F^V_{\rho\sigma} + a_1'\varepsilon^{\mu\nu\rho\sigma}F^A_{\mu\nu}F^A_{\rho\sigma} \tag{1.3.31}$$

where $a_1$ and $a_1'$ are, in principle, distinct coefficients. In future discussions, we will avoid such redundancies where possible. It is important to note that in our specific case—where the currents are defined as in eq. (1.3.3) with the same coupling constant $g$—the anomalous coefficients are related by

$$a_1' = \frac{a_1}{3} \tag{1.3.32}$$

6. As we will discuss in the next chapters, the chiral anomaly has important implications in condensed matter physics, particularly in the study of Weyl semimetals and topological insulators.

[1] We are neglecting mass effects.

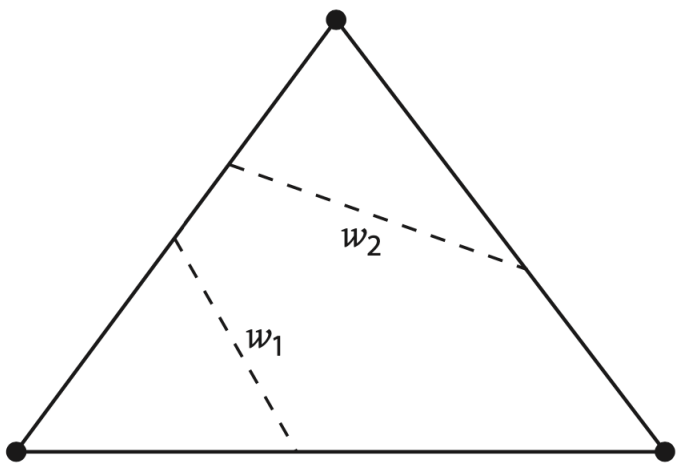

Figure 1.3: Three-loop diagram

7. When computing our diagrams, we have different types of regularization at our disposal. For example, in the Pauli–Villars regularization scheme, we subtract from the integrand what the integrand would have been if the electron mass were replaced by some regulator mass. We can then take the limit where the mass tends to infinity. In other words, we define

$$\begin{aligned}\Delta^{\lambda\mu\nu}(k_1,k_2) = -g^3 \int \frac{d^4p}{(2\pi)^4}\,\mathrm{tr}\Bigg(&\gamma^\lambda\gamma^5\frac{1}{\not p-\not q}\gamma^\nu\frac{1}{\not p-\not k_1}\gamma^\mu\frac{1}{\not p}\\ &-\gamma^\lambda\gamma^5\frac{1}{\not p-\not q-M}\gamma^\nu\frac{1}{\not p-\not k_1-M}\gamma^\mu\frac{1}{\not p-M}\Bigg)+\{\mu,k_1\leftrightarrow\nu,k_2\}.\end{aligned} \tag{1.3.33}$$

Now, the integral appears to be logarithmically divergent, and we can shift the integration variable $p$ freely. So, how does the chiral anomaly arise? By introducing the regulator mass $M$, we have explicitly broken axial current conservation. The anomaly is the statement that this breaking persists even in the limit where $M$ tends to infinity.
Another common regularization method is the Breitenlohner-Maison-'t Hooft-Veltman scheme [48], which we will employ in the following chapters.

8. Consider now the non-Abelian theory described by the Lagrangian

$$\mathcal{L} = \bar\psi i\gamma^\mu\left(\partial_\mu + igV^a_\mu T^a\right)\psi \tag{1.3.34}$$

In computing the Feynman amplitude, we simply include a factor of $T^a$ in each vertex. The procedure follows as in the Abelian case, except that when summing over all fermions running in the loop, we obtain the additional factor

$$D_{abc} = \frac{1}{2}\,\mathrm{Tr}[T_a\{T_b,T_c\}] \tag{1.3.35}$$

This immediately reveals that, in a non-Abelian gauge theory, the divergence of the axial current satisfies

$$\partial_\mu\langle J^\mu_{A\,a}\rangle = \frac{ig^3}{(4\pi)^2}D_{abc}\,\varepsilon^{\mu\nu\rho\sigma}F^b_{\mu\nu}F^c_{\rho\sigma} \tag{1.3.36}$$

Remarkably, non-Abelian symmetry implies that the anomaly includes not only terms quadratic in the gauge field $V$, but also cubic and quartic terms. As a result, chiral anomalies can involve two, three or even four gauge bosons at a time. These anomalies are commonly referred to as the triangle, square, and pentagon anomalies, respectively.

9. We previously computed the chiral anomaly in the free theory, described by the Lagrangian $\mathcal{L} = \bar\psi\left(i\gamma^\mu\partial_\mu - m\right)\psi$. Now, consider coupling the fermion to a scalar field by introducing the interaction term $f\varphi\bar{\psi}\psi$, or alternatively, coupling it to the electromagnetic field. In such cases, computing the anomaly requires evaluating higher-order diagrams, such as the three-loop diagram shown in Fig. 1.3. A natural expectation would be that the right-hand side of eq. (1.3.23) acquires a multiplicative correction of the form $1 + h(f, g, \ldots)$, where $h$ is some function of the couplings in the theory. However, in a remarkable result, Adler and Bardeen proved that $h = 0$. This nontrivial fact, known as the nonrenormalization of the anomaly, can be understood heuristically as follows. Before integrating over the momenta of the scalar propagators in Fig. 1.3 (denoted $w_1$ and $w_2$), the Feynman integrand contains seven fermion propagators. This ensures that the integral is sufficiently convergent, allowing shifts in integration variables without introducing ambiguities. Consequently, prior to integrating over $w_1$ and $w_2$, all relevant Ward identities hold, including

$$q_\lambda \Delta^{\lambda\mu\nu}_{3\text{ loops}}(k_1, k_2; w_1, w_2) = 0. \tag{1.3.37}$$

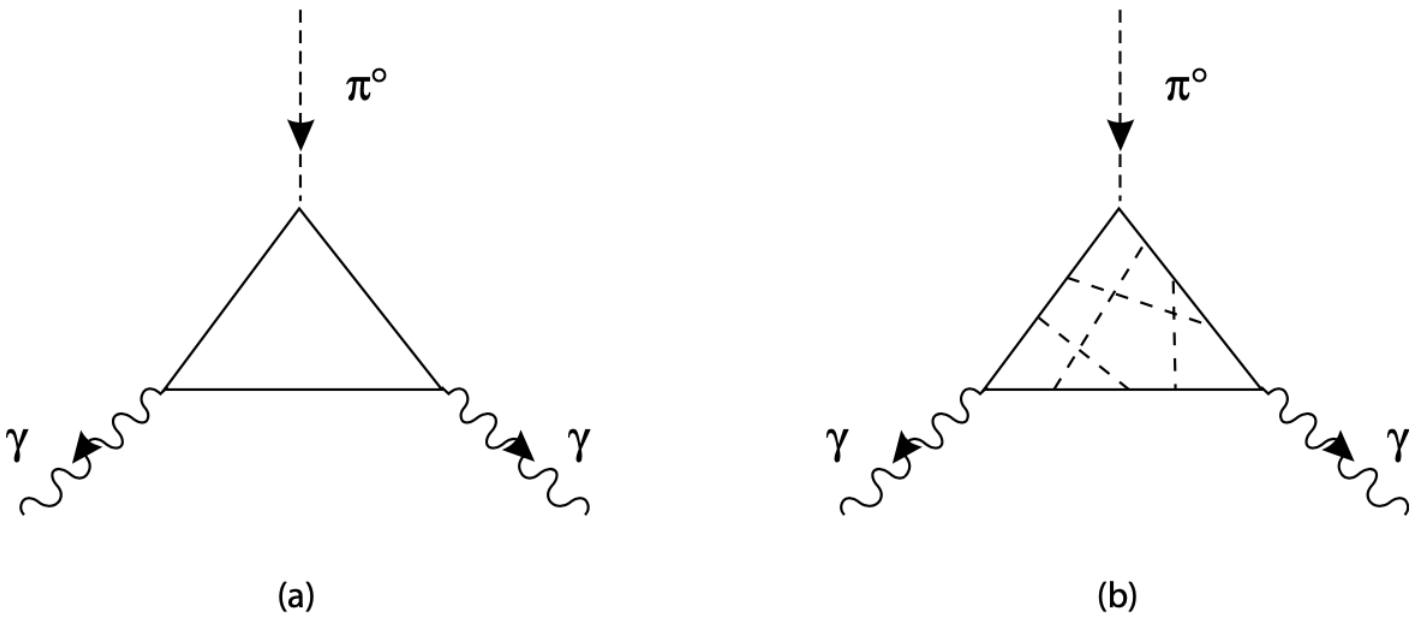

Figure 1.4: Decay of the Pion

We will return to this topic in Chapter 4.

10. The preceding point was of great importance in the history of particle physics, as it led directly to the notion of color. The nonrenormalization of the anomaly allowed for the decay amplitude of $\pi^0 \to \gamma + \gamma$ to be calculated with confidence in the late 1960s. In the quark model of that time, the amplitude was given by an infinite number of Feynman diagrams, as shown in Fig. 1.4 (with a quark running around the fermion loop). However, the nonrenormalization of the anomaly indicates that only diagram (a) contributes. In other words, the amplitude does not depend on the details of the strong interaction. Consequently, the fact that the result was a factor of 3 too small suggested that quarks exist in three copies.

11. It is natural to speculate as to whether quarks and leptons are composites of yet more fundamental fermions known as preons. The nonrenormalization of the chiral anomaly provides a powerful tool for this sort of theoretical speculation. No matter how complicated the relevant interactions might be, as long as they are described by field theory as we know it, the anomaly at the preon level must be the same as the anomaly at the quark-lepton level. This so-called anomaly matching condition severely constrains the possible preon theories [49, 50].

## 1.4 The Fujikawa method

Historically, field theorists were skeptical of the path integral formulation, favoring the canonical approach. When the chiral anomaly was discovered, some even argued that its existence demonstrated a fundamental flaw in the path integral. These critics contended that the path integral fails to reflect the non-invariance of the theory under the chiral transformation $\psi \to e^{i\theta\gamma^5}\psi$.
Fujikawa resolved this issue by demonstrating that the path integral correctly accounts for the anomaly: under the chiral transformation, the measure $D\bar{\psi}D\psi$ acquires a nontrivial Jacobian. While the action itself may be invariant, the path integral is not. In this section, we explore this aspect. Our result will be the same as the one we derived in the previous section.
Consider once again a massless fermion in a gauge field. For the following discussion, the gauge field can be treated as an external field, but the same applies to the quantized case as well. In fact, for an external gauge field, we define the generating functional as

$$Z = \mathcal{N} \int \mathcal{D}\psi\mathcal{D}\bar{\psi} e^{i\int d^4x\, \bar{\psi} i\gamma^\mu D_\mu \psi} \tag{1.4.1}$$

where

$$D_\mu = \partial_\mu + igV_\mu \tag{1.4.2}$$

As we will be interested only in checking the current conservation we can put to zero the external sources for the fermion fields. If the gauge field $V_\mu$ is quantized we have to insert a further functional integration in $V_\mu$ but this will not play any role in the following, so for simplicity we will take $V_\mu$ as an external field.
As previously discussed, this Lagrangian is invariant under global chiral transformations, leading to the classically conserved current $J_A$. We will once again demonstrate that quantum corrections violate this conservation law using the path integral formulation. To this end, we proceed by performing a change of variables within $Z$

$$\begin{aligned} \psi(x) &\to \psi'(x) = \left(1 + i\theta(x)\gamma^5\right)\psi(x) \\ \bar{\psi}(x) &\to \bar{\psi}'(x) = \bar{\psi}(x)\left(1 + i\theta(x)\gamma^5\right). \end{aligned} \tag{1.4.3}$$

The variation of the action is given by

$$\delta S = \int d^4x\, \theta(x)\partial_\mu\left(\bar{\psi}\gamma^\mu\gamma^5\psi\right) \tag{1.4.4}$$

In order to evaluate the change of the integration measure it is convenient to expand the fermion field in a basis of eigenvectors of $\gamma^\mu D_\mu$

$$\begin{aligned} i\gamma^\mu D_\mu \phi_m(x) &= \lambda_m \phi_m(x) \\ -iD_\mu \tilde{\phi}_m(x)\gamma^\mu &= \lambda_m \tilde{\phi}_m(x) \end{aligned} \tag{1.4.5}$$

where we have used a discrete notation for the eigenvalues. In the case of a zero gauge field $V_\mu$ the eigenvalues $\lambda_m$ are given by

$$\lambda_m^2 = k_0^2 - |\vec{k}|^2 \tag{1.4.6}$$

Notice that they become negative definite after being Wick rotated

$$\lambda_m^2 \to -k_4^2 - |\vec{k}|^2 = -k_E^2 \tag{1.4.7}$$

By hypothesis the eigenvectors $\phi_m$ span an orthonormal basis and therefore we can expand $\psi$ and $\bar{\psi}$ as

$$\psi(x) = \sum_m a_m \phi_m(x), \quad \bar{\psi}(x) = \sum_m b_m \tilde{\phi}_m(x) \tag{1.4.8}$$

where $a_m$ and $b_m$ are Grassmann coefficients. Then, except for a possible constant we have

$$\mathcal{D}\psi\mathcal{D}\bar{\psi} = \prod_m da_m db_m \tag{1.4.9}$$

We can evaluate the effect of the change of variables on the Grassmann coefficients by starting from

$$\psi'(x) = \left(1 + i\theta(x)\gamma^5\right)\psi(x) = \sum_m a'_m \phi_m(x) \tag{1.4.10}$$

Then, from the orthogonality relation

$$\int d^4x\, \phi_m^\dagger \phi_n = \delta_{mn} \tag{1.4.11}$$

we get

$$a'_m = \int d^4x\, \phi_m^\dagger(x)\psi'(x) = a_m + \sum_n \int d^4x\, \phi_m^\dagger(x) i\theta(x)\gamma^5\phi_n(x) a_n = a_m + \sum_n C_{mn} a_n \tag{1.4.12}$$

with

$$C_{mn} = \int d^4x \phi_m^\dagger(x) i\theta(x)\gamma^5\phi_n(x) \tag{1.4.13}$$

A similar transformation holds for $b_m$. Therefore, from the transformation rules for the Grassmann measure we get

$$\mathcal{D}\psi'\mathcal{D}\bar{\psi}' = \frac{1}{\det|\mathcal{I}|^2}\mathcal{D}\psi\mathcal{D}\bar{\psi} \tag{1.4.14}$$

where $\mathcal{I} = 1 + C$ is the jacobian of the transformation. Since we are interested in the infinitesimal transformation, we have

$$\det|\mathcal{I}| = e^{\mathrm{Tr}\log(1+C)} \approx e^{\mathrm{Tr}\, C} \tag{1.4.15}$$

that is

$$\log\det|\mathcal{I}| \approx \mathrm{Tr}\, C = i\int d^4x\, \theta(x) \sum_n \phi_n^\dagger(x)\gamma^5\phi_n(x) \tag{1.4.16}$$

The coefficient of $\theta(x)$ is the trace of $\gamma^5$ taken over all the Hilbert space. Although the trace of $\gamma^5$ is zero on the spinor space it gets multiplied by an infinite factor coming from the total Hilbert space. Therefore this expression is ill defined and it needs to be regularized. Furthermore, we would like to maintain the gauge invariance, and therefore we should regulate the expression in a gauge invariant way. A natural choice for the regularization is the following

$$\sum_n \phi_n^\dagger(x)\gamma^5\phi_n(x) = \lim_{M\to\infty}\sum_n \phi_n^\dagger(x)\gamma^5\phi_n(x) e^{\lambda_n^2/M^2} \tag{1.4.17}$$

This is in fact a convergence factor in the euclidean space. The previous expression can be rewritten as

$$\begin{aligned}\sum_n \phi_n^\dagger(x)\gamma^5\phi_n(x) &= \lim_{M\to\infty}\sum_n \phi_n^\dagger(x)\gamma^5 e^{(i\gamma^\mu D_\mu)^2/M^2}\phi_n(x) \\ &= \lim_{M\to\infty}\langle x|\mathrm{tr}\left[\gamma^5 e^{(i\gamma^\mu D_\mu)^2/M^2}\right]|x\rangle\end{aligned} \tag{1.4.18}$$

where tr is the trace over the Dirac matrices. The evaluation of $(\gamma_\mu D^\mu)^2$ gives

$$\begin{aligned}(\gamma_\mu D^\mu)^2 = \gamma_\mu\gamma_\nu D^\mu D^\nu = \frac{1}{2}\left[\gamma_\mu,\gamma_\nu\right]_+ D^\mu D^\nu + \frac{1}{2}\left[\gamma_\mu,\gamma_\nu\right]_- D^\mu D^\nu \\ = D^2 - \frac{i}{2}\sigma_{\mu\nu}\left[D^\mu, D^\nu\right]_- = D^2 + \frac{g}{2}\sigma_{\mu\nu}F^{\mu\nu}\end{aligned} \tag{1.4.19}$$

In order to get a contribution from the trace over the Dirac indices we need at least four $\gamma$ matrices. Therefore the leading term is obtained by expanding the exponential up to the order $\left(\sigma_{\mu\nu}F^{\mu\nu}\right)^2$ and neglecting $V_\mu$ in all the other places. Therefore we get

$$\sum_n \phi_n^\dagger(x)\gamma^5\phi_n(x) = \lim_{M\to\infty} \mathrm{tr}\left[\gamma^5\frac{1}{2!}\left(-\frac{g}{2M^2}\sigma_{\mu\nu}F^{\mu\nu}\right)^2\right]\langle x|e^{-\Box/M^2}|x\rangle \tag{1.4.20}$$

The matrix element can be easily evaluated by inserting momentum eigenstates and performing a Wick's rotation

$$\langle x|e^{-\Box/M^2}|x\rangle = \lim_{x\to y}\int\frac{d^4k}{(2\pi)^4}e^{k^2/M^2}e^{-ik(x-y)} = i\int\frac{d^4k_E}{(2\pi)^4}e^{-k_E^2/M^2} = i\frac{M^4}{16\pi^2} \tag{1.4.21}$$

Therefore

$$\sum_n \phi_n^\dagger\gamma^5\phi_n = \frac{ig^2}{8\cdot 16\pi^2}\mathrm{tr}\left[\gamma^5\sigma_{\mu\nu}\sigma_{\rho\lambda}\right]F^{\mu\nu}F^{\rho\lambda} \tag{1.4.22}$$

Recalling that

$$\mathrm{tr}\left[\gamma^5\sigma_{\mu\nu}\sigma_{\rho\lambda}\right] = 4i\epsilon_{\mu\nu\rho\lambda} \tag{1.4.23}$$

we can write

$$\sum_n \phi_n^\dagger\gamma^5\phi_n = -\frac{g^2}{32\pi^2}\epsilon_{\mu\nu\rho\lambda}F^{\mu\nu}F^{\rho\lambda} \tag{1.4.24}$$

Therefore the determinant we are looking for is given by

$$\det|\mathcal{I}| = e^{-i\int d^4x\,\theta(x)\frac{g^2}{32\pi^2}\epsilon_{\mu\nu\rho\lambda}F^{\mu\nu}F^{\rho\lambda}} \tag{1.4.25}$$

and after the change of variables the generating functional is given by

$$Z = \mathcal{N}\int\mathcal{D}\psi\mathcal{D}\bar{\psi}e^{i\int d^4x\bar{\psi}i\hat{D}\psi}e^{i\int d^4x\,\theta(x)\left(\partial_\mu(\bar{\psi}\gamma^\mu\gamma^5\psi)+\frac{g^2}{16\pi^2}\epsilon_{\mu\nu\rho\lambda}F^{\mu\nu}F^{\rho\lambda}\right)} \tag{1.4.26}$$

Since a change of variables should not alter the path integral, one can write the anomalous conservation law as

$$\partial_\mu\langle J_A^\mu\rangle = \frac{ig^3}{16\pi^2}\epsilon^{\mu\nu\rho\sigma}F_{\mu\nu}F_{\rho\sigma} \tag{1.4.27}$$

in accordance with the result of the previous section.

## 1.5 Parity-odd anomalies

In this section, we are going to discuss some general features of anomalies. Besides the chiral Abelian anomaly we have just examined, we will introduce additional ones, such as the gravitational chiral anomaly and trace anomalies. Our main focus is on parity-odd anomalies.

### 1.5.1 Chiral anomalies

Of all the anomalies, the chiral one is among the most widely discussed, both in its global and gauged forms. Historically, chiral symmetry was one of the first contexts where an anomaly emerged, playing a crucial role in explaining the observed decay rate of a neutral pion into two photons.
As we have shown, the chiral anomaly appears in parity-odd correlators involving axial vector currents, such as $\langle J_V J_V J_A\rangle$ and $\langle J_A J_A J_A\rangle$. We have examined the simplest case within an Abelian framework and in flat spacetime. This anomaly can be generalized to include gravitational contributions, which also violate the conservation of the axial current. Such violations can be analyzed through the parity-odd correlator $\langle T T J_A\rangle$, constructed with an axial vector current and two energy-momentum tensors. This discussion is presented in Chapter 5.
The anomalous conservation Ward identity can be expressed in its general form as

$$\nabla_\mu \langle J_A^\mu\rangle = a_1\, \varepsilon^{\mu\nu\rho\sigma} F_{\mu\nu} F_{\rho\sigma} + a_2\, \varepsilon^{\mu\nu\rho\sigma} R^{\alpha\beta}{}_{\mu\nu} R_{\alpha\beta\rho\sigma}. \tag{1.5.1}$$

where the anomalous terms on the right-hand side are known as Pontryagin densities. As we have mentioned, the first term ($F\tilde{F}$) appears in the analysis of the $\langle J_V J_V J_A\rangle$ correlator, while the second term ($R\tilde{R}$) arises in the study of the $\langle T T J_A\rangle$ correlator.
These results can also be extended to the non-Abelian case. For a general axial current $J^\mu_{A\,i}$, the anomaly takes the form

$$\nabla_\mu \langle J^\mu_{A\,i}\rangle = a_1\, D_{ijk}\, \varepsilon^{\mu\nu\rho\sigma} F^j_{\mu\nu} F^k_{\rho\sigma} + a_2\, D_i\, \varepsilon^{\mu\nu\rho\sigma} R^{\alpha\beta}{}_{\mu\nu} R_{\alpha\beta\rho\sigma}, \tag{1.5.2}$$

where we have introduced the tensors

$$D_{ijk} = \frac{1}{2}\mathrm{Tr}[\{T_i, T_j\}\, T_k], \qquad D_i = \mathrm{Tr}[T_i]. \tag{1.5.3}$$

constructed with the non-Abelian generators of the theory.
The approach developed in this thesis is general and adaptable to various contexts; however, the impact of an anomaly can range from benign to problematic, depending on the circumstances. Consider, for example, a scenario involving a Dirac fermion interacting with a vector potential $V_\mu$. Kimura, Delbourgo, and Salam were the first to compute the anomaly in this case, observed in the divergence of $J_A$ [51, 52]. This specific anomaly poses no threat and may even hold phenomenological interest. For convenience, we can also introduce an axial-vector field $A_\mu$, which couples to $J_A$, but only as an external source. Allowing $A_\mu$ to act as a dynamical gauge field would introduce an anomalous gauge symmetry, undermining both unitarity and renormalizability. When the axial current couples directly to a dynamical gauge field $A_\mu$, gauge invariance demands the absence of the anomaly. In such cases, for gauge currents, the following condition must hold

$$D_{ijk} = 0 \tag{1.5.4}$$

Another example of potentially harmful anomalies arises in chiral models involving a Weyl fermion $\psi_{L/R}$ interacting with both gravity and a gauge field. Here, the anomaly appears in the divergence of $J_{L/R}$, posing a threat to unitarity and renormalization unless it is properly canceled [53–55].
Gravitational anomalies are also significant in the context of the Standard Model. Consider the $D_i$ term within the Standard Model's symmetry group, $SU(3)\times SU(2)\times U(1)_Y$. In this framework, only the hypercharge ($U(1)_Y$) contribution, represented by $\langle T T J_Y\rangle$, is relevant because the $SU(2)$ and $SU(3)$ generators are traceless. The anomaly cancels out when summing over each generation of particles, a result of considerable importance. This cancellation ensures the consistency of coupling the Standard Model to gravity, guaranteeing that gauge currents remain conserved in a gravitational background.

Correlators influenced by chiral anomalies, as well as discrete anomalies, play a vital role in condensed matter theory, particularly in the context of topological materials [9–15, 15–18]. The gravitational anomaly has been investigated, in the same context, in other interesting works [56, 57]. Crucial, in this analysis, is the correspondence between thermal stresses and gravity, as summarized by Luttinger's relation [58] connecting a gravitational potential to a thermal gradient [59, 60].
Understanding these phenomena is essential for unraveling the intricate properties of such materials. Since their dynamical contribution in the evolution of topological matter, in the realistic case, is characterized by both thermal effecs and by Fermi surfaces, which break the charge conjugation $C$ invariance of the vacuum, the quantification of such corrections becomes crucial for phenomenology.

### 1.5.2 Trace anomalies

As we have seen in Section 1.2.2, the invariance of a field theory under Weyl transformations classically implies the tracelessness of the energy-momentum tensor. At the quantum level, the energy-momentum tensor becomes an operator, whose renormalized expectation value may develop a non-vanishing trace value; in such a case, it is said that we are in the presence of a trace anomaly and the Weyl symmetry is broken.
Trace anomalies have some distinctive features that make them more complex compared to the chiral ones, due to the presence of both topological and non-topological terms in the anomaly functional. As was first found by Capper and Duff on dimensional grounds and by requiring covariance [61, 62], the general structure of the trace anomaly in four dimensions is given by

$$g_{\mu\nu}\langle T^{\mu\nu}\rangle = b_1\,E_4 + b_2\,C^{\mu\nu\rho\sigma}C_{\mu\nu\rho\sigma} + b_3\,\nabla^2 R + b_4\,F^{\mu\nu}F_{\mu\nu} + f_1\,\varepsilon^{\mu\nu\rho\sigma}R_{\alpha\beta\mu\nu}R^{\alpha\beta}{}_{\rho\sigma} + f_2\,\varepsilon^{\mu\nu\rho\sigma}F_{\mu\nu}F_{\rho\sigma}, \tag{1.5.5}$$

where $C_{\mu\nu\rho\sigma}$ is the Weyl tensor and $E_4$ is the Gauss-Bonnet term

$$\begin{aligned} C^{\mu\nu\rho\sigma}C_{\mu\nu\rho\sigma} &= R^{\mu\nu\rho\sigma}R_{\mu\nu\rho\sigma} - 2R^{\mu\nu}R_{\mu\nu} + \frac{1}{3}R^2, \\ E_4 &= R^{\mu\nu\rho\sigma}R_{\mu\nu\rho\sigma} - 4R^{\mu\nu}R_{\mu\nu} + R^2. \end{aligned} \tag{1.5.6}$$

The trace anomaly can include both parity-even and parity-odd terms. The parity-even terms are parameterized by the coefficients $b_i$, while the parity-odd terms are parameterized by $f_i$.
In recent years, the parity-odd terms have been the subject of debate, particularly regarding whether the coefficients $f_i$ vanish in the case of Weyl fermions. Some argue that they do, while others maintain that they do not. From a cohomological perspective, all terms are consistent and should not be dismissed. Nevertheless, the question of whether $f_i$ vanishes for Weyl fermions remains open to discussion.
Without engaging in the debate over whether free field theory, and the Standard Model in particular, are affected by parity-odd anomalies, one can investigate, on more general grounds, whether conformal field theories allow such parity-odd terms when the coefficients $f_i$ are generic, as well as the constraints these terms impose on the structure of the correlators. We will explore this topic in greater depth in Chapter 6.

### 1.5.3 The role of anomalies

As we will demonstrate in the following chapters, it is highly advantageous to decompose a correlator into its distinct components: longitudinal, trace and transverse-traceless parts. This decomposition proves to be particularly useful, as it allows for a clearer understanding of how different parts of the interaction are affected by various physical phenomena, such as anomalies.

As can be seen in equations (1.5.1) and (1.5.5), anomalies specifically influence the longitudinal and trace components of the interactions, leading to significant modifications in these parts. Further analysis of parity-odd correlators uncovers a key feature: the presence of an anomaly—whether it is chiral or conformal—is invariably linked to the appearance of either a longitudinal or a trace structure, marked by the inclusion of an anomaly pole of the form $1/q^2$ [63, 64].
In contrast, the transverse-traceless component of the interaction remains unaffected by the anomaly and, in the most general settings, is left undetermined. However, additional physical principles, such as conformal invariance, can be imposed to further restrict or determine the form of this component. This approach will be explored in the first part of this thesis, where we apply such constraints to derive the structure of entire correlators.
Alternatively, one can opt for a more explicit route by calculating the interaction perturbatively within a given Lagrangian framework. This approach allows for a more direct and detailed determination of the full structure of the correlator, providing a comprehensive understanding of the underlying interactions.

## 1.6 Worked example: the Ward identities of the $\langle TTJ_A\rangle$ correlator

In this section we consider the following correlator

$$\left\langle T^{\mu_1\nu_1}(x_1)\,T^{\mu_2\nu_2}(x_2)J_A^{\mu_3}(x_3)\right\rangle \equiv \frac{2^2}{\sqrt{-g(x_1)}\sqrt{-g(x_2)}\sqrt{-g(x_3)}}\frac{\delta^3\mathcal{S}}{\delta g_{\mu_1\nu_1}(x_1)\delta g_{\mu_2\nu_2}(x_2)\delta A_{\mu_3}(x_3)}\Bigg|_{\substack{g=\delta;\\ A=0;}}. \tag{1.6.1}$$

Our aim is to derive the Ward identity for the correlator in momentum-space

$$\begin{aligned}(2\pi)^4\delta^4\big(p_1+p_2+p_3\big)\Big\langle T^{\mu_1\nu_1}(p_1)\,T^{\mu_2\nu_2}(p_2)J_A^{\mu_3}(p_3)\Big\rangle =\\ \int d^4x_1d^4x_2d^4x_3\,e^{-ip_1x_1-ip_2x_2-ip_3x_3}\left\langle T^{\mu_1\nu_1}(x_1)\,T^{\mu_2\nu_2}(x_2)J_A^{\mu_3}(x_3)\right\rangle\end{aligned} \tag{1.6.2}$$

We choose the $\langle TTJ_A\rangle$ because it represents a sufficiently complex and diverse example to illustrate the derivation process effectively. A similar approach can be applied to any other correlator.

- **Diffeomorphism invariance**

We begin by imposing diffeomorphism invariance. Following the procedure outlined in Section 1.2.1, we obtain the following Ward identity

$$\nabla^\mu\left\langle T_{\mu\nu}\right\rangle - F^A_{\mu\nu}\left\langle J_A^\mu\right\rangle + A_\nu\nabla_\mu\left\langle J_A^\mu\right\rangle = 0. \tag{1.6.3}$$

where $F^A_{\mu\nu}$ is the field strength tensor of the axial field.
To derive a Ward identity for the correlator $\langle TTJ_A\rangle$, we apply the functional derivatives $\frac{\delta^2}{\delta g\,\delta A}$ to the equation above. In the limit $g_{\mu\nu}\to\delta_{\mu\nu}$ and $A_\mu\to 0$, all 1-point and 2-point functions involving scalars, currents, and the energy-momentum tensor vanish in $4d$. Going to momentum-space, this procedure generates the equation

$$0 = p_{i\mu_i}\Big\langle T^{\mu_1\nu_1}(p_1)\,T^{\mu_2\nu_2}(p_2)J_A^{\mu_3}(p_3)\Big\rangle, \qquad\qquad i=1,2. \tag{1.6.4}$$

- **Gauge invariance**

Since we are dealing with an axial current, gauge invariance is spoiled by the presence of anomalies. We recall that

$$\nabla_\mu \langle J_A^\mu \rangle = a_1\, \varepsilon^{\mu\nu\rho\sigma} F_{\mu\nu} F_{\rho\sigma} + a_2\, \varepsilon^{\mu\nu\rho\sigma} R_{\alpha\beta\mu\nu} R^{\alpha\beta}{}_{\rho\sigma}, \tag{1.6.5}$$

where the right-hand side accounts for gauge and gravitational anomalies. We now apply two functional derivatives with respect to the metric to such equation, and perform the limit $g_{\mu\nu} \to \delta_{\mu\nu}$ and $A_\mu \to 0$. The second term on the right-hand side contributes non-trivially. Moving to momentum-space, we derive the anomalous relation

$$p_{3\mu_3} \langle T^{\mu_1\nu_1}(p_1) T^{\mu_2\nu_2}(p_2) J_A^{\mu_3}(p_3)\rangle = 4\, i\, a_2\, (p_1\cdot p_2) \left\{ \left[ \varepsilon^{\nu_1\nu_2 p_1 p_2} \left( g^{\mu_1\mu_2} - \frac{p_1^{\mu_2} p_2^{\mu_1}}{p_1\cdot p_2} \right) + (\mu_1 \leftrightarrow \nu_1) \right] + (\mu_2 \leftrightarrow \nu_2) \right\}. \tag{1.6.6}$$

This constraint is satisfied by the inclusion of an anomaly pole in the correlator (see Section 5.2).

- **Weyl invariance**

We start from the anomalous equation

$$g_{\mu\nu}\langle T^{\mu\nu}\rangle = b_1\, E_4 + b_2\, C^{\mu\nu\rho\sigma} C_{\mu\nu\rho\sigma} + b_3 \nabla^2 R + b_4\, F^{\mu\nu} F_{\mu\nu} + f_1\, \varepsilon^{\mu\nu\rho\sigma} R_{\alpha\beta\mu\nu} R^{\alpha\beta}{}_{\rho\sigma} + f_2\, \varepsilon^{\mu\nu\rho\sigma} F_{\mu\nu} F_{\rho\sigma}, \tag{1.6.7}$$

and we derive once with respect to the metric and once with respect to the axial field. After setting all the remaing sources (the metric and the axial field) to the vacuum state, we find only one term. Specifically, all the anomalous terms do not contribute to the correlator in consideration and all the appearing two point function can be set to zero since as we have already mentioned their parity-odd contribution vanish in $4d$. In momentum-space we can write

$$g_{\mu_i\nu_i} \langle T^{\mu_1\nu_1}(p_1) T^{\mu_2\nu_2}(p_2) J_A^{\mu_3}(p_3)\rangle = 0, \qquad i = 1, 2 \tag{1.6.8}$$

# Chapter 2

# Conformal field theories

Conformal field theory is a cornerstone of theoretical physics, providing a powerful framework for systems that remain invariant under transformations preserving angles but not necessarily distances. CFTs emerge naturally across diverse areas, from the study of critical phenomena to the exploration of quantum field theories and string theory in high-energy physics.
The conformal symmetry extends the Poincaré group by including scale and special conformal transformations, imposing stringent structural constraints on CFTs that set them apart within quantum field theory. These additional symmetries significantly restrict physical observables, especially correlation functions, which encapsulate the theory's fundamental dynamics and interactions. Due to these constraints, CFTs often allow for the exact calculation of correlation functions, revealing universal properties and behaviors that apply across many different models.
In addition to applications in condensed matter and particle physics, CFT plays a central role in the AdS/CFT correspondence, where conformal field theories in $d$ dimensions are linked to gravitational theories in $(d+1)$-dimensional anti-de Sitter (AdS) space. This correspondence has opened new avenues for understanding quantum gravity and holography.
In this chapter, we introduce the framework of conformal field theories, with a particular emphasis on the constraints imposed on correlation functions. We begin by examining these constraints in coordinate-space and then transition to an analysis in momentum-space, which will also be the primary focus of subsequent chapters. Our analysis is set in $d$-dimensional Euclidean space.
A key distinction arises between two-dimensional CFTs and those in higher dimensions. In two dimensions, the theory is governed by an infinite-dimensional conformal algebra, which imposes exceptionally strong constraints on correlation functions. In contrast, in higher dimensions we just have the SO(d,2) Lie algebra, providing fewer but still significant constraints. Despite this difference, several core features are shared across dimensions. In this thesis, our focus will be specifically on the case $d > 2$.

## 2.1 Conformal transformations

By definition, a conformal transformation of the coordinates is an invertible mapping $x \to x'$, which leaves the metric tensor invariant up to a scale factor

$$g'_{\mu\nu}(x') = \Lambda(x') g_{\mu\nu}(x'). \tag{2.1.1}$$

The set of conformal transformations forms a group, known as the conformal group, which includes the Poincaré group as a subgroup. The Poincaré transformations correspond to the special case where $\Lambda(x) = 1$, meaning the metric remains exactly invariant without any scaling. The term conformal derives from the property that the transformation preserves the angles between two arbitrary curves crossing each other at some point, even though it may change the distances.
Given an infinitesimal arbitrary transformation $x^\mu \to x'^\mu = x^\mu + \epsilon^\mu(\boldsymbol{x})$, the metric changes as follows

$$g_{\mu\nu} \to g_{\mu\nu} - \left(\nabla_\mu \epsilon_\nu + \nabla_\nu \epsilon_\mu\right). \tag{2.1.2}$$

where $\nabla_\mu$ denotes the covariant derivative. The requirement for the transformation to be conformal implies that the change in the metric is proportional to the metric itself

$$\nabla_\mu \epsilon_\nu + \nabla_\nu \epsilon_\mu = f(x) g_{\mu\nu} \tag{2.1.3}$$

with $f(x) = 1 - \Lambda(x)$. The above equation is known as the conformal Killing equation, and the vectors $\epsilon^\mu$ that satisfy it are called conformal Killing vectors (CKV). The factor $f(x)$ can be determined by taking the trace on both sides of the equation

$$f(x) = \frac{2}{d} \nabla_\mu \epsilon^\mu. \tag{2.1.4}$$

In the flat spacetime limit, the conformal killing equation is given by

$$\partial_\mu \epsilon_\nu + \partial_\nu \epsilon_\mu = \frac{2}{d} (\partial \cdot \epsilon) \delta_{\mu\nu}. \tag{2.1.5}$$

For $d > 2$, the general solution to the above equation can then be written as

$$\epsilon_\mu = a_\mu + \omega_{\mu\nu} x^\nu + \lambda x_\mu + b_\mu x^2 - 2(b \cdot x) x_\mu \tag{2.1.6}$$

where:

- $a_\mu$ is a constant vector corresponding to translations,
- $\omega_{\mu\nu} = -\omega_{\nu\mu}$ represents the antisymmetric parameters for Lorentz transformations (rotations and boosts),
- $\lambda$ is a scalar parameter for dilatations,
- $b_\mu$ is a constant vector for special conformal transformations.

The finite form of the transformations are

$$\begin{aligned}
&\text{(translation)} && x'^\mu = x^\mu + a^\mu \\
&\text{(dilatation)} && x'^\mu = \alpha x^\mu \\
&\text{(Lorentz transformation)} && x'^\mu = M^\mu{}_\nu x^\nu \\
&\text{(special conformal transformation)} && x'^\mu = \frac{x^\mu - b^\mu x^2}{1 - 2b \cdot x + b^2 x^2}
\end{aligned} \tag{2.1.7}$$

Notice that, by considering the definition of the inversion

$$x_\mu \to x'_\mu = \frac{x_\mu}{x^2}, \tag{2.1.8}$$

the special conformal transformations can be realized as a translation preceded and followed by an inversion.

## 2.2 Conformal group

Considering the infinitesimal transformations described in the previous section, the generators of the conformal group acting on a function can be expressed as

$$\begin{aligned}
&\text{(translation)} && P_\mu = \partial_\mu, \\
&\text{(dilatations)} && D = x^\mu \partial_\mu, \\
&\text{(Lorentz transformation)} && L_{\mu\nu} = x_\nu \partial_\mu - x_\mu \partial_\nu, \\
&\text{(special conformal transformation)} && K_\mu = 2x_\mu x^\nu \partial_\nu - x^2 \partial_\mu.
\end{aligned} \tag{2.2.1}$$

These generators satisfy the following commutation rules that define the conformal algebra

$$\begin{aligned}
&[D, K_\mu] = K_\mu, \\
&[D, P_\mu] = -P_\mu, \\
&[P_\mu, K_\nu] = 2\left(\delta_{\mu\nu} D + L_{\mu\nu}\right), \\
&[P_\mu, L_{\nu\rho}] = \delta_{\mu\rho} P_\nu - \delta_{\mu\nu} P_\rho, \\
&[K_\mu, K_\nu] = [P_\mu, P_\nu] = [D, D] = [D, L_{\mu\nu}] = 0, \\
&[K_\mu, L_{\rho\sigma}] = \delta_{\mu\rho} K_\sigma - \delta_{\mu\sigma} K_\rho, \\
&[L_{\mu\nu}, L_{\rho\sigma}] = \delta_{\mu\rho} L_{\nu\sigma} + \delta_{\nu\sigma} L_{\mu\rho} - \delta_{\nu\rho} L_{\mu\sigma} - \delta_{\mu\sigma} L_{\nu\rho}.
\end{aligned} \tag{2.2.2}$$

It is important to note that translations, Lorentz transformations and dilatations form a subgroup of the full conformal group. It is a subtle question whether theories can exhibit scale invariance and Poincaré symmetry without also possessing special conformal invariance. Extensive studies suggest that examples of such theories, while mathematically possible, are not physically compelling. They typically suffer from issues such as nonunitarity, an unbounded spectrum or lack of interactions, making them less relevant in physical applications.
We now define the generators $J_{ab}$, where $J_{ba} = -J_{ab}$ and $a, b \in -1, 0, 1, \ldots, d$, as

$$\begin{aligned}
&J_{\mu\nu} = L_{\mu\nu}, \quad J_{-1,\mu} = \frac{1}{2}\left(P_\mu - K_\mu\right), \\
&J_{-1,0} = D, \quad J_{0,\mu} = \frac{1}{2}\left(P_\mu + K_\mu\right),
\end{aligned} \tag{2.2.3}$$

These relations allow the commutation rules to be rewritten in a compact form

$$[J_{ab}, J_{cd}] = \eta_{ac} J_{bd} + \eta_{bd} J_{ac} - \eta_{ad} J_{bc} - \eta_{bc} J_{ad} \tag{2.2.4}$$

where $\eta_{ab} = \text{diag}(-1, 1, 1, \ldots, 1)$. The commutation relations (2.2.4) define the algebra of the group $SO(1, d+1)$, demonstrating the isomorphism between the conformal group in $d$-dimensional Euclidean space and the group $SO(1, d+1)$ with $(d+2)(d+1)/2$ parameters.

## 2.3 Representations of the conformal group

In this section, we illustrate how a general field $\mathcal{O}$ transforms under conformal transformations. We begin by examining the transformation properties of the field at $x = 0$. Then, using the momentum vector $P_\mu$, we shift the argument of the field to an arbitrary point $x$ to derive the general transformation rule.
We start by introducing a matrix representation $S_{\mu\nu}$ which define the action of infinitesimal Lorentz transformations on $\mathcal{O}(0)$ as follows

$$[L_{\mu\nu}, \mathcal{O}(0)] = \mathcal{S}_{\mu\nu}\mathcal{O}(0), \tag{2.3.1}$$

where $S_{\mu\nu}$ determines the spin of the field $\mathcal{O}(0)$. For the conformal algebra, we also establish the commutation relation with the dilatation operator $D$

$$[D, \mathcal{O}(0)] = \Delta_{\mathcal{O}}\mathcal{O}(0). \tag{2.3.2}$$

This relation implies that $\mathcal{O}$ has the scaling dimension $\Delta_{\mathcal{O}}$. Consequently, under dilatations $x \mapsto x' = \lambda x$ the field transforms as

$$\mathcal{O}(x) \mapsto \mathcal{O}'(x') = \lambda^{-\Delta_{\mathcal{O}}}\mathcal{O}(x). \tag{2.3.3}$$

A field $\mathcal{O}$, which transforms covariantly under an irreducible representation of the conformal algebra, has a fixed scaling dimension and it is therefore an eigenstate of the dilatation operator $D$. In particular, in a unitary CFT, there is a lower bound on the scaling dimensions of the fields. In $d$ dimensions, the bound for fields of spin $s$ takes the form

$$\Delta \geq \frac{d-2}{2}, \qquad s = 0, \tag{2.3.4}$$

$$\Delta \geq \frac{d-1}{2}, \qquad s = 1/2, \tag{2.3.5}$$

$$\Delta \geq d + s - 2, \qquad s \geq 1. \tag{2.3.6}$$

These bounds are saturated by a free scalar field ($s = 0$), a conserved current ($s = 1$) and a conserved symmetric traceless tensor ($s = 2$).
Consindering the commutation relations of $D$ with $P_\mu$ and $K_\mu$, one observes that the operator $P_\mu$ raises the scaling dimension , while $K_\mu$ lowers it

$$[D, [P_\mu, \mathcal{O}(0)]] = [P_\mu, [D, \mathcal{O}(0)]] + [[D, P_\mu], \mathcal{O}(0)] = (\Delta_{\mathcal{O}} + 1)[P_\mu, \mathcal{O}(0)], \tag{2.3.7}$$

$$[D, [K_\mu, \mathcal{O}(0)]] = [K_\mu, [D, \mathcal{O}(0)]] + [[D, K_\mu], \mathcal{O}(0)] = (\Delta_{\mathcal{O}} - 1)[K_\mu, \mathcal{O}(0)]. \tag{2.3.8}$$

Since in a unitary CFT there is a lower bound on the scaling dimensions of fields, every conformal representation must contain operators of the lowest dimension, which are annihilated by $K_\mu$ at $x = 0$

$$[K_\mu, \mathcal{O}(0)] = 0. \tag{2.3.9}$$

In a given irreducible multiplet of the conformal algebra, the *conformal primary fields* are those fields of lowest scaling dimension, as specified by (2.3.9). All the other fields, known as *conformal descendants*, can be obtained by acting with $P_\mu$ on such conformal primary fields.
We now introduce the operator $U(x) = \exp\left(\hat{P}_\mu x^\mu\right)$ that, when acting on $O(0)$, gives

$$U(x)\mathcal{O}(0)\,U^{-1}(x) = \mathcal{O}(x). \tag{2.3.10}$$

Using such operator, we can derive the commutation relations for a conformal primary field $\mathcal{O}(x)$, taking into account the relations of the conformal algebra. To illustrate this process, we examine the case of $[P_\mu, \mathcal{O}(x)]$. Expanding the operator $U(x)$ and using the Haussdorf formula

$$U(x)\mathcal{O}(0)\,U^{-1}(x) = \sum_{n=0}^{\infty} \frac{1}{n!} x^{\nu_1} \dots x^{\nu_n} [P_{\nu_1}, [\dots [P_{\nu_n}, \mathcal{O}(0)] \dots]], \tag{2.3.11}$$

one obtains

$$[P_\mu, U(x)\mathcal{O}(0)\,U^{-1}(x)] = \left[P_\mu\,,\, \sum_{n=0}^{\infty}\frac{1}{n!}x^{\nu_1}\dots x^{\nu_n}[P_{\nu_1},[\dots[P_{\nu_n},\mathcal{O}(0)]\dots]]\right] = \partial_\mu\mathcal{O}(x), \tag{2.3.12}$$

By proceeding in this manner, we can derive the commutation relations

$$\begin{aligned}
[P_\mu,\mathcal{O}(x)] &= \partial_\mu\mathcal{O}(x),\\
[D,\mathcal{O}(x)] &= (\Delta_{\mathcal{O}} + x^\mu\partial_\mu)\mathcal{O}(x),\\
[L_{\mu\nu},\mathcal{O}(x)] &= \left(\mathcal{S}_{\mu\nu} + x_\nu\partial_\mu - x_\mu\partial_\nu\right)\mathcal{O}(x),\\
[K_\mu,\mathcal{O}(x)] &= \left[2x_\mu(x\cdot\partial) - x^2\partial_\mu + 2\Delta_{\mathcal{O}}x_\mu - 2x^\nu\mathcal{S}_{\mu\nu}\right]\mathcal{O}(x).
\end{aligned} \tag{2.3.13}$$

Correlation functions involving primary fields transform covariantly. In particular, in the case of scalar conformal primaries, we have

$$\langle\mathcal{O}_1(x_1')\dots\mathcal{O}_n(x_n')\rangle = \left|\frac{\partial x_1'}{\partial x_1}\right|^{-\Delta_1/d}\dots\left|\frac{\partial x_n'}{\partial x_n}\right|^{-\Delta_n/d}\langle\mathcal{O}_1(x_1)\dots\mathcal{O}_n(x_n)\rangle. \tag{2.3.14}$$

where $\left|\frac{\partial x_i'}{\partial x_i}\right|$ is the Jacobian of the conformal transformation of the coordinates.
When a theory is coupled to a curved background metric $g_{\mu\nu}$, it is natural to consider the action of the Weyl group, under which the metric and the scalar primary fields transform as

$$\begin{aligned}
g'_{\mu\nu}(x) &= e^{2\sigma(x)}g_{\mu\nu}(x),\\
\mathcal{O}'_i(x) &= e^{-\sigma(x)\Delta_i}\mathcal{O}_i(x).
\end{aligned} \tag{2.3.15}$$

The function $\sigma(x)$ allows to define a subgroup of the local Weyl transformation group, induced by conformal transformations, known as the conformal Weyl group. The explicit form of $\sigma(x)$ is given by

$$\sigma(x) = \log\left(\frac{1}{1-2b\cdot x + b^2x^2}\right), \tag{2.3.16}$$

with $b$ any constant vector. For $b$ infinitesimal and using the expression of $\sigma(x)$, the transformation (2.3.15) corresponds exactly to the conformal transformations given in (2.1.6). Every conformal field theory is Weyl invariant, and the scalar correlation function of primary fields given above transforms under the Weyl group as

$$\langle\mathcal{O}_1(x_1)\dots\mathcal{O}_n(x_n)\rangle_{e^{2\sigma}g_{\mu\nu}} = e^{-\sigma(x_1)\Delta_1}\dots e^{-\sigma(x_n)\Delta_n}\langle\mathcal{O}_1(x_1)\dots\mathcal{O}_n(x_n)\rangle_{g_{\mu\nu}}. \tag{2.3.17}$$

The relationship between conformal and Weyl symmetry can be understood in physical terms through the Weyl gauging procedure, which leads to Weyl gravity if a specific metric is treated as a dynamical field rather than a compensator field. In this case, in each free-falling frame identified within a local region of spacetime, the local flat symmetry in the tangent space is governed by the generators of the conformal group rather than those of the Poincaré group.
It is worth noting that (2.3.17) is valid at non-coincident point only, i.e. for $x_i \neq x_j$, $i \neq j$. This fact is due to the choice of the regularisation scheme that one has to consider in order to have a defined quantum field theory. The (dimensionful) regulator used in the scheme chosen, may break some of the symmetries of the theory that cannot be restored once the regulator is removed. For instance, dimensional regularisation, which preserves Lorentz invariance, breaks Weyl invariance, and it is manifest in a local manner, i.e., it affects the correlation functions at coincident points only. The effects of the breaking of Weyl symmetry is manifest in the trace or Weyl anomaly. In quantum field theory scale and conformal invariance are treated as equivalent, in the sense that all the realistic field theories which are scale invariant are also conformal invariant. Such enhancement of the symmetry (from dilatation invariance to conformal), whether it holds in general or not, has been discussed at length in the literature [65], with counterexamples which are quite unrealistic from the point of view of a local, covariant, relativistic field theory. In our case scale and conformal invariance will be taken as equivalent, as is indeed the case in ordinary Lagrangian field theories.

## 2.4 Correlation functions in CFT

The presence of conformal symmetry in a quantum field theory imposes significant constraints on the structure of correlation functions. Conformal invariance fixes the form of correlation functions up to three points, leaving only multiplicative constants undetermined. For simplicity, this section will focus on scalar correlators.
Let us examine the simplest scenario: one-point functions constructed from a scalar operator $\mathcal{O}$. Due to translational invariance, these functions must be constant. When we then impose scale invariance, it follows that this constant must equal zero, yielding the condition

$$\langle \mathcal{O}(x) \rangle = 0, \qquad \text{if } \Delta \neq 0. \tag{2.4.1}$$

This requirement on the conformal dimension $\Delta \neq 0$ aligns with the unitarity bounds in a unitary conformal field theory, where all operators (except for the identity) possess strictly positive conformal dimensions. Thus, in a conformal theory, we conclude $\langle \mathcal{O}(x) \rangle = 0$, assuming $\mathcal{O}$ is not proportional to the identity operator.
Moving on to the next non trivial case, we consider 2-point functions of two scalar operators $\mathcal{O}_1$ and $\mathcal{O}_2$ of conformal dimensions $\Delta_1$ and $\Delta_2$, respectively. Due to Poincaré invariance, these functions can only depend on the difference of coordinates and imposing scaling invariance one finds

$$\langle \mathcal{O}_1(x_1) \mathcal{O}_2(x_2) \rangle = \frac{C_{12}}{|x_1 - x_2|^{\Delta_1 + \Delta_2}}. \tag{2.4.2}$$

To ensure invariance under the special conformal transformations, we recall that

$$\left| \frac{\partial x_i'}{\partial x_i} \right| = \frac{1}{(1 - 2b \cdot x_i + b^2 x_i^2)^d}. \tag{2.4.3}$$

Under this transformation the 2-point function transforms as

$$\langle \mathcal{O}_1'(x_1') \mathcal{O}_2'(x_2') \rangle = \left| \frac{\partial x_1'}{\partial x_1} \right|^{\Delta_1/d} \left| \frac{\partial x_2'}{\partial x_2} \right|^{\Delta_2/d} \langle \mathcal{O}_1(x_1') \mathcal{O}_2(x_2') \rangle = \frac{(\gamma_1 \gamma_2)^{(\Delta_1 + \Delta_2)/2}}{\gamma_1^{\Delta_1} \gamma_2^{\Delta_2}} \frac{C_{12}}{|x_1 - x_2|^{\Delta_1 + \Delta_2}} \tag{2.4.4}$$

where $\gamma_i = (1 - 2b \cdot x_i + b^2 x_i^2)$. The invariance of the correlation function under special conformal transformations requires that

$$\frac{(\gamma_1 \gamma_2)^{(\Delta_1 + \Delta_2)/2}}{\gamma_1^{\Delta_1} \gamma_2^{\Delta_2}} = 1 \tag{2.4.5}$$

which is identically satisfied only if $\Delta_1 = \Delta_2$. This means that two primary fields are correlated only if they have identical scaling dimensions

$$\langle \mathcal{O}_1(x_1) \mathcal{O}_2(x_2) \rangle = \frac{C_{12}}{|x_1 - x_2|^{2\Delta_1}} \delta_{\Delta_1 \Delta_2}. \tag{2.4.6}$$

A similar analysis applies to three-point functions. Poincaré invariance requires that

$$\langle \mathcal{O}_1(x_1)\mathcal{O}_2(x_2)\mathcal{O}_3(x_3)\rangle = f\Big(|x_1-x_2|, |x_1-x_3|, |x_2-x_3|\Big). \tag{2.4.7}$$

Scale invariance further constrains this function to

$$f\Big(|x_1-x_2|, |x_1-x_3|, |x_2-x_3|\Big) = \lambda^{\Delta_1+\Delta_2+\Delta_3} f\Big(\lambda|x_1-x_2|, \lambda|x_1-x_3|, \lambda|x_2-x_3|\Big). \tag{2.4.8}$$

This implies that the three-point function can be expressed as

$$\langle \mathcal{O}_1(x_1)\mathcal{O}_2(x_2)\mathcal{O}_3(x_3)\rangle = \frac{C_{123}}{x_{12}^a\, x_{23}^b\, x_{13}^c} \tag{2.4.9}$$

where $x_{ij} = |x_i - x_j|$ and $C_{123}$ is a constant that captures the strength of the correlation. The coefficients $a$, $b$, $c$ must satisfy the relation

$$a+b+c = \Delta_1+\Delta_2+\Delta_3. \tag{2.4.10}$$

The invariance of (2.4.9) under special conformal transformations leads to

$$\frac{C_{123}}{\gamma_1^{\Delta_1}\gamma_2^{\Delta_2}\gamma_3^{\Delta_3}}\frac{(\gamma_1\gamma_2)^{a/2}(\gamma_2\gamma_3)^{b/2}(\gamma_1\gamma_3)^{c/2}}{x_{12}^a\, x_{23}^b\, x_{13}^c} = \frac{C_{123}}{x_{12}^a\, x_{23}^b\, x_{13}^c} \tag{2.4.11}$$

Such equation is satisfied if

$$\gamma_1^{a/2+c/2-\Delta_1} = 1, \qquad \gamma_2^{a/2+b/2-\Delta_2} = 1, \qquad \gamma_3^{b/2+c/2-\Delta_3} = 1, \tag{2.4.12}$$

which imply

$$a = \Delta_1+\Delta_2-\Delta_3, \qquad b = \Delta_2+\Delta_3-\Delta-1, \qquad c = \Delta_3+\Delta_1-\Delta_2. \tag{2.4.13}$$

Therefore, the generic three-point function can be expressed as

$$\langle \mathcal{O}_1(x_1)\mathcal{O}_2(x_2)\mathcal{O}_3(x_3)\rangle = \frac{C_{123}}{x_{12}^{\Delta_1+\Delta_2-\Delta_3}\, x_{23}^{\Delta_2+\Delta_3-\Delta-1}\, x_{13}^{\Delta_3+\Delta_1-\Delta_2}}. \tag{2.4.14}$$

$C_{123}$ and $\Delta_i$'s are usually referred to as "the conformal data".
While the conformal constraints on two- and three-point functions provide significant information, they do not uniquely determine the structure of four-point functions or higher. For these cases, one can construct scalar conformal invariants from the external coordinates $\{x_j\}$. Poincaré invariance implies the correlation function must depend on the distances $x_{ij} = |x_i - x_j|$. To maintain scale invariance, one has to construct ratios of these distances. We recall that, under the special conformal transformation, the distance separating two points $x_{ij}$ is mapped to

$$|x_i' - x_j'| = \frac{|x_i - x_j|}{(1-2b\cdot x_i + b^2 x_i^2)^{1/2}(1-2b\cdot x_j + b^2 x_j^2)^{1/2}}. \tag{2.4.15}$$

Therefore, the simplest objects that are conformally invariant are

$$u_{ijkl} \equiv \frac{x_{ij}\,x_{kl}}{x_{ik}\,x_{jl}}, \qquad v_{ijkl} = \frac{x_{ij}\,x_{kl}}{x_{il}\,x_{jk}}, \tag{2.4.16}$$

known as anharmonic ratios or cross-ratios. These cross-ratios are well-defined provided the points are distinct. Consequently, any function of these cross-ratios will be conformally invariant. For this reason 4- and higher-point functions are determined up to an arbitrary function of the cross-ratios, and are not uniquely fixed by conformal invariance. For instance, the 4-point function can generally be written as

$$\langle \mathcal{O}_1(x_1)\mathcal{O}_2(x_2)\mathcal{O}_3(x_3)\mathcal{O}_4(x_4)\rangle = f\left(\frac{x_{12}x_{34}}{x_{13}x_{24}}, \frac{x_{12}x_{34}}{x_{23}x_{14}}\right) \prod_{1\leq i<j\leq 4}^{4} x_{ij}^{\Delta_t/3-\Delta_i-\Delta_j} \tag{2.4.17}$$

where $\Delta_t = \sum_{i=1}^4 \Delta_i$ and $f$ is an undetermined function. In the case of 4-point functions there are two independent conformal ratios. In the case of $n$-point functions there are $n(n-3)/2$ independent conformal ratios. In conclusion, the correlation functions for four-point and higher-order interactions remain significantly less constrained and can take on a broader variety of forms.

## 2.5 Conformal Ward identities in coordinate-space

In the previous section, we examined how correlation functions are determined by conformal symmetry. In this section, we derive the conformal Ward identities that these correlators must satisfy. In general, the Ward identities for a correlation function that is invariant under a symmetry transformation $g$ are given by

$$\sum_{i=1}^{n} \langle \mathcal{O}_1(x_1)\dots\delta_g\mathcal{O}_i(x_i)\dots\mathcal{O}_n(x_n)\rangle = 0, \tag{2.5.1}$$

where $\mathcal{O}_i$ are arbitrary conformal primary operators. In the case of conformal transformations, the variations $\delta_g\mathcal{O}_i$ are determined by the transformations rules (2.3.13). Therefore, the translations Ward identity takes the form

$$0 = \sum_{j=1}^{n} \frac{\partial}{\partial x_j^\mu} \langle \mathcal{O}_1(x_1)\dots\mathcal{O}_n(x_n)\rangle, \tag{2.5.2}$$

implying that the correlation function depends only on the differences $x_i - x_j$. To derive the Ward identity associated with Lorentz transformations, we first consider the $n$-point function of scalar operators, yielding

$$\sum_{j=1}^{n}\left(x_j^\nu \frac{\partial}{\partial x_{j\mu}} - x_j^\mu \frac{\partial}{\partial x_{j\nu}}\right)\langle \mathcal{O}_1(x_1)\dots\mathcal{O}_n(x_n)\rangle = 0. \tag{2.5.3}$$

For tensor operators one needs to add to the left-hand side of the previous equation the contribution of $\mathcal{S}_{\mu\nu}$ in (2.3.13), the finite-dimensional representation of the rotations determining the spin for the field. If we assume that the tensor $\mathcal{O}_j$ has $r_j$ Lorentz indices, i.e. $\mathcal{O}_j = \mathcal{O}_j^{\mu_{j_1}\dots\mu_{j_{r_j}}}$, for $j = 1, 2, \dots, n$, in this case we need to add the extra term

$$\sum_{j=1}^{n}\sum_{h=1}^{r_j}\left[\delta_{\nu\alpha_{j_h}}\delta_\mu^{\mu_{j_h}} - \delta_{\mu\alpha_{j_h}}\delta_\nu^{\mu_{j_h}}\right]\langle \mathcal{O}^{\mu_{1_1}\mu_{1_2}\dots\mu_{1_{r_1}}}(x_1)\dots\ \mathcal{O}^{\mu_{j_1}\dots\alpha_{j_h}\dots\mu_{j_{r_j}}}(x_j)\ \dots\ \mathcal{O}^{\mu_{n_1}\mu_{n_2}\dots\mu_{n_{r_n}}}(x_n)\rangle. \tag{2.5.4}$$

Next, the dilatation Ward identities can be constructed using the dilatation generator, yielding

$$\left[\sum_{j=1}^{n}\Delta_j + \sum_{j=1}^{n} x_j^\alpha \frac{\partial}{\partial x_j^\alpha}\right]\langle \mathcal{O}_1(x_1)\dots\mathcal{O}_n(x_n)\rangle = 0. \tag{2.5.5}$$

where here we indicate with $\Delta_1,\ldots,\Delta_n$ the dimensions of the conformal primaries operators $\mathcal{O}_1,\ldots,\mathcal{O}_n$. Lastly, the special conformal Ward identity for the scalar case is given by

$$\sum_{j=1}^{n}\left(2\Delta_j\, x_j^k + 2x_j^\kappa\, x_j^\alpha \frac{\partial}{\partial x_{j\alpha}} - x_j^2 \frac{\partial}{\partial x_{jk}}\right)\langle \mathcal{O}_1(x_1)\ldots\mathcal{O}_n(x_n)\rangle = 0, \tag{2.5.6}$$

where $k$ is a free Lorentz index. In the case of tensor operators one needs to add the additional term

$$2\sum_{j=1}^{n}\sum_{h=1}^{r_j}\left[(x_j)_{\alpha_{j_h}}\delta^{\kappa\mu_{j_h}} - x_j^{\mu_{j_h}}\delta^{\kappa}_{\alpha_{j_h}}\right]\langle \mathcal{O}^{\mu_{1_1}\mu_{1_2}\cdots\mu_{1_{r_1}}}(x_1)\ldots\ \mathcal{O}^{\mu_{j_1}\cdots\alpha_{j_h}\cdots\mu_{j_{r_j}}}(x_j)\ \ldots\ \mathcal{O}^{\mu_{n_1}\mu_{n_2}\cdots\mu_{n_{r_n}}}(x_n)\rangle \tag{2.5.7}$$

to the left-hand side of (2.5.6). In Appendix E, we present an alternative derivation of the CWIs using the $\langle TTJ_A\rangle$ correlator as a working example.

## 2.6 Conformal Ward identities in momentum-space

The original approach to identify correlation functions in conformal field theories was traditionally formulated using coordinate-space methods, both for scalar and tensor correlators. The correlators examined in previous sections serve as basic examples, devoid of anomalies, with solutions in coordinate-space valid when computed at distinct points.
In the presence of anomalies, the solutions of the conformal Ward identities can be obtained by partitioning the domain of definition of each correlator into nonlocal and contact contributions. The equations are initially solved for the regions in which the external coordinate points of the correlators are all noncoincident. Anomalous corrections, which arise when all the points coincide, are manually added by including additional local terms with support defined by products of delta functions. This methodology was pioneered in groundbreaking works [66, 67], and it was successfully applied to correlators containing the energy-momentum tensor $T$ and conserved vector currents $J_V$, such as the $\langle TTT\rangle$ and the $\langle TJ_VJ_V\rangle$. Given the intricate nature of chiral and conformal anomalies, which are related to contact interactions, the coordinate-space approach becomes unwieldy and the hierarchical character of the CWIs certainly becomes rather involved.
As we move to momentum-space, due to the Fourier transform acting on a correlation function's spacetime points, the integration also includes domains in which two or more coordinates coalesce. Therefore the anomaly contribution is naturally included in the expression of a correlator in momentum-space [68–71].
Furthermore, one of the main advantages of the momentum-space approach to the determination of CFT correlators is to establish a link with the ordinary perturbative Feynman expansion. In particular, it allows to compare general results with explicit realizations of CFT's, where many methods are available. While the latter is directly connected with a specific Lagrangian realization, the analysis of the conformal Ward identities (CWI's) in momentum-space, on the other hand, allows to investigate the operatorial content of a CFT in the most general way, whenever this is possible.
The use of conformal Ward identities to determine the structure of three-point functions in momentum-space was independently introduced in [28] and [29], with the latter extending the methodology to tensor cases. Renormalized parity even correlators, type-A and type-B Weyl anomalies have been extensively discussed [25–27, 72]. Furthermore, perturbative analysis of some important correlators such as the $\langle TTT\rangle$ and $\langle TJ_VJ_V\rangle$, in the parity-even sector, have been conducted in the conformal and nonconformal limits in previous works in QED and QCD [64, 73–77]. Lastly, the analysis of 4-point functions in both generic CFTs and in free field realizations has been explored in [78, 79].

These studies have primarily focused on the parity-even sector, incorporating contributions from the conformal anomaly. In contrast, investigations into parity-odd correlators have only recently commenced, and they will be the primary focus of this thesis.
We now proceed with the reformulation of the conformal constraints, specifically the conformal Ward identities, in momentum-space. To simplify the expressions, we will employ a condensed notation for transforms in momentum-space, using the following conventions

$$\begin{aligned} &\Phi(\underline{x}) \equiv \langle \mathcal{O}_1(x_1)\mathcal{O}_2(x_2)\dots\mathcal{O}_n(x_n)\rangle \qquad e^{i\underline{px}} \equiv e^{i(p_1x_1+p_2x_2+\dots p_nx_n)} \\ &\underline{dp} \equiv dp_1dp_2\dots dp_n \qquad \Phi(\underline{p}) \equiv \Phi(p_1,p_2,\dots,p_n). \end{aligned} \tag{2.6.1}$$

where $\Phi$ represents an $n$-point correlation functions of primary operators $\mathcal{O}_i$. It will also be useful to introduce the total momentum $P = \sum_{j=1}^{n} p_j$.
The momentum constraint is enforced by a delta function $\delta(P)$ under integration. Indeed, the translation invariance of $\Phi(\underline{x})$ leads to

$$\Phi(\underline{x}) = \int \underline{dp}\,\delta(P)\,e^{i\underline{px}}\Phi(\underline{p}). \tag{2.6.2}$$

In general, for an $n$-point function $\Phi(x_1,x_2,\dots,x_n) = \langle \mathcal{O}_1(x_1)\mathcal{O}_2(x_2)\dots\mathcal{O}_n(x_n)\rangle$, the condition of translation invariance

$$\langle \mathcal{O}_1(x_1)\mathcal{O}_2(x_2)\dots\mathcal{O}_n(x_n)\rangle = \langle \mathcal{O}_1(x_1+a)\mathcal{O}_2(x_2+a)\dots\mathcal{O}_n(x_n+a)\rangle \tag{2.6.3}$$

leads to the momentum-space expression in the form of (2.6.2), allowing us to remove one of the momenta, conventionally the last one, $p_n$, by replacing it with $\overline{p}_n = -(p_1+p_2+\dots p_{n-1})$. This yields

$$\Phi(x_1,x_2,\dots,x_n) = \int dp_1dp_2...dp_{n-1}e^{i(p_1x_1+p_2x_2+\dots p_{n-1}x_{n-1}+\overline{p}_nx_n)}\Phi(p_1,p_2,\dots,\overline{p}_n). \tag{2.6.4}$$

### 2.6.1 The dilatation equation

We begin by considering the dilatation Ward identity. The condition of scale covariance for fields $\mathcal{O}_i$ with scale dimensions $\Delta_i$ is

$$\Phi(\lambda x_1,\lambda x_2,\dots,\lambda x_n) = \lambda^{-\Delta}\Phi(x_1,x_2,\dots,x_n), \qquad \Delta = \Delta_1+\Delta_2+\dots\Delta_n \tag{2.6.5}$$

which, by setting $\lambda = 1+\epsilon$ and expanding to the first order in $\epsilon$, results in the scaling relation

$$(D_n+\Delta)\Phi \equiv \sum_{j=1}^{n}\left(x_j^\alpha\frac{\partial}{\partial x_j^\alpha}+\Delta_j\right)\Phi(x_1,x_2,\dots,x_n) = 0. \tag{2.6.6}$$

with

$$D_n = \sum_{j=1}^{n} x_j^\alpha\frac{\partial}{\partial x_j^\alpha}. \tag{2.6.7}$$

The momentum-space form of the dilatation equation can be derived by Fourier transforming (2.6.6) or, more directly, by using (2.6.5). Utilizing translation invariance and removing the delta function constraint, we obtain

$$\begin{aligned} \Phi(\lambda x_1,\lambda x_2,\dots,\lambda x_n) &= \int d^dp_1d^dp_2\dots d^dp_{n-1}e^{i\lambda(p_1x_1+p_2x_2+\dots p_{n-1}x_{n-1}+\overline{p}_nx_n)}\Phi(p_1,p_2,\dots,\overline{p}_n) \\ &= \lambda^{-\Delta}\int d^dp_1d^dp_2\dots d^dp_{n-1}e^{i(p_1x_1+p_2x_2+\dots p_{n-1}x_{n-1}+\overline{p}_nx_n)}\Phi(p_1,p_2,\dots,\overline{p}_n). \end{aligned} \tag{2.6.8}$$

We now perform the change of variables $p_i = p_i'/\lambda$ on the right-hand-side of the equation above (first line) with $dp_1...dp_{n-1} = (1/\lambda)^{d(n-1)} d^d p_1' \dots d^d p_{n-1}'$ to derive the relation

$$\frac{1}{\lambda^{d(n-1)}}\Phi\left(\frac{p_1}{\lambda},\frac{p_2}{\lambda},\dots,\frac{\overline{p}_n}{\lambda}\right) = \lambda^{-\Delta}\Phi(p_1,p_2,\dots,\overline{p}_n). \tag{2.6.9}$$

Setting $\lambda = 1/s$, we can write

$$s^{(n-1)d-\Delta}\Phi(sp_1,sp_2,\dots,s\overline{p}_n) = \Phi(p_1,p_2,\dots,\overline{p}_n). \tag{2.6.10}$$

Expanding to the first order in $s \sim 1+\epsilon$, we obtain

$$\left[\sum_{j=1}^{n}\Delta_j - (n-1)d - \sum_{j=1}^{n-1} p_j^{\alpha}\frac{\partial}{\partial p_j^{\alpha}}\right]\Phi(p_1,p_2,\dots,\overline{p}_n) = 0. \tag{2.6.11}$$

Notice that the differentiation is restricted to the $n-1$ independent momenta due to momentum conservation. This point has been illustrated at length in [75], to which we refer for further details.

### 2.6.2 The special conformal equations

We now turn to the analysis of the special conformal transformations in momentum-space. We discuss both the symmetric and the asymmetric forms of the equations with respect to momenta, focusing our attention first on the scalar case. The CWI in coordinate-space for a scalar correlator is expressed as

$$\sum_{j=1}^{n}\left(-x_j^2\frac{\partial}{\partial x_j^{\kappa}} + 2x_j^{\kappa}x_j^{\alpha}\frac{\partial}{\partial x_j^{\alpha}} + 2\Delta_j x_j^{\kappa}\right)\Phi(x_1,x_2,\dots,x_n) = 0. \tag{2.6.12}$$

To transition to momentum-space, we use the substitutions

$$x_j^{\alpha} \to -i\frac{\partial}{\partial p_j^{\alpha}} \qquad \frac{\partial}{\partial x_j^{\kappa}} \to i p_j^{\kappa}. \tag{2.6.13}$$

The CWI then becomes

$$\sum_{j=1}^{n}\int \underline{d^d p}\,\delta^d(P)\Phi(\underline{p})\left(p_j^{\kappa}\frac{\partial^2}{\partial p_j^{\alpha}\partial p_j^{\kappa}} - 2p_j^{\alpha}\frac{\partial^2}{\partial p_j^{\alpha}\partial p_j^{\kappa}} - 2\Delta_j\frac{\partial}{\partial p_j^{\kappa}}\right)e^{i\underline{p\cdot x}} = 0, \tag{2.6.14}$$

After integrating by parts, we can write

$$\int \underline{d^d p} e^{i\underline{px}}\mathcal{K}_s^k\Phi(\underline{p})\delta^d(P) + \delta'_{\text{term}} = 0 \tag{2.6.15}$$

where we have introduced the following differential operator acting on a scalar correlator in a symmetric form

$$\mathcal{K}_s^k = \sum_{j=1}^{n}\left(p_j^{\kappa}\frac{\partial^2}{\partial p_j^{\alpha}\partial p_j^{\alpha}} + 2(\Delta_j - d)\frac{\partial}{\partial p_j^{\kappa}} - 2p_j^{\alpha}\frac{\partial^2}{\partial p_j^{\kappa}\partial p_j^{\alpha}}\right). \tag{2.6.16}$$

Using some distributional identities derived in [75]

$$\begin{aligned}\mathcal{K}_s^\kappa \delta^d(P) &= \left(P^k \frac{\partial^2}{\partial P^\alpha \partial P_\alpha} - 2P^\alpha \frac{\partial^2}{\partial P^\alpha \partial P^k} + 2(\Delta - n\,d)\frac{\partial}{\partial P^k}\right)\delta^d(P)\\ &= 2d(d\,n - d - \Delta)P^k \frac{\delta^d(P)}{P^2}\\ &= -2(d\,n - d - \Delta)\frac{\partial}{\partial P_k}\delta^d(P), \qquad \Delta = \sum_{j=1}^n \Delta_j,\end{aligned} \tag{2.6.17}$$

we obtain

$$\begin{aligned}\delta'_{\text{term}} = \int \underline{d^d p}\, e^{\underline{ip\cdot x}}\Bigg[&\frac{\partial}{\partial P^\alpha}\delta^d(P)\sum_{j=1}^n\left(p_j^\alpha \frac{\partial}{\partial p_j^\kappa} - p_j^\kappa \frac{\partial}{\partial p_j^\alpha}\right)\Phi(\underline{p})\\ &. + 2\frac{\partial}{\partial P^\kappa}\delta^d(P)\left(\sum_{j=1}^n\left(\Delta_j - p_j^\alpha \frac{\partial}{\partial p_j^\alpha}\right) - (n-1)d\right)\Phi(\underline{p})\Bigg].\end{aligned} \tag{2.6.18}$$

Such terms vanish as shown in [75], due to the symmetric scaling relation and the rotational invariance of the scalar correlator

$$\sum_{j=1}^n\left(p_j^\alpha \frac{\partial}{\partial p_j^\kappa} - p_j^\kappa \frac{\partial}{\partial p_j^\alpha}\right)\Phi(\underline{p}) = 0, \tag{2.6.19}$$

which follows from the $SO(d)$ symmetry condition

$$\sum_{j=1}^n L_{\mu\nu}(x_j)\langle \mathcal{O}(x_1)\ldots\mathcal{O}(x_n)\rangle = 0, \tag{2.6.20}$$

with

$$L_{\mu\nu}(x) = i\left(x_\mu \partial_\nu - x_\nu \partial_\mu\right), \tag{2.6.21}$$

Using (2.6.17) and the vanishing of the $\delta'_{\text{term}}$ terms, we find that the structure of the CWI on the correlator $\Phi(p)$ assumes the symmetric form

$$\sum_{j=1}^n \int \underline{d^d p}\, e^{\underline{ip\cdot x}}\left(p_j^\kappa \frac{\partial^2}{\partial p_{j\alpha}\partial p_j^\alpha} - 2p_j^\alpha \frac{\partial^2}{\partial p_j^\alpha \partial p_j^\kappa} + 2(\Delta_j - d)\frac{\partial}{\partial p_j^\kappa}\right)\Phi(\underline{p})\delta^d(P) = 0. \tag{2.6.22}$$

This symmetric expression is the starting point in order to proceed with the elimination of one of the momenta, say $p_n$. One can proceed by following the same procedure used in the derivation of the dilatation identity, dropping the contribution coming from the dependent momentum $p_n$, thereby obtaining the final form of the equation

$$\sum_{j=1}^{n-1}\left(p_j^\kappa \frac{\partial^2}{\partial p_j^\alpha \partial p_j^\alpha} + 2(\Delta_j - d)\frac{\partial}{\partial p_j^\kappa} - 2p_j^\alpha \frac{\partial^2}{\partial p_j^\kappa \partial p_j^\alpha}\right)\Phi(p_1,\ldots p_{n-1},\bar{p}_n) = 0. \tag{2.6.23}$$

This equation can be extended to the tensorial case, where $O_j$ has $r_j$ Lorentz indices, i.e. $O_j = O_j^{\mu_1\ldots\mu_{r_j}}$ for $j = 1,2,\ldots n$. In this case we need to add the following term to the left-hand side of the equation

$$2\sum_{j=1}^{n-1}\sum_{h=1}^{n_j}\left[\delta^{\kappa\mu_{j_h}}\frac{\partial}{\partial p_j^{\alpha_{j_h}}} - \delta^\kappa_{\alpha_{j_h}}\frac{\partial}{\partial p_{j\mu_{j_h}}}\right]\left\langle O_1^{\mu_{11}\ldots\mu_{1r_1}}(p_1)\ldots O_j^{\mu_{j1}\ldots\alpha_{j_h}\ldots\mu_{jr_j}}\left(p_j\right)\ldots O_n^{\mu_{n1}\ldots\mu_{nr_n}}(p_n)\right\rangle \tag{2.6.24}$$

### 2.6.3 Conformal Ward identities from the vector to the scalar form

Before concluding this chapter, we outline the procedure for deriving conformal constraints as partial differential equations with respect to scalar invariants.
As an example, let us consider the parity-even three-point function of scalar primary operators. Due to translation invariance, this correlator can be expressed solely in terms of three independent scalar invariants, specifically the magnitudes of the momenta, defined as $p_i = \sqrt{p_i^2}$

$$\langle \mathcal{O}_1(p_1)\mathcal{O}_2(p_2)\mathcal{O}_3(\bar{p}_3)\rangle = \Phi(p_1, p_2, p_3). \tag{2.6.25}$$

All the conformal Ward identities (WIs) can be reformulated in scalar form using the chain rule

$$\frac{\partial \Phi}{\partial p_i^\mu} = \frac{p_i^\mu}{p_i}\frac{\partial \Phi}{\partial p_i} - \frac{\bar{p}_3^\mu}{p_3}\frac{\partial \Phi}{\partial p_3}, \qquad i = 1, 2 \tag{2.6.26}$$

where $\bar{p}_3^\mu = -p_1^\mu - p_2^\mu$ and $p_3 = \sqrt{(p_1 + p_2)^2}$. Using this expression, the dilatation operator can be written as

$$\sum_{i=1}^{2} p_i^\mu \frac{\partial}{\partial p_i{}^\mu}\Phi(p_1, p_2, p_3) = \left(p_1\frac{\partial}{\partial p_1} + p_2\frac{\partial}{\partial p_2} + p_3\frac{\partial}{\partial p_3}\right)\Phi(p_1, p_2, p_3). \tag{2.6.27}$$

Therefore, the scale equation becomes

$$\left(\sum_{i=1}^{3}\Delta_i - 2d - \sum_{i=1}^{3} p_i\frac{\partial}{\partial p_i}\right)\Phi(p_1, p_2, \bar{p}_3) = 0. \tag{2.6.28}$$

Through a straightforward yet lengthy computation, it can be shown that the special conformal Ward identities in $d$ dimensions take the form

$$\begin{aligned} 0 = K^\kappa_{scalar}\Phi(p_1, p_2, p_3) &= \left(p_1^\kappa K_1 + p_2^\kappa K_2 + \bar{p}_3^\kappa K_3\right)\Phi(p_1, p_2, p_3) \\ &= p_1^\kappa\left(K_1 - K_3\right)\Phi(p_1, p_2, p_3) + p_2^\kappa\left(K_2 - K_3\right)\Phi(p_1, p_2, p_3), \end{aligned} \tag{2.6.29}$$

where we have used the conservation of the total momentum and we have introduced the following operator

$$K_i \equiv \frac{\partial^2}{\partial p_i \partial p_i} + \frac{d + 1 - 2\Delta_i}{p_i}\frac{\partial}{\partial p_i} \tag{2.6.30}$$

The equation (2.6.29) is satisfied if every coefficients of the independent four-momenta $p_1^\kappa$, $p_2^\kappa$ are equal to zero. Therefore, the special conformal equations can be written in the coincise form

$$K_{13}\,\Phi(p_1, p_2, p_3) = 0 \qquad \text{and} \qquad K_{23}\,\Phi(p_1, p_2, p_3) = 0. \tag{2.6.31}$$

with

$$K_{ij} \equiv K_i - K_j \tag{2.6.32}$$

These kinds of equations can be solved using a special class of functions known as 3K integrals. We will explore such solutions for parity-odd correlators in later chapters.

# Chapter 3

# Parity-violating interactions with scalars in CFTs

In this chapter, we examine parity-odd 3-point correlators involving currents, energy-momentum tensors and at least one scalar operator in momentum-space CFTs. In particular, we illustrate how the conformal Ward identities completely determine the structure of such correlators in $d = 4$.
From the perturbative point of view, parity-odd correlators can be realized, for example, by considering a loop interaction with chiral fermions or by directly inserting a $\gamma^5$ into an operator, as in the case of an axial fermionic current $J_A^\mu = \bar{\psi}\gamma^\mu\gamma^5\psi$. Here, we study parity-odd correlators independently of their possible realizations. For example, the $\langle TJO\rangle_{\text{odd}}$ can be realized using a pseudoscalar operator, an axial current, or a pseudo-tensor $T$. The behavior under parity is a global property of the correlator and not necessarily of a single operator in the triple.
The correlators we examine in this chapter play a key role in the interactions between (pseudo)scalar fields—for example, an axion-like field—and gravity in a conformal symmetric phase. We will comment on the possible implications of this analysis from the phenomenological point of view. Indeed, our results constrain the interactions of axions or dilatons in a conformal phase of the early universe.
The nature of the scalar operator introduced in our investigation is general and can have any scaling dimension $\Delta$. The most interesting cases, from our perspective, are for $\Delta = 4$, where $O = \nabla \cdot J_A$ or $O = g_{\mu\nu}T^{\mu\nu}$, which carry direct physical implications and are directly linked to the chiral and conformal anomalies.

## 3.1 Three-point functions with multiple scalars

We start our discussion by considering correlators with at least two scalar operators $O(p_i)$. Before examining the conformal constraints, we write the most general form of the correlators in terms of form factors and tensorial structures. In this respect, the vanishing of such correlators in the parity-odd sector is rather straightforward, due to symmetry. Indeed, since we are dealing with four-dimensional parity-odd correlators, the tensorial structures need to include an $\varepsilon^{\alpha_1\alpha_2\alpha_3\alpha_4}$ tensor. It is to Figure out that this condition cannot be satisfied in this case, thereby giving the vanishing relations

$$\begin{aligned}
&\langle O(p_1)O(p_2)O(p_3)\rangle_{odd} = 0,\\
&\langle J^\mu(p_1)O(p_2)O(p_3)\rangle_{odd} = 0, \qquad (3.1.1)\\
&\langle T^{\mu\nu}(p_1)O(p_2)O(p_3)\rangle_{odd} = 0.
\end{aligned}$$

We emphasize that these equations are valid without imposing full conformal invariance and even when considering nonconserved currents and energy-momentum tensors. All possible two-point functions involving scalar operators, currents, and energy-momentum tensors are zero in 4d for the same reason.
Conversely, if we consider a three-point correlator with only one scalar operator, it is possible to construct at least one parity-odd tensorial structure with an $\varepsilon^{\alpha_1\alpha_2\alpha_3\alpha_4}$ tensor and the correlator does not necessarily vanish. In the following sections, we will examine how conformal invariance constrains such correlators by a direct application of the full methodology of CFT in momentum-space.

## 3.2 The $\langle JJO\rangle_{odd}$ correlator in CFT

In this section we study the conformal constraints on the $\langle JJO\rangle_{odd}$ correlator where each $J$ can be a conserved vector current $J_V$ or an anomalous axial-vector current $J_A$. Our result is valid for every parity-odd $\langle JJO\rangle_{odd}$ correlator that is constructed with two potentially different (axial and/or conserved) currents and a scalar/pseudoscalar operator. As we will see, the solution of the CWIs of the correlator can be written in terms of 3K integrals that needs a regularization. Therefore, we work in the general scheme $\{u,v\}$ where $d = 4+2u\epsilon$ and the conformal dimension $\Delta_i$ of the operators is shifted by $(u+v_i)\epsilon$ with arbitrary $u$ and $v_i$.
We start by decomposing the currents $J$ in terms of their transverse part and longitudinal ones (also termed "local") [29]

$$J^{\mu_i}(p_i) \equiv j^{\mu_i}(p_i) + j^{\mu_i}_{loc}(p_i), \tag{3.2.1}$$

where

$$j^{\mu_i}(p_i) = \pi^{\mu_i}_{\alpha_i}(p_i)J^{\alpha_i}(p_i), \qquad\qquad j^{\mu_i}_{loc}(p_i) = \frac{p_i^{\mu_i}\,p_{i\,\alpha_i}}{p_i^2}J^{\alpha_i}(p_i), \tag{3.2.2}$$

having introduced the transverse projector

$$\pi^{\mu}_{\alpha} = \delta^{\mu}_{\alpha} - \frac{p^{\mu}p_{\alpha}}{p^2}. \tag{3.2.3}$$

We then consider the following conservation Ward identities

$$\nabla_\mu\langle J^\mu_V\rangle = 0, \qquad\qquad \nabla_\mu\langle J^\mu_A\rangle = a\,\varepsilon^{\mu\nu\rho\sigma}F_{\mu\nu}F_{\rho\sigma} \tag{3.2.4}$$

of the expectation value of the vector $J^\mu_V$ and anomalous $J^\mu_A$ currents.
The vector current is coupled to the vector source $V_\mu$ and the axial-vector current to the source $A_\mu$. The operator $O$ in the $\langle JJO\rangle_{odd}$ is coupled to a scalar field source $\phi$. Applying multiple functional derivatives to (3.2.4) with respect to the sources, after a Fourier transform, we find the conservation Ward identities related to the entire correlator

$$p_{i\mu_i}\langle J^{\mu_1}(p_1)J^{\mu_2}(p_2)O(p_3)\rangle_{odd} = 0. \qquad i = 1,2 \tag{3.2.5}$$

Such equation is satisfied independently of the fact that $J$'s are conserved vector or axial-vector currents. Indeed, the chiral anomaly does not contribute to the $\langle JJO\rangle_{odd}$. Due to the identities in (3.2.5), the longitudinal part of the correlator vanishes. On the other hand, the transverse part can be formally expressed in terms of the following tensor structure

$$\langle J^{\mu_1}(p_1)J^{\mu_2}(p_2)O(p_3)\rangle_{odd} = \langle j^{\mu_1}(p_1)j^{\mu_2}(p_2)O(p_3)\rangle_{odd} = \pi^{\mu_1}_{\alpha_1}(p_1)\pi^{\mu_2}_{\alpha_2}(p_2)\Big[A(p_1,p_2,p_3)\varepsilon^{\alpha_1\alpha_2p_1p_2}\Big] \tag{3.2.6}$$

where $\varepsilon^{\alpha_1\alpha_2 p_1 p_2} \equiv \varepsilon^{\alpha_1\alpha_2\rho\sigma} p_{1\rho} p_{2\sigma}$. Notice that in this case one can omit the projectors $\pi^{\mu_i}_{\alpha_i}(p_i)$ since they act as an identity on the tensorial structure in the brackets.
Therefore, the entire analysis of the correlator can be reduced to determining $A(p_1,p_2,p_3)$. In the following, we will examine how the conformal constraints fix such form factor. These constraints will take the form of differential equations for $A(p_1,p_2,p_3)$. Specifically, we are going to solve the constraints from both the dilatation and the special conformal transformations on the parity-odd structure (3.2.6) admitted by Lorentz covariance.

### 3.2.1 Dilatation and special conformal Ward identities

We denote by $\Delta_i$ the conformal dimension of the operators in our correlator. Specifically, since we are working in four-dimensional space-time, the conformal dimensions of the conserved/axial currents are

$$\Delta_1 = \Delta_2 = 3. \tag{3.2.7}$$

while we do not fix the conformal dimension $\Delta_3$ of the scalar operator $O$. The invariance of the correlator under dilatation is reflected in the equation

$$\left(\sum_{i=1}^{3}\Delta_i - 2d - \sum_{i=1}^{2} p_i^\mu \frac{\partial}{\partial p_i^\mu}\right)\langle J^{\mu_1}(p_1)J^{\mu_2}(p_2)O(p_3)\rangle_{odd} = 0. \tag{3.2.8}$$

It is useful to introduce the invariants $p_i = |\sqrt{p_i^2}|$ as variables. Then, by using the chain rule

$$\frac{\partial}{\partial p_i^\mu} = \sum_{j=1}^{3} \frac{\partial p_j}{\partial p_i^\mu}\frac{\partial}{\partial p_j}, \tag{3.2.9}$$

and by considering the decomposition (3.2.6), we can rewrite the dilatation equation as a constraint on the form factor $A$

$$\sum_{i=1}^{3} p_i \frac{\partial A}{\partial p_i} - \left(\sum_{i=1}^{3}\Delta_i - 2d - N\right)A = 0, \tag{3.2.10}$$

with $N = 2$, the number of momenta that the form factor $A$ multiply in the decomposition.
The invariance of the correlator with respect to the special conformal transformations is encoded in the special conformal Ward identities

$$\begin{aligned}
0 = &\sum_{j=1}^{2}\left[-2\frac{\partial}{\partial p_{j\kappa}} - 2p_j^\alpha \frac{\partial^2}{\partial p_j^\alpha \partial p_{j\kappa}} + p_j^\kappa \frac{\partial^2}{\partial p_j^\alpha \partial p_{j\alpha}}\right]\langle J^{\mu_1}(p_1)J^{\mu_2}(p_2)O(p_3)\rangle_{odd} \\
&+ 2\left(\delta^{\mu_1\kappa}\frac{\partial}{\partial p_1^{\alpha_1}} - \delta^\kappa_{\alpha_1}\frac{\partial}{\partial p_{1\mu_1}}\right)\langle J^{\alpha_1}(p_1)J^{\mu_2}(p_2)O(p_3)\rangle_{odd} \\
&+ 2\left(\delta^{\mu_2\kappa}\frac{\partial}{\partial p_2^{\alpha_2}} - \delta^\kappa_{\alpha_2}\frac{\partial}{\partial p_{2\mu_2}}\right)\langle J^{\mu_1}(p_1)J^{\alpha_2}(p_2)O(p_3)\rangle_{odd} \equiv \mathcal{K}^\kappa \langle J^{\mu_1}(p_1)J^{\mu_2}(p_2)O(p_3)\rangle_{odd}.
\end{aligned} \tag{3.2.11}$$

The special conformal operator $\mathcal{K}^\kappa$ acts as an endomorphism on the transverse sector of the entire correlator. We can perform a transverse projection on all the indices in order to identify a set of partial differential equations

$$\pi^{\lambda_1}_{\mu_1}(p_1)\pi^{\lambda_2}_{\mu_2}(p_2)\left(\mathcal{K}^\kappa \langle J^{\mu_1}(p_1)J^{\mu_2}(p_2)O(p_3)\rangle\right) = 0. \tag{3.2.12}$$

We then decompose the action of the special conformal operator on the transverse sector in the following way

$$
\begin{aligned}
0 =&\pi_{\mu_1}^{\lambda_1}(p_1)\pi_{\mu_2}^{\lambda_2}(p_2)\mathcal{K}^k\langle J^{\mu_1}(p_1)J^{\mu_2}(p_2)O(p_3)\rangle = \pi_{\mu_1}^{\lambda_1}(p_1)\pi_{\mu_2}^{\lambda_2}(p_2)\Big[\\
&C_{11}\varepsilon^{p_1p_2\mu_1\mu_2}p_1^\kappa + C_{21}\varepsilon^{p_1p_2\mu_1\mu_2}p_2^\kappa + C_{31}\varepsilon^{p_1\kappa\mu_1\mu_2} + C_{32}\varepsilon^{p_2\kappa\mu_1\mu_2} + C_{33}p_2^{\mu_1}\varepsilon^{p_1p_2\kappa\mu_2} + C_{34}p_3^{\mu_2}\varepsilon^{p_1p_2\kappa\mu_1}\Big],
\end{aligned}
\tag{3.2.13}
$$

where $C_{ij}$ are scalar functions of the form factor $A$ and its derivatives. The tensor structures we have written are not all independent and can be simplified in order to find the minimal decomposition, using the following Schouten identities

$$
\begin{aligned}
&\varepsilon^{[p_1p_2\mu_1\mu_2}p_1^{\kappa]} = 0,\\
&\varepsilon^{[p_1p_2\mu_1\mu_2}p_2^{\kappa]} = 0,
\end{aligned}
\tag{3.2.14}
$$

according to which we can eliminate $C_{33}$ and $C_{34}$

$$
\begin{aligned}
&\pi_{\mu_1}^{\lambda_1}(p_1)\pi_{\mu_2}^{\lambda_2}(p_2)\left(\varepsilon^{p_1p_2\kappa\mu_1}p_3^{\mu_2}\right) = \pi_{\mu_1}^{\lambda_1}(p_1)\pi_{\mu_2}^{\lambda_2}(p_2)\left(\frac{1}{2}\varepsilon^{p_1\kappa\mu_1\mu_2}(p_1^2+p_2^2-p_3^2) + \varepsilon^{p_1p_2\mu_1\mu_2}p_1^\kappa + \varepsilon^{p_2\kappa\mu_1\mu_2}p_1^2\right)\\
&\pi_{\mu_1}^{\lambda_1}(p_1)\pi_{\mu_2}^{\lambda_2}(p_2)\left(\varepsilon^{p_1p_2\kappa\mu_2}p_2^{\mu_1}\right) = \pi_{\mu_1}^{\lambda_1}(p_1)\pi_{\mu_2}^{\lambda_2}(p_2)\left(-\frac{1}{2}\varepsilon^{p_2\kappa\mu_1\mu_2}(p_1^2+p_2^2-p_3^2) + \varepsilon^{p_1p_2\mu_1\mu_2}p_2^\kappa - \varepsilon^{p_1\kappa\mu_1\mu_2}p_2^2\right).
\end{aligned}
\tag{3.2.15}
$$

Therefore we can rewrite eq. (3.2.13) in the minimal form

$$
\begin{aligned}
0 = &\pi_{\mu_1}^{\lambda_1}(p_1)\pi_{\mu_2}^{\lambda_2}(p_2)\mathcal{K}^k\langle J^{\mu_1}(p_1)J^{\mu_2}(p_2)O(p_3)\rangle =\\
&\pi_{\mu_1}^{\lambda_1}(p_1)\pi_{\mu_2}^{\lambda_2}(p_2)\Big[C_{11}\varepsilon^{p_1p_2\mu_1\mu_2}p_1^\kappa + C_{21}\varepsilon^{p_1p_2\mu_1\mu_2}p_2^\kappa + C_{31}\varepsilon^{p_1\kappa\mu_1\mu_2} + C_{32}\varepsilon^{p_2\kappa\mu_1\mu_2}\Big],
\end{aligned}
\tag{3.2.16}
$$

where we have redefined the form factors $C_{ij}$ in order to include the contribution of the old $C_{33}$ and $C_{34}$. Due to the independence of the tensor structures listed above, the special conformal equations can be written as

$$
C_{ij} = 0. \tag{3.2.17}
$$

In particular, $C_{11} = 0$ and $C_{21} = 0$ are equations of the second order and therefore they are called primary equations [29]. All the others are first order differential equations and are called secondary equations. The explicit form of the primary equations is

$$
\begin{aligned}
&K_{31}A = 0,\\
&K_{32}A = 0,
\end{aligned}
\tag{3.2.18}
$$

where we have defined

$$
K_i = \frac{\partial^2}{\partial p_i^2} + \frac{(d+1-2\Delta_i)}{p_i}\frac{\partial}{\partial p_i}, \qquad K_{ij} = K_i - K_j. \tag{3.2.19}
$$

The secondary equations are

$$
\begin{aligned}
&0 = p_2\frac{\partial A}{\partial p_2} + (d-1-\Delta_2)A,\\
&0 = p_1\frac{\partial A}{\partial p_1} + (d-1-\Delta_1)A.
\end{aligned}
\tag{3.2.20}
$$

### 3.2.2 Solution of the CWIs

The most general solution of the conformal Ward identities of the $\langle JJO\rangle$ can be written in terms of integrals involving a product of three Bessel functions, namely 3K integrals. For a review on the properties of such integrals, see appendix A and [25, 29, 80]. We recall the definition of the general 3K integral

$$I_{\alpha\{\beta_1\beta_2\beta_3\}}(p_1,p_2,p_3)=\int dx x^{\alpha}\prod_{j=1}^{3}p_j^{\beta_j}K_{\beta_j}\left(p_jx\right), \tag{3.2.21}$$

where $K_\nu$ is a modified Bessel function of the second kind

$$K_\nu(x)=\frac{\pi}{2}\frac{I_{-\nu}(x)-I_\nu(x)}{\sin(\nu\pi)},\qquad \nu\notin\mathbb{Z}\qquad\qquad I_\nu(x)=\left(\frac{x}{2}\right)^{\nu}\sum_{k=0}^{\infty}\frac{1}{\Gamma(k+1)\Gamma(\nu+1+k)}\left(\frac{x}{2}\right)^{2k} \tag{3.2.22}$$

with the property

$$K_n(x)=\lim_{\epsilon\to0}K_{n+\epsilon}(x),\quad n\in\mathbb{Z} \tag{3.2.23}$$

The triple-K integral depends on four parameters: the power $\alpha$ of the integration variable $x$, and the three Bessel function indices $\beta_j$ . The arguments of the 3K integral are magnitudes of momenta $p_j$ with $j=1,2,3$. We will also use the reduced version of the 3K integral defined as

$$J_{N\{k_j\}}=I_{\frac{d}{2}-1+N\{\Delta_j-\frac{d}{2}+k_j\}}, \tag{3.2.24}$$

where we introduced the condensed notation $\{k_j\}=\{k_1,k_2,k_3\}$. The 3K integral satisfies an equation analogous to the dilatation equation with scaling degree

$$\deg\left(J_{N\{k_j\}}\right)=\Delta_t+k_t-2d-N, \tag{3.2.25}$$

where

$$k_t=k_1+k_2+k_3,\qquad\qquad \Delta_t=\Delta_1+\Delta_2+\Delta_3. \tag{3.2.26}$$

From this analysis, it is simple to relate the form factors to the 3K integrals. Indeed, the dilatation Ward identities tell us that the form factor $A$ needs to be written as a combination of integrals of the following type

$$J_{N+k_t,\{k_1,k_2,k_3\}}, \tag{3.2.27}$$

with $N=2$, the number of momenta that the form factor multiplies in the decomposition (3.2.6). The special conformal Ward identities fix the remaining indices $k_1$, $k_2$ and $k_3$. Indeed, recalling the following property of the 3K integrals

$$K_{nm}J_{N\{k_j\}}=-2k_nJ_{N+1\{k_j-\delta_{jn}\}}+2k_mJ_{N+1\{k_j-\delta_{jm}\}}, \tag{3.2.28}$$

we can write the most general solution of the primary equations (3.2.18) as

$$A=c_1J_{2\{0,0,0\}}, \tag{3.2.29}$$

where $c_1$ is an arbitrary constant. Before moving on to the secondary equations, we need to discuss the possible divergences in 3K integrals which, in this case, can occur for some specific values of $\Delta_3$. In general, it can be shown that the 3K integral $I_{\alpha\{\beta_1,\beta_2,\beta_3\}}$ diverges if

$$\alpha+1\pm\beta_1\pm\beta_2\pm\beta_3=-2k\quad,\quad k=0,1,2,\dots \tag{3.2.30}$$

For a more detailed review of the topic, see appendix A and [25, 29, 80]. If the above condition is satisfied, we need to regularize the integrals

$$d \to 4 + 2u\epsilon \qquad \Delta_i \to \Delta_i + (u + v_i)\epsilon. \tag{3.2.31}$$

In general, the regularisation parameters $u$ and $v_i$ are arbitrary. Here for simplicity we will choose the same $v_i = v$ for each operator. If a 3K integral in our solution diverges, we can expand the coefficient in front of such integral in the solution in powers of $\epsilon$

$$c_i = \sum_{j=-\infty}^{\infty} c_i^{(j)} \epsilon^j, \tag{3.2.32}$$

and then we can require that our entire solution is finite for $\epsilon \to 0$ by constraining the coefficients $c_i^{(j)}$. Looking at our solution we can see that $J_{2\{0,0,0\}} \equiv I_{3+u\epsilon\{1+v\epsilon,1+v\epsilon,-2+\Delta_3+v\epsilon\}}$ diverges for $\epsilon \to 0$ when

$$\Delta_3 = 0, 4, 6, 8, 10, \ldots \tag{3.2.33}$$

Let us now look at the secondary conformal equations

$$\begin{aligned} 0 &= p_2 \frac{\partial A}{\partial p_2} - (\Delta_2 - d + 1)A \\ 0 &= p_1 \frac{\partial A}{\partial p_1} - (\Delta_1 - d + 1)A. \end{aligned} \tag{3.2.34}$$

We can solve such equation by performing the limit $p_i \to 0$ for various values of $\Delta_3$ (see appendix A for a review of the procedure). If the eq. (3.2.33) is not satisfied and the 3K integral is finite, the secondary equations lead to the condition $c_1 = \mathcal{O}(\epsilon)$ and therefore the entire correlator vanishes. However, in the case $\Delta_3 = 4$, the secondary equations still lead to $c_1 = c_1^{(1)}\epsilon + \mathcal{O}(\epsilon^2)$ but since the following 3K integral has a pole [29, 80, 81]

$$I_{3+u\epsilon\{1+v\epsilon,1+v\epsilon,2+v\epsilon\}} = \frac{2}{(u-3v)\epsilon} + \mathcal{O}(\epsilon^0), \tag{3.2.35}$$

the $\epsilon$ in the solution cancels out and we end up with a finite nonzero solution. The coefficient $c_1^{(1)}$ remains unconstrained. A similar story occurs when $\Delta_3 = 0$ since

$$I_{3+u\epsilon\{1+v\epsilon,1+v\epsilon,-2+v\epsilon\}} = \frac{2}{(u-v)\,p_3^4\,\epsilon} + \mathcal{O}(\epsilon^0). \tag{3.2.36}$$

Note that such value of $\Delta_3$ is not physical and violates unitarity bounds. However, sometimes this may not constitute a problem if, for example, we are working in a holographic contest. One can check that for $\Delta_3 = 6, 8, 10\ldots$ and so on, the correlator vanishes. Indeed, in this cases the secondary equations require the coefficient $c_1$ to scale with high powers of $\epsilon$ that can not be compensated by the poles of the 3K integral $I_{3+u\epsilon\{1+v\epsilon,1+v\epsilon,-2+\Delta_3+v\epsilon\}}$. In the end we have

$$\begin{aligned} \left\langle J^{\mu_1}(p_1) J^{\mu_2}(p_2) O_{(\Delta_3\neq 0,4)}(p_3)\right\rangle_{odd} &= 0 \\ \left\langle J^{\mu_1}(p_1) J^{\mu_2}(p_2) O_{(\Delta_3=4)}(p_3)\right\rangle_{odd} &= c_1^{(1)} \varepsilon^{p_1 p_2 \mu_1 \mu_2} \\ \left\langle J^{\mu_1}(p_1) J^{\mu_2}(p_2) O_{(\Delta_3=0)}(p_3)\right\rangle_{odd} &= \frac{c_1^{(1)}}{p_3^4} \varepsilon^{p_1 p_2 \mu_1 \mu_2}, \end{aligned} \tag{3.2.37}$$

where we have absorbed a factor $2/(u-3v)$ or $2/(u-v)$ in the constant $c_1^{(1)}$. Therefore, excluding the unphysical case $\Delta_3 = 0$, the only other case where the correlator does not vanish is $\Delta_3 = 4$, which is, for example, the dimension of the scalar operators $O = \nabla_\mu J_A^\mu$ and $O = T_\mu^\mu$. Furthermore, it is important to note that the most general solution that we have found for the $\langle JJO\rangle_{odd}$ with $\Delta_3 = 4$ can be written in terms of functional derivatives $F\tilde{F}$

$$(2\pi)^4\delta^4(p_1+p_2+p_3)\Big\langle J^{\mu_1}(p_1)J^{\mu_2}(p_2)O_{(\Delta_3=4)}(p_3)\Big\rangle_{odd} = \int dx_1\,dx_2\,dx_3\,e^{-i(p_1x_1+p_2x_2+p_3x_3)}\,\frac{\delta^2\Big[c_1\,\varepsilon^{\mu\nu\rho\sigma}F_{\mu\nu}(x_3)F_{\rho\sigma}(x_3)\Big]}{\delta A_{\mu_1}(x_1)\,\delta A_{\mu_2}(x_2)}, \tag{3.2.38}$$

in accord with the chiral anomaly formula (1.5.1) for $O_{(\Delta_3=4)} = \nabla\cdot J_A$, and potentially a parity-odd trace anomaly $F\tilde{F}$ for the case $O_{(\Delta_3=4)} = T_\mu^\mu$.

## 3.3 The $\langle TJO\rangle_{odd}$ correlator in CFT

In this section we study the conformal constraints on the $\langle TJO\rangle_{odd}$ correlator where $T$ is the energy momentum tensor, $J$ can be a conserved vector current $J_V$ or an anomalous axial-vector current $J_A$ and $O$ is a scalar or pseudoscalar operator.
It is quite immediate to realize that the correlator cannot exhibit any anomaly content and its longitudinal and trace part vanishes

$$\begin{aligned}
0 &= \delta_{\mu_1\nu_1}\,\langle T^{\mu_1\nu_1}(p_1)J^{\mu_2}(p_2)O(p_3)\rangle_{odd}, \\
0 &= p_{1\,\mu_1}\,\langle T^{\mu_1\nu_1}(p_1)J^{\mu_2}(p_2)O(p_3)\rangle_{odd}, \\
0 &= p_{2\,\mu_2}\,\langle T^{\mu_1\nu_1}(p_1)J^{\mu_2}(p_2)O(p_3)\rangle_{odd}.
\end{aligned} \tag{3.3.1}$$

Indeed, the absence of a mixed anomaly in this correlator - specifically, the lack of contractions involving the Riemann or Weyl tensors together with the field strength of the Abelian current $J$ at scaling dimension four - will be crucial for determining this correlation function. As we are going to find out, the result of this procedure in the parity-odd sector of a generic 3-point function, is directly linked with the presence or absence of parity-odd anomalies.
We start by decomposing the energy-momentum tensor $T^{\mu\nu}$ and the current $J^\mu$ in terms of their transverse-traceless part and longitudinal-trace ones (also called "local")

$$T^{\mu_i\nu_i}(p_i) = t^{\mu_i\nu_i}(p_i) + t_{loc}^{\mu_i\nu_i}(p_i), \tag{3.3.2}$$

$$J^{\mu_i}(p_i) = j^{\mu_i}(p_i) + j_{loc}^{\mu_i}(p_i), \tag{3.3.3}$$

where

$$\begin{aligned}
t^{\mu_i\nu_i}(p_i) &= \Pi^{\mu_i\nu_i}_{\alpha_i\beta_i}(p_i)\,T^{\alpha_i\beta_i}(p_i), &\qquad t_{loc}^{\mu_i\nu_i}(p_i) &= \Sigma^{\mu_i\nu_i}_{\alpha_i\beta_i}(p)\,T^{\alpha_i\beta_i}(p_i), \\
j^{\mu_i}(p_i) &= \pi^{\mu_i}_{\alpha_i}(p_i)\,J^{\alpha_i}(p_i), &\qquad j_{loc}^{\mu_i}(p_i) &= \frac{p_i^{\mu_i}\,p_{i\,\alpha_i}}{p_i^2}\,J^{\alpha_i}(p_i),
\end{aligned} \tag{3.3.4}$$

having introduced the transverse-traceless ($\Pi$), transverse ($\pi$) and longitudinal-trace ($\Sigma$) projectors, given respectively by

$$\pi^{\mu}_{\alpha} = \delta^{\mu}_{\alpha} - \frac{p^{\mu}p_{\alpha}}{p^2}, \tag{3.3.5}$$

$$\Pi^{\mu\nu}_{\alpha\beta} = \frac{1}{2}\left(\pi^{\mu}_{\alpha}\pi^{\nu}_{\beta} + \pi^{\mu}_{\beta}\pi^{\nu}_{\alpha}\right) - \frac{1}{d-1}\pi^{\mu\nu}\pi_{\alpha\beta}, \tag{3.3.6}$$

$$\Sigma^{\mu_i\nu_i}_{\alpha_i\beta_i} = \frac{p_{i\beta_i}}{p_i^2}\left[2\delta^{(\nu_i}_{\alpha_i}p_i^{\mu_i)} - \frac{p_{i\alpha_i}}{(d-1)}\left(\delta^{\mu_i\nu_i} + (d-2)\frac{p_i^{\mu_i}p_i^{\nu_i}}{p_i^2}\right)\right] + \frac{\pi^{\mu_i\nu_i}(p_i)}{(d-1)}\delta_{\alpha_i\beta_i} \tag{3.3.7}$$

Because of eq. (3.3.1), the correlator is composed only of a transverse-traceless part which can be written as

$$\langle T^{\mu_1\nu_1}(p_1)J^{\mu_2}(p_2)O(p_3)\rangle_{odd} = \Pi^{\mu_1\nu_1}_{\alpha_1\beta_1}(p_1)\,\pi^{\mu_2}_{\alpha_2}(p_2)\,X^{\alpha_1\beta_1\alpha_2}. \tag{3.3.8}$$

Here, $X^{\alpha_1\beta_1\alpha_2}$ is a parity-odd tensor that can be expressed in its most general form as

$$X^{\alpha_1\beta_1\alpha_2} = A(p_1,p_2,p_3)\varepsilon^{p_1p_2\alpha_1\alpha_2}p_2^{\beta_1} \tag{3.3.9}$$

where $A(p_1,p_2,p_3)$ is an arbitrary form factor.

### 3.3.1 Dilatation and special conformal Ward identities

We denote by $\Delta_i$ the conformal dimension of the operators in our correlator. Specifically, since we are working in four-dimensional space-time, the conformal dimensions of the energy-momentum tensor and a conserved/axial current are, respectively,

$$\Delta_1 = 4, \qquad \Delta_2 = 3. \tag{3.3.10}$$

The invariance of the correlator under dilatation in momentum-space is reflected in the equation

$$\left(\sum_{i=1}^{3}\Delta_i - 2d - \sum_{i=1}^{2}p_i^{\mu}\frac{\partial}{\partial p_i^{\mu}}\right)\langle T^{\mu_1\nu_1}(p_1)J^{\mu_2}(p_2)O(p_3)\rangle_{odd} = 0. \tag{3.3.11}$$

By using the chain rule

$$\frac{\partial}{\partial p_i^{\mu}} = \sum_{j=1}^{3}\frac{\partial p_j}{\partial p_i^{\mu}}\frac{\partial}{\partial p_j} \tag{3.3.12}$$

we can then express the derivatives with respect to 4-vectors in term of the invariants $p_i = |\sqrt{p_i^2}|$. Furthermore, using eq. (3.3.8), (3.3.9) and (3.3.10), we can rewrite the dilatations equations as a constraint on the form factor

$$\sum_{i=1}^{3}p_i\frac{\partial A}{\partial p_i}(p_1,p_2,p_3) + (4-\Delta_3)A(p_1,p_2,p_3) = 0. \tag{3.3.13}$$

On the other hand, the invariance of the correlator under special conformal transformations is encoded in the following equation

$$\begin{aligned}
0 &= \mathcal{K}^\kappa \langle T^{\mu_1\nu_1}(p_1)J^{\mu_2}(p_2)O(p_3)\rangle_{odd} \\
&\equiv \sum_{j=1}^{2}\left(2\left(\Delta_j-d\right)\frac{\partial}{\partial p_{j\kappa}}-2p_j^\alpha\frac{\partial}{\partial p_j^\alpha}\frac{\partial}{\partial p_{j\kappa}}+\left(p_j\right)^\kappa\frac{\partial}{\partial p_j^\alpha}\frac{\partial}{\partial p_{j\alpha}}\right)\langle T^{\mu_1\nu_1}(p_1)J^{\mu_2}(p_2)O(p_3)\rangle_{odd} \\
&+4\left(\delta^{\kappa(\mu_1}\frac{\partial}{\partial p_1^{\alpha_1}}-\delta^\kappa_{\alpha_1}\delta^{(\mu_1}_\lambda\frac{\partial}{\partial p_{1\lambda}}\right)\langle T^{\nu_1)\alpha_1}(p_1)J^{\mu_2}(p_2)O(p_3)\rangle_{odd} \\
&+2\left(\delta^{\kappa\mu_2}\frac{\partial}{\partial p_2^{\alpha_2}}-\delta^\kappa_{\alpha_2}\delta^{\mu_2}_\lambda\frac{\partial}{\partial p_{2\lambda}}\right)\langle T^{\mu_1\nu_1}(p_1)J^{\alpha_2}(p_2)O(p_3)\rangle_{odd}\,.
\end{aligned} \tag{3.3.14}$$

We then perform a transverse projection on all the indices in order to identify a set of partial differential equations

$$0=\Pi^{\rho_1\sigma_1}_{\mu_1\nu_1}(p_1)\pi^{\rho_2}_{\mu_2}(p_2)\mathcal{K}^k\langle T^{\mu_1\nu_1}(p_1)J^{\mu_2}(p_2)O(p_3)\rangle_{odd} \tag{3.3.15}$$

and decompose the action of the special conformal operator on the correlator in the following way

$$\begin{aligned}
0 &= \Pi^{\rho_1\sigma_1}_{\mu_1\nu_1}(p_1)\pi^{\rho_2}_{\mu_2}(p_2)\mathcal{K}^k\langle T^{\mu_1\nu_1}(p_1)J^{\mu_2}(p_2)O(p_3)\rangle_{odd} \\
&= \Pi^{\rho_1\sigma_1}_{\mu_1\nu_1}(p_1)\pi^{\rho_2}_{\mu_2}(p_2)\Big[C_1\varepsilon^{p_1p_2\mu_1\mu_2}p_2^{\nu_1}p_1^\kappa+C_2\varepsilon^{p_1p_2\mu_1\mu_2}p_2^{\nu_1}p_2^\kappa+C_3\varepsilon^{p_1\kappa\mu_1\mu_2}p_2^{\nu_1}+C_4\varepsilon^{p_2\kappa\mu_1\mu_2}p_2^{\nu_1} \\
&\qquad +C_5\varepsilon^{p_1p_2\kappa\mu_2}p_2^{\mu_1}p_2^{\nu_1}+C_6\varepsilon^{p_1p_2\kappa\mu_1}p_3^{\mu_2}p_2^{\nu_1}+C_7\varepsilon^{p_1p_2\kappa\mu_1}\delta^{\mu_2\nu_1}+C_8\varepsilon^{p_1p_2\mu_1\mu_2}\delta^{\kappa\nu_1}\Big]
\end{aligned} \tag{3.3.16}$$

where $C_i$ are scalar function that depend on the form factor $A$ and its derivatives with respect to the momenta. The tensor structures listed in the equation above are not all independent and can be simplified in order to find a minimal decomposition, using the following Schouten identities

$$\begin{aligned}
0 &= \varepsilon^{[p_1p_2\mu_1\mu_2}p_1^{\kappa]}, \\
0 &= \varepsilon^{[p_1p_2\mu_1\mu_2}p_2^{\kappa]}, \\
0 &= \varepsilon^{[p_1p_2\mu_1\mu_2}\delta^{\kappa]\nu_1}.
\end{aligned} \tag{3.3.17}$$

where the square brackets indicate the antisymmetrization with respect to the enclosed indices. After applying the transverse-traceless projectors to these identities, we can use them to eliminate the tensorial structures corresponding to the form factors $C_3$, $C_5$ and $C_6$

$$\begin{aligned}
&\Pi^{\rho_1\sigma_1}_{\mu_1\nu_1}(p_1)\pi^{\rho_2}_{\mu_2}(p_2)\left(\varepsilon^{p_1p_2\kappa\mu_1}p_3^{\mu_2}\right)=\Pi^{\rho_1\sigma_1}_{\mu_1\nu_1}(p_1)\pi^{\rho_2}_{\mu_2}(p_2)\left(\frac{1}{2}\varepsilon^{p_1\kappa\mu_1\mu_2}(p_1^2+p_2^2-p_3^2)+\varepsilon^{p_1p_2\mu_1\mu_2}p_1^\kappa+\varepsilon^{p_2\kappa\mu_1\mu_2}p_1^2\right), \\
&\Pi^{\rho_1\sigma_1}_{\mu_1\nu_1}(p_1)\pi^{\rho_2}_{\mu_2}(p_2)\left(\varepsilon^{p_1p_2\kappa\mu_2}p_2^{\mu_1}\right)=\Pi^{\rho_1\sigma_1}_{\mu_1\nu_1}(p_1)\pi^{\rho_2}_{\mu_2}(p_2)\left(-\frac{1}{2}\varepsilon^{p_2\kappa\mu_1\mu_2}(p_1^2+p_2^2-p_3^2)+\varepsilon^{p_1p_2\mu_1\mu_2}p_2^\kappa-\varepsilon^{p_1\kappa\mu_1\mu_2}p_2^2\right), \\
&\Pi^{\rho_1\sigma_1}_{\mu_1\nu_1}(p_1)\pi^{\rho_2}_{\mu_2}(p_2)\left(\varepsilon^{\mu_1\mu_2\kappa p_1}p_2^{\nu_1}\right)=\Pi^{\rho_1\sigma_1}_{\mu_1\nu_1}(p_1)\pi^{\rho_2}_{\mu_2}(p_2)\left(-\varepsilon^{\kappa p_1p_2\mu_1}\delta^{\mu_2\nu_1}-\varepsilon^{p_1p_2\mu_1\mu_2}\delta^{\kappa\nu_1}\right).
\end{aligned} \tag{3.3.18}$$

Therefore, we can rewrite eq. (3.3.16) in the minimal form

$$\begin{aligned}
0 &= \Pi^{\rho_1\sigma_1}_{\mu_1\nu_1}(p_1)\pi^{\rho_2}_{\mu_2}(p_2)\mathcal{K}^k\langle T^{\mu_1\nu_1}(p_1)J^{\mu_2}(p_2)O(p_3)\rangle_{odd}=\Pi^{\rho_1\sigma_1}_{\mu_1\nu_1}(p_1)\pi^{\rho_2}_{\mu_2}(p_2)\Big[C_1\varepsilon^{\mu_1\mu_2p_1p_2}p_2^{\nu_1}p_1^\kappa \\
&\qquad +C_2\varepsilon^{\mu_1\mu_2p_1p_2}p_2^{\nu_1}p_2^\kappa+C_3\varepsilon^{p_2\kappa\mu_1\mu_2}p_2^{\nu_1}+C_4\varepsilon^{p_1p_2\kappa\mu_1}\delta^{\mu_2\nu_1}+C_5\varepsilon^{p_1p_2\mu_1\mu_2}\delta^{\kappa\nu_1}\Big]
\end{aligned} \tag{3.3.19}$$

where we have redefined the function $C_i$. Due to the independence of the tensorial structures listed in the equation above, now all the coefficients $C_i$ need to vanish

$$C_i = 0. \tag{3.3.20}$$

In particular, $C_1 = 0$ and $C_2 = 0$ are differential equation of the second order, called primary equations. Their explicit form is given by

$$\begin{aligned} K_{31} A &= 0, \\ K_{32} A &= 0 \end{aligned} \tag{3.3.21}$$

The remaining special CWIs ($C_i = 0$ with $i = \{3,4,5\}$) are differential equations of the first order, i.e. the secondary equations. Their explicit expressions are given by

$$\begin{aligned} 0 &= A - p_1 \frac{\partial A}{\partial p_1}, \\ 0 &= \frac{\partial A}{\partial p_2}, \\ 0 &= -2\,\frac{p_1^2 - 2p_2^2 + 2p_3^2}{p_1^2} A - \frac{p_1^2 + p_2^2 - p_3^2}{p_1} \frac{\partial A}{\partial p_1} - 4p_2 \frac{\partial A}{\partial p_2}. \end{aligned} \tag{3.3.22}$$

### 3.3.2 Solution of the CWIs

The solution of the conformal Ward identities for the $\langle TJO\rangle_{odd}$ can be written in terms of integrals involving a product of three Bessel functions, namely 3K integrals [29, 80], as illustrated Appendix A. The 3K integrals satisfy an equation analogous to the dilatation equation with scaling degree [29]

$$\deg\left(J_{N\{k_1,k_2,k_3\}}\right) = \Delta_t + k_t - 2d - N, \tag{3.3.23}$$

where

$$k_t = k_1 + k_2 + k_3, \qquad \Delta_t = \Delta_1 + \Delta_2 + \Delta_3. \tag{3.3.24}$$

From this analysis, it is simple to relate the form factor $A$ to the 3K integrals. Indeed, the dilatation WI (3.3.13) tells us that the form factor $A$ can to be written as a combination of integrals of the following type

$$J_{3+k_t,\{k_1,k_2,k_3\}}. \tag{3.3.25}$$

The special CWIs fix the remaining indices $k_1$, $k_2$ and $k_3$. Recalling the following property of the 3K integrals

$$K_{nm} J_{N\{k_j\}} = -2k_n J_{N+1\{k_j-\delta_{jn}\}} + 2k_m J_{N+1\{k_j-\delta_{jm}\}}, \tag{3.3.26}$$

we can write the most general solution of the primary equations (3.3.21) as

$$A = c_1 J_{3\{0,0,0\}} \equiv c_1 I_{4\{2,1,\Delta_3-2\}}, \tag{3.3.27}$$

where $c_1$ is an arbitrary constant. When dealing with such a 3K integral, one needs to be careful as $J_{3\{0,0,0\}}$ may diverge. Depending on the value of $\Delta_3$, a regularization may be necessary. In general, it can be shown that the 3K integral $I_{\alpha\{\beta_1,\beta_2,\beta_3\}}$ diverges if

$$\alpha + 1 \pm \beta_1 \pm \beta_2 \pm \beta_3 = -2k \quad , \quad k = 0, 1, 2, \ldots \tag{3.3.28}$$

For a more detailed review of the topic, see appendix A and [25, 29, 80]. If the above condition is satisfied, we need to regularize the integral. This can be done by shifting the parameters of the 3K integrals as

$$I_{\alpha\{\beta_1,\beta_2,\beta_3\}} \to I_{\alpha+u\epsilon\{\beta_1+v_1\epsilon,\beta_2+v_2\epsilon,\beta_3+v_3\epsilon\}}, \qquad J_{N\{k_1,k_2,k_3\}} \to J_{N+u\epsilon\{k_1+v_1\epsilon,k_2+v_2\epsilon,k_3+v_3\epsilon\}} \tag{3.3.29}$$

or equivalently

$$d \to 4+2u\epsilon \qquad \Delta_i \to \Delta_i + (u+v_i)\epsilon. \tag{3.3.30}$$

In general, the regularization parameters $u$ and $v_i$ are arbitrary. However, in certain specific cases, there can be some constraints on them.
In any case, regardless of the implementation of a regularization, after inserting our solution (3.3.27) back into the secondary equations (3.3.22), one finds $c_1 = 0$. Therefore, we have shown that the conformal constraints require the parity-odd sector of the correlator to be zero

$$\langle T^{\mu_1\nu_1}(p_1)J^{\mu_2}(p_2)O(p_3)\rangle_{odd} = 0. \tag{3.3.31}$$

Considering the fact that the parity-even sector of the $\langle TJO\rangle$ vanishes as well [29], we come to the conclusion that conformal symmetry prohibits the off-shell interaction of a graviton with a photon and a scalar/pseudoscalar, such as a dilaton/axion. In other words, a gravitational field cannot induce an axion to a spin-1 transition in the presence of conformal symmetry.

## 3.4 The $\langle TTO\rangle_{odd}$ correlator in CFT

In this section we study the conformal constraints on the $\langle TTO\rangle_{odd}$ correlator. Our result is valid for every parity-odd $\langle TTO\rangle_{odd}$ correlator constructed with standard energy-momentum tensors and/or $T_5$ (obtained by inserting a $\gamma^5$ in the standard $T$) and a scalar/pseudoscalar operator $O$. As we will see, the solution of the CWIs of the correlator can be written in terms of 3K integrals that needs a regularization. Therefore, we work in the general scheme $\{u,v\}$ where $d = 4+2u\epsilon$, with the conformal dimensions $\Delta_i$ of the operators is shifted by $(u+v_i)\epsilon$ with arbitrary $u$ and $v_i$.
We can procede in a manner similar to the previous correlators by decomposing the $\langle TTO\rangle_{odd}$ into its longitudinal and transverse-traceless parts. In this case, the conservation and trace Ward identities for the energy-momentum tensor lead to the condition[1]

$$p_{i\mu_i}\langle T^{\mu_1\nu_1}(p_1)\,T^{\mu_2\nu_2}(p_2)O(p_3)\rangle_{odd} = 0, \qquad \delta_{\mu_i\nu_i}\langle T^{\mu_1\nu_1}(p_1)\,T^{\mu_2\nu_2}(p_2)O(p_3)\rangle_{odd} = 0, \tag{3.4.1}$$

for $i = \{1,2\}$. Due to the equations (3.4.1), the longitudinal and trace part of the correlator vanishes. Therefore, the correlator only consists of a transverse-traceless part that can be formally expressed in terms of the following form factors

$$\begin{aligned}\langle T^{\mu_1\nu_1}(p_1)\,T^{\mu_2\nu_2}(p_2)O(p_3)\rangle_{odd} &= \langle t^{\mu_1\nu_1}(p_1)\,t^{\mu_2\nu_2}(p_2)O(p_3)\rangle_{odd} = \\ \Pi^{\mu_1\nu_1}_{\alpha_1\beta_1}(p_1)\Pi^{\mu_2\nu_2}_{\alpha_2\beta_2}(p_2)&\Big[A_1(p_1,p_2,p_3)\varepsilon^{\alpha_1\alpha_2p_1p_2}p_2^{\beta_1}p_3^{\beta_2} + A_2(p_1,p_2,p_3)\varepsilon^{\alpha_1\alpha_2p_1p_2}\delta^{\beta_1\beta_2}\Big].\end{aligned} \tag{3.4.2}$$

[1] Note that such equations remain valid even if we allow an parity-odd term $\varepsilon^{\mu\nu\rho\sigma}R^{\alpha\beta}{}_{\mu\nu}R_{\alpha\beta\rho\sigma}$ in the trace anomaly. Indeed if we consider the case $O = T^\mu_\mu$ and we trace again over one of the other energy-momentum tensors, we get a vanishing result. We will show this more in detail when examining the $\langle TTT\rangle_{odd}$ correlator.

### 3.4.1 Dilatation and special conformal Ward identities

We can now analyse the conformal constraints on the form factors $A_1$ and $A_2$. We proceed in a manner similar to the previous correlators. The invariance of the $\langle TTO\rangle_{odd}$ under dilatation is reflected in the following constraints on the form factors

$$\begin{aligned}&\sum_{i=1}^{3} p_i \frac{\partial A_1}{\partial p_i} - \left(\sum_{i=1}^{3}\Delta_i - 2d - 4\right) A_1 = 0,\\ &\sum_{i=1}^{3} p_i \frac{\partial A_2}{\partial p_i} - \left(\sum_{i=1}^{3}\Delta_i - 2d - 2\right) A_2 = 0.\end{aligned} \tag{3.4.3}$$

The invariance of the correlator with respect to the special conformal transformations is instead encoded in the special conformal Ward identities

$$\begin{aligned}0 = \mathcal{K}^\kappa \langle T^{\mu_1\nu_1}(p_1)T^{\mu_2\nu_2}(p_2)O(p_3)\rangle_{odd} = \\ \sum_{j=1}^{2}\left(2\left(\Delta_j - d\right)\frac{\partial}{\partial p_{j\kappa}} - 2p_j^\alpha \frac{\partial}{\partial p_j^\alpha}\frac{\partial}{\partial p_{j\kappa}} + \left(p_j\right)^\kappa \frac{\partial}{\partial p_j^\alpha}\frac{\partial}{\partial p_{j\alpha}}\right)\langle T^{\mu_1\nu_1}(p_1)T^{\mu_2\nu_2}(p_2)O(p_3)\rangle_{odd}\\ + 4\left(\delta^{\kappa(\mu_1}\frac{\partial}{\partial p_1^{\alpha_1}} - \delta^\kappa_{\alpha_1}\delta^{(\mu_1}_\lambda \frac{\partial}{\partial p_{1\lambda}}\right)\left\langle T^{\nu_1)\alpha_1}(p_1)T^{\mu_2\nu_2}(p_2)O(p_3)\right\rangle_{odd}\\ + 4\left(\delta^{\kappa(\mu_2}\frac{\partial}{\partial p_2^{\alpha_2}} - \delta^\kappa_{\alpha_2}\delta^{(\mu_2}_\lambda \frac{\partial}{\partial p_{2\lambda}}\right)\left\langle T^{\nu_2)\alpha_2}(p_2)T^{\mu_1\nu_1}(p_1)O(p_3)\right\rangle_{odd}.\end{aligned} \tag{3.4.4}$$

We can perform a transverse projection on all the indices in order to identify a set of partial differential equations, using the following minimal decomposition

$$\begin{aligned}0 =&\Pi^{\rho_1\sigma_1}_{\mu_1\nu_1}(p_1)\Pi^{\rho_2\sigma_2}_{\mu_2\nu_2}(p_2)\mathcal{K}^k\langle T^{\mu_1\nu_1}(p_1)T^{\mu_2\nu_2}(p_2)O(p_3)\rangle_{odd} = \Pi^{\rho_1\sigma_1}_{\mu_1\nu_1}(p_1)\Pi^{\rho_2\sigma_2}_{\mu_2\nu_2}(p_2)\Big[\\ &C_{11}\varepsilon^{\mu_1\mu_2p_1p_2}p_2^{\nu_1}p_3^{\nu_2}p_1^\kappa + C_{12}\varepsilon^{\mu_1\mu_2p_1p_2}\delta_3^{\nu_1\nu_2}p_1^\kappa + C_{21}\varepsilon^{\mu_1\mu_2p_1p_2}p_2^{\nu_1}p_3^{\nu_2}p_2^\kappa + C_{22}\varepsilon^{\mu_1\mu_2p_1p_2}\delta_3^{\nu_1\nu_2}p_2^\kappa\\ &+ C_{31}\varepsilon^{p_1\kappa\mu_1\mu_2}\delta^{\nu_1\nu_2} + C_{32}\varepsilon^{p_2\kappa\mu_1\mu_2}\delta^{\nu_1\nu_2} + C_{41}\varepsilon^{p_1p_2\mu_1\mu_2}p_3^{\nu_2}\delta^{\kappa\nu_1} + C_{51}\varepsilon^{p_1p_2\mu_1\mu_2}p_2^{\nu_1}\delta^{\kappa\nu_2}\Big].\end{aligned} \tag{3.4.5}$$

All the tensor structures in this formula are independent. Indeed, we have not considered the following tensors

$$\begin{aligned}&\varepsilon^{p_1\kappa\mu_1\mu_2}p_2^{\nu_1}p_3^{\nu_2}, && \varepsilon^{p_2\kappa\mu_1\mu_2}p_2^{\nu_1}p_3^{\nu_2}, && \varepsilon^{p_1p_2\kappa\mu_2}p_2^{\mu_1}p_2^{\nu_1}p_3^{\nu_2},\\ &\varepsilon^{p_1p_2\kappa\mu_1}p_3^{\mu_2}p_2^{\nu_1}p_3^{\nu_2}, && \varepsilon^{p_1p_2\kappa\mu_1}p_3^{\nu_2}\delta^{\mu_2\nu_1}, && \varepsilon^{p_1p_2\kappa\mu_2}p_2^{\nu_1}\delta^{\mu_1\nu_2},\end{aligned} \tag{3.4.6}$$

which can be rewritten in terms of the ones present in our decomposition, using the Schouten identities

$$\begin{aligned}\varepsilon^{[p_1p_2\mu_1\mu_2}p_1^{\kappa]} &= 0\\ \varepsilon^{[p_1p_2\mu_1\mu_2}p_2^{\kappa]} &= 0\\ \varepsilon^{[p_1p_2\mu_1\mu_2}\delta^{\kappa]\nu_1} &= 0\\ \varepsilon^{[p_1p_2\mu_1\mu_2}\delta^{\kappa]\nu_2} &= 0,\end{aligned} \tag{3.4.7}$$

according to which we have

$$
\begin{aligned}
&\Pi^{\rho_1\sigma_1}_{\mu_1\nu_1}(p_1)\Pi^{\rho_2\sigma_2}_{\mu_2\nu_2}(p_2)\Big(\varepsilon^{p_1p_2\kappa\mu_1}p_3^{\mu_2}\Big)=\Pi^{\rho_1\sigma_1}_{\mu_1\nu_1}(p_1)\Pi^{\rho_2\sigma_2}_{\mu_2\nu_2}(p_2)\Big(\frac{1}{2}\varepsilon^{p_1\kappa\mu_1\mu_2}(p_1^2+p_2^2-p_3^2)+\varepsilon^{p_1p_2\mu_1\mu_2}p_1^\kappa+\varepsilon^{p_2\kappa\mu_1\mu_2}p_1^2\Big)\\
&\Pi^{\rho_1\sigma_1}_{\mu_1\nu_1}(p_1)\Pi^{\rho_2\sigma_2}_{\mu_2\nu_2}(p_2)\Big(\varepsilon^{p_1p_2\kappa\mu_2}p_2^{\mu_1}\Big)=\Pi^{\rho_1\sigma_1}_{\mu_1\nu_1}(p_1)\Pi^{\rho_2\sigma_2}_{\mu_2\nu_2}(p_2)\Big(-\frac{1}{2}\varepsilon^{p_2\kappa\mu_1\mu_2}(p_1^2+p_2^2-p_3^2)+\varepsilon^{p_1p_2\mu_1\mu_2}p_2^\kappa-\varepsilon^{p_1\kappa\mu_1\mu_2}p_2^2\Big)\\
&\Pi^{\rho_1\sigma_1}_{\mu_1\nu_1}(p_1)\Pi^{\rho_2\sigma_2}_{\mu_2\nu_2}(p_2)\Big(\varepsilon^{\mu_1\mu_2\kappa p_1}p_2^{\nu_1}\Big)=\Pi^{\rho_1\sigma_1}_{\mu_1\nu_1}(p_1)\Pi^{\rho_2\sigma_2}_{\mu_2\nu_2}(p_2)\Big(-\varepsilon^{\kappa p_1p_2\mu_1}\delta^{\mu_2\nu_1}-\varepsilon^{p_1p_2\mu_1\mu_2}\delta^{\kappa\nu_1}\Big)\\
&\Pi^{\rho_1\sigma_1}_{\mu_1\nu_1}(p_1)\Pi^{\rho_2\sigma_2}_{\mu_2\nu_2}(p_2)\Big(\varepsilon^{p_2\mu_1\mu_2\kappa}p_1^{\nu_2}\Big)=\Pi^{\rho_1\sigma_1}_{\mu_1\nu_1}(p_1)\Pi^{\rho_2\sigma_2}_{\mu_2\nu_2}(p_2)\Big(-\varepsilon^{\mu_2\kappa p_1p_2}\delta^{\mu_1\nu_2}-\varepsilon^{p_1p_2\mu_1\mu_2}\delta^{\kappa\nu_2}\Big).
\end{aligned}
\tag{3.4.8}
$$

We remark that we can always change ($\mu_1\leftrightarrow\nu_1$) and/or ($\mu_2\leftrightarrow\nu_2$) in order to obtain new Schouten identities. Due to the independence of tensor structures in our decomposition, the special conformal equations are written, from (3.4.5), as

$$
C_{ij}=0\qquad i=1,\dots 4\qquad j=1,2. \tag{3.4.9}
$$

The equations with $i=\{1,2\}$ are second order differential equations called primary equations. Their explicit form is

$$
\begin{aligned}
&K_{31}A_1=0, && K_{32}A_1=0,\\
&K_{31}A_2-2p_2\frac{\partial A_1}{\partial p_2}+4A_1=0, && K_{32}A_2-2p_1\frac{\partial A_1}{\partial p_1}+4A_1=0.
\end{aligned}
\tag{3.4.10}
$$

The other equations are of the first order and are called secondary ones. Their explicit form is

$$
\begin{aligned}
&0=(p_1^2-p_2^2-p_3^2)A_1+2A_2+2p_1p_2^2\frac{\partial A_1}{\partial p_1}-p_2(p_1^2+p_2^2-p_3^2)\frac{\partial A_1}{\partial p_2}-2p_2\frac{\partial A_2}{\partial p_2},\\
&0=(p_1^2-p_2^2+p_3^2)A_1-2A_2-2p_1^2p_2\frac{\partial A_1}{\partial p_2}+p_1(p_1^2+p_2^2-p_3^2)\frac{\partial A_1}{\partial p_1}+2p_1\frac{\partial A_2}{\partial p_1},\\
&0=2\left(\frac{p_1^2+2p_2^2-2p_3^2}{p_1^2}\right)A_1+\frac{8}{p_1^2}A_2-\frac{p_1^2+p_2^2-p_3^2}{p_1}\frac{\partial A_1}{\partial p_1}-\frac{2}{p_1}\frac{\partial A_2}{\partial p_1}-4p_2\frac{\partial A_1}{\partial p_2},\\
&0=-2\left(\frac{2p_1^2+p_2^2-2p_3^2}{p_2^2}\right)A_1-\frac{8}{p_2^2}A_2+4p_1\frac{\partial A_1}{\partial p_1}+\frac{p_1^2+p_2^2-p_3^2}{p_2}\frac{\partial A_1}{\partial p_2}+\frac{2}{p_2}\frac{\partial A_2}{\partial p_2}.
\end{aligned}
\tag{3.4.11}
$$

### 3.4.2 Solutions of the CWIs

In order to solve the primary eqs. (3.4.10), we rewrite them as a set of homogeneous equations by repeatedly applying the operator $K_{ij}$ on them

$$
\begin{aligned}
&K_{31}A_1=0, && K_{32}A_1=0,\\
&K_{31}K_{31}A_2=0, && K_{32}K_{32}A_2=0.
\end{aligned}
\tag{3.4.12}
$$

The most general solution of these equations can be written in terms of the following combinations of 3K integrals

$$
\begin{aligned}
&A_1=c_1J_{4\{0,0,0\}},\\
&A_2=c_2J_{3\{1,0,0\}}+c_3J_{3\{0,1,0\}}+c_4J_{3\{0,0,1\}}+c_5J_{4\{1,1,0\}}+c_6J_{2\{0,0,0\}}.
\end{aligned}
\tag{3.4.13}
$$

We then insert these solutions back into the nonhomogeneous primary eqs. (3.4.10) and the secondary eqs. (3.4.11) in order to fix the constants $c_i$. We can solve such constraints for different values of the conformal dimensions $\Delta_3$. As seen in the computations of the previous correlators, the procedure may involve a regularization, the use of the properties of 3K integrals and their limits $p_i \to 0$ described in the Appendix A. For odd values of $\Delta_3$, no regularization is needed, and we found only vanishing solutions. We also considered different examples with even values of $\Delta_3$. In particular, we found

$$
\begin{aligned}
&\Big\langle T^{\mu_1\nu_1}(p_1)\,T^{\mu_2\nu_2}(p_2)\,O_{(\Delta_3=-2)}(p_3)\Big\rangle_{odd} = \Pi^{\mu_1\nu_1}_{\alpha_1\beta_1}(p_1)\,\Pi^{\mu_2\nu_2}_{\alpha_2\beta_2}(p_2)\times\\
&\quad\frac{c_1}{p_3^8}\bigg[\Big(3\big(p_1^2-p_2^2\big)^2-2\big(p_1^2+p_2^2\big)p_3^2-p_3^4\Big)\varepsilon^{\alpha_1\alpha_2p_1p_2}\delta^{\beta_1\beta_2}-2\big(3p_1^2+3p_2^2+p_3^2\big)\varepsilon^{\alpha_1\alpha_2p_1p_2}p_2^{\beta_1}p_3^{\beta_2}\bigg],\\
&\Big\langle T^{\mu_1\nu_1}(p_1)\,T^{\mu_2\nu_2}(p_2)\,O_{(\Delta_3=0)}(p_3)\Big\rangle_{odd} =\\
&\qquad\Pi^{\mu_1\nu_1}_{\alpha_1\beta_1}(p_1)\,\Pi^{\mu_2\nu_2}_{\alpha_2\beta_2}(p_2)\,\frac{c_1}{p_3^4}\bigg[-\frac{p_1^2+p_2^2-p_3^2}{2}\varepsilon^{\alpha_1\alpha_2p_1p_2}\delta^{\beta_1\beta_2}+\varepsilon^{\alpha_1\alpha_2p_1p_2}p_2^{\beta_1}p_3^{\beta_2}\bigg],\\
&\Big\langle T^{\mu_1\nu_1}(p_1)\,T^{\mu_2\nu_2}(p_2)\,O_{(\Delta_3=2)}(p_3)\Big\rangle_{odd} = 0,\\
&\Big\langle T^{\mu_1\nu_1}(p_1)\,T^{\mu_2\nu_2}(p_2)\,O_{(\Delta_3=4)}(p_3)\Big\rangle_{odd} =\\
&\qquad\Pi^{\mu_1\nu_1}_{\alpha_1\beta_1}(p_1)\,\Pi^{\mu_2\nu_2}_{\alpha_2\beta_2}(p_2)\,c_1\bigg[-\frac{p_1^2+p_2^2-p_3^2}{2}\varepsilon^{\alpha_1\alpha_2p_1p_2}\delta^{\beta_1\beta_2}+\varepsilon^{\alpha_1\alpha_2p_1p_2}p_2^{\beta_1}p_3^{\beta_2}\bigg],\\
&\Big\langle T^{\mu_1\nu_1}(p_1)\,T^{\mu_2\nu_2}(p_2)\,O_{(\Delta_3=6)}(p_3)\Big\rangle_{odd} = \Pi^{\mu_1\nu_1}_{\alpha_1\beta_1}(p_1)\,\Pi^{\mu_2\nu_2}_{\alpha_2\beta_2}(p_2)\times\\
&\quad c_1\bigg[\Big(3\big(p_1^2-p_2^2\big)^2-2\big(p_1^2+p_2^2\big)p_3^2-p_3^4\Big)\varepsilon^{\alpha_1\alpha_2p_1p_2}\delta^{\beta_1\beta_2}-2\big(3p_1^2+3p_2^2+p_3^2\big)\varepsilon^{\alpha_1\alpha_2p_1p_2}p_2^{\beta_1}p_3^{\beta_2}\bigg],\\
&\Big\langle T^{\mu_1\nu_1}(p_1)\,T^{\mu_2\nu_2}(p_2)\,O_{(\Delta_3=8)}(p_3)\Big\rangle_{odd} = 0.
\end{aligned}
\tag{3.4.14}
$$

Note that we have also considered examples with $\Delta_3 \leq 0$. Although these cases violate unitarity and are nonphysical, they can be relevant in particular contexts such as holography. It is important to note that the conformal equations do not require these nonphysical solutions to vanish.
We now focus on the solution with $\Delta_3 = 4$, which is satisfied for example if $O = \nabla\cdot J_A$ or $O = T^\mu_\mu$. After contracting the indices of the transverse-traceless projectors with the argument in the square brackets, we have

$$
\begin{aligned}
&\Big\langle T^{\mu_1\nu_1}(p_1)\,T^{\mu_2\nu_2}(p_2)\,O_{(\Delta_3=4)}(p_3)\Big\rangle_{odd} =\\
&\qquad\frac{c_1}{4}\bigg[\varepsilon^{\nu_1\nu_2p_1p_2}\Big((p_1\cdot p_2)\delta^{\mu_1\mu_2}-p_1^{\mu_2}p_2^{\mu_1}\Big)+(\mu_1\leftrightarrow\nu_1)+(\mu_2\leftrightarrow\nu_2)+\begin{pmatrix}\mu_1\leftrightarrow\nu_1\\ \mu_2\leftrightarrow\nu_2\end{pmatrix}\bigg].
\end{aligned}
\tag{3.4.15}
$$

This solution can be rewritten as functional derivatives of $R\tilde{R}$

$$
\begin{aligned}
&(2\pi)^4\delta^4(p_1+p_2+p_3)\Big\langle T^{\mu_1\nu_1}(p_1)\,T^{\mu_2\nu_2}(p_2)\,O_{(\Delta_3=4)}(p_3)\Big\rangle_{odd} =\\
&\qquad\int dx_1\,dx_2\,dx_3\,e^{-i(p_1x_1+p_2x_2+p_3x_3)}\,\frac{\delta^2\Big[c_1\,\varepsilon^{\mu\nu\rho\sigma}R^{\alpha\beta}{}_{\mu\nu}(x_3)R_{\alpha\beta\rho\sigma}(x_3)\Big]}{\delta g_{\mu_1\nu_1}(x_1)\;\delta g_{\mu_2\nu_2}(x_2)}
\end{aligned}
\tag{3.4.16}
$$

in agreement with the chiral anomaly formula (1.5.1) in the case $O_{(\Delta_3=4)} = \nabla\cdot J_A$ and potentially a parity-odd trace anomaly $R\tilde{R}$ for the case $O_{(\Delta_3=4)} = T^\mu_\mu$.

## 3.5 Connections with the anomalies

In this section we discuss on the nonvanishing solutions we found untill now. In particular, we will show how such solutions are related to chiral and conformal anomalies.

### 3.5.1 The $\Delta_3 = 4$ solutions

As we have seen in the previous sections, specifically in eq. (3.2.37) and (3.4.14), the $\langle JJO\rangle_{odd}$ and $\langle TTO\rangle_{odd}$ do not always vanish. Indeed, such correlators are protected by chiral and conformal anomalies. Specifically, we can justify the existence of the nonvanishing solutions with $\Delta_3 = 4$ by providing an example of a scalar operator $O$ which is linked to the chiral anomaly

$$O = \nabla_\mu \langle J_A^\mu\rangle = \mathcal{A}_{chiral} = a_1\, \varepsilon^{\mu\nu\rho\sigma} F_{\mu\nu}F_{\rho\sigma} + a_2\, \varepsilon^{\mu\nu\rho\sigma} R^\alpha_{\beta\mu\nu} R^\beta_{\alpha\rho\sigma}. \tag{3.5.1}$$

Choosing such scalar operator, then we can write

$$\begin{aligned} \langle J^{\mu_1}(x_1) J^{\mu_2}(x_2) O(x_3)\rangle_{odd} &= \frac{\delta\mathcal{A}_{chiral}(x_3)}{\delta A_{\mu_1}(x_1)\delta A_{\mu_2}(x_2)}\bigg|_{g_{\mu\nu}=\eta_{\mu\nu}, A_\mu=0}, \\ \langle T^{\mu_1\nu_1}(x_1) T^{\mu_2\nu_2}(x_2) O(x_3)\rangle_{odd} &= 4\frac{\delta\mathcal{A}_{chiral}(x_3)}{\delta g_{\mu_1\nu_1}(x_1)\delta g_{\mu_2\nu_2}(x_2)}\bigg|_{g_{\mu\nu}=\eta_{\mu\nu}, A_\mu=0} \end{aligned} \tag{3.5.2}$$

in the flat limit. By Fourier transforming such expressions, one obtains exactly the solutions in eq. (3.2.37) and (3.4.14) with $\Delta_3 = 4$. Note that $O = \nabla\cdot J_A$ is a primary operator since, by acting on it with the special conformal operator $\mathcal{K}$, we obtain a vanishing result

$$\mathcal{K}_\nu P_\mu \left|J_A^\mu\right\rangle = \left[\mathcal{K}_\nu, P_\mu\right]\left|J_A^\mu\right\rangle = 2\left(\mathcal{D}\,\delta_{\mu\nu} - M_{\nu\mu}\right)\left|J_A^\mu\right\rangle = 2(\Delta - d + 1)|J_{A\,\nu}\rangle = 0 \tag{3.5.3}$$

where $\mathcal{D}$ is the dilatation operator and in the last passage we used the fact that the conformal dimension of $J_A$ is $\Delta = d-1$.
Besides $O = \nabla\cdot J_A$, one could alternatively consider $O = g_{\mu\nu}T^{\mu\nu}$ as scalar operator with $\Delta_3 = 4$. The correlators would then be similarly protected by the possible existence of parity-odd trace anomalies in CFT

$$g_{\mu\nu}\left\langle T^{\mu\nu}\right\rangle_{odd} = f_1 \varepsilon^{\mu\nu\rho\sigma} R_{\alpha\beta\mu\nu} R^{\alpha\beta}_{\rho\sigma} + f_2 \varepsilon^{\mu\nu\rho\sigma} F_{\mu\nu}F_{\rho\sigma}. \tag{3.5.4}$$

This last hypothesis has been analyzed in depth in [35].

### 3.5.2 The shadow transforms and the $\Delta_3 \le 0$ solutions

In the previous section, we justified the nonvanishing solutions of the $\langle JJO\rangle_{odd}$ and $\langle TTO\rangle_{odd}$ correlators with $\Delta_3 = 4$ through their connection to the chiral and conformal anomalies. However, these are not the only nonvanishing solutions that we have discussed. Indeed, such correlators can also be nonzero when $\Delta_3 = 0$.
The general solution of these correlators with $\Delta_3 = 4$ and $\Delta_3 = 0$ differ by a factor $p_3^4$ . Although the condition $\Delta_3 = 0$ is nonphysical and violate unitarity, we can now ask ourself if there are any justifications for the existance of such nonvanishing solutions in CFT. As we will see, the conformal and chiral anomalies can still be responsible for such solutions. To demonstrate this, we need to introduce the concept of shadow transform, which we will briefly review [82–84].

Given a primary operator $O_{l,\Delta}$ of spin $l$ and scaling dimension $\Delta$, we can construct a *shadow* primary field in $d$ dimensions in the following way

$$\tilde{O}_{l,\bar{\Delta}}(x) = \int d^d y\, G_{l,\bar{\Delta}}(x-y) O_{l,\Delta}(y) \tag{3.5.5}$$

with spin $l$ and conjugate scaling dimension

$$\bar{\Delta} = d - \Delta. \tag{3.5.6}$$

The kernel $G_{l,\bar{\Delta}}(x-y)$ takes the form of a 2-point function, with spin $l$ and dimension $\bar{\Delta}$ operators, in a $d$-dimensional CFT. The integral defines the shadow transform. In particular, for a scalar operator we have

$$\tilde{O}_{\bar{\Delta}}(x) = \int d^d y \frac{c_{\bar{\Delta}}}{|x-y|^{2\bar{\Delta}}} O_{\Delta}(y). \tag{3.5.7}$$

The constant $c_{\bar{\Delta}}$ is a normalization factor, which we leave arbitrary for now. What is special about this particular choice of integration kernel - as opposed to say a kernel of the form $|x-y|^{-2\Delta'}$ for some generic $\Delta'$ - is that the resulting object $\tilde{O}(x)$ transforms again as a local primary field under the conformal group.
The inverse of a shadow transform is again a shadow transform. For example, the scalar relation (3.5.7) may be inverted as

$$O_{\Delta}(x) = \int d^d y \frac{\tilde{c}_{\Delta}}{|x-y|^{2\Delta}} \tilde{O}_{\bar{\Delta}}(y). \tag{3.5.8}$$

This becomes obvious in the momentum-space description. The Fourier transform of the expression (3.5.7) can be computed by using the following relation

$$\int d^d x\, x^{-2\bar{\Delta}} e^{-ik\cdot x} = \frac{\pi^{d/2}\, 2^{d-2\bar{\Delta}}\, \Gamma\left[\frac{d-2\bar{\Delta}}{2}\right]}{\Gamma[\bar{\Delta}]} k^{2\bar{\Delta}-d}. \tag{3.5.9}$$

In particular, for a suitable choice of $c_{\bar{\Delta}}$ we end up with

$$\tilde{O}_{\bar{\Delta}}(k) = k^{2\bar{\Delta}-d} O_{\Delta}(k). \tag{3.5.10}$$

Given this definition, we now reexamine the solutions of the $\langle JJO\rangle_{odd}$ and $\langle TTO\rangle_{odd}$ in eq. (3.2.37) and (3.4.14). For both correlators, the solution with $\Delta_3 = 0$ is connected to the one with $\Delta_3 = 4$ through a shadow transformation of the scalar operator $O$.
In the previous sections, we have provided examples of nonvanishing $\langle JJO\rangle_{odd}$ and $\langle TTO\rangle_{odd}$ correlators, using $\nabla \cdot J_A$ or $g_{\mu\nu}T^{\mu\nu}$ as scalar operators with $\Delta_3 = 4$. Applying a shadow transform to these specific scalar operators yields nonvanishing correlators that satisfy eqs. (3.2.37) and (3.4.14) for $\Delta_3 = 0$. This demonstrates that the chiral and conformal anomaly content of the quantum expectation values of these operators, can explain the nonvanishing nature of these other solutions too.
To conclude, it is worth noting that for the $\langle TTO\rangle_{odd}$ correlator, we have identified additional nonzero solutions, specifically for $\Delta_3 = -2$ and 6. These solutions exhibit a more complex structure compared to those with $\Delta_3 = 0$ and 4, and they do not seem to be directly related to the functional derivatives of the anomalies that we have previously discussed. It would be interesting to find a justification for the existence of these solutions as well. Notably, the solutions with $\Delta_3 = -2$ and 6 are also related to each other by a shadow transform.

## 3.6 Physical implications for chiral/gravitational backgrounds

Our analysis has been set to explore the possible implication of conformal symmetry on 3-point functions whenever scalar operators appear in mixed correlators which have a direct physical relevance. Obviously, the use of conformal symmetry limits the generality of the result, for being surely specific, and it is then natural to ask under which conditions these results can be applied.
One possibility, but surely not the only one, is encountered in early universe cosmology, when conformal symmetry is expected to have played a significant role before that any physical scale appeared in the dynamics.
Indeed, scalar and pseudoscalar fields are considered potential components of dark matter in current cosmological models, although definitive evidence for their existence has yet to be collected. Other important applications are natural inflation models [85], which involve a pseudoscalar field (the inflaton) with a periodic potential inspired by axion-like particles. Mechanisms for baryogenesis also involve scalar fields carrying baryon or lepton number. If these fields are pseudoscalars, they can exhibit CP-violating interactions that generate a baryon asymmetry [86].
Pseudoscalar fields, in general, can generate distinctive signatures if they couple to gravity, detectable through their imprints on the stochastic background of gravitational waves. In the case of axion-like fields and their coupling to gauge fields, they can induce spin-1 helicity asymmetries before any spontaneously broken phase intervenes. In this context, a correlator such as the $\langle TTO\rangle_{odd}$ for example, where $O$ is a pseudoscalar operator coupled to an axion-like field $\phi$, is directly connected with a gravitational anomaly. The local effective action describing this interaction in the infrared, once a conformal symmetry breaking scale ($\Lambda$) is present, is of the form

$$\mathcal{L}_{axion} \supset \frac{\phi}{\Lambda} R_{\mu\nu} \tilde{R}^{\mu\nu} \tag{3.6.1}$$

where $R_{\mu\nu}$ is the Ricci tensor. This correlator mediates an interaction between the pseudoscalar and two gravitational waves at semiclassical level and is part of the effective action generated by integrating out conformal matter in quantum corrections. The interaction is the analog of the the $(\phi/f)F\tilde{F}$ interaction of ordinary axion physics, which is associated with the $\langle JJO\rangle_{odd}$ ($\Delta_3 = 4$) in (3.2.37). This process, referred to as *conformal backreaction* [87], is based on the hypothesis that the universe underwent a conformal phase before any ordinary mechanism based on spontaneous symmetry breaking took place.
This assumption is supported by the fact that ordinary gauge theories in their exact phases are classically scale invariant and manifest conformal and chiral anomalies in their fermion sectors. Both the $\langle TTO\rangle_{odd}$ and the $\langle JJO\rangle_{odd}$ correlators are part of the gravitational effective action, where semiclassical corrections modify the interaction of gravity with other fields and are essential for investigating the chiral behavior in the spectrum of gravitational waves.
The $\langle JJO\rangle_{odd}$, where $J$ is a spin-1 current to gauge fields like the hypercharge $Y$ and axion-like fields, can also be linked to gravity nonperturbatively. Indeed, $O$ can represent the trace of the energy-momentum tensor, including its parity-odd anomalous contribution. This point has come to the attention of several groups in the recent literature. The apperance of this anomaly with Standard Model fermions has been attributed to the implementation of regularization procedure in the perturbative analysis of such contributions with conflicting results [21, 88–94].
In a previous analysis [35], we showed that the Conformal Ward Identities (CWIs) permit a minimal form of this interaction. However, such approach is nonperturbative. In this context, the axion/dilaton field would interact according to the form described in (3.6.1).
Another possibility is to consider $O \sim \partial \cdot J_{CS}$, representing the divergence of a Chern-Simons (CS) current. This idea was discussed in [95], where the current generates an anomaly for a spin-1 particle

[4, 5, 96]. This anomaly is associated with a CS current of the form

$$J^{\lambda}_{CS} = \varepsilon^{\lambda\mu\nu\rho} V_\mu \partial_\nu V_\rho. \tag{3.6.2}$$

The $J_{CS}$ current mediates the gravitational chiral anomaly with spin-1 virtual particles in the loops, creating an asymmetry between their two circular modes and inducing optical helicity [97].
A final comment concerns the $\langle TJO\rangle_{odd}$ that vanished identically as an off-shell correlator in a CFT, both in its even and odd components. This result can be interpreted as a constrain on the absence of an axion/dilaton to a spin-1 mixing in the presence of an external gravitational field. This can be viewed as a constrain similar to the axion to photon transition in the presence of an external magnetic field, which is the basic process for axion detection with helioscopes [98].

## 3.7 Conclusions

We have analyzed all possible 3-point functions constructed with the energy-momentum tensor, currents and at least one scalar operator in CFT. Most of the correlators are constrained to be zero, except some specific case which are protected from vanishing by quantum anomalies.
In particular, all the correlators built with at least two scalar operator vanish because of Lorentz invariance. There is no need to require invariance under the full conformal group, as it is not even possible to construct tensorial structures for them.
The analysis of the $\langle TJO\rangle_{odd}$ is more intricate and requires the full conformal group in order to determine the correlator. Specifically, we have shown that such correlator vanishes in a CFT. The even part of $\langle TJO\rangle_{odd}$ has been investigated in [29] and it was found to be zero as well.
Among all the correlators we have examined, the only ones not constrained to vanish are the $\langle JJO\rangle_{odd}$ and $\langle TTO\rangle_{odd}$. Interestingly, this effect is not visible in coordinate-space, as these correlators were found to be zero in that context [32]. However, in momentum-space, such correlators cannot always vanish since they are protected by the chiral and conformal anomalies when $\Delta_3 = 0$ and 4. For instance, one can consider as scalar operators $O = \nabla \cdot J_A$ or $O = g_{\mu\nu} T^{\mu\nu}$ and therefore the chiral/conformal anomaly prevent the correlators from vanishing when $\Delta_3 = 4$. Furthermore, the nonzero solutions we found for the $\langle JJO\rangle_{odd}$ and $\langle TTO\rangle_{odd}$ with $\Delta_3 = 0$ can be obtained by selecting the shadow transform of $\nabla \cdot J_A$ or $g_{\mu\nu} T^{\mu\nu}$ as scalar operator $O$.
Moving forward, one might wonder if there are other examples of scalar operators with $\Delta_3 = 0$ or 4 which generate nonvanishing $\langle JJO\rangle_{odd}$ and $\langle TTO\rangle_{odd}$ correlators. Moreover, it would also be interesting to determine if there is any sort of explanation for the nonvanishing solutions of the $\langle TTO\rangle_{odd}$ with $\Delta_3 = -2$ and 6. We hope to revisit this topic in the future.

# Chapter 4

# The chiral anomaly in CFT

> *"On the other hand, quite frankly, just between us friends, we won't get too upset if quantum fluctuation violates axial current conservation. Who cares if the axial charge is not constant in time?"*
>
> A. Zee, *Quantum Field Theory in a Nutshell*

In this chapter, we illustrate how the Conformal Ward Identities determine the structure of a chiral anomaly interaction, taking the example of the $\langle J_V J_V J_A\rangle$ (vector/vector/axial-vector) and $\langle J_A J_A J_A\rangle$ correlators in momentum-space. We employ a longitudinal/transverse decomposition of tensor structures and form factors. The longitudinal component is fixed by the anomaly content and the anomaly pole, while the transverse sector is determined by imposing conformal invariance. As a result, the entire correlator depends solely on the anomaly coefficient. For both the $\langle J_V J_V J_A\rangle$ and $\langle J_A J_A J_A\rangle$, our CFT result matches the one-loop perturbative expression, as expected. The CWIs for correlators of mixed chirality $\langle J_L J_V J_R\rangle$ generate solutions in agreement with the all-orders nonrenormalization theorems of perturbative QCD in the chiral limit.

## 4.1 Introduction

Parity even correlators in $d = 4$ play a central role in CFT and have been investigated in coordinate-space [66, 67] for quite some time. Parity-odd ones, instead, have attracted less attention, except for $\langle J_V J_V J_A\rangle$ and $\langle J_A J_A J_A\rangle$, due to the role of the chiral anomaly

$$\partial_\mu \langle J_A^\mu\rangle = a_1 \varepsilon^{\mu\nu\rho\sigma} F_{\mu\nu} F_{\rho\sigma} \tag{4.1.1}$$

The conformal properties of the $\langle J_V J_V J_A\rangle$ diagram, in the coordinate-space, have been studied since the 70's (see for instance [99] and [100]). In more recent years, its nonrenormalization property, which affects its anomaly (longitudinal) part, as shown by Adler and Bardeen [101], has been redrawn in a more general context, given its relevance in the analysis of $g-2$ of the muon [102]. The numerical value of $a_\mu = (g-2)_\mu = F_M(0)$, which measures the muon anomaly, is given by the Pauli form factor at zero momentum transfer, and involves a soft photon limit on one of the two vector currents of the $\langle J_V J_V J_A\rangle$ diagram, which interpolates with $F_M$ via electroweak corrections.
In such special kinematic configuration, the correlator is described by the longitudinal and transverse components $w_L$ and $w_T$ of the diagram. If we adopt such a $L/T$ separation of the vertex, $w_L$ does not receive radiative corrections, as shown in [101]. Therefore, any constraining relation involving both $w_L$ and $w_T$ is bound to identify combinations of form factors in the $T$ sector, which are protected against radiative corrections. The conditions under which such nonrenormalization constraints hold, may involve a special kinematics. For this reason, the analysis of the correlator requires moving into momentum-space.
It has been argued that in the chiral limit of QCD and with a soft $V$ line, the relation

$$w_L = 2\, w_T \tag{4.1.2}$$

holds to all orders in the strong coupling constant $\alpha_s$ [102]. This constraint is indeed reproduced by the bare fermion loop, which is conformal and is a byproduct of our analysis of the CWIs in momentum-space as well. Notice that, given the complete overlap of the CFT result with the perturbative one at the lowest order, identities such as the Crewther-Broadhurst-Kataev (CBK) relation [103–105], are exactly satisfied in our case as well (see [106]). The CBK identity connects the Adler function for electron-positron annihilation with the Bjorken function of deep-inelastic sum rules. Away from the conformal limit, these relations acquire corrections proportional to the QCD $\beta$ function at higher orders.
A perturbative analysis of the radiative corrections of the $\langle J_V J_V J_A\rangle$ correlator in QCD at two-loop level ($O(\alpha_s)$, showed that the entire vertex is also non renormalized [107], in agreement with (4.1.2). However, at higher orders in $\alpha_s$ ($O(\alpha_s^2)$), the nonrenormalization properties of the entire diagram are violated and conformality is lost. In particular (4.1.2) is only limited to the soft photon limit discussed in [102], as shown in explicit computations [108].
In our analysis, the constraint in (4.1.2) can be viewed also as a result of conformal symmetry, without resorting to a perturbative picture, since in the unique solution of the CWIs from momentum-space, such constraint is verified. While the result is expected, the procedure is novel.
One interesting aspect of the identification of the $\langle J_V J_V J_A\rangle$ or $\langle J_A J_A J_A\rangle$ diagrams from the CWIs in momentum-space, is the centrality of the anomaly pole in the longitudinal sector, that allows to identify every part of the correlator. The link between such massless state and the nonlocal anomaly effective action has been stressed in the past in several works [63, 64, 109]. More recently, these analysis have been extended in the investigation of chiral density waves, with local actions [18, 110]. These actions play a key role in very different contexts where either chiral or conformal anomalies are involved, such as in topological materials [12, 111].

### 4.1.1 Massless intermediate states in the anomaly: the pivot

The construction of the entire correlator proceeds from the anomaly pole, that plays a key role in any anomaly diagram. This takes the role of a pivot in the procedure, and it allows to solve for the longitudinal anomalous WI quite straightforwardly.
The tensorial expansion of a chiral vertex is not unique, due to the presence of Schouten relations among its tensor components. For instance, an anomaly pole in the virtuality of the axial-vector current - denoted as $1/p_3^2$ in our notations - can be inserted or removed from a given tensorial decomposition, just by the use of these relations.
These identities connect two of the most common representation of this vertex, the first of them introduced long ago by Rosenberg [112], expressed in terms of 6 tensor structures and form factors, and the second one [113], more recent, introduced in the context of the analysis of the $g-2$ anomalous magnetic moment of the muon. The latter, which is the most valuable from the physical point of view, allows to attribute the anomaly to the exchange of a pole in the longitudinal channel [64, 73, 114].
In this second parameterization of the vertex, the decomposition identifies longitudinal and transverse components

$$\langle J_V^{\mu_1}(p_1) J_V^{\mu_2}(p_2) J_A^{\mu_3}(p_3)\rangle = \frac{1}{8\pi^2}\left(W_L^{\mu_1\mu_2\mu_3} - W_T^{\mu_1\mu_2\mu_3}\right) \tag{4.1.3}$$

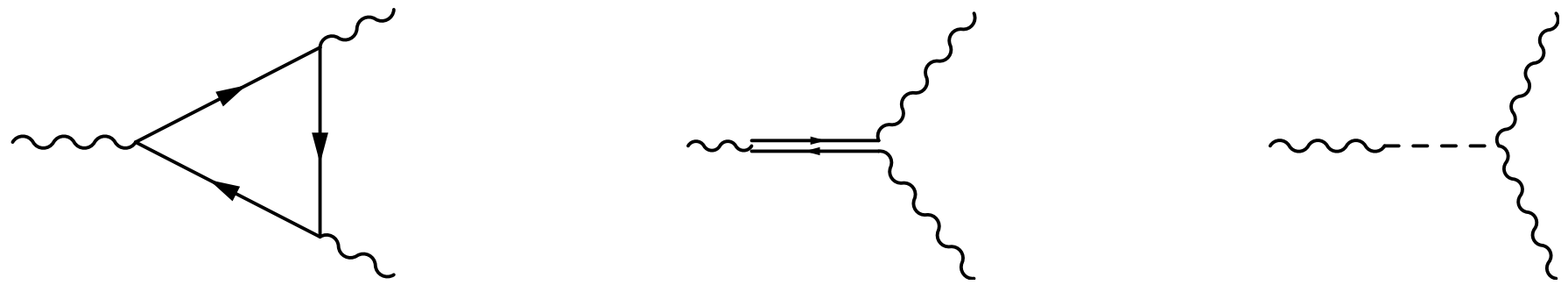

Figure 4.1: The fermion loop (a); the collinear region (b); the effective pseudoscalar exchange (c).

where $W_T$ is the transverse part, while the longitudinal tensor structure is given by

$$W_L^{\mu_1\mu_2\mu_3} = w_L\, p_3^{\mu_3}\, \varepsilon^{\mu_1\mu_2\rho\sigma} p_{1\rho} p_{2\sigma} \equiv w_L\, p_3^{\mu_3}\, \varepsilon^{\mu_1\mu_2 p_1 p_2} \tag{4.1.4}$$

$w_L$ is the anomaly form factor, that in the massless (chiral or conformal) has a $1/p_3^2$ pole. In the case of gauge anomalies, the cancelation of the anomaly poles is entirely connected with the particle content of the theory and defines the condition for the elimination of such massless interactions. The total residue at the pole then identifies the total anomaly of a certain fermion multiplet.
A similar behavior holds for conformal anomaly correlators with energy-momentum tensors, where the residue at the pole coincides with the $\beta$-function of the Lagrangian field theory, and is determined by the number of massless degrees of freedom included in the corresponding anomaly vertex, at the scale at which the perturbative prescription holds [109].

As illustrated in Fig. 4.1, the pole emerges from the region of integration in momentum-space in the $\langle J_V J_V J_A\rangle$ loop when an effective interaction, mediated by the fermion/antifermion pair is generated. The two collinear (on-shell) particles describe an effective pseudoscalar interaction interpolating between the incoming axial vector and the two outgoing vector currents.
Away from the conformal limit, when the $\langle J_V J_V J_A\rangle$ diagram is recomputed with the inclusion of a fermion of mass $m$, one discovers the presence of a sum rule satisfied by the spectral density of the form factor $w_L$ of the longitudinal channel. Such form factor ($w_L$) is characterized by a spectral density $\rho(s)$, whose integration in the region $4m^2 < s < \infty$ in the dispersive variable $s$, related to the virtuality of the axial-vector current, is mass-independent and given by

$$\int_{4m^2}^{\infty} ds\, \rho(s, m^2) = a_1 \tag{4.1.5}$$

We will explore this aspect in greater detail in Chapter 9.
For on-shell vector lines the perturbative vertex reduces only to the longitudinal component $W_L$. Our analysis shows the centrality of the pole in determining the general off-shell solution of the CWIs, that is fixed only by the anomaly content.
In the case of the $\langle J_A J_A J_A\rangle$ diagram, as we are going to see, we encounter contributions which are proportional to Chern-Simons (CS) forms. These are terms linear in two of the three momenta and allow to move the anomaly form one vertex to the other. For instance, in a perturbative evaluation of the $\langle J_V J_V J_A\rangle$ and $\langle J_A J_A J_A\rangle$ diagrams, in the chiral limit, the anomaly can be moved around the vertices by the inclusion of such CS terms [115]

$$\langle J_V^{\mu_1} J_V^{\mu_2} J_A^{\mu_3}\rangle = \langle J_A^{\mu_1} J_A^{\mu_2} J_A^{\mu_3}\rangle + \frac{8\, a_1\, i}{3}\varepsilon^{\alpha\mu_1\mu_2\mu_3}(p_1 - p_2)_\alpha. \tag{4.1.6}$$

where the last term is a Chern-Simons contribution.
In our analysis we proceed from the $\langle J_V J_V J_A\rangle$ case, then turning to the $\langle J_A J_A J_A\rangle$ and show in both cases how primary and secondary CWI can be solved quite efficiently in momentum-space. As shown in Fig. 4.2, the structure of the $\langle J_A J_A J_A \rangle$ is dictated by symmetry. Our solutions are then compared with the two most common representations of such diagrams, the Rosenberg and the L/T one. The latter is particularly useful for the role that provides for the longitudinal component of the correlator. The same representation has been used in the past for the derivation of new nonrenormalization theorems in perturbative QCD, in the chiral limit. This result has been derived in [113] and is confirmed by the formal analysis of the CWIs.

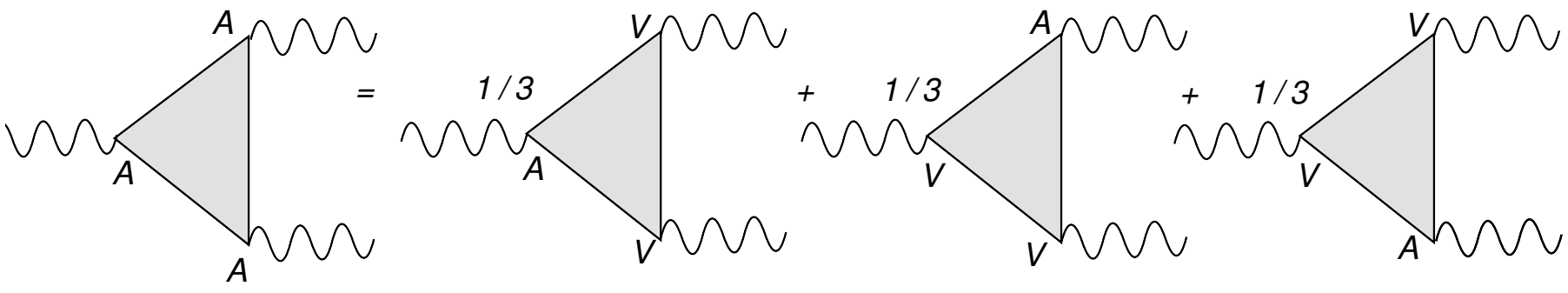

Figure 4.2: Distribution of the axial anomaly for the $\langle J_A J_A J_A \rangle$ diagram

## 4.2 The conformal $\langle J_V J_V J_A \rangle$ correlator

### 4.2.1 Longitudinal/transverse decomposition

Our analysis of the conformal constraints for the $\langle J_V J_V J_A \rangle$ is performed by applying the L/T decomposition to the correlator. We focus on the $d = 4$ case, where the conformal dimensions of the currents are $\Delta_i = 3$ and the tensorial structures of the correlator will involve the antisymmetric tensor in four dimensions $\varepsilon^{\mu\nu\alpha\beta}$. The procedure to obtain the general structure of the correlator starts from the conservation Ward identities in flat space-time

$$\partial_\mu \langle J_V^\mu \rangle = 0, \qquad \partial_\mu \langle J_A^\mu \rangle = a_1\, \varepsilon^{\mu\nu\rho\sigma} F_{\mu\nu} F_{\rho\sigma} \tag{4.2.1}$$

with $F_{\mu\nu}$ the field stregth tensor of the vector field. We recall that the vector current is coupled to the vector field $V_\mu$ and the axial-vector current to the source $A_\mu$. Applying multiple functional derivatives to (4.2.1) with respect to the source $V_\mu$, after a Fourier transform, we find the conservation Ward identities related to the entire correlator

$$\begin{aligned} p_{i\mu_i}\, \langle J_V^{\mu_1}(p_1) J_V^{\mu_2}(p_2) J_A^{\mu_3}(p_3) \rangle &= 0, \qquad i = 1,2 \\ p_{3\mu_3}\, \langle J_V^{\mu_1}(p_1) J_V^{\mu_2}(p_2) J_A^{\mu_3}(p_3) \rangle &= -8\, a_1\, i\, \varepsilon^{p_1 p_2 \mu_1 \mu_2} \end{aligned} \tag{4.2.2}$$

where $\varepsilon^{p_1 p_2 \mu_1 \mu_2} \equiv \varepsilon^{\alpha\beta\mu_1\mu_2} p_{1\alpha} p_{2\beta}$ and the momenta are all incoming. From this relations we construct the general form of the correlator, splitting the operators into a transverse and a longitudinal part as

$$\begin{aligned} J^\mu(p) &= j^\mu(p) + j^\mu_{loc}(p), \\ j^\mu = \pi^\mu_\alpha(p) J^\alpha(p), &\qquad j^\mu_{loc}(p) = \frac{p^\mu}{p^2}\, p \cdot J(p), \end{aligned} \tag{4.2.3}$$

where the transverse projector is

$$\pi^\mu_\alpha(p) \equiv \delta^\mu_\alpha - \frac{p_\alpha p^\mu}{p^2}. \tag{4.2.4}$$

Due to (4.2.2), the correlator is purely transverse in the vector currents. Therefore, the correlator can be decomposed in the following way

$$\langle J_V^{\mu_1}(p_1) J_V^{\mu_2}(p_2) J_A^{\mu_3}(p_3) \rangle = \langle j_V^{\mu_1}(p_1) j_V^{\mu_2}(p_2) j_A^{\mu_3}(p_3) \rangle + \langle j_V^{\mu_1}(p_1) j_V^{\mu_2}(p_2)\, j^{\mu_3}_{A\,\text{loc}}(p_3) \rangle \tag{4.2.5}$$

where the first term is completely transverse with respect to all the momenta and the second term is the longitudinal part that is proper of the anomaly contribution. Using the anomaly constraint, one obtains

$$\langle j_V^{\mu_1}(p_1) j_V^{\mu_2}(p_2) j_{A\,\text{loc}}^{\mu_3}(p_3)\rangle = \frac{p_3^{\mu_3}}{p_3^2}\, p_{3\,\alpha_3}\, \langle j_V^{\mu_1}(p_1) j_V^{\mu_2}(p_2) J_A^{\alpha_3}(p_3)\rangle = -\frac{8\,a_1\,i}{p_3^2}\,\varepsilon^{p_1 p_2 \mu_1 \mu_2}\, p_3^{\mu_3} \tag{4.2.6}$$

On the other hand, the transverse part can be formally written as

$$\begin{aligned}\langle j_V^{\mu_1}(p_1) j_V^{\mu_2}(p_2) j_A^{\mu_3}(p_3)\rangle = \pi^{\mu_1}_{\alpha_1}(p_1)\pi^{\mu_2}_{\alpha_2}(p_2)\pi^{\mu_3}_{\alpha_3}(p_3)\Big[&A_1(p_1,p_2,p_3)\,\varepsilon^{p_1 p_2 \alpha_1 \alpha_2} p_1^{\alpha_3} + A_2(p_1,p_2,p_3)\,\varepsilon^{p_1 p_2 \alpha_1 \alpha_3} p_3^{\alpha_2} \\ &+ A_3(p_1,p_2,p_3)\,\varepsilon^{p_1 p_2 \alpha_2 \alpha_3} p_2^{\alpha_1} + A_4(p_1,p_2,p_3)\,\varepsilon^{p_1 \alpha_1 \alpha_2 \alpha_3} + A_5(p_1,p_2,p_3)\,\varepsilon^{p_2 \alpha_1 \alpha_2 \alpha_3}\Big]\end{aligned} \tag{4.2.7}$$

where we have made a choice on which independent momenta to consider for each index, and in particular

$$\alpha_1 \leftrightarrow p_1, p_2, \qquad \alpha_2 \leftrightarrow p_2, p_3, \qquad \alpha_3 \leftrightarrow p_3, p_1 \tag{4.2.8}$$

The correlator has to be symmetric under the exchange of the two vector currents and this fact is reflected in the symmetry constraints

$$A_3(p_1,p_2,p_3) = -A_2(p_1,p_2,p_3), \quad A_5(p_1,p_2,p_3) = -A_4(p_1,p_2,p_3), \quad A_1(p_1,p_2,p_3) = -A_1(p_1,p_2,p_3) \tag{4.2.9}$$

reducing by two the number of independent form factors. Furthermore, in $d = 4$ a class of tensor identities has to be considered, for instance the Schouten identity

$$\delta^{\beta_3[\alpha_1}\varepsilon^{\alpha_2\alpha_3\beta_1\beta_2]} = 0. \tag{4.2.10}$$

From this type of tensor identities we find that

$$\pi^{\mu_1}_{\alpha_1}\pi^{\mu_2}_{\alpha_2}\pi^{\mu_3}_{\alpha_3}\left(p_2^{\alpha_1}\,\varepsilon^{p_1 p_2 \alpha_2 \alpha_3}\right) = \pi^{\mu_1}_{\alpha_1}\pi^{\mu_2}_{\alpha_2}\pi^{\mu_3}_{\alpha_3}\left(-\left(p_2\cdot p_1\right)\varepsilon^{p_2\alpha_1\alpha_2\alpha_3} + p_2^2\,\varepsilon^{p_1\alpha_1\alpha_2\alpha_3} - p_2^{\alpha_3}\,\varepsilon^{p_1 p_2 \alpha_1 \alpha_2}\right), \tag{4.2.11}$$

$$\pi^{\mu_1}_{\alpha_1}\pi^{\mu_2}_{\alpha_2}\pi^{\mu_3}_{\alpha_3}\left(p_3^{\alpha_2}\,\varepsilon^{p_1 p_2 \alpha_1 \alpha_3}\right) = \pi^{\mu_1}_{\alpha_1}\pi^{\mu_2}_{\alpha_2}\pi^{\mu_3}_{\alpha_3}\left(-p_1^2\,\varepsilon^{p_2\alpha_1\alpha_2\alpha_3} + \left(p_1\cdot p_2\right)\varepsilon^{p_1\alpha_1\alpha_2\alpha_3} - p_1^{\alpha_3}\,\varepsilon^{p_1 p_2 \alpha_1 \alpha_2}\right), \tag{4.2.12}$$

reducing the number of independent form factors just to two. We conclude that the general structure of the transverse part is given by

$$\begin{aligned}\langle j_V^{\mu_1}(p_1) j_V^{\mu_2}(p_2) j_A^{\mu_3}(p_3)\rangle = \pi^{\mu_1}_{\alpha_1}(p_1)\pi^{\mu_2}_{\alpha_2}(p_2)\pi^{\mu_3}_{\alpha_3}(p_3)\Big[&A_1(p_1,p_2,p_3)\,\varepsilon^{p_1 p_2 \alpha_1 \alpha_2} p_1^{\alpha_3} \\ &+ A_2(p_1,p_2,p_3)\,\varepsilon^{p_1\alpha_1\alpha_2\alpha_3} - A_2(p_2,p_1,p_3)\,\varepsilon^{p_2\alpha_1\alpha_2\alpha_3}\Big]\end{aligned} \tag{4.2.13}$$

where $A_1(p_1,p_2,p_3) = -A_1(p_2,p_1,p_3)$.

### 4.2.2 Dilatation Ward identities

We now start to analyse the conformal constraints on the form factors. In this and the next sections, we closely follows the methodology adopted in [29]. The invariance of the correlator under dilatation is reflected in the equation

$$\left(\sum_{i=1}^{3}\Delta_i-2d-\sum_{i=1}^{2}p_i^\mu\frac{\partial}{\partial p_i^\mu}\right)\langle J_V^{\mu_1}(p_1)J_V^{\mu_2}(p_2)J_A^{\mu_3}(p_3)\rangle=0. \tag{4.2.14}$$

Considering the decomposition (4.2.5), the transverse part of the correlator has to satisfy the same equation

$$\left(\sum_{i=1}^{3}\Delta_i-2d-\sum_{i=1}^{2}p_i^\mu\frac{\partial}{\partial p_i^\mu}\right)\langle j_V^{\mu_1}(p_1)j_V^{\mu_2}(p_2)j_A^{\mu_3}(p_3)\rangle=0. \tag{4.2.15}$$

Then, by using the chain rule

$$\frac{\partial}{\partial p_i^\mu}=\sum_{j=1}^{3}\frac{\partial p_j}{\partial p_i^\mu}\frac{\partial}{\partial p_j} \tag{4.2.16}$$

in term of the invariants $p_i=|\sqrt{p_i^2}|$, we can rewrite the dilatation equation as the following constraints on the form factors

$$\sum_{i=1}^{3}p_i\frac{\partial A_j}{\partial p_i}-\Big(\Delta_3-2-N_j\Big)A_j=0, \tag{4.2.17}$$

where $N_j$ is the number of momenta that the form factors multiply in the decomposition (4.2.13), i.e. $N_1=3$ and $N_2=1$.

### 4.2.3 Special conformal Ward identities

The invariance of the correlator with respect to the special conformal transformations is encoded in the special conformal Ward identities

$$\begin{aligned}0=&\sum_{j=1}^{2}\left[-2\frac{\partial}{\partial p_{j\kappa}}-2p_j^\alpha\frac{\partial^2}{\partial p_j^\alpha\,\partial p_{j\kappa}}+p_j^\kappa\frac{\partial^2}{\partial p_j^\alpha\,\partial p_{j\alpha}}\right]\langle J_V^{\mu_1}(p_1)J_V^{\mu_2}(p_2)J_A^{\mu_3}(p_3)\rangle\\&+2\left(\delta^{\mu_1\kappa}\frac{\partial}{\partial p_1^{\alpha_1}}-\delta^\kappa_{\alpha_1}\frac{\partial}{\partial p_{1\mu_1}}\right)\langle J_V^{\alpha_1}(p_1)J_V^{\mu_2}(p_2)J_A^{\mu_3}(p_3)\rangle\\&+2\left(\delta^{\mu_2\kappa}\frac{\partial}{\partial p_2^{\alpha_2}}-\delta^\kappa_{\alpha_2}\frac{\partial}{\partial p_{2\mu_2}}\right)\langle J_V^{\mu_1}(p_1)J_V^{\alpha_2}(p_2)J_A^{\mu_3}(p_3)\rangle\equiv\mathcal{K}^\kappa\langle J_V^{\mu_1}(p_1)J_V^{\mu_2}(p_2)J_A^{\mu_3}(p_3)\rangle.\end{aligned} \tag{4.2.18}$$

We can perform a transverse projection ($3-\pi$ projection) on all the indices in order to identify a set of partial differential equations [29]

$$\pi^{\lambda_1}_{\mu_1}(p_1)\pi^{\lambda_2}_{\mu_2}(p_2)\pi^{\lambda_3}_{\mu_3}(p_3)\Big(\mathcal{K}^\kappa\langle J_V^{\mu_1}(p_1)J_V^{\mu_2}(p_2)J_A^{\mu_3}(p_3)\rangle\Big)=0 \tag{4.2.19}$$

splitting the correlator into its transverse and longitudinal parts. The contribution given by the transverse part of the correlator can be written as

$$\pi^{\lambda_1}_{\mu_1}(p_1)\pi^{\lambda_2}_{\mu_2}(p_2)\pi^{\lambda_3}_{\mu_3}(p_3)\bigg(\mathcal{K}^\kappa \langle j_V^{\mu_1}(p_1) j_V^{\mu_2}(p_2) j_A^{\mu_3}(p_3)\rangle\bigg) = \pi^{\lambda_1}_{\mu_1}(p_1)\pi^{\lambda_2}_{\mu_2}(p_2)\pi^{\lambda_3}_{\mu_3}(p_3)\, X^{\kappa\mu_1\mu_2\mu_3}, \tag{4.2.20}$$

where the tensor $X^{\kappa\mu_1\mu_2\mu_3}$ can be constructed using a set of possible tensor structures, of the form

$$\varepsilon^{\mu_1\mu_2\mu_3 p_1} p_1^\kappa, \quad \varepsilon^{\mu_1\mu_2\mu_3 p_2} p_1^\kappa, \quad \varepsilon^{\mu_1\mu_2 p_1 p_2} p_1^{\mu_3} p_1^\kappa, \quad \varepsilon^{\mu_1\mu_3 p_1 p_2} p_3^{\mu_2} p_1^\kappa, \quad \varepsilon^{\mu_2\mu_3 p_1 p_2} p_2^{\mu_1} p_1^\kappa, \tag{4.2.21}$$

whose complete list is given in Appendix C.1. These tensor structures are not all independent, and are simplified using some Schouten identites in order to find the minimal number of tensor structures in which $X$ can be expanded. The first two identities are

$$\varepsilon^{[\mu_2\mu_3\kappa p_1}\delta^{\mu_1]\alpha} = 0, \tag{4.2.22}$$

$$\varepsilon^{[\mu_2\mu_3\kappa p_2}\delta^{\mu_1]\alpha} = 0, \tag{4.2.23}$$

that can be contracted with $p_{1\alpha}$ and $p_{2\alpha}$ to generate, after the projection, tensor identities of the form

$$\pi^{\lambda_1}_{\mu_1}\pi^{\lambda_2}_{\mu_2}\pi^{\lambda_3}_{\mu_3}\bigg(\varepsilon^{p_1\kappa\mu_1\mu_3} p_3^{\mu_2}\bigg) = \pi^{\lambda_1}_{\mu_1}\pi^{\lambda_2}_{\mu_2}\pi^{\lambda_3}_{\mu_3}\bigg(-p_1^2\varepsilon^{\kappa\mu_1\mu_2\mu_3} + \varepsilon^{p_1\mu_1\mu_2\mu_3}p_1^\kappa - \varepsilon^{p_1\kappa\mu_1\mu_2}p_1^{\mu_3}\bigg) \tag{4.2.24}$$

$$\pi^{\lambda_1}_{\mu_1}\pi^{\lambda_2}_{\mu_2}\pi^{\lambda_3}_{\mu_3}\bigg(\varepsilon^{p_1\kappa\mu_2\mu_3} p_2^{\mu_1}\bigg) = \pi^{\lambda_1}_{\mu_1}\pi^{\lambda_2}_{\mu_2}\pi^{\lambda_3}_{\mu_3}\bigg(\frac{1}{2}\left(p_1^2+p_2^2-p_3^2\right)\varepsilon^{\kappa\mu_1\mu_2\mu_3} + \varepsilon^{p_1\kappa\mu_1\mu_2}p_1^{\mu_3} + \varepsilon^{p_1\mu_1\mu_2\mu_3}p_2^\kappa\bigg). \tag{4.2.25}$$

More technical details and a full list of such tensor relations is given in the Appendix C.1. Combining all the expressions, we obtain the following projected decomposition

$$\begin{aligned}
&\pi^{\lambda_1}_{\mu_1}(p_1)\pi^{\lambda_2}_{\mu_2}(p_2)\pi^{\lambda_3}_{\mu_3}(p_3)\bigg(\mathcal{K}^\kappa \langle j_V^{\mu_1}(p_1) j_V^{\mu_2}(p_2) j_A^{\mu_3}(p_3)\rangle\bigg) = \\
&= \pi^{\lambda_1}_{\mu_1}(p_1)\pi^{\lambda_2}_{\mu_2}(p_2)\pi^{\lambda_3}_{\mu_3}(p_3)\bigg[p_1^\kappa\bigg(C_{11}\,\varepsilon^{\mu_1\mu_2\mu_3p_1} + C_{12}\,\varepsilon^{\mu_1\mu_2\mu_3p_2} + C_{13}\,\varepsilon^{\mu_1\mu_2p_1p_2}p_1^{\mu_3}\bigg) \\
&\quad + p_2^\kappa\bigg(C_{21}\,\varepsilon^{\mu_1\mu_2\mu_3p_1} + C_{22}\,\varepsilon^{\mu_1\mu_2\mu_3p_2} + C_{23}\,\varepsilon^{\mu_1\mu_2p_1p_2}p_1^{\mu_3}\bigg) + C_{31}\varepsilon^{\kappa\mu_1\mu_2\mu_3} + C_{32}\varepsilon^{\kappa\mu_1\mu_2p_1}p_1^{\mu_3} \\
&\quad + C_{33}\varepsilon^{\kappa\mu_1\mu_2p_2}p_1^{\mu_3} + C_{34}\varepsilon^{\kappa\mu_1p_1p_2}\delta^{\mu_2\mu_3} + C_{35}\varepsilon^{\kappa\mu_2p_1p_2}\delta^{\mu_1\mu_3} + C_{36}\varepsilon^{\kappa\mu_3p_1p_2}\delta^{\mu_1\mu_2}\bigg],
\end{aligned} \tag{4.2.26}$$

where all the tensor structures are independent. Here, the coefficients $C_{ij}$ can be expressed in terms of the form factors $A_1$ and $A_2$ and their derivatives. In particular, $C_{1j}$ and $C_{2j}$ contains second order derivatives, while all the others coefficients contain at most first order derivatives. On the other hand, the action of $\mathcal{K}^\kappa$ on the longitudinal part is given by

$$\begin{aligned}
&\pi^{\lambda_1}_{\mu_1}(p_1)\pi^{\lambda_2}_{\mu_2}(p_2)\pi^{\lambda_3}_{\mu_3}(p_3)\bigg(\mathcal{K}^\kappa \langle j_V^{\mu_1}(p_1) j_V^{\mu_2}(p_2) j_{A\,loc}^{\mu_3}(p_3)\rangle\bigg) = \pi^{\lambda_1}_{\mu_1}(p_1)\,\pi^{\lambda_2}_{\mu_2}(p_2)\,\pi^{\lambda_3}_{\mu_3}(p_3)\Big[\mathcal{A}\,\delta^{\mu_3\kappa}\varepsilon^{\mu_1\mu_2p_1p_2}\Big] \\
&= \pi^{\lambda_1}_{\mu_1}(p_1)\,\pi^{\lambda_2}_{\mu_2}(p_2)\,\pi^{\lambda_3}_{\mu_3}(p_3)\bigg[\mathcal{A}\bigg(\varepsilon^{\kappa\mu_2p_1p_2}\delta^{\mu_1\mu_3} - \varepsilon^{\kappa\mu_1p_1p_2}\delta^{\mu_2\mu_3} + \varepsilon^{\kappa\mu_1\mu_2p_1}p_1^{\mu_3} + \varepsilon^{\kappa\mu_1\mu_2p_2}p_1^{\mu_3}\bigg)\bigg],
\end{aligned} \tag{4.2.27}$$

where

$$\mathcal{A} \equiv -\frac{16\,a_1\,i\,(\Delta_3-1)}{p_3^2} \tag{4.2.28}$$

is related to the chiral anomaly. Due to the independence of the tensor structures listed in Eq. (4.2.26), the special conformal equations can be written as

$$C_{ij} = 0, \quad i = 1,2, \quad j = 1,2,3 \tag{4.2.29}$$
$$C_{31} = 0, \tag{4.2.30}$$
$$C_{3j} + \mathcal{A} = 0, \quad j = 2,3,5 \tag{4.2.31}$$
$$C_{34} - \mathcal{A} = 0, \tag{4.2.32}$$
$$C_{36} = 0. \tag{4.2.33}$$

The explicit form of the primary equations (4.2.29) is

$$\begin{aligned}
&K_{31} A_1 = 0, && K_{32} A_1 = 0, \\
&K_{31} A_2 = 0, && K_{32} A_2 = \left(\frac{4}{p_1^2} - \frac{2}{p_1}\frac{\partial}{\partial p_1}\right) A_2(p_1 \leftrightarrow p_2) + 2A_1, \\
&K_{31} A_2(p_1 \leftrightarrow p_2) = \left(\frac{4}{p_2^2} - \frac{2}{p_2}\frac{\partial}{\partial p_2}\right) A_2 - 2A_1, && K_{32} A_2(p_1 \leftrightarrow p_2) = 0,
\end{aligned} \tag{4.2.34}$$

where we have defined

$$K_i = \frac{\partial^2}{\partial p_i^2} + \frac{(d+1-2\Delta_i)}{p_i}\frac{\partial}{\partial p_i}, \qquad K_{ij} = K_i - K_j. \tag{4.2.35}$$

These equations can also be reduced to a set of homogenous equations by repeatedly applying the operator $K_{ij}$

$$\begin{aligned}
&K_{31} A_1 = 0, && K_{32} A_1 = 0, \\
&K_{31} A_2 = 0, && K_{32} K_{32} A_2 = 0.
\end{aligned} \tag{4.2.36}$$

### 4.2.4 Solutions of the CWIs

The most general solution of the conformal Ward identities of the $\langle J_V J_V J_A \rangle$ can be written in terms of 3K integrals. Indeed, recalling the following property of the 3K integrals

$$K_{nm} J_{N\{k_j\}} = -2k_n J_{N+1\{k_j - \delta_{jn}\}} + 2k_m J_{N+1\{k_j - \delta_{jm}\}} \tag{4.2.37}$$

we can write the most general solution of the homogeneous equations (4.2.36) as

$$A_1 = \alpha_1 J_{3\{0,0,0\}}, \tag{4.2.38}$$
$$A_2 = \alpha_2 J_{1\{0,0,0\}} + \alpha_3 J_{2\{0,1,0\}}. \tag{4.2.39}$$

Inserting these solutions to the nonhomogenous equations (4.2.34), we find the constraint

$$\alpha_2 = -2\alpha_3 \tag{4.2.40}$$

and the solution reduces to

$$A_1 = \alpha_1 J_{3\{0,0,0\}}, \tag{4.2.41}$$
$$A_2 = -2\alpha_3 J_{1\{0,0,0\}} + \alpha_3 J_{2\{0,1,0\}}. \tag{4.2.42}$$

We now consider the first order differential equations (4.2.30), (4.2.31), (4.2.32) and (4.2.33) in their explicit form, ignoring the trivial ones. We start with

$$C_{36} = \left(\frac{4}{p_2^2} - \frac{2}{p_2}\frac{\partial}{\partial p_2}\right)A_2 - \left(\frac{4}{p_1^2} - \frac{2}{p_1}\frac{\partial}{\partial p_1}\right)A_2(p_1 \leftrightarrow p_2) - 2A_1 = 0. \tag{4.2.43}$$

Inserting our solution in the equation above and using the property of the 3K integral

$$\frac{\partial}{\partial p_n} J_{N\{k_j\}} = -p_n J_{N+1\{k_j-\delta_{jn}\}}, \tag{4.2.44}$$

we have

$$C_{36} = 2\alpha_1 J_{3\{0,0,0\}} = 0, \tag{4.2.45}$$

obtaining the constraint

$$\alpha_1 = 0. \tag{4.2.46}$$

Taking into account this constraint, we consider the other secondary Ward identity as

$$\begin{aligned} C_{31} = 2p_2\frac{\partial}{\partial p_2}A_2(p_1 \leftrightarrow p_2) - 2p_1\frac{\partial}{\partial p_1}A_2 + \frac{2(p_1^2+p_2^2-p_3^2)}{p_1^2 p_2^2}\left(p_1^2 A_2 - p_2^2 A_2(p_1 \leftrightarrow p_2)\right) + \\ \frac{2p_1\cdot p_2}{p_2}\frac{\partial}{\partial p_2}A_2 - \frac{2p_1\cdot p_2}{p_1}\frac{\partial}{\partial p_1}A_2(p_1 \leftrightarrow p_2) = 0. \end{aligned} \tag{4.2.47}$$

This equation with the constraints obtained before on the coefficients $\alpha_i$ is trivially satisfied by our solution. The last equation we have to consider is the one related to the anomaly that takes the form

$$-\frac{2}{p_3^2}\left(p_2\frac{\partial}{\partial p_2} + p_1\frac{\partial}{\partial p_1} + 2\right)\left(A_2 + A_2(p_1 \leftrightarrow p_2)\right) - \left(\frac{4}{p_1^2} - \frac{2}{p_1}\frac{\partial}{\partial p_1}\right)A_2(p_1 \leftrightarrow p_2) - \mathcal{A} = 0. \tag{4.2.48}$$

This equation can be easily solved by substituting our solution and taking the limit $p_3 \to 0$. Setting $d = 4$ and $\Delta_3 = 3$, we obtain

$$\alpha_3 = 8i\, a_1. \tag{4.2.49}$$

In summary, once the conformal constraints are solved, we find the solution of the transverse part in terms of one coefficient, proportional to the anomaly and in particular we have

$$\begin{aligned} \langle j_V^{\mu_1}(p_1) j_V^{\mu_2}(p_2) j_A^{\mu_3}(p_3)\rangle = \pi^{\mu_1}_{\alpha_1}(p_1)\pi^{\mu_2}_{\alpha_2}(p_2)\pi^{\mu_3}_{\alpha_3}(p_3)\bigg[8ia_1\Big(-2J_{1\{0,0,0\}} + J_{2\{0,1,0\}}\Big)\varepsilon^{p_1\alpha_1\alpha_2\alpha_3} \\ - 8ia_1\Big(-2J_{1\{0,0,0\}} + J_{2\{1,0,0\}}\Big)\varepsilon^{p_2\alpha_1\alpha_2\alpha_3}\bigg] \end{aligned} \tag{4.2.50}$$

or in terms of the simplified version of the $3K$ integrals as

$$J_{1\{0,0,0\}} = I_{2\{1,1,1\}}, \quad J_{2\{0,1,0\}} = I_{3\{1,2,1\}}, \quad J_{2\{1,0,0\}} = I_{3\{2,1,1\}}. \tag{4.2.51}$$

Explicitly we have

$$\langle j_V^{\mu_1}(p_1) j_V^{\mu_2}(p_2) j_A^{\mu_3}(p_3)\rangle = 8i a_1\, \pi^{\mu_1}_{\alpha_1}(p_1)\pi^{\mu_2}_{\alpha_2}(p_2)\pi^{\mu_3}_{\alpha_3}(p_3)\bigg[\bigg(-2\,I_{2\{1,1,1\}} + I_{3\{1,2,1\}}\bigg)\varepsilon^{p_1\alpha_1\alpha_2\alpha_3} - \bigg(-2\,I_{2\{1,1,1\}} + I_{3\{2,1,1\}}\bigg)\varepsilon^{p_2\alpha_1\alpha_2\alpha_3}\bigg]. \tag{4.2.52}$$

Furthermore, this expression can be reduced by noticing that

$$-2\,I_{2\{1,1,1\}} + I_{3\{1,2,1\}} = p_2^2\, I_{3\{1,0,1\}}, \tag{4.2.53}$$

$$-2\,I_{2\{1,1,1\}} + I_{3\{1,2,1\}} = p_1^2\, I_{3\{0,1,1\}}, \tag{4.2.54}$$

finally giving

$$\langle j_V^{\mu_1}(p_1) j_V^{\mu_2}(p_2) j_A^{\mu_3}(p_3)\rangle = 8i a_1\, \pi^{\mu_1}_{\alpha_1}(p_1)\pi^{\mu_2}_{\alpha_2}(p_2)\pi^{\mu_3}_{\alpha_3}(p_3)\bigg[p_2^2\, I_{3\{1,0,1\}}\,\varepsilon^{p_1\alpha_1\alpha_2\alpha_3} - p_1^2\, I_{3\{0,1,1\}}\,\varepsilon^{p_2\alpha_1\alpha_2\alpha_3}\bigg] \tag{4.2.55}$$

## 4.3 Reducing the 3K integral and the perturbative solution

Considering the reduction relations presented in [80,81], we find that the solution obtained for the $\langle J_V J_V J_A\rangle$ in the previous section is finite and can be reduced to the standard perturbation results as

$$I_{3\{1,0,1\}} = p_1\, p_3 \frac{\partial^2}{\partial p_1 \partial p_3} I_{1\{0,0,0\}} \tag{4.3.1}$$

where $I_{1\{0,0,0\}}$ is the master integral related to the three-point function of the operator $\varphi^2$ in the theory of free massless scalar $\varphi$ in $d = 4$. Indeed, the relation between the master integrals and the massless scalar 1-loop 3-point momentum-space integral is

$$I_{1\{0,0,0\}} = (2\pi)^2 K_{4,\{1,1,1\}} = (2\pi)^2 \int \frac{d^4k}{(2\pi)^4} \frac{1}{k^2\,(k-p_1)^2\,(k+p_2)^2} = \frac{1}{4} C_0(p_1^2, p_2^2, p_3^2) \tag{4.3.2}$$

where

$$K_{d\{\delta_1\delta_2\delta_3\}} = \int \frac{d^dk}{(2\pi)^d} \frac{1}{(k^2)^{\delta_3}\,((k-p_1)^2)^{\delta_2}\,((k+p_2)^2)^{\delta_1}}. \tag{4.3.3}$$

By using the relations of the derivative acting on the master integral presented in [75], and analytically continuing $C_0$ to $d$ dimensions we find that

$$p_3 \frac{\partial}{\partial p_3} C_0(p_1^2, p_2^2, p_3^2) = \frac{1}{\lambda}\bigg\{2(d-3)\Big[(p_1^2 - p_2^2 + p_3^2)B_0(p_1^2) + (p_2^2 + p_3^2 - p_1^2)B_0(p_2^2) - 2p_3^2 B_0(p_3^2)\Big] + \Big[(d-4)(p_1^2 - p_2^2)^2 - (d-2)p_3^4 + 2p_3^2(p_1^2 + p_2^2)\Big] C_0(p_1^2, p_2^2, p_3^2)\bigg\} \tag{4.3.4}$$

where $\lambda$ is the Källen $\lambda$-function

$$\lambda \equiv (p_1 - p_2 - p_3)(p_1 + p_2 - p_3)(p_1 - p_2 + p_3)(p_1 + p_2 + p_3). \tag{4.3.5}$$

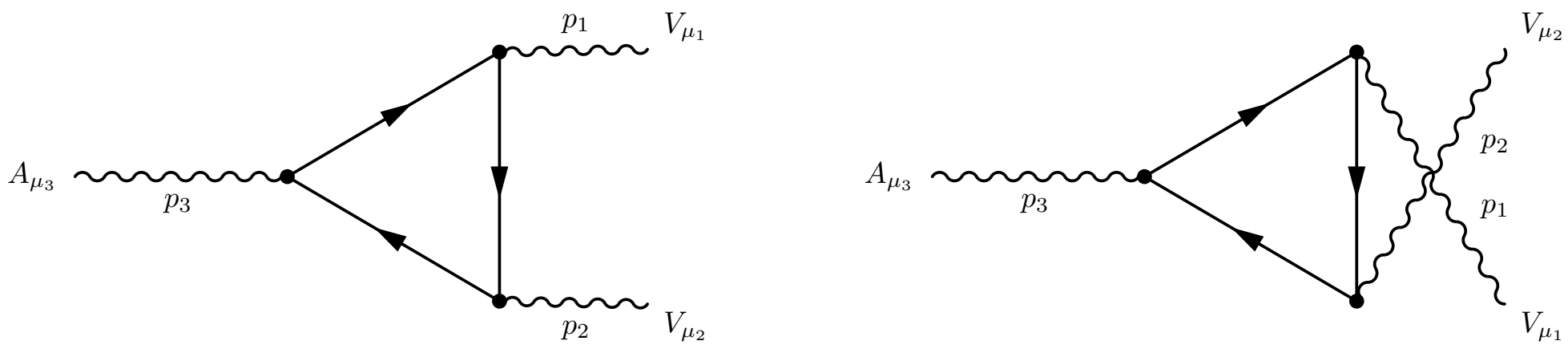

Figure 4.3: Feynman diagrams for the $\langle J_V J_V J_A\rangle$ interaction

Then

$$\begin{aligned}
4 I_{3\{1,0,1\}} = {} & p_1 \frac{\partial}{\partial p_1} \frac{1}{\lambda} \bigg\{ 2(d-3)\Big[(p_1^2-p_2^2+p_3^2)B_0(p_1^2)+(p_2^2+p_3^2-p_1^2)B_0(p_2^2)-2p_3^2B_0(p_3^2)\Big] \\
& + \Big[(d-4)(p_1^2-p_2^2)^2-(d-2)p_3^4+2p_3^2(p_1^2+p_2^2)\Big]C_0(p_1^2,p_2^2,p_3^2)\bigg\} \\
= {} & -\frac{4p_1^2(p_1^2-p_2^2-p_3^2)}{\lambda^2}\bigg\{2(d-3)\Big[(p_1^2-p_2^2+p_3^2)B_0(p_1^2)+(p_2^2+p_3^2-p_1^2)B_0(p_2^2)-2p_3^2B_0(p_3^2)\Big] \\
& + \Big[-(d-2)p_3^4+2p_3^2(p_1^2+p_2^2)\Big]C_0(p_1^2,p_2^2,p_3^2)\bigg\} \\
& + \frac{1}{\lambda}\bigg\{2(d-3)\Big[2p_1^2\,B_0(p_1^2)+(p_1^2-p_2^2+p_3^2)\,(d-4)B_0(p_1^2)-2p_1^2\,B_0(p_2^2)\Big]+4p_1^2p_3^2\,C_0(p_1^2,p_2^2,p_3^2) \\
& + \frac{1}{\lambda}\Big[-(d-2)p_3^2+2p_3^2(p_1^2+p_2^2)\Big]\bigg[2(d-3)\Big((p_1^2+p_2^2-p_3^2)B_0(p_2^2)+(p_1^2-p_2^2+p_3^2)B_0(p_3^2)-2p_1^2B_0(p_1^2)\Big) \\
& + \Big(-(d-2)p_1^4+2p_1^2(p_2^2+p_3^2)\Big)C_0(p_1^2,p_2^2,p_3^2)\bigg]\bigg\} + O(d-4).
\end{aligned} \tag{4.3.6}$$

This expression is finite in the expansion around $d=4$ and the result simplifies to the form

$$\begin{aligned}
I_{3\{1,0,1\}}(p_1^2,p_2^2,p_3^2) = {} & \frac{1}{\lambda^2}\bigg\{-2p_1^2p_3^2\Big[p_1^2\left(p_2^2-2p_3^2\right)+p_1^4+p_2^2p_3^2-2p_2^4+p_3^4\Big]C_0\left(p_1^2,p_2^2,p_3^2\right) \\
& + p_1^2\left(\left(p_1^2-p_2^2\right)^2+4p_2^2p_3^2-p_3^4\right)\log\left(\frac{p_1^2}{p_2^2}\right)+4p_1^2p_3^2\left(p_1^2-p_3^2\right)\log\left(\frac{p_1^2}{p_3^2}\right) \\
& - p_3^2\Big((p_2^2-p_3^2)^2+4p_1^2p_2^2-p_1^4\Big)\log\left(\frac{p_2^2}{p_3^2}\right)-\lambda(p_1^2-p_2^2+p_3^2)\bigg\}
\end{aligned} \tag{4.3.7}$$

with the form factor given by

$$A_2^{(CFT)}(p_1,p_2,p_3) = 8ia_1\,p_2^2\,I_{3\{1,0,1\}}(p_1^2,p_2^2,p_3^2). \tag{4.3.8}$$

### 4.3.1 Perturbative realization of the correlator

The $\langle J_V J_V J_A\rangle$ correlator can be investigated perturbatively within the framework of free field theory involving a chiral fermion. Alternatively, one can also consider working in QED with an external field $A_\mu$ coupled to the axial current.

Here, we choose to work in Minkowski space, using the Breitenlohner-Maison-'t Hooft-Veltman scheme [48]. The triangle diagrams depicted in Fig. 4.3 can be expressed as follows

$$\langle J_V^{\mu_1}(p_1) J_V^{\mu_2}(p_2) J_A^{\mu_3}(p_3)\rangle = \\ -i^3 \int \frac{d^4 l}{(2\pi)^4} \frac{\mathrm{Tr}\Big[(-ig\gamma^{\mu_1})(\not{l}-\not{p}_1)(-ig\gamma^{\mu_3}\gamma^5)(\not{l}+\not{p}_2)(-ig\gamma^{\mu_2})\not{l}\Big]}{(l-p_1)^2(l+p_2)^2 l^2} + [(p_1,\mu_1)\leftrightarrow(p_2,\mu_2)] \tag{4.3.9}$$

After computing the integral, one can verify that the vector currents are conserved, while the axial current is not. Specifically, the perturbative results satisfy eq. (4.2.2), with the anomaly coefficient given by

$$a_1 = \frac{i g^3}{16\pi^2}. \tag{4.3.10}$$

We can now employ the decomposition of the correlator derived in Section 4.2.1. In particular, the transverse part of the correlator can be expressed using eq. (4.2.13). The perturbative result yields a single non-vanishing form factor

$$\begin{aligned} A_1^{(Pert)} =& 0 \\ A_2^{(Pert)} =& \frac{-g^3 p_2^2}{2\pi^2\lambda^2}\Bigg\{-\lambda(p_1^2-p_2^2+p_3^2)+p_1^2\Big[\left(p_1^2-p_2^2\right)^2+4p_3^2p_2^2-p_3^4\Big]\log\left(\frac{p_1^2}{p_2^2}\right) \\ &+p_3^2\Big[p_1^4-4p_1^2p_2^2-\left(p_2^2-p_3^2\right)^2\Big]\log\left(\frac{p_2^2}{p_3^2}\right)+4p_1^2p_3^2\left(p_1^2-p_3^2\right)\log\left(\frac{p_1^2}{p_3^2}\right) \\ &-2p_1^2p_3^2\Big[p_1^2\left(p_2^2-2p_3^2\right)+p_1^4+p_2^2p_3^2-2p_2^4+p_3^4\Big]C_0\left(p_1^2,p_2^2,p_3^2\right)\Bigg\}. \end{aligned} \tag{4.3.11}$$

One can directly check the complete agreement between the perturbative and CFT results, provided the anomaly coefficient is set as in eq. (4.3.10).

## 4.4 The conformal $\langle J_A J_A J_A\rangle$ correlator

In this section we illustrate how conformal invariance completely determines the structure of the $\langle J_A J_A J_A\rangle$ correlator. Differently from the $\langle J_V J_V J_A\rangle$ correlator, a 3K integral regularization is needed in this case. Therefore, we will work in $d = 4+\epsilon$ and then perform the limit $\epsilon \to 0$.

### 4.4.1 Longitudinal/transverse decomposition

First of all, we consider the anomalous Ward identity

$$\partial_\mu J_A^\mu = a_1' \varepsilon^{\mu\nu\rho\sigma} F^A_{\mu\nu} F^A_{\rho\sigma} \tag{4.4.1}$$

where $F^A_{\mu\nu}$ is the axial field strength tensor. We impose such identity symmetrically on the all the three external axial-vector currents of the $\langle J_A J_A J_A\rangle$ correlator, leading to the following equations

$$\begin{aligned} p_{1\mu_1}\Big\langle J_A^{\mu_1}(p_1)J_A^{\mu_2}(p_2)J_A^{\mu_3}(p_3)\Big\rangle &= -8\,a_1'\,i\,\varepsilon^{p_1p_2\mu_2\mu_3} \\ p_{2\mu_2}\Big\langle J_A^{\mu_1}(p_1)J_A^{\mu_2}(p_2)J_A^{\mu_3}(p_3)\Big\rangle &= 8\,a_1'\,i\,\varepsilon^{p_1p_2\mu_1\mu_3} \\ p_{3\mu_3}\Big\langle J_A^{\mu_1}(p_1)J_A^{\mu_2}(p_2)J_A^{\mu_3}(p_3)\Big\rangle &= -8\,a_1'\,i\,\varepsilon^{p_1p_2\mu_1\mu_2}. \end{aligned} \tag{4.4.2}$$

Notice that the equations above are symmetric in the three momenta, but for technical reasons we prefer to express them only in terms of $p_1$ and $p_2$. Note that, if we contract the correlator with multiple momenta at the same time, the result is zero.
We can then decompose the correlator into the following transverse and longitudinal parts

$$\left\langle J_A^{\mu_1} J_A^{\mu_2} J_A^{\mu_3} \right\rangle = \left\langle j_A^{\mu_1} j_A^{\mu_2} j_A^{\mu_3} \right\rangle + \left\langle j_{A\,\mathrm{loc}}^{\mu_1} j_A^{\mu_2} j_A^{\mu_3} \right\rangle + \left\langle j_A^{\mu_1} j_{A\,\mathrm{loc}}^{\mu_2} j_A^{\mu_3} \right\rangle + \left\langle j_A^{\mu_1} j_A^{\mu_2} j_{A\,\mathrm{loc}}^{\mu_3} \right\rangle \tag{4.4.3}$$

The longitudinal parts are completely fixed by the anomalous Ward identities above in the form

$$\begin{aligned}
\left\langle j_{A\,\mathrm{loc}}^{\mu_1}(p_1) j_A^{\mu_2}(p_2) j_A^{\mu_3}(p_3) \right\rangle &= -\frac{8\,a_1' i}{p_1^2} \varepsilon^{p_1 p_2 \mu_2 \mu_3} p_1^{\mu_1} \\
\left\langle j_A^{\mu_1}(p_1) j_{A\,\mathrm{loc}}^{\mu_2}(p_2) j_A^{\mu_3}(p_3) \right\rangle &= \frac{8\,a_1' i}{p_2^2} \varepsilon^{p_1 p_2 \mu_1 \mu_3} p_2^{\mu_2} \\
\left\langle j_A^{\mu_1}(p_1) j_A^{\mu_2}(p_2) j_{A\,\mathrm{loc}}^{\mu_3}(p_3) \right\rangle &= -\frac{8\,a_1' i}{p_3^2} \varepsilon^{p_1 p_2 \mu_1 \mu_2} p_3^{\mu_3}.
\end{aligned} \tag{4.4.4}$$

On the other hand, the transverse part can be expressed as

$$\begin{aligned}
\left\langle j_A^{\mu_1}(p_1) j_A^{\mu_2}(p_2) j_A^{\mu_3}(p_3) \right\rangle = \pi^{\mu_1}_{\alpha_1}(p_1)\pi^{\mu_2}_{\alpha_2}(p_2)\pi^{\mu_3}_{\alpha_3}(p_3)\Big[ \\
\tilde{A}(p_1,p_2,p_3)\varepsilon^{p_1 p_2 \alpha_1 \alpha_2} p_1^{\alpha_3} + A(p_1,p_2,p_3)\varepsilon^{p_1 \alpha_1 \alpha_2 \alpha_3} - A(p_2,p_1,p_3)\varepsilon^{p_2 \alpha_1 \alpha_2 \alpha_3}\Big].
\end{aligned} \tag{4.4.5}$$

Differently from the $\langle J_V J_V J_A \rangle$ correlator, there is an additional Bose symmetry condition we have to consider: $\{p_1,\mu_1\} \leftrightarrow \{p_3,\mu_3\}$ and $\{p_2,\mu_2\} \leftrightarrow \{p_3,\mu_3\}$. Therefore, in this case, there is only one independent form factor contributing to the transverse part of the correlator. Indeed we have

$$\tilde{A}(p_1,p_2,p_3) = 2\,\frac{A(p_1,p_2,p_3) + A(p_2,p_3,p_1) + A(p_3,p_1,p_2)}{p_1^2 + p_2^2 + p_3^2} \tag{4.4.6}$$

and moreover

$$A(p_1,p_2,p_3) = \frac{(p_1^2 - p_2^2 + p_3^2)A(p_3,p_2,p_1) - 2p_2^2 A(p_1,p_3,p_2) - 2p_2^2 A(p_2,p_1,p_3)}{p_1^2 + p_2^2 + p_3^2} \tag{4.4.7}$$

For now on we will ignore such relations and we will treat $\tilde{A}$ and $A$ as independent quantities. We can still check later that our final result is invariant under the exchange of the currents.

### 4.4.2 Dilatation and special conformal Ward identities

The dilatation Ward identities of the transverse part are not affected by the longitudinal terms. Therefore the constraints are the same of the $\langle J_V J_V J_A \rangle$ correlator

$$\begin{aligned}
\sum_{i=1}^{3} p_i \frac{\partial \tilde{A}}{\partial p_i} - (\Delta_3 - 5)\tilde{A} = 0 \\
\sum_{i=1}^{3} p_i \frac{\partial A}{\partial p_i} - (\Delta_3 - 3)A = 0.
\end{aligned} \tag{4.4.8}$$

The invariance of the correlator with respect to the special conformal transformations is encoded in the following relation

$$0 = \pi^{\lambda_1}_{\mu_1}(p_1)\,\pi^{\lambda_2}_{\mu_2}(p_2)\,\pi^{\lambda_3}_{\mu_3}(p_3)\,\mathcal{K}^k\Big[\Big\langle j_A^{\mu_1} j_A^{\mu_2} j_A^{\mu_3}\Big\rangle + \Big\langle j_{A\,\text{loc}}^{\mu_1} j_A^{\mu_2} j_A^{\mu_3}\Big\rangle + \Big\langle j_A^{\mu_1} j_{A\,\text{loc}}^{\mu_2} j_A^{\mu_3}\Big\rangle + \Big\langle j_A^{\mu_1} j_A^{\mu_2} j_{A\,\text{loc}}^{\mu_3}\Big\rangle\Big]. \tag{4.4.9}$$

We procede in a manner similar to the $\langle J_V J_V J_A\rangle$ correlator, using the same Schouten identities. We can then decompose the contribution of the transverse part into the following form factors

$$\begin{aligned}
&\pi^{\lambda_1}_{\mu_1}(p_1)\pi^{\lambda_2}_{\mu_2}(p_2)\pi^{\lambda_3}_{\mu_3}(p_3)\Big(\mathcal{K}^\kappa\langle j_A^{\mu_1}(p_1) j_A^{\mu_2}(p_2) j_A^{\mu_3}(p_3)\rangle\Big) = \\
&= \pi^{\lambda_1}_{\mu_1}(p_1)\pi^{\lambda_2}_{\mu_2}(p_2)\pi^{\lambda_3}_{\mu_3}(p_3)\Big[p_1^\kappa\Big(C_{11}\,\varepsilon^{\mu_1\mu_2\mu_3 p_1} + C_{12}\,\varepsilon^{\mu_1\mu_2\mu_3 p_2} + C_{13}\,\varepsilon^{\mu_1\mu_2 p_1 p_2} p_1^{\mu_3}\Big) \\
&\qquad + p_2^\kappa\Big(C_{21}\,\varepsilon^{\mu_1\mu_2\mu_3 p_1} + C_{22}\,\varepsilon^{\mu_1\mu_2\mu_3 p_2} + C_{23}\,\varepsilon^{\mu_1\mu_2 p_1 p_2} p_1^{\mu_3}\Big) + C_{31}\varepsilon^{\kappa\mu_1\mu_2\mu_3} + C_{32}\varepsilon^{\kappa\mu_1\mu_2 p_1} p_1^{\mu_3} \\
&\qquad\qquad + C_{33}\varepsilon^{\kappa\mu_1\mu_2 p_2} p_1^{\mu_3} + C_{34}\varepsilon^{\kappa\mu_1 p_1 p_2}\delta^{\mu_2\mu_3} + C_{35}\varepsilon^{\kappa\mu_2 p_1 p_2}\delta^{\mu_1\mu_3} + C_{36}\varepsilon^{\kappa\mu_3 p_1 p_2}\delta^{\mu_1\mu_2}\Big],
\end{aligned}$$

where the explicit expression for the form factor is the same of the $\langle J_V J_V J_A\rangle$. However, both the primary and secondary equations will contain anomalous terms from the longitudinal parts of the correlators. Indeed, the primary Ward identities are given by

$$\begin{aligned}
&C_{11} = 0 && C_{21} = -\frac{16(d-2)a_1' i}{p_1^2} \\
&C_{12} = \frac{16(d-2)a_1' i}{p_2^2} && C_{22} = 0 \\
&C_{13} = 0 && C_{23} = 0
\end{aligned} \tag{4.4.10}$$

while the secondary equations can be written as

$$\begin{aligned}
&C_{31} = -\frac{8ia_1'(d-2)(p_1-p_2)(p_1+p_2)\left(p_1^2+p_2^2-p_3^2\right)}{p_1^2 p_2^2} && C_{32} = \frac{32ia_1'}{p_3^2} - \frac{16ia_1'(d-2)}{p_1^2} \\
&C_{33} = \frac{32ia_1'}{p_3^2} - \frac{16ia_1'(d-2)}{p_2^2} && C_{34} = -\frac{32ia_1'}{p_3^2} + \frac{16ia_1'(d-2)}{p_2^2} \\
&C_{35} = -\frac{16ia_1'(d-2)}{p_1^2} + \frac{32ia_1'}{p_3^2} && C_{36} = -\frac{16ia_1'(d-2)(p_1-p_2)(p_1+p_2)}{p_1^2 p_2^2}
\end{aligned} \tag{4.4.11}$$

The explicit form of the primary special conformal Ward identity is

$$\begin{aligned}
&K_{31}\tilde{A} = 0, && K_{32}\tilde{A} = 0, \\
&K_{31}A = 0, && K_{32}A = 2\left(\frac{d-2}{p_1^2} - \frac{1}{p_1}\frac{\partial}{\partial p_1}\right)A(p_1\leftrightarrow p_2) + 2\tilde{A} + \frac{16a_1' i(d-2)}{p_1^2} \\
&K_{31}A(p_1\leftrightarrow p_2) = 2\left(\frac{d-2}{p_2^2} - \frac{1}{p_2}\frac{\partial}{\partial p_2}\right)A - 2\tilde{A} + \frac{16a_1' i(d-2)}{p_2^2}, && K_{32}A(p_1\leftrightarrow p_2) = 0
\end{aligned} \tag{4.4.12}$$

These equations can also be reduced to a set of homogenous equations by repeatedly applying the operator $K_{ij}$

$$K_{31}\tilde{A} = 0, \qquad K_{32}\tilde{A} = 0, \tag{4.4.13}$$
$$K_{31}A = 0, \qquad K_{32}K_{32}A = 0, \tag{4.4.14}$$
$$K_{32}A(p_1 \leftrightarrow p_2) = 0, \qquad K_{31}K_{31}A(p_1 \leftrightarrow p_2) = 0 \tag{4.4.15}$$

As we can see, the presence of anomalous terms in the primary equations does not affect the structure of the homogenous equations which are exactly the same of the $\langle J_V J_V J_A \rangle$ correlator.

### 4.4.3 Solutions of the CWIs

The solutions of the primary homogeneous equations can be written in terms of the following 3K integrals, extending the previous approach of the $\langle J_V J_V J_A \rangle$

$$\tilde{A} = \alpha_1 J_{3\{0,0,0\}} = \alpha_1 I_{\frac{d}{2}+2\{\frac{d}{2}-1,\frac{d}{2}-1,\frac{d}{2}-1\}}, \tag{4.4.16}$$
$$A = \alpha_2 J_{1\{0,0,0\}} + \alpha_3 J_{2\{0,1,0\}} = \alpha_2 I_{\frac{d}{2}\{\frac{d}{2}-1,\frac{d}{2}-1,\frac{d}{2}-1\}} + \alpha_3 I_{\frac{d}{2}+1,\{\frac{d}{2}-1,\frac{d}{2},\frac{d}{2}-1\}}. \tag{4.4.17}$$

Inserting our solutions back to the nonhomogeneous equations, in the limit $p_3 \to 0$ we find

$$\left(\frac{d-4}{3}\right)\alpha_1 = -\frac{i a_1' \, 2^{6-\frac{d}{2}}(d-2)\sin\left(\frac{\pi d}{2}\right)}{\pi\Gamma\left(\frac{d}{2}+1\right)} + \alpha_2 + \alpha_3(d-2). \tag{4.4.18}$$

We then focus on the first two secondary equations in (4.4.11). The explicit form of the first equation is

$$\begin{aligned} 0 =& \frac{p_1^2+p_2^2-p_3^2}{p_1^2 p_2^2}\left\{-p_1^2 p_2^2\tilde{A} + (d-2)\left[p_1^2 A - p_2^2 A(p_2 \leftrightarrow p_1) + 8 i a_1'(p_1^2-p_2^2)\right]\right\} \\ &+ \left(p_1^2+p_2^2-p_3^2\right)\left(\frac{1}{p_1}\frac{\partial}{\partial p_1}A(p_2 \leftrightarrow p_1) - \frac{1}{p_2}\frac{\partial}{\partial p_2}A\right) - 2p_1\frac{\partial}{\partial p_1}A + 2p_2\frac{\partial}{\partial p_2}A(p_2 \leftrightarrow p_1) \end{aligned} \tag{4.4.19}$$

which leads to the condition

$$\alpha_2 = (d-2)\left(-\alpha_3 + \frac{i a_1' \, 2^{6-\frac{d}{2}}\sin\left(\frac{\pi d}{2}\right)}{\pi\Gamma\left(\frac{d}{2}+1\right)}\right). \tag{4.4.20}$$

The explicit form of the second equation in (4.4.11) is

$$\begin{aligned} 0 =& -2\,\frac{2p_1^2+(d-2)p_3^2}{p_1^2 p_3^2}A(p_2 \leftrightarrow p_1) + 2\left(1-\frac{p_1^2}{p_3^2}\right)\frac{1}{p_1}\frac{\partial}{\partial p_1}A(p_2 \leftrightarrow p_1) - \frac{2p_1}{p_3^2}\frac{\partial}{\partial p_1}A \\ &- \frac{2p_2}{p_3^2}\frac{\partial}{\partial p_2}A - \frac{2p_2}{p_3^2}\frac{\partial}{\partial p_2}A(p_2 \leftrightarrow p_1) - \frac{4A}{p_3^2} + \frac{p_1\left(-p_1^2+p_2^2+p_3^2\right)}{p_3^2}\frac{\partial}{\partial p_1}\tilde{A} \\ &+ \frac{p_2\left(-p_1^2+p_2^2-3p_3^2\right)}{p_3^2}\frac{\partial}{\partial p_2}\tilde{A} - \frac{2\left(2p_1^2-2p_2^2+p_3^2\right)\tilde{A}}{p_3^2} - 16\,a_1'\,i\left(\frac{d-2}{p_1^2} - \frac{2}{p_3^2}\right) \end{aligned} \tag{4.4.21}$$

which leads to the constraint

$$\alpha_3 = \frac{i\,a_1' \, 2^{7-\frac{d}{2}}(d-1)\sin\left(\frac{\pi d}{2}\right)}{\pi(d-4)\Gamma\left(\frac{d}{2}+1\right)}. \tag{4.4.22}$$

The other secondary equations don't impose any other constraints. We insert such conditions into our solution and use the following property of the 3K integral

$$I_{\frac{d}{2}+1\{\frac{d}{2}-1,\frac{d}{2},\frac{d}{2}-1\}} = p_2^2 I_{\frac{d}{2}+1\{\frac{d}{2}-1,\frac{d}{2}-2,\frac{d}{2}-1\}} + (d-2) I_{\frac{d}{2}\{\frac{d}{2}-1,\frac{d}{2}-1,\frac{d}{2}-1\}} \tag{4.4.23}$$

in order to arrive to the following expression in $d = 4+\epsilon$

$$\begin{aligned} &\tilde{A} = 0, \\ &A = 24\, i\, a_1' p_2^2 I_{3\{1,0,1\}} + 8\, i\, a_1' \epsilon\, I_{2+\frac{\epsilon}{2}\{1+\frac{\epsilon}{2},1+\frac{\epsilon}{2},1+\frac{\epsilon}{2}\}}. \end{aligned} \tag{4.4.24}$$

Note that we are keeping the second term of $A$ because the 3K integral has a pole in $\epsilon$. In the end we have

$$\begin{aligned} \langle j_A^{\mu_1}(p_1) j_A^{\mu_2}(p_2) j_A^{\mu_3}(p_3)\rangle &= \pi^{\mu_1}_{\alpha_1}(p_1)\pi^{\mu_2}_{\alpha_2}(p_2)\pi^{\mu_3}_{\alpha_3}(p_3)\, 24\, i\, a_1' \Big[ \\ &\Big(I_{3\{1,0,1\}} p_2^2 + \frac{\epsilon}{3} I_{2+\frac{\epsilon}{2}\{1+\frac{\epsilon}{2},1+\frac{\epsilon}{2},1+\frac{\epsilon}{2}\}}\Big)\varepsilon^{p_1\alpha_1\alpha_2\alpha_3} - \Big(I_{3\{0,1,1\}} p_1^2 + \frac{\epsilon}{3} I_{2+\frac{\epsilon}{2}\{1+\frac{\epsilon}{2},1+\frac{\epsilon}{2},1+\frac{\epsilon}{2}\}}\Big)\varepsilon^{p_2\alpha_1\alpha_2\alpha_3}\Big]. \end{aligned} \tag{4.4.25}$$

### 4.4.4 Connection with the $\langle J_V J_V J_A\rangle$ correlator

If the charge associated to the vector and axial field is the same, the chiral anomalous coefficients satisfy the following condition

$$a_1' = \frac{a_1}{3} \tag{4.4.26}$$

In this section, we are going to show that our conformal results are in accordance with the perturbative formula

$$\left\langle J_A^{\mu_1} J_A^{\mu_2} J_A^{\mu_3}\right\rangle = \frac{1}{3}\left[\left\langle J_A^{\mu_1} J_V^{\mu_2} J_V^{\mu_3}\right\rangle + \left\langle J_V^{\mu_1} J_A^{\mu_2} J_V^{\mu_3}\right\rangle + \left\langle J_V^{\mu_1} J_V^{\mu_2} J_A^{\mu_3}\right\rangle\right] \tag{4.4.27}$$

The proof of this statement is trivial for the longitudinal parts of the correlators; hence, we now examine the transverse parts. In order to prove such formula we use the relation

$$\left\langle j_V^{\mu_1}(p_1) j_V^{\mu_2}(p_2) j_A^{\mu_3}(p_3)\right\rangle = \pi^{\mu_1}_{\alpha_1}(p_1)\pi^{\mu_2}_{\alpha_2}(p_2)\pi^{\mu_3}_{\alpha_3}(p_3)\, 24\, i\, a_1' \left[I_{3\{1,0,1\}} p_2^2\, \varepsilon^{p_1\alpha_1\alpha_2\alpha_3} - I_{3\{0,1,1\}} p_1^2\, \varepsilon^{p_2\alpha_1\alpha_2\alpha_3}\right]. \tag{4.4.28}$$

Then, exchanging the second current with the third one, we have

$$\begin{aligned} \left\langle j_V^{\mu_1}(p_1) j_A^{\mu_2}(p_2) j_V^{\mu_3}(p_3)\right\rangle =& \pi^{\mu_1}_{\alpha_1}(p_1)\pi^{\mu_2}_{\alpha_2}(p_2)\pi^{\mu_3}_{\alpha_3}(p_3)\, 24\, i\, a_1' \Big[ \\ & -\left(p_1^2 I_{3\{0,1,1\}} + p_3^2 I_{3\{1,1,0\}}\right)\varepsilon^{p_1\alpha_1\alpha_2\alpha_3} - p_1^2 I_{3\{0,1,1\}}\, \varepsilon^{p_2\alpha_1\alpha_2\alpha_3}, \Big] \end{aligned} \tag{4.4.29}$$

while, exchanging the first current with the third one, we obtain

$$\begin{aligned} \left\langle j_A^{\mu_1}(p_1) j_V^{\mu_2}(p_2) j_V^{\mu_3}(p_3)\right\rangle =& \pi^{\mu_1}_{\alpha_1}(p_1)\pi^{\mu_2}_{\alpha_2}(p_2)\pi^{\mu_3}_{\alpha_3}(p_3)\, 24\, i\, a_1' \Big[ \\ & p_2^2 I_{3\{1,0,1\}}\, \varepsilon^{p_1\alpha_1\alpha_2\alpha_3} + \left(p_2^2 I_{3\{1,0,1\}} + p_3^2 I_{3\{1,1,0\}}\right)\varepsilon^{p_2\alpha_1\alpha_2\alpha_3} \Big] \end{aligned} \tag{4.4.30}$$

Summing the last three equations and using the following identity of the 3K integrals

$$p_1^2 I_{3\{0,1,1\}} + p_2^2 I_{3\{1,0,1\}} + p_3^2 I_{3\{1,1,0\}} = -\epsilon\, I_{2+\frac{\epsilon}{2}\{1+\frac{\epsilon}{2},1+\frac{\epsilon}{2},1+\frac{\epsilon}{2}\}} \tag{4.4.31}$$

we obtain exactly Eq. (4.4.25).

## 4.5 Comparison with other parameterizations

Having established the agreement between the perturbative (lowest order) and the nonperturbative computation of the $\langle J_V J_V J_A \rangle$, we try to relate the result of the expansion introduced in (4.2.13) with the two most popular parameterizations of the same vertex.
As we have already mentioned in the introduction, the original parameterization of the $\langle J_V J_V J_A \rangle$ was presented in [112]. Lorentz symmetry and parity fix the correlation function in the form

$$\begin{aligned}\langle J_V^{\mu_1}(p_1) J_V^{\mu_2}(p_2) J_A^{\mu_3}(p_3)\rangle &= B_1(p_1,p_2)\varepsilon^{p_1\mu_1\mu_2\mu_3} + B_2(p_1,p_2)\varepsilon^{p_2\mu_1\mu_2\mu_3} + B_3(p_1,p_2)\varepsilon^{p_1p_2\mu_1\mu_3}{p_1}^{\mu_2} \\ &+ B_4(p_1,p_2)\varepsilon^{p_1p_2\mu_1\mu_3}p_2^{\mu_2} + B_5(p_1,p_2)\varepsilon^{p_1p_2\mu_2\mu_3}p_1^{\mu_1} + B_6(p_1,p_2)\varepsilon^{p_1p_2\mu_2\mu_3}p_2^{\mu_1},\end{aligned} \tag{4.5.1}$$

with $B_1$ and $B_2$ divergent by power counting. If we use a diagrammatic evaluation of the correlator, the four invariant amplitudes $B_i$ for $i \geq 3$ are given by explicit parametric integrals [112]

$$\begin{aligned} B_3(p_1,p_2) &= -B_6(p_2,p_1) = 16\pi^2 I_{11}(p_1,p_2), \\ B_4(p_1,p_2) &= -B_5(p_2,p_1) = -16\pi^2\left[I_{20}(p_1,p_2) - I_{10}(p_1,p_2)\right], \end{aligned} \tag{4.5.2}$$

where the general massive $I_{st}$ integral is defined by

$$I_{st}(p_1,p_2) = \int_0^1 dw \int_0^{1-w} dz w^s z^t \left[z(1-z)p_1^2 + w(1-w)p_2^2 + 2wz(p_1\cdot p_2) - m^2\right]^{-1}. \tag{4.5.3}$$

Both $B_1$ and $B_2$ can be rendered finite by imposing the Ward identities on the two vector lines, giving

$$B_1(p_1,p_2) = p_1\cdot p_2\, B_3(p_1,p_2) + p_2^2\, B_4(p_1,p_2), \tag{4.5.4}$$

$$B_2(p_1,p_2) = p_1^2\, B_5(p_1,p_2) + p_1\cdot p_2\, B_6(p_1,p_2), \tag{4.5.5}$$

which allow to reexpress the formally divergent amplitudes in terms of the convergent ones. The Bose symmetry on the two vector vertices is fulfilled thanks to the relations

$$\begin{aligned} B_5\,(p_1,p_2) &= -B_4\,(p_2,p_1) \\ B_6\,(p_1,p_2) &= -B_3\,(p_2,p_1). \end{aligned} \tag{4.5.6}$$

Using the conservation WIs for the vector currents, we obtain the convergent expansion [116]

$$\begin{aligned}\langle J_V^{\mu_1} J_V^{\mu_2} J_A^{\mu_3}\rangle &= B_3(p_1\cdot p_2\varepsilon^{p_1\mu_1\mu_2\mu_3} + p_1^{\mu_2}\varepsilon^{p_1p_2\mu_1\mu_3}) + B_4(p_2\cdot p_2\varepsilon^{p_1\mu_1\mu_2\mu_3} + p_2^{\mu_2}\varepsilon^{p_1p_2\mu_1\mu_3}) \\ &+ B_5(p_1\cdot p_1\varepsilon^{p_2\mu_1\mu_2\mu_3} + p_1^{\mu_1}\varepsilon^{p_1p_2\mu_2\mu_3}) + B_6(p_1\cdot p_2\varepsilon^{p_2\mu_1\mu_2\mu_3} + p_2^{\mu_1}\varepsilon^{p_1p_2\mu_2\mu_3}) \\ &\equiv B_3\,\eta_3^{\mu_1\mu_2\mu_3}(p_1,p_2) + B_4\,\eta_4^{\mu_1\mu_2\mu_3}(p_1,p_2) + B_5\,\eta_5^{\mu_1\mu_2\mu_3}(p_1,p_2) + B_6\,\eta_6^{\mu_1\mu_2\mu_3}(p_1,p_2),\end{aligned} \tag{4.5.7}$$

where in the last step we have introduced four tensor structures $\eta_i$ that are mapped into one another under the Bose symmetry of the two vector lines. One can identifies six of them, as indicated in Table 4.1, but two of them

$$\eta_1^{\mu_1\mu_2\mu_3}(p_1,p_2) = p_1^{\mu_3}\,\varepsilon^{p_1p_2\mu_1\mu_2}, \qquad \eta_2^{\mu_1\mu_2\mu_3}(p_1,p_2) = p_2^{\mu_3}\,\varepsilon^{p_1p_2\mu_1\mu_2}, \tag{4.5.8}$$

are related by the Schouten relations to the other four, $\eta_3, \ldots \eta_6$. Indeed one has

$$\eta_1^{\mu_1\mu_2\mu_3}(p_1,p_2) = \eta_3^{\mu_1\mu_2\mu_3}(p_1,p_2) - \eta_5^{\mu_1\mu_2\mu_3}(p_1,p_2), \tag{4.5.9}$$

$$\eta_2^{\mu_1\mu_2\mu_3}(p_1,p_2) = \eta_4^{\mu_1\mu_2\mu_3}(p_1,p_2) - \eta_6^{\mu_1\mu_2\mu_3}(p_1,p_2). \tag{4.5.10}$$

| | |
|---|---|
| $\eta_1$ | $p_1^{\mu_3}\,\varepsilon^{p_1 p_2 \mu_1 \mu_2}$ |
| $\eta_2$ | $p_2^{\mu_3}\,\varepsilon^{p_1 p_2 \mu_1 \mu_2}$ |
| $\eta_3$ | $p_1 \cdot p_2 \varepsilon^{p_1 \mu_1 \mu_2 \mu_3} + p_1^{\mu_2}\,\varepsilon^{p_1 p_2 \mu_1 \mu_3}$ |
| $\eta_4$ | $p_2 \cdot p_2 \varepsilon^{p_1 \mu_1 \mu_2 \mu_3} + p_2^{\mu_2}\,\varepsilon^{p_1 p_2 \mu_1 \mu_3}$ |
| $\eta_5$ | $p_1 \cdot p_1 \varepsilon^{p_2 \mu_1 \mu_2 \mu_3} + p_1^{\mu_1}\,\varepsilon^{p_1 p_2 \mu_2 \mu_3}$ |
| $\eta_6$ | $p_1 \cdot p_2 \varepsilon^{p_2 \mu_1 \mu_2 \mu_3} + p_2^{\mu_1}\,\varepsilon^{p_1 p_2 \mu_2 \mu_3}$ |

Table 4.1: Tensor structures of odd parity in the expansion of the $\langle J_V J_V J_A \rangle$ with conserved vector currents.

The remaining tensor structures are inter-related by the Bose symmetry

$$\eta_3^{\mu_1\mu_2\mu_3}(p_1,p_2) = -\eta_6^{\mu_2\mu_1\mu_3}(p_2,p_1) \qquad \eta_4^{\mu_1\mu_2\mu_3}(p_1,p_2) = -\eta_5^{\mu_2\mu_1\mu_3}(p_2,p_1). \tag{4.5.11}$$

The correct counting of the independent form factors/tensor structures can be done only after we split each of them into their symmetric and antisymmetric components

$$\eta_i^{\mu_1\mu_2\mu_3} = \eta_i^{S\,\mu_1\mu_2\mu_3} + \eta_i^{A\,\mu_1\mu_2\mu_3} \qquad \eta_i^{S/A\,\mu_1\mu_2\mu_3} \equiv \frac{1}{2}\Big(\eta_i^{\mu_1\mu_2\mu_3}(p_1,p_2) \pm \eta_i^{\mu_2\mu_1\mu_3}(p_2,p_1)\Big) \equiv \eta_i^{\pm\,\mu_1\mu_2\mu_3} \tag{4.5.12}$$

with $i \geq 3$, giving

$$\eta_3^{+}(p_1,p_2) = -\eta_6^{+}(p_1,p_2) \qquad \eta_3^{-}(p_1,p_2) = \eta_6^{-}(p_1,p_2) \tag{4.5.13}$$

$$\eta_4^{+}(p_1,p_2) = -\eta_5^{+}(p_1,p_2) \qquad \eta_4^{-}(p_1,p_2) = \eta_5^{-}(p_1,p_2). \tag{4.5.14}$$

where we omitted all the tensorial indices which are in the order $\mu_1\,\mu_2\,\mu_3$. We can then reexpress the correlator as

$$\langle J_V J_V J_A \rangle = B_3^{+}\eta_3^{+} + B_3^{-}\eta_3^{-} + B_4^{+}\eta_4^{+} + B_4^{-}\eta_4^{-} \tag{4.5.15}$$

in terms of four tensor structures of definite symmetry times 4 independent form factors.

### 4.5.1 L/T decomposition

An alternative parameterization of the $\langle J_V J_V J_A \rangle$ correlator, which allows to set a direct comparison with the one that we have introduced in the previous sections is given by [113]

$$\langle J_V^{\mu_1}(p_1) J_V^{\mu_2}(p_2) J_A^{\mu_3}(p_3) \rangle = \frac{1}{8\pi^2}\Big(W^{L\,\mu_1\mu_2\mu_3} - W^{T\,\mu_1\mu_2\mu_3}\Big) \tag{4.5.16}$$

where the longitudinal component is specified in eq. (4.1.4), while the transverse component is given by

$$\begin{aligned} W^{T\,\mu_1\mu_2\mu_3}(p_1,p_2,p_3^2) = {} & w_T^{(+)}\left(p_1^2,p_2^2,p_3^2\right) t^{(+)\,\mu_1\mu_2\mu_3}(p_1,p_2) + w_T^{(-)}\left(p_1^2,p_2^2,p_3^2\right) t^{(-)\,\mu_1\mu_2\mu_3}(p_1,p_2) \\ & + \widetilde{w}_T^{(-)}\left(p_1^2,p_2^2,p_3^2\right) \widetilde{t}^{(-)\,\mu_1\mu_2\mu_3}(p_1,p_2), \end{aligned} \tag{4.5.17}$$

This decomposition automatically account for all the symmetries of the correlator with the transverse tensors given by

$$
\begin{aligned}
t^{(+)\mu_1\mu_2\mu_3}(p_1,p_2) &= p_1^{\mu_2}\,\varepsilon^{\mu_1\mu_3 p_1 p_2} - p_2^{\mu_1}\,\varepsilon^{\mu_2\mu_3 p_1 p_2} - (p_1\cdot p_2)\,\varepsilon^{\mu_1\mu_2\mu_3(p_1-p_2)} + \frac{p_1^2+p_2^2-p_3^2}{p_3^2}\,(p_1+p_2)^{\mu_3}\,\varepsilon^{\mu_1\mu_2 p_1 p_2},\\
t^{(-)\mu_1\mu_2\mu_3}(p_1,p_2) &= \left[(p_1-p_2)^{\mu_3} - \frac{p_1^2-p_2^2}{p_3^2}\,(p_1+p_2)^{\mu_3}\right]\varepsilon^{\mu_1\mu_2 p_1 p_2},\\
\tilde{t}^{(-)\mu_1\mu_2\mu_3}(p_1,p_2) &= p_1^{\mu_2}\,\varepsilon^{\mu_1\mu_3 p_1 p_2} + p_2^{\mu_1}\,\varepsilon^{\mu_2\mu_3 p_1 p_2} - (p_1\cdot p_2)\,\varepsilon^{\mu_1\mu_2\mu_3(p_1+p_2)}.
\end{aligned}
\tag{4.5.18}
$$

The map between the Rosenberg representation and the current one is given by the relations

$$
B_3(p_1,p_2) = \frac{1}{8\pi^2}\left[w_L - \tilde{w}_T^{(-)} - \frac{p_1^2+p_2^2}{p_3^2}w_T^{(+)} - 2\,\frac{p_1\cdot p_2+p_2^2}{p_3^2}w_T^{(-)}\right], \tag{4.5.19}
$$

$$
B_4(p_1,p_2) = \frac{1}{8\pi^2}\left[w_L + 2\,\frac{p_1\cdot p_2}{p_3^2}w_T^{(+)} + 2\,\frac{p_1\cdot p_2+p_1^2}{k^2}w_T^{(-)}\right], \tag{4.5.20}
$$

and viceversa

$$
w_L(p_1^2,p_2^2,p_3^2) = \frac{8\pi^2}{p_3^2}\,[B_1 - B_2] \tag{4.5.21}
$$

or, after the imposition of the Ward identities in Eqs.(4.5.4,4.5.5),

$$
\begin{aligned}
w_L(p_1^2,p_2^2,p_3^2) &= \frac{8\pi^2}{p_3^2}\left[(B_3-B_6)p_1\cdot p_2 + B_4\,p_2^2 - B_5\,p_1^2\right], && (4.5.22)\\
w_T^{(+)}(p_1^2,p_2^2,p_3^2) &= -4\pi^2\,(B_3-B_4+B_5-B_6), && (4.5.23)\\
w_T^{(-)}(p_1^2,p_2^2,p_3^2) &= 4\pi^2\,(B_4+B_5), && (4.5.24)\\
\tilde{w}_T^{(-)}(p_1^2,p_2^2,p_3^2) &= -4\pi^2\,(B_3+B_4+B_5+B_6). && (4.5.25)
\end{aligned}
$$

As already mentioned, (4.5.21) is a special relation, since it shows that the pole is not affected by Chern-Simons forms, showing us of the physical character of this part of the interaction.
Also in this case, the counting of the form factor is four, one for the longitudinal pole part and 3 for the transverse part. Notice that all of them are either symmetric or antisymmetric by construction.

$$
\begin{aligned}
w_L(p_1^2,p_2^2,p_3^2) &= w_L(p_2^2,p_1^2,p_3^2)\\
w_T^{(+)}(p_1^2,p_2^2,p_3^2) &= w_T^{(+)}(p_2^2,p_1^2,p_3^2)\\
w_T^{(-)}(p_1^2,p_2^2,p_3^2) &= -w_T^{(-)}(p_2^2,p_1^2,p_3^2)\\
\tilde{w}_T^{(-)}(p_1^2,p_2^2,p_3^2) &= -\tilde{w}_T^{(-)}(p_2^2,p_1^2,p_3^2).
\end{aligned}
\tag{4.5.26}
$$

To relate this decomposition to our, we apply the transverse projectors and obtain

$$
\begin{aligned}
&\pi^{\lambda_1}_{\mu_1}(p_1)\pi^{\lambda_2}_{\mu_2}(p_2)\pi^{\lambda_3}_{\mu_3}(p_3)\Big(W^{T\,\mu_1\mu_2\mu_3}(p_1,p_2,p_3)\Big) =\\
&= \pi^{\lambda_1}_{\mu_1}(p_1)\pi^{\lambda_2}_{\mu_2}(p_2)\pi^{\lambda_3}_{\mu_3}(p_3)\bigg[\Big(w_T^{(+)}+\tilde{w}_T^{(-)}\Big)p_3^{\alpha_2}\,\varepsilon^{\alpha_1\alpha_3 p_1 p_2} - 2\,w_T^{(-)}\,p_1^{\alpha_3}\,\varepsilon^{\alpha_1\alpha_2 p_1 p_2}\\
&\quad + \Big(w_T^{(+)}-\tilde{w}_T^{(-)}\Big)p_2^{\alpha_1}\,\varepsilon^{\alpha_2\alpha_3 p_1 p_2} + \Big(w_T^{(+)}+\tilde{w}_T^{(-)}\Big)(p_1\cdot p_2)\varepsilon^{\alpha_1\alpha_2\alpha_3 p_1} - \Big(w_T^{(+)}-\tilde{w}_T^{(-)}\Big)(p_1\cdot p_2)\varepsilon^{\alpha_1\alpha_2\alpha_3 p_2}\bigg],
\end{aligned}
\tag{4.5.27}
$$

and by using the Schouten identities

$$\pi^{\mu_1}_{\alpha_1}\pi^{\mu_2}_{\alpha_2}\pi^{\mu_3}_{\alpha_3}\left(p_2^{\alpha_1}\,\varepsilon^{p_1p_2\alpha_2\alpha_3}\right)=\pi^{\mu_1}_{\alpha_1}\pi^{\mu_2}_{\alpha_2}\pi^{\mu_3}_{\alpha_3}\left(-\left(p_2\cdot p_1\right)\varepsilon^{p_2\alpha_1\alpha_2\alpha_3}+p_2^2\,\varepsilon^{p_1\alpha_1\alpha_2\alpha_3}+p_1^{\alpha_3}\,\varepsilon^{p_1p_2\alpha_1\alpha_2}\right),\tag{4.5.28}$$

$$\pi^{\mu_1}_{\alpha_1}\pi^{\mu_2}_{\alpha_2}\pi^{\mu_3}_{\alpha_3}\left(p_3^{\alpha_2}\,\varepsilon^{p_1p_2\alpha_1\alpha_3}\right)=\pi^{\mu_1}_{\alpha_1}\pi^{\mu_2}_{\alpha_2}\pi^{\mu_3}_{\alpha_3}\left(-p_1^2\,\varepsilon^{p_2\alpha_1\alpha_2\alpha_3}+\left(p_1\cdot p_2\right)\varepsilon^{p_1\alpha_1\alpha_2\alpha_3}-p_1^{\alpha_3}\,\varepsilon^{p_1p_2\alpha_1\alpha_2}\right),\tag{4.5.29}$$

we obtain

$$\begin{aligned}\pi^{\lambda_1}_{\mu_1}(p_1)\pi^{\lambda_2}_{\mu_2}(p_2)\pi^{\lambda_3}_{\mu_3}(p_3)\left(W^{T\,\mu_1\mu_2\mu_3}(p_1,p_2,p_3)\right)=\pi^{\lambda_1}_{\mu_1}(p_1)\pi^{\lambda_2}_{\mu_2}(p_2)\pi^{\lambda_3}_{\mu_3}(p_3)\Big\{-p_1^2\left(w_T^{(+)}+\widetilde{w}_T^{(-)}\right)\varepsilon^{p_2\alpha_1\alpha_2\alpha_3}\\+p_2^2\left(w_T^{(+)}-\widetilde{w}_T^{(-)}\right)\varepsilon^{p_1\alpha_1\alpha_2\alpha_3}-2\left(\widetilde{w}_T^{(-)}+w_T^{(-)}\right)p_1^{\alpha_3}\,\varepsilon^{p_1p_2\alpha_1\alpha_2}\Big\}.\end{aligned}\tag{4.5.30}$$

We then identify the form factors of our decomposition and the current L/T one in the form

$$\begin{aligned}A_1&=\frac{1}{4\pi^2}\left(\widetilde{w}_T^{(-)}+w_T^{(-)}\right),\\A_2&=-\frac{1}{8\pi^2}\,p_2^2\left(w_T^{(+)}-\widetilde{w}_T^{(-)}\right).\end{aligned}\tag{4.5.31}$$

Notice that $A_1$ is antisymmetric in the exchange of the two vector lines and counts for one independent form factor, while $A_2$ contains both symmetric and antisymmetric components and counts as two. Combined with $w_L$, we again find that our form factors are four in the general case, before enforcing the conformal WIs on the parameterization. One can check from the solution of the CWIs that this number is reduced by one in both representations, since in this case

$$\widetilde{w}_T^{(-)}=-w_T^{(-)}\tag{4.5.32}$$

for the L/T one. In our case the form factor $A_1$ vanishes

$$A_1=0,\tag{4.5.33}$$

and we are left with three form factors in both cases. Proceeding in a similar manner, we can also map the Rosenberg parametrization into the one we worked in. The results are

$$\begin{aligned}A_1&=B_3-B_6\\A_2&=p_2^2(B_6+B_4).\end{aligned}\tag{4.5.34}$$

## 4.6 Nonrenormalization theorems

In this section we establish a connection between the conformal solutions of correlators of mixed chirality and the nonrenormalization theorems presented in [113]. nonrenormalization theorems identify constraints between $w_L$, the anomalous pole part, and combinations of transverse form factors, which are not affected by radiative corrections. The authors of [113] examine in particular the $\langle J_LJ_VJ_R\rangle$ correlator, where we have defined the left and right chiral currents in the following way

$$J_L\equiv\frac{1}{2}(J_V-J_A),\qquad J_R\equiv\frac{1}{2}(J_V+J_A)\tag{4.6.1}$$

As we are going to show, the building block of the $\langle J_L J_V J_R\rangle$ is the $\langle J_V J_V J_A\rangle$ correlator. All the other diagrams, in the chiral limit, such as the $\langle J_A J_V J_V\rangle$ or $\langle J_V J_A J_V\rangle$ or $\langle J_A J_A J_A\rangle$, are trivially related to the $\langle J_V J_V J_A\rangle$ due to the anticommuting property of $\gamma^5$ and the symmetry constrain

$$\langle J_A J_A J_A\rangle = \frac{1}{3}\left(\langle J_A J_V J_V\rangle + \langle J_V J_A J_V\rangle + \langle J_V J_V J_A\rangle\right) \tag{4.6.2}$$

as illustrated in Fig. 4.2. In perturbation theory, the anomaly ($a_1'$) content of the $\langle J_A J_A J_A\rangle$, for instance, can be determined on the basis of symmetry, assuming an equal sharing ($a_1/3$) of the anomaly for each external axial-vector line. The constraint can be used as a starting point for moving the anomaly around the vertices, by the inclusion of appropriate Chern-Simons terms.
From the definition of the chiral currents, one can derive

$$\langle J_L J_V J_R\rangle = \frac{1}{4}\Big[\langle J_V J_V J_V\rangle - \langle J_A J_V J_V\rangle - \langle J_A J_V J_A\rangle + \langle J_V J_V J_A\rangle\Big]. \tag{4.6.3}$$

Then, using the charge conjugation invariance, the authors of [113] set the parity-even contribution $\langle J_V J_V J_V\rangle$ and $\langle J_A J_V J_A\rangle$ to zero. Alternatively, one can assume that the correlator is conformally invariant and arrive to the same conclusion [29]. We are going to prove this point in the next section

### 4.6.1 The abelian $\langle JJJ\rangle_{even}$ in CFT

Conformal invariance requires the abelian correlator $\langle JJJ\rangle_{even}$ to be zero in any dimension. In this section, we will briefly demonstrate this statement. This correlator can be realized, for example, with three vector currents $J_V$ or with one vector current and two axial currents $J_A$. Note that, in the latter case, the axial anomaly does not appear in the correlator which is parity-even. We start our derivation by imposing the following conservation Ward identities

$$p_{i\mu_i}\,\langle J^{\mu_1}(p_1)J^{\mu_2}(p_2)J^{\mu_3}(p_3)\rangle = 0, \qquad i = 1,2,3 \tag{4.6.4}$$

Therefore the correlator is purely transverse and can be expressed as

$$\begin{aligned}\langle J^{\mu_1}(p_1)J^{\mu_2}(p_2)J^{\mu_3}(p_3)\rangle = \pi^{\mu_1}_{\alpha_1}(p_1)\pi^{\mu_2}_{\alpha_2}(p_2)\pi^{\mu_3}_{\alpha_3}(p_3)\Big[A_1(p_1,p_2,p_3)\,p_2^{\alpha_1}p_3^{\alpha_2}p_1^{\alpha_3}\\ +A_2(p_1,p_2,p_3)\,\delta^{\alpha_1\alpha_2}p_1^{\alpha_3} + A_2(p_3,p_1,p_2)\,\delta^{\alpha_1\alpha_3}p_3^{\alpha_2} + A_2(p_2,p_3,p_1)\,\delta^{\alpha_2\alpha_3}p_2^{\alpha_1}\Big].\end{aligned} \tag{4.6.5}$$

The $A_1$ form factor is completely antisymmetric for any permutation of the momenta while the form factor $A_2$ is antisymmetric under ($p_1 \leftrightarrow p_2$). We now list the conformal constraints on the form factors. The dilatation Ward identities are

$$\begin{aligned}&\sum_{i=1}^{3} p_i\frac{\partial A_1}{\partial p_i} - (d-6)A_1 = 0,\\ &\sum_{i=1}^{3} p_i\frac{\partial A_2}{\partial p_i} - (d-4)A_2 = 0,\end{aligned} \tag{4.6.6}$$

while, the primary special conformal Ward identities are given by

$$\begin{aligned}&K_{12}A_1 = 0, \qquad && K_{13}A_1 = 0,\\ &K_{12}A_2 = 0, && K_{13}A_2 = 2A_1.\end{aligned} \tag{4.6.7}$$

The solution to such equations can be written in terms of the following 3K integrals

$$\begin{aligned} A_1 &= \alpha_1 J_{3\{000\}} \\ A_2 &= \alpha_1 J_{2\{001\}} + \alpha_2 J_{1\{000\}} \end{aligned} \tag{4.6.8}$$

Lastly, we need to impose the secondary special conformal Ward identities, which are given by

$$\begin{aligned} &L_3[A_1(p_1,p_2,p_3)] + 2R[A_2(p_1,p_2,p_3) - A_2(p_3,p_1,p_2)] = 0 \\ &L_1[A_2(p_2,p_3,p_1)] + 2p_1^2 A_2(p_3,p_1,p_2) - 2p_1^2 A_2(p_1,p_2,p_3) = 0 \end{aligned} \tag{4.6.9}$$

where we have defined the following operators

$$\begin{aligned} \mathrm{L}_N &= p_1\left(p_1^2+p_2^2-p_3^2\right)\frac{\partial}{\partial p_1} + 2p_1^2 p_2 \frac{\partial}{\partial p_2} \\ &\quad + \left[(2d-\Delta_1-2\Delta_2+N)p_1^2 + (2\Delta_1-d)\left(p_3^2-p_2^2\right)\right] \\ \mathrm{R} &= p_1\frac{\partial}{\partial p_1} - (2\Delta_1-d) \end{aligned} \tag{4.6.10}$$

Inserting our solution (4.6.8) into such equations, we arrive to the conditions $\alpha_1 = \alpha_2 = 0$. Therefore, the correlator vanishes.

### 4.6.2 Nonrenormalization theorems from conformal invariance

As we have just proven by imposing conformal invariance, two of the four terms in Eq. (4.6.3) vanish. Therefore, one can write

$$\langle J_L J_V J_R\rangle = \frac{1}{4}\Big(\langle J_V J_V J_A\rangle - \langle J_A J_V J_V\rangle\Big). \tag{4.6.11}$$

where we used Eq. (4.6.2).

The authors of [113] assumed that the $\langle J_L J_V J_R\rangle$ correlator is simply a Chern-Simons term, in order to prove the nonrenormalization theorems. Using our conformal solution, we can directly write the expression for the $\langle J_L J_V J_R\rangle$ correlator and prove their statement. Indeed for the longitudinal part we can write

$$\langle J_V J_V J_A\rangle_{loc} - \langle J_A J_V J_V\rangle_{loc} = -8\, i\, a_1\left[\frac{p_3^{\mu_3}}{p_3^2}\varepsilon^{p_1 p_2 \mu_1 \mu_2} - \frac{p_1^{\mu_1}}{p_1^2}\varepsilon^{p_1 p_2 \mu_2 \mu_3}\right] \tag{4.6.12}$$

while for the transverse part, after contracting the projectors in our solution, we have

$$\begin{aligned} \langle J_V J_V J_A\rangle_{transv} - \langle J_A J_V J_V\rangle_{transv} &= 8\, i\, a_1\, \epsilon\, I_{2+\frac{\epsilon}{2},\{1+\frac{\epsilon}{2},1+\frac{\epsilon}{2},1+\frac{\epsilon}{2}\}}\left[\varepsilon^{p_2\mu_1\mu_2\mu_3} - \frac{p_3^{\mu_3}}{p_3^2}\varepsilon^{p_1p_2\mu_1\mu_2} + \frac{p_1^{\mu_1}}{p_1^2}\varepsilon^{p_1p_2\mu_2\mu_3}\right] \\ &= -8\, i\, a_1\left[\varepsilon^{p_2\mu_1\mu_2\mu_3} - \frac{p_3^{\mu_3}}{p_3^2}\varepsilon^{p_1p_2\mu_1\mu_2} + \frac{p_1^{\mu_1}}{p_1^2}\varepsilon^{p_1p_2\mu_2\mu_3}\right] \end{aligned} \tag{4.6.13}$$

where in the last passage we used the explicit expression of the 3K integral. Adding the contributions together, we arrive to

$$\langle J_L J_V J_R\rangle = \langle J_V J_V J_A\rangle - \langle J_A J_V J_V\rangle = -8\, i\, a_1\, \varepsilon^{p_2\mu_1\mu_2\mu_3}. \tag{4.6.14}$$

which tells us that the $\langle J_L J_V J_R\rangle$ correlator is simply given by a Chern-Simons term, as expected. Note that, proceeding in a similar manner, one can prove Eq. (4.1.6) too.

From this consideration, as shown in [113], one can derive the following nonrenormalization theorems, which were originally obtained in the chiral limit of perturbative QCD

$$0 = \left\{ \left[ w_T^{(+)} + w_T^{(-)} \right] \left( q_1^2, q_2^2, (q_1+q_2)^2 \right) - \left[ w_T^{(+)} + w_T^{(-)} \right] \left( (q_1+q_2)^2, q_2^2, q_1^2 \right) \right\}_{\text{pQCD}} \tag{4.6.15}$$

$$0 = \left\{ \left[ \tilde{w}_T^{(-)} + w_T^{(-)} \right] \left( q_1^2, q_2^2, (q_1+q_2)^2 \right) + \left[ \tilde{w}_T^{(-)} + w_T^{(-)} \right] \left( (q_1+q_2)^2, q_2^2, q_1^2 \right) \right\}_{\text{pQCD}} \tag{4.6.16}$$

and

$$\begin{aligned} &\left\{ \left[ w_T^{(+)} + \tilde{w}_T^{(-)} \right] \left( q_1^2, q_2^2, (q_1+q_2)^2 \right) + \left[ w_T^{(+)} + \tilde{w}_T^{(-)} \right] \left( (q_1+q_2)^2, q_2^2, q_1^2 \right) \right\}_{\text{pQCD}} - w_L \left( (q_1+q_2)^2, q_2^2, q_1^2 \right) \\ &\qquad = - \left\{ \frac{2(q_2^2 + q_1 \cdot q_2)}{q_1^2} w_T^{(+)} \left( (q_1+q_2)^2, q_2^2, q_1^2 \right) - 2 \frac{q_1 \cdot q_2}{q_1^2} w_T^{(-)} \left( (q_1+q_2)^2, q_2^2, q_1^2 \right) \right\}_{\text{pQCD}} \end{aligned} \tag{4.6.17}$$

Naturally, one can also directly substitute our conformal solution into these equations to confirm their validity. In the end, the nonrenormalization theorems follow directly from conformal invariance. Moreover, invariance under conformal transformations imposes an even stronger condition than the nonrenormalization theorems stated above. Specifically, as we have shown, the entire transverse part is determined in terms of the longitudinal part, which, as we know, is protected from loop corrections. Consequently, one might also conclude that the transverse part is similarly protected from such corrections in a CFT. Indeed, the nonvanishing loop corrections to the transverse part, computed in [108], explicitly break conformal invariance and are proportional to the $\beta$ function.

## 4.7 Conclusions

In this work we have illustrated how the CWIs in momentum-space can be used to determine the structure of chiral anomaly diagrams. This shows that anomalies can be treated consistently in this specific CFT framework, that allows to establish a link between such correlators and the ordinary perturbative amplitudes. Parity-odd correlators are important in many physical context, and in this case we have shown how the conformal properties of previous perturbative analysis, performed in free field theory in the chiral limit, are the result of conformal symmetry and of its constraints. Our derivation does not rely on any Lagrangian realization. The analysis of parity-odd correlators, especially for 4-point functions, is still under investigation, given its complexity, and the inclusion of the anomaly content is for certainly an interesting aspect that deserves close attention.

# Chapter 5

# The gravitational anomaly in CFT

In this chapter, we illustrate how the conformal Ward identities and the gravitational chiral anomaly completely determine the structure of the $\langle TTJ_A\rangle$ correlator in momentum-space. Our approach is nonperturbative and not Lagrangian-based, requiring the inclusion of a single anomaly pole in the solution of the anomaly constraint. The pole and its residue, along with the CWIs, determine the entire correlator in all of its sectors (longitudinal/transverse), all of which are proportional to the same anomaly coefficient. The method does not rely on a specific expression of the CP-odd anomalous current, which in free field theory can be represented either by a bilinear fermion current or by a gauge-dependent Chern-Simons current; the analysis relies solely on the symmetry constraints. The $\langle TTJ_A\rangle$ plays a fundamental role in the study of the conformal backreaction in early universe cosmology, affecting the particle content and the evolution of the primordial plasma. Furthermore, we compute the correlator perturbatively at one-loop in free field theory and verify its exact agreement with the nonperturbative result. A comparison with the perturbative analysis confirms the presence of a sum rule satisfied by the correlator, similar to the parity-even $\langle TJJ\rangle$ and the chiral $\langle J_VJ_VJ_A\rangle$.

## 5.1 Introduction

### 5.1.1 The gravitational anomaly and the $\langle TTJ_A\rangle$

Gravitational anomalies generated by spin 1/2 and spin 3/2 particles have been extensively studied in several works since the 70's, due to their connection with ordinary gauge theories [51, 52], supergravity [117] and self-dual antisymmetric fields in string theory [53], just to mention a few (see also [118] for a recent study on the properties of chiral anomalies in the context of black holes)

$$\nabla_\mu\langle J_A^\mu\rangle = a_1\,\varepsilon^{\mu\nu\rho\sigma}F_{\mu\nu}F_{\rho\sigma} + a_2\,\varepsilon^{\mu\nu\rho\sigma}R^\alpha_{\ \beta\mu\nu}R^\beta_{\ \alpha\rho\sigma} \tag{5.1.1}$$

The gravitational anomaly $R\tilde{R}$ can appear in different settings. The nonperturbative approach developed in this chapter is general and adaptable to many contexts, yet the anomaly's impact can vary from benign to dangerous, based on the circumstances. Consider, for example, a scenario involving a Dirac fermion interacting with gravity and a vector potential $V_\mu$. Kimura, Delbourgo and Salam were the first to compute the anomaly in this case, observed in the divergence of $J_A$ [51, 52]. This specific anomaly poses no threat and may be of interest in phenomenology. For convenience, we can also introduce an axial-vector field $A_\mu$, which couples to $J_A$, but only as an external source since an anomalous gauge symmetry for $A_\mu$ would spoil unitarity and renormalizability.
Another instance involves a chiral model incorporating a Weyl fermion $\psi_{L/R}$ interacting with gravity and a gauge field. In this case, the anomaly emerges in the divergence of $J_{L/R}$, potentially endangering unitarity and renormalization, unless it is canceled [53] (see [54] for a detailed account on the types of chiral anomalies).
In general, in perturbation theory, the evaluation of a chiral trace of Dirac matrices hinges on the choice of a specific regularization and the related treatment of the antisymmetric $\varepsilon$ tensor in the loop. In the case of the Breitenlohner-Maison-'t Hooft-Veltman scheme [48], for example, the anomaly of parity-odd correlators is present only on the Ward identity of the chiral current, while the energy-momentum tensor and the vector currents are conserved. In other regularizations, one can potentially find a violation of the latter as well.
The correlator under scrutiny in this work is the $\langle TTJ_A\rangle$, reinvestigated using CFT in momentum-space. This correlator involves two energy-momentum tensors and one parity-odd current, denoted as $J_A$. A study of this correlator was previously discussed in [119] using coordinate-space methods. In the Standard Model, when $J_A$ is the non-Abelian $SU(2)$ gauge current or the hypercharge gauge current, this anomaly cancels out by summing over the chiral spectrum of each fermion generation. This feature is usually interpreted as an indication of the compatibility of the Standard Model when coupled to a gravitational background, providing an essential constraint on its possible extensions. The correlator plays a crucial role in mediating anomalies of global currents associated with baryon ($B$) and lepton ($L$) numbers in the presence of gravity.
In condensed matter theory, correlators affected both by chiral and conformal anomalies, as well as by discrete anomalies, play an important role in the context of topological materials [10–14]. The gravitational anomaly has been investigated, in the same context, in other interesting works [56, 57]. In our analysis, we demonstrate how investigating CFT in momentum-space allows us to independently reproduce previous results found in coordinate-space [119].
The solution is uniquely constructed by assuming the exchange of a single anomaly pole in the longitudinal sector of this correlator when we proceed with its sector decomposition. We will show that the momentum-space solution, derived from the CFT constraints, is unique and depends on a single constant: the anomaly coefficient at the pole.
This result appears to be a common feature in correlation functions that are finite and affected by parity-odd anomalies, complementing our previous analysis of similar correlation functions such as the $\langle J_V J_V J_A\rangle$.
In this chapter we decompose the correlator into all of its sectors and derive the conformal equations for its reconstruction [29]. We will first proceed with a parametrizaton of the form factors and tensors structures of the transverse-traceless sector. We introduce a form factor in the longitudinal part, in the form of an anomaly pole, and proceed with a complete determination of the entire correlation function by solving the equations of all the remaining sectors. We follow the steps introduced in [29], extended to the parity-odd case, and split the equations into primary and secondary CWIs. The solution, as we are going to show, will coincide with the perturbative one and will depend on a single constant, the coefficient of the anomaly. The off-shell parametrization of the vertex that results from this construction is quite economical, and is expressed in terms of only two form factors in the transverse-traceless sector, plus the anomaly form factor that takes the form of a $1/p_3^2$ anomaly pole. The anomaly, in this formulation, is the residue at the pole.

### 5.1.2 Fermionic and Chern-Simons currents

In free field theory, two realizations of currents have been discussed for the $\langle TTJ_A\rangle$ correlator: the bilinear axial fermion current (($J_A = J_{5f}$) and the bilinear gauge-dependent Chern-Simons (CS) current ($J_A = J_{CS}$) [95,120]. We recall that in previous analysis it has been shown that both currents of the form

$$J^{\lambda}_{5f} = -ig\bar{\psi}\gamma^5\gamma^\lambda\psi \tag{5.1.2}$$

or of the Chern-Simons form

$$J^{\lambda}_{CS} = \varepsilon^{\lambda\mu\nu\rho}V_\mu\partial_\nu V_\rho, \tag{5.1.3}$$

could be considered in a perturbative realization of the $\langle TTJ_A\rangle$ correlator and generate a gravitational anomaly. Notice that this second version of the current can be incorporated into an ordinary partition function - in an ordinary Lagrangian realization by a path integral - only in the presence of a coupling to an axial-vector gauge field ($A_\lambda$) via an interaction of the form

$$\mathcal{S}_{AVF} \equiv \int d^4x\sqrt{g}A_\lambda J^{\lambda}_{CS}. \tag{5.1.4}$$

The term, usually denoted as $A\,V \wedge F_V$, is the Abelian Chern-Simons form that allows to move the anomaly from one vertex to another in the usual $\langle J_V J_V J_A\rangle$ diagram. Details on these point can be found in [115,121].

$J_{CS}$ is responsible for mediating the gravitational chiral anomaly with spin-1 virtual photons in the loops, resulting in a difference between their two circular modes and inducing an optical helicity. This interaction is relevant in early universe cosmology and has an impact on the polarization of the Cosmic Microwave Background (CMB) [97].

In this case, the classical symmetry to be violated is the discrete duality invariance ($E \to B$, $B \to -E$) of the Maxwell equations in the vacuum (see [96,122]). The $\langle TTJ_A\rangle$ correlator induces similar effects on gravitational waves [4,5]. Spectral asymmetries induced by chiral anomalies, particularly the ordinary chiral anomaly (the $F\tilde{F} \sim E\cdot B$ term), have been investigated for their impact on the evolution of the primordial plasma, affecting the magneto-hydrodynamical equations and the generation of cosmological magnetic fields [6–8].

As previously mentioned, our method exclusively exploits the correlator's symmetries to identify its structure, which remains identical for a generic parity-odd $J_A$. In both cases ($J_{5f}$ and $J_{CS}$), the solution is entirely centered around the anomaly pole, serving as a pivot for the complete reconstruction of the corresponding correlators.

Both realizations of the $\langle TTJ_A\rangle$ correlator - using a $J_{CS}$ current or a $J_{5f}$ current - have been shown in [95, 120] to reduce to the exchange of an anomaly pole for on-shell gluons and photons for the unique form factors present in the diagrams.

In these works, the authors introduced a mass deformation of the propagators in the loops and showed the emergence of the pole as the mass was sent to zero. The method relies on the spectral density of the amplitude and it has been used also more recently in [64] and [109], in studies of the parity-even $\langle TJJ\rangle$ and in supersymmetric variants.

We comment on this point in Section 5.6 and illustrate, by a simple computation, that the spectral densities of the only surviving form factors in the on-shell $\langle TTJ_A\rangle$, with $J_A \equiv J_{5f}$ and $J_A \equiv J_{CS}$, satisfy two (mass-independent) sum rules.

## 5.2 Decomposition of the correlator

In this section we find the most general expression of the $\langle TTJ_A\rangle$ correlator, satisfying the anomalous conservation WI and trace WI. The analysis is performed by applying the L/T decomposition to the correlator. We focus on a parity-odd four-dimensional correlator, therefore its tensorial structure will involve the antisymmetric tensor $\varepsilon^{\mu\nu\rho\sigma}$.

We start by decomposing the energy-momentum tensor $T^{\mu\nu}$ and the current $J_A^\mu$ in terms of their transverse-traceless part and longitudinal ones (also called "local")

$$T^{\mu_i\nu_i}(p_i) = t^{\mu_i\nu_i}(p_i) + t_{loc}^{\mu_i\nu_i}(p_i), \tag{5.2.1}$$

$$J_A^{\mu_i}(p_i) = j_A^{\mu_i}(p_i) + j_{A\,loc}^{\mu_i}(p_i), \tag{5.2.2}$$

where

$$t^{\mu_i\nu_i}(p_i) = \Pi^{\mu_i\nu_i}_{\alpha_i\beta_i}(p_i)\,T^{\alpha_i\beta_i}(p_i), \qquad t_{loc}^{\mu_i\nu_i}(p_i) = \Sigma^{\mu_i\nu_i}_{\alpha_i\beta_i}(p)\,T^{\alpha_i\beta_i}(p_i),$$
$$j_A^{\mu_i}(p_i) = \pi^{\mu_i}_{\alpha_i}(p_i)\,J_A^{\alpha_i}(p_i), \qquad j_{A\,loc}^{\mu_i}(p_i) = \frac{p_i^{\mu_i}\,p_{i\,\alpha_i}}{p_i^2}\,J_A^{\alpha_i}(p_i), \tag{5.2.3}$$

having introduced the transverse-traceless ($\Pi$), transverse ($\pi$) and longitudinal ($\Sigma$) projectors, given respectively by

$$\pi^\mu_\alpha = \delta^\mu_\alpha - \frac{p^\mu p_\alpha}{p^2}, \tag{5.2.4}$$

$$\Pi^{\mu\nu}_{\alpha\beta} = \frac{1}{2}\left(\pi^\mu_\alpha\pi^\nu_\beta + \pi^\mu_\beta\pi^\nu_\alpha\right) - \frac{1}{d-1}\pi^{\mu\nu}\pi_{\alpha\beta}, \tag{5.2.5}$$

$$\Sigma^{\mu_i\nu_i}_{\alpha_i\beta_i} = \frac{p_{i\,\beta_i}}{p_i^2}\left[2\delta^{(\nu_i}_{\alpha_i}p_i^{\mu_i)} - \frac{p_{i\alpha_i}}{(d-1)}\left(\delta^{\mu_i\nu_i} + (d-2)\frac{p_i^{\mu_i}p_i^{\nu_i}}{p_i^2}\right)\right] + \frac{\pi^{\mu_i\nu_i}(p_i)}{(d-1)}\delta_{\alpha_i\beta_i}. \tag{5.2.6}$$

Such decomposition allows to split our correlation function into the following terms

$$\begin{aligned}\left\langle T^{\mu_1\nu_1}T^{\mu_2\nu_2}J_A^{\mu_3}\right\rangle =& \left\langle t^{\mu_1\nu_1}t^{\mu_2\nu_2}j_A^{\mu_3}\right\rangle + \left\langle T^{\mu_1\nu_1}T^{\mu_2\nu_2}j_{A\,loc}^{\mu_3}\right\rangle + \left\langle T^{\mu_1\nu_1}t_{loc}^{\mu_2\nu_2}J_A^{\mu_3}\right\rangle + \left\langle t_{loc}^{\mu_1\nu_1}T^{\mu_2\nu_2}J_A^{\mu_3}\right\rangle \\ &- \left\langle T^{\mu_1\nu_1}t_{loc}^{\mu_2\nu_2}j_{A\,loc}^{\mu_3}\right\rangle - \left\langle t_{loc}^{\mu_1\nu_1}t_{loc}^{\mu_2\nu_2}J_A^{\mu_3}\right\rangle - \left\langle t_{loc}^{\mu_1\nu_1}T^{\mu_2\nu_2}j_{A\,loc}^{\mu_3}\right\rangle + \left\langle t_{loc}^{\mu_1\nu_1}t_{loc}^{\mu_2\nu_2}j_{A\,loc}^{\mu_3}\right\rangle.\end{aligned} \tag{5.2.7}$$

Using the conservation and trace WIs derived in Section 1.6, it is then possible to completely fix all the longitudinal parts, i.e. the terms containing at least one $t_{loc}^{\mu\nu}$ or $j_{A\,loc}^{\mu}$. We start by considering the non-anomalous equations

$$\begin{aligned}&\delta_{\mu_i\nu_i}\langle T^{\mu_1\nu_1}(p_1)T^{\mu_2\nu_2}(p_2)J_A^{\mu_3}(p_3)\rangle = 0, && i = \{1,2\}\\ &p_{i\mu_i}\langle T^{\mu_1\nu_1}(p_1)T^{\mu_2\nu_2}(p_2)J_A^{\mu_3}(p_3)\rangle = 0, && i = \{1,2\}.\end{aligned} \tag{5.2.8}$$

Thanks to these WIs, we can eliminate most of terms on the right-hand side of equation (5.2.7), ending up only with two terms

$$\left\langle T^{\mu_1\nu_1}T^{\mu_2\nu_2}J_A^{\mu_3}\right\rangle = \left\langle t^{\mu_1\nu_1}t^{\mu_2\nu_2}j_A^{\mu_3}\right\rangle + \left\langle T^{\mu_1\nu_1}T^{\mu_2\nu_2}j_{A\,loc}^{\mu_3}\right\rangle = \left\langle t^{\mu_1\nu_1}t^{\mu_2\nu_2}j_A^{\mu_3}\right\rangle + \left\langle t^{\mu_1\nu_1}t^{\mu_2\nu_2}j_{A\,loc}^{\mu_3}\right\rangle. \tag{5.2.9}$$

The remaining local term is then fixed by the anomalous WI of $J_A$. First, we construct the most general expression in terms of tensorial structures and form factors

$$\left\langle t^{\mu_1\nu_1}t^{\mu_2\nu_2}j_{A\,loc}^{\mu_3}\right\rangle = p_3^{\mu_3}\,\Pi^{\mu_1\nu_1}_{\alpha_1\beta_1}(p_1)\Pi^{\mu_2\nu_2}_{\alpha_2\beta_2}(p_2)\,\varepsilon^{\alpha_1\alpha_2p_1p_2}\left(F_1\,g^{\beta_1\beta_2} + F_2\,p_1^{\beta_2}p_2^{\beta_1}\right) \tag{5.2.10}$$

where, due to the Bose symmetry, both $F_1$ and $F_2$ are symmetric under the exchange $(p_1 \leftrightarrow p_2)$. Then, recalling the definition of $j_{A\,loc}$ and the anomalous WI

$$p_{3\mu_3}\langle T^{\mu_1\nu_1}(p_1)T^{\mu_2\nu_2}(p_2)J_A^{\mu_3}(p_3)\rangle = 4\,i\,a_2\,(p_1\cdot p_2)\left\{\left[\varepsilon^{\nu_1\nu_2p_1p_2}\left(g^{\mu_1\mu_2} - \frac{p_1^{\mu_2}p_2^{\mu_1}}{p_1\cdot p_2}\right) + (\mu_1\leftrightarrow\nu_1)\right] + (\mu_2\leftrightarrow\nu_2)\right\}, \tag{5.2.11}$$

we can write

$$\left\langle t^{\mu_1\nu_1} t^{\mu_2\nu_2} j^{\mu_3}_{A\,loc}\right\rangle = 4ia_2 \frac{p_3^{\mu_3}}{p_3^2}(p_1\cdot p_2)\left\{\left[\varepsilon^{\nu_1\nu_2 p_1 p_2}\left(g^{\mu_1\mu_2} - \frac{p_1^{\mu_2}p_2^{\mu_1}}{p_1\cdot p_2}\right) + (\mu_1\leftrightarrow\nu_1)\right] + (\mu_2\leftrightarrow\nu_2)\right\}. \tag{5.2.12}$$

One can show that this formula coincides with eq. (5.2.10) after contracting the projectors' indices and fixing the form factors in the following way

$$F_1 = \frac{16ia_2(p_1\cdot p_2)}{p_3^2}, \qquad\qquad F_2 = -\frac{16ia_2}{p_3^2}. \tag{5.2.13}$$

Therefore, all the local terms of the $\langle TTJ_A\rangle$ are fixed. The only remaining term to be studied in order to reconstruct the entire correlator is the transverse-traceless part $\left\langle t^{\mu_1\nu_1}t^{\mu_2\nu_2}j_A^{\mu_3}\right\rangle$. Its explicit form is given by

$$\left\langle t^{\mu_1\nu_1}(p_1)\, t^{\mu_2\nu_2}(p_2)\, j_A^{\mu_3}(p_3)\right\rangle = \Pi^{\mu_1\nu_1}_{\alpha_1\beta_1}(p_1)\Pi^{\mu_2\nu_2}_{\alpha_2\beta_2}(p_2)\pi^{\mu_3}_{\alpha_3}(p_3)X^{\alpha_1\beta_1\alpha_2\beta_2\alpha_3} \tag{5.2.14}$$

where $X^{\alpha_1\beta_1\alpha_2\beta_2\alpha_3}$ is a general rank five tensor built by products of metric tensors, momenta and the Levi-Civita symbol with the appropriate choice of indices. Indeed, as a consequence of the projectors in (5.2.14), $X^{\alpha_1\beta_1\alpha_2\beta_2\alpha_3}$ can not be constructed by using $g_{\alpha_i\beta_i}$, nor by $p_{i\,\alpha_i}$ with $i=\{1,2,3\}$. We also must keep in mind that, due to symmetries of the correlator, form factors associated with structures linked by a $(1\leftrightarrow 2)$ transformation (the gravitons exchange) are dependent. Then, the transverse-traceless part can be written as

$$\begin{aligned}
\langle t^{\mu_1\nu_1}(p_1)t^{\mu_2\nu_2}(p_2)j_A^{\mu_3}(p_3)\rangle = \Pi^{\mu_1\nu_1}_{\alpha_1\beta_1}(p_1)\Pi^{\mu_2\nu_2}_{\alpha_2\beta_2}(p_2)\pi^{\mu_3}_{\alpha_3}(p_3)\Big[ \\
& A_1\varepsilon^{p_1\alpha_1\alpha_2\alpha_3}p_2^{\beta_1}p_3^{\beta_2} - A_1(p_1\leftrightarrow p_2)\varepsilon^{p_2\alpha_1\alpha_2\alpha_3}p_2^{\beta_1}p_3^{\beta_2} \\
& + A_2\varepsilon^{p_1\alpha_1\alpha_2\alpha_3}\delta^{\beta_1\beta_2} - A_2(p_1\leftrightarrow p_2)\varepsilon^{p_2\alpha_1\alpha_2\alpha_3}\delta^{\beta_1\beta_2} \\
& + A_3\varepsilon^{p_1p_2\alpha_1\alpha_2}p_2^{\beta_1}p_3^{\beta_2}p_1^{\alpha_3} + A_4\varepsilon^{p_1p_2\alpha_1\alpha_2}\delta^{\beta_1\beta_2}p_1^{\alpha_3} \\
& + A_5\varepsilon^{p_1p_2\alpha_1\alpha_3}p_2^{\beta_1}p_3^{\alpha_2}p_3^{\beta_2} + A_5(p_1\leftrightarrow p_2)\varepsilon^{p_1p_2\alpha_2\alpha_3}p_3^{\beta_2}p_2^{\alpha_1}p_2^{\beta_1} \\
& + A_6\varepsilon^{p_1p_2\alpha_1\alpha_3}p_3^{\alpha_2}\delta^{\beta_1\beta_2} + A_6(p_1\leftrightarrow p_2)\varepsilon^{p_1p_2\alpha_2\alpha_3}p_2^{\alpha_1}\delta^{\beta_1\beta_2} \\
& + A_7\varepsilon^{p_1p_2\alpha_1\alpha_2}p_2^{\beta_1}\delta^{\beta_2\alpha_3} - A_7(p_1\leftrightarrow p_2)\varepsilon^{p_1p_2\alpha_1\alpha_2}p_3^{\beta_2}\delta^{\beta_1\alpha_3}\Big]
\end{aligned} \tag{5.2.15}$$

where $A_3$ and $A_4$ are antisymmetric under the exchange $(p_1\leftrightarrow p_2)$ and we have made a choice on which independent momenta to consider for each index

$$\{\alpha_1,\beta_1\}\leftrightarrow p_2, \qquad \{\alpha_2,\beta_2\}\leftrightarrow p_3, \qquad \{\alpha_3\}\leftrightarrow p_1. \tag{5.2.16}$$

Since we are working in $d=4$ the form factors in eq. (5.2.15) are not all independent and the decomposition is not minimal. Indeed, one needs to consider the following class of tensor identities

$$0 = \varepsilon^{[p_1p_2\alpha_1\alpha_2}\delta^{\alpha_3]}_{\alpha} \tag{5.2.17}$$

If we set $\alpha=\beta_1$ or $\alpha=\beta_2$ and apply the projectors, we have

$$\begin{aligned}
&\Pi^{\mu_1\nu_1}_{\alpha_1\beta_1}(p_1)\Pi^{\mu_2\nu_2}_{\alpha_2\beta_2}(p_2)\pi^{\mu_3}_{\alpha_3}(p_3)\left[\varepsilon^{p_1p_2\alpha_1\alpha_2}\delta^{\alpha_3\beta_1}\right]=\\
&\qquad\Pi^{\mu_1\nu_1}_{\alpha_1\beta_1}(p_1)\Pi^{\mu_2\nu_2}_{\alpha_2\beta_2}(p_2)\pi^{\mu_3}_{\alpha_3}(p_3)\left[\varepsilon^{p_1\alpha_1\alpha_2\alpha_3}p_2^{\beta_1}+\varepsilon^{p_1p_2\alpha_1\alpha_3}\delta^{\alpha_2\beta_1}\right],\\
&\Pi^{\mu_1\nu_1}_{\alpha_1\beta_1}(p_1)\Pi^{\mu_2\nu_2}_{\alpha_2\beta_2}(p_2)\pi^{\mu_3}_{\alpha_3}(p_3)\left[\varepsilon^{p_1p_2\alpha_1\alpha_2}\delta^{\alpha_3\beta_2}\right]=\\
&\qquad\Pi^{\mu_1\nu_1}_{\alpha_1\beta_1}(p_1)\Pi^{\mu_2\nu_2}_{\alpha_2\beta_2}(p_2)\pi^{\mu_3}_{\alpha_3}(p_3)\left[\varepsilon^{p_2\alpha_1\alpha_2\alpha_3}p_3^{\beta_2}-\varepsilon^{p_1p_2\alpha_2\alpha_3}\delta^{\alpha_1\beta_2}\right]
\end{aligned}\tag{5.2.18}$$

according to which we can rewrite the tensorial structures multiplying $A_7$ in terms of the others. If we instead contract the identity (5.2.17) with $p_{1\alpha}$ and $p_{2\alpha}$, we arrive to

$$\begin{aligned}
&\Pi^{\mu_1\nu_1}_{\alpha_1\beta_1}(p_1)\Pi^{\mu_2\nu_2}_{\alpha_2\beta_2}(p_2)\pi^{\mu_3}_{\alpha_3}(p_3)\left[\varepsilon^{p_1p_2\alpha_1\alpha_3}p_3^{\alpha_2}\right]=\\
&\qquad\Pi^{\mu_1\nu_1}_{\alpha_1\beta_1}(p_1)\Pi^{\mu_2\nu_2}_{\alpha_2\beta_2}(p_2)\pi^{\mu_3}_{\alpha_3}(p_3)\left[\varepsilon^{p_1\alpha_1\alpha_2\alpha_3}(p_1\cdot p_2)-\varepsilon^{p_1p_2\alpha_1\alpha_2}p_1^{\alpha_3}-\varepsilon^{p_2\alpha_1\alpha_2\alpha_3}p_1^2\right],\\
&\Pi^{\mu_1\nu_1}_{\alpha_1\beta_1}(p_1)\Pi^{\mu_2\nu_2}_{\alpha_2\beta_2}(p_2)\pi^{\mu_3}_{\alpha_3}(p_3)\left[\varepsilon^{p_1p_2\alpha_2\alpha_3}p_2^{\alpha_1}\right]=\\
&\qquad\Pi^{\mu_1\nu_1}_{\alpha_1\beta_1}(p_1)\Pi^{\mu_2\nu_2}_{\alpha_2\beta_2}(p_2)\pi^{\mu_3}_{\alpha_3}(p_3)\left[\varepsilon^{p_1\alpha_1\alpha_2\alpha_3}p_2^2+\varepsilon^{p_1p_2\alpha_1\alpha_2}p_1^{\alpha_3}-\varepsilon^{p_2\alpha_1\alpha_2\alpha_3}(p_1\cdot p_2)\right]
\end{aligned}\tag{5.2.19}$$

according to which we can rewrite the form factors $A_5$ and $A_6$ in terms of the first four. We conclude that the general structure of the transverse-traceless part is given by

$$\begin{aligned}
\langle t^{\mu_1\nu_1}(p_1)t^{\mu_2\nu_2}(p_2)j_A^{\mu_3}(p_3)\rangle=\Pi^{\mu_1\nu_1}_{\alpha_1\beta_1}(p_1)\Pi^{\mu_2\nu_2}_{\alpha_2\beta_2}(p_2)\pi^{\mu_3}_{\alpha_3}(p_3)\Big[&\\
A_1\varepsilon^{p_1\alpha_1\alpha_2\alpha_3}p_2^{\beta_1}p_3^{\beta_2}-A_1(p_1\leftrightarrow p_2)\varepsilon^{p_2\alpha_1\alpha_2\alpha_3}p_2^{\beta_1}p_3^{\beta_2}&\\
+A_2\varepsilon^{p_1\alpha_1\alpha_2\alpha_3}\delta^{\beta_1\beta_2}-A_2(p_1\leftrightarrow p_2)\varepsilon^{p_2\alpha_1\alpha_2\alpha_3}\delta^{\beta_1\beta_2}&\\
+A_3\varepsilon^{p_1p_2\alpha_1\alpha_2}p_2^{\beta_1}p_3^{\beta_2}p_1^{\alpha_3}+A_4\varepsilon^{p_1p_2\alpha_1\alpha_2}\delta^{\beta_1\beta_2}p_1^{\alpha_3}\Big]&
\end{aligned}\tag{5.2.20}$$

where we have redefined the form factors $A_1,\ldots,A_4$. Once again, $A_3$ and $A_4$ are antisymmetric under the exchange $(p_1\leftrightarrow p_2)$.

## 5.3 The conformal analysis of the $\langle TTJ_A\rangle$

In the previous section we have seen that the conservation and trace WIs fix the longitudinal part of the correlator. In this section we examine the conformal constraints on the $\langle TTJ_A\rangle$, following closely the methodology adopted in [29]. We will see that the transverse-traceless part of the correlator is completely determined by conformal invariance together with the $R\tilde{R}$ part of the boundary condition coming from the anomaly relation (1.5.1), corresponding to the anomalous coefficient $a_2$.

### 5.3.1 Dilatation Ward identities

The invariance of the transverse-traceless part of the correlator under dilatation is reflected in the equation

$$\left(\sum_{i=1}^{3}\Delta_i - 2d - \sum_{i=1}^{2} p_i^\mu \frac{\partial}{\partial p_i^\mu}\right)\langle t^{\mu_1\nu_1}(p_1)t^{\mu_2\nu_2}(p_2)j_A^{\mu_3}(p_3)\rangle = 0. \tag{5.3.1}$$

By using the chain rule

$$\frac{\partial}{\partial p_i^\mu} = \sum_{j=1}^{3}\frac{\partial p_j}{\partial p_i^\mu}\frac{\partial}{\partial p_j} \tag{5.3.2}$$

to express the derivatives respect to 4-vectors in term of the invariants $p_i = |\sqrt{p_i^2}|$, we rewrite (5.3.1) as a constraint on the form factors

$$\sum_{i=1}^{3} p_i \frac{\partial A_j}{\partial p_i} - \left(\sum_{i=1}^{3}\Delta_i - 2d - N_j\right)A_j = 0 \tag{5.3.3}$$

with $N_j$ the number of momenta that the form factors multiply in the decomposition of eq. (5.2.20)

$$N_1 = 3, \qquad N_2 = 1, \qquad N_3 = 5, \qquad N_4 = 3, \tag{5.3.4}$$

### 5.3.2 Special conformal Ward identities

The invariance of the correlator with respect to the special conformal transformations is encoded in the following equation

$$\begin{aligned}
0 =& \mathcal{K}^\kappa\left\langle T^{\mu_1\nu_1}(p_1)T^{\mu_2\nu_2}(p_2)J_A^{\mu_3}(p_3)\right\rangle \\
\equiv& \sum_{j=1}^{2}\left(2\left(\Delta_j - d\right)\frac{\partial}{\partial p_{j\kappa}} - 2p_j^\alpha\frac{\partial}{\partial p_j^\alpha}\frac{\partial}{\partial p_{j\kappa}} + \left(p_j\right)^\kappa\frac{\partial}{\partial p_j^\alpha}\frac{\partial}{\partial p_{j\alpha}}\right)\left\langle T^{\mu_1\nu_1}(p_1)T^{\mu_2\nu_2}(p_2)J_A^{\mu_3}(p_3)\right\rangle \\
&+4\left(\delta^{\kappa(\mu_1}\frac{\partial}{\partial p_1^{\alpha_1}} - \delta^\kappa_{\alpha_1}\delta^{(\mu_1}_\lambda\frac{\partial}{\partial p_{1\lambda}}\right)\left\langle T^{\nu_1)\alpha_1}(p_1)T^{\mu_2\nu_2}(p_2)J_A^{\mu_3}(p_3)\right\rangle \\
&+4\left(\delta^{\kappa(\mu_2}\frac{\partial}{\partial p_2^{\alpha_2}} - \delta^\kappa_{\alpha_2}\delta^{(\mu_2}_\lambda\frac{\partial}{\partial p_{2\lambda}}\right)\left\langle T^{\nu_2)\alpha_2}(p_2)T^{\mu_1\nu_1}(p_1)J_A^{\mu_3}(p_3)\right\rangle.
\end{aligned} \tag{5.3.5}$$

The special conformal operator $\mathcal{K}^\kappa$ acts as an endomorphism on the transverse-traceless sector of the entire correlator. Therefore we can perform a transverse-traceless projection on all the indices in order to identify a set of partial differential equations

$$\begin{aligned}
0 =& \Pi^{\rho_1\sigma_1}_{\mu_1\nu_1}(p_1)\Pi^{\rho_2\sigma_2}_{\mu_2\nu_2}(p_2)\pi^{\rho_3}_{\mu_3}(p_3)\mathcal{K}^\kappa\left\langle T^{\mu_1\nu_1}(p_1)T^{\mu_2\nu_2}(p_2)J_A^{\mu_3}(p_3)\right\rangle = \\
&\Pi^{\rho_1\sigma_1}_{\mu_1\nu_1}(p_1)\Pi^{\rho_2\sigma_2}_{\mu_2\nu_2}(p_2)\pi^{\rho_3}_{\mu_3}(p_3)\mathcal{K}^\kappa\left(\left\langle t^{\mu_1\nu_1}t^{\mu_2\nu_2}j_A^{\mu_3}\right\rangle + \left\langle t^{\mu_1\nu_1}t^{\mu_2\nu_2}j_{A\,loc}^{\mu_3}\right\rangle\right),
\end{aligned} \tag{5.3.6}$$

splitting the correlator into its transverse and longitudinal parts. The action of the special conformal operator $\mathcal{K}^\kappa$ on the longitudinal part of the correlator is given by

$$\Pi^{\rho_1\sigma_1}_{\mu_1\nu_1}(p_1)\Pi^{\rho_2\sigma_2}_{\mu_2\nu_2}(p_2)\pi^{\rho_3}_{\mu_3}(p_3)\left[\mathcal{K}^\kappa\left\langle t^{\mu_1\nu_1}t^{\mu_2\nu_2}j^{\mu_3}_{A\,loc}\right\rangle\right]= \Pi^{\rho_1\sigma_1}_{\mu_1\nu_1}(p_1)\Pi^{\rho_2\sigma_2}_{\mu_2\nu_2}(p_2)\pi^{\rho_3}_{\mu_3}(p_3)\left[2\frac{(\Delta_3-1)}{p_3^2}\delta^{\kappa\mu_3}p_{3\alpha}\left\langle t^{\mu_1\nu_1}t^{\mu_2\nu_2}j^{\alpha}_{A\,loc}\right\rangle\right]. \tag{5.3.7}$$

Using the eq. (5.2.11) together with the Schouten identities mentioned in the Appendix C.2, we can write

$$\begin{aligned}\Pi^{\rho_1\sigma_1}_{\mu_1\nu_1}(p_1)\Pi^{\rho_2\sigma_2}_{\mu_2\nu_2}(p_2)\pi^{\rho_3}_{\mu_3}(p_3)\left[\mathcal{K}^\kappa\left\langle t^{\mu_1\nu_1}t^{\mu_2\nu_2}j^{\mu_3}_{A\,loc}\right\rangle\right]&=\Pi^{\rho_1\sigma_1}_{\mu_1\nu_1}(p_1)\Pi^{\rho_2\sigma_2}_{\mu_2\nu_2}(p_2)\pi^{\rho_3}_{\mu_3}(p_3)\frac{16\,i\,a_2(\Delta_3-1)}{p_3^2}\Bigg[\\
&p_1^\kappa\varepsilon^{p_2\mu_1\mu_2\mu_3}\Big(-2p_2^{\nu_1}p_3^{\nu_2}+(p_1^2+p_2^2-p_3^2)\delta^{\nu_1\nu_2}\Big)+p_2^\kappa\varepsilon^{p_1\mu_1\mu_2\mu_3}\Big(2p_2^{\nu_1}p_3^{\nu_2}-(p_1^2+p_2^2-p_3^2)\delta^{\nu_1\nu_2}\Big)\\
&\delta^{\kappa\mu_1}\Big(+(p_1^2+p_2^2-p_3^2)\varepsilon^{p_1p_2\mu_2\mu_3}\delta^{\nu_1\nu_2}-2p_2^2\varepsilon^{p_1\mu_2\mu_3\nu_1}p_3^{\nu_2}-(p_1^2+p_2^2-p_3^2)\varepsilon^{p_2\mu_2\mu_3\nu_1}p_3^{\nu_2}+2\varepsilon^{p_1p_2\mu_2\nu_1}p_1^{\mu_3}p_3^{\nu_2}\Big)\\
&\delta^{\kappa\mu_2}\Big(-(p_1^2+p_2^2-p_3^2)\varepsilon^{p_1p_2\mu_1\mu_3}\delta^{\nu_1\nu_2}+2p_1^2\varepsilon^{p_2\mu_1\mu_3\nu_2}p_2^{\nu_1}+(p_1^2+p_2^2-p_3^2)\varepsilon^{p_1\mu_1\mu_3\nu_2}p_2^{\nu_1}-2\varepsilon^{p_1p_2\mu_1\nu_2}p_1^{\mu_3}p_2^{\nu_1}\Big)\Bigg].\end{aligned} \tag{5.3.8}$$

Using the Schouten identities reported in Appendix C.2, we can then decompose the action of the special conformal operator on the entire correlator in the following minimal expression

$$\begin{aligned}0=\Pi^{\rho_1\sigma_1}_{\mu_1\nu_1}(p_1)\Pi^{\rho_2\sigma_2}_{\mu_2\nu_2}(p_2)\pi^{\rho_3}_{\mu_3}(p_3)\Big(\mathcal{K}^\kappa\left\langle T^{\mu_1\nu_1}(p_1)T^{\mu_2\nu_2}(p_2)J^{\mu_3}_A(p_3)\right\rangle\Big)&=\Pi^{\rho_1\sigma_1}_{\mu_1\nu_1}(p_1)\Pi^{\rho_2\sigma_2}_{\mu_2\nu_2}(p_2)\pi^{\rho_3}_{\mu_3}(p_3)\Bigg[\\
&p_1^\kappa\Big(C_{11}\varepsilon^{p_1\mu_1\mu_2\mu_3}p_2^{\nu_1}p_3^{\nu_2}+C_{12}\varepsilon^{p_2\mu_1\mu_2\mu_3}p_2^{\nu_1}p_3^{\nu_2}+C_{13}\varepsilon^{p_1\mu_1\mu_2\mu_3}\delta^{\nu_1\nu_2}+C_{14}\varepsilon^{p_2\mu_1\mu_2\mu_3}\delta^{\nu_1\nu_2}\\
&+C_{15}\varepsilon^{p_1p_2\mu_1\mu_2}p_2^{\nu_1}p_3^{\nu_2}p_1^{\mu_3}+C_{16}\varepsilon^{p_1p_2\mu_1\mu_2}\delta^{\nu_1\nu_2}p_1^{\mu_3}\Big)\\
&+p_2^\kappa\Big(C_{21}\varepsilon^{p_1\mu_1\mu_2\mu_3}p_2^{\nu_1}p_3^{\nu_2}+C_{22}\varepsilon^{p_2\mu_1\mu_2\mu_3}p_2^{\nu_1}p_3^{\nu_2}+C_{23}\varepsilon^{p_1\mu_1\mu_2\mu_3}\delta^{\nu_1\nu_2}+C_{24}\varepsilon^{p_2\mu_1\mu_2\mu_3}\delta^{\nu_1\nu_2}\\
&+C_{25}\varepsilon^{p_1p_2\mu_1\mu_2}p_2^{\nu_1}p_3^{\nu_2}p_1^{\mu_3}+C_{26}\varepsilon^{p_1p_2\mu_1\mu_2}\delta^{\nu_1\nu_2}p_1^{\mu_3}\Big)\\
&+\delta^{\kappa\mu_1}\Big(C_{31}\varepsilon^{p_1\mu_2\mu_3\nu_1}p_3^{\nu_2}+C_{32}\varepsilon^{p_2\mu_2\mu_3\nu_1}p_3^{\nu_2}+C_{33}\varepsilon^{p_1p_2\mu_2\nu_1}p_1^{\mu_3}p_3^{\nu_2}+C_{34}\varepsilon^{p_1p_2\mu_2\mu_3}\delta^{\nu_1\nu_2}\Big)\\
&+\delta^{\kappa\mu_2}\Big(C_{41}\varepsilon^{p_1\mu_1\mu_3\nu_2}p_2^{\nu_1}+C_{42}\varepsilon^{p_2\mu_1\mu_3\nu_2}p_2^{\nu_1}+C_{43}\varepsilon^{p_1p_2\mu_1\nu_2}p_1^{\mu_3}p_2^{\nu_1}+C_{44}\varepsilon^{p_1p_2\mu_1\mu_3}\delta^{\nu_1\nu_2}\Big)\\
&+C_{51}\varepsilon^{\kappa\mu_1\mu_2\mu_3}\delta^{\nu_1\nu_2}+C_{52}\varepsilon^{\kappa\mu_1\mu_2\mu_3}p_2^{\nu_1}p_3^{\nu_2}+C_{53}\varepsilon^{p_1\kappa\mu_1\mu_2}p_1^{\mu_3}\delta^{\nu_1\nu_2}+C_{54}\varepsilon^{p_2\kappa\mu_1\mu_2}p_1^{\mu_3}\delta^{\nu_1\nu_2}\Bigg]\end{aligned} \tag{5.3.9}$$

where the coefficients $C_{ij}$ depend on the gravitational anomalous coefficient $a_2$, the form factors $A_i$ and their derivatives with respect to the momenta.

Due to the independence of the tensorial structures listed in the equation above, all the coefficients $C_{ij}$ need to vanish. In particular the primary equations are

$$0=C_{ij}\qquad\qquad i=\{1,2\},\quad j=\{1,\dots 6\} \tag{5.3.10}$$

They correspond to second order differential equations. The secondary equations are instead given by

$$0=C_{ij}\qquad\qquad i=\{3,4,5\},\quad j=\{1,\dots 4\} \tag{5.3.11}$$

and they are differential equations of the first order.

### 5.3.3 Solving the CWIs

The most general solution of the CWIs of the $\langle TTJ_A\rangle$ can be written in terms of integrals involving a product of three Bessel functions, namely 3K integrals [29, 80]. We start by considering the explicit form of the primary equations (5.3.10) involving the form factor $A_3$

$$K_{31}A_3 = 0, \qquad\qquad K_{32}A_3 = 0 \tag{5.3.12}$$

Recalling the following property of the 3K integrals

$$K_{nm}J_{N\{k_j\}} = -2k_n J_{N+1\{k_j-\delta_{jn}\}} + 2k_m J_{N+1\{k_j-\delta_{jm}\}}, \tag{5.3.13}$$

we can write the most general solution of the primary equations as

$$A_3 = \zeta_1 J_{\{5,0,0,0\}} \tag{5.3.14}$$

where $\zeta_1$ is an arbitrary constant. Note that this solution is symmetric under the exchange of momenta $p_1 \leftrightarrow p_2$. Indeed, from the definition of the 3K integral, it follows that for any permutation $\sigma$ of the set $\{1,2,3\}$ we have

$$J_{N\{k_{\sigma(1)},k_{\sigma(2)},k_{\sigma(3)}\}}(p_1,p_2,p_3) = J_{N\{k_1,k_2,k_3\}}\left(p_{\sigma^{-1}(1)},p_{\sigma^{-1}(2)},p_{\sigma^{-1}(3)}\right) \tag{5.3.15}$$

However, due to the Bose symmetry, the form factor $A_3$ needs to be antisymmetric under the exchange of momenta $p_1 \leftrightarrow p_2$. This leads to

$$\zeta_1 = 0 \quad \Longrightarrow \quad A_3 = 0 \tag{5.3.16}$$

After setting $A_3 = 0$, the explicit form of the primary equations involving the form factor $A_4$ can be written as

$$K_{31}A_4 = 0, \qquad\qquad K_{32}A_4 = 0 \tag{5.3.17}$$

Their solution is given by

$$A_4 = \zeta_2 J_{\{3,0,0,0\}}, \tag{5.3.18}$$

where $\zeta_2$ is an arbitrary constant. Once again, due to the Bose symmetry, $A_4$ needs to be antisymmetric under the exchange of momenta $p_1 \leftrightarrow p_2$. This leads to

$$\zeta_2 = 0 \quad \Longrightarrow \quad A_4 = 0. \tag{5.3.19}$$

After setting $A_4 = 0$, we can write the remaining primary equations as[1]

$$\begin{aligned} &0 = K_{31}A_1, && 0 = K_{32}A_1 + \frac{2}{p_1^2}\left(p_1\frac{\partial}{\partial p_1} - 4\right)A_1(p_1 \leftrightarrow p_2) \\ &0 = K_{31}A_2 + 4A_1, && 0 = K_{32}A_2 + \frac{2}{p_1^2}\left(p_1\frac{\partial}{\partial p_1} - 4\right)A_2(p_1 \leftrightarrow p_2) + 4A_1 \end{aligned} \tag{5.3.20}$$

These equations can also be reduced to a set of homogenous equations by repeatedly applying the operator $K_{ij}$

$$\begin{aligned} &0 = K_{31}A_1, && 0 = K_{32}K_{32}A_1 \\ &0 = K_{31}K_{31}A_2, && 0 = K_{32}K_{32}K_{32}A_2. \end{aligned} \tag{5.3.21}$$

[1] For simplicity, we are actually considering on the right side two equations that are obtained by a combination of primary and secondary equations: $0 = C_{21} - C_{33}$ and $0 = C_{23} - C_{34}$. The contribution of the anomalous term coming from eq. (5.3.8) does not appear in such combinations of equations.

The most general solution of such homogenous equations can be written in terms of the following 3K integrals

$$\begin{aligned} A_1 =& \eta_1 J_{3\{0,0,0\}} + \eta_2 J_{4\{0,1,0\}}, \\ A_2 =& \theta_1 J_{4\{1,2,0\}} + \theta_2 J_{3\{0,2,0\}} + \theta_3 J_{3\{1,1,0\}} \\ & + \theta_4 J_{2\{0,1,0\}} + \theta_5 J_{2\{1,0,0\}} + \theta_6 J_{1\{0,0,0\}} + \theta_7 J_{3\{0,1,1\}} + \theta_8 J_{2\{0,0,1\}}, \end{aligned} \tag{5.3.22}$$

where $\eta_i$ and $\theta_i$ are arbitrary constants. The explicit form of such 3K integrals can be determined by following the procedure in [29, 80, 81]. Before moving on, we need to examine the divergences in the 3K integrals. For a more detailed review of the topic, see Appendix A and [25, 29, 80]. In general, it can be shown that a 3K integral $I_{\alpha\{\beta_1,\beta_2,\beta_3\}}$ diverges if

$$\alpha + 1 \pm \beta_1 \pm \beta_2 \pm \beta_3 = -2k \quad , \quad k = 0, 1, 2, \ldots \tag{5.3.23}$$

If the above condition is satisfied, we need to regularize the integral. Therefore, we shift its parameters by small amounts proportional to a regulator $\epsilon$ according to the formula

$$\begin{aligned} I_{\alpha\{\beta_1,\beta_2,\beta_3\}} &\longmapsto I_{\alpha+u\epsilon\{\beta_1+v_1\epsilon,\beta_2+v_2\epsilon,\beta_3+v_3\epsilon\}} \\ J_{N\{k_1,k_2,k_3\}} &\longmapsto J_{N+u\epsilon\{k_1+v_1\epsilon,k_2+v_2\epsilon,k_3+v_3\epsilon\}}. \end{aligned} \tag{5.3.24}$$

The arbitrary numbers $u$, $v_1$, $v_2$ and $v_3$ specify the direction of the shift. In general the regulated integral exists, but exhibits singularities when $\epsilon$ is taken to zero. If a 3K integral in our solution diverges, we can expand the coefficient in front of such integral in the solution in powers of $\epsilon$

$$\eta_i = \sum_{j=-\infty}^{\infty} \eta_i^{(j)} \epsilon^j \qquad \theta_i = \sum_{j=-\infty}^{\infty} \theta_i^{(j)} \epsilon^j, \tag{5.3.25}$$

and then we can require that our entire solution is finite for $\epsilon \to 0$ by constraining the coefficients $\eta_i^{(j)}$ and $\theta_i^{(j)}$. Both of the 3K integrals appearing in the eq. (5.3.22) in the solution for $A_1$ diverge like $1/\epsilon$. Therefore, we require

$$\eta_1 = \eta_1^{(0)} + \eta_1^{(1)}\epsilon, \qquad \eta_2 = \eta_2^{(0)} + \eta_2^{(1)}\epsilon. \tag{5.3.26}$$

Higher order terms do not contribute to the solution and therefore they can be neglected. In the case of $A_2$, since some of the 3K integrals diverge like $1/\epsilon^2$, we need to set

$$\begin{aligned} &\theta_1 = \theta_1^{(0)} + \theta_1^{(1)}\epsilon + \theta_1^{(2)}\epsilon^2, \quad \theta_2 = \theta_2^{(0)} + \theta_2^{(1)}\epsilon + \theta_2^{(2)}\epsilon^2, \quad \theta_3 = \theta_3^{(0)} + \theta_3^{(1)}\epsilon + \theta_3^{(2)}\epsilon^2, \\ &\theta_4 = \theta_4^{(0)} + \theta_4^{(1)}\epsilon + \theta_4^{(2)}\epsilon^2, \quad \theta_5 = \theta_5^{(0)} + \theta_5^{(1)}\epsilon + \theta_5^{(2)}\epsilon^2, \quad \theta_6 = \theta_6^{(0)} + \theta_6^{(1)}\epsilon + \theta_6^{(2)}\epsilon^2, \\ &\qquad \theta_7 = \theta_7^{(0)} + \theta_7^{(1)}\epsilon, \qquad \theta_8 = \theta_8^{(0)} + \theta_8^{(1)}\epsilon. \end{aligned} \tag{5.3.27}$$

The last step consists in analyzing all the conformal constraints on the numerical coefficients $\eta_i^{(j)}$ and $\theta_i^{(j)}$. In order to do that, we insert our solution back into the primary nonhomogenous equations

(5.3.20) and into the secondary equations. The explicit form of the secondary equations is given by[2]

$$
\begin{aligned}
0 &= -2p_1 \frac{\partial}{\partial p_1} A_1 + 2p_2 \frac{\partial}{\partial p_2} A_1(p_1 \leftrightarrow p_2) \\
0 &= -\left(p_1^2 - p_2^2 + p_3^2\right) A_1 + \left(-p_1^2 + p_2^2 + p_3^2\right) A_1(p_1 \leftrightarrow p_2) - 2p_1 \frac{\partial}{\partial p_1} A_2 + 2p_2 \frac{\partial}{\partial p_2} A_2(p_1 \leftrightarrow p_2) \\
&\quad + 2A_2 - 2A_2(p_1 \leftrightarrow p_2) \\
0 &= -\frac{2p_2^3}{p_3^2} \frac{\partial}{\partial p_2} A_1(p_1 \leftrightarrow p_2) - 2\left(\frac{p_2^2 + p_3^2}{p_3^2}\right) p_2 \frac{\partial}{\partial p_2} A_1 + \left(-\frac{2p_2^2}{p_3^2} + \frac{p_3^2 - p_2^2 - p_1^2}{p_1^2}\right) p_1 \frac{\partial}{\partial p_1} A_1 \\
&\quad + 2p_2^2 \left(\frac{p_3^2 - p_1^2}{p_3^2 p_1}\right) \frac{\partial}{\partial p_1} A_1(p_1 \leftrightarrow p_2) - 4p_2^2 \left(\frac{2}{p_1^2} + \frac{1}{p_3^2}\right) A_1(p_1 \leftrightarrow p_2) + 4\left(\frac{p_1^2 + p_2^2 - p_3^2}{p_1^2} - \frac{p_2^2}{p_3^2}\right) A_1 \\
&\quad - \frac{2}{p_1} \frac{\partial}{\partial p_1} A_2 + \frac{8}{p_1^2} A_2 - \frac{64 i a_2 p_2^2}{p_3^2} \\
0 &= -\left(\frac{p_1^2 + p_2^2 - p_3^2}{p_3^2}\right) p_1 \frac{\partial}{\partial p_1} A_1 - \frac{\left(p_1^2 - 2p_3^2\right)\left(p_1^2 + p_2^2 - p_3^2\right)}{p_1 p_3^2} \frac{\partial}{\partial p_1} A_1(p_1 \leftrightarrow p_2) - \left(\frac{p_1^2 + p_2^2 - p_3^2}{p_3^2}\right) p_2 \frac{\partial}{\partial p_2} A_1 \\
&\quad - \left(\frac{p_1^2 + p_2^2 - 3p_3^2}{p_3^2}\right) p_2 \frac{\partial}{\partial p_2} A_1(p_1 \leftrightarrow p_2) - 2\left(\frac{p_1^2 + p_2^2 - 2p_3^2}{p_3^2} + 4\frac{p_1^2 + p_2^2 - p_3^2}{p_1^2}\right) A_1(p_1 \leftrightarrow p_2) \\
&\quad - 2\left(\frac{p_1^2 + p_2^2 - 2p_3^2}{p_3^2}\right) A_1 + \frac{2}{p_1} \frac{\partial}{\partial p_1} A_2(p_1 \leftrightarrow p_2) - \frac{8}{p_1^2} A_2(p_1 \leftrightarrow p_2) - \frac{32 i a_2 \left(p_1^2 + p_2^2 - p_3^2\right)}{p_3^2} \\
0 &= \frac{2p_1}{p_3^2} \frac{\partial}{\partial p_1} A_1 + 2\left(\frac{p_1^2 - p_3^2}{p_3^2 p_1}\right) \frac{\partial}{\partial p_1} A_1(p_1 \leftrightarrow p_2) + \frac{2p_2}{p_3^2} \frac{\partial}{\partial p_2} A_1 + \frac{2p_2}{p_3^2} \frac{\partial}{\partial p_2} A_1(p_1 \leftrightarrow p_2) \\
&\quad + 4\left(\frac{2}{p_1^2} + \frac{1}{p_3^2}\right) A_1(p_1 \leftrightarrow p_2) + \frac{4}{p_3^2} A_1 + \frac{64 i a_2}{p_3^2} \\
0 &= -\frac{2p_1}{p_3^2} \frac{\partial}{\partial p_1} A_2 + 2\left(\frac{p_3^2 - p_1^2}{p_3^2 p_1}\right) \frac{\partial}{\partial p_1} A_2(p_1 \leftrightarrow p_2) - \frac{2p_2}{p_3^2} \frac{\partial}{\partial p_2} A_2 - \frac{2p_2}{p_3^2} \frac{\partial}{\partial p_2} A_2(p_1 \leftrightarrow p_2) - \frac{8}{p_1^2} A_2(p_1 \leftrightarrow p_2) \\
&\quad + \frac{32 i a_2 \left(p_1^2 + p_2^2 - p_3^2\right)}{p_3^2}
\end{aligned}
\tag{5.3.28}
$$

We can solve all these equations by performing the limit $p_i \to 0$, as explained in the Appendix A.2. After some lengthy computations, using all the properties of the 3K integral listed in the Appendix A, we find that all the nonvanishing coefficients $\eta_i^{(j)}$ and $\theta_i^{(j)}$ depend on the anomaly coefficient $a_2$ of eq. (1.5.1). In particular the final solution can be written in the compact form

$$
\begin{aligned}
A_1 &= -4\, i\, a_2\, p_2^2\, I_{5\{2,1,1\}} \\
A_2 &= -8\, i\, a_2\, p_2^2 \left(p_3^2\, I_{4\{2,1,0\}} - 1\right) \\
A_3 &= 0 \\
A_4 &= 0.
\end{aligned}
\tag{5.3.29}
$$

[2]Not of the secondary equations are independent from each other. Here we listed only the relevant ones.

## 5.4 Perturbative realization

In this section we compute the $\langle TTJ_A\rangle$ perturbatevely at one-loop, working in the Breitenlohner-Maison scheme. For this analysis, we work in Minkowski space, where the effective action is defined as

$$e^{i\mathcal{S}} = \mathcal{N}\int d\chi\, e^{iS_0} \tag{5.4.1}$$

We consider the following action with a fermionic field in a gravitational and axial gauge field background

$$S_0 = \int d^dx\,\frac{e}{2}\,e^\mu_a\Big[i\bar\psi\gamma^a\big(D_\mu\psi\big) - i\big(D_\mu\bar\psi\big)\gamma^a\psi\Big] \tag{5.4.2}$$

where $e^\mu_a$ is the vielbein, $e$ is its determinant and $D_\mu$ is the covariant derivative defined as

$$\begin{aligned} D_\mu\psi &= \big(\nabla_\mu + ig\gamma^5A_\mu\big)\psi = \Big(\partial_\mu + ig\gamma^5A_\mu + \frac{1}{2}\omega_{\mu ab}\Sigma^{ab}\Big)\psi,\\ D_\mu\bar\psi &= \big(\nabla_\mu - ig\gamma^5A_\mu\big)\bar\psi = \Big(\partial_\mu - ig\gamma^5A_\mu - \frac{1}{2}\omega_{\mu ab}\Sigma^{ab}\Big)\bar\psi. \end{aligned} \tag{5.4.3}$$

$\Sigma^{ab}$ are the generators of the Lorentz group in the case of a spin $1/2$-field, while the spin connection is given by

$$\omega_{\mu ab} \equiv e^\nu_a\left(\partial_\mu e_{\nu b} - \Gamma^\lambda_{\mu\nu}e_{\lambda b}\right). \tag{5.4.4}$$

The Latin and Greek indices are related to the (locally) flat basis and the curved background respectively. Using the explicit expression of the generators of the Lorentz group one can reexpress the action $S_0$ as follows

$$S_0 = \int d^dx\, e\Big[\frac{i}{2}\bar\psi e^\mu_a\gamma^a\big(\partial_\mu\psi\big) - \frac{i}{2}\big(\partial_\mu\bar\psi\big)e^\mu_a\gamma^a\psi - gA_\mu\bar\psi e^\mu_a\gamma^a\gamma^5\psi + \frac{i}{4}\omega_{\mu ab}e^\mu_c\bar\psi\gamma^{abc}\psi\Big] \tag{5.4.5}$$

with

$$\gamma^{abc} = \{\Sigma^{ab},\gamma^c\}. \tag{5.4.6}$$

Taking a first variation of the action with respect to the metric one can construct the energy momentum tensor as

$$T^{\mu\nu} = -\frac{i}{2}\left[\bar\psi\gamma^{(\mu}\nabla^{\nu)}\psi - \nabla^{(\mu}\bar\psi\gamma^{\nu)}\psi - g^{\mu\nu}\left(\bar\psi\gamma^\lambda\nabla_\lambda\psi - \nabla_\lambda\bar\psi\gamma^\lambda\psi\right)\right] - g\bar\psi\left(g^{\mu\nu}\gamma^\lambda A_\lambda - \gamma^{(\mu}A^{\nu)}\right)\gamma^5\psi. \tag{5.4.7}$$

The computation of the vertices can be done by taking functional derivatives of the action with respect to the metric and the gauge field and Fourier transforming to momentum-space. Their explicit expressions is reported in Appendix I.

### 5.4.1 Feynman diagrams

The $\langle TTJ_A\rangle$ correlator around flat space is extracted by taking three functional derivatives of the effective action with respect to the metric and the gauge field, evaluated when the sources are turned off

$$\left\langle T^{\mu_1\nu_1}(x_1)\,T^{\mu_2\nu_2}(x_2)\,J_A^{\mu_3}(x_3)\right\rangle \equiv 4\,\frac{\delta^3\mathcal{S}}{\delta g_{\mu_1\nu_1}(x_1)\delta g_{\mu_2\nu_2}(x_2)\delta A_{\mu_3}(x_3)}\Bigg|_{\substack{g=\eta;\\ A=0;}} \tag{5.4.8}$$

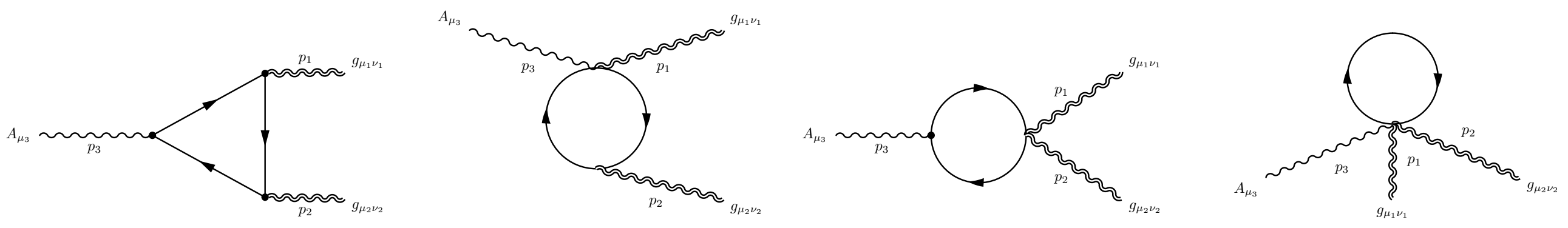

Figure 5.1: Feynman diagrams of the three different topologies appearing in the perturbative computation.

Having denoted with $S_0$ the conformal invariant classical action, recalling eq. (5.4.1), we can write

$$\langle T^{\mu_1\nu_1}(x_1)\,T^{\mu_2\nu_2}(x_2)J_A^{\mu_3}(x_3)\rangle = \\ 4\left\{-i\left\langle\frac{\delta S_0}{\delta g_1}\frac{\delta S_0}{\delta g_2}\frac{\delta S_0}{\delta A_3}\right\rangle-\left\langle\frac{\delta^2 S_0}{\delta g_1\delta g_2}\frac{\delta S_0}{\delta A_3}\right\rangle-\left\langle\frac{\delta^2 S_0}{\delta g_1\delta A_3}\frac{\delta S_0}{\delta g_2}\right\rangle-\left\langle\frac{\delta^2 S_0}{\delta g_2\delta A_3}\frac{\delta S_0}{\delta g_1}\right\rangle+i\left\langle\frac{\delta^3 S_0}{\delta g_1\delta g_2\delta A_3}\right\rangle\right\} \tag{5.4.9}$$

where for the sake of simplicity we have used the notation $g_i = g_{\mu_i\nu_i}(x_i)$ and $A_i = A_{\mu_i}(x_i)$. The angle brackets denote the vacuum expectation value and each of the terms correspond to a Feynman diagram of specific topology. In particular, the first term has a triangle topology while the others are all bubble diagrams, except for the last one, which is a tadpole (see Fig. 5.1). The contribution of the triangle diagrams is given by

$$V_1^{\mu_1\nu_1\mu_2\nu_2\mu_3} = -i^3\int\frac{d^d l}{(2\pi)^d}\,\frac{\text{tr}\Big[V^{\mu_1\nu_1}_{g\bar\psi\psi}(l-p_1,l)(\not l-\not p_1)V^{\mu_3}_{A\bar\psi\psi}(\not l+\not p_2)V^{\mu_2\nu_2}_{g\bar\psi\psi}(l,l+p_2)\not l\Big]}{(l-p_1)^2\,(l+p_2)^2\,l^2}+\text{exchange} \tag{5.4.10}$$

while the bubble diagrams are

$$V_2^{\mu_1\nu_1\mu_2\nu_2\mu_3} = -i^2\int\frac{d^d l}{(2\pi)^d}\,\frac{\text{tr}\Big[V^{\mu_1\nu_1\mu_3}_{gA\bar\psi\psi}(\not l+\not p_2)V^{\mu_2\nu_2}_{g\bar\psi\psi}(l,l+p_2)\not l\Big]}{(l+p_2)^2\,l^2}+\text{exchange} \tag{5.4.11}$$

and

$$V_3^{\mu_1\nu_1\mu_2\nu_2\mu_3} = -i^2\int\frac{d^d l}{(2\pi)^d}\,\frac{\text{tr}\Big[V^{\mu_1\nu_1\mu_2\nu_2}_{gg\bar\psi\psi}(p_1,p_2,l-p_1-p_2,l)(\not l-\not p_1-\not p_2)V^{\mu_3}_{A\bar\psi\psi}\not l\Big]}{(l-p_1-p_2)^2\,l^2}. \tag{5.4.12}$$

After performing the integration, one can verify that $V_2^{\mu_1\nu_1\mu_2\nu_2\mu_3}$ vanishes. Lastly, the tadpole diagram is given by

$$V_4^{\mu_1\nu_1\mu_2\nu_2\mu_3} = -i\int\frac{d^d l}{(2\pi)^d}\,\frac{\text{tr}\Big[V^{\mu_1\nu_1\mu_2\nu_2\mu_3}_{ggA\bar\psi\psi}\,\not l\Big]}{l^2}. \tag{5.4.13}$$

This last diagram vanishes since it contains the trace of two $\gamma$'s and a $\gamma^5$. The perturbative realization of the correlator will be written down as the sum of the amplitudes, formally given by the expression

$$\left\langle T^{\mu_1\nu_1}T^{\mu_2\nu_2}J_A^{\mu_3}\right\rangle = 4\sum_{i=1}^{4}V_i^{\mu_1\nu_1\mu_2\nu_2\mu_3}. \tag{5.4.14}$$

### 5.4.2 Reconstruction of the correlator

The perturbative realization of the $\langle TTJ_A\rangle$ satisfies the (anomalous) conservation and trace WIs. Therefore, the correlator can be decomposed as described in section 5.2. In particular, it is comprised of two terms

$$\left\langle T^{\mu_1\nu_1}T^{\mu_2\nu_2}J_A^{\mu_3}\right\rangle = \left\langle t^{\mu_1\nu_1}t^{\mu_2\nu_2}j_A^{\mu_3}\right\rangle + \left\langle t^{\mu_1\nu_1}t^{\mu_2\nu_2}j_{A\,loc}^{\mu_3}\right\rangle. \tag{5.4.15}$$

The anomalous pole is given by

$$\left\langle t^{\mu_1\nu_1}t^{\mu_2\nu_2}j_{A\,loc}^{\mu_3}\right\rangle = \frac{g}{96\pi^2}\frac{p_3^{\mu_3}}{p_3^2}(p_1\cdot p_2)\left\{\left[\varepsilon^{\nu_1\nu_2 p_1 p_2}\left(g^{\mu_1\mu_2}-\frac{p_1^{\mu_2}p_2^{\mu_1}}{p_1\cdot p_2}\right)+(\mu_1\leftrightarrow\nu_1)\right]+(\mu_2\leftrightarrow\nu_2)\right\}, \tag{5.4.16}$$

which corresponds to eq. (5.2.12) with

$$a_2 = -\frac{ig}{384\pi^2}. \tag{5.4.17}$$

The transverse-traceless part $\left\langle t^{\mu_1\nu_1}t^{\mu_2\nu_2}J_A^{\mu_3}\right\rangle$ can be expressed in terms of four form factors as described in eq. (5.2.20). The perturbative calculation in four dimensions gives

$$\begin{aligned}
A_1 &= \frac{gp_2^2}{24\pi^2\lambda^4}\left\{A_{11}+A_{12}\log\left(\frac{p_1^2}{p_2^2}\right)+A_{13}\log\left(\frac{p_1^2}{p_3^2}\right)+A_{14}\,C_0(p_1^2,p_2^2,p_3^2)\right\},\\
A_2 &= \frac{gp_2^2}{48\pi^2\lambda^3}\left\{A_{21}+A_{22}\log\left(\frac{p_1^2}{p_2^2}\right)+A_{23}\log\left(\frac{p_1^2}{p_3^2}\right)+A_{24}\,C_0(p_1^2,p_2^2,p_3^2)\right\},\\
A_3 &= 0,\\
A_4 &= 0,
\end{aligned} \tag{5.4.18}$$

where $C_0$ in Minkowski space is the master integral

$$C_0(p_1^2,p_2^2,p_3^2) \equiv \frac{1}{i\pi^2}\int d^d l\,\frac{1}{l^2(l-p_1)^2(l+p_2)^2} \tag{5.4.19}$$

and we have introduced the following quantities

$$
\begin{aligned}
A_{11} =& -\lambda\Big[2p_1^{10} - p_1^8(p_2^2+p_3^2) - 2p_1^6(5p_2^4 - 48p_2^2p_3^2 + 5p_3^4) + 4p_1^4(p_2^2+p_3^2)(4p_2^4 - 23p_2^2p_3^2 + 4p_3^4) \\
&\qquad - 8p_1^2(p_2^2-p_3^2)^2(p_2^4 + 4p_2^2p_3^2 + p_3^4) + (p_2^2-p_3^2)^4(p_2^2+p_3^2)\Big] \\
A_{12} =& +2p_2^2\Big[p_3^2(p_3^2-p_2^2)^5 + p_1^{10}(38p_3^2 - 12p_2^2) + p_1^8(18p_2^4 + 41p_2^2p_3^2 - 121p_3^4) - 4p_1^6(3p_2^6 + 46p_2^4p_3^2 - 38p_2^2p_3^4 - 26p_3^6) \\
&\qquad + p_1^4(p_2-p_3)(p_2+p_3)(3p_2^6 + 95p_2^4p_3^2 + 215p_2^2p_3^4 + 11p_3^6) + 14p_1^2p_3^2(p_2^2-p_3^2)^3(p_2^2+p_3^2) + 3p_1^{12}\Big] \\
A_{13} =& +2p_3^2\Big[3p_1^{12} + 2p_1^{10}(19p_2^2 - 6p_3^2) + p_1^8(-121p_2^4 + 41p_2^2p_3^2 + 18p_3^4) + 4p_1^6(26p_2^6 + 38p_2^4p_3^2 - 46p_2^2p_3^4 - 3p_3^6) \\
&\qquad - 14p_1^2p_2^2(p_2^2-p_3^2)^3(p_2^2+p_3^2) - p_1^4(p_2-p_3)(p_2+p_3)(11p_2^6 + 215p_2^4p_3^2 + 95p_2^2p_3^4 + 3p_3^6) + p_2^2(p_2^2-p_3^2)^5\Big] \\
A_{14} =& -24p_1^4p_2^2p_3^2\Big[(p_1^2-p_2^2)^3(2p_1^2+3p_2^2) - 3p_3^4(p_1^4 + 4p_1^2p_2^2 - 4p_2^4) - 3p_3^2(p_1^6 - 6p_1^4p_2^2 + 4p_1^2p_2^4 + p_2^6) - 3p_3^8p_3^6(7p_1^2 - 3p_2^2)\Big] \\
A_{21} =& -\lambda\Big[2p_3^6(3p_1^2+p_2^2) + 4p_1^2p_3^4(3p_2^2 - 2p_1^2) + (p_1^2-p_2^2)^4 + 2p_3^2(p_1-p_2)(p_1+p_2)(p_1^4 + 8p_1^2p_2^2 + p_2^4) - p_3^8\Big] \\
A_{22} =& -2p_2^2p_3^2\Big[-17p_1^8 + p_1^6(28p_2^2 + 26p_3^2) - 4p_1^4(p_2^4 + 15p_2^2p_3^2) + (p_2^2-p_3^2)^4 - 2p_1^2(p_2^2-p_3^2)^2(4p_2^2+5p_3^2)\Big] \\
A_{23} =& +2p_3^2\Big[2p_1^{10} - p_1^8(p_2^2 + 6p_3^2) + p_1^6(-10p_2^4 + 46p_2^2p_3^2 + 6p_3^4) - 2p_1^2(4p_2^2+5p_3^2)(p_2^3 - p_2p_3^2)^2 + p_2^2(p_2^2-p_3^2)^4 \\
&\qquad + 2p_1^4(8p_2^6 - 21p_2^4p_3^2 - 18p_2^2p_3^4 - p_3^6)\Big] \\
A_{24} =& -12p_1^4p_2^2p_3^2\Big[p_3^4(3p_2^2 - 5p_1^2) + (p_1^2-p_2^2)^3 + p_3^2(p_1^4 + 4p_1^2p_2^2 - 5p_2^4) + 3p_3^6\Big]
\end{aligned}
\tag{5.4.20}
$$

with the Källen $\lambda$-function given by

$$
\lambda \equiv \lambda(p_1,p_2,p_3) = (p_1-p_2-p_3)(p_1+p_2-p_3)(p_1-p_2+p_3).(p_1+p_2+p_3) \tag{5.4.21}
$$

## 5.5 Matching the perturbative solution

In this section, we verify the matching between the perturbative form factors in Eq. (5.4.18) and the nonperturbative ones in Eq. (5.3.29). First of all, one can immediately see that $A_3$ and $A_4$ vanish in both calculations. On the other hand, in order to verify the matching between the first two form factors, we will need to rewrite the 3K integrals in the conformal solution in terms of the master integral $C_0$. For this purpose, we recall the reduction relations presented in [80, 81]

$$
I_{\alpha\{\beta_1\beta_2\beta_3\}} = (-1)^{\beta_t}\, \mathrm{K}_{j,\beta_j}^{|n_0|-1}\left[p_1^{2\beta_1}p_2^{2\beta_2}p_3^{2\beta_3}\left(\frac{1}{p_1}\frac{\partial}{\partial p_1}\right)^{\beta_1}\left(\frac{1}{p_2}\frac{\partial}{\partial p_2}\right)^{\beta_2}\left(\frac{1}{p_3}\frac{\partial}{\partial p_3}\right)^{\beta_3} I_{1\{000\}}\right] \tag{5.5.1}
$$

Moreover, the integral $I_{1\{0,0,0\}}$ is related to the massless scalar 1-loop 3-point momentum-space integral

$$
I_{1\{0,0,0\}} = (2\pi)^2 K_{4,\{1,1,1\}} = (2\pi)^2 \int \frac{d^4k}{(2\pi)^4}\frac{1}{k^2\,(k-p_1)^2\,(k+p_2)^2} = \frac{1}{4}C_0(p_1^2,p_2^2,p_3^2) \tag{5.5.2}
$$

where

$$K_{d\{\delta_1\delta_2\delta_3\}} \equiv \int \frac{d^dk}{(2\pi)^d} \frac{1}{(k^2)^{\delta_3}\,((k-p_1)^2)^{\delta_2}\,((k+p_2)^2)^{\delta_1}}. \tag{5.5.3}$$

Hence, it follows that the 3K integrals in our conformal solutions (5.3.29) are finite and can be reduced to

$$\begin{aligned} I_{5\{2,1,1\}} &= \frac{i}{4} p_1 p_2 p_3 \left(p_1 \frac{\partial}{\partial p_1} - 1\right) \frac{\partial^3}{\partial p_1 \partial p_2 \partial p_3} C_0(p_1^2, p_2^2, p_3^2) \\ I_{4\{2,1,0\}} &= -\frac{i}{4} p_1 p_2 \left(p_1 \frac{\partial}{\partial p_1} - 1\right) \frac{\partial^2}{\partial p_1 \partial p_2} C_0(p_1^2, p_2^2, p_3^2) \end{aligned} \tag{5.5.4}$$

By using the relations of the derivative acting on the master integral in Appendix B and setting the anomalous coefficient as in eq. (5.4.17), one can then verify the matching between the perturbative and nonperturbative form factors.

## 5.6 The anomaly pole of the gravitational anomaly and the sum rules

As we have already mentioned in the previous sections, it is clear that our conformal result does not depend on the specific expression of the current $J_A$ appearing in the correlator, since we have been using only the general symmetry properties of this 3-point function and its anomaly content in order to solve the conformal constraints.

Being the result unique and expressed in terms of a single constant, it shows that in a parity-odd CFT the gravitational anomaly vertex is generated by the exchange of an anomaly pole, with the entire correlator built around such massless pole and the value of its residue. Since this massless exchange was also present in the perturbative analysis of [120], we are now going to elaborate on those previous findings under the light of our current result.

### 5.6.1 Duality symmetry

The Maxwell equations in the absence of charges and currents satisfy the duality symmetry ($E \to B$ and $B \to -E$). The symmetry can be viewed as a special case of a continuous symmetry

$$\delta F_V^{\mu\nu} = \beta \tilde{F}_V^{\mu\nu} \tag{5.6.1}$$

where $\delta\beta$ is an infinitesimal $SO(2)$ rotation and $\tilde{F}^{\mu\nu} = \varepsilon^{\mu\nu\rho\sigma} F_{\rho\sigma}/2$. Its finite form

$$\begin{pmatrix} E \\ B \end{pmatrix} = \begin{pmatrix} \cos\beta & \sin\beta \\ -\sin\beta & \cos\beta \end{pmatrix} \begin{pmatrix} E \\ B \end{pmatrix} \tag{5.6.2}$$

is indeed a symmetry of the equations of motion, but not of the Maxwell action. Notice that the action

$$\mathcal{S} = \int d^4x F_V^{\mu\nu} F_{V\,\mu\nu} \tag{5.6.3}$$

is invariant under an infinitesmal transformation modulo a total derivative. For $\beta = \pi/2$, the discrete case, then the action flips sign since ($F_V^2 \to -\tilde{F}_V^2$).

In general, the infinitesimal variation of the action takes the form

$$\delta_\beta S_0 = -\beta \int d^4x \partial_\mu \left(\tilde{F}_V^{\mu\nu} V_\nu\right). \tag{5.6.4}$$

Due to the equivalence (dual Bianchi identity)

$$\partial_\nu F_V^{\mu\nu} = 0 \leftrightarrow \varepsilon^{\mu\nu\rho\sigma}\partial_\nu \tilde{F}_{V\rho\sigma} = 0, \tag{5.6.5}$$

we can introduce the dual gauge field $\tilde{V}^\mu$

$$\tilde{F}_V^{\mu\nu} = \partial^\mu \tilde{V}^\nu - \partial^\nu \tilde{V}^\mu, \tag{5.6.6}$$

which is related to the original $V_\mu$ one by

$$\partial^\mu \tilde{V}^\nu - \partial^\nu \tilde{V}^\mu = \varepsilon^{\mu\nu\rho\sigma}\partial_\rho V_\sigma. \tag{5.6.7}$$

The current corresponding to the infinitesimal symmetry (5.6.4) can be expressed in the form

$$J^\mu = \tilde{F}_V^{\mu\nu} V_\nu - F_V^{\mu\nu}\tilde{V}_\nu, \tag{5.6.8}$$

whose conserved charge is gauge invariant

$$Q_5 = \int d^3x \Big( V\cdot\nabla\times V - \tilde{V}\cdot\nabla\times\tilde{V}\Big) \tag{5.6.9}$$

after an integration by parts. Notice that the two terms on the equation above count the linking number of magnetic and electric lines respectively. In fluid mechanics, helicity is the volume integral of the scalar product of the velocity field with its curl given by

$$\mathcal{H}_{fluid} = \int d^3x\, \vec{v}\cdot\nabla\times\vec{v} \tag{5.6.10}$$

and one recognizes in (5.6.9) the expression

$$Q_5 = \int d^3x \Big( B\cdot V - E\cdot\tilde{V}\Big) \tag{5.6.11}$$

with $B = \nabla\times V$ and $E = -\nabla\times\tilde{V}$, that coincides with the optical helicity of the electromagnetic field [4]. As already mentioned, a perturbative analysis of $\langle TTJ_{CS}\rangle$ has been presented long ago in [95].
The presence of anomaly poles in this correlator can indeed be extracted from [95], in agreement with our result. Indeed, for on-shell gravitons ($g$) and photons ($\gamma$), the authors obtain, with the inclusion of mass effects in the $\langle J_V J_V J_A\rangle$, $\langle TTJ_f\rangle$ and $\langle TTJ_{CS}\rangle$ the following expressions for the matrix elements

$$\langle 0|J^\mu_{5f}|\gamma\gamma\rangle = f_1(q^2)\frac{q^\mu}{q^2}F_{\kappa\lambda}\tilde{F}^{\kappa\lambda} \tag{5.6.12}$$

$$\langle 0|J^\mu_{5f}|gg\rangle = f_2(q^2)\frac{q^\mu}{q^2}R_{\kappa\lambda\rho\sigma}\tilde{R}^{\kappa\lambda\rho\sigma} \tag{5.6.13}$$

$$\langle 0|J^\mu_{CS}|gg\rangle = f_3(q^2)\frac{q^\mu}{q^2}R_{\kappa\lambda\rho\sigma}\tilde{R}^{\kappa\lambda\rho\sigma}, \tag{5.6.14}$$

where $q$ is the momentum of the chiral current. The anomaly poles are extracted by including a mass $m$ in the propagators of the loop corrections, in the form of either a fermion mass for the $\langle J_V J_V J_A\rangle$ and the $\langle TTJ_{5f}\rangle$, or working with a Proca spin-1 in the case of $\langle TTJ_{CS}\rangle$, and then taking the limit for $m\to 0$. A dispersive analysis gives for the corresponding spectral densities [95]

$$\begin{aligned}
\Delta_{J_VJ_VJ_A}(q^2,m)\equiv \mathrm{Im} f_1(q^2) &= \frac{d_{J_VJ_VJ_A}}{q^2}(1-v^2)\log\frac{1+v}{1-v}\\
\Delta_{TTJ_{5f}}(q^2,m)\equiv \mathrm{Im} f_2(q^2) &= \frac{d_{TTJ_{5f}}}{q^2}(1-v^2)^2\log\frac{1+v}{1-v}\\
\Delta_{TTJ_{CS}}(q^2,m)\equiv \mathrm{Im} f_3(q^2) &= \frac{d_{TTJ_{CS}}}{q^2}v^2(1-v^2)^2\log\frac{1+v}{1-v},
\end{aligned} \tag{5.6.15}$$

with $v=\sqrt{1-4m^2/q^2}$ and $d_{J_VJ_VJ_A}=-1/2\,\alpha_{em}$, $d_{TTJ_{5f}}=1/(192\pi)$ and $d_{TTJ_{CS}}=1/(96\pi)$ being the corresponding anomaly coefficients in the normalization of the currents of [95], with $\alpha_{em}$ the electromagnetic coupling.
Notice the different functional forms of $\Delta_{TTJ_{5f}}(q^2,m)$ and $\Delta_{TTJ_{CS}}(q^2,m)$ away from the conformal limit, when the mass $m$ is nonzero. One can easily check that in the massless limit the branch cut present in the previous spectral densities at $q^2=4m^2$ turns into a pole

$$\lim_{m\to 0}\Delta(q^2,m)\propto\delta(q^2) \tag{5.6.16}$$

in all the three cases. Beside, one can easily show that the same spectral densities satisfy three sum rules

$$\int_{4m^2}^{\infty} ds\,\Delta_{J_VJ_VJ_A}(s,m) = 2\,d_{J_VJ_VJ_A} \tag{5.6.17}$$

$$\int_{4m^2}^{\infty} ds\,\Delta_{TTJ_{5f}}(s,m) = \frac{2}{3}d_{TTJ_{5f}} \tag{5.6.18}$$

$$\int_{4m^2}^{\infty} ds\,\Delta_{TTJ_{CS}}(s,m) = \frac{14}{45}d_{TTJ_{CS}}, \tag{5.6.19}$$

indicating that for any deformation $m$ from the conformal limit, the integral under $\Delta(s,m)$ is mass independent. Therefore, the numerical value of the area equals the value of the anomaly coefficient in each case. We will explore this topic in greater depth in Chapter 9.
One can verify, from equation (5.3.29), by taking the on-shell photon/graviton limit, that the transverse sector of $\langle TTJ_A\rangle$, corresponding to the form factors $A_1$ and $A_2$, vanishes, since these two form factors are zero, limiting each of these matrix elements to only single form factors, as indicated in (5.6.13) and (5.6.14). Then it is clear that, in general, the structure of the anomaly action responsible for the generation of the gravitational chiral anomaly can be expressed in the form

$$\mathcal{S}_{anom}\sim\int d^4x\,d^4y\,\partial_\lambda A^\lambda\,\frac{1}{\square}(x,y)R\tilde{R}(y)+\ldots \tag{5.6.20}$$

where the ellipses stand for the transverse sector, and $A_\lambda$ is a spin-1 external source. For on-shell gravitons, as remarked above, this action summarizes the effect of the entire chiral gravitational anomaly vertex, being exactly given by the exchange of a single anomaly pole.

## 5.7 Summary of the results

Before coming to our comments and conclusions, for the reader's convenience, we briefly summarize our findings.

We have shown that in a general CFT the $\langle TTJ_A\rangle$ correlator can be written as a sum of two terms

$$\left\langle T^{\mu_1\nu_1}T^{\mu_2\nu_2}J_A^{\mu_3}\right\rangle = \left\langle t^{\mu_1\nu_1}t^{\mu_2\nu_2}j_A^{\mu_3}\right\rangle + \left\langle t^{\mu_1\nu_1}t^{\mu_2\nu_2}j_{A\,loc}^{\mu_3}\right\rangle, \tag{5.7.1}$$

the first term being the transverse component and the second, the longitudinal one, expressed in terms of a single anomaly form factor and tensor structure. This is characterized by an interpolating anomaly pole.
The anomaly part is given by the expression

$$\left\langle t^{\mu_1\nu_1}t^{\mu_2\nu_2}j_{A\,loc}^{\mu_3}\right\rangle = 4ia_2\frac{p_3^{\mu_3}}{p_3^2}(p_1\cdot p_2)\left\{\left[\varepsilon^{\nu_1\nu_2p_1p_2}\left(g^{\mu_1\mu_2}-\frac{p_1^{\mu_2}p_2^{\mu_1}}{p_1\cdot p_2}\right)+(\mu_1\leftrightarrow\nu_1)\right]+(\mu_2\leftrightarrow\nu_2)\right\} \tag{5.7.2}$$

while the transverse-traceless part is

$$\begin{aligned}\langle t^{\mu_1\nu_1}(p_1)t^{\mu_2\nu_2}(p_2)j_A^{\mu_3}(p_3)\rangle &= \Pi^{\mu_1\nu_1}_{\alpha_1\beta_1}(p_1)\Pi^{\mu_2\nu_2}_{\alpha_2\beta_2}(p_2)\pi^{\mu_3}_{\alpha_3}(p_3)\Big[\\ &A_1\varepsilon^{p_1\alpha_1\alpha_2\alpha_3}p_2^{\beta_1}p_3^{\beta_2}-A_1(p_1\leftrightarrow p_2)\varepsilon^{p_2\alpha_1\alpha_2\alpha_3}p_2^{\beta_1}p_3^{\beta_2}+A_2\varepsilon^{p_1\alpha_1\alpha_2\alpha_3}\delta^{\beta_1\beta_2}-A_2(p_1\leftrightarrow p_2)\varepsilon^{p_2\alpha_1\alpha_2\alpha_3}\delta^{\beta_1\beta_2}\Big]\end{aligned} \tag{5.7.3}$$

with $A_1$ and $A_2$ given by eq. (5.3.29).
The entire correlator is therefore determined *only by the anomalous coefficient* $a_2$ in (5.7.2).
We have also computed the correlator perturbatively at one-loop in free field theory and verified the agreement of the expression with the nonperturbative results obtained by imposing the conformal symmetry. The explicit expressions of the form factors $A_1$ and $A_2$ have been given in (5.4.18).
The solutions of the conformal constraints, expressed in terms of 3K integrals $I_{5\{2,1,1\}}$ and $I_{4\{2,1,0\}}$, can be related to the ordinary one-loop master integrals $C_0$ and $B_0$ by (5.5.4). They can be reconstructed using recursively the relations included in Appendix B.

## 5.8 Comments: nonrenormalization of the $\langle J_V J_V J_A\rangle$ and $\langle TTJ_A\rangle$

Before coming to our conclusions, we pause for few comments on the results of our paper, in relation to our previous study of the $\langle J_V J_V J_A\rangle$ chiral anomaly vertex, in a general CP-violating CFT [34].
In the case of the $\langle J_V J_V J_A\rangle$ vertex, the Adler-Bardeen theorem shows that the longitudinal part of the interaction is not affected by renormalization and therefore can be computed exactly just from the one-loop triangle diagram, being protected from perturbative corrections at higher orders. This is not true for the transverse part of the same diagram. However, in [102] it was pointed out that, in the kinematic limit where the momentum of one of the vector currents is vanishingly small, another nonrenormalization theorem is valid. Indeed, in that limit just two form factors are needed to fully describe the $\langle J_V J_V J_A\rangle$ correlator. One of these form factors is related to the axial anomaly and, therefore, it is not renormalized. The other form factor belongs to the transverse sector. In [102] it was shown that, due to helicity conservation in massless QCD, the two form factors are in fact proportional to each other, and so the nonrenormalization of one of them implies that of the other. If the anomalous behavior is identified with the exchange of an anomaly pole, that result relates the anomaly pole to the transverse part of the diagram, when one of the photons becomes soft.
In the most general kinematics, perturbative analysis of the diagram showed that at two-loops the entire diagram is nonrenormalized [107], a feature that disappears at higher perturbative orders. Indeed, the authors of [108] found nonvanishing corrections to the correlator at $O(\alpha_S^2)$.
In the most general kinematics, the nonrenormalization to all orders of a specific combination of the transverse form factors of the $\langle J_V J_V J_A \rangle$ was shown to hold in [113], in the chiral limit of QCD
In the previous chapter, we have shown that such results follow from conformal symmetry, once the conformal constraints are solved either in the most general kinematics or in the specific one required by Vainshtein's conjecture [102]. Therefore, the breaking of the nonrenormalizaton theorem for the entire vertex in QCD must originate from terms breaking conformal invariance and must be proportional to the QCD $\beta$ function.
In this chapter, we have verified that a similar connection between the longitudinal and the transverse part is present in the case of the $\langle TTJ_A \rangle$ correlator in the conformal limit, being both sectors proportional to the $a_2$ anomalous coefficient.
With these new indications, that follow quite closely the $\langle J_V J_V J_A \rangle$ case previously discussed by us, it would be interesting to test, at the perturbative level, if in the soft graviton limit a similar result holds for the $\langle TTJ_A \rangle$ at all orders in perturbative QCD. We do expect that the higher order corrections will be proportional to the QCD $\beta$ function, therefore breaking the conformal symmetry.

## 5.9 Conclusions

We have presented an analysis of the gravitational anomaly vertex from the perspective of CFT in momentum-space. We have shown how the vertex can be completely defined by the inclusion of a single anomaly pole together with the CWIs. This explicit analysis shows that reconstruction method formulated in the parity-even sector in the case of conformal anomaly correlators can be extended quite naturally to the parity-odd sector. This provides a different and complementary perspective on the origin of anomalies and their related effective actions, which may account for such phenomena. This extension highlights the intrinsic connection between these seemingly distinct sectors and suggests a unified framework for comprehending the origin of anomalies. It underscores the notion that anomalies, whether chiral, conformal, or supersymmetric, share a common underlying structure characterized by the presence of a single (anomaly) form factor, together with a specific tensor structure responsible for generating the anomaly.
The approach does not rely on the explicit structure of the parity-odd current appearing in the correlator but, rather, on its symmetry properties. We have also shown that, similarly to previous dispersive analysis of the anomalous form factors for the $\langle TJJ \rangle$ and $\langle J_V J_V J_A \rangle$ diagrams, the spectral density of the anomalous form factor of the $\langle TTJ_A \rangle$ satisfies a sum rule. The numerical value of the sum rule is fixed by the anomaly.

# Chapter 6

# CP-Violating trace anomalies

In this chapter, we explore the role of parity-odd trace anomalies in shaping correlators within momentum-space CFT, focusing on the $\langle JJO\rangle_{odd}$, $\langle TTO\rangle_{odd}$, $\langle JJT\rangle_{odd}$, and $\langle TTT\rangle_{odd}$. Here, $O$ represents either a scalar or pseudoscalar operator, which may correspond to the trace of the energy-momentum tensor. We specifically examine how all these correlators are constrained by conformal Ward identities and investigate the impact of parity-odd trace anomalies on the equations governing the correlators. The $\langle JJO\rangle_{odd}$ and $\langle TTO\rangle_{odd}$ can be different from zero in a CFT. This can occur, for example, when the conformal dimension of the scalar operator is $\Delta_3 = 4$, as in the case where $O = T_\mu^\mu$. Moreover, assuming the presence of parity-odd trace anomalies, both $\langle JJT\rangle_{\text{odd}}$ and $\langle TTT\rangle_{\text{odd}}$ become nonzero. In the case of $\langle JJT\rangle_{\text{odd}}$, the transverse-traceless component vanishes, leaving the correlator entirely determined by its trace part, characterized by an anomaly pole.
This study is motivated by its relevance to holography, early universe cosmology, and ongoing debates about parity-odd trace anomalies in chiral fermion theories.

## 6.1 Introduction

Conformal anomalies have some distinctive features that make them more complex compared to the chiral ones, due to the presence of both topological and non-topological terms in the anomaly functional. As was first found by Capper and Duff on dimensional grounds and by requiring covariance [61, 62], the structure of the trace anomaly in four dimensions is given by

$$g_{\mu\nu}\langle T^{\mu\nu}\rangle = \mathcal{A} = b_1\, E_4 + b_2\, C^{\mu\nu\rho\sigma}C_{\mu\nu\rho\sigma} + b_3\,\nabla^2 R + b_4\, F^{\mu\nu}F_{\mu\nu} + f_1\,\varepsilon^{\mu\nu\rho\sigma}R_{\alpha\beta\mu\nu}R^{\alpha\beta}{}_{\rho\sigma} + f_2\,\varepsilon^{\mu\nu\rho\sigma}F_{\mu\nu}F_{\rho\sigma}, \quad (6.1.1)$$

where $C_{\mu\nu\rho\sigma}$ is the Weyl tensor and $E_4$ is the Gauss-Bonnet term

$$\begin{aligned} C^{\mu\nu\rho\sigma}C_{\mu\nu\rho\sigma} &= R^{\mu\nu\rho\sigma}R_{\mu\nu\rho\sigma} - 2R^{\mu\nu}R_{\mu\nu} + \frac{1}{3}R^2, \\ E_4 &= R^{\mu\nu\rho\sigma}R_{\mu\nu\rho\sigma} - 4R^{\mu\nu}R_{\mu\nu} + R^2. \end{aligned} \quad (6.1.2)$$

The trace anomaly can include both parity-even and parity-odd terms. All the terms in the trace anomaly (6.1.1) are parity-even except for the last two of them which are parametrized by the coefficient $f_1$ and $f_2$. All the coefficients $b_i$ in the parity-even terms have been computed and their values strictly depend on the number and type of fields entering in the perturbative quantum corrections, but they are all real. Indeed, if we consider that the energy-momentum tensor is a fundamental composite operator of the Standard Model, the presence of imaginary coefficients would endanger the consistency of the theory.
Concerning the parity-odd terms in eq. (6.1.1), several computations using different regularization schemes indicate that the coefficients $f_i$ vanish for Weyl fermions. Specifically, the computation of parity-odd terms has been performed in free field theory using various regularization schemes, ranging from dimensional reduction [88] to Pauli-Villars [123,124] and the 't Hooft-Veltman Breitenlohner-Mason scheme [89], all consistently showing that these terms are zero. This result has been further corroborated by several other studies [21, 24, 92, 125, 126].
In recent years, however, following the work of Nakayama [127], Bonora and collaborators have claimed the existence of nonvanishing parity-odd terms in the trace anomaly of Weyl fermions. They employed both diagrammatic approaches and the heat-kernel method to reach this conclusion [20, 22, 55, 90, 91, 128, 129]. Moreover, new independent studies [130, 131] (see also [132, 133] for the supersymmetric version) have recently supported these alternative conclusions. Notably, in these computations, the coefficients $f_i$ are found to be nonzero and imaginary, raising critical concerns about unitarity. If these results are confirmed, they would have significant phenomenological implications. Such anomalies would indicate that the chiral spectrum of the Standard Model would require modifications for consistent coupling to gravity—possibly involving the addition of extra chiral fermions. For instance, accepting these findings would immediately imply that the correlator $\langle TJ_YJ_Y\rangle$, where $J_Y$ represents the hypercharge current of the Standard Model, does not vanish, necessitating additional chiral matter.
The study of parity-odd trace anomalies also intersects with holographic considerations. In [127,134], the author presented a holographic model predicting a Pontryagin density in the trace of the energy-momentum tensor, though this approach also encounters unitarity issues.
Without delving into the debate of whether free field theory, and the Standard Model in particular, are affected by parity-odd anomalies, one can investigate, on more general grounds, whether conformal field theories permit such parity-odd terms, when the coefficients $f_i$ are generic, as well as the constraints on the structure of the correlators. For our purposes, the real or imaginary nature of these coefficients is secondary and independent of the fact that a free field theory realization is possible. As we are going to see, significant constraints on the structure of correlators can be derived from the solution of their CWIs in the presence and absence of this kind of anomalies.
Concerning the methodology, we recall that the general solutions of the CWIs have been successfully analyzed in momentum-space in the presence of ordinary (parity-even) trace anomalies [26, 27]. The conformal analysis for the $\langle TTT\rangle$ and $\langle TJJ\rangle$ correlators, in the parity even case, has been matched exactly with one-loop perturbation theory, providing a direct way to investigate the renormalization of the corresponding correlators using ordinary Feynman diagrams. The three constants, appearing in the solution of the corresponding CWIs, are uniquely associated with the multiplicities of scalars ($n_S$), fermions ($n_f$) and spin-1 gauge fields ($n_V$) of the free field theory realizations [75,77]. A general review of these methods can be found in [72].
In this chapter, we investigate the conformal constraints of the $\langle JJO\rangle_{odd}$, $\langle TTO\rangle_{odd}$, $\langle JJT\rangle_{odd}$ and $\langle TTT\rangle_{odd}$ correlators. Specifically, we examine the implications of parity-odd trace anomalies on the structure of these correlators and the corresponding conformal equations. Our findings reveal interesting and unique features of the correlators which we decompose as usual into transverse-traceless, trace and longitudinal sectors.
As we are going to discuss in Section 6.3, no renormalization counterterms are introduced in the analysis of the parity-odd correlators mentioned above. In this sense, the analysis presented in this chapter closely parallels that of the $\langle J_VJ_VJ_A\rangle$ chiral anomaly correlator, which satisfies the standard dilatation and special CWIs. Indeed, in the perturbative analysis of the $\langle J_VJ_VJ_A\rangle$, no counterterms is required: the chiral anomaly is topological and the corresponding interaction become finite solely by enforcing the conservation Ward identities on the vector currents.

In this chapter, we demonstrate that, despite the potential existence of parity-odd trace anomalies, the correlators $\langle JJO\rangle_{odd}$, $\langle TTO\rangle_{odd}$ and $\langle JJT\rangle_{odd}$ always satisfy non-anomalous ordinary CWIs. The only correlator that exhibits anomalous CWIs is the $\langle TTT\rangle_{odd}$, limitedly to the special conformal transformation.
Furthermore, we show that in the case of the $\langle JJT\rangle_{odd}$, the correlator is nonvanishing only if we allow an anomalous trace. The anomalous-trace solution is particularly intriguing, as all other components of the correlator (transverse-traceless and longitudinal) are constrained to vanish under the conformal equations.

## 6.2 Defining the conformal anomaly

The violation of conformal symmetry at the quantum level is manifested in the energy-momentum tensor's failure to remain traceless. Thus, the conformal anomaly is generally defined through the trace of the energy-momentum tensor, as in eq. (6.1.1). However, this anomaly can also be interpreted as the failure of the trace operation to commute with the quantum average. Specifically, the anomaly may be defined as the difference between two trace operations on the energy-momentum tensor, one performed before the quantum average and the other after

$$\mathcal{A} = g^{\mu\nu}\left\langle T_{\mu\nu}(x)\right\rangle - \left\langle T^{\mu}_{\mu}(x)\right\rangle, \tag{6.2.1}$$

This perspective was introduced by Duff [62,135] and has also been adopted by Bonora [20] to demonstrate the existence of parity-odd trace anomalies. The significance of the equation above lies in its ability to eliminate contributions to the trace that arise from classically non-invariant terms in non-conformal theories, such as mass-dependent terms. These contributions are removed by the subtraction in (6.2.1). A distinctive feature of Bonora's analysis is that the second term in (6.2.1), initially introduced to cancel classical contributions in non-conformal theories, generate parity-odd trace anomalies. Indeed, as shown in [20], this term also contributes to the trace anomaly even within a conformal theory.
We now turn our attention to the possible gauge anomaly $F\tilde{F}$. To examine the contribution from the first term on the right-hand side of (6.2.1) in $d = 4$, we focus on the correlator $\langle JJT\rangle_{odd}$. By computing this correlator and tracing over the energy-momentum tensor, one can potentially obtain the anomaly. The parity-odd nature of this anomaly emerges in a perturbative analysis through the insertion of a $\gamma^5$ in one or all of the three operators of the correlator. Alternatively, one can consider a theory coupled to chiral fermions. From a CFT perspective, all these cases are unified under the classification $\langle JJT\rangle_{odd}$.
The second term in (6.2.1) involves a trace taken inside the quantum average, yielding the expectation value of a scalar operator. The correlator best associated with this term is $\langle JJO\rangle_{odd}$, where $O = T^{\mu}_{\mu}$. Once more, this correlator can be realized perturbatively by inserting an odd number of operators with a $\gamma^5$. According to Bonora's analysis, both terms in (6.2.1) contribute to the $F\tilde{F}$ anomaly.
A similar analysis applies to the trace anomaly $R\tilde{R}$. For the first term in (6.2.1), one examines the $\langle TTT\rangle_{odd}$ correlator, while for the second term, the relevant correlator is $\langle TTO\rangle_{odd}$. However, Bonora's results indicate that only the second term contributes to the $R\tilde{R}$ anomaly, implying that taking the external trace of $\langle TTT\rangle_{odd}$ yields zero.
In this chapter, we will concentrate on solving the CFT constraints for the correlators mentioned above, both in the presence and absence of parity-odd trace anomalies.

## 6.3 Anomalies and renormalization

The correlation functions in $d = 4$ dimensions can be fixed by imposing anomalous CWIs. The general approach involves considering the ordinary (non-anomalous) CWIs satisfied by the theory in general $d$ dimensions and then introducing a renormalization procedure that allows to remove the divergences generated by the solution in the $d \to 4$ limit. This method has been developed, independently of any free field theory realization, in [26, 27, 29] by introducing suitable counterterms for each type of correlator. In the case of parity-even trace anomalies, correlators such as the $\langle TJJ\rangle_{even}$ satisfy ordinary CWIs in $d$ dimensions and are renormalized by the inclusion of a $1/(d-4)F^2$ counterterm. Obviously, the types of counterterms needed in the renormalization of a generic CFT are those that appear in the expression of the trace anomaly in the parity even case. For correlators such as the $\langle TTT\rangle_{even}$ two counterterms are needed, $V_E/\epsilon$ and $V_{C^2}/\epsilon$, with

$$\begin{aligned} V_E(g,d) &\equiv \mu^\epsilon \int d^dx\,\sqrt{-g}\,E, \qquad\qquad \epsilon = d-4 \\ V_{C^2}(g,d) &\equiv \mu^\epsilon \int d^dx\,\sqrt{-g}\,C^2. \end{aligned} \tag{6.3.1}$$

The inclusion of $E$, despite its topological nature in $d = 4$, ensures consistency with the Wess-Zumino condition for the anomaly effective action. The generation of the anomaly, in DR, is associated with the non-invariance of the two counterterms under Weyl variations

$$\begin{aligned} 2g_{\mu\nu}\frac{\delta}{\delta g_{\mu\nu}}\left(\frac{V_E}{\epsilon}\right) &= \sqrt{g}E, \\ 2g_{\mu\nu}\frac{\delta}{\delta g_{\mu\nu}}\left(\frac{V_{C^2}}{\epsilon}\right) &= \sqrt{g}\left[C^2+\frac{2}{3}\Box R\right] \end{aligned} \tag{6.3.2}$$

When considering parity-odd contributions, a distinct issue arises: it is not possible to define $d$-dimensional counterterms for terms like $R\tilde{R}$ and $F\tilde{F}$ and then take the limit $\epsilon = d-4 \to 0$ to induce a finite renormalization. This situation is in stark contrast to the parity-even case, where the inclusion of counterterms is necessary to handle divergences. For the parity-odd interactions, no counterterms are required. Perturbative computations confirm that these contributions do not exhibit divergences and therefore do not necessitate renormalization. This holds true regardless of whether the results yield parity-odd trace anomalies. Additionally, it is worth noting that topological anomalies can be generated without any renormalization procedure, as demonstrated in the $\langle J_V J_V J_A\rangle$ chiral anomaly diagram.

## 6.4 Anomalous conformal Ward identities

In this section, we examine the anomalous CWIs using an alternative derivation to the one commonly employed. The equations we present apply to both the parity-even and parity-odd sectors of correlators. Moreover, these equations can be naturally extended to correlators of a non-Lagrangian CFT. For clarity and simplicity, we adopt a direct approach by deriving the CWIs starting from the effective action $\mathcal{S}$ in euclidean space with

$$e^{-\mathcal{S}} \equiv \mathcal{N}\int D\chi\, e^{-S_0} \tag{6.4.1}$$

where $\chi$ is a collection of conformal fields, including scalars and fermions. We recall the definition of the following one-point functions

$$\langle T^{\mu\nu}(x)\rangle = \frac{2}{\sqrt{-g(x)}}\frac{\delta\mathcal{S}}{\delta g_{\mu\nu}(x)}\bigg|_0, \qquad \langle O(x)\rangle = \frac{1}{\sqrt{-g(x)}}\frac{\delta\mathcal{S}}{\delta\phi(x)}\bigg|_0,$$
$$\left\langle J_A^{\mu}(x)\right\rangle = \frac{1}{\sqrt{-g(x)}}\frac{\delta\mathcal{S}}{\delta A_\mu(x)}\bigg|_0, \qquad \left\langle J_V^{\mu}(x)\right\rangle = \frac{1}{\sqrt{-g(x)}}\frac{\delta\mathcal{S}}{\delta V_\mu(x)}\bigg|_0, \tag{6.4.2}$$

where the subscript zero indicates switching off the sources and the metric is taken flat. If we focus specifically on the scalar operator $O = T^{\mu}_{\mu}$, the field $\phi$ in the equation above corresponds to the dilaton and one can write

$$\frac{\delta}{\delta\phi} = 2g_{\mu\nu}\frac{\delta}{\delta g_{\mu\nu}} \tag{6.4.3}$$

as described by the Weyl-gauging procedure [136]. By applying multiple functional derivative to the action $\mathcal{S}$, one can obtain higher-point functions.
In order to derive the dilatation and the special CWIs, we need first to introduce the following conformal current

$$J^{\mu}_{(K)} \equiv K_\nu(x)\, T^{\mu\nu}(x) \tag{6.4.4}$$

expressed in terms of $K_\mu(x)$, which are the Conformal Killing Vectors (CKVs) satisfying the equation

$$\nabla_{(\mu}K_{\nu)} \equiv \frac{1}{2}\left(\nabla_\mu K_\nu + \nabla_\nu K_\mu\right) = \frac{1}{d}\, g_{\mu\nu}\,(\nabla\cdot K) \tag{6.4.5}$$

We recall that the CKVs in flat space for the dilatations are given by

$$K^{(D)}_\mu \equiv x_\mu, \qquad \partial\cdot K^{(D)} = d, \tag{6.4.6}$$

while the special CKVs are

$$K^{(S)\kappa}_{\ \ \mu} \equiv 2x^\kappa x_\mu - x^2\delta^\kappa_{\ \mu}, \qquad \partial\cdot K^{(S)\kappa} = (2d)\,x^\kappa, \quad \kappa = 1,\dots,d \tag{6.4.7}$$

The conservation of $J^{\mu}_{(K)}$ is violated by the presence of the conformal anomaly. Indeed, one can write

$$\nabla_\mu J^{\mu}_{(K)} = K_\nu\nabla_\mu T^{\mu\nu} + \left(\nabla_\mu K_\nu\right)T^{\mu\nu} = K_\nu\nabla_\mu T^{\mu\nu} + \frac{(\nabla\cdot K)}{d}\, g_{\mu\nu}T^{\mu\nu} \tag{6.4.8}$$

where we have used (6.4.5). The first term on the right-hand side is related to the conservation of the energy-momentum tensor, while the conformal anomaly appears in the second term of the equation. As we are going to see, the current (6.4.4) can be inserted in the $n$-point correlator to derive its anomalous CWIs. The derivation of these equations can be found in [71] for the $\langle TTT\rangle$, while the explicit derivation for the $\langle JJT\rangle$ will be worked out below.
Before starting our derivation, we introduce, for convenience, the following quantity, which will be used in subsequent sections

$$\begin{aligned}\tilde{\mathcal{A}} &\equiv \sqrt{g}\mathcal{A} = 2g_{\mu\nu}\frac{\delta\mathcal{S}}{\delta g_{\mu\nu}} = \sqrt{g}\, g_{\mu\nu}T^{\mu\nu}\\ &= \sqrt{g}\Big(b_1\, E_4 + b_2\, C^{\mu\nu\rho\sigma}C_{\mu\nu\rho\sigma} + b_3\,\nabla^2 R + b_4\, F^{\mu\nu}F_{\mu\nu} + f_1\,\varepsilon^{\mu\nu\rho\sigma}R_{\alpha\beta\mu\nu}R^{\alpha\beta}_{\ \ \rho\sigma} + f_2\,\varepsilon^{\mu\nu\rho\sigma}F_{\mu\nu}F_{\rho\sigma}\Big).\end{aligned} \tag{6.4.9}$$

The quantity $\tilde{\mathcal{A}}$ differs from the anomaly $\mathcal{A}$, as defined in earlier sections, by a factor of $\sqrt{g}$. Let us now examine the final term in the equation, which influences the $\langle JJT\rangle_{\text{odd}}$ correlator and is parametrized by the coefficient $f_2$. Since we employ the Levi-Civita pseudotensor $\varepsilon^{\mu\nu\rho\sigma}$, defined with $\varepsilon^{0123}=\frac{1}{\sqrt{g}}$, the $\sqrt{g}$ dependence in Eq. (6.4.9) cancels out for this term. Consequently, it becomes independent of the metric. This metric-independence will prove crucial in the analysis that follows.

### 6.4.1 $\langle JJT\rangle$

In this section, we derive the anomalous CWIs applicable to both the parity-even and parity-odd sectors of the $\langle JJT\rangle$ correlator. We begin by assuming that the following surface term vanishes due to the rapid fall-off behavior of the correlation function at infinity

$$
\begin{aligned}
0=\int dx\, \partial^{(x)}_\mu \langle J^\mu_{(K)}(x) J^{\mu_1}(x_1) J^{\mu_2}(x_2) T^{\mu_3\nu_3}(x_3)\rangle = \int dx\, \partial_\mu \Big[K_\nu \langle T^{\mu\nu}(x) J^{\mu_1}(x_1) J^{\mu_2}(x_2) T^{\mu_3\nu_3}(x_3)\rangle\Big] = \\
\int dx \left(\partial_\mu K_\nu\right)\langle T^{\mu\nu}(x) J^{\mu_1}(x_1) J^{\mu_2}(x_2) T^{\mu_3\nu_3}(x_3)\rangle + K_\nu \partial_\mu \langle T^{\mu\nu}(x) J^{\mu_1}(x_1) J^{\mu_2}(x_2) T^{\mu_3\nu_3}(x_3)\rangle.
\end{aligned}
\tag{6.4.10}
$$

Recalling the conformal Killing vector equation (6.4.5), we can then write

$$
0=\int dx \left(\frac{\partial\cdot K}{d}\right)\eta_{\mu\nu}\langle T^{\mu\nu}(x) J^{\mu_1}(x_1) J^{\mu_2}(x_2) T^{\mu_3\nu_3}(x_3)\rangle + K_\nu \partial_\mu \langle T^{\mu\nu}(x) J^{\mu_1}(x_1) J^{\mu_2}(x_2) T^{\mu_3\nu_3}(x_3)\rangle. \tag{6.4.11}
$$

On the right-hand side of the equation, we encounter the trace and the divergence of a four-point correlator function. These terms can be rewritten using the anomalous trace equation and the conservation of the energy-momentum tensor. We show this below.
The invariance under diffeomorphism leads to

$$
\nabla^\mu\langle T_{\mu\nu}\rangle - F^A_{\mu\nu}\langle J^\mu_A\rangle + A_\nu \nabla_\mu \langle J^\mu_A\rangle - F^V_{\mu\nu}\langle J^\mu_V\rangle + V_\nu\nabla_\mu\langle J^\mu_V\rangle = 0. \tag{6.4.12}
$$

Applying functional derivatives to this equation and going to the flat limit, we obtain

$$
\begin{aligned}
0 =& \partial_\mu \langle T^{\mu\nu}(x) J^{\mu_1}(x_1) J^{\mu_2}(x_2) T^{\mu_3\nu_3}(x_3)\rangle - \eta^{\mu_3\nu_3}\left(\partial_\mu \delta_{xx_3}\right)\langle J^{\mu_1}(x_1) J^{\mu_2}(x_2) T^{\mu\nu}(x)\rangle + \\
&\Big[-\delta^{\mu_1}_\mu \partial_\nu \delta_{xx_1} + \delta^{\mu_1}_\nu \partial_\mu \delta_{xx_1}\Big]\langle J^{\mu}(x) J^{\mu_2}(x_2) T^{\mu_3\nu_3}(x_3)\rangle + \Big[-\delta^{\mu_2}_\mu \partial_\nu \delta_{xx_2} + \delta^{\mu_2}_\nu \partial_\mu \delta_{xx_2}\Big]\langle J^{\mu_1}(x_1) J^{\mu}(x) T^{\mu_3\nu_3}(x_3)\rangle \\
&+\frac{1}{2}\langle J^{\mu_1}(x_1) J^{\mu_2}(x_2) T^{\mu\lambda}(x)\rangle\Big[\delta^{\nu_3}_\lambda \delta^{\nu\mu_3}\partial_\mu \delta_{xx_3} + \delta^{\nu_3}_\mu \delta^{\nu\mu_3}\partial_\lambda \delta_{xx_3} - \delta^{\nu_3}_\lambda \delta^{\mu_3}_\mu \partial^\nu \delta_{xx_3} + \\
&\delta^{\mu_3}_\lambda \delta^{\nu\nu_3}\partial_\mu \delta_{xx_3} + \delta^{\mu_3}_\mu \delta^{\nu\nu_3}\partial_\lambda \delta_{xx_3} - \delta^{\mu_3}_\lambda \delta^{\nu_3}_\mu \partial^\nu \delta_{xx_3}\Big] + \delta^{\mu_3\nu_3}\left(\partial_\lambda \delta_{xx_3}\right)\langle J^{\mu_1}(x_1) J^{\mu_2}(x_2) T^{\lambda\nu}(x)\rangle,
\end{aligned}
\tag{6.4.13}
$$

where we have denoted with $\delta_{xy}$ the Dirac delta function $\delta^4(x-y)$. Note that, in the equation above, $J$ may represent either a vector or an axial-vector current. The anomalous trace equation is instead given by

$$
0 = 2g_{\mu\nu}(x)\frac{\delta\mathcal{S}[g]}{\delta g_{\mu\nu}(x)} - \tilde{\mathcal{A}}(x). \tag{6.4.14}
$$

Applying once again functional derivatives to the equation above and going to the flat limit, we find

$$
\begin{aligned}
0 =& \eta_{\mu\nu}\langle T^{\mu\nu}(x) J^{\mu_1}(x_1) J^{\mu_2}(x_2) T^{\mu_3\nu_3}(x_3)\rangle \\
&+ 2\delta_{xx_3}\langle J^{\mu_1}(x_1) J^{\mu_2}(x_2) T^{\mu_3\nu_3}(x_3)\rangle - 2\frac{\delta^3\tilde{\mathcal{A}}(x)}{\delta A_{\mu_1}(x_1)\delta A_{\mu_2}(x_2)\delta g_{\mu_3\nu_3}(x_3)}.
\end{aligned}
\tag{6.4.15}
$$

We can now insert the equations (6.4.13) and (6.4.15) into (6.4.11) and integrate by parts. In the case of dilatations, using the conformal Killing vectors in eq. (6.4.6), we find

$$\sum_{i=1}^{3}\left(\Delta_i + x_i^\mu \frac{\partial}{\partial x_i^\mu}\right)\langle J^{\mu_1}(x_1)J^{\mu_2}(x_2)T^{\mu\nu}(x_3)\rangle = 2\int dx \frac{\delta^3 \tilde{\mathcal{A}}(x)}{\delta A_{\mu_1}(x_1)\delta A_{\mu_2}(x_2)\delta g_{\mu_3\nu_3}(x_3)}. \tag{6.4.16}$$

If, instead, we consider the special conformal transformations, the conformal Killing vectors are given in (6.4.7) and we find

$$\begin{aligned}
&\sum_{i=1}^{3}\left[2x_i^\kappa\left(\Delta_i + x_i^\alpha \frac{\partial}{\partial x_i^\alpha}\right) - x_i^2\delta^{\kappa\alpha}\frac{\partial}{\partial x_i^\alpha}\right]\langle J^{\mu_1}(x_1)J^{\mu_2}(x_2)T^{\mu_3\nu_3}(x_3)\rangle \\
&+2\left[\delta^{\kappa\mu_1}x_{1\alpha} - \delta^\kappa_\alpha x_1^{\mu_1}\right]\langle J^{\alpha}(x_1)J^{\mu_2}(x_2)T^{\mu_3\nu_3}(x_3)\rangle + 2\left[\delta^{\kappa\mu_2}x_{2\alpha} - \delta^\kappa_\alpha x_2^{\mu_2}\right]\langle J^{\mu_1}(x_1)J^{\alpha}(x_2)T^{\mu_3\nu_3}(x_3)\rangle \\
&+2\left[\delta^{\kappa\mu_3}x_{3\alpha} - \delta^\kappa_\alpha x_3^{\mu_3}\right]\langle J^{\mu_1}(x_1)J^{\mu_2}(x_2)T^{\alpha\nu_3}(x_3)\rangle + 2\left[\delta^{\kappa\nu_3}x_{3\alpha} - \delta^\kappa_\alpha x_3^{\nu_3}\right]\langle J^{\mu_1}(x_1)J^{\mu_2}(x_2)T^{\mu_3\alpha}(x_3)\rangle \\
&= 2^2\int dx\, x^\kappa \frac{\delta^3 \tilde{\mathcal{A}}(x)}{\delta A_{\mu_1}(x_1)\delta A_{\mu_2}(x_2)\delta g_{\mu_3\nu_3}(x_3)}.
\end{aligned} \tag{6.4.17}$$

### 6.4.2 $\langle TTT\rangle$

Proceeding in a manner similar to the previous section, one can derive the anomalous conformal Ward identities of other correlators. The case of the $\langle TTT\rangle$ has been worked out in [71]. The dilatation equation is given by

$$\sum_{i=1}^{3}\left(\Delta_i + x_i^\mu \frac{\partial}{\partial x_i^\mu}\right)\langle T^{\mu_1\nu_1}(x_1)T^{\mu_2\nu_2}(x_2)T^{\mu_3\nu_3}(x_3)\rangle = 2^3\int dx \frac{\delta^3 \tilde{\mathcal{A}}(x)}{\delta g_{\mu_1\nu_1}(x_1)\delta g_{\mu_2\nu_2}(x_2)\delta g_{\mu_3\nu_3}(x_3)}, \tag{6.4.18}$$

while the special conformal equations are

$$\begin{aligned}
&\sum_{i=1}^{3}\left[2x_i^\kappa\left(\Delta_i + x_i^\alpha \frac{\partial}{\partial x_i^\alpha}\right) - x_i^2\delta^{\kappa\alpha}\frac{\partial}{\partial x_i^\alpha}\right]\langle T^{\mu_1\nu_1}(x_1)T^{\mu_2\nu_2}(x_2)T^{\mu_3\nu_3}(x_3)\rangle \\
&+2\left[\delta^{\kappa\mu_1}x_{1\alpha} - \delta^\kappa_\alpha x_1^{\mu_1}\right]\langle T^{\alpha\nu_1}(x_1)T^{\mu_2\nu_2}(x_2)T^{\mu_3\nu_3}(x_3)\rangle + 2\left[\delta^{\kappa\nu_1}x_{1\alpha} - \delta^\kappa_\alpha x_1^{\nu_1}\right]\langle T^{\mu_1\alpha}(x_1)T^{\mu_2\nu_2}(x_2)T^{\mu_3\nu_3}(x_3)\rangle \\
&+2\left[\delta^{\kappa\mu_2}x_{2\alpha} - \delta^\kappa_\alpha x_2^{\mu_2}\right]\langle T^{\mu_1\nu_1}(x_1)T^{\alpha\nu_2}(x_2)T^{\mu_3\nu_3}(x_3)\rangle + 2\left[\delta^{\kappa\nu_2}x_{2\alpha} - \delta^\kappa_\alpha x_2^{\nu_2}\right]\langle T^{\mu_1\nu_1}(x_1)T^{\mu_2\alpha}(x_2)T^{\mu_3\nu_3}(x_3)\rangle \\
&+2\left[\delta^{\kappa\mu_3}x_{3\alpha} - \delta^\kappa_\alpha x_3^{\mu_3}\right]\langle T^{\mu_1\nu_1}(x_1)T^{\mu_2\nu_2}(x_2)T^{\alpha\nu_3}(x_3)\rangle + 2\left[\delta^{\kappa\nu_3}x_{3\alpha} - \delta^\kappa_\alpha x_3^{\nu_3}\right]\langle T^{\mu_1\nu_1}(x_1)T^{\mu_2\nu_2}(x_2)T^{\mu_3\alpha}(x_3)\rangle \\
&= 2^4\int dx\, x^\kappa \frac{\delta^3 \tilde{\mathcal{A}}(x)}{\delta g_{\mu_1\nu_1}(x_1)\delta g_{\mu_2\nu_2}(x_2)\delta g_{\mu_3\nu_3}(x_3)}.
\end{aligned} \tag{6.4.19}$$

### 6.4.3 $\langle JJO\rangle$ and $\langle TTO\rangle$

As discussed earlier in this chapter, the $\langle JJO\rangle$ and $\langle TTO\rangle$ correlators are pivotal in studying trace anomalies. In Chapter 3, we have determined the general structure of these correlators within the framework of CFT, under the assumption that no trace anomaly is present. A natural question then arises: what happens if we relax this assumption and allow for the presence of anomalies? In this section, we derive the general anomalous conformal Ward identities for the $\langle JJO\rangle$ and $\langle TTO\rangle$ correlators. In the subsequent section, we then demonstrate that the parity-odd trace anomalies, $F\tilde{F}$ and $R\tilde{R}$, do not contribute to these CWIs. This result confirms the validity of the analysis presented in Chapter 3 in all cases. We also recall that our analysis considers a scalar operator $O$ coupled to an external field $\phi$. Specifically, when $O = T_\mu^\mu$, the field $\phi$ corresponds to the dilaton.
For the $\langle JJO\rangle$ correlator, the dilatation Ward identity is expressed as

$$\sum_{i=1}^{3}\left(\Delta_i + x_i^\mu \frac{\partial}{\partial x_i^\mu}\right)\langle J^{\mu_1}(x_1)J^{\mu_2}(x_2)O(x_3)\rangle = \int dx \frac{\delta^3\tilde{\mathcal{A}}(x)}{\delta A_{\mu_1}(x_1)\delta A_{\mu_2}(x_2)\delta\phi(x_3)} \tag{6.4.20}$$

while the special CWIs are

$$\begin{aligned}\sum_{i=1}^{3}&\left[2x_i^\kappa\left(\Delta_i + x_i^\alpha\frac{\partial}{\partial x_i^\alpha}\right) - x_i^2\delta^{\kappa\alpha}\frac{\partial}{\partial x_i^\alpha}\right]\langle J^{\mu_1}(x_1)J^{\mu_2}(x_2)O(x_3)\rangle + \\ &2\left[\delta^{\kappa\mu_1}x_{1\alpha} - \delta^\kappa_\alpha x_1^{\mu_1}\right]\langle J^\alpha(x_1)J^{\mu_2}(x_2)O(x_3)\rangle + 2\left[\delta^{\kappa\mu_2}x_{2\alpha} - \delta^\kappa_\alpha x_2^{\mu_2}\right]\langle J^{\mu_1}(x_1)J^\alpha(x_2)O(x_3)\rangle \\ &= 2\int dx\, x^\kappa \frac{\delta^3\tilde{\mathcal{A}}(x)}{\delta A_{\mu_1}(x_1)\delta A_{\mu_2}(x_2)\delta\phi(x_3)}\end{aligned} \tag{6.4.21}$$

The dilatation WI of the $\langle TTO\rangle$ is

$$\sum_{i=1}^{3}\left(\Delta_i + x_i^\mu \frac{\partial}{\partial x_i^\mu}\right)\langle T^{\mu_1\nu_1}(x_1)T^{\mu_2\nu_2}(x_2)O(x_3)\rangle = 2^2\int dx \frac{\delta^3\tilde{\mathcal{A}}(x)}{\delta g_{\mu_1\nu_1}(x_1)\delta g_{\mu_2\nu_2}(x_2)\delta\phi(x_3)}, \tag{6.4.22}$$

and the special CWIs are

$$\begin{aligned}\sum_{i=1}^{3}&\left[2x_i^\kappa\left(\Delta_i + x_i^\alpha\frac{\partial}{\partial x_i^\alpha}\right) - x_i^2\delta^{\kappa\alpha}\frac{\partial}{\partial x_i^\alpha}\right]\langle T^{\mu_1\nu_1}(x_1)T^{\mu_2\nu_2}(x_2)O(x_3)\rangle \\ &+ 2\left[\delta^{\kappa\mu_1}x_{1\alpha} - \delta^\kappa_\alpha x_1^{\mu_1}\right]\langle T^{\alpha\nu_1}(x_1)T^{\mu_2\nu_2}(x_2)O(x_3)\rangle + 2\left[\delta^{\kappa\nu_1}x_{1\alpha} - \delta^\kappa_\alpha x_1^{\nu_1}\right]\langle T^{\mu_1\alpha}(x_1)T^{\mu_2\nu_2}(x_2)O(x_3)\rangle \\ &+ 2\left[\delta^{\kappa\mu_2}x_{2\alpha} - \delta^\kappa_\alpha x_2^{\mu_2}\right]\langle T^{\mu_1\nu_1}(x_1)T^{\alpha\nu_2}(x_2)O(x_3)\rangle + 2\left[\delta^{\kappa\nu_2}x_{2\alpha} - \delta^\kappa_\alpha x_2^{\nu_2}\right]\langle T^{\mu_1\nu_1}(x_1)T^{\mu_2\alpha}(x_2)O(x_3)\rangle \\ &= 2^3\int dx\, x^\kappa \frac{\delta^3\tilde{\mathcal{A}}(x)}{\delta g_{\mu_1\nu_1}(x_1)\delta g_{\mu_2\nu_2}(x_2)\delta\phi(x_3)}.\end{aligned} \tag{6.4.23}$$

The anomalous contribution is present, naturally, in all the CWIs, but in some cases it can vanishes, as we are going to show in the next section.

### 6.4.4 The momentum-space analysis: the $\langle TTT\rangle$ correlator

The structure of the anomalous CWIs in momentum-space is defined via Fourier transformation. Here, we focus on the general structure of the anomalous CWIs for the $\langle TTT\rangle$, as other correlators follow a similar pattern. Furthermore, as we will show in the next section, when we restrict ourselves to the parity-odd sector, the $\langle TTT\rangle$ is the only correlator exhibiting anomalous CWIs.
We begin by introducing the Fourier transform of the functional derivatives of the anomaly in the form

$$(2\pi)^4\,\delta^4(p_1+\dots+p_{n+1})\,\tilde{\mathcal{A}}_n^{\mu_2\nu_2\dots\mu_{n+1}\nu_{n+1}}(p_1,\dots,p_{n+1}) \equiv \int d^4x_1\dots d^4x_{n+1}\; e^{ip_1\cdot x_1+\dots+ip_{n+1}\cdot x_{n+1}}\, \frac{\delta^n\tilde{\mathcal{A}}(x_1)}{\delta g_{\mu_2\nu_2}(x_2)\dots\delta g_{\mu_{n+1}\nu_{n+1}}(x_{n+1})}\bigg|_{flat}. \tag{6.4.24}$$

We can then write the anomalous dilatation WI in momentum-space as

$$\left[4-p_1\cdot\frac{\partial}{\partial p_1}-p_2\cdot\frac{\partial}{\partial p_2}\right]\langle T^{\mu_1\nu_1}(p_1)T^{\mu_2\nu_2}(p_2)T^{\mu_3\nu_3}(-p_1-p_2)\rangle = 8\,\tilde{\mathcal{A}}_3^{\mu_1\nu_1\mu_2\nu_2\mu_3\nu_3}(p_1,p_2,-p_1-p_2), \tag{6.4.25}$$

where we have set $\Delta_i = d = 4$. As we are going to see, the $R\tilde{R}$ term does not contribute to the anomaly on the right-hand side of the dilatation equation. However, this equation also applies to the parity-even sector of $\langle TTT\rangle$, where the anomaly contribution remains.
The special CWIs, on the other hand, are given by

$$\begin{aligned}
&\sum_{j=1}^{2}\left[-2p_{j\alpha}\frac{\partial^2}{\partial p_{j\alpha}\partial p_{j\kappa}}+p_j^\kappa\frac{\partial^2}{\partial p_{j\alpha}\partial p_j^\alpha}\right]\langle T^{\mu_1\nu_1}(p_1)T^{\mu_2\nu_2}(p_2)T^{\mu_3\nu_3}(-p_1-p_2)\rangle\\
&\quad+2\left(\eta^{\kappa\mu_1}\frac{\partial}{\partial p_1^{\alpha_1}}-\delta^\kappa{}_{\alpha_1}\frac{\partial}{\partial p_{1\mu_1}}\right)\langle T^{\mu_1\nu_1}(p_1)T^{\mu_2\nu_2}(p_2)T^{\mu_3\nu_3}(-p_1-p_2)\rangle\\
&\quad+2\left(\eta^{\kappa\nu_1}\frac{\partial}{\partial p_1^{\beta_1}}-\delta^\kappa{}_{\beta_1}\frac{\partial}{\partial p_{1\nu_1}}\right)\langle T^{\mu_1\nu_1}(p_1)T^{\mu_2\nu_2}(p_2)T^{\mu_3\nu_3}(-p_1-p_2)\rangle\\
&\quad+2\left(\eta^{\kappa\mu_2}\frac{\partial}{\partial p_2^{\alpha_2}}-\delta^\kappa{}_{\alpha_2}\frac{\partial}{\partial p_{2\mu_2}}\right)\langle T^{\mu_1\nu_1}(p_1)T^{\mu_2\nu_2}(p_2)T^{\mu_3\nu_3}(-p_1-p_2)\rangle\\
&\quad+2\left(\eta^{\kappa\nu_2}\frac{\partial}{\partial p_2^{\beta_2}}-\delta^\kappa{}_{\nu_2}\frac{\partial}{\partial p_{2\nu_2}}\right)\langle T^{\mu_1\nu_1}(p_1)T^{\mu_2\nu_2}(p_2)T^{\mu_3\nu_3}(-p_1-p_2)\rangle\\
&\quad= -16\left[\frac{\partial}{\partial p_{3\kappa}}\tilde{\mathcal{A}}_3^{\mu_1\nu_1\mu_2\nu_2\mu_3\nu_3}(p_1,p_2,p_3)\right]_{p_3=-p_1-p_2}.
\end{aligned} \tag{6.4.26}$$

Notice that the differentiation of $\mathcal{A}_3$ with respect to one of the momenta on the r.h.s. of (6.4.26) is what makes the anomalous term nonzero in the case of a topological anomaly. This differentiation arises due to the presence of the factor $x^\kappa$ in the integration of (6.4.19).

### 6.4.5 The parity-odd case: the non-anomalous character of the CWIs

In this section, we delve into the anomaly terms within the CWIs introduced earlier, with a focus on the parity-odd sector of the correlators. Specifically, we analyze the parity-odd anomaly contributions, $R\tilde{R}$ and $F\tilde{F}$, which are topological. Our findings demonstrate that the anomalous contributions vanish in nearly all the CWIs. In other words, the majority of these equations correspond to ordinary (i.e., non-anomalous) CWIs.
First of all, since we are considering topological anomalies, the term on the right-hand sides of (6.4.16), (6.4.18), (6.4.20) and (6.4.22) always vanishes. Topological anomalies are scale-invariant: they do not break dilatations. Indeed, we have

$$\frac{\delta}{\delta g_{\mu_1\nu_1}(x_1)}\int dx\tilde{\mathcal{A}}(x) = 0. \tag{6.4.27}$$

This result follows from the fact that the integrand is a total derivative. When integrated, it yields a constant term that vanishes under functional differentiation. However, special conformal transformations can potentially be broken by a topological anomaly. Specifically, the anomalous terms on the right-hand sides of (6.4.19), (6.4.17), (6.4.21), (6.4.23) may be non-vanishing due to the presence of $x^\kappa$ in the integral. We will now illustrate this point case by case.
We start by considering the anomouls CWIs of the $\langle JJT\rangle_{odd}$ correlator. In this case, we need to examine the anomaly contribution $\tilde{\mathcal{A}}(x) = \sqrt{g}\, f\, \varepsilon^{\mu\nu\rho\sigma} F_{\mu\nu}F_{\rho\sigma}$. It is clear that such anomaly does not depend on the metric since

$$\varepsilon^{\mu\nu\rho\sigma} = \frac{\epsilon^{\mu\nu\rho\sigma}}{\sqrt{g}} \tag{6.4.28}$$

with $\epsilon^{0123} = 1$. Therefore, one finds

$$\frac{\delta^3 \tilde{\mathcal{A}}(x)}{\delta A_{\mu_1}(x_1)\delta A_{\mu_2}(x_2)\delta g_{\mu_3\nu_3}(x_3)} = 0. \tag{6.4.29}$$

This shows that (6.4.16) and (6.4.17) are homogeneous and henceforth they are ordinary CWIs. A similar result holds for the $\langle JJO\rangle_{odd}$ correlator in eq. (6.4.20) and (6.4.21). Here, the relevant case is given by $O = T^\mu_\mu$ and the functional derivative with respect to the dilaton field can be expressed as

$$\frac{\delta}{\delta\phi} = 2g_{\mu\nu}\frac{\delta}{\delta g_{\mu\nu}}. \tag{6.4.30}$$

Once again, $\tilde{\mathcal{A}}(x) = \sqrt{g}\, f\, \varepsilon^{\mu\nu\rho\sigma} F_{\mu\nu}F_{\rho\sigma}$ vanishes when functionally differentiated with respect to the metric. Therefore, the dilatations and special conformal transformations are not broken by anomalies in the case of the $\langle JJT\rangle_{odd}$ and $\langle JJO\rangle_{odd}$.
Next, we consider the conformal equations of the $\langle TTO\rangle_{\text{odd}}$ correlator. The anomalous term in this case is $R\tilde{R}$, which is Weyl-invariant. Thus, applying the operator $\frac{\delta}{\delta\phi} = 2g_{\mu\nu}\frac{\delta}{\delta g_{\mu\nu}}$ to it yields a vanishing result. Consequently, the anomalous term vanishes in (6.4.22) and (6.4.23), and the correlator satisfies ordinary CWIs.
Finally, we examine the $\langle TTT\rangle_{\text{odd}}$ correlator. As previously noted, dilatations are not broken by a topological anomaly. However, in this case, the anomalous term does not vanish in the special conformal Ward identities.
In conclusion, almost all of the CWIs considered in this section are ordinary and non-anomalous. The only anomalous CWI is given by (6.4.19) in the presence of an anomaly of the form $R\tilde{R}$.

## 6.5 Review of the $\langle JJO\rangle_{odd}$ and $\langle TTO\rangle_{odd}$ correlators in CFT

In this section, we briefly summarize the key results regarding the $\langle JJO\rangle_{odd}$ and $\langle TTO\rangle_{odd}$ correlators. As detailed in Section 6.2, these correlators play a central role in the study of trace anomalies. In Chapter 3, we have determined the general structure of these correlators within the framework of CFT, focusing specifically on their connection to parity-odd anomalies.
As emphasized earlier, the $\langle JJO\rangle_{odd}$ and $\langle TTO\rangle_{odd}$ correlators are deeply tied to the study of the $F\tilde{F}$ and $R\tilde{R}$ anomalies, respectively. Despite this, as demonstrated in Section 6.4.5, the conformal Ward identities associated with these correlators remain non-anomalous. This observation reinforces the robustness of the CWIs under such circumstances.
For our analysis, we focus on the specific case where $\Delta_3 = 4$, corresponding to the operator choice $O = T^\mu_\mu$. Within this setting, the solutions to the CWIs for the $\langle JJO\rangle_{odd}$ and $\langle TTO\rangle_{odd}$ correlators are precisely the structures induced by the aforementioned parity-odd anomalies. Explicitly, these structures can be expressed as follows:

$$
\begin{aligned}
&(2\pi)^4\delta^4(p_1+p_2+p_3)\Big\langle J^{\mu_1}(p_1)J^{\mu_2}(p_2)O_{(\Delta_3=4)}(p_3)\Big\rangle_{odd} = \\
&\qquad\int dx_1\,dx_2\,dx_3\,e^{-i(p_1x_1+p_2x_2+p_3x_3)}\,\frac{\delta^2\Big[f_2\,\varepsilon^{\mu\nu\rho\sigma}F_{\mu\nu}(x_3)F_{\rho\sigma}(x_3)\Big]}{\delta A_{\mu_1}(x_1)\,\delta A_{\mu_2}(x_2)}, \\
&(2\pi)^4\delta^4(p_1+p_2+p_3)\Big\langle T^{\mu_1\nu_1}(p_1)T^{\mu_2\nu_2}(p_2)O_{(\Delta_3=4)}(p_3)\Big\rangle_{odd} = \\
&\qquad\int dx_1\,dx_2\,dx_3\,e^{-i(p_1x_1+p_2x_2+p_3x_3)}\,\frac{\delta^2\Big[f_1\,\varepsilon^{\mu\nu\rho\sigma}R^{\alpha\beta}{}_{\mu\nu}(x_3)R_{\alpha\beta\rho\sigma}(x_3)\Big]}{\delta g_{\mu_1\nu_1}(x_1)\,\delta g_{\mu_2\nu_2}(x_2)}.
\end{aligned}
\tag{6.5.1}
$$

It is worth highlighting that these expressions were derived in Chapter 3 without invoking any specific assumptions regarding the trace anomaly. Instead, they emerge naturally as solutions to the CWIs. This connection underscores the compatibility of the conformal symmetry framework with the inclusion of $F\tilde{F}$ and $R\tilde{R}$ terms in the trace anomaly.
These results not only provide a deeper understanding of the structure of parity-odd contributions to correlation functions in CFT but also pave the way for further exploration of their implications for anomaly-induced phenomena in quantum field theory.

## 6.6 The $\langle JJT\rangle_{odd}$ correlator in CFT without the anomaly

In this section we explore the conformal constraints of the $\langle JJT\rangle_{odd}$ correlator. Assuming the correlator is non-anomalous, we impose the condition that tracing over the energy-momentum tensor yields a vanishing result. This assumption will be relaxed in a following section. We start our analysis by recalling the trace and conservation Ward identities in the presence of a field $A_\mu$ and $V_\mu$

$$
\begin{aligned}
&\nabla\cdot\langle J_V\rangle = 0, && \nabla\cdot\langle J_A\rangle = a\,\varepsilon^{\mu\nu\rho\sigma}F_{\mu\nu}F_{\rho\sigma}, \\
&\nabla^\mu\langle T_{\mu\nu}\rangle - F^A_{\mu\nu}\langle J^\mu_A\rangle + A_\nu\nabla_\mu\langle J^\mu_A\rangle - F^V_{\mu\nu}\langle J^\mu_V\rangle + V_\nu\nabla_\mu\langle J^\mu_V\rangle = 0, && g_{\mu\nu}\langle T^{\mu\nu}\rangle = 0.
\end{aligned}
\tag{6.6.1}
$$

By applying multiple functional derivatives to these identities with respect to the field sources and performing a Fourier transform, we obtain

$$
p_{i\mu_i}\langle J^{\mu_1}(p_1)J^{\mu_2}(p_2)T^{\mu_3\nu_3}(p_3)\rangle_{odd} = 0, \qquad i=1,2,3 \tag{6.6.2}
$$

and

$$
\delta_{\mu_3\nu_3}\langle J^{\mu_1}(p_1)J^{\mu_2}(p_2)T^{\mu_3\nu_3}(p_3)\rangle_{odd} = 0. \tag{6.6.3}
$$

Such equations are satisfied independently of the fact that $J$ are conserved vector or axial-vector currents. Notice that the right-hand side of the equations contains no 2-point functions, as they all vanish in the parity-odd sector.
Due to these Ward identities, all the longitudinal and trace components of the correlator vanish. On the other hand, the transverse-traceless part can be formally expressed in terms of the following independent tensor structures and form factors

$$
\begin{aligned}
&\langle J^{\mu_1}(p_1)J^{\mu_2}(p_2)T^{\mu_3\nu_3}(p_3)\rangle_{odd} = \langle j^{\mu_1}(p_1)j^{\mu_2}(p_2)t^{\mu_3\nu_3}(p_3)\rangle_{odd} = \\
&\qquad \pi^{\mu_1}_{\alpha_1}(p_1)\pi^{\mu_2}_{\alpha_2}(p_2)\Pi^{\mu_3\nu_3}_{\alpha_3\beta_3}(p_3)\Big[A_1\varepsilon^{p_1p_2\alpha_1\alpha_2}p_1^{\alpha_3}p_1^{\beta_3} + A_2\varepsilon^{p_1\alpha_1\alpha_2\alpha_3}p_1^{\beta_3} + \\
&\qquad\qquad A_3\varepsilon^{p_2\alpha_1\alpha_2\alpha_3}p_1^{\beta_3} + A_4\varepsilon^{p_1p_2\alpha_2\alpha_3}\delta^{\alpha_1\beta_3}\Big].
\end{aligned}
\tag{6.6.4}
$$

Note that the following tensor structures are not included in our decomposition

$$p_2^{\alpha_1}p_1^{\beta_3}\varepsilon^{p_1p_2\alpha_2\alpha_3}, \qquad p_3^{\alpha_2}p_1^{\beta_3}\varepsilon^{p_1p_2\alpha_1\alpha_3}, \qquad \delta^{\beta_3\alpha_2}\varepsilon^{p_1p_2\alpha_1\alpha_3}. \tag{6.6.5}$$

These tensor structures can be related to those written in Eq. (6.6.4), by using the Schouten identities

$$\varepsilon^{[p_1p_2\alpha_1\alpha_2}\delta_\alpha^{\alpha_3]} = 0 \tag{6.6.6}$$

which can be contracted with $p_{1\alpha}$, $p_{2\alpha}$ and $\delta_\alpha^{\beta_3}$ and with the transverse-traceless projectors, yielding

$$\begin{aligned}
&\pi^{\mu_1}_{\alpha_1}\pi^{\mu_2}_{\alpha_2}\Pi^{\mu_3\nu_3}_{\alpha_3\beta_3}\left(\varepsilon^{p_1p_2\alpha_1\alpha_3}p_3^{\alpha_2}\right) = \pi^{\mu_1}_{\alpha_1}\pi^{\mu_2}_{\alpha_2}\Pi^{\mu_3\nu_3}_{\alpha_3\beta_3}\left(-\frac{p_1^2+p_2^2-p_3^2}{2}\varepsilon^{p_1\alpha_1\alpha_2\alpha_3} - p_1^2\varepsilon^{p_2\alpha_1\alpha_2\alpha_3} - \varepsilon^{p_1p_2\alpha_1\alpha_2}p_1^{\alpha_3}\right)\\
&\pi^{\mu_1}_{\alpha_1}\pi^{\mu_2}_{\alpha_2}\Pi^{\mu_3\nu_3}_{\alpha_3\beta_3}\left(\varepsilon^{p_1p_2\alpha_2\alpha_3}p_2^{\alpha_1}\right) = \pi^{\mu_1}_{\alpha_1}\pi^{\mu_2}_{\alpha_2}\Pi^{\mu_3\nu_3}_{\alpha_3\beta_3}\left(\frac{p_1^2+p_2^2-p_3^2}{2}\varepsilon^{p_2\alpha_1\alpha_2\alpha_3} + p_2^2\varepsilon^{p_1\alpha_1\alpha_2\alpha_3} + \varepsilon^{p_1p_2\alpha_1\alpha_2}p_1^{\alpha_3}\right)\\
&\pi^{\mu_1}_{\alpha_1}\pi^{\mu_2}_{\alpha_2}\Pi^{\mu_3\nu_3}_{\alpha_3\beta_3}\left(\varepsilon^{p_1p_2\alpha_1\alpha_3}\delta^{\alpha_2\beta_3}\right) = \pi^{\mu_1}_{\alpha_1}\pi^{\mu_2}_{\alpha_2}\Pi^{\mu_3\nu_3}_{\alpha_3\beta_3}\left(\varepsilon^{p_1\alpha_1\alpha_2\alpha_3}p_1^{\beta_3} + \varepsilon^{p_2\alpha_1\alpha_2\alpha_3}p_1^{\beta_3} + \varepsilon^{p_1p_2\alpha_2\alpha_3}\delta^{\alpha_1\beta_3}\right)
\end{aligned} \tag{6.6.7}$$

### 6.6.1 Dilatation and special conformal Ward Identities

In this section we begin to examine the conformal constraints on the $\langle JJT\rangle$. The invariance of the correlator under dilatation is reflected in the equation

$$\left(\sum_{i=1}^{3}\Delta_i - 2d - \sum_{i=1}^{2}p_i^\mu\frac{\partial}{\partial p_i^\mu}\right)\langle J^{\mu_1}(p_1)J^{\mu_2}(p_2)T^{\mu_3\nu_3}(p_3)\rangle_{odd} = 0. \tag{6.6.8}$$

By using the chain rule

$$\frac{\partial}{\partial p_i^\mu} = \sum_{j=1}^{3}\frac{\partial p_j}{\partial p_i^\mu}\frac{\partial}{\partial p_j}, \tag{6.6.9}$$

in term of the invariants $p_i = |\sqrt{p_i^2}|$ and by considering the decomposition (6.6.4), we can rewrite the dilatation equation as a constraint on the form factors

$$\begin{aligned}
&\sum_{i=1}^{3}p_i\frac{\partial A_1}{\partial p_i}(p_1,p_2,p_3) - \left(\sum_{i=1}^{3}\Delta_i - 2d - 4\right)A_1(p_1,p_2,p_3) = 0\\
&\sum_{i=1}^{3}p_i\frac{\partial A_2}{\partial p_i}(p_1,p_2,p_3) - \left(\sum_{i=1}^{3}\Delta_i - 2d - 2\right)A_2(p_1,p_2,p_3) = 0\\
&\sum_{i=1}^{3}p_i\frac{\partial A_3}{\partial p_i}(p_1,p_2,p_3) - \left(\sum_{i=1}^{3}\Delta_i - 2d - 2\right)A_3(p_1,p_2,p_3) = 0\\
&\sum_{i=1}^{3}p_i\frac{\partial A_4}{\partial p_i}(p_1,p_2,p_3) - \left(\sum_{i=1}^{3}\Delta_i - 2d - 2\right)A_4(p_1,p_2,p_3) = 0.
\end{aligned} \tag{6.6.10}$$

The invariance of the correlator with respect to the special conformal transformations is instead encoded in the special conformal Ward identities

$$\begin{aligned}
0=&\sum_{j=1}^{2}\left[-2\frac{\partial}{\partial p_{j\kappa}}-2p_j^{\alpha}\frac{\partial^2}{\partial p_j^{\alpha}\partial p_{j\kappa}}+p_j^{\kappa}\frac{\partial^2}{\partial p_j^{\alpha}\partial p_{j\alpha}}\right]\langle J^{\mu_1}(p_1)J^{\mu_2}(p_2)T^{\mu_3\nu_3}(p_3)\rangle_{odd}\\
&+2\left(\delta^{\mu_1\kappa}\frac{\partial}{\partial p_1^{\alpha_1}}-\delta^{\kappa}_{\alpha_1}\frac{\partial}{\partial p_{1\mu_1}}\right)\langle J^{\alpha_1}(p_1)J^{\mu_2}(p_2)T^{\mu_3\nu_3}(p_3)\rangle_{odd}\\
&+2\left(\delta^{\mu_2\kappa}\frac{\partial}{\partial p_2^{\alpha_2}}-\delta^{\kappa}_{\alpha_2}\frac{\partial}{\partial p_{2\mu_2}}\right)\langle J^{\mu_1}(p_1)J^{\alpha_2}(p_2)T^{\mu_3\nu_3}(p_3)\rangle_{odd}\equiv\mathcal{K}^{\kappa}\langle J^{\mu_1}(p_1)J^{\mu_2}(p_2)T^{\mu_3\nu_3}(p_3)\rangle_{odd}. \qquad (6.6.11)
\end{aligned}$$

We can perform a transverse projection on all the indices in order to identify a set of independent partial differential equations

$$\begin{aligned}
&0=\pi^{\alpha_1}_{\mu_1}(p_1)\pi^{\alpha_2}_{\mu_2}(p_2)\Pi^{\alpha_3\beta_3}_{\mu_3\nu_3}(p_3)\mathcal{K}^{k}\langle J^{\mu_1}(p_1)J^{\mu_2}(p_2)T^{\mu_3\nu_3}(p_3)\rangle_{odd}=\\
&\pi^{\alpha_1}_{\mu_1}(p_1)\pi^{\alpha_2}_{\mu_2}(p_2)\Pi^{\alpha_3\beta_3}_{\mu_3\nu_3}(p_3)\Big[\Big(C_{11}\varepsilon^{p_1\alpha_1\alpha_2\alpha_3}p_1^{\beta_3}+C_{12}\varepsilon^{p_2\alpha_1\alpha_2\alpha_3}p_1^{\beta_3}+C_{13}\varepsilon^{p_1p_2\alpha_1\alpha_2}p_1^{\alpha_3}p_1^{\beta_3}+C_{14}\varepsilon^{p_1p_2\alpha_2\alpha_3}\delta^{\alpha_1\beta_3}\Big)p_1^{\kappa}+\\
&\quad\Big(C_{21}\varepsilon^{p_1\alpha_1\alpha_2\alpha_3}p_1^{\beta_3}+C_{22}\varepsilon^{p_2\alpha_1\alpha_2\alpha_3}p_1^{\beta_3}+C_{23}\varepsilon^{p_1p_2\alpha_1\alpha_2}p_1^{\alpha_3}p_1^{\beta_3}+C_{24}\varepsilon^{p_1p_2\alpha_2\alpha_3}\delta^{\alpha_1\beta_3}\Big)p_2^{\kappa}+\\
&\quad C_{31}\varepsilon^{\kappa\mu_1\mu_2\mu_3}p_1^{\nu_3}+C_{32}\varepsilon^{p_1\kappa\mu_2\mu_3}\delta^{\mu_1\nu_3}+C_{33}\varepsilon^{p_2\kappa\mu_1\mu_3}\delta^{\mu_2\nu_3}+C_{34}\varepsilon^{p_1p_2\kappa\mu_3}\delta^{\mu_1\mu_2}p_1^{\nu_3}+\\
&\quad C_{41}\delta^{\mu_1\kappa}\varepsilon^{\mu_2\mu_3p_1p_2}p_1^{\nu_3}+C_{51}\delta^{\mu_2\kappa}\varepsilon^{\mu_1\mu_3p_1p_2}p_1^{\nu_3}+C_{61}\delta^{\mu_3\kappa}\varepsilon^{p_1\mu_1\mu_2\nu_3}+C_{62}\delta^{\mu_3\kappa}\varepsilon^{p_2\mu_1\mu_2\nu_3}\Big],
\end{aligned} \qquad (6.6.12)$$

where $C_{ij}$ are scalar functions of the form factors $A$ and their derivatives. The decomposition (6.6.12) is obtained by writing all the possibile tensor structures and then identifying all the independent ones. The full procedure is shown in Appendix C.3. Due to the independence of the tensor structures of eq. (6.6.12), the special conformal constraints can then be written as

$$C_{ij}=0 \qquad i=1,\dots 6 \qquad j=1,\dots 4. \qquad (6.6.13)$$

### 6.6.2 Solution of the CWIs

In order to solve the CWIs, it is easier to first analyze the equations $C_{ij}=0$ which involve only the $A_1$ form factor. Their explicit form is

$$K_{31}A_1=0, \qquad K_{32}A_1=0, \qquad \left(\frac{\partial}{\partial p_3}+4-\Delta_3\right)A_1=0. \qquad (6.6.14)$$

The first two equations are the primary ones. Their solution can be expressed in terms of the following 3K integral

$$A_1=b_1J_{4\{0,0,0\}}, \qquad (6.6.15)$$

which does not diverge in $d=4$ (see Appendix A). Therefore, we can solve the last equation in (6.6.14) directly without the need of a regularization. We set $\Delta_3=d=4$ and take the limit $p_1\to 0$ in the equation, as outlined in Appendix A. This procedure yields the condition $b_1=0$ which leads to

$$A_1=0. \qquad (6.6.16)$$

We now examine the other conformal equations $C_{ij} = 0$. After setting $A_1 = 0$, we can write the remaining primary special conformal Ward indentities as

$$\begin{aligned}
&0 = K_{31}A_2, && 0 = K_{32}A_2 - \left(\frac{2}{p_3}\frac{\partial}{\partial p_3} - \frac{2\Delta_3}{p_3^2}\right)(A_4 + A_2 - A_3),\\
&0 = K_{31}A_3 + \left(\frac{2}{p_3}\frac{\partial}{\partial p_3} - \frac{2\Delta_3}{p_3^2}\right)(A_4 + A_2 - A_3), && 0 = K_{32}A_3,\\
&0 = K_{31}A_4, && 0 = K_{32}A_4.
\end{aligned} \tag{6.6.17}$$

These equations can be reduced to a set of homogenous equations by repeatedly applying the operator $K_{ij}$ on them

$$\begin{aligned}
&0 = K_{31}A_2, && 0 = K_{21}K_{32}A_2, \quad ,\\
&0 = K_{21}K_{31}A_3, && 0 = K_{32}A_3,\\
&0 = K_{31}A_4, && 0 = K_{32}A_4,
\end{aligned} \tag{6.6.18}$$

which can be easly derived by noting that $K_{21}A_4 = 0$ and $K_{21}(A_2 - A_3) = 0$.
The solutions of the homogenous equations can then be written in terms of the following 3K integrals

$$A_4 = c_1 J_{2,\{0,0,0\}}, \qquad A_2 = c_2 J_{3\{0,1,0\}} + c_3 J_{2,\{0,0,0\}}, \qquad A_3 = c_4 J_{3\{1,0,0\}} + c_5 J_{2,\{0,0,0\}}. \tag{6.6.19}$$

In $d = 4 + 2u\epsilon$, these integrals diverge like $1/\epsilon$ and therefore we should perform a regularization as we did for the other correlators. However here for simplicity we will avoid such procedure since such divergences don't particularly spoil our equations and we can arrive to the same conclusions with or without the regularization. Therefore, we will set

$$d = 4, \qquad \Delta_1 = 3, \qquad \Delta_2 = 3, \qquad \Delta_3 = 4. \tag{6.6.20}$$

Inserting our solutions back into the nonhomogeneous equations we find

$$c_1 = -4c_2 - c_3 + c_5, \qquad c_4 = -c_2. \tag{6.6.21}$$

The secondary equations are given by $C_{ij} = 0$ with $i \geq 3$. Their explicit expression is

$$\begin{aligned}
0 &= p_2\frac{\partial A_2}{\partial p_2} + \frac{p_2^2 + p_3^2}{p_3}\frac{\partial A_2}{\partial p_3}\\
0 &= p_2\frac{\partial A_3}{\partial p_2} - p_2\frac{\partial A_4}{\partial p_2} + \frac{p_2^2 + p_3^2}{p_3}\frac{\partial A_3}{\partial p_3}\\
0 &= -\frac{8}{p_3^2}A_2 + \left(\frac{4}{p_1^2} + \frac{8}{p_3^2}\right)A_3 - \left(\frac{4}{p_1^2} + \frac{8}{p_3^2}\right)A_4 - \frac{2p_1}{p_3^2}\frac{\partial A_2}{\partial p_1}\\
&\quad + \left(-\frac{2}{p_1} + \frac{2p_1}{p_3^2}\right)\frac{\partial A_3}{\partial p_1} + \left(\frac{2}{p_1} - \frac{2p_1}{p_3^2}\right)\frac{\partial A_4}{\partial p_1} - \frac{2p_2}{p_3^2}\frac{\partial A_2}{\partial p_2} + \frac{2p_2}{p_3^2}\frac{\partial A_3}{\partial p_2} - \frac{2p_2}{p_3^2}\frac{\partial A_4}{\partial p_2},\\
0 &= \left(\frac{4}{p_2^2} + \frac{8}{p_3^2}\right)A_2 - \frac{8}{p_3^2}A_3 + \frac{8}{p_3^2}A_4 - \frac{2}{p_2}\frac{\partial A_2}{\partial p_2} - \frac{2}{p_3}\frac{\partial A_2}{\partial p_3} + \frac{2}{p_3}\frac{\partial A_3}{\partial p_3} - \frac{2}{p_3}\frac{\partial A_4}{\partial p_3}\\
0 &= A_2 + A_3 - A_4 - p_2\frac{\partial A_2}{\partial p_2} + p_2\frac{\partial A_3}{\partial p_2} + p_2\frac{\partial A_4}{\partial p_2} - p_3\frac{\partial A_2}{\partial p_3} + p_3\frac{\partial A_4}{\partial p_3}\\
0 &= A_3 - A_4 + p_2\frac{\partial A_4}{\partial p_2} + p_3\frac{\partial A_4}{\partial p_3}\\
0 &= A_2 + p_2\frac{\partial A_4}{\partial p_2}.
\end{aligned} \tag{6.6.22}$$

Working in the limit $p_3 \to 0$, the first and second equation lead to the following constraints

$$c_3 = -2c_2, \qquad c_5 = 4c_2. \tag{6.6.23}$$

Inserting such conditions into the last equation and using the properties (A.8) of the 3K integrals, we arrive to $c_2 = 0$. Therefore, we have

$$\langle J^{\mu_1}(p_1)J^{\mu_2}(p_2)T^{\mu_3\nu_3}(p_3)\rangle_{odd} = 0. \tag{6.6.24}$$

We conclude that in the absence of a parity-odd trace anomaly, the $\langle JJT\rangle_{odd}$ vanishes.

## 6.7 The $\langle JJT\rangle_{odd}$ correlator in CFT with trace anomalies

In the previous section we have assumed that there is no parity-odd trace anomaly. In this section we relax such assumption by considering a term of the following form

$$g_{\mu\nu}\langle T^{\mu\nu}\rangle_{odd} = f_2\,\varepsilon^{\mu\nu\rho\sigma}F_{\mu\nu}F_{\rho\sigma}, \tag{6.7.1}$$

where $f_2$ is an arbitrary constant. In this case, the trace part of the $\langle JJT\rangle_{odd}$ no longer vanishes. By applying multiple functional derivatives with respect to the field sources to the equation above and performing a Fourier transform, we find

$$\delta_{\mu\nu}\langle J^{\mu_1}(p_1)J^{\mu_2}(p_2)T^{\mu\nu}(p_3)\rangle_{odd} = 8\,f_2\,\varepsilon^{p_1p_2\mu_1\mu_2}. \tag{6.7.2}$$

Therefore, there will be a nonvanishing term of the form $\left\langle j^{\mu_1}j^{\mu_2}t_{loc}^{\mu_3\nu_3}\right\rangle_{odd}$ contributing to the correlator. Note that we are still assuming the conservation of the currents and the energy-momentum tensor as in [22]

$$p_{i\mu_i}\langle J^{\mu_1}(p_1)J^{\mu_2}(p_2)T^{\mu_3\nu_3}(p_3)\rangle_{odd} = 0, \qquad i = 1,2,3. \tag{6.7.3}$$

We can now decompose the correlator in terms of its longitudinal-trace and transverse-traceless parts

$$\langle J^{\mu_1}(p_1)J^{\mu_2}(p_2)T^{\mu_3\nu_3}(p_3)\rangle_{odd} = \left\langle j^{\mu_1}(p_1)j^{\mu_2}(p_2)t_{loc}^{\mu_3\nu_3}(p_3)\right\rangle_{odd} + \langle j^{\mu_1}(p_1)j^{\mu_2}(p_2)t^{\mu_3\nu_3}(p_3)\rangle_{odd}, \tag{6.7.4}$$

where we recall the definition of the local term

$$t_{loc}^{\mu\nu}(p) \equiv \left(I^{\mu\nu}_{\alpha\beta} + \frac{1}{d-1}\pi^{\mu\nu}\delta_{\alpha\beta}\right)T^{\alpha\beta}(p) \tag{6.7.5}$$

Inside $I^{\mu\nu}_{\alpha\beta}$ there is a momentum that, when contracted with $T^{\alpha\beta}$, leads to a vanishing result due to the conservation of the energy-momentum tensor. Therefore, only the second term on the right-hand side of the equation survives, yielding

$$\left\langle j^{\mu_1}(p_1)j^{\mu_2}(p_2)t_{loc}^{\mu_3\nu_3}(p_3)\right\rangle_{odd} = \frac{8}{3}\,f_2\,\pi^{\mu_3\nu_3}(p_3)\varepsilon^{p_1p_2\mu_1\mu_2}. \tag{6.7.6}$$

In order to fix the transverse-traceless part of the correlator, we need to analyze the conformal constraints in the presence of an anomaly. As we have seen Section 6.4.5, the conformal equations of the $\langle JJT\rangle_{odd}$ are not modified by the presence of a trace anomaly $F\tilde{F}$. Moreover, the trace term (6.7.6) does not affect the conformal constraints written before for the transverse-traceless part since

$$\pi^{\rho_1}_{\mu_1}(p_1)\,\pi^{\rho_2}_{\mu_2}(p_2)\,\Pi^{\rho_3\sigma_3}_{\mu_3\nu_3}(p_3)\,K^{\kappa}\left\langle j^{\mu_1}j^{\mu_2}t_{loc}^{\mu_3\nu_3}\right\rangle_{odd} = 0. \tag{6.7.7}$$

We can prove such equation by using the conservation of the energy-momentum tensor together with the properties of the projectors. Therefore, our analysis of the transverse-traceless part in the previous section still applies. The transverse-traceless part remains zero since it is unaffected by the presence of the anomaly. Consequently, the correlator reduces entirely to its trace component

$$\langle J^{\mu_1}(p_1)J^{\mu_2}(p_2)T^{\mu_3\nu_3}(p_3)\rangle_{odd} = \Big\langle J^{\mu_1}(p_1)J^{\mu_2}(p_2)t^{\mu_3\nu_3}_{loc}(p_3)\Big\rangle_{odd} = \frac{8}{3}f_2\,\pi^{\mu_3\nu_3}(p_3)\varepsilon^{p_1p_2\mu_1\mu_2}. \tag{6.7.8}$$

The correlator $\langle JJT\rangle_{odd}$ exhibits notable differences compared to $\langle J_VJ_VJ_A\rangle$ and $\langle TTJ_A\rangle$. While the latter cases include both longitudinal and transverse-traceless sectors, the $\langle JJT\rangle_{odd}$ comprises only an anomaly term. This distinction arises because in the $\langle JJT\rangle_{odd}$ the anomaly part of the correlator is decoupled from the transverse-traceless part in the conformal equations, leaving the trace sector as the sole surviving component after imposing the conformal constraints.

## 6.8 The $\langle TTT\rangle_{odd}$ correlator

The $\langle TTT\rangle_{odd}$ correlator has been extensively studied in coordinate-space CFT (see, for example, [137]). All studies agree that the correlator vanishes. Coordinate-space analyses typically avoid contact terms, where the external points of the correlator coalesce, as these are the source of the conformal anomaly.[1] Moreover, these investigations consistently assume the absence of parity-odd trace terms

$$\delta_{\mu_i\nu_i}\langle T^{\mu_1\nu_1}T^{\mu_2\nu_2}T^{\mu_3\nu_3}\rangle_{odd} = 0. \tag{6.8.1}$$

Perturbative analyses in momentum-space confirm the vanishing of the correlator under various regularization schemes for Weyl fermions. Additionally, Bonora's claim in [20] is that the gravity contributions to the parity-odd trace anomaly originate entirely from the second term of (6.2.1), so (6.8.1) remains valid.

Nonetheless, one may wonder what happens in a general CFT if we allow a parity-odd trace anomaly of this kind, where the trace is performed after the quantum average of $T$. In general, such anomalies are admitted, as predicted by Capper and Duff [61,62]. Under these assumptions, the $\langle TTT\rangle_{odd}$ would necessarily be nonvanishing, containing at least a trace anomalous term. If we admit the Pontryagin density $f_1\,\varepsilon^{\mu\nu\rho\sigma}R_{\alpha\beta\mu\nu}R^{\alpha\beta}_{\ \ \rho\sigma}$ as an anomalous trace term for the correlator, we can write

$$\delta_{\mu_3\nu_3}\langle T^{\mu_1\nu_1}T^{\mu_2\nu_2}T^{\mu_3\nu_3}\rangle_{odd} = -16f_1\left\{\left[\varepsilon^{\nu_1\nu_2p_1p_2}\left((p_1\cdot p_2)\delta^{\mu_1\mu_2}-p_1^{\mu_2}p_2^{\mu_1}\right)+(\mu_1\leftrightarrow\nu_1)\right]+(\mu_2\leftrightarrow\nu_2)\right\}. \tag{6.8.2}$$

Analogous expressions can be derived for the trace over $T^{\mu_1\nu_1}$ or $T^{\mu_2\nu_2}$ by exchanging the momenta and indices in the equation above. Moreover, from this equation, one can see that if we trace over two energy-momentum tensors simultaneously, the result vanishes. Therefore, terms constructed with more than one $t_{loc}$ do not need to be included in the decomposition of the $\langle TTT\rangle_{odd}$ correlator. Proceeding in a manner similar to previous correlators, we can decompose the $\langle TTT\rangle_{odd}$ as

$$\langle T^{\mu_1\nu_1}T^{\mu_2\nu_2}T^{\mu_3\nu_3}\rangle_{odd} = \langle t^{\mu_1\nu_1}t^{\mu_2\nu_2}t^{\mu_3\nu_3}\rangle_{odd} + \langle t^{\mu_1\nu_1}_{loc}t^{\mu_2\nu_2}t^{\mu_3\nu_3}\rangle_{odd} + \langle t^{\mu_1\nu_1}t^{\mu_2\nu_2}_{loc}t^{\mu_3\nu_3}\rangle_{odd} + \langle t^{\mu_1\nu_1}t^{\mu_2\nu_2}t^{\mu_3\nu_3}_{loc}\rangle_{odd} \tag{6.8.3}$$

with

$$\langle t^{\mu_1\nu_1}t^{\mu_2\nu_2}t^{\mu_3\nu_3}_{loc}\rangle_{odd} = -\frac{16f_1}{3}\pi^{\mu_3\nu_3}(p_3)\left\{\left[\varepsilon^{\nu_1\nu_2p_1p_2}\left((p_1\cdot p_2)\delta^{\mu_1\mu_2}-p_1^{\mu_2}p_2^{\mu_1}\right)+(\mu_1\leftrightarrow\nu_1)\right]+(\mu_2\leftrightarrow\nu_2)\right\} \tag{6.8.4}$$

[1] The inclusion of such contact terms in the parity-even case was discussed long ago in [66].

Analogous expressions can also be derived for $\langle t_{loc}^{\mu_1\nu_1} t^{\mu_2\nu_2} t^{\mu_3\nu_3}\rangle_{odd}$ and $\langle t^{\mu_1\nu_1} t_{loc}^{\mu_2\nu_2} t^{\mu_3\nu_3}\rangle_{odd}$ by appropriately exchanging momenta and indices. In order to fix the transverse-traceless part of the correlator, we need to examine the conformal constraints. The situation here is notably different from the $\langle JJT\rangle_{odd}$ case. Since $R\tilde{R}$ is a topological anomaly, dilatations remain unbroken. However, in this case, special conformal transformations are broken. Thus, if we assume such anomaly in the $\langle TTT\rangle_{odd}$ correlator, we also expect a nonvanishing transverse-traceless part. We hope to address this point in a future work.

## 6.9 Summary of the results and conclusions

In this section, we summarize the result of our analysis. We have investigated parity-odd terms in the trace anomaly

$$\mathcal{A}_{odd} = f_1\, \varepsilon^{\mu\nu\rho\sigma} R_{\alpha\beta\mu\nu} R^{\alpha\beta}_{\ \ \rho\sigma} + f_2\, \varepsilon^{\mu\nu\rho\sigma} F_{\mu\nu} F_{\rho\sigma}, \tag{6.9.1}$$

If we consider the abelian gauge contribution to the trace anomaly, i.e. the Chern-Pontryagin density $F\tilde{F}$, the first term of the anomaly (6.2.1) can be evaluated by computing the $\langle JJT\rangle_{odd}$ correlator. We have shown that conformal invariance requires the $\langle JJT\rangle_{odd}$ correlator to be purely given by a trace term. It is possible to include such trace part in the correlator without breaking the CWIs. In momentum-space this is summarized by the following expression

$$\langle J^{\mu_1}(p_1) J^{\mu_2}(p_2) T^{\mu_3\nu_3}(p_3)\rangle_{odd} = \frac{8}{3} f_2\, \pi^{\mu_3\nu_3}(p_3) \varepsilon^{p_1 p_2 \mu_1 \mu_2} \tag{6.9.2}$$

On the other hand, in order to analyze the second term of eq. (6.2.1) in CFTs, we need to consider the $\langle JJO\rangle_{odd}$ correlator. Since $O = T_\mu^\mu$, the conformal dimension of the scalar operator is fixed to $\Delta_3 = 4$. Remarkably, as we have seen in Chapter 3, such value of $\Delta_3$ is the only physical case where the correlator can be different from zero. Furthermore, conformal invariance fixes the $\langle JJO\rangle_{odd}$ to assume the required form induced by the anomaly (6.9.1)

$$\langle J^{\mu_1} J^{\mu_2} O_{(\Delta_3=4)}\rangle_{odd} = \mathcal{F}\left[\frac{\delta^2 \mathcal{A}_{odd}(x_3)}{\delta A_{\mu_1}(x_1)\, \delta A_{\mu_2}(x_2)}\right] \tag{6.9.3}$$

where we denoted with $\mathcal{F}[\cdot]$ the Fourier transform with respect to all the coordinates $(x_1, x_2, x_3)$. In the formula, we are also ignoring the delta $\delta^4(\sum p_i)$ appearing after Fourier transforming since it is not included in the definition of the correlator.

We have also analyzed the gravitational contribution to the parity-odd trace anomaly, i.e. the Pontryagin density $R\tilde{R}$. In this case, the first term in eq. (6.2.1), can be evaluated by computing the $\langle TTT\rangle_{odd}$ correlator. Contrary to popular belief, such correlator doesn't need to vanish if we allow the presence of a trace term. However, although the regularization and prescriptions adopted by Bonora lead to an anomalous $\langle JJT\rangle_{odd}$, in the case of the $\langle TTT\rangle_{odd}$, the correlator vanishes. Indeed, according to Bonora, the gravitational contribution to the trace anomaly comes entirely from the second term of eq. (6.2.1). Therefore, in this case we don't need to impose the presence of anomalies for the $\langle TTT\rangle_{odd}$. The study of the complete uncontracted anomalous $\langle TTT\rangle_{odd}$ is less relevant for the purpose of a comparison with the perturbative prediction and the ongoing debate over whether the parity-odd components of this correlator vanish.

In order to analyze the second term in the anomaly (6.2.1), we need to considere the $\langle TTO\rangle_{odd}$ correlator. Conformal invariance requires the correlator to either vanish or not depending on the value of $\Delta_3$. A nonvanishing case occurs when $\Delta_3 = 4$, which is exactly what we need for $O = T_\mu^\mu$. As we demonstrated in Chapter 3, the structure assumed by such a correlator in a CFT precisely corresponds to that induced by the trace anomaly (6.9.1)

$$\langle T^{\mu_1\nu_1}T^{\mu_2\nu_2}O_{(\Delta_3=4)}\rangle_{odd} = \mathcal{F}\left[\frac{\delta^2\mathcal{A}_{odd}(x_3)}{\delta g_{\mu_1\nu_1}(x_1)\,\delta g_{\mu_2\nu_2}(x_2)}\right] \tag{6.9.4}$$

The implications for the structure of the effective action in the external gauge and gravity fields can be summarized by rather simple nonlocal functionals. Our analysis shows that in terms of the parity-odd contribution to the effective action, in the $\langle JJT\rangle_{odd}$ case, the correlator is represented by the term

$$\mathcal{S}_{JJT} = f_2\int d^4x'\sqrt{g(x')}\int d^4x\sqrt{g(x)}R(x)\Box^{-1}_{x,x'}F\tilde{F}(x') \tag{6.9.5}$$

The entire $\langle JJT\rangle_{odd}$ correlator can be obtained from (6.9.5). Other parity-odd terms my be taken into account by inserting additional energy-momentum tensors, constrained by external conservation and CWIs.
In the case of the $\langle TTT\rangle_{odd}$, the Weyl-variant part of the correlator is generated instead by the functional

$$\mathcal{S}_{TTT} = f_1\int d^4x'\sqrt{g(x')}\int d^4x\sqrt{g(x)}R(x)\Box^{-1}_{x,x'}R\tilde{R}(x'). \tag{6.9.6}$$

It would be interesting to see what the conformal constraints predict in more general cases. However, the pattern for 3-point functions is the usual one, as in the case of the parity-even trace anomalies and ordinary chiral anomalies

$$\mathcal{S}_{J_VJ_VJ_A} = a_1\int d^4x'\sqrt{g(x')}\int d^4x\sqrt{g(x)}\partial\cdot A\,\Box^{-1}_{x,x'}F\tilde{F}(x'), \tag{6.9.7}$$

In all the cases, the trace and longitudinal sectors are characterized by the presence of the nonlocal $1/\Box$ interaction. The form is that of a bilinear mixing between a spin 1 (for the chiral current, $\partial\cdot A$) or spin 2 external field (R) and an intermediate scalar or pseudoscalar interpolating state, times the anomaly contributions ($F\tilde{F}$ or $R\tilde{R}$). These interactions play a role in the cosmological context, in the analysis of the conformal backreaction on gravity, and provide a cosistent basis for nonlocal cosmological models for the dark energy (see for instance [138, 139]). Furthermore, if conformal symmetry plays a significant role in the very early universe, it is reasonable to consider scenarios where the sources of CP violation are directly linked to gravity.

# Chapter 7

# The chiral anomaly at finite density and temperature

In this chapter, we investigate the general structure of chiral anomaly vertices in the presence of a temperature $T_{\text{emp}}$ and a chemical potential $\mu$ within perturbation theory. This study finds applications in anomalous transport, particularly in scenarios involving chirally unbalanced matter with propagating external currents that are classically conserved. Examples include topological materials and the chiral magnetic effect in the plasma state of the early universe.
We classify the minimal number of form factors in the parametrization of $\langle J_V J_V J_A \rangle$ through a complete analysis of Schouten identities and symmetries, considering a heat-bath at finite temperature and density. We then proceed with a perturbative computation of the correlator, focusing on the case of finite density and zero temperature. We demonstrate that the longitudinal (anomaly) sector in the axial-vector channel remains protected against corrections arising from the insertion of a chemical potential in the fermion loop. The cancellation of the $\mu$-dependence extends to any chiral current within the Standard Model, including examples such as $B$ (baryon), $L$ (lepton) and $B-L$.
When the photons are on-shell, we prove that the transverse sector of the $\langle J_V J_V J_A \rangle$ is also $\mu$-independent and vanishes. Consequently, the correlator reduces to the anomalous longitudinal component, as in the case of vanishing chemical potentials. The associated effective action is shown to be consistently described by the exchange of a massless anomaly pole. This pole is interpreted as an interpolating axion-like quasiparticle generated by the anomaly. In the off-shell case, the $\mu$-dependent contribution to $\langle J_V J_V J_A \rangle$ resides within its transverse sector, for which we provide an explicit expression.

## 7.1 Introduction

The interest in exploring anomalies in chiral matter, specifically the Adler-Bell-Jackiw (ABJ) anomaly, has witnessed substantial growth over the past two decades. This surge extends across both condensed matter theory [9, 12–16, 57, 140–149] and high-energy physics, particularly in the theoretical and experimental study of matter under high densities and in transport [47, 110, 150–156]. Experiments involving heavy ion collisions have tested and confirmed the anomalous behavior of matter in the presence of finite density chiral asymmetric backgrounds and strong fields [157].
The chiral interaction implies that, for massless fermions coupled to electric and magnetic fields $(\vec{E}, \vec{B})$ the chiral fermion number $N_5$, if nonzero, is not conserved, but is modified by the anomaly

$$\frac{dN_5}{dt} = \frac{g^2}{2\pi^2}\mathbf{E}\cdot\mathbf{B}. \tag{7.1.1}$$

A similar effect was pointed out to be possible in a crystal subjected to the same external fields, more than a decade later after the discovery of the ABJ anomaly, by Nielsen and Ninomiya [158].
In condensed matter physics, the more recent identification of Dirac and Weyl semimetals has paved the way for the exploration of analogous anomaly-related phenomena [159–162]. These materials exhibit distinctive features, notably the presence of "Dirac-points" where the conduction and valence bands come into contact. In Weyl metals, these Dirac points are concealed within the Fermi surface [163].
In topological materials, the relativistic description is considered an analog rather than a true relativistic scenario. This is because the relativistic dynamics in these systems do not arise from a genuine relativistic dispersion relation for the propagation of fundamental states (whether real or virtual). Instead, the behavior is reflected in the Fermi velocity, which behaves in a relativistic-like manner. This results in the manifestation of the Weyl and Dirac equations in these systems. As a consequence, topological materials offer a unique table-top platform for investigating exotic phenomena typically observed in high-energy particle physics.
Eq. (7.1.1) clearly shows that if chiral asymmetries are present in the background in the form of propagating external currents, with a net nonzero chiral charge, the response of the matter system may be characterized by a dynamical evolution that, as it has been pointed out, can even lead to instabilities. The general feature of the phenomenon is such that it can be incorporated into the magnetohydrodynamic equations of the early universe plasma as well [8], generating a chiral magnetic cascade. Indeed, a notable example of a similar phenomenon is the chiral magnetic effect (CME) in the quark gluon plasma [164,165]. The CME, for example, is characterized by the generation of an electric charge separation along an external magnetic field, due to a chirality unbalance that is driven by the anomaly. In this case one introduces the chiral current

$$\vec{J}_A = \mu_5 \vec{B}, \tag{7.1.2}$$

with $\mu_5 = \mu_L - \mu_R$ being the chiral chemical potential of the left ($L$) and right ($R$) Weyl fermions present in the initial state, and $\vec{B}$ a magnetic field. $\vec{J}_A$ is the external source triggering the anomalous response at quantum level. The result of the chiral anomaly interaction is the generation of an electric field $\vec{E}$ parallel to the magnetic field.
This macroscopic phenomenon gives rise to a collective motion within the Dirac sea. The topological nature of the unbalance ($\mu_5 \neq 0$) imparts a unique property to the CME current, ensuring a nondissipative behavior even with the inclusion of radiative corrections. Therefore, in matter under extreme conditions, as found, for example, at neutron star densities [166] or in the primordial plasma [7, 8], chiral anomalies induce a nonequilibrium phase that deserves a close attention. Given the different values of the external chiral asymmetries and of the related chemical potentials, the evaluation of the finite density corrections is expected to be of remarkable phenomenological relevance.
These interactions may leave an imprint in the stochastic background(s) of gravitational waves via the gravitational chiral anomaly [5, 167] as well as contributing to the generation of primordial magnetic fields [1, 168]. This may clearly occur at any phase of the early universe, and impact the production of gravitational waves in strongly first order phase transitions. Similar spectral asymmetries for spin-1 particles in the anomaly loop can be generated by Chern-Simons currents (see the discussion and references in [36]).
Both chiral and conformal anomalies have been associated with the presence of interpolating massless states. In the chiral case, they are clearly recognized in the on-shell effective action, since the anomaly diagram is entirely represented by a pole. However, they appear in combination with transverse sectors as soon as in vector/vector/axial-vector ($\langle J_V J_V J_A \rangle$) interactions one moves off-shell in the vector lines.

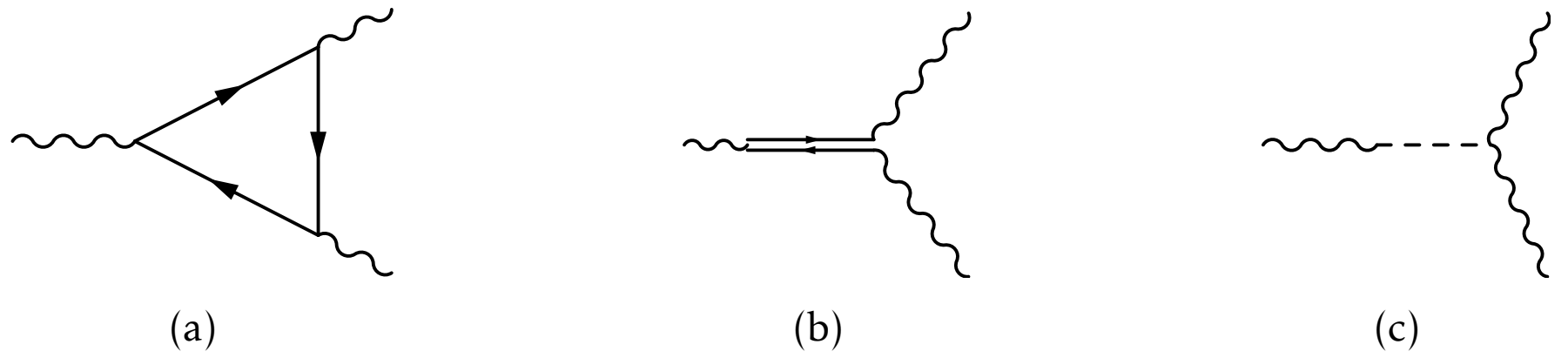

Figure 7.1: The fermion loop (a); the collinear region in the loop integration (b); the effective pseudoscalar exchange as an effective axion (c).

The possibility that future experiments in Weyl semimetals may shed light on this phenomenon is the motivation of our analysis. In this chapter we show how the entire anomaly interaction is dominated also at finite density by the exchange of a massless pole, that can be interpreted by an interpolating, on-shell fermion/antifermion pair in the axial vector channel. This intermediate state can be classified as an axion-like quasiparticle excitation of the medium. Specifically, we demonstrate that the pole remains unaffected by any correction and reaches saturation in its interaction when external electromagnetic fields are on-shell. Systems with these characteristics exhibit a topological response, and conducting a table-top experiment provides a viable means to investigate this interpolating state.

### 7.1.1 Topological protection in the Ward identity

The Ward identity associated with the chiral interaction has been shown to be protected by the quantum corrections in a specific way. For example, at zero (massless) fermion density and temperature, the anomaly is not modified by radiative corrections [101]. This is the content of the Adler-Bardeen theorem. On the other end, finite fermion mass corrections are shown to modify the anomaly which in the flat limit has the following form

$$\partial_\mu \langle J_A^\mu \rangle = a_1 \varepsilon^{\mu\nu\lambda\sigma} F_{\mu\nu} F_{\lambda\sigma} + 2m \langle J_A \rangle \tag{7.1.3}$$

where

$$J_A = -ig\bar{\psi}\gamma^5\psi \tag{7.1.4}$$

The mass term appears together with a pseudoscalar interaction inserted in the anomaly loop. The second contribution in (7.1.3), in axion physics, is responsible for the finite mass corrections of the axion-anomaly interaction. The corrections are obtained by coupling the axion field, $\varphi$, treated as an asymptotic state, to (7.1.3) in the standard form

$$\frac{\varphi}{f_a} \partial \cdot \langle J_A \rangle \tag{7.1.5}$$

with the inclusion of a scale $f_a$ in order to preserve the quartic mass dimensions of the interaction. The $\varphi \to \gamma\gamma$ decay of an axion is then obtained by differentiating twice (7.1.3) with respect to the background electromagnetic gauge field.
Concerning the generalization of the chiral Ward identity at finite density, previous studies have shown that the anomaly is preserved in this case [46, 169, 170].
The goal of the present analysis is to identify the structure of the three-point functions $\langle J_V J_V J_A \rangle$ at finite density within perturbation theory. We will also consider the $\langle J_A J_A J_A \rangle$ interaction and correlators involving chiral fermions, such as $\langle J_L J_L J_L \rangle$ and $\langle J_R J_R J_R \rangle$, which are defined by the inclusion of left- and right-chirality projectors. Our analysis applies to any chiral current associated with the global symmetries of the Standard Model. Examples include the baryon ($B$) and lepton ($L$) currents, which correspond to the conservation of baryon and lepton numbers in the Standard Model. While both are conserved at the classical level, they become anomalous at the quantum level. The $B-L$ chiral current cancels the mixed $(B-L)Y^2$ anomaly, where $Y$ is the hypercharge, but the cancellation of the $(B-L)^3$ anomaly requires the inclusion of a singlet right-handed neutrino. Our results hold for all these cases as well, since, in the absence of a spontaneously broken phase, chiral diagrams remain unaffected by density corrections on-shell.

### 7.1.2 Content of the chapter

The $\langle J_V J_V J_A \rangle$ vertex provides the simplest (free-field) realization of the correlators mentioned above, which can be studied at finite density, either perturbatively or, at least in principle, using modified conformal Ward identities. Notice that, in CFTs at zero density, the structure of such correlators is determined by the inclusion of the anomaly contribution in the solutions of the conformal Ward identities (CWIs), as shown in [34] and in Chapter 4, without resorting to perturbation theory. The results follow from the conformal symmetry of the interaction at $\mu = 0$, which remains preserved for on-shell photons even at finite $\mu$, as we will demonstrate. However, in the off-shell case, conformal symmetry is broken by the new scale ($\mu$).
We will proceed with an explicit identification of all sectors of a chiral anomaly interaction, classifying the tensor structures and form factors involved. Our representation of the full $\langle J_V J_V J_A \rangle$ vertex differs from that in [169] in both the number of form factors and the tensor structures used, and we provide a detailed derivation. Our goal is to illustrate explicitly how the reduction of the general parametrization of the vertex proceeds. The original parametrization, valid at finite temperature and density, involves a large number of tensor structures—sixty—due to the presence of three independent 4-vectors (the momenta $p_1$, $p_2$, and the heat-bath vector $\eta$). This number is significantly reduced by imposing Bose symmetry, vector current conservation, and the use of Schouten identities. We work in a fully covariant manner, which is crucial for extracting an effective action that captures the essential features of the anomaly phenomenon.
We then proceed with a perturbative analysis of the correlator for both on-shell and off-shell photons at finite density and zero temperature. We consider the simplest case of an Abelian theory, such as QED coupled to external axial-vector fields. Some interesting implications emerge when extending this analysis to the full spectrum of Standard Model particles, rather than restricting it to QED, as we do in this work. The specialization of these results to fermion families and gauge currents in the Standard Model is an aspect that we hope to explore in the future, particularly in relation to axion physics in dense astrophysical environments.
By requiring that the anomaly interaction occurs on a null surface, we will show that the only surviving contribution in the chiral vertex is the anomalous pole. From a perturbative perspective, this implies that when the two vector currents in $\langle J_V J_V J_A \rangle$ are on-shell, the entire interaction— in the $m = 0$ limit—is simply described by the exchange of a pole in the longitudinal component of the correlator. This protection has significant ramifications, particularly in influencing the hydrodynamic and transport properties of systems containing chiral fermions.

## 7.2 The action and its solutions

We start by considering the massless Dirac Lagrangian

$$\mathcal{L}_0 = \bar{\psi} i \partial\!\!\!/ \psi \tag{7.2.1}$$

to which we add the contribution of a vector chemical potential

$$\mathcal{L}_V^{Dens} = -\mu\bar{\psi}\gamma^0\psi. \tag{7.2.2}$$

or a chiral chemical potential

$$\mathcal{L}_A^{Dens} = -\mu\psi^\dagger\gamma^5\psi \tag{7.2.3}$$

The Dirac fermion $\psi$ can be split into its left-handed and right-handed chiral components. The chemical potential $\mu$ accounts for the presence of a nonvanishing background density of particles ($\mu > 0$), of total charge $Q$ by the inclusion of a term $-\mu Q$ in the action.
The action has a $U(1)_L \times U(1)_R$ global symmetry. Both the chiral charges of the left and right components,

$$Q_{L/R} = \int d^3x\, n^0_{L/R} \qquad\qquad Q = -g(Q_L + Q_R) \tag{7.2.4}$$

are classically conserved, where

$$n_R = \mu_R\bar{\psi}_R\gamma^0\psi_R, \qquad\qquad n_L = \mu_L\bar{\psi}_L\gamma^0\psi_L \tag{7.2.5}$$

are the density of the $L$ and $R$ fermionic modes.
The chiral anomaly vertex render $J_A$ not conserved at quantum level with the anomalous Ward identity

$$\partial_\mu\langle J^\mu_A\rangle = a_1 F\tilde{F} = a_1 \mathbf{E}\cdot\mathbf{B}. \tag{7.2.6}$$

The chiral charge

$$Q_5 \equiv Q_L - Q_R, \tag{7.2.7}$$

acquire a nonzero time dependence

$$\dot{Q}_5 = \dot{N}_L - \dot{N}_R = \int d^3x\, \mathbf{E}\cdot\mathbf{B}. \tag{7.2.8}$$

In this chapter, we use the Dirac matrices in the chiral base, given by

$$\gamma^\mu = \begin{pmatrix} 0 & \sigma^\mu \\ \bar{\sigma}^\mu & 0 \end{pmatrix} \tag{7.2.9}$$

where $\sigma^\mu = (\mathbb{I}, \boldsymbol{\sigma})$, $\bar{\sigma}^\mu = (\mathbb{I}, -\boldsymbol{\sigma})$ and $\boldsymbol{\sigma} = (\sigma_1, \sigma_2, \sigma_3)$ are the Pauli matrices. The chiral matrix $\gamma^5$ in this basis is defined as

$$\gamma^5 = \begin{pmatrix} -\mathbb{I} & 0 \\ 0 & \mathbb{I} \end{pmatrix}. \tag{7.2.10}$$

Left- and right-handed spinors are defined by the projectors

$$P_R = \frac{1+\gamma^5}{2} \qquad P_L = \frac{1-\gamma^5}{2}. \tag{7.2.11}$$

We define Dirac spinors as sum of two 4 components chiral spinors $\psi_L$ and $\psi_R$

$$\psi = \psi_L + \psi_R \qquad \psi_L = \begin{pmatrix} \chi_L \\ 0 \end{pmatrix}\ \psi_R = \begin{pmatrix} 0 \\ \chi_R \end{pmatrix}. \tag{7.2.12}$$

### 7.2.1 Vector chemical potential

If the chemical potential is vector-like, such that $\mu_L = \mu_R = \mu$, its contribution to the Lagrangian takes the form

$$\mathcal{L}_V^{Dens} = -\mu\bar{\psi}\gamma^0\psi. \tag{7.2.13}$$

The corresponding equations of motions are

$$i\gamma^\mu\partial_\mu\psi - \mu\gamma^0\psi = 0, \tag{7.2.14}$$

which we analyze using the ansatz

$$\psi \sim e^{-ipx}u(p) = e^{-ipx}\begin{pmatrix}\chi_L\\ \chi_R\end{pmatrix}. \tag{7.2.15}$$

Therefore, we can write

$$\begin{pmatrix}0 & \sigma\cdot p\\ \bar{\sigma}\cdot p & 0\end{pmatrix}\begin{pmatrix}\chi_L\\ \chi_R\end{pmatrix} - \mu\begin{pmatrix}0 & \mathbb{1}\\ \mathbb{1} & 0\end{pmatrix}\begin{pmatrix}\chi_L\\ \chi_R\end{pmatrix} = 0, \tag{7.2.16}$$

obtaining

$$\begin{aligned}\sigma\cdot\mathbf{p}\,\chi_R &= (E-\mu)\,\chi_R\\ \sigma\cdot\mathbf{p}\,\chi_L &= -(E-\mu)\,\chi_L.\end{aligned} \tag{7.2.17}$$

Iterating these equations, we get the constraint

$$|\mathbf{p}|^2\,\chi_{L/R} = (E-\mu)^2\chi_{L/R}, \tag{7.2.18}$$

giving

$$E^{(\pm)} \equiv E^{(\pm)}_{L/R} = \mu \pm |\mathbf{p}|. \tag{7.2.19}$$

Both equations in (7.2.17) are of the form

$$\sigma\cdot\mathbf{n}\chi = \lambda\chi \tag{7.2.20}$$

with

$$\mathbf{n} = \frac{1}{|\mathbf{p}|}(p_1,p_2,p_3) = (\sin\theta\cos\varphi, \sin\theta\sin\varphi, \cos\theta) \tag{7.2.21}$$

with $\lambda = \pm 1$. The two Weyl spinors that solve the equations are classified by their helicities

$$\chi^{(\pm)} = \begin{pmatrix}\left(\frac{1\pm n_3}{2}\right)^{1/2}\\ \pm\left(\frac{1\pm n_3}{2}\right)^{1/2}\frac{1\mp n_3}{n_1-in_2}\end{pmatrix}. \tag{7.2.22}$$

The equation for $\chi_R$ becomes

$$\sigma\cdot\mathbf{n}\chi_R = \frac{E-\mu}{|\mathbf{p}|}\chi_R \tag{7.2.23}$$

If $E = E^{(+)}$, we can write

$$\sigma\cdot\mathbf{n}\,\chi_R^{(+)} = +\chi_R^{(+)}, \tag{7.2.24}$$

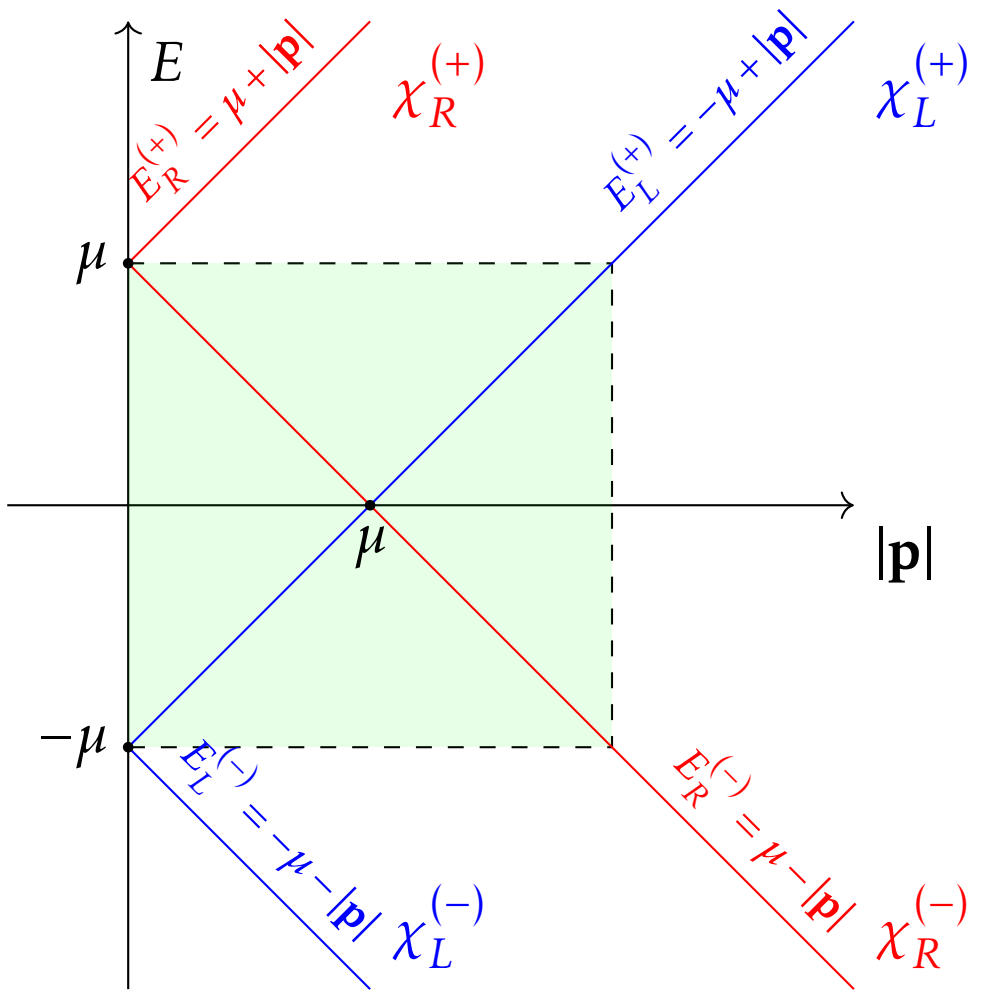

Figure 7.2: The four folds of the dispersion relation in the $E$, $|\mathbf{p}|$ plane in the case of a chiral chemical potential. In the case of a vector chemical potential we have a two-folds solution.

with

$$\chi_R^{(+)}=\begin{pmatrix}(\frac{1+n_3}{2})^{1/2}\\ (\frac{1+n_3}{2})^{1/2}\frac{1-n_3}{n_1-in_2}\end{pmatrix}, \tag{7.2.25}$$

of helicity +1. If $E=E^{(-)}$, we have

$$\sigma\cdot\mathbf{n}\chi_R^{(-)}=-\chi_R^{(-)} \tag{7.2.26}$$

with

$$\chi_R^{(-)}=\begin{pmatrix}(\frac{1-n_3}{2})^{1/2}\\ -(\frac{1-n_3}{2})^{1/2}\frac{1+n_3}{n_1-in_2}\end{pmatrix}. \tag{7.2.27}$$

The same procedure applies to $\chi_L$. This leads to the solution $\chi_L^{(-)}$ with helicity $-1$ for $E=E^{(+)}$. Similarly, for $E=E^{(-)}$, we obtain the solution $\chi_L=\chi_L^{(+)}$ with helicity +1.

### 7.2.2 Chiral chemical potential

We now consider the case of a chiral chemical potential

$$\mathcal{L}_A^{Dens}=-\mu\psi^\dagger\gamma^5\psi \tag{7.2.28}$$

where $\mu_L=-\mu$ and $\mu_R=\mu$. The equation of motion for the fermion is

$$i\gamma^\mu\partial_\mu\psi-\mu\bar{\psi}\gamma^0\gamma^5\psi=0, \tag{7.2.29}$$

or equivalently

$$\sigma\cdot\mathbf{p}\chi_R=(E-\mu)\chi_R, \tag{7.2.30}$$

and

$$\sigma\cdot\mathbf{p}\chi_L=-(E+\mu)\chi_L. \tag{7.2.31}$$

Iterating them, we get

$$\mathbf{p}^2\chi_R = (E-\mu)^2\chi_R, \tag{7.2.32}$$

giving two dispersion relations $E_R^{(\pm)} = \mu \pm |\mathbf{p}|$ for the $R$ component. For the $L$ component we get two different dispersion relations $E_L^{(\pm)} = -\mu \pm |\mathbf{p}|$.

In this case, we encounter equations similar to those in the vector case. For $E_R = E_R^{(+)}$, we have $\chi_R^{(+)}$ with helicity +1, while for $E_R = E_R^{(-)}$, the chiral solution $\chi_R^{(-)}$ has helicity −1. Similar results hold for the left components, with $\chi_L^{(-)}$ corresponding to energy $E_L^{(+)}$ and $\chi_L^{(+)}$ corresponding to energy $E_L^{(-)}$. The fourfold solutions are shown in Fig. 7.2.

### 7.2.3 The massive fermion case

The generalization to the massive case can be worked out in a similar way. We choose a vector chemical potential. The equations of motion

$$i\gamma^\mu\partial_\mu\psi - m\psi - \mu\gamma^0\psi = 0, \tag{7.2.33}$$

can be rewritten in the Hamiltonian form

$$H\psi = E\psi. \tag{7.2.34}$$

We have the Hamiltonian equations

$$\begin{aligned} E\chi_L &= -\sigma\cdot\mathbf{p}\chi_L + \mu\chi_L + m\chi_R \\ E\chi_R &= \sigma\cdot\mathbf{p}\chi_R + \mu\chi_R + m\chi_L, \end{aligned} \tag{7.2.35}$$

that we rewrite as before in the form

$$\begin{aligned} \sigma\cdot\mathbf{p}\chi_L &= -E'\chi_L + m\chi_R \\ \sigma\cdot\mathbf{p}\chi_R &= E'\chi_R - m\chi_L, \end{aligned} \tag{7.2.36}$$

with $E' = E - \mu$. The eigenvalues can be easily found by iterating the equations, yielding the eigenvalues

$$E_1 = \mu - E(m), \quad E_2 = \mu + E(m), \tag{7.2.37}$$

where

$$E(m) = \sqrt{\mathbf{p}^2 + m^2}. \tag{7.2.38}$$

The equations for the $L$ and $R$ modes are obviously coupled. One can check that the solutions can be obtained in each case by selecting for $\chi_R$ the two options of spin up $\chi_R^{\uparrow} = (1,0)$ and down $\chi_R^{\downarrow} = (0,1)$ and then solving the coupled equations. For the eigenvalue $E_1 = \mu - E(m)$ we derive the two degenerate eigenfunctions

$$\begin{aligned} \psi_{11} &= \left[-\frac{-p_1 + ip_2}{m}, -\frac{-p_3 + E(m)}{m}, 0, 1\right] \\ \psi_{12} &= \left[-\frac{-p_3 + E(m)}{m}, -\frac{-p_1 + ip_2}{m}, 1, 0\right], \end{aligned} \tag{7.2.39}$$

while, for eigenvalue $E_2 = \mu + E(m)$, they are given by

$$\psi_{21} = \left[ -\frac{-p_1 + ip_2}{m}, -\frac{p_3 - E(m)}{m}, 0, 1 \right] \tag{7.2.40}$$
$$\psi_{22} = \left[ -\frac{-p_3 - E(m)}{m}, -\frac{-p_1 - ip_2}{m}, 1, 0 \right].$$

Our analysis in the following sections will be limited to the massless case. In the massive case, the corrections to the anomaly vertex arising from nonzero $m$ generate an axial-vector Ward identity that includes, in addition to the anomaly, an extra $\mu$ and $m$-dependent contribution.

## 7.3 Thermal propagator of a fermion

In this section we proceed with an analysis of the thermal propagator of a fermion in the Furry picture (see also [171]). Computations at finite temperature have been traditionally performed both in the real and in the imaginary time formalisms [172–174]. While the imaginary time formalism is very efficient in the computation of vacuum diagrams, such as in the derivation of the equations of state of QED [175, 176] and QCD [177] to rather large orders, and in the resummation program of hard thermal loops [178, 179], the real time formalism has the advantage of providing direct access to time-dependent quantities directly [180].
We start by considering a Dirac fermion in the vacuum. Using static energy solutions we may represent the second quantized fermion field in the Furry picture [181]

$$\Psi(\mathbf{x},t) = \sum_{\lambda,\kappa} b_{\lambda\kappa}\psi^{(+)}_{\lambda\kappa}(\mathbf{x},t) + d^\dagger_{\lambda\kappa}\psi^{(-)}_{\lambda\kappa}(\mathbf{x},t), \tag{7.3.1}$$

where $\lambda$ is a polarization index, $\kappa$ denotes the energy and momentum (or other) quantum numbers (discrete and/or continuous) needed in order to completely characterize the solutions, and $(\pm)$ denotes positive and negative energy solutions of the corresponding Dirac equation

$$(i\,\partial\!\!\!/ - m)\psi^{(\pm)}_{\lambda\kappa}(\mathbf{x},t) = 0, \tag{7.3.2}$$

The creation and annihilation operators satisfy the canonical anti-commutation relations

$$\{d_{\lambda'\kappa'}, d^\dagger_{\lambda\kappa}\} = \delta_{\lambda'\lambda}\delta_{\kappa'\kappa} = \{b_{\lambda'\kappa'}, b^\dagger_{\lambda\kappa}\}, \tag{7.3.3}$$

while other anti-commutators are zero. The completeness relation

$$\sum_{\lambda,\kappa} \psi^{(+)\dagger}_{\lambda\kappa,a}(\mathbf{x}',t)\psi^{(+)}_{\lambda\kappa,b}(\mathbf{x},t) + \psi^{(-)\dagger}_{\lambda\kappa,a}(\mathbf{x}',t)\psi^{(-)}_{\lambda\kappa,b}(\mathbf{x},t) = \delta_{ab}\delta^3(\mathbf{x}'-\mathbf{x}), \tag{7.3.4}$$

where $\psi^{(\pm)}_{\lambda\kappa,a}$ denotes the $a$-component of the Dirac spinor $\psi^{(\pm)}_{\lambda\kappa}$, leads to the canonical anti-commutation relations for the fields

$$\{\Psi_a(\mathbf{x}',t), \Psi^\dagger_b(\mathbf{x},t)\} = \delta_{ab}\delta^3(\mathbf{x}'-\mathbf{x}). \tag{7.3.5}$$

In the vacuum, the fermion propagator $iS_0(x';x|m)$ is then defined by

$$iS_0(x';x|m) = \langle 0|\mathbf{T}\left(\Psi(\mathbf{x}',t')\overline{\Psi}(\mathbf{x},t)\right)|0\rangle = \theta(t'-t)\sum_{\lambda\kappa}\psi^{(+)}_{\lambda\kappa}(\mathbf{x}',t')\overline{\psi}^{(+)}_{\lambda\kappa}(\mathbf{x},t) - \theta(t-t')\sum_{\lambda\kappa}\psi^{(-)}_{\lambda\kappa}(\mathbf{x}',t')\overline{\psi}^{(-)}_{\lambda\kappa}(\mathbf{x},t). \tag{7.3.6}$$

Since $\psi^{(\pm)}_{\lambda\kappa}(\mathbf{x},t)$ satisfies the Dirac equation, only the time derivative acting on the step functions gives a non-zero contribution, so one finds that

$$(i\,\partial\!\!\!/ - m)S_0(x';x|m) = \mathbb{1}\cdot\delta^4(x'-x). \tag{7.3.7}$$

The real-time propagator at finite temperature and density, denoted by $\langle iS_F(x';x|m)\rangle_{\beta,\mu}$, can now be obtained by the following reasoning. First, we introduce the Fermi-Dirac thermodynamical distribution function

$$f_F^+(E) = \frac{1}{e^{\beta(E-\mu)}+1}, \tag{7.3.8}$$

where $\beta$ is the inverse temperature. A particle can propagate forward in time in a state which is unoccupied by thermal particles, whereas a hole in the occupied states can propagate backwards in time. We can therefore write

$$\begin{aligned}\langle iS_F(x';x|m)\rangle_{\beta,\mu} &= \\ \sum_{\lambda,\kappa}\Big[\theta(t'-t)\Big([1-f_F^+(E_\kappa)]\psi^{(+)}_{\lambda\kappa}(\mathbf{x}',t')\overline{\psi}^{(+)}_{\lambda\kappa}(\mathbf{x},t) &+ [1-f_F^+(-E_\kappa)]\psi^{(-)}_{\lambda\kappa}(\mathbf{x}',t')\overline{\psi}^{(-)}_{\lambda\kappa}(\mathbf{x},t)\Big) \\ -\theta(t-t')\Big(f_F^+(-E_\kappa)\psi^{(-)}_{\lambda\kappa}(\mathbf{x}',t')\overline{\psi}^{(-)}_{\lambda\kappa}(\mathbf{x},t) &+ f_F^+(E_\kappa)\psi^{(+)}_{\lambda\kappa}(\mathbf{x}',t')\overline{\psi}^{(+)}_{\lambda\kappa}(\mathbf{x},t)\Big)\Big].\end{aligned} \tag{7.3.9}$$

We can now extract the vacuum part of the propagator eq. (7.3.6) and write

$$\langle iS_F(x';x|m)\rangle_{\beta,\mu} = iS_0(x';x|m) + iS_1(x';x|m), \tag{7.3.10}$$

where the thermal part is defined by

$$S_1(x';x|m) = i\sum_{\lambda,\kappa}\Big(f_F^+(E_\kappa)\psi^{(+)}_{\lambda\kappa}(\mathbf{x}',t')\overline{\psi}^{(+)}_{\lambda\kappa}(\mathbf{x},t) - f_F^-(E_\kappa)\psi^{(-)}_{\lambda\kappa}(\mathbf{x}',t')\overline{\psi}^{(-)}_{\lambda\kappa}(\mathbf{x},t)\Big), \tag{7.3.11}$$

and where we have introduced the distribution

$$f_F^-(E_\kappa) = 1 - f_F^+(-E_\kappa). \tag{7.3.12}$$

Notice that there is no time-ordering in $S_1(x';x|m)$ despite the fact that the time-ordering in eq. (7.3.9) is non-trivial. The thermal propagator eq. (7.3.10) therefore also trivially satisfies eq. (7.3.7). These considerations can, of course, easily be extended to treat particles with Bose-Einstein statistics as well. The result eq. (7.3.11) can also be derived from an explicit calculation using the second-quantized field operators and appropriate thermal averages, i.e. we use Wicks theorem

$$\mathbf{T}\Big(\Psi(\mathbf{x}',t')\overline{\Psi}(\mathbf{x},t)\Big) = iS_F(x';x|m) + :\Psi(\mathbf{x}',t')\overline{\Psi}(\mathbf{x},t): \tag{7.3.13}$$

where the last term corresponds to a normal ordering. We then obtain

$$\langle\mathbf{T}\Big(\Psi(\mathbf{x}',t')\overline{\Psi}(\mathbf{x},t)\Big)\rangle_{\beta,\mu} = iS_0(x';x|m) + iS_1(x';x|m) \tag{7.3.14}$$

where we have used the only non-zero bilinear thermal averages

$$\begin{aligned}\langle b^\dagger_{\lambda\kappa}b_{\lambda'\kappa'}\rangle_{\beta,\mu} &= f_F^+(E_\kappa)\delta_{\lambda\lambda'}\delta_{\kappa\kappa'} \\ \langle d^\dagger_{\lambda\kappa}d_{\lambda'\kappa'}\rangle_{\beta,\mu} &= f_F^-(E_\kappa)\delta_{\lambda\lambda'}\delta_{\kappa\kappa'}\end{aligned} \tag{7.3.15}$$

In principle we do not have to restrict ourselves to thermal distributions as given by eq. (7.3.8). In fact, we can allow for *any* such one-particle distribution function $f_F^+$ and the definition in eq. (7.3.12). In this chapter, we are particularly interested in the limit of finite density and $T_{emp}\to 0$ for which

$$f_F^+(E) \to \theta(\mu - E) \tag{7.3.16}$$

### 7.3.1 The propagator in momentum-space

As we will compute Feynman diagrams in the following sections, it is necessary to transition to the propagator in momentum-space. Below, we present the general expression for the case of a massive particle at finite temperature and density. For the in-depth derivation, we refer to [182] and its references. Specifically, we have

$$S_F(k,\beta,\mu,m) = (\not{k}+m)\left\{\frac{1}{k^2-m^2} + 2\pi i\delta\left(k^2-m^2\right)\left[\frac{\theta(k_0)}{e^{\beta(E-\mu)}+1} + \frac{\theta(-k_0)}{e^{\beta(E+\mu)}+1}\right]\right\}. \tag{7.3.17}$$

In the limits $T_{emp} \to 0$ and $m \to 0$, this result reduces to

$$\begin{aligned} S_F(k,\mu) &= S_0(k) + S_1(k,\mu), \\ S_0(k) = \frac{\not{k}}{k^2}, &\qquad S_1(k,\mu) = 2\pi i \not{k}\theta(k_0)\theta(\mu-k_0), \end{aligned} \tag{7.3.18}$$

where we recall that $S_1$ represents the thermal contribution. More generally, this result can be formulated covariantly in the form

$$S_F(k,\mu) = \frac{1}{\not{k}} + 2\pi i\not{k}\,\delta(k^2)\,\theta(\eta\cdot k)\,\theta(\mu-\eta\cdot k), \tag{7.3.19}$$

where the final expression is written in terms of a four-vector $\eta^\mu$, representing the velocity of the heat-bath. One can then eventualy set $\eta = (1,\mathbf{0})$.
The propagator for chiral fermions can be obtained from (7.3.17) by the inclusion of the appropriate chiral projectors $P_L$ and $P_R$ and the corresponding chemical potential $\mu \to \mu_{L/R}$, as in previous analyses performed within the Standard Model [183].

## 7.4 General parametrization of the correlator

The correlation function

$$\Gamma^{\lambda\mu\nu} \equiv \frac{i}{g^3}\langle J_V^\mu(p_1) J_V^\nu(p_2) J_A^\lambda(q)\rangle \tag{7.4.1}$$

at finite temperature and density is realized in free field theory by the simple triangle Feynman diagram. The analysis of this correlator will be performed both on general grounds, by the classification of the minimal number of form factors appearing in its expression and, at the same time, perturbatively, by using the expression of the propagator at finite density. The general dependence of $\Gamma$ on the external momenta of the outgoing vextor lines ($p_1$ and $p_2$) and the incoming axial vector line ($q = p_1 + p_2$) is extended by the inclusion of a four-vector $\eta^\mu$ characterizing the velocity of the heat-bath in a covariant formulation.
As already mentioned, the analysis of these interactions has been performed in the previous literature by focusing the attention on the Ward identities satisfied by the correlator at perturbative level, showing that the anomaly is protected against thermal corrections [46]. Our goal is to extend such previous studies and proceed with an analysis of the complete vertex, proving the emergence of a $1/q^2$ interaction in its anomaly form factor, which is the signature of the new effective degree of freedom associated with the chiral anomaly. When the two vector lines (photons) are on-shell, we are going to show explicitly that the interaction reduces just to the anomaly pole, as in the vacuum case. For off-shell photons, the transverse sector of the correlator will be shown to be $\mu$-dependent, but the longitudinal contribution will exhibit the same massless pole structure, deprived of any finite density correction.
The minimal decomposition of $\Gamma$ differs from the standard vacuum vertex at zero $\mu$, which is rather elementary, since it requires fewer form factors. The expansion of the correlator at finite density is complicated by the presence of the four-vector $\eta$ characterizing the heat-bath, and the number of allowed tensor structures increases quite significantly. The complete list of 60 tensor structures $\tau^i$ and form factors $B_i$ formally written as

$$\Gamma^{\lambda\mu\nu} = \sum_{i=1}^{60} B_i \tau_i^{\lambda\mu\nu} \tag{7.4.2}$$

is reported in Appendix D.2. The general dependence on the external momenta in each of the form factors $B_i$ can be expressed in terms of the scalar products $(p_1^2, p_2^2, q^2, p_1\cdot\eta, p_2\cdot\eta)$.
We now pause for a moment and define our conventions for the vacuum contributions and the finite density parts. We are going to denote with $\Gamma$ the entire contribution to the vertex, that we split into a vacuum part and a finite density part

$$\Gamma = \Gamma^{(\text{Vac})} + \Gamma^{(\text{Dens})}. \tag{7.4.3}$$

Moreover, the action of the longitudinal and transverse projectors defined in the axial-vector channel allows us to separate the entire vertex $\Gamma$ as

$$\pi^{\lambda}_{\rho}\Gamma^{\rho\mu\nu} \equiv \Gamma_T^{\lambda\mu\nu}, \qquad \Sigma^{\lambda}_{\rho}\Gamma^{\rho\mu\nu} \equiv \Gamma_L^{\lambda\mu\nu}, \qquad \pi^{\lambda\rho} = g^{\lambda\rho} - \frac{q^\lambda q^\rho}{q^2}, \qquad \Sigma^{\lambda\rho} = \frac{q^\lambda q^\rho}{q^2}. \tag{7.4.4}$$

Such projections can be applied to both $\Gamma^{(0)}$ and $\Gamma^{(1)}$, thereby identifying four distinct components in the decomposition of the correlator. This decomposition simplifies once we impose the Ward identity on the free indices $\mu$ and $\nu$. Specifically, the correlator does not include a longitudinal component with respect to the photon momenta. Moreover, using the Schouten identities, we can significantly lower the number of form factors. We will consider their reduction, first in the general case and then, at a second stage, assuming a simplifying kinematical condition, in which both vector lines (i.e. the two photons) have equal projection on the 4-vector of the heat-bath $(p_1\cdot\eta = p_2\cdot\eta)$.

### 7.4.1 Schouten relations

The analysis of the Schouten relation is rather involved. In $d=4$ spacetime dimensions, the complete antisymmetrization of a tensor over five of its indices must vanish. These tensors can have free indices as well as indices contracted with the two momenta $p_1$, $p_2$, and $\eta$.
We start by considering tensors with three free indices $(\lambda\mu\nu)$ and three indices $(\alpha\beta\gamma)$ that we are going to contract with the momenta and $\eta$. The possible tensors are

$$\varepsilon^{[\lambda\mu\nu\alpha}\delta^{\beta]\gamma}, \quad \varepsilon^{[\mu\nu\alpha\beta}\delta^{\gamma]\lambda}, \quad \varepsilon^{[\lambda\mu\alpha\beta}\delta^{\gamma]\nu}, \quad \varepsilon^{[\lambda\nu\alpha\beta}\delta^{\gamma]\mu}. \tag{7.4.5}$$

Contracting these with $p_1$,$p_2$, and $\eta$ yields 12 equations, 11 of which are independent

$$\begin{aligned}
\varepsilon^{\mu\nu p_1 p_2}{p_1}^\lambda &= \varepsilon^{\lambda\nu p_1 p_2}{p_1}^\mu - \varepsilon^{\lambda\mu p_1 p_2}{p_1}^\nu + \varepsilon^{\lambda\mu\nu p_2}{p_1}^2 - \varepsilon^{\lambda\mu\nu p_1}(p_1\cdot p_2)\\
\varepsilon^{\mu\nu p_1 \eta}{p_1}^\lambda &= \varepsilon^{\lambda\nu p_1 \eta}{p_1}^\mu - \varepsilon^{\lambda\mu p_1 \eta}{p_1}^\nu + \varepsilon^{\lambda\mu\nu \eta}{p_1}^2 - \varepsilon^{\lambda\mu\nu p_1}(p_1\cdot \eta)\\
\varepsilon^{\mu\nu p_1 p_2}{p_2}^\lambda &= \varepsilon^{\lambda\nu p_1 p_2}{p_2}^\mu - \varepsilon^{\lambda\mu p_1 p_2}{p_2}^\nu + \varepsilon^{\lambda\mu\nu p_2}(p_1\cdot p_2) - \varepsilon^{\lambda\mu\nu p_1}{p_2}^2\\
\varepsilon^{\mu\nu p_1 p_2}\eta^\lambda &= \varepsilon^{\lambda\nu p_1 p_2}\eta^\mu - \varepsilon^{\lambda\mu p_1 p_2}\eta^\nu + \varepsilon^{\lambda\mu\nu p_2}(p_1\cdot \eta) - \varepsilon^{\lambda\mu\nu p_1}(p_2\cdot\eta)\\
\varepsilon^{\mu\nu p_1 \eta}{p_2}^\lambda &= \varepsilon^{\lambda\nu p_1 \eta}{p_2}^\mu - \varepsilon^{\lambda\mu p_1 \eta}{p_2}^\nu + \varepsilon^{\lambda\mu\nu \eta}(p_1\cdot p_2) - \varepsilon^{\lambda\mu\nu p_1}(p_2\cdot\eta)
\end{aligned}$$

$$\begin{aligned}
\varepsilon^{\mu\nu p_2\eta}{p_1}^{\lambda} &= \varepsilon^{\lambda\nu p_2\eta}{p_1}^{\mu}-\varepsilon^{\lambda\mu p_2\eta}{p_1}^{\nu}+\varepsilon^{\lambda\mu\nu\eta}(p_1\cdot p_2)-\varepsilon^{\lambda\mu\nu p_2}(p_1\cdot\eta)\\
\varepsilon^{\mu\nu p_1\eta}\eta^{\lambda} &= \varepsilon^{\lambda\nu p_1\eta}\eta^{\mu}-\varepsilon^{\lambda\mu p_1\eta}\eta^{\nu}+\varepsilon^{\lambda\mu\nu\eta}(p_1\cdot\eta)-\varepsilon^{\lambda\mu\nu p_1}\eta^2\\
\varepsilon^{\mu\nu p_2\eta}{p_2}^{\lambda} &= \varepsilon^{\lambda\nu p_2\eta}{p_2}^{\mu}-\varepsilon^{\lambda\mu p_2\eta}{p_2}^{\nu}+\varepsilon^{\lambda\mu\nu\eta}{p_2}^2-\varepsilon^{\lambda\mu\nu p_2}(p_2\cdot\eta)\\
\varepsilon^{\mu\nu p_2\eta}\eta^{\lambda} &= \varepsilon^{\lambda\nu p_2\eta}\eta^{\mu}-\varepsilon^{\lambda\mu p_2\eta}\eta^{\nu}+\varepsilon^{\lambda\mu\nu\eta}(p_2\cdot\eta)-\varepsilon^{\lambda\mu\nu p_2}\eta^2\\
\varepsilon^{\mu p_1 p_2\eta}\delta^{\lambda\nu} &= \varepsilon^{\lambda p_1 p_2\eta}\delta^{\mu\nu}-\varepsilon^{\lambda\mu p_2\eta}{p_1}^{\nu}+\varepsilon^{\lambda\mu p_1\eta}{p_2}^{\nu}-\varepsilon^{\lambda\mu p_1 p_2}\eta^{\nu}\\
\varepsilon^{\nu p_1 p_2\eta}\delta^{\lambda\mu} &= \varepsilon^{\lambda p_1 p_2\eta}\delta^{\mu\nu}-\varepsilon^{\lambda\nu p_2\eta}{p_1}^{\mu}+\varepsilon^{\lambda\nu p_1\eta}{p_2}^{\mu}-\varepsilon^{\lambda\nu p_1 p_2}\eta^{\mu}.
\end{aligned} \tag{7.4.6}$$

A second set of Schouten identities can be obtained by considering tensors with two free indices and four contracted ones

$$\varepsilon^{[\mu\nu\alpha\beta}\delta^{\gamma]\delta},\quad \varepsilon^{[\lambda\mu\alpha\beta}\delta^{\gamma]\delta},\quad \varepsilon^{[\lambda\nu\alpha\beta}\delta^{\gamma]\delta}. \tag{7.4.7}$$

After contracting such structures with the three available four-vectors, we multiply the result for $p_1^{(\cdot)}$, $p_2^{(\cdot)}$ or $\eta^{(\cdot)}$ , where $(\cdot)$ denote the remaining free index, thereby obtaining a rank-3 tensor. Using (7.4.6) to eliminate redundancies, we get 21 new independent relations, that we can use to eliminate the following structures

$$\begin{array}{ccccccc}
{p_1}^{\lambda}{p_1}^{\mu}\varepsilon^{\nu p_1 p_2\eta} & {p_1}^{\mu}{p_2}^{\lambda}\varepsilon^{\nu p_1 p_2\eta} & \eta^{\lambda}{p_1}^{\mu}\varepsilon^{\nu p_1 p_2\eta} & {p_1}^{\lambda}{p_2}^{\mu}\varepsilon^{\nu p_1 p_2\eta} & {p_2}^{\lambda}{p_2}^{\mu}\varepsilon^{\nu p_1 p_2\eta} & \eta^{\lambda}{p_2}^{\mu}\varepsilon^{\nu p_1 p_2\eta} & \eta^{\mu}{p_1}^{\lambda}\varepsilon^{\nu p_1 p_2\eta}\\
\eta^{\mu}{p_2}^{\lambda}\varepsilon^{\nu p_1 p_2\eta} & \eta^{\lambda}\eta^{\mu}\varepsilon^{\nu p_1 p_2\eta} & {p_1}^{\lambda}{p_1}^{\nu}\varepsilon^{\mu p_1 p_2\eta} & {p_1}^{\lambda}{p_2}^{\nu}\varepsilon^{\mu p_1 p_2\eta} & \eta^{\nu}{p_1}^{\lambda}\varepsilon^{\mu p_1 p_2\eta} & {p_1}^{\nu}{p_2}^{\lambda}\varepsilon^{\mu p_1 p_2\eta} & {p_2}^{\lambda}{p_2}^{\nu}\varepsilon^{\mu p_1 p_2\eta}\\
\eta^{\nu}{p_2}^{\lambda}\varepsilon^{\mu p_1 p_2\eta} & \eta^{\lambda}{p_1}^{\nu}\varepsilon^{\mu p_1 p_2\eta} & \eta^{\lambda}{p_2}^{\nu}\varepsilon^{\mu p_1 p_2\eta} & \eta^{\lambda}\eta^{\nu}\varepsilon^{\mu p_1 p_2\eta} & \eta^{\mu}{p_1}^{\nu}\varepsilon^{\lambda p_1 p_2\eta} & {p_1}^{\nu}{p_2}^{\mu}\varepsilon^{\lambda p_1 p_2\eta} & \eta^{\nu}{p_2}^{\mu}\varepsilon^{\lambda p_1 p_2\eta}.
\end{array} \tag{7.4.8}$$

All possible relations derived from other types of antisymmetric tensors have been verified to depend on Eqs. (7.4.6) and (7.4.7). Consequently, there are a total of 32 Schouten identities, which reduce the number of independent tensors from the original 60 to 28. At this stage, the vertex can be simplified into the form

$$\begin{aligned}
\Gamma^{\lambda\mu\nu}(p_1,p_2,\eta) = {} & B_1(p_1,p_2,\eta)\varepsilon^{\lambda\mu\nu p_1}+B_2(p_1,p_2,\eta)\varepsilon^{\lambda\mu\nu p_2}+B_3(p_1,p_2,\eta)\varepsilon^{\lambda\mu\nu\eta}\\
&+{p_1}^{\mu}B_4(p_1,p_2,\eta)\varepsilon^{\lambda\nu p_1 p_2}+{p_1}^{\mu}B_5(p_1,p_2,\eta)\varepsilon^{\lambda\nu p_1\eta}+{p_1}^{\mu}B_6(p_1,p_2,\eta)\varepsilon^{\lambda\nu p_2\eta}\\
&+{p_2}^{\mu}B_7(p_1,p_2,\eta)\varepsilon^{\lambda\nu p_1 p_2}+{p_2}^{\mu}B_8(p_1,p_2,\eta)\varepsilon^{\lambda\nu p_1\eta}+{p_2}^{\mu}B_9(p_1,p_2,\eta)\varepsilon^{\lambda\nu p_2\eta}\\
&+\eta^{\mu}B_{10}(p_1,p_2,\eta)\varepsilon^{\lambda\nu p_1 p_2}+\eta^{\mu}B_{11}(p_1,p_2,\eta)\varepsilon^{\lambda\nu p_1\eta}+\eta^{\mu}B_{12}(p_1,p_2,\eta)\varepsilon^{\lambda\nu p_2\eta}\\
&+{p_1}^{\nu}B_{13}(p_1,p_2,\eta)\varepsilon^{\lambda\mu p_1 p_2}+{p_1}^{\nu}B_{14}(p_1,p_2,\eta)\varepsilon^{\lambda\mu p_1\eta}+{p_1}^{\nu}B_{15}(p_1,p_2,\eta)\varepsilon^{\lambda\mu p_2\eta}\\
&+{p_2}^{\nu}B_{16}(p_1,p_2,\eta)\varepsilon^{\lambda\mu p_1 p_2}+{p_2}^{\nu}B_{17}(p_1,p_2,\eta)\varepsilon^{\lambda\mu p_1\eta}+{p_2}^{\nu}B_{18}(p_1,p_2,\eta)\varepsilon^{\lambda\mu p_2\eta}\\
&+\eta^{\nu}B_{19}(p_1,p_2,\eta)\varepsilon^{\lambda\mu p_1 p_2}+\eta^{\nu}B_{20}(p_1,p_2,\eta)\varepsilon^{\lambda\mu p_1\eta}+\eta^{\nu}B_{21}(p_1,p_2,\eta)\varepsilon^{\lambda\mu p_2\eta}\\
&+{p_1}^{\mu}{p_1}^{\nu}B_{22}(p_1,p_2,\eta)\varepsilon^{\lambda p_1 p_2\eta}+{p_1}^{\mu}{p_2}^{\nu}B_{23}(p_1,p_2,\eta)\varepsilon^{\lambda p_1 p_2\eta}\\
&+\eta^{\nu}{p_1}^{\mu}B_{24}(p_1,p_2,\eta)\varepsilon^{\lambda p_1 p_2\eta}+{p_2}^{\mu}{p_2}^{\nu}B_{25}(p_1,p_2,\eta)\varepsilon^{\lambda p_1 p_2\eta}\\
&+\eta^{\mu}{p_2}^{\nu}B_{26}(p_1,p_2,\eta)\varepsilon^{\lambda p_1 p_2\eta}+\eta^{\mu}\eta^{\nu}B_{27}(p_1,p_2,\eta)\varepsilon^{\lambda p_1 p_2\eta}\\
&+\delta^{\mu\nu}B_{28}(p_1,p_2,\eta)\varepsilon^{\lambda p_1 p_2\eta}
\end{aligned} \tag{7.4.9}$$

where we have redefined the form factors $B_i$. In the next sections, we are going to simplify this decomposition even further by imposing the Bose symmetry of the two photon lines.

### 7.4.2 Bose symmetry

Due to the Bose symmetry, the $\Gamma$ vertex is symmetric under the exchange $\{p_1,\mu_1\}\leftrightarrow\{p_2,\mu_2\}$. The relations among the form factors obtained by imposing the Bose symmetry are presented in Appendix

D.3. Using those relations the expression of the amplitude takes the form

$$
\begin{aligned}
\Gamma^{\lambda\mu\nu}(p_1,p_2,\eta) = {} & B_1(p_1,p_2,\eta)\varepsilon^{\lambda\mu\nu p_1} - B_1(p_2,p_1,\eta)\varepsilon^{\lambda\mu\nu p_2} + B_3(p_1,p_2,\eta)\varepsilon^{\lambda\mu\nu\eta} \\
& + {p_1}^{\mu} B_4(p_1,p_2,\eta)\varepsilon^{\lambda\nu p_1 p_2} - {p_2}^{\nu} B_4(p_2,p_1,\eta)\varepsilon^{\lambda\mu p_1 p_2} + {p_1}^{\mu} B_5(p_1,p_2,\eta)\varepsilon^{\lambda\nu p_1 \eta} \\
& + {p_2}^{\nu} B_5(p_2,p_1,\eta)\varepsilon^{\lambda\mu p_2 \eta} + {p_1}^{\mu} B_6(p_1,p_2,\eta)\varepsilon^{\lambda\nu p_2 \eta} + {p_2}^{\nu} B_6(p_2,p_1,\eta)\varepsilon^{\lambda\mu p_1 \eta} \\
& + {p_2}^{\mu} B_7(p_1,p_2,\eta)\varepsilon^{\lambda\nu p_1 p_2} - {p_1}^{\nu} B_7(p_2,p_1,\eta)\varepsilon^{\lambda\mu p_1 p_2} + {p_2}^{\mu} B_8(p_1,p_2,\eta)\varepsilon^{\lambda\nu p_1 \eta} \\
& + {p_1}^{\nu} B_8(p_2,p_1,\eta)\varepsilon^{\lambda\mu p_2 \eta} + {p_2}^{\mu} B_9(p_1,p_2,\eta)\varepsilon^{\lambda\nu p_2 \eta} + {p_1}^{\nu} B_9(p_2,p_1,\eta)\varepsilon^{\lambda\mu p_1 \eta} \\
& + \eta^{\mu} B_{10}(p_1,p_2,\eta)\varepsilon^{\lambda\nu p_1 p_2} - \eta^{\nu} B_{10}(p_2,p_1,\eta)\varepsilon^{\lambda\mu p_1 p_2} + \eta^{\mu} B_{11}(p_1,p_2,\eta)\varepsilon^{\lambda\nu p_1 \eta} \\
& + \eta^{\nu} B_{11}(p_2,p_1,\eta)\varepsilon^{\lambda\mu p_2 \eta} + \eta^{\mu} B_{12}(p_1,p_2,\eta)\varepsilon^{\lambda\nu p_2 \eta} + \eta^{\nu} B_{12}(p_2,p_1,\eta)\varepsilon^{\lambda\mu p_1 \eta} \\
& + {p_1}^{\mu}{p_1}^{\nu} B_{22}(p_1,p_2,\eta)\varepsilon^{\lambda p_1 p_2 \eta} - {p_2}^{\mu}{p_2}^{\nu} B_{22}(p_2,p_1,\eta)\varepsilon^{\lambda p_1 p_2 \eta} \\
& + {p_1}^{\mu}{p_2}^{\nu} B_{23}(p_1,p_2,\eta)\varepsilon^{\lambda p_1 p_2 \eta} + \eta^{\nu}{p_1}^{\mu} B_{24}(p_1,p_2,\eta)\varepsilon^{\lambda p_1 p_2 \eta} \\
& - \eta^{\mu}{p_2}^{\nu} B_{24}(p_2,p_1,\eta)\varepsilon^{\lambda p_1 p_2 \eta} + \eta^{\mu}\eta^{\nu} B_{27}(p_1,p_2,\eta)\varepsilon^{\lambda p_1 p_2 \eta} \\
& + \delta^{\mu\nu} B_{28}(p_1,p_2,\eta)\varepsilon^{\lambda p_1 p_2 \eta},
\end{aligned}
\tag{7.4.10}
$$

which, at this stage, is expressed in terms of 16 independent form factors. Therefore, we have reduced the structures according to the sequence **60** → **28** → **16**.

### 7.4.3 Vector Ward identities

The last step involves the implementation of the Ward identities on the vector lines

$$
{p_1}_{\mu}\Gamma^{\lambda\mu\nu}(p_1,p_2,\eta) = {p_2}_{\nu}\Gamma^{\lambda\mu\nu}(p_1,p_2,\eta) = 0. \tag{7.4.11}
$$

By imposing these conditions, we get additional relations, reported in the Appendix D.3, that will be essential for further simplifications. In the end, we can express the final form of the amplitude as follows

$$
\begin{aligned}
\Gamma^{\lambda\mu\nu}(p_1,p_2,\eta) = {} & \chi_1^{\lambda\mu\nu}(p_1,p_2,\eta) B_1(p_1,p_2,\eta) + \chi_1^{\lambda\mu\nu}(p_2,p_1,\eta) B_1(p_2,p_1,\eta) \\
& + \chi_2^{\lambda\mu\nu}(p_1,p_2,\eta) B_2(p_1,p_2,\eta) + \chi_2^{\lambda\mu\nu}(p_2,p_1,\eta) B_2(p_2,p_1,\eta) \\
& + \chi_3^{\lambda\mu\nu}(p_1,p_2,\eta) B_3(p_1,p_2,\eta) + \chi_3^{\lambda\mu\nu}(p_2,p_1,\eta) B_3(p_2,p_1,\eta) \\
& + \chi_4^{\lambda\mu\nu}(p_1,p_2,\eta) B_4(p_1,p_2,\eta) + \chi_4^{\lambda\mu\nu}(p_2,p_1,\eta) B_4(p_2,p_1,\eta) \\
& + \chi_5^{\lambda\mu\nu}(p_1,p_2,\eta) B_5(p_1,p_2,\eta) + \chi_5^{\lambda\mu\nu}(p_2,p_1,\eta) B_5(p_2,p_1,\eta) \\
& + \chi_6^{\lambda\mu\nu}(p_1,p_2,\eta) B_6(p_1,p_2,\eta) + \chi_6^{\lambda\mu\nu}(p_2,p_1,\eta) B_6(p_2,p_1,\eta) \\
& + \chi_7^{\lambda\mu\nu}(p_1,p_2,\eta) B_7(p_1,p_2,\eta) + \chi_7^{\lambda\mu\nu}(p_2,p_1,\eta) B_7(p_2,p_1,\eta) \\
& + \chi_8^{\lambda\mu\nu}(p_1,p_2,\eta) B_8(p_1,p_2,\eta) + \chi_8^{\lambda\mu\nu}(p_2,p_1,\eta) B_8(p_2,p_1,\eta) \\
& + \chi_9^{\lambda\mu\nu}(p_1,p_2,\eta) B_9(p_1,p_2,\eta) + \chi_{10}^{\lambda\mu\nu}(p_1,p_2,\eta) B_{10}(p_1,p_2,\eta),
\end{aligned}
\tag{7.4.12}
$$

where we have redefined the form factors $B_i$ and we have introduced new tensor structures

$$
\begin{aligned}
\chi_1^{\lambda\mu\nu}(p_1,p_2,\eta) &= \left(\varepsilon^{\lambda\mu\nu p_2} p_1^2 + \varepsilon^{\lambda\nu p_1 p_2}{p_1}^{\mu}\right) \\
\chi_2^{\lambda\mu\nu}(p_1,p_2,\eta) &= \left(\varepsilon^{\lambda\nu p_2 \eta}{p_1}^{\mu} - \frac{\varepsilon^{\lambda\nu p_2 \eta}\eta^{\mu} p_1^2}{\eta\cdot p_1}\right)
\end{aligned}
$$

$$
\begin{aligned}
\chi_3^{\lambda\mu\nu}(p_1,p_2,\eta) &= \left(\varepsilon^{\lambda\nu p_1 p_2}{p_2}^{\mu} + \varepsilon^{\lambda\mu\nu p_2}(p_1\cdot p_2)\right)\\
\chi_4^{\lambda\mu\nu}(p_1,p_2,\eta) &= \left(\varepsilon^{\lambda\nu p_2\eta}{p_2}^{\mu} - \frac{\varepsilon^{\lambda\nu p_2\eta}\eta^{\mu}(p_1\cdot p_2)}{p_1\cdot\eta}\right)\\
\chi_5^{\lambda\mu\nu}(p_1,p_2,\eta) &= \left(\varepsilon^{\lambda\nu p_1 p_2}\eta^{\mu} + \varepsilon^{\lambda\mu\nu p_2}(p_1\cdot\eta)\right)\\
\chi_6^{\lambda\mu\nu}(p_1,p_2,\eta) &= \Bigg(\varepsilon^{\lambda p_1 p_2\eta}{p_1}^{\mu}{p_1}^{\nu} - \frac{1}{2}\varepsilon^{\lambda\mu p_2\eta}{p_1}^{2}{p_1}^{\nu} - \frac{1}{2}\varepsilon^{\lambda p_1 p_2\eta}g^{\mu\nu}{p_1}^{2} - \frac{1}{2}\varepsilon^{\lambda\nu p_1\eta}{p_2}^{\mu}{p_1}^{2}\\
&\quad + \varepsilon^{\lambda\nu p_1\eta}{p_1}^{\mu}(p_1\cdot p_2) + \frac{1}{2}\varepsilon^{\lambda\mu\nu\eta}{p_1}^{2}(p_1\cdot p_2)\Bigg)\\
\chi_7^{\lambda\mu\nu}(p_1,p_2,\eta) &= \left(\frac{1}{2}\varepsilon^{\lambda p_1 p_2\eta}{p_1}^{\mu}{p_2}^{\nu} - \frac{1}{2}\varepsilon^{\lambda\mu p_2\eta}{p_1}^{2}{p_2}^{\nu} + \frac{1}{2}\varepsilon^{\lambda\nu p_1\eta}{p_1}^{\mu}{p_2}^{2} + \frac{1}{2}\varepsilon^{\lambda\mu\nu\eta}{p_1}^{2}{p_2}^{2}\right)\\
\chi_8^{\lambda\mu\nu}(p_1,p_2,\eta) &= \Bigg(\varepsilon^{\lambda p_1 p_2\eta}{p_1}^{\mu}\eta^{\nu} - \frac{1}{2}\varepsilon^{\lambda\mu p_2\eta}{p_1}^{2}\eta^{\nu} - \frac{\varepsilon^{\lambda p_1 p_2\eta}\eta^{\mu}{p_1}^{2}\eta^{\nu}}{2(p_1\cdot\eta)}\\
&\quad + \varepsilon^{\lambda\nu p_1\eta}{p_1}^{\mu}(p_2\cdot\eta) + \frac{1}{2}\varepsilon^{\lambda\mu\nu\eta}{p_1}^{2}(p_2\cdot\eta) - \frac{\varepsilon^{\lambda\nu p_1\eta}\eta^{\mu}{p_1}^{2}(p_2\cdot\eta)}{2(p_1\cdot\eta)}\Bigg)\\
\chi_9^{\lambda\mu\nu}(p_1,p_2,\eta) &= \left(\varepsilon^{\lambda p_1 p_2\eta}g^{\mu\nu} + \varepsilon^{\lambda\nu p_1\eta}{p_2}^{\mu} - \varepsilon^{\lambda\mu p_2\eta}{p_1}^{\nu} + \varepsilon^{\lambda\mu\nu\eta}(p_1\cdot p_2)\right)\\
\chi_{10}^{\lambda\mu\nu}(p_1,p_2,\eta) &= \left(\varepsilon^{\lambda\mu\nu\eta} + \frac{\varepsilon^{\lambda\nu p_1\eta}\eta^{\mu}}{p_1\cdot\eta} - \frac{\varepsilon^{\lambda\mu p_2\eta}\eta^{\nu}}{p_2\cdot\eta} + \frac{\varepsilon^{\lambda p_1 p_2\eta}\eta^{\mu}\eta^{\nu}}{(p_1\cdot\eta)(p_2\cdot\eta)}\right).
\end{aligned}
\tag{7.4.13}
$$

Therefore, using the Schouten relations and imposing the symmetry of the interaction, we have simplified the off-shell vertex from its original expression in terms of 60 form factors in (7.4.2), down to 10. The procedure can be summarized according to this sequence: **60** initial form factors → **28** form factors after using the 32 Schouten relations → **16** after using the Bose symmetry → **10** after using the vector WIs.

## 7.5 The symmetric decomposition of the correlator

In the last sections, we have derived the most general decomposition of the Γ vertex at finite density. In this section, we choose to impose some kinematic constraints in order to further simplify the vertex. Specifically, we consider the following additional conditions

$$
\begin{aligned}
p_1^2 = p_2^2 &= p^2\\
p_1\cdot\eta = p_2\cdot\eta &= p\cdot\eta,
\end{aligned}
\tag{7.5.1}
$$

which state that the two photons have equal invariant mass as well as equal projections on the four-vector of the heat-bath. Note that we are still working with off-shell photons. The relations above constrain the form factors to depend only on three scalar products, namely $q^2$, $p^2$ and $p\cdot\eta$. Due to this restriction, the form factors are automatically invariant under the exchange $(p_1 \leftrightarrow p_2)$.

We now adopt an approach analogous to the decomposition outlined in the previous section, starting with the same initial set of 60 form factors. We then apply the same Schouten identities, which remain largely unaffected by the symmetric constraint (7.5.1). The primary difference lies in a simple relabeling of the scalar products. From the 60 structures listed in Appendix (D.2), we reduce the set to 28 form factors and tensor structures by employing the Schouten relations, as in the previous section.

We then examine the Bose symmetry constraints. These conditions are more restrictive in this context,

reducing the number of form factors to 12, as opposed to 16 in the previous section. The constraints are expressed as

$$
\begin{aligned}
B_2 &= -B_1 & B_3 &= -B_3 \\
B_{13} &= -B_7 & B_{14} &= B_9 \\
B_{15} &= B_8 & B_{16} &= -B_4 \\
B_{17} &= B_6 & B_{18} &= B_5 \\
B_{19} &= -B_{10} & B_{20} &= B_{12} \\
B_{21} &= B_{11} & B_{23} &= -B_{23} \\
B_{25} &= -B_{22} & B_{26} &= -B_{24} \\
B_{27} &= -B_{27} & B_{28} &= -B_{28}
\end{aligned}
\tag{7.5.2}
$$

Given these relations and the symmetric constraints (7.5.1), it follows that the antisymmetric form factors $B_3$, $B_{23}$, $B_{27}$ and $B_{28}$ must vanish. Therefore, we can express the correlator in terms of 12 form factors

$$
\begin{aligned}
\Gamma^{\lambda\mu\nu}(p_1,p_2,\eta) = {} & B_5\left(p_1{}^{\mu}\varepsilon^{\lambda\nu p_1\eta} + p_2{}^{\nu}\varepsilon^{\lambda\mu p_2\eta}\right) + B_6\left(p_1{}^{\mu}\varepsilon^{\lambda\nu p_2\eta} + p_2{}^{\nu}\varepsilon^{\lambda\mu p_1\eta}\right) + B_8\left(p_2{}^{\mu}\varepsilon^{\lambda\nu p_1\eta} + p_1{}^{\nu}\varepsilon^{\lambda\mu p_2\eta}\right) \\
& + B_9\left(p_1{}^{\nu}\varepsilon^{\lambda\mu p_1\eta} + p_2{}^{\mu}\varepsilon^{\lambda\nu p_2\eta}\right) + B_{10}\left(\eta^{\mu}\varepsilon^{\lambda\nu p_1 p_2} - \eta^{\nu}\varepsilon^{\lambda\mu p_1 p_2}\right) + B_{11}\left(\eta^{\mu}\varepsilon^{\lambda\nu p_1\eta} + \eta^{\nu}\varepsilon^{\lambda\mu p_2\eta}\right) \\
& + B_{12}\left(\eta^{\nu}\varepsilon^{\lambda\mu p_1\eta} + \eta^{\mu}\varepsilon^{\lambda\nu p_2\eta}\right) + B_{22}\left(p_1{}^{\mu}p_1{}^{\nu}\varepsilon^{\lambda p_1 p_2\eta} - p_2{}^{\mu}p_2{}^{\nu}\varepsilon^{\lambda p_1 p_2\eta}\right) \\
& + B_4\left(p_1{}^{\mu}\varepsilon^{\lambda\nu p_1 p_2} - p_2{}^{\nu}\varepsilon^{\lambda\mu p_1 p_2}\right) + B_7\left(p_2{}^{\mu}\varepsilon^{\lambda\nu p_1 p_2} - p_1{}^{\nu}\varepsilon^{\lambda\mu p_1 p_2}\right) \\
& + B_{24}\left(\eta^{\nu}p_1{}^{\mu}\varepsilon^{\lambda p_1 p_2\eta} - \eta^{\mu}p_2{}^{\nu}\varepsilon^{\lambda p_1 p_2\eta}\right) + B_1\left(\varepsilon^{\lambda\mu\nu p_1} - \varepsilon^{\lambda\mu\nu p_2}\right).
\end{aligned}
\tag{7.5.3}
$$

At this stage, we finally need to impose the conservation Ward identity

$$
\begin{aligned}
p_{2\nu}\Gamma^{\lambda\mu\nu}(p_1,p_2,\eta) = {} & \varepsilon^{\lambda\mu p_1 p_2}\Big(-B_{10}\,(p\cdot\eta) - B_7\,(p_1\cdot p_2) - B_1 - B_4\,p^2\Big) \\
& + \varepsilon^{\lambda\mu p_2\eta}\Big(B_{11}\,(p\cdot\eta) + B_8\,(p_1\cdot p_2) + B_5\,p^2\Big) + \varepsilon^{\lambda\mu p_1\eta}\Big(B_{12}\,(p\cdot\eta) + B_9\,(p_1\cdot p_2) + B_6\,p^2\Big) \\
& + \varepsilon^{\lambda p_1 p_2\eta}\Big(-(B_{11} + B_{24}\,p^2)\eta^{\mu} + p_1{}^{\mu}\left[B_{24}\,(p\cdot\eta) + B_{22}\,(p_1\cdot p_2) - B_5\right] - (B_8 + B_{22}\,p^2)p_2{}^{\mu}\Big) = 0,
\end{aligned}
\tag{7.5.4}
$$

obtaining the following set of relations

$$
\begin{aligned}
B_1 &= -B_7\,(p_1\cdot p_2) - B_{10}(p\cdot\eta) - B_4\,p^2 \\
B_5 &= B_{22}\,(p_1\cdot p_2) + B_{24}(p\cdot\eta) \\
B_8 &= -B_{22}\,p^2 \\
B_{11} &= -B_{24}\,p^2 \\
B_{12} &= -\frac{B_9\,(p_1\cdot p_2) + B_6\,p^2}{p\cdot\eta}.
\end{aligned}
\tag{7.5.5}
$$

The final decomposition can thus be expressed in terms of a reduced set of 7 form factors, compared to the 10 in the general decomposition of the previous section. The symmetric decomposition can be written as

$$
\Gamma^{\lambda\mu\nu}(p_1,p_2,\eta) = B_1\chi_1^{\lambda\mu\nu} + B_2\chi_2^{\lambda\mu\nu} + B_3\chi_3^{\lambda\mu\nu} + B_4\chi_4^{\lambda\mu\nu} + B_5\chi_5^{\lambda\mu\nu} + B_6\chi_6^{\lambda\mu\nu} + B_7\chi_7^{\lambda\mu\nu},
\tag{7.5.6}
$$

where we have redefined the form factors $B_i$ and we have introduced new tensor structures

$$\chi_1^{\lambda\mu\nu} = \left(-\frac{p^2\eta^\nu\varepsilon^{\lambda\mu p_1\eta}}{p\cdot\eta} - \frac{p^2\eta^\mu\varepsilon^{\lambda\nu p_2\eta}}{p\cdot\eta} + p_1{}^\mu\varepsilon^{\lambda\nu p_2\eta} + p_2{}^\nu\varepsilon^{\lambda\mu p_1\eta}\right)$$
$$\chi_2^{\lambda\mu\nu} = \left(-\frac{\eta^\mu(p_1\cdot p_2)\varepsilon^{\lambda\nu p_2\eta}}{p\cdot\eta} - \frac{\eta^\nu(p_1\cdot p_2)\varepsilon^{\lambda\mu p_1\eta}}{p\cdot\eta} + p_1{}^\nu\varepsilon^{\lambda\mu p_1\eta} + p_2{}^\mu\varepsilon^{\lambda\nu p_2\eta}\right)$$
$$\chi_3^{\lambda\mu\nu} = \left(-(p\cdot\eta)\varepsilon^{\lambda\mu\nu p_1} + (p\cdot\eta)\varepsilon^{\lambda\mu\nu p_2} + \eta^\mu\varepsilon^{\lambda\nu p_1 p_2} - \eta^\nu\varepsilon^{\lambda\mu p_1 p_2}\right)$$
$$\chi_4^{\lambda\mu\nu} = \left(-p^2 p_1{}^\nu\varepsilon^{\lambda\mu p_2\eta} - p^2 p_2{}^\mu\varepsilon^{\lambda\nu p_1\eta} + p_1{}^\mu p_1{}^\nu\varepsilon^{\lambda p_1 p_2\eta} - p_2{}^\mu p_2{}^\nu\varepsilon^{\lambda p_1 p_2\eta} + p_1{}^\mu(p_1\cdot p_2)\varepsilon^{\lambda\nu p_1\eta} + p_2{}^\nu(p_1\cdot p_2)\varepsilon^{\lambda\mu p_2\eta}\right)$$
$$\chi_5^{\lambda\mu\nu} = \left(-p^2\eta^\mu\varepsilon^{\lambda\nu p_1\eta} + p_1{}^\mu(p\cdot\eta)\varepsilon^{\lambda\nu p_1\eta} - p^2\eta^\nu\varepsilon^{\lambda\mu p_2\eta} + p_2{}^\nu(p\cdot\eta)\varepsilon^{\lambda\mu p_2\eta} - \eta^\mu p_2{}^\nu\varepsilon^{\lambda p_1 p_2\eta} + \eta^\nu p_1{}^\mu\varepsilon^{\lambda p_1 p_2\eta}\right)$$
$$\chi_6^{\lambda\mu\nu} = \left(-p^2\varepsilon^{\lambda\mu\nu p_1} + p^2\varepsilon^{\lambda\mu\nu p_2} + p_1{}^\mu\varepsilon^{\lambda\nu p_1 p_2} - p_2{}^\nu\varepsilon^{\lambda\mu p_1 p_2}\right)$$
$$\chi_7^{\lambda\mu\nu} = \left(p_2{}^\mu\varepsilon^{\lambda\nu p_1 p_2} - p_1{}^\nu\varepsilon^{\lambda\mu p_1 p_2} - (p_1\cdot p_2)\varepsilon^{\lambda\mu\nu p_1} + (p_1\cdot p_2)\varepsilon^{\lambda\mu\nu p_2}\right). \tag{7.5.7}$$

## 7.6 The perturbative realization

In this section, we begin by examining the perturbative realization of the Γ vertex at finite density and zero temperature. The perturbative expression of the vertex will be compared with the parametrization in terms of form factors of the Γ vertex derived in the previous sections. We therefore write

$$\Gamma^{\lambda\mu\nu}_{pert} = \Gamma^{\lambda\mu\nu}, \tag{7.6.1}$$

where the left-hand side represents the perturbative realization of the Γ vertex, expressed in terms of Feynman integrals at finite density, and the right-hand side represents the Γ vertex expressed in terms of tensor structures and form factors, as parametrized in (7.4.12) for the general case and (7.5.6) for the symmetric case. The equation is tensorial in nature.

To explicitly identify the form factors in our parametrization, we need to contract both sides of (7.6.1) with a set of tensor structures to saturate the tensorial indices. This yields scalar equations that relate the form factors in the parametrization to the Feynman integrals at finite density. The list of tensor structures required for the contractions is extensive, and the resulting equations must be carefully analyzed to extract the independent ones. In the final step, solving the derived linear system of equations allows us to determine the perturbative expressions of the form factors on the right-hand side of (7.6.1).

While this procedure may appear intricate, it can first be tested against known results for the perturbative correlator at zero density. For instance, it is well established that in this specific case, the perturbative vertex reduces to its longitudinal sector when the two vector lines are on-shell. In our case, this check is performed by taking the $\eta\to 0$ limit, combined with the $p^2\to 0$ condition and the transversality of the photon polarization vectors ($\varepsilon_1\cdot p_1 = 0$, $\varepsilon_2\cdot p_2 = 0$). We will illustrate this method in detail in a future section.

Let us now focus on the perturbative expression of the interaction. The Γ vertex receives contributions from the direct and exchange triangle diagrams. These contributions can be written as

$$\Gamma^{\lambda\mu\nu}_{pert} = \int\frac{d^4k}{(2\pi)^4}\mathrm{Tr}[(\not{k}-\not{q})\gamma^\nu(\not{k}-\not{p}_1)\gamma^\mu\not{k}\gamma^\lambda\gamma^5]\,G_F(k)\,G_F(k-p_1)\,G_F(k-q) + \begin{pmatrix} p_1\leftrightarrow p_2 \\ \mu\leftrightarrow\nu\end{pmatrix} \tag{7.6.2}$$

where we have separated the scalar and tensor components of the fermion propagator $S_F(k)$, together with their finite density and vacuum parts, as

$$
\begin{aligned}
S_F(k) &= \not{k}\, G_F(k) \\
G_F(k) &= G_0(k) + G_1(k) \\
G_0(k) &= \frac{1}{k^2} \\
G_1(k) &= 2\pi i\, \delta(k^2)\, \theta(\eta\cdot k)\, \theta(\mu - \eta\cdot k).
\end{aligned}
\tag{7.6.3}
$$

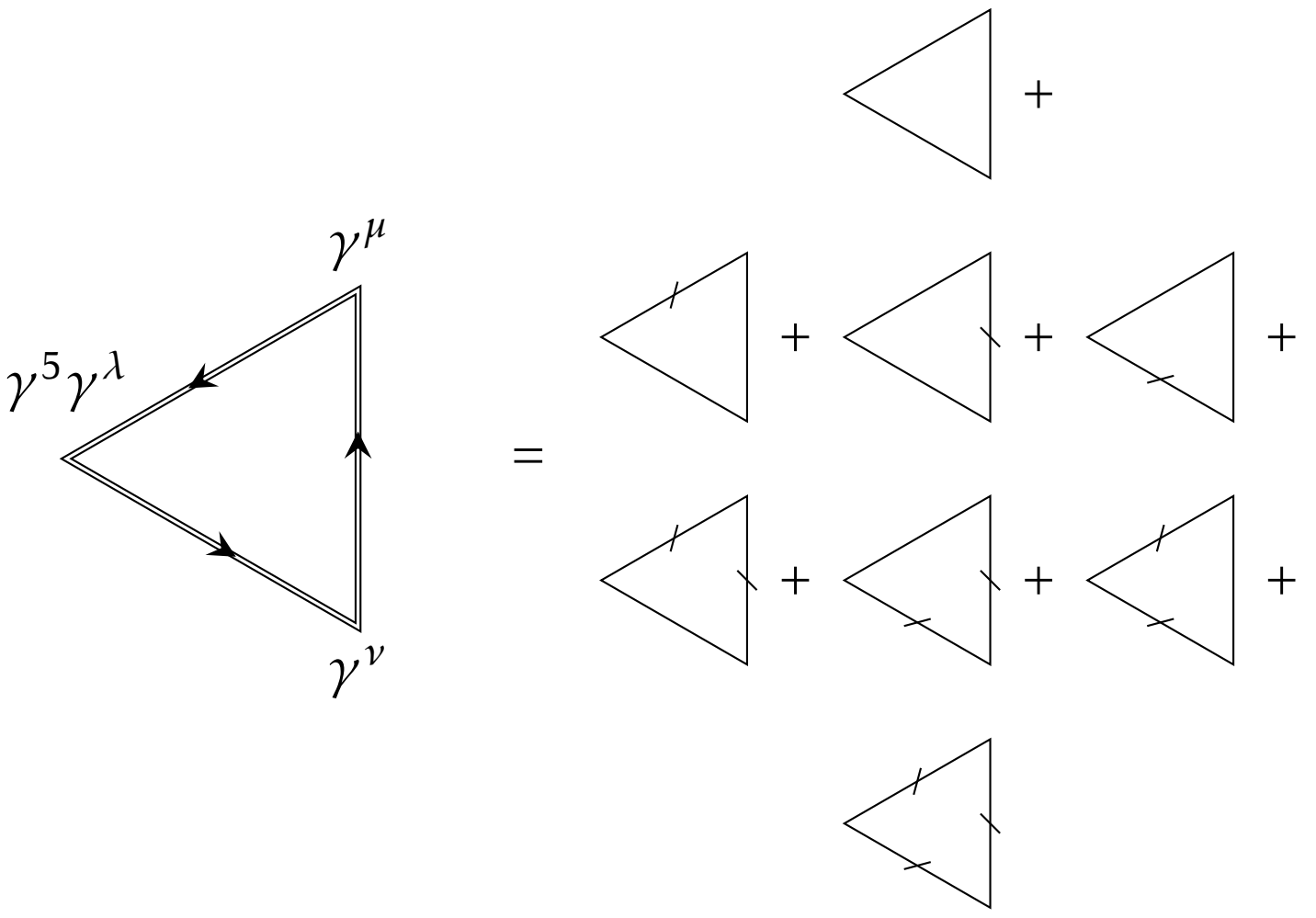

Figure 7.3: List of the vacuum and finite density contributions in the expansion of the chiral anomaly interaction. The finite density insertions are indicated with cut lines.

Expanding the trace in (7.6.2), we can write

$$
\mathrm{Tr}[(\not{k}-\not{q})\gamma^\nu(\not{k}-\not{p}_1)\gamma^\mu \not{k}\gamma^\lambda\gamma^5] = iD_{1\ \alpha}{}^{\lambda\mu\nu}(p_1,p_2)k^\alpha + iD_{2\ \alpha\beta}{}^{\lambda\mu\nu}(p_1,p_2)k^\alpha k^\beta + iD_{3\ \alpha\beta\gamma}{}^{\lambda\mu\nu}(p_1,p_2)k^\alpha k^\beta k^\gamma
\tag{7.6.4}
$$

where we have defined the tensors

$$
\begin{aligned}
D_{1\ \alpha}{}^{\lambda\mu\nu}(p_1,p_2) &= 4g^{\alpha\mu}\varepsilon^{\lambda\nu p_1 q} - 4g^{\lambda\nu}\varepsilon^{\alpha\mu p_1 q} + 4p_1{}^\alpha\varepsilon^{\lambda\mu\nu q} - 4q^\lambda\varepsilon^{\alpha\mu\nu p_1} + 4p_1{}^\mu\varepsilon^{\alpha\lambda\nu q} + 4q^\nu\varepsilon^{\alpha\lambda\mu p_1} \\
D_{2\ \alpha\beta}{}^{\lambda\mu\nu}(p_1,p_2) &= 4g^{\beta\lambda}\varepsilon^{\alpha\mu\nu p_1} + 4g^{\beta\mu}\varepsilon^{\alpha\lambda\nu p_1} - 4g^{\beta\nu}\varepsilon^{\alpha\lambda\mu p_1} - 4g^{\alpha\beta}\varepsilon^{\lambda\mu\nu q} - 8g^{\beta\mu}\varepsilon^{\alpha\lambda\nu q} + 4p_1{}^\beta\varepsilon^{\alpha\lambda\mu\nu} \\
D_{3\ \alpha\beta\gamma}{}^{\lambda\mu\nu}(p_1,p_2) &= 4g^{\beta\gamma}\varepsilon^{\lambda\mu\nu\alpha}.
\end{aligned}
\tag{7.6.5}
$$

The perturbative integral can then be written as

$$
\begin{aligned}
\Gamma^{\lambda\mu\nu}_{pert} &= iD_{1\ \alpha}{}^{\lambda\mu\nu}(p_1,p_2)\int \frac{d^4k}{(2\pi)^4}\, k^\alpha\, G_F(k)\, G_F(k-p_1)\, G_F(k-q) \\
&\quad + iD_{2\ \alpha\beta}{}^{\lambda\mu\nu}(p_1,p_2)\int \frac{d^4k}{(2\pi)^4}\, k^\alpha\, k^\beta\, G_F(k)\, G_F(k-p_1)\, G_F(k-q) \\
&\quad + iD_{3\ \alpha\beta\tau}{}^{\lambda\mu\nu}(p_1,p_2)\int \frac{d^4k}{(2\pi)^4 k}\, k^\alpha\, k^\beta\, k^\tau\, G_F(k)\, G_F(k-p_1)\, G_F(k-q) \\
&\quad + \begin{pmatrix} p_1 \leftrightarrow p_2 \\ \mu \leftrightarrow \nu \end{pmatrix}.
\end{aligned}
\tag{7.6.6}
$$

As discussed at the beginning of this section, we need to contract both sides of equation (7.6.1) with a specific set of parity-odd tensor structures to extract a perturbative expression for the form factors.

Contracting equation (7.6.6) with a parity-odd tensor structure yields integrals of the following type

$$\begin{aligned}
J[f(k,p_1,p_2,\eta)] &= \int \frac{d^4k}{(2\pi)^4} f(k,p_1,p_2,\eta)\, G_F(k)\, G_F(k-p_1)\, G_F(k-q) \\
H[f(k,p_1,p_2,\eta)] &= \int \frac{d^4k}{(2\pi)^4} f(k,p_1,p_2,\eta)\, G_F(k)\, G_F(k-p_2)\, G_F(k-q),
\end{aligned} \tag{7.6.7}$$

$f(k,p_1,p_2,\eta)$ indicates a general scalar function obtained by contracting $p_1, p_2$ and $\eta$ with the loop momentum $k$. The integrals $J[\dots]$ and the $H[\dots]$ we have just defined refer to the direct and the exchanged triangle graph respectively.

The next step is to expand the propagators $G_F(k)$, $G_F(k-p_i)$ and $G_F(k-q)$ into their vacuum ($G_0$) and thermal ($G_1$) components. This results in each integral splitting into eight terms that correspond to different combinations of vacuum and thermal propagators. To manage these combinations, we introduce superscripts $(l_1, l_2, l_3)$ with $l_i = 0/1$ to label the specific contributions associated to each propagator. For example

$$\begin{aligned}
J^{(1,0,0)}[f(k,p_1,p_2,\eta)] &= \int \frac{d^4k}{(2\pi)^4} f(k,p_1,p_2,\eta)\, G_1(k)\, G_0(k-p_1)\, G_0(k-q), \\
J^{(0,1,0)}[f(k,p_1,p_2,\eta)] &= \int \frac{d^4k}{(2\pi)^4} f(k,p_1,p_2,\eta)\, G_0(k)\, G_1(k-p_1)\, G_0(k-q), \\
J^{(1,0,1)}[f(k,p_1,p_2,\eta)] &= \int \frac{d^4k}{(2\pi)^4} f(k,p_1,p_2,\eta)\, G_1(k)\, G_0(k-p_1)\, G_1(k-q).
\end{aligned}$$

In most cases, the integration at finite density cannot be performed covariantly. Specifically, such integrals can only be computed in a specific frame of reference. This represents the final step of our computation.

The potential UV divergences in the (0,0,0) term, corresponding to the standard diagram at zero density, are removed by enforcing the vector Ward identities. Other terms are automatically free of UV divergences due to the presence of a cutoff imposed by the chemical potential. However, there may still be other types of divergences, such as IR or collinear ones, which can arise in the on-shell photon case. These divergences will be handled using dimensional regularization (DR). While such divergences may appear in specific contributions of the diagram, we will show that they ultimately cancel out in the complete amplitude.

### 7.6.1 The zero-density limit

Before proceeding with our computations, we verify that, when the chemical potential is set to zero, our results align with the standard zero-density case described in Chapter 4. Specifically, we perform this verification within the symmetric decomposition (7.5.6), as it will be the primary framework used in subsequent sections. By setting $\eta \to 0$, all tensors in this decomposition vanish except for two

$$\lim_{\eta\to 0} \Gamma^{\lambda\mu\nu} = \bar{B}_6 \chi_6^{\lambda\mu\nu} + \bar{B}_7 \chi_7^{\lambda\mu\nu}. \tag{7.6.8}$$

This parametrization coincides the Rosenberg one, as described in eq. (4.5.15), after the antisymmetric terms are eliminated. Indeed, as previously noted, due to the symmetric conditions

$$p_1^2 = p_2^2, \quad p_1 \cdot \eta = p_2 \cdot \eta, \tag{7.6.9}$$

it is impossible to construct a scalar form factor that is antisymmetric under the exchange $p_1 \leftrightarrow p_2$. We then equate eq. (7.6.8) to the perturbative computation, evaluated in the same zero-density limit. The perturbative analysis of the two form factors yields

$$\begin{aligned}
\bar{B}_6 &= \frac{i}{4\pi^2 (p^2 - p_1 \cdot p_2)^2 (p^2 + p_1 \cdot p_2)} \Bigg\{ \Bigg[ (p_1 \cdot p_2)^2 \left( \log\left( \frac{2\left(p^2 + p_1 \cdot p_2\right)}{p^2} \right) \right) \\
&\quad + p^2 (p_1 \cdot p_2) \left( 2 \log\left( \frac{2(p^2 + p_1 \cdot p_2)}{p^2} \right) - 1 \right) + p^4 \Bigg] \\
&\quad - p^2 (p_1 \cdot p_2) \left(p^2 + 2(p_1 \cdot p_2)\right) C_0\left(p^2, p^2, 2p^2 + 2p_1 \cdot p_2\right) \Bigg\}, \\
\bar{B}_7 &= \frac{i}{4\pi^2 (p^2 - p_1 \cdot p_2)^2 (p^2 + p_1 \cdot p_2)} \Bigg\{ p^4 \left(p^2 + 2(p_1 \cdot p_2)\right) C_0\left(p^2, p^2, 2p^2 + 2p_1 \cdot p_2\right) \\
&\quad + (p_1 \cdot p_2)^2 - 2p^4 \left( \log\left( \frac{2\left(p^2 + p_1 \cdot p_2\right)}{p^2} \right) \right) - p^2 (p_1 \cdot p_2) \left( \log\left( \frac{2\left(p^2 + p_1 \cdot p_2\right)}{p^2} \right) + 1 \right) \Bigg\},
\end{aligned} \tag{7.6.10}$$

These results are in complete agreement with the zero-density analysis presented in Chapter 4, specifically under the kinematical conditions defined by eq. (7.6.9).

## 7.7 The chiral anomaly with on-shell photons

The case of the decay of the axial-vector line into two physical on-shell photons is surely one of the most interesting, since the computations are slightly simplified and one discovers, as we have already anticipated, that the anomaly pole is not affected by any $\mu$ dependent correction. In this section, we are going to illustrate this point by an explicit computation. We work in the symmetric case, using the decomposition in (7.5.6). For physical photons, we take the limit $p_1^2 = p_2^2 = p^2 = 0$ and discard all tensors containing $p_1^\mu$ or $p_2^\nu$, as these terms vanish upon contraction with the physical polarization vectors of the photons

$$\epsilon_1 \cdot p_1 = 0, \qquad \epsilon_2 \cdot p_2 = 0. \tag{7.7.1}$$

Accounting for this consideration, the tensors (7.5.7) simplifies significantly.

### 7.7.1 The longitudinal component

In order to analyze the anomaly in the correlator, we focus on the longitudinal component with respect to the axial momentum $q$. Therefore, we employ the $\Sigma$ projector, we have defined in eq. (7.4.4). Using the decomposition (7.5.6), in the case of physical on-shell photons, one obtains

$$\Sigma^\lambda_\alpha(q)\Gamma^{\alpha\mu\nu} = \frac{q^\lambda}{q^2} \left\{ \Big[ 2(p\cdot\eta) B_3 + 2(p_1 \cdot p_2) B_7 \Big] \varepsilon^{\mu\nu p_1 p_2} + B_2 \left[ \frac{p_1 \cdot p_2}{p \cdot \eta} \eta^\mu \varepsilon^{\nu p_1 p_2 \eta} - \frac{p_1 \cdot p_2}{p \cdot \eta} \eta^\nu \varepsilon^{\mu p_1 p_2 \eta} + p_1^\nu \varepsilon^{\mu p_1 p_2 \eta} - p_2^\mu \varepsilon^{\nu p_1 p_2 \eta} \right] \right\} \tag{7.7.2}$$

Comparing this expression with the perturbative computation, one discovers that $B_2$ vanishes, leaving the correlator with only one tensor structure, which can be rewritten in the following form

$$\Sigma_\alpha^\lambda(q)\,\Gamma^{\alpha\mu\nu}(p_1,p_2,\eta) = \omega_L \varepsilon^{\mu\nu p_1 p_2}\frac{q^\lambda}{q^2}. \tag{7.7.3}$$

This new form factor contains a vacuum and a finite density part

$$\omega_L = \omega_L^{(\mathrm{Vac})} + \omega_L^{(\mathrm{Dens})}, \tag{7.7.4}$$

The analysis of the zero density part is quite direct. By applying the longitudinal projector to $\Gamma_{pert}^{(\mathrm{Vac})}$ and proceeding as discussed in Section 7.6, one finds

$$\begin{aligned}\omega_L^{(\mathrm{Vac})} &= 8i(p_1\cdot p_2)H^{(0,0,0)}[p_1\cdot k] + 8i(p_1\cdot p_2)J^{(0,0,0)}[p_2\cdot k] - 8iH^{(0,0,0)}[(p_1\cdot k)(p_2\cdot k)] + 4iH^{(0,0,0)}[(k\cdot k)(p_1\cdot k)]\\ &- 4iH^{(0,0,0)}[(k\cdot k)(p_2\cdot k)] + 8iH^{(0,0,0)}[(p_2\cdot k)(p_2\cdot k)] - 8iJ^{(0,0,0)}[(p_1\cdot k)(p_2\cdot k)] - 4iJ^{(0,0,0)}[(k\cdot k)(p_1\cdot k)]\\ &+ 8iJ^{(0,0,0)}[(p_1\cdot k)(p_1\cdot k)] + 4iJ^{(0,0,0)}[(k\cdot k)(p_2\cdot k)] - 8iJ^{(0,0,0)}[(p_2\cdot k)(p_2\cdot k)] - 8iH^{(0,0,0)}[(p_1\cdot k)(p_1\cdot k)].\end{aligned} \tag{7.7.5}$$

Such integrals are easy to compute, being ordinary Feynman integral in the vacuum case. The result is

$$\omega_L^{(\mathrm{Vac})} = -\frac{i}{2\pi^2}, \tag{7.7.6}$$

As expected, this yields the standard anomaly pole for the longitudinal component of the correlator. In contrast, the finite-density part arises from the contributions

$$\begin{aligned}\omega_L^{(\mathrm{Dens})} &= 8i(p_1\cdot p_2)H^{(1,0,0)}[p_1\cdot k] + 8i(p_1\cdot p_2)J^{(1,0,0)}[p_2\cdot k] - 8iH^{(1,0,0)}[(p_1\cdot k)(p_2\cdot k)]\\ &+ 4iH^{(1,0,0)}[(k\cdot k)(p_1\cdot k)] - 4iH^{(1,0,0)}[(k\cdot k)(p_2\cdot k)] + 8iH^{(1,0,0)}[(p_2\cdot k)(p_2\cdot k)]\\ &- 8iJ^{(1,0,0)}[(p_1\cdot k)(p_2\cdot k)] - 4iJ^{(1,0,0)}[(k\cdot k)(p_1\cdot k)] + 8iJ^{(1,0,0)}[(p_1\cdot k)(p_1\cdot k)]\\ &+ 4iJ^{(1,0,0)}[(k\cdot k)(p_2\cdot k)] - 8iJ^{(1,0,0)}[(p_2\cdot k)(p_2\cdot k)] - 8iH^{(1,0,0)}[(p_1\cdot k)(p_1\cdot k)]\\ &+ (0,1,0) + (0,0,1) + (1,0,1)\end{aligned} \tag{7.7.7}$$

One can easily see that other contributions, such as $(1,1,1)$, $(0,1,1)$, and $(1,1,0)$, vanish due to the Dirac delta function contained within the finite-density part $G_1$ of the fermion propagator. In the next section, we will demonstrate through direct computation that the contributions written in the equation above vanish as well.

### 7.7.2 Evaluation of the integrals in a special frame

The explicit evaluation of (7.7.7) is performed by choosing a frame of reference in which the $J[\dots]$ and $H[\dots]$ integrals simplify. We choose a frame where the particle of momentum $q$ and invariant mass $4p_0^2$ decays into two on-shell back-to-back photons. The heat-bath is assumed to be at rest in this frame, with $\eta^\mu = (1,0,0,0)$. Together with (7.5.1) (and $p^2 = 0$), this choice translates into the following parametrizations

$$\begin{aligned} p_1^\mu &= (p_0,0,0,p_0) & p_2^\mu &= (p_0,0,0,-p_0)\\ q^\mu &= (2p_0,0,0,0) & \eta^\mu &= (1,0,0,0).\end{aligned} \tag{7.7.8}$$

As it will become apparent soon, integrals in this frame are plagued by collinear divergences as well as IR ones, that will cancel out in the overall sum (7.7.7). As an explicit example, we sketch the evaluation of the $H^{(1,0,0)}[p_1 \cdot k]$ integral, treated in DR

$$\begin{aligned} H^{(1,0,0)}[p_1 \cdot k] &= i \int \frac{d^d k}{(2\pi)^{d-1}} k \cdot p_1 \Big( \theta(\mu - \eta \cdot k)\theta(\eta \cdot k)\delta(k^2) \Big) \frac{1}{(k-q)^2} \frac{1}{(k-p_2)^2} \\ &= i \int \frac{d^d k}{(2\pi)^{d-1}} (k_0 p_0 - |\mathbf{k}| p_0 \cos\theta) \Big( \theta(\mu - k_0) \frac{\delta(k_0 - |\mathbf{k}|)}{2|\mathbf{k}|} \Big) \frac{1}{4p_0^2 + k^2 - 4k_0 p_0} \frac{1}{k^2 - 2k_0 p_0 + 2|\mathbf{k}| p_0 \cos\theta}. \end{aligned} \tag{7.7.9}$$

After performing the energy $dk^0$ integration, we have

$$H^{(1,0,0)}[p_1 \cdot k] = \frac{i}{16 p_0} \int \frac{d^{d-1}k}{(2\pi)^{d-3}} \frac{1}{|\mathbf{k}|(|\mathbf{k}| - p^0)} \frac{1 - \cos\theta}{1 + \cos\theta} \theta(\mu - |\mathbf{k}|). \tag{7.7.10}$$

We then work with spherical coordinates in $d-1$ dimensions and write

$$H^{(1,0,0)}[p_1 \cdot k] = \frac{i}{16 p_0} \int \frac{d|\mathbf{k}|}{(2\pi)^{d-3}} \theta(\mu - |\mathbf{k}|) \frac{|\mathbf{k}|^{d-3}}{|\mathbf{k}| - p_0} \int d\theta_1 d\theta_2 ... d\theta_{d-2} \frac{1 - \cos\theta_1}{1 + \cos\theta_1} \sin^{d-3}\theta_1 \sin^{d-4}\theta_2 ... \sin\theta_{d-3}. \tag{7.7.11}$$

In the frame we have chosen, there is a factorization of the integrals, as the angular and radial integrations do not intertwine. This occurs in every integral computed in this special frame and is one of the main reasons we can evaluate the result analytically. However, if we take the photons to be off-shell, this factorization is no longer present. The radial integration is limited by the theta function. The angular part can be integrated for all variables except $\theta_1$, yielding a $\Omega(d-3)$ volume. The integral can then be reexpressed in the form

$$H^{(1,0,0)}[p_1 \cdot k] = \frac{i\Omega(d-3)}{16 p_0 (2\pi)^{d-1}} \int_0^\mu d|\mathbf{k}| \frac{|\mathbf{k}|^{d-3}}{|\mathbf{k}| - p_0} \int_{-1}^1 dt \frac{1-t}{1+t} (1-t^2)^{\frac{d}{2}-1}. \tag{7.7.12}$$

Using

$$\begin{aligned} \int_{-1}^1 dt \frac{1-t}{1+t} (1-t^2)^{\frac{d}{2}-1} &= \frac{\sqrt{\pi}(d-2)}{2} \frac{\Gamma(d/2-1)}{\Gamma(d/2-1/2)} \\ \int_0^\mu d|\mathbf{k}| \frac{|\mathbf{k}|^{d-3}}{|\mathbf{k}| - p^0} &= \frac{\mu^{d-2}}{(d-2)p_0} {}_2F_1(1, d-2, d-1, \mu/p_0) \\ \Omega(d-3) &= \frac{\pi^{d/2-3/2}}{\Gamma(d/2-1/2)} \end{aligned} \tag{7.7.13}$$

where ${}_2F_1$ is an Hypergeometric function, we finally obtain

$$H^{(1,0,0)}[p_1 \cdot k] = -\frac{2^{-d-4} \pi^{-d/2} \mu^{d-2} \Gamma\left(\frac{d}{2} - 2\right) {}_2F_1\left(1, d-2; d-1; \frac{\mu}{p_0}\right)}{p_0^2 \Gamma\left(\frac{d-1}{2}\right)^2}. \tag{7.7.14}$$

This integral is singular in the limit $d \to 4$ due to collinear divergences. Notice that we need to require that $p_0 > \mu$ in order to guarantee the convergence of all the contributions. This requirement, in this case, is necessary only at the intermediate stage, since all the contributions, as already mentioned, add up to zero.
The result of each integral appearing in eq. (7.7.7) is reported in Appendix D.4. One could argue, looking at the table of integrals, that the possibility of residual IR divergences cannot be ruled out a priori. We now demonstrate how in (7.7.7) such divergences do not intervene in the overall sum.
From (D.12), it is apparent that the only cases in which IR divergences could be present are for $J^{(0,1,0)}[p_2 \cdot k]$, $H^{(0,1,0)}[p_1 \cdot k]$, $J^{(0,1,0)}[(p_2 \cdot k)^2]$ and $H^{(0,1,0)}[(p_1 \cdot k)^2]$. The IR divergent part of (7.7.7) thus can be identified in the contributions

$$\omega_{IR} = 8i\,(p_1 \cdot p_2)(H_{IR}^{(0,1,0)}[p_1 \cdot k] + J_{IR}^{(0,1,0)}[p_2 \cdot k]) - 8i\,(H_{IR}^{(0,1,0)}[(p_1 \cdot k)^2] + J_{IR}^{(0,1,0)}[(p_2 \cdot k)^2]). \tag{7.7.15}$$

In our frame, $p_1 \cdot p_2 = 2p_0^2$ and $J^{(0,1,0)}[p_2 \cdot k] = H^{(0,1,0)}[p_1 \cdot k]$, $J^{(0,1,0)}[(p_2 \cdot k)^2] = H^{(0,1,0)}[(p_1 \cdot k)^2]$. The IR divergent part of these integrals takes the form

$$\omega_{IR} = \frac{8i}{(2\pi)^{d-1}} \left( \underbrace{-\frac{2p_0^2}{4} \int_0^\mu d|\mathbf{k}||\mathbf{k}|^{d-5} \int_{-1}^1 dt\, \frac{(1-t^2)^{\frac{d}{2}-1}}{1-t^2}}_{2\,(p_1 \cdot p_2)\,H^{(0,1,0)}[p_1 \cdot k]} + \underbrace{\frac{p_0^2}{2} \int_0^\mu d|\mathbf{k}||\mathbf{k}|^{d-5} \int_{-1}^1 dt\, \frac{(1-t^2)^{\frac{d}{2}-1}}{1-t^2}}_{2\,H^{(0,1,0)}[(p_1 \cdot k)^2]} \right) \tag{7.7.16}$$

and therefore vanishes.
Using the integrals in the Appendix D.4, one is able to show that the whole (7.7.7) combination vanishes in the limit $d \to 4$, leading to

$$\omega_L = \omega_L^{(\mathrm{Vac})} + \omega_L^{(\mathrm{Dens})} = \omega_L^{(\mathrm{Vac})} = -\frac{i}{2\pi^2}, \tag{7.7.17}$$

The longitudinal part of the correlator Γ, at least in the on-shell photon case, is determined by an anomaly pole that remains uncorrected at finite fermion density. By covariance, this result is valid in any reference frame. Therefore, one can write

$$\langle \partial_\mu J_A^\mu \rangle = a_1 F\tilde{F} \tag{7.7.18}$$

in flat space, with no extra corrections due to density effects.
As we will demonstrate, for on-shell photons, the entire diagram is uniquely characterized by the anomaly pole, since the transverse part vanishes even at finite density. The solution of the correlator in this case coincides with the one we found in Chapter 4, which satisfies conformal symmetry. However, in the case of off-shell photons, finite density corrections will appear in the transverse part of the correlator.

## 7.8 Finite density corrections and the transverse sector of Γ

As we have shown in the previous section by an explicit computation in the on-shell case, the chiral anomaly - and hence the longitudinal component of Γ - is protected from finite density corrections. This statement has been also proven in the offshell case by other analyses at finite temperature and density [46, 169]. Therefore, we are now going to impose this condition on the off-shell expression (7.5.6) of the chiral interaction, in order to reduce the number of form factors in our parametrization. Specifically, we impose the following axial Ward identity

$$q_\lambda \Gamma^{\lambda\mu\nu} = q_\lambda \Gamma^{(\mathrm{Vac})\,\lambda\mu\nu} = \omega_L^{(\mathrm{Vac})} \varepsilon^{\mu\nu p_1 p_2}. \tag{7.8.1}$$

This implies that the pure finite density corrections contained in $\Gamma^{(\text{Dens})}$ are transverse with respect to the axial-vector channel, leading to

$$\Gamma^{\lambda\mu\nu} = \omega_L \frac{q^\lambda}{q^2} \varepsilon^{\mu\nu p_1 p_2} + \Gamma_T^{\lambda\mu\nu} = \omega_L^{(\text{Vac})} \frac{q^\lambda}{q^2} \varepsilon^{\mu\nu p_1 p_2} + \Gamma_T^{(\text{Vac})\,\lambda\mu\nu} + \Gamma_T^{(\text{Dens})\,\lambda\mu\nu}. \tag{7.8.2}$$

We now focus on the transverse part $\Gamma_T^{\lambda\mu\nu}$ which can also be decomposed as in eq. (7.5.6) but with the following additional constraint

$$\begin{aligned} 0 = q_\lambda \Gamma_T^{\lambda\mu\nu} &= \Big(2B_6 p^2 + 2B_3 (p\cdot\eta) + 2B_7 (p_1\cdot p_2)\Big)\varepsilon^{\mu\nu p_1 p_2} \\ &+ \varepsilon^{\mu p_1 p_2 \eta}\bigg[\Big(B_4 p^2 + B_2\Big) {p_1}^\nu + {p_2}^\nu \left(-B_5 (p\cdot\eta) - B_4\, p_1\cdot p_2 + B_1\right) \\ &\qquad + \eta^\nu \Big( - B_1 \frac{p^2}{p\cdot\eta} + B_5 p^2 - B_2 \frac{p_1\cdot p_2}{p\cdot\eta}\Big)\bigg] \\ &+ \varepsilon^{\nu p_1 p_2 \eta}\bigg[{p_1}^\mu \left(B_5\, p\cdot\eta + B_4 p_1\cdot p_2 - B_1\right) + \Big(-B_4 p^2 - B_2\Big) {p_2}^\mu \\ &\qquad + \eta^\mu \Big(B_1 \frac{p^2}{p\cdot\eta} - B_5 p^2 + B_2 \frac{p_1\cdot p_2}{p\cdot\eta}\Big)\bigg]. \end{aligned} \tag{7.8.3}$$

The standard procedure, that we have used for the vector WIs, in this case leads us to the relations

$$\begin{aligned} B_1 &= B_5\, p\cdot\eta + B_4\, p_1\cdot p_2, \\ B_2 &= -B_4 p^2, \\ B_3 &= -\frac{1}{p\cdot\eta}\Big(B_6 p^2 + B_7\, p_1\cdot p_2\Big). \end{aligned} \tag{7.8.4}$$

The final form of the decomposition (for $p_1^2 = p_2^2 \equiv p^2$, $p_1\cdot\eta = p_2\cdot\eta$) can then be written as

$$\Gamma_{\text{sym}}^{\lambda\mu\nu} = \Gamma_L^{\lambda\mu\nu} + \Gamma_T^{\lambda\mu\nu} = \omega_L^{(\text{Vac})} \frac{q^\lambda}{q^2} \varepsilon^{\mu\nu p_1 p_2} + \hat{B}_1 \chi_1^{\lambda\mu\nu} + \hat{B}_2 \chi_2^{\lambda\mu\nu} + \hat{B}_3 \chi_3^{\lambda\mu\nu} + \hat{B}_4 \chi_4^{\lambda\mu\nu}, \tag{7.8.5}$$

where we have defined new form factors ($\hat{B}$) corresponding to the recombined structures

$$\begin{aligned} \chi_1^{\lambda\mu\nu} &= \bigg(\frac{{p_2}^\mu {p_2}^\nu \varepsilon^{\lambda p_1 p_2 \eta}}{p^2} - \frac{{p_1}^\mu {p_1}^\nu \varepsilon^{\lambda p_1 p_2 \eta}}{p^2} - \frac{{p_1}^\mu (p_1\cdot p_2)\, \varepsilon^{\lambda\nu p_1 \eta}}{p^2} - \frac{{p_1}^\mu (p_1\cdot p_2)\, \varepsilon^{\lambda\nu p_2 \eta}}{p^2} \\ &\quad - \frac{{p_2}^\nu (p_1\cdot p_2)\, \varepsilon^{\lambda\mu p_1 \eta}}{p^2} - \frac{{p_2}^\nu (p_1\cdot p_2)\, \varepsilon^{\lambda\mu p_2 \eta}}{p^2} + {p_2}^\mu \varepsilon^{\lambda\nu p_1 \eta} + {p_1}^\nu \varepsilon^{\lambda\mu p_2 \eta} + {p_1}^\nu \varepsilon^{\lambda\mu p_1 \eta} + {p_2}^\mu \varepsilon^{\lambda\nu p_2 \eta}\bigg), \\ \chi_2^{\lambda\mu\nu} &= \bigg(-\frac{{p_1}^\mu (k\cdot\eta)\, \varepsilon^{\lambda\nu p_1 p_2}}{p^2} + \frac{{p_2}^\nu (k\cdot\eta)\, \varepsilon^{\lambda\mu p_1 p_2}}{p^2} + \eta^\mu \varepsilon^{\lambda\nu p_1 p_2} - \eta^\nu \varepsilon^{\lambda\mu p_1 p_2}\bigg), \\ \chi_3^{\lambda\mu\nu} &= \bigg({p_2}^\nu (k\cdot\eta)\, \varepsilon^{\lambda\mu p_1 \eta} + {p_1}^\mu (k\cdot\eta)\, \varepsilon^{\lambda\nu p_2 \eta} - p^2 \eta^\mu \varepsilon^{\lambda\nu p_1 \eta} - p^2 \eta^\nu \varepsilon^{\lambda\mu p_1 \eta} + {p_1}^\mu (k\cdot\eta)\, \varepsilon^{\lambda\nu p_1 \eta} - p^2 \eta^\mu \varepsilon^{\lambda\nu p_2 \eta} \\ &\quad - p^2 \eta^\nu \varepsilon^{\lambda\mu p_2 \eta} + {p_2}^\nu (k\cdot\eta)\, \varepsilon^{\lambda\mu p_2 \eta} - \eta^\mu {p_2}^\nu \varepsilon^{\lambda p_1 p_2 \eta} + \eta^\nu {p_1}^\mu \varepsilon^{\lambda p_1 p_2 \eta}\bigg), \\ \chi_4^{\lambda\mu\nu} &= \bigg(-\frac{{p_1}^\mu (p_1\cdot p_2)\, \varepsilon^{\lambda\nu p_1 p_2}}{p^2} + \frac{{p_2}^\nu (p_1\cdot p_2)\, \varepsilon^{\lambda\mu p_1 p_2}}{p^2} + {p_2}^\mu \varepsilon^{\lambda\nu p_1 p_2} - {p_1}^\nu \varepsilon^{\lambda\mu p_1 p_2}\bigg). \end{aligned} \tag{7.8.6}$$

In summary, eq. (7.8.5) describes the entire off-shell vertex at nonzero $\mu$ in the symmetric case

### 7.8.1 General expressions of the form factors in the transverse sector

Although the explicit computation of the scalar integrals in the off-shell case cannot be performed analytically, it is still worthwhile to analyze the structure of the form factors in the parametrization outlined in (7.8.5). Accordingly, we present the expressions for all the form factors in terms of integrals that can be evaluated numerically, as all the integrals are finite. This finiteness is ensured by the nonzero virtualities of the external vector lines, which are sufficient to eliminate both collinear and infrared divergences. As demonstrated in the previous sections, such divergences occur only in certain integrals appearing in the on-shell case, where they cancel out in the complete expression. In the off-shell case, however, all integrals are finite. The perturbative results are

$$
\begin{aligned}
\hat{B}_1 &= -\frac{8i}{M^2(M^4-4p_0^4)}\Big[(M^6+4M^2p_0^4+16p_0^6)J[k\cdot\eta]+(4M^4p_0-16p_0^5)J[(k\cdot\eta)^2] \\
&+(-M^4-6M^2p_0^2+8p_0^4)J[k\cdot\eta\, k\cdot p_1]+(-M^4-6M^2p_0^2+8p_0^4)J[k\cdot\eta\, k\cdot p_2] \\
&+(-3M^4p_0-2M^2p_0^3-8p_0^5)J[k\cdot p_1]-3M^2p_0J[k^2\, k\cdot p_1]+7M^2p_0J[(k\cdot p_1)^2] \\
&-4M^2p_0J[k\cdot p_1\, k\cdot p_2]+(3M^4p_0+2M^2p_0^3-8p_0^5)J[k\cdot p_2]+M^2p_0J[(k\cdot p_2)^2]\Big], \\
\hat{B}_2 &= \frac{8i}{M^2(2p_0^2-M^2)}\Big[(4p_0^4-M^4)J[k\cdot\eta](+(M^2-2p_0^2)J[k\cdot\eta\, k\cdot p_1]+(M^2-2p_0^2)J[k\cdot\eta\, k\cdot p_2] \\
&+(M^2p_0+2p_0^3)J[k\cdot p_1]+p_0J[k^2\, k\cdot p_1]-p_0J[(k\cdot p_1)^2] \\
&+(-M^2p_0-2p_0^3)J[k\cdot p_2]+p_0J[(k\cdot p_2)^2])\Big], \\
\hat{B}_3 &= \frac{8i}{M^4\left(M^2-2p_0^2\right)}\Big[\left(4M^4p_0-16p_0^5\right)J[k\cdot\eta]+\left(16p_0^4-4M^4\right)J[(k\cdot\eta)^2]+\left(4M^2p_0-8p_0^3\right)J[k\cdot\eta\, k\cdot p_1] \\
&+\left(4M^2p_0-8p_0^3\right)J[k\cdot\eta\, k\cdot p_2]+8p_0^4J[k\cdot p_1]+M^2J[k^2\, k\cdot p_2] \\
&-2M^2J[(k\cdot p_1)^2]+2M^2J[k\cdot p_1\, k\cdot p_2]+\left(-2M^4-4M^2p_0^2+8p_0^4\right)J[k\cdot p_2]\Big)\Big], \\
\hat{B}_4 &= -\frac{8i}{M^2(M^2-2p_0^2)^2}\Big[(3M^4p_0-8M^2p_0^3+4p_0^5)J[k\cdot\eta]+(4M^2p_0^2-2M^4)J[(k\cdot\eta)^2] \\
&+(M^2p_0-2p_0^3)J[k\cdot\eta\, k\cdot p_1]+(M^2p_0-2p_0^3)J[k\cdot\eta\, k\cdot p_2] \\
&+(-M^4+M^2p_0^2+2p_0^4)J[k\cdot p_1]+(p_0^2-M^2)J[k^2\, k\cdot p_1] \\
&+(M^2-p_0^2)J[(k\cdot p_1)^2]+(-M^4+3M^2p_0^2-2p_0^4)J[k\cdot p_2]+(p_0^2-M^2)J[(k\cdot p_2)^2]\Big].
\end{aligned}
\tag{7.8.7}
$$

Notice that, in the expressions above, the integrals $H[\dots]$ are not present due to relations such as $H[p_1\cdot k]=J[p_2\cdot k]$. Some additional details on the manipulations of these integrals are given in Appendix D.5.

### 7.8.2 The on-shell case: $\mu$ independence of the amplitude $\Gamma_T$

In this section, we compute the integrals in eq. (7.8.7) in the case of on-shell photons in the frame of reference (7.7.8). Performing the $p^2\to 0$ limit, we obtain

$$
\hat{B}_2 = -\frac{4i}{(p_1\cdot p_2)(p_1\cdot p_2-2(k\cdot\eta)^2)}\Big(-(H[\eta\cdot k]+J[\eta\cdot k])(p_1\cdot p_2)^2
$$

$$
\begin{aligned}
&+(p_1\cdot p_2)\Big(-(k\cdot\eta)(-H[p_1\cdot k]+H[p_2\cdot k]+H[k\cdot k]+J[p_1\cdot k]-J[p_2\cdot k]+J[k\cdot k])\\
&+H[(p_1\cdot k)(\eta\cdot k)]+H[(p_2\cdot k)(\eta\cdot k)]+J[(p_1\cdot k)(\eta\cdot k)]+J[(p_2\cdot k)(\eta\cdot k)]\Big)\\
&+(k\cdot\eta)\,(H[(k\cdot k)(p_1\cdot k)]-H[(p_1\cdot k)^2]-H[(k\cdot k)(p_2\cdot k)]+H[(p_2\cdot k)^2]\\
&-J[(k\cdot k)(p_1\cdot k)]+J[(p_1\cdot k)^2]+J[(k\cdot k)(p_2\cdot k)]-J[(p_2\cdot k)^2]\Big),
\end{aligned}
\tag{7.8.8}
$$

$$
\hat{B}_3=0. \tag{7.8.9}
$$

$$
\hat{B}_4=\hat{B}_2, \tag{7.8.10}
$$

$\hat{B}_1$ does not play any role in the case of physical photons, given that $t_1^{\lambda\mu\nu}$ is always perpendicular to the photon polarization $\epsilon^\mu$ and $\epsilon^\nu$.
In the reference frame (7.7.8), the denominator of the equation (7.8.8) happens to be singular, since

$$
p_1\cdot p_2=2\,(p\cdot\eta)^2. \tag{7.8.11}
$$

To overcome this singularity we slightly modify the heat-bath velocity by setting $\eta=(1+\epsilon,0,0,0)$, with $\epsilon=d-4$. The denominator in (7.8.8) is now proportional to $1/(2p_0^2\epsilon)$. This allows us to proceed with the analytic computation of the integrals. For example, in this new reference frame a typical integral takes the form

$$
\begin{aligned}
H^{(1,0,0)}[p_1\cdot k] &= i\int\frac{d^dk}{(2\pi)^{d-1}}\,k\cdot p_1\Big(\theta(\mu-\eta\cdot k)\,\theta(\eta\cdot k)\,\delta(k^2)\Big)\frac{1}{(k-q)^2}\frac{1}{(k-p_2)^2}\\
&= i\int\frac{d^dk}{(2\pi)^{d-1}}\,(k_0\,p_0-|\mathbf{k}|p_0\cos\theta)\Big(\theta(\mu-k_0[1+\epsilon])\,\frac{\delta(k_0-|\mathbf{k}|)}{2|\mathbf{k}|}\Big)\\
&\quad\times\frac{1}{4p_0^2+k^2-4k_0p_0}\,\frac{1}{k^2-2k_0p_0+2|\mathbf{k}|p_0\cos\theta}.
\end{aligned}
\tag{7.8.12}
$$

The next step, in this example, is to perform the $k^0$ integration

$$
H^{(1,0,0)}[p_1\cdot k]=\frac{i}{16\,p_0}\int\frac{d^{d-1}k}{(2\pi)^{d-3}}\,\frac{1}{|\mathbf{k}|\,(|\mathbf{k}|-p^0)}\,\frac{1-\cos\theta}{1+\cos\theta}\,\theta(\mu-|\mathbf{k}|[1+\epsilon]) \tag{7.8.13}
$$

and use spherical coordinates in $d-1$ dimensions. The integral can be re-expressed in the form

$$
H^{(1,0,0)}[p_1\cdot k]=\frac{i\Omega(d-3)}{16\,p_0\,(2\pi)^{d-1}}\int_0^{\mu/[1+\epsilon]}d|\mathbf{k}|\frac{|\mathbf{k}|^{d-3}}{|\mathbf{k}|-p_0}\int_{-1}^{1}dt\,\frac{1-t}{1+t}\,(1-t^2)^{\frac{d}{2}-1}. \tag{7.8.14}
$$

If we redefine $\tilde{\mu}=\mu/[1+\epsilon]$, the computation is identical to the one encountered in eq. (7.7.12). Therefore we get

$$
H^{(1,0,0)}[p_1\cdot k]=-\frac{2^{-d-4}\,\pi^{-d/2}\,\tilde{\mu}^{d-2}\,\Gamma\left(\frac{d}{2}-2\right)\,{}_2F_1\left(1,d-2;d-1;\frac{\tilde{\mu}}{p_0}\right)}{p_0^2\,\Gamma\left(\frac{d-1}{2}\right)^2}. \tag{7.8.15}
$$

It is easy to show that the linear combinations of integrals in (7.8.8) scale as $\sim O((d-4)^2)$, giving

$$\tilde{B}_2 = \frac{1}{2p_0^2\epsilon} O((d-4)^2). \tag{7.8.16}$$

This demonstrates that the transverse part of the Γ interaction at finite density remains unaltered by the chemical potential in the on-shell limit.
As shown in a previous section, the longitudinal part of Γ corresponds to its vacuum part as well. Consequently, the entire on-shell interaction does not receive any finite-density corrections

$$\Gamma^{(\text{Dens})} = 0. \tag{7.8.17}$$

The entire interaction is therefore determined by its vacuum contribution, which reduces to the anomalous part in the on-shell limit

$$\Gamma^{\lambda\mu\nu} = \Gamma^{(\text{Vac})\,\lambda\mu\nu} = -\frac{i}{2\pi^2}\varepsilon^{\mu\nu p_1 p_2}\frac{q^\lambda}{q^2}. \tag{7.8.18}$$

At the level of the on-shell effective action, the implication is that it retains its exact form

$$\mathcal{S}_{eff} \sim \frac{e^2}{2\pi}\int d^4x\, d^4y\, \partial\cdot A\, \square^{-1}(x,y)\, F\tilde{F}(y), \tag{7.8.19}$$

where $A_\lambda$ denotes the external axial-vector source field.

### 7.8.3 The off-shell case: scaling violations in $\mu$

We now examine the off-shell case. One might wonder whether the dependence on the chemical potential cancels out in the off-shell perturbative form factors (7.8.7), as it does in the on-shell case. In this section, we show that this is not the case. Our approach involves applying the operator $\partial/\partial\mu$ to the transverse form factors. We will show that the result does not vanish. The derivative operator acts solely on the hot propagator term, specifically affecting one of the step functions

$$\theta(\mu-\eta\cdot k), \qquad \theta(\mu-\eta\cdot(k-p_1)), \qquad \theta(\mu-\eta\cdot(k-q)). \tag{7.8.20}$$

By applying a derivative on these functions, we obtain Dirac delta functions. In the $(1,0,0)$ case, this leads to a further saturation of the integration variables, as can be seen in

$$\frac{\partial}{\partial\mu}\Big(\delta(k^2)\,\theta(\eta\cdot k)\,\theta(\mu-\eta\cdot k)\Big) = \delta(k^2)\,\theta(\eta\cdot k)\,\delta(\mu-\eta\cdot k) = \frac{1}{2|\mathbf{k}|}\,\delta(k_0-|\mathbf{k}|)\,\delta(\mu-|\mathbf{k}|) \tag{7.8.21}$$

The first delta saturates the time integral while the second delta eliminates the $|\mathbf{k}|$ one, leaving us with only angular integrals. This process happens in all integrals with only one hot propagator, while for $(1,1,0)$ cases and permutations, the additional delta generates conflicting conditions that cancel the integral altogether. We take the following case as an example

$$\begin{aligned}&\frac{\partial}{\partial\mu}\Big(\delta(k^2)\,\theta(\mu-\eta\cdot k)\,\delta\big((k-q)^2\big)\,\theta\big(\mu-\eta\cdot(k-q)\big)\,\theta(\eta\cdot k)\,\theta(\eta\cdot(k-q))\Big) = \\ &\delta(k^2)\,\delta\big((k-q)^2\big)\,\theta(\eta\cdot k)\,\theta\big(\eta\cdot(k-q)\big)\Big[\delta(\mu-\eta\cdot k)\,\theta\big(\mu-\eta\cdot(k-q)\big)+\theta(\mu-\eta\cdot k)\,\delta\big(\mu-\eta\cdot(k-q)\big)\Big]\end{aligned} \tag{7.8.22}$$

Looking at the first term of the product

$$\delta(k^2)\,\delta(\mu-\eta\cdot k)\,\delta((k-q)^2)\,\theta(\eta\cdot k)\,\theta(\eta\cdot(k-q)) \propto \delta(k^0-|\mathbf{k}|)\,\delta(\mu-|\mathbf{k}|)\,\delta\Big(-4p_0(|\mathbf{k}|-p_0)\Big) \tag{7.8.23}$$

It becomes apparent that the deltas cannot be satisfied simultaneously, and thus the entire contribution vanishes. A similar reasoning applies to the other term in equation (7.8.22), as well as to integrals of the type $(0,1,1)$, $(1,1,0)$ and $(1,1,1)$.

Using these principles, one can evaluate the derivative of the form factors (7.8.7). We present here some examples of integrals that appear in the evaluation

$$\begin{aligned}
&\frac{\partial}{\partial\mu}J^{(1,0,0)}[p_1\cdot k] = \frac{\mu}{16p_0(p_0+\mu)}\int_{-1}^{1} dt\,\frac{A-t}{B_+ + t} + \frac{p_0}{8(p_0+\mu)\sqrt{p_0^2-M^2}}\int_{-1}^{1} dt\,\frac{1}{B_+ + t},\\
&\frac{\partial}{\partial\mu}J^{(0,1,0)}[p_1\cdot k] = \frac{1}{8\sqrt{p_0^2-M^2}}\int_{-1}^{1} dt\,\frac{1}{B_- - t}\\
&\frac{\partial}{\partial\mu}J^{(0,0,1)}[p_1\cdot k] = \frac{\mu}{p_0(p_0-\mu)}\int_{-1}^{1} dt\,\frac{A-t}{B_- + t}\\
&\frac{\partial}{\partial\mu}J^{(1,0,0)}[(p_1\cdot k)^2] = \frac{\mu^2\sqrt{p_0^2-M^2}}{16p_0(p_0-\mu)}\int_{-1}^{1} dt\,\frac{(A-t)^2}{B_- + t}\\
&\frac{\partial}{\partial\mu}J^{(0,1,0)}[(p_1\cdot k)^2] = \frac{\mu}{8}\int_{-1}^{1} dt\,\frac{(A-t+\frac{M^2}{\mu\sqrt{p_0^2-M^2}})^2}{(B_+ - t)(B_- - t)}\\
&\frac{\partial}{\partial\mu}J^{(0,0,1)}[(p_1\cdot k)^2] = \frac{\mu^2\sqrt{p_0^2-M^2}}{16p_0(p_0-\mu)}\int_{-1}^{1} dt\,\frac{(A-t+\frac{2p_0^2}{\mu\sqrt{p_0^2-M^2}})^2}{B_+ + t}
\end{aligned} \tag{7.8.24}$$

where

$$\begin{aligned}
A &= \frac{p_0}{\sqrt{p_0^2-M^2}}\\
B_- &= \frac{M^2}{2\mu\sqrt{p_0^2-M^2}} - A\\
B_+ &= \frac{M^2}{2\mu\sqrt{p_0^2-M^2}} + A
\end{aligned} \tag{7.8.25}$$

Performing the full computation of the derivative with respect to the chemical potential for the $\hat{B}_2$ form factor, one obtains

$$\frac{\partial}{\partial\mu}\hat{B}_2 \neq 0. \tag{7.8.26}$$

Specifically, the log independent part reads

$$\frac{\partial}{\partial\mu}\hat{B}_2 \propto \frac{\mu(-\mu M^2 + 4p_0^3 + 2\mu p_0^2)}{4\pi^3 p_0(M^2-2p_0^2)(\mu+p_0)} + \text{Log terms} \tag{7.8.27}$$

This result clearly means that the chemical potential plays an important role in the off-shell case, although its contribution is limited to the transverse sector of the interaction.

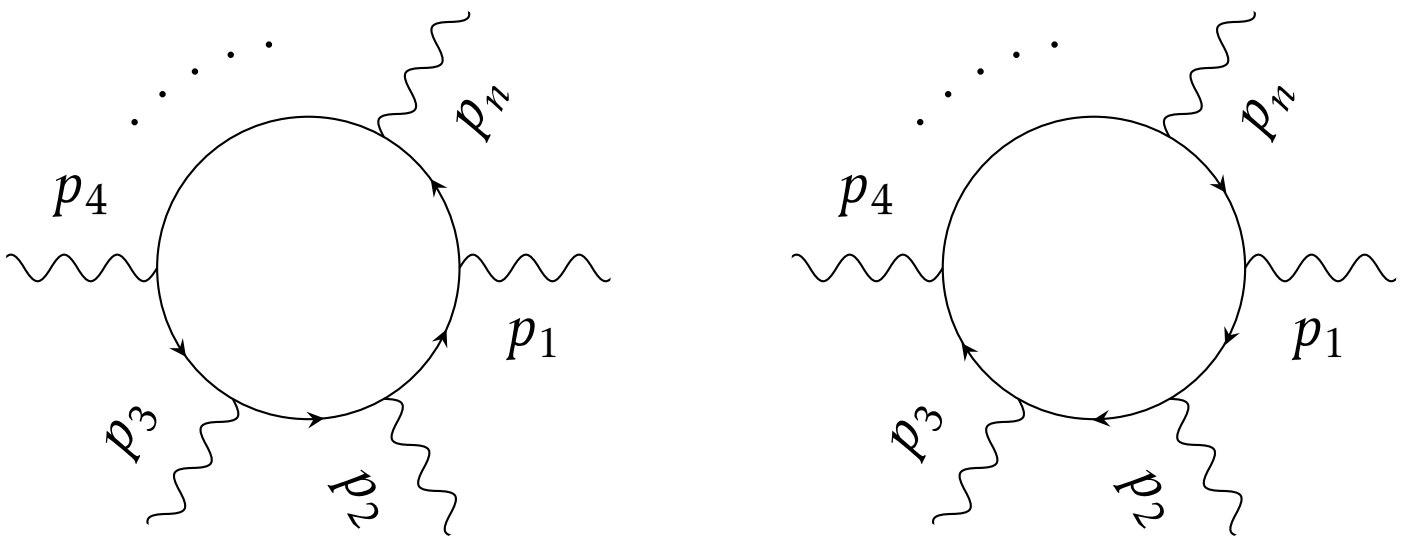

Figure 7.4: Diagrams with reversed fermion flow and an odd number of external photon lines

## 7.9 Furry's theorem and chiral current correlators

The relation between the $\langle J_V J_V J_A\rangle$ diagram and other diagrams involving $J_L$ and $J_R$ currents requires analyzing the role of charge conjugation symmetry in the presence of both chiral and vector currents. This includes examining the $\langle J_V J_V J_V\rangle$ correlator of three vector currents, which vanishes due to $C$ invariance in the ordinary vacuum sector but becomes nonzero in the presence of finite-density backgrounds.

### 7.9.1 Furry's theorem

According to Furry's theorem, Feynman diagrams containing a closed electron loop with an odd number of photon vertices vanish. In a closed loop, both electrons and positrons can "circle around." These particles interact with the electromagnetic field with opposite signs of charge, and their contributions cancel each other for an odd number of vertices. Denoting the diagrams in Fig. 7.4 by $M_a$ and $M_b$, one can show that

$$M_{\text{tot}} = M_a + M_b = 0. \tag{7.9.1}$$

However, as we will see, Furry's theorem no longer holds in the presence of a chemical potential, as the background breaks charge conjugation invariance.
We start by writing the expression of the diagrams displayed in Fig. 7.4 in the finite temperature and density case

$$\begin{aligned}
M_a(\beta,\mu) &= \int \frac{d^4k}{(2\pi)^4}\text{Tr}\Big\{S_F(k,\beta,\mu)\,\gamma_{\mu_n}\ldots S_F(k+p_1+p_2,\beta,\mu)\,\gamma_{\mu_2}\,S_F(k+p_1,\beta,\mu)\,\gamma_{\mu_1}\Big\}\\
M_b(\beta,\mu) &= \int \frac{d^4k}{(2\pi)^4}\text{Tr}\Big\{\gamma_{\mu_1}\,S_F(k-p_1,\beta,\mu)\,\gamma_{\mu_2}\,S_F(k-p_1-p_2,\beta,\mu)\ldots\gamma_{\mu_n}\,S_F(k,\beta,\mu)\Big\}
\end{aligned} \tag{7.9.2}$$

Here, we have explicitly retained the dependence of the fermion propagators on the inverse temperature $\beta$ and the chemical potential $\mu$. These diagrams are closely related to each other. To see this we make use of the charge conjugation matrix $\hat{C} = i\gamma^2\gamma^0$ with the property

$$\hat{C}\gamma_\mu\hat{C}^{-1} = -\gamma_\mu^T \tag{7.9.3}$$

Applied to the Feynman propagator at finite temperature and density, this transformation yields

$$\begin{aligned}
\hat{C}\,S_F(k,\beta,\mu)\,\hat{C}^{-1} &= \Big(k^\mu\hat{C}\,\gamma_\mu\,\hat{C}^{-1}+m\Big)\,G_F(k,\beta,\mu) = \Big(-k^\mu\gamma_\mu^T+m\Big)\,G_F(k,\beta,\mu) = \\
&\Big(-k^\mu\gamma_\mu^T+m\Big)\,G_F(-k,\beta,-\mu) = S_F(-k,\beta,-\mu)^T
\end{aligned} \tag{7.9.4}$$

Now, inserting multiple factors of $\hat{C}^{-1}\hat{C}$ into the exchanged diagram $M_b$, we get

$$M_b(\beta,\mu) = \int \frac{d^4k}{(2\pi)^4} \text{Tr}\Big\{\hat{C}^{-1}\hat{C}\,\gamma_{\mu_1}\,\hat{C}^{-1}\hat{C}\,S_F(k-p_1,\beta,\mu)\,\hat{C}^{-1}\hat{C}\,\gamma_{\mu_2}\,\hat{C}^{-1}\hat{C}\,S_F(k-p_1-p_2,\beta,\mu)\dots \\ \times \hat{C}^{-1}\hat{C}\,\gamma_{\mu_n}\,\hat{C}^{-1}\hat{C}\,S_F(k,\beta,\mu)\Big\} \tag{7.9.5}$$

Using the properties (7.9.3) and (7.9.4), we can then write

$$M_b(\beta,\mu) = \\ (-1)^n \int \frac{d^4k}{(2\pi)^4} \text{Tr}\Big\{\gamma^T_{\mu_1}\,S_F(-k+p_1,\beta,-\mu)^T\,\gamma^T_{\mu_2}\,S_F(-k+p_1+p_2,\beta,-\mu)^T \dots \gamma^T_{\mu_n}\,S_F(-k,\beta,-\mu)^T\Big\} = \\ (-1)^n \int \frac{d^4k}{(2\pi)^4} \text{Tr}\Big\{S_F(-k,\beta,-\mu)\,\gamma_{\mu_n} \dots S_F(-k+p_1+p_2,\beta,-\mu)\,\gamma_{\mu_2}\,S_F(-k+p_1,\beta,-\mu)\,\gamma_{\mu_1}\Big\}^T \tag{7.9.6}$$

Applying the change of variable $k \to -k$ in the integral, we arrive to

$$M_b(\beta,\mu) = (-1)^n M_a(\beta,-\mu) \tag{7.9.7}$$

For an odd number $n$ of vertices, this condition reduces to the Furry's theorem for any value of the temperature when $\mu = 0$

$$M_{\text{tot}}(\beta,\mu=0) = M_a(\beta,0) - M_a(\beta,0) = 0 \tag{7.9.8}$$

However in general the sum of the two diagrams $M_{\text{tot}}$ will not vanish

$$M_{\text{tot}}(\beta,\mu) = M_a(\beta,\mu) - M_a(\beta,-\mu) \tag{7.9.9}$$

The presence of a chemical potential manifestly breaks charge conjugation invariance and leads to the appearance of new processes in a perturbative expansion. Note that there is still a case where $\mu \neq 0$ and Furry's theorem is satisfied. Indeed, in the limit $\beta \to 0$ (infinite temperature) the fermion propagator (7.3.17) does no longer depend on the value of the chemical potential and therefore $M_{\text{tot}} = 0$.

### 7.9.2 Left- and right-handed chiral correlators

Using the previous results, one can extend the analysis in this chapter to correlators constructed with left- and right-handed currents, defined as

$$J_L = -ig\bar{\psi}_L\gamma^\mu\psi_L, \qquad J_R = -ig\bar{\psi}_R\gamma^\mu\psi_R, \tag{7.9.10}$$

which, as we recall, are related to the axial and vector currents in the following way:

$$J_L = \frac{1}{2}\Big(J_V - J_A\Big), \qquad J_R = \frac{1}{2}\Big(J_V + J_A\Big). \tag{7.9.11}$$

The propagator of a chiral fermion can be obtained by inserting a chiral projector $P_L$ or $P_R$ into the usual fermionic propagator

$$S_{L/R} = \int d^4k e^{-ik(x'-x)}\,P_{L/R}\left[\frac{\not{k}}{k^2} + 2i\pi\delta(k^2)\theta(k\cdot\eta)\theta(\mu - k\cdot\eta)\right] \tag{7.9.12}$$

We now consider a correlator constructed with three left-handed currents, which can be related to correlators built with vector and axial currents in the following way

$$\Delta_L \equiv \langle J_L J_L J_L \rangle = \frac{1}{8}\Big( \langle J_V J_V J_V \rangle - \langle J_V J_V J_A \rangle - \langle J_V J_A J_V \rangle - \langle J_A J_V J_V \rangle + \langle J_V J_A J_A \rangle + \langle J_A J_V J_A \rangle + \langle J_A J_A J_V \rangle - \langle J_A J_A J_A \rangle \Big) \tag{7.9.13}$$

Recalling the following relations

$$\begin{aligned} \langle J_V J_V J_V \rangle &= \frac{1}{3}\Big( \langle J_A J_A J_V \rangle + \langle J_A J_V J_A \rangle + \langle J_V J_A J_A \rangle \Big) \\ \langle J_A J_A J_A \rangle &= \frac{1}{3}\Big( \langle J_V J_V J_A \rangle + \langle J_V J_A J_V \rangle + \langle J_A J_V J_V \rangle \Big) \end{aligned} \tag{7.9.14}$$

one can re-express $\Delta_L$ in a simpler form

$$\Delta_L \equiv \frac{1}{2}\Big( \langle J_V J_V J_V \rangle - \langle J_A J_A J_A \rangle \Big) \tag{7.9.15}$$

Proceeding in a similar manner with right-handed currents, one obtains

$$\Delta_R = \frac{1}{2}\Big( \langle J_V J_V J_V \rangle + \langle J_A J_A J_A \rangle \Big) \tag{7.9.16}$$

As noted in the previous section, since we are working at finite density and temperature, the $\langle J_V J_V J_V \rangle$ correlator does not vanish in this case.
In this chapter, we have investigated the $\langle J_V J_V J_A \rangle$ and pointed out that its anomalous sector is protected from thermal corrections. A similar argument must also apply to a correlator built with three axial currents which can be written as

$$\langle J_A J_A J_A \rangle^{\lambda_1\lambda_2\lambda_3} = \frac{\omega_L}{3}\left( \frac{p_1^{\lambda_1}}{p_1^2}\varepsilon^{\lambda_2\lambda_3 p_2 p_3} + \frac{p_2^{\lambda_2}}{p_2^2}\varepsilon^{\lambda_1\lambda_3 p_1 p_3} + \frac{p_3^{\lambda_3}}{p_3^2}\varepsilon^{\lambda_1\lambda_2 p_1 p_2} \right) + \Delta_T^{\lambda_1\lambda_2\lambda_3} \tag{7.9.17}$$

where $\Delta_T^{\lambda_1\lambda_2\lambda_3}$ denotes the transverse part of the correlator, which includes all the thermal corrections. The three anomaly poles in this expression distribute the anomaly equally across the three axial-vector vertices.
The equation above is a natural consequences of (7.9.14), since each permutation of the $\langle J_V J_V J_A \rangle$ in (7.9.14) is characterized by a $\mu$-independent longitudinal sector in the axial vector channel. By symmetry, the result can be extended to the $\langle J_A J_A J_A \rangle$ case.
Similarly, recalling eq. (7.9.15) and (7.9.16), one can verify that the longitudinal sector of $\Delta_L$ and $\Delta_R$ is $\mu$-independent and we can write

$$\begin{aligned} \Delta_L &= -\frac{\omega_L}{6}\left( \frac{p_1^{\lambda_1}}{p_1^2}\varepsilon^{\lambda_2\lambda_3 p_2 p_3} + \frac{p_2^{\lambda_2}}{p_2^2}\varepsilon^{\lambda_1\lambda_3 p_1 p_3} + \frac{p_3^{\lambda_3}}{p_3^2}\varepsilon^{\lambda_1\lambda_2 p_1 p_2} \right) + \Delta_{L,T}, \\ \Delta_R &= \frac{\omega_L}{6}\left( \frac{p_1^{\lambda_1}}{p_1^2}\varepsilon^{\lambda_2\lambda_3 p_2 p_3} + \frac{p_2^{\lambda_2}}{p_2^2}\varepsilon^{\lambda_1\lambda_3 p_1 p_3} + \frac{p_3^{\lambda_3}}{p_3^2}\varepsilon^{\lambda_1\lambda_2 p_1 p_2} \right) + \Delta_{R,T} \end{aligned} \tag{7.9.18}$$

where $\Delta_{L,T}$ and $\Delta_{R,T}$ are transverse in all momenta and include the thermal corrections.

## 7.10 Conclusions

In this chapter, we have examined the chiral anomaly interaction at finite density and temperature, classifying all the tensor structures involved. The parametrization we presented is minimal, as it accounts for all relevant symmetries, as well as the Schouten and Ward identities. For the case of finite density and zero temperature, we have provided a direct perturbative identification of the corresponding form factors.

We have shown that the on-shell vertex, where the photons are physical, reduces to a massless anomaly pole, even at finite density, in a manner similar to the ordinary vacuum case. Therefore, from the perspective of the 1PI effective action, our main conclusion is that this action takes the following form

$$\mathcal{S}_{eff} \sim \frac{g^2}{2\pi} \int d^4x\, d^4y\, \partial \cdot A\, \Box^{-1}(x,y)\, F\tilde{F}(y). \tag{7.10.1}$$

This result is independent of the chemical potential. Nonlocal actions of such type cover, at least in the vacuum case, all the chiral and conformal anomaly interactions, thereby characterizing a unique trend.

In the off-shell case, the longitudinal part of the vertex is not modified by the chemical potential, while the transverse part is. In this chapter, we have provided the expression for the finite density corrections.

From the theoretical perspective, in recent years, it has become increasingly evident that chiral anomalies, associated with the breaking of classical global symmetries by quantum corrections, play a crucial role in the dynamics of fundamental interactions, not only in the vacuum case, but also in the presence of chiral chemical potentials, as shown in the case of the chiral magnetic effect. The possibility of performing experimental tests of these interactions, that could identify the nature of the pseudoparticle emerging from the virtual corrections, as predicted by the current and previous analyses [64, 73, 74, 184] should be taken very seriously at experimental level.

# Chapter 8

# The gravitational chiral anomaly at finite density and temperature

Considerable attention has been dedicated to examining the effects of finite temperature and density on the axial gauge anomaly in several contexts [1,42–47,182]. Despite the diverse approaches taken by the various authors to address this issue, a unanimous consensus emerges: such anomaly remains insensitive to corrections from finite temperature and density.
In the previous chapter, we have investigated the general structure of the chiral anomaly vertex $\langle J_V J_V J_A \rangle$ in the presence of chemical potentials in perturbation theory. We have classified the minimal number of tensorial structures for the parameterization of the correlator and we have provided a direct perturbative identification of their corresponding form factors. Moreover, when the photons are on-shell, we have shown that the entire correlator reduces to the longitudinal anomalous sector.
The gravitational axial anomaly has yet to be explored in the context of finite temperature and density. In this chapter, we aim to fill this gap by investigating this interaction. Our approach involves utilizing the real-time Green's function method [185]. Through this technique, we will compute the $\langle TTJ_A \rangle$ with thermal Feynman diagrams. Our objective is to demonstrate that the gravitational axial anomaly remains unchanged, with no contributions from either density or temperature effects, through a direct analysis of perturbative contributions. These results find application in several contexts, from topological materials to the early universe plasma. They affect the decay of any axion or axion-like particle into gravitational waves, in very dense and hot environments.
Correlators influenced by chiral and conformal anomalies, as well as discrete anomalies, play a vital role in condensed matter theory, particularly in the context of topological materials [9–15,15–18]. The gravitational anomaly has been investigated, in the same context, in other interesting works [56, 57]. Crucial, in this analysis, is the correspondence between thermal stresses and gravity, as summarized by Luttinger's relation [58] connecting a gravitational potential to a thermal gradient [59, 60].
Understanding these phenomena is essential for unraveling the intricate properties of such materials. Since their dynamical contribution in the evolution of topological matter, in the realistic case, is characterized by both thermal effects and by Fermi surfaces, which break the charge conjugation $C$ invariance of the vacuum, the quantification of such corrections becomes crucial for phenomenology.

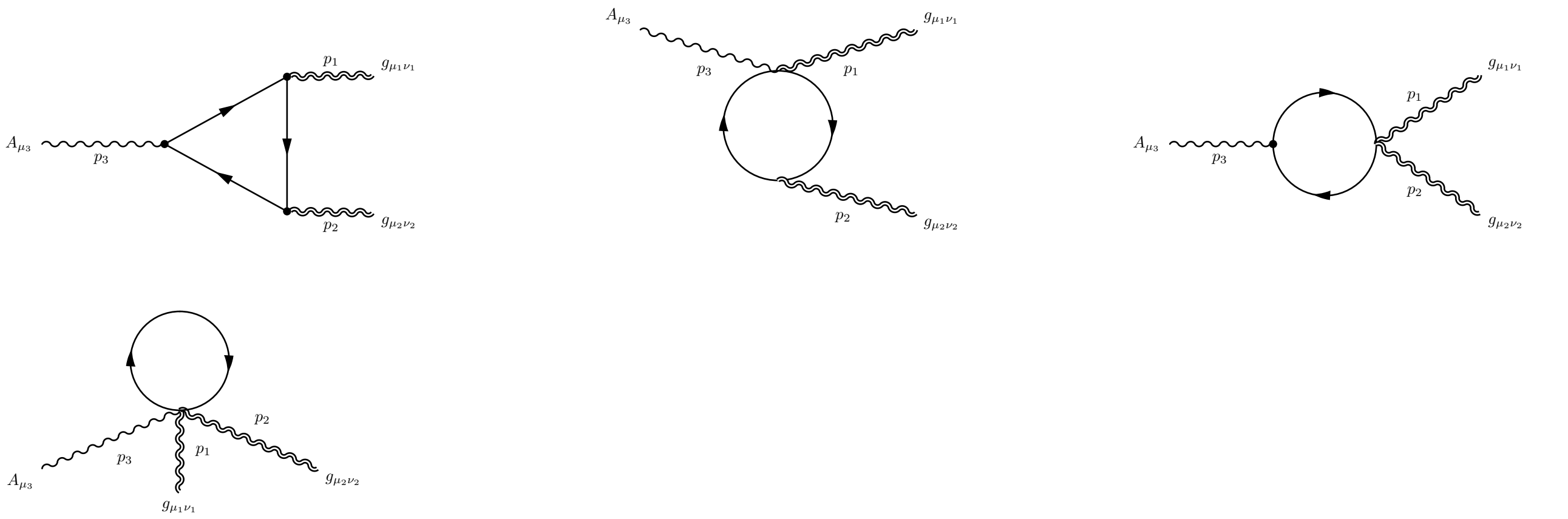

Figure 8.1: Feynman Diagrams of the three different topologies appearing in the perturbative computation.

## 8.1 The perturbative realization

In order to compute the finite density and temperature corrections to the gravitational anomaly we need to consider the same Feynman diagrams defined in the vacuum case in Section 5.4. However, now we have to replace the usual fermionic propagator with its generalization in a hot medium in the real time formulation. We use the expression

$$S_F \equiv (\not{k}+m)\,G_F = (\not{k}+m)\left\{\frac{1}{k^2-m^2} + 2\pi i\delta\left(k^2-m^2\right)\left[\frac{\theta(k_0)}{e^{\beta(E-\mu)}+1} + \frac{\theta(-k_0)}{e^{\beta(E+\mu)}+1}\right]\right\} \tag{8.1.1}$$

with the fermion mass $m$ set to zero. As we saw in the previous chapter, such expression can also be formulated covariantly by introducing the four-vector $\eta$, defining the velocity of the heat-bath.
The contribution of the triangle diagrams in Fig. 8.1 is given by

$$\begin{aligned} &V_1^{\mu_1\nu_1\mu_2\nu_2\mu_3} = \\ &-i^3\int\frac{d^d l}{(2\pi)^d}\,\mathrm{tr}\left[V^{\mu_1\nu_1}_{g\bar\psi\psi}(l-p_1,l)(\not{l}-\not{p}_1)V^{\mu_3}_{A\bar\psi\psi}(\not{l}+\not{p}_2)V^{\mu_2\nu_2}_{g\bar\psi\psi}(l,l+p_2)\not{l}\right]G_F(l-p_1)G_F(l+p_2)G_F(l) + \text{exchange}\end{aligned} \tag{8.1.2}$$

while the bubble diagrams are

$$V_2^{\mu_1\nu_1\mu_2\nu_2\mu_3} = -i^2\int\frac{d^d l}{(2\pi)^d}\,\mathrm{tr}\left[V^{\mu_1\nu_1\mu_3}_{gA\bar\psi\psi}(\not{l}+\not{p}_2)V^{\mu_2\nu_2}_{g\bar\psi\psi}(l,l+p_2)\not{l}\right]G_F(l+p_2)G_F(l) + \text{exchange} \tag{8.1.3}$$

and

$$V_3^{\mu_1\nu_1\mu_2\nu_2\mu_3} = -i^2\int\frac{d^d l}{(2\pi)^d}\,\mathrm{tr}\left[V^{\mu_1\nu_1\mu_2\nu_2}_{gg\bar\psi\psi}(p_1,p_2,l-p_1-p_2,l)(\not{l}-\not{p}_1-\not{p}_2)V^{\mu_3}_{A\bar\psi\psi}\not{l}\right]G_F(l-p_1-p_2)G_F(l). \tag{8.1.4}$$

As previously mentioned, in the case of zero temperature and density, $V_2$ vanishes. The proof relies on Lorentz symmetry, which is violated by temperature and density effects. Consequently, $V_2$ now exhibits nonvanishing thermal contributions that cannot be discarded. Furthermore, the presence of $V_2$ is essential to demonstrate the cancellation of all diagram contributions to the gravitational anomaly. Lastly, similar to the scenario of zero temperature and density, the tadpole diagram vanishes as it contains the trace of two $\gamma$'s and a $\gamma^5$.
We now decompose $G_F$ into a standard contribution to the Fermi propagator $G_0$ and the finite density and temperature corrections $G_1$

$$G_F = G_0 + G_1, \qquad G_0 = \frac{1}{k^2 - m^2}, \qquad G_1 = 2\pi i \delta\left(k^2 - m^2\right)\left[\frac{\theta(k_0)}{e^{\beta(E-\mu)}+1} + \frac{\theta(-k_0)}{e^{\beta(E+\mu)}+1}\right] \tag{8.1.5}$$

Then, we can split the $\langle TTJ_A\rangle$ correlator into four different parts depending on the number of $G_1$ contained in the loop integrals

$$\langle TTJ_A\rangle = \langle TTJ_A\rangle^{(0)} + \langle TTJ_A\rangle^{(1)} + \langle TTJ_A\rangle^{(2)} + \langle TTJ_A\rangle^{(3)} \tag{8.1.6}$$

$\langle TTJ_A\rangle^{(0)}$ represents the zero density and temperature part that was computed in the previous section, $\langle TTJ_A\rangle^{(1)}$ contains only one $G_1$ and so on. Note that the triangles diagrams contribute to all the four terms on the right-hand side of eq. (8.1.6), while the bubble diagrams to not contribute to $\langle TTJ_A\rangle^{(3)}$ since they contain two fermionic propagators $G_F$.

## 8.2 The gravitational anomaly at finite temperature and density

In this section, we examine the longitudinal anomalous sector of $J_A$, showing that it is protected from finite density and temperature effects. To achieve this, we dissect the terms associated with finite density and temperature corrections in eq. (8.1.6) separately. We will illustrate that each of these terms vanishes upon contraction with the momentum of the axial current, $p_3^{\mu_3}$. The sole surviving term is the zero density and temperature one which showcases the effect of the gravitational anomaly

$$\begin{aligned} p_{3\mu_3}\left\langle T^{\mu_1\nu_1}T^{\mu_2\nu_2}J_A^{\mu_3}\right\rangle &= p_{3\mu_3}\left\langle T^{\mu_1\nu_1}T^{\mu_2\nu_2}J_A^{\mu_3}\right\rangle^{(0)} = \\ &= 4\,i\,a_2\,(p_1\cdot p_2)\left\{\left[\varepsilon^{\nu_1\nu_2 p_1 p_2}\left(g^{\mu_1\mu_2} - \frac{p_1^{\mu_2}p_2^{\mu_1}}{p_1\cdot p_2}\right) + (\mu_1 \leftrightarrow \nu_1)\right] + (\mu_2 \leftrightarrow \nu_2)\right\}. \end{aligned} \tag{8.2.1}$$

The proof of such statement may appear miraculous, as exceedingly long expressions cancel out. Furthermore, in order to achieve this, we do not need to specify the explicit form of $G_1$, except for the fact that it contains a Dirac delta. There's also no requirement to perform the loop integral since the cancellations occur within the integrand itself.
This procedure has been previously applied to the more simple case of the gauge chiral anomaly [42, 46]. As we will see, in the case of the gravitational chiral anomaly, the computations are significantly lengthier and require the use of Schouten identities, which relate different tensor structures. We now proceed with our proof.

### 8.2.1 $\langle \mathbf{TTJ_A}\rangle^{(1)}$

We start by considering $\langle TTJ_A\rangle^{(1)}$. The terms contributing to $\langle TTJ_A\rangle^{(1)}$ contain only one thermal correction $G_1$ but with different momenta as arguments

$$\begin{aligned} &G_1(l), \quad G_1(l+p_1), \quad G_1(l-p_1), \quad G_1(l+p_2), \quad G_1(l-p_2), \\ &G_1(l+p_3), \quad G_1(l-p_3), \quad G_1(l-p_1+p_2), \quad G_1(l+p_1-p_2) \end{aligned} \tag{8.2.2}$$

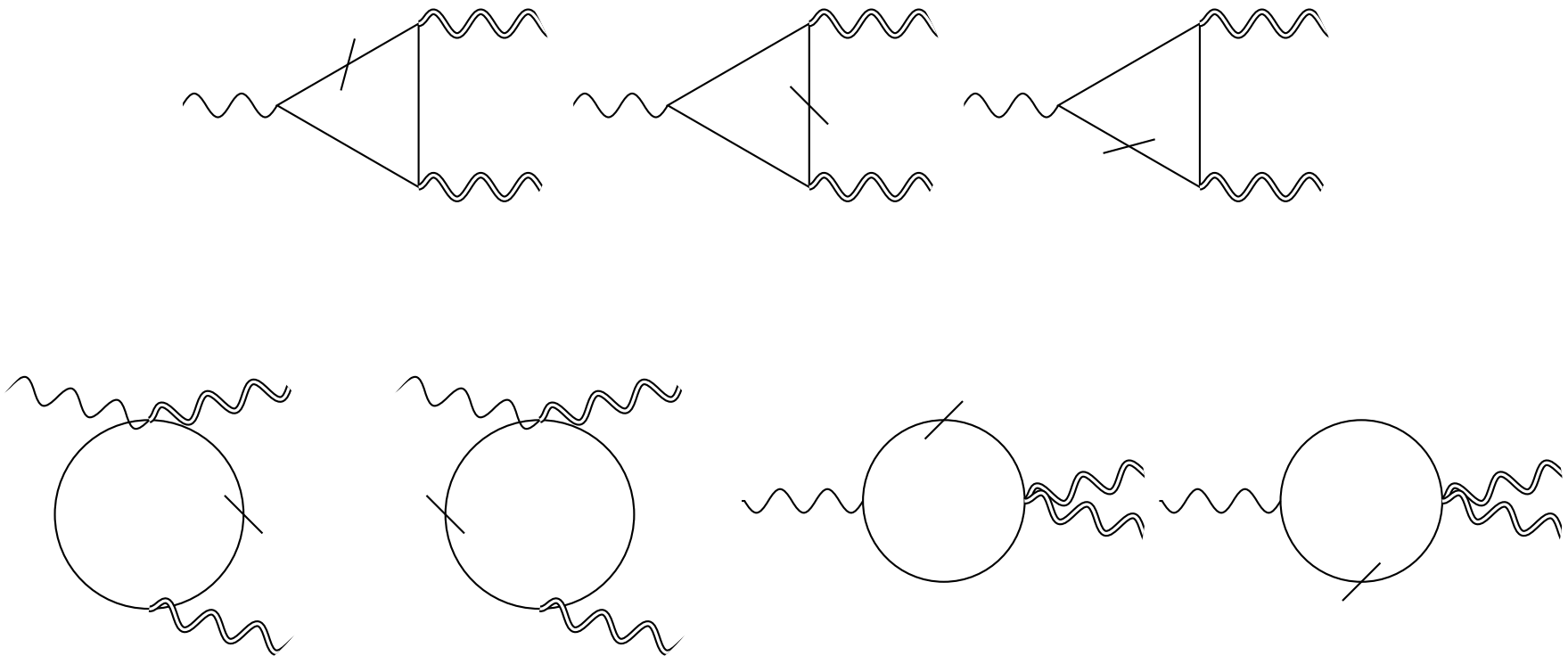

Figure 8.2: Topologies of diagrams contributing to $\langle TTJ_A\rangle^{(1)}$. The bar on the fermions' lines denotes the insertion of a hot propagator $G_1$

However, since we are dealing with finite integrals, we can perform multiple shifts in the loop momentum to ensure that the terms inside $\langle TTJ_A\rangle^{(1)}$ only have one single dependence, for example $G_1(l)$. We can then write the contracted correlator in the following form

$$p_{3\mu_3}\left\langle T^{\mu_1\nu_1}T^{\mu_2\nu_2}J_A^{\mu_3}\right\rangle^{(1)} = \int \frac{d^d l}{(2\pi)^d}\, W_1^{\mu_1\nu_1\mu_2\nu_2}(p_1,p_2,l)\,G_1(l) \tag{8.2.3}$$

The dependence on the temperature and chemical potential in the integral above is contained only in $G_1(l)$. $W_1^{\mu_1\nu_1\mu_2\nu_2}$ is a parity-odd tensor constructed with the momenta $p_1$ $p_2$ and $l$. It is symmetric under the exchange $\{\mu_1 \leftrightarrow \nu_1\}$ and $\{\mu_2 \leftrightarrow \nu_2\}$ and $\{(\mu_1,\nu_1,p_1) \leftrightarrow (\mu_2,\nu_2,p_2)\}$. The explicit initial expression for $W_1^{\mu_1\nu_1\mu_2\nu_2}$ is extremely long but one can significantly simplify it by utilizing a set of tensorial relations, known as Schouten identities, which are detailed in Appendix C.4. These identities arise from the dimensional degeneracies of tensor structures, given that we are working in $d = 4$.
Surprisingly, by applying all the Schouten identities reported in the Appendix C.4 and then setting $l^2 = 0$ due to the $\delta(l^2)$ contained in $G_1(l)$, one is able to prove that

$$0 = W_1^{\mu_1\nu_1\mu_2\nu_2}\Big|_{l^2=0}. \tag{8.2.4}$$

Therefore

$$p_{3\mu_3}\left\langle T^{\mu_1\nu_1}T^{\mu_2\nu_2}J_A^{\mu_3}\right\rangle^{(1)} = 0 \tag{8.2.5}$$

which means that the $\langle TTJ_A\rangle^{(1)}$ contributions do not modify the axial anomalous Ward identity.

### 8.2.2 $\langle \mathbf{TTJ_A}\rangle^{(2)}$

The procedure for $\langle TTJ_A\rangle^{(2)}$ is very similar to the one followed in the previous subsection. Considering our parametrization of the loop integral, the combinations in which the thermal corrections $G_1$ appear in $\langle TTJ_A\rangle^{(2)}$ are

$$\begin{gathered} G_1(l)G_1(l+p_1), \qquad G_1(l)G_1(l+p_2), \qquad G_1(l)G_1(l+p_3), \\ G_1(l)G_1(l-p_2), \qquad G_1(l)G_1(l-p_1), \qquad G_1(l+p_1)G_1(l-p_2), \qquad G_1(l-p_1)G_1(l+p_2). \end{gathered} \tag{8.2.6}$$

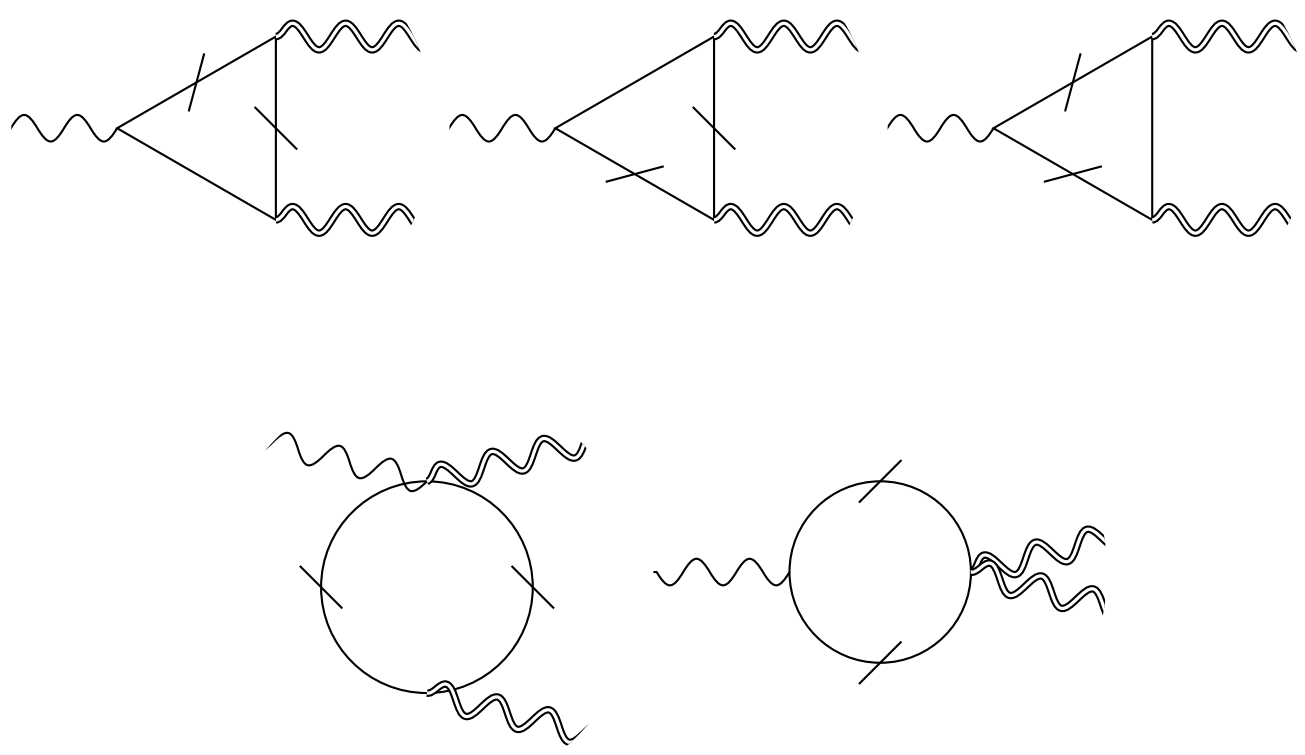

Figure 8.3: Topologies of diagrams contributing to $\langle TTJ_A\rangle^{(2)}$. They are characterized by the insertion of two hot propagators $G_1$

Since the integrals are finite, we can perform multiple shifts to the loop momentum, reducing all the combinations above to only three terms

$$p_{3\mu_3}\left\langle T^{\mu_1\nu_1}T^{\mu_2\nu_2}J_A^{\mu_3}\right\rangle^{(2)} = \int \frac{d^d l}{(2\pi)^d}\, W_{2,1}^{\mu_1\nu_1\mu_2\nu_2}(p_1,p_2,l)\, G_1(l)G_1(l+p_1) \\ + W_{2,2}^{\mu_1\nu_1\mu_2\nu_2}(p_1,p_2,l)\, G_1(l)G_1(l+p_2) + W_{2,3}^{\mu_1\nu_1\mu_2\nu_2}(p_1,p_2,l)\, G_1(l)G_1(l+p_3) \tag{8.2.7}$$

The dependence on the temperature and chemical potential in the integrals above is contained in $G_1$. The tensors $W_{2,i}^{\mu_1\nu_1\mu_2\nu_2}$ are constructed with the momenta $p_1$ $p_2$ and $l$. They are parity-odd and symmetric under the exchange $\{\mu_1 \leftrightarrow \nu_1\}$ and $\{\mu_2 \leftrightarrow \nu_2\}$. Moreover, due to the Bose symmetry, we can write

$$W_{2,1}^{\mu_1\nu_1\mu_2\nu_2}(p_1,p_2,l) = W_{2,2}^{\mu_2\nu_2\mu_1\nu_1}(p_2,p_1,l), \qquad W_{2,3}^{\mu_1\nu_1\mu_2\nu_2}(p_1,p_2,l) = W_{2,3}^{\mu_2\nu_2\mu_1\nu_1}(p_2,p_1,l) \tag{8.2.8}$$

The explicit initial expression for $W_{2,i}^{\mu_1\nu_1\mu_2\nu_2}$ is extremely long but one can significantly simplify it by utilizing the Schouten identities. Indeed, by applying all the identities reported in the Appendix C.4 and using the fact that $G_1$ contains delta functions, we can prove that all three terms in eq. (8.2.7) vanish individually

$$0 = W_{2,1}\big|_{l^2=(l+p_1)^2=0}, \qquad 0 = W_{2,2}\big|_{l^2=(l+p_2)^2=0}, \qquad 0 = W_{2,3}\big|_{l^2=(l+p_3)^2=0}. \tag{8.2.9}$$

Therefore, we have

$$p_{3\mu_3}\left\langle T^{\mu_1\nu_1}T^{\mu_2\nu_2}J_A^{\mu_3}\right\rangle^{(2)} = 0 \tag{8.2.10}$$

which means that the $\langle TTJ_A\rangle^{(2)}$ contributions do not modify the axial anomalous Ward identity.

### 8.2.3 $\langle \mathbf{TTJ_A}\rangle^{(3)}$

Only triangle diagrams contribute to $\langle TTJ_A\rangle^{(3)}$ since the bubble diagrams have only two fermionic propagators. The combination of momenta that appear as arguments of $G_1$ are

$$G_1(l)G_1(l-p_1)G_1(l+p_2), \qquad G_1(l)G_1(l+p_1)G_1(l-p_2). \tag{8.2.11}$$

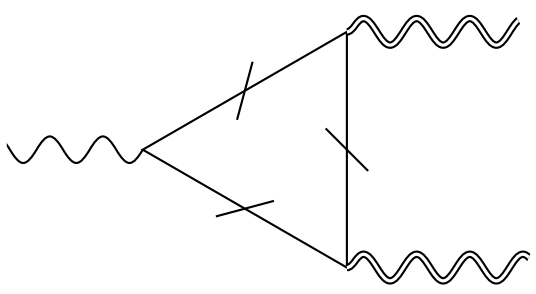

Figure 8.4: Topology of diagram contributing to $\langle TTJ_A\rangle^{(3)}$. Such diagram is characterized by the insertion of three hot propagators $G_1$

Such combination of $G_1$ can not be further reduced by shift in the loop momentum as in the previous case. Therefore, we can write

$$\begin{aligned} p_{3\mu_3}\left\langle T^{\mu_1\nu_1}T^{\mu_2\nu_2}J_A^{\mu_3}\right\rangle^{(3)} = \int \frac{d^d l}{(2\pi)^d}\, & W_{3,1}^{\mu_1\nu_1\mu_2\nu_2}(p_1,p_2,l)\, G_1(l)G_1(l-p_1)G_1(l+p_2) \\ & + W_{3,2}^{\mu_1\nu_1\mu_2\nu_2}(p_1,p_2,l)\, G_1(l)G_1(l+p_1)G_1(l-p_2) \end{aligned} \tag{8.2.12}$$

$W_{3,i}^{\mu_1\nu_1\mu_2\nu_2}$ are parity-odd tensors that depend on the momenta $p_1$ $p_2$ and $l$. They are symmetric under the exchange $\{\mu_1 \leftrightarrow \nu_1\}$ and $\{\mu_2 \leftrightarrow \nu_2\}$. Moreover, they are related to each other due to the Bose symmetry

$$W_{3,1}^{\mu_1\nu_1\mu_2\nu_2}(p_1,p_2,l) = W_{3,2}^{\mu_2\nu_2\mu_1\nu_1}(p_2,p_1,l), \tag{8.2.13}$$

Once again, we can use the Schouten identities and the fact that $G_1$ contains delta functions in order to prove that

$$0 = W_{3,1}\Big|_{l^2=(l-p_1)^2=(l+p_2)^2=0}, \qquad 0 = W_{3,2}\Big|_{l^2=(l+p_1)^2=(l-p_2)^2=0} \tag{8.2.14}$$

Therefore, we have

$$p_{3\mu_3}\left\langle T^{\mu_1\nu_1}T^{\mu_2\nu_2}J_A^{\mu_3}\right\rangle^{(3)} = 0 \tag{8.2.15}$$

This completes the proof that the longitudinal axial WI is not modified w.r.t. the vacuum part and the solution of the longitudinal equation is still given by the exchange of an anomaly pole, as shown in (8.2.1).

## 8.3 Conclusions

In this chapter, we have examined the gravitational anomaly vertex $\langle TTJ_A\rangle$ in a hot and dense medium. We have shown that the anomaly is protected from finite density and temperature corrections. It would be worthwhile to investigate how dilatations and special conformal transformations are broken in this context [186]. Moreover, it would be interesting to examine the effect of finite temperature and density on the transverse-traceless part of the $\langle TTJ_A\rangle$.
Our result has application to the decay of an axion or any axion-like particle into gravitational waves, as well as in the production of chiral currents from gravitational waves, in very dense and hot environments. Furthermore, the protection of the gravitational axial anomaly against finite density and temperature corrections presents intriguing experimental opportunities within condensed matter theory, particularly in the context of topological materials.

# Chapter 9

# Sum rules of the anomaly interactions

In this chapter, we explore the structure of the off-shell $\langle J_V J_V J_A \rangle$ and $\langle TTJ_A \rangle$ interactions, which are influenced by chiral and gravitational anomalies. These interactions are vital for understanding anomaly-driven dynamics in perturbative quantum field theory. To move beyond the conformal limit, we introduce massive fermions into our framework. We decompose the correlators into longitudinal and transverse components and provide their minimal form factor representation. Additionally, we compute the mass corrections explicitly and examine their impact on the anomaly structure.
Anomalies are inherently linked to interactions that, within the perturbative regime, manifest as poles or branch cuts depending on the external invariants, and are related to specific sum rules. Sum rules are fundamental area laws governing the absorptive amplitudes of anomaly vertices and are properties of the longitudinal form factors of these vertices.
In the case of the $\langle J_V J_V J_A \rangle$ interaction, sum rules were discussed in previous works, including [187], building on earlier analyses [114] and [188], which were formulated in specific kinematic settings. A broader exploration of these features in the effective action for both chiral and conformal anomalies can be found in [64]. Other perturbative studies, such as the proof of sum rule existence in the superconformal anomaly multiplet of the $\mathcal{N} = 1$ theory and the Konishi anomaly, are presented in [109].
Several prior studies overlap with our analysis of the $\langle J_V J_V J_A \rangle$ correlator [114,188–192], most of which use the Rosenberg representation [112], whose dispersive analysis requires polynomial subtractions in the dispersion relations. Other approaches have explored different decompositions of the $\langle J_V J_V J_A \rangle$ correlator, such as the longitudinal/transverse sector decomposition ($L/T_K$) introduced in [193] for the analysis of the muon's $g-2$ anomaly. These studies address the problem of defining the couplings of intermediate states [64,116].
In our current investigation, we extend previous analyses by examining how universal sum rules regulate the anomaly correlators across all kinematic conditions. Moving away from the conformal point, our dispersive analysis focuses on the longitudinal sectors, where the anomalous Ward identity arises. Importantly, the sum rules apply exclusively to the form factors in this sector.
We perform a detailed spectral density analysis of the relevant anomaly form factors, identifying the distinct components involved, analyzing their cancellation patterns before and after evaluating the dispersive integral, and observing their self-similar behavior. Through explicit computations, we examine the spectral flow of the anomaly form factors, revealing an "area law" for the absorptive part of the form factors as one deviates from or returns to the conformal point.
The study of the gravitational anomaly, encapsulated in the $\langle TTJ_A \rangle$ correlator, mirrors that of the chiral anomaly in the $\langle JJJ_A \rangle$ case, revealing similar structures in their sum rules and spectral flows.
By analyzing the intricate relationship between anomaly form factors, spectral densities, and sum rules, this chapter provides a comprehensive understanding of the dynamics governing anomaly in-teractions. These results have implications not only for high-energy physics, including the study of axion-like particles, but also for condensed matter systems, where similar anomaly-driven processes arise in materials with topological phases.
We will use the notations $s_1 = p_1^2$ and $s_2 = p_2^2$ interchangeably to denote the momenta of the external particles. The momentum of the axial-vector line will be represented by $q \equiv p_1 + p_2 = -p_3$.

## 9.1 The perturbative $\langle J_V J_V J_A \rangle$ in the massive case

In this section, we compute the $\langle J_V J_V J_A \rangle$ correlator. To extend our analysis from Chapter 4 beyond the conformal limit, we employ ordinary perturbation theory and we introduce a fermion with mass $m$ into the framework. The chiral anomaly interaction is obtained via the fermion loop

$$\langle J_V^{\mu_1}(p_1) J_V^{\mu_2}(p_2) J_A^{\mu_3}(p_3)\rangle = \\ -i^3 \int \frac{d^4l}{(2\pi)^4} \text{Tr}\left[(-ig\gamma^{\mu_1})\frac{i}{\slashed{l}-\slashed{p}_1-m}(-ig\gamma^{\mu_3}\gamma^5)\frac{i}{\slashed{l}+\slashed{p}_2-m}(-ig\gamma^{\mu_2})\frac{i}{\slashed{l}-m}\right] + [(p_1,\mu_1)\leftrightarrow(p_2,\mu_2)] \tag{9.1.1}$$

Regardless of the regularization used in the computation, the result must satisfy the following conservation Ward identities

$$p_{1\mu_1}\langle J_V^{\mu_1}(p_1) J_V^{\mu_2}(p_2) J_A^{\mu_3}(p_3)\rangle = 0, \qquad p_{2\mu_2}\langle J_V^{\mu_1}(p_1) J_V^{\mu_2}(p_2) J_A^{\mu_3}(p_3)\rangle = 0. \tag{9.1.2}$$

Furthermore, the divergence of the axial-vector current is given by

$$p_{3\mu_3}\langle J_V^{\mu_1}(p_1) J_V^{\mu_2}(p_2) J_A^{\mu_3}(p_3)\rangle = \frac{g^3}{2\pi^2}\left[1 + 2m^2 C_0\left(p_1^2, p_2^2, p_3^2, m^2\right)\right]\epsilon^{\mu_1\mu_2 p_1 p_2} \tag{9.1.3}$$

In the massive case, an additional term explicitly breaks chiral symmetry. Indeed, the first term on the right-hand side of the equation above corresponds to the anomaly, while the second one accounts for the explicit breaking of chiral symmetry.
A direct computation in the on-shell case ($p_1^2 = p_2^2 = 0$) shows that the entire chiral interaction simplifies to

$$\langle J_V^{\mu_1}(p_1) J_V^{\mu_2}(p_2) J_A^{\mu_3}(p_3)\rangle = \frac{g^3}{2\pi^2}\left[1 + 2m^2 C_0\left(0, 0, p_3^2, m^2\right)\right]\frac{p_3^{\mu_3}}{p_3^2}\epsilon^{\mu_1\mu_2 p_1 p_2} \tag{9.1.4}$$

Here, we have used Schouten identities and eliminated all terms containing $p_1^{\mu_1}$ and $p_2^{\mu_2}$,which vanish after contraction with the polarization vectors $\epsilon_1^{\mu_1}$ and $\epsilon_2^{\mu_2}$, respectively.
The general off-shell interaction is more involved. Using the parametrization introduced in Chapter 4, we decompose the correlator as follows

$$\langle J_V^{\mu_1}(p_1) J_V^{\mu_2}(p_2) J_A^{\mu_3}(p_3)\rangle = \langle j_V^{\mu_1}(p_1) j_V^{\mu_2}(p_2) j_A^{\mu_3}(p_3)\rangle + \langle j_V^{\mu_1}(p_1) j_V^{\mu_2}(p_2) j_{A\,\text{loc}}^{\mu_3}(p_3)\rangle \tag{9.1.5}$$

where the first term is completely transverse with respect to all momenta, and the second term corresponds to the longitudinal part associated with the anomaly contribution. The longitudinal part is given by

$$\langle j_V^{\mu_1}(p_1) j_V^{\mu_2}(p_2) j_{A\,\text{loc}}^{\mu_3}(p_3)\rangle = \Phi_0\, g^3\, \varepsilon^{p_1 p_2 \mu_1 \mu_2}\, p_3^{\mu_3} \tag{9.1.6}$$

where

$$\Phi_0 = \frac{m^2}{\pi^2}\frac{1}{p_3^2}C_0\left(p_1^2, p_2^2, p_3^2, m^2\right) + \frac{1}{2\pi^2}\frac{1}{p_3^2}, \tag{9.1.7}$$

On the other hand, the transverse part can be formally written in the following minimal form

$$\langle j_V^{\mu_1}(p_1) j_V^{\mu_2}(p_2) j_A^{\mu_3}(p_3)\rangle = \pi^{\mu_1}_{\alpha_1}(p_1)\pi^{\mu_2}_{\alpha_2}(p_2)\pi^{\mu_3}_{\alpha_3}(p_3)\Big[A_1(p_1,p_2,p_3)\,\varepsilon^{p_1p_2\alpha_1\alpha_2}p_1^{\alpha_3} + A_2(p_1,p_2,p_3)\,\varepsilon^{p_1\alpha_1\alpha_2\alpha_3} - A_2(p_2,p_1,p_3)\,\varepsilon^{p_2\alpha_1\alpha_2\alpha_3}\Big] \tag{9.1.8}$$

Defining $s_1 = p_1^2$, $s_2 = p_2^2$ and $s = p_3^2$, the explicit expressions of the form factors obtained from perturbative computations are given by

$$A_1(s_1,s_2,s,m^2) = 0. \tag{9.1.9}$$

and

$$\begin{aligned} A_2(s_1,s_2,s,m^2) = \frac{g^3 s_2}{2\pi^2\lambda^2}\Big\{ & 2\Big[\lambda m^2(s+s_1-s_2) + \lambda s_1(s+3s_2) + 3ss_1s_2(3s_1+s_2) - 3s_1s_2(s_1-s_2)^2\Big]C_0(s_1,s_2,s,m^2) + \\ & Q(s_1)s_1\Big[-5\lambda - 6s(s_1+s_2) + 6(s_1-s_2)^2\Big] - Q(s_2)\Big[\lambda(s+s_1) + 12ss_1s_2\Big] + \\ & Q(s)\Big[s\left(\lambda + 6s_1(s_1+3s_2)\right) + 6s_1(\lambda-(s_1-s_2)^2)\Big] + \lambda\,(s+s_1-s_2)\Big\}, \end{aligned} \tag{9.1.10}$$

where $\lambda$ is the Källen function and

$$Q(s) \equiv b(s)\log\left[s\frac{b(s)-1}{2m^2}+1\right] \qquad b(s) \equiv \sqrt{1-\frac{4m^2}{s}}, \tag{9.1.11}$$

### 9.1.1 Pole cancellation in $\Phi_0$ for $m \neq 0$ for on-shell photons

Let us consider the longitudinal form factor $\Phi_0$, which is the only non-vanishing contribution, in the case of on-shell photons

$$\Phi_0 = \frac{m^2}{\pi^2}\frac{1}{p_3^2}C_0\left(0,0,p_3^2,m^2\right) + \frac{1}{2\pi^2}\frac{1}{p_3^2}, \tag{9.1.12}$$

The right-hand side of this equation exhibits an apparent $1/p_3^2$ contribution, often interpreted as an anomaly pole interaction. However, $\Phi_0$ actually contains a branch cut rather than a genuine pole, as we will clarify in the following discussion. The analysis in this section aims to demonstrate the cancellation of the anomaly pole in $\Phi_0$. Specifically, one can show that there is a cancellation between the second and first terms in (9.1.12). This behavior arises from a sum rule of clear topological origin, which remains satisfied by the anomaly form factor as we move away from the conformal point.
To illustrate this, consider the explicit expression of $C_0$ in the form

$$C_0(0,0,p_3^2,m^2) \equiv C_0(p_3^2,m^2) = -\frac{1}{2m^2}\,{}_3F_2\left(1,1,1;2,\frac{3}{2};x\right), \tag{9.1.13}$$

where the ${}_3F_2$ hypergeometric function can be expanded as a series for $p_3^2/4m^2 < 1$ as

$${}_3F_2\left(1,1,1;2,\frac{3}{2};x\right) = \sum_{k=0}^{\infty}\frac{(1)_k(1)_k(1)_k}{(2)_k\left(\frac{3}{2}\right)_k}\frac{x^k}{k!}, \tag{9.1.14}$$

with $x = \frac{p_3^2}{4m^2}$. Here, $(a)_k$ denotes the Pochhammer symbol, defined as

$$(a)_k = a(a+1)(a+2)\cdots(a+k-1), \quad (a)_0 = 1. \tag{9.1.15}$$

Substituting the Pochhammer symbols for the given parameters

$$
\begin{aligned}
(1)_k &= 1\cdot 2\cdot 3\cdots(k) = k!,\\
(2)_k &= 2\cdot 3\cdot 4\cdots(k+1) = (k+1)k!,\\
\left(\frac{3}{2}\right)_k &= \frac{3}{2}\cdot\frac{5}{2}\cdot\frac{7}{2}\cdots\left(\frac{3}{2}+k-1\right).
\end{aligned}
\tag{9.1.16}
$$

we obtain the expansion

$$
C_0(p_3^2,m^2) = -\frac{1}{2m^2}\left(1+\frac{p_3^2}{12m^2}+\frac{p_3^4}{180m^4}+\mathcal{O}\left(\frac{p_3^6}{m^6}\right)\right),
\tag{9.1.17}
$$

which leads to the anomaly form factor

$$
\Phi_0 = -\frac{1}{4\pi^2}\frac{1}{12m^2} - \frac{1}{4\pi^2}\frac{p_3^2}{180m^4} + \dots
\tag{9.1.18}
$$

Notice that the $1/p_3^2$ term in (9.1.12) cancels with a similar term arising from the scalar loop $C_0$. Given the condition $p_3^2 < 4m^2$, the expansion is valid below the cut. Furthermore, the form factor vanishes as the fermion mass approaches infinity, indicating the decoupling of the heavy fermion in the perturbative loop.

## 9.2 Chiral sum rule

In this section, we begin our investigation of the sum rules governing the anomaly form factors, starting with the case of the $\langle J_V J_V J_A\rangle$ correlator. The original derivation of the sum rule for $\langle J_V J_V J_A\rangle$ has been discussed for some specific cases. Horejsi investigated the case $p_1^2 = p_2^2 = 0$ with $m \neq 0$ [189]. Other previous analyses derived the sum rule for the scenario where only a single vector line is off-shell $(p_1^2 = 0, p_2^2 \neq 0, m \neq 0)$ [114] and for $(p_1^2 = p_2^2 < 0, m = 0)$ [188].
The area law implicit in this specific feature of anomaly interactions is, at least in part, a result of the analyticity property of the vertex. It also manifests as a correlation between the longitudinal and transverse sectors of the interaction, a point that was observed in [109].
The analyticity conditions of $\Phi_0$ in the complex $q^2$ plane, where $q = -p_3 = p_1 + p_2$, are summarized by the following two formulas

$$
\oint_C \Phi_0(s,s_1,s_2,m^2)\,ds \;=\; 0,
\tag{9.2.1}
$$

and

$$
\Phi_0(q^2,s_1,s_2,m^2) \;=\; \frac{1}{2\pi i}\oint_C \frac{\Phi_0(s,s_1,s_2,m^2)}{s-q^2}\,ds,
\tag{9.2.2}
$$

where, by Jordan's lemma, the contour $C$ can be chosen to encircle the positive real axis $s \geq 0$ axis, where the function, as we are going to show, exhibits poles and a cut for $s > 4m^2$. We start by defining the general discontinuity of $\Phi_0$ for $q^2 > 0$, as

$$
\text{disc}\,\Phi_0 \;=\; \Phi_0(q^2+i\epsilon,p_1^2,p_2^2,m^2) - \Phi_0(q^2-i\epsilon,p_1^2,p_2^2,m^2)
\tag{9.2.3}
$$

and its imaginary part, the spectral density $\Delta\Phi_0$ in the $s$ variable as

$$\Delta\Phi_0(s,s_1,s_2,m^2) \equiv \mathrm{Im}\Phi_0(s,s_1,s_2,m^2) = \frac{1}{2i}\mathrm{disc}\,\Phi_0. \tag{9.2.4}$$

The vanishing of the boundary integral on the large radius allows to rewrite the dispersion relation (9.2.2) as

$$\Phi_0(q^2,p_1^2,p_2^2,m^2) = \frac{1}{\pi}\int_0^\infty \frac{\Delta\Phi_0(s,p_1^2,p_2^2,m^2)}{s-q^2}ds. \tag{9.2.5}$$

This relation holds in the complex $q^2$ plane at fixed positive values of $s_1$ and $s_2$. The computation of the spectral density of the anomaly form factors is carried out using the principal value prescription for the $1/q^2$ pole appearing in their explicit expressions

$$\mathrm{disc}\,\frac{1}{q^2} \equiv \frac{1}{q^2+i\epsilon} - \frac{1}{q^2-i\epsilon} = -2\pi i\delta(q^2). \tag{9.2.6}$$

Recalling eq. (9.1.7), we can write

$$\mathrm{disc}\,\Phi_0 \;=\; \frac{m^2}{\pi^2}\mathrm{disc}\left(\frac{C_0(p_1^2,p_2^2,q^2,m^2)}{q^2}\right) + \frac{1}{2\pi^2}\mathrm{disc}\frac{1}{q^2}, \tag{9.2.7}$$

Then, using eq. (9.2.6), we can separate the pole and continuum contributions as follows

$$\mathrm{disc}\,\Phi_0 = \frac{1}{2\pi^2}\left(-4\pi i m^2\,C_0(p_1^2,p_2^2,q^2,m^2)\,\delta(q^2) - 2\pi i\,\delta(q^2) + \frac{2m^2}{q^2}\mathrm{disc}\,C_0(p_1^2,p_2^2,q^2,m^2)\right), \tag{9.2.8}$$

showing that the spectral density exhibits two distinct localized terms at $s=0$ ($\delta$ spikes) and a continuum contribution for $q^2>4m^2$, with an interplay among these three components, as we are going to show, which is related to the property of analiticity of $\Phi_0$. We will refer to the localized components as

$$\begin{aligned} \mathrm{spk}_1 &\equiv 2\pi i\,\delta(q^2), \\ \mathrm{spk}_2 &\equiv 4\pi i m^2\,C_0(p_1^2,p_2^2,q^2,m^2)\,\delta(q^2). \end{aligned} \tag{9.2.9}$$

As we shall demonstrate, either before or after integration, these features exhibit cancellation patterns tied to the external kinematics of the vertex. Such cancellations arise either within the two spikes themselves or through the interaction of $\mathrm{spk}_2$ with the continuum. The continuum corresponds to the discontinuity in the scalar loop, $\mathrm{disc}\,C_0$.

### 9.2.1 The sum rule in the on-shell case

In this section, we derive the chiral sum rule in the simple setting in which the two vector lines are on-shell ($p_1^2=p_2^2=0$). We have already shown that the anomaly form factor, in this case, is characterized by a branch cut without any pole. This point can be reinvestigated using the dispersive approach. For this purpose, we start from the expression of the chiral interaction

$$\langle J_V^{\mu_1}(p_1)J_V^{\mu_2}(p_2)J_A^{\mu_3}(p_3)\rangle = g^3\,\Phi_0\,p_3^{\mu_3}\,\epsilon^{\mu_1\mu_2 p_1 p_2} \tag{9.2.10}$$

where $\Phi_0$ is given by

$$\Phi_0 = \frac{1}{2\pi^2 q^2}\left(1+2m^2 C_0\left(q^2,m^2\right)\right) = \frac{1}{2\pi^2 q^2}\left(1+\frac{m^2}{q^2}\log^2\left(\frac{\sqrt{\tau(q^2,m^2)}+1}{\sqrt{\tau(q^2,m^2)}-1}\right)\right), \qquad q^2>0, \tag{9.2.11}$$

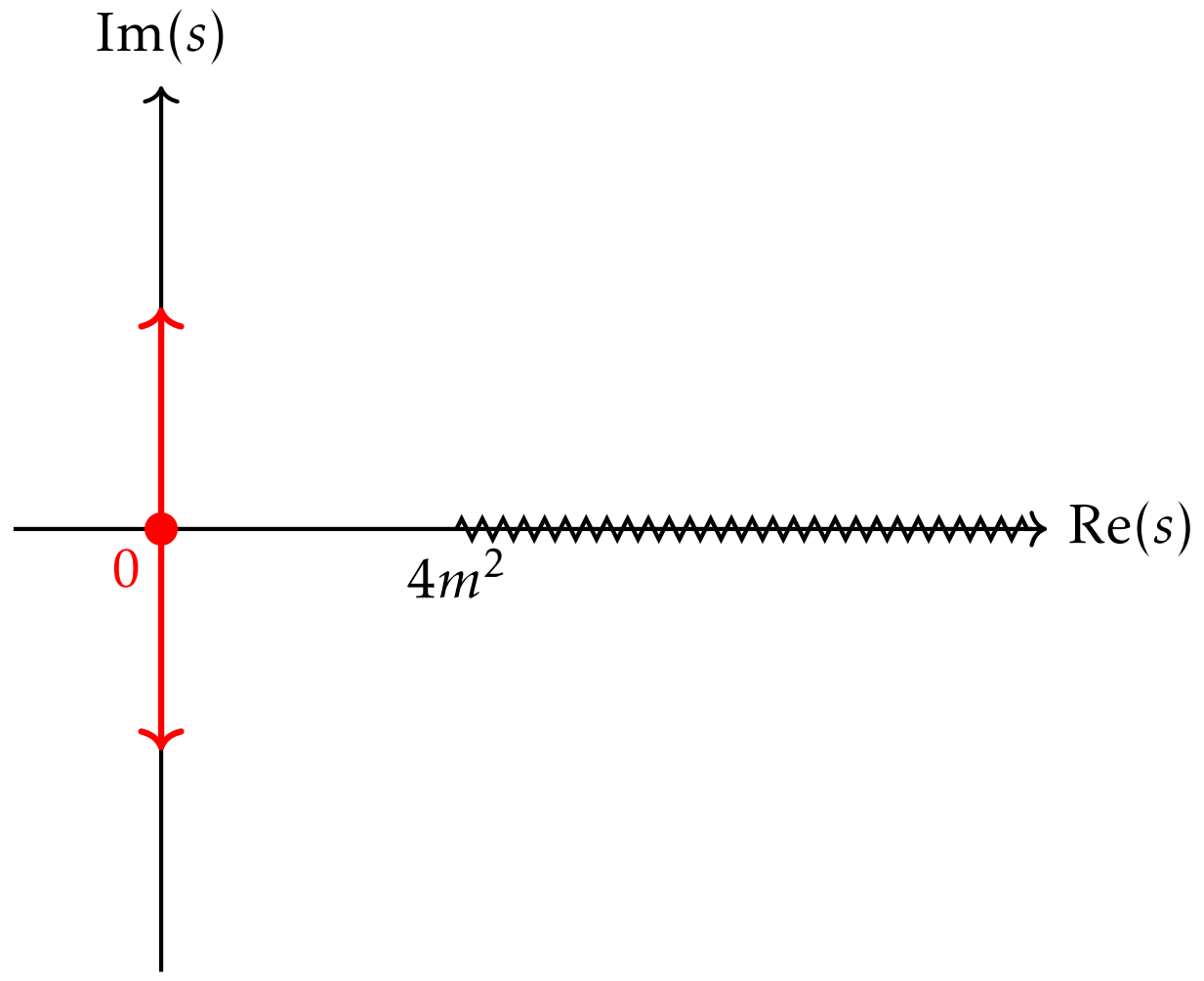

Figure 9.1: Analyticity region of the on-shell anomaly form factor in the complex $s \equiv q^2$-plane, with two bold spikes at $s = 0$ of different colour, for visual clarity. The sum rule of the anomaly form factor of the $AVV$ involves a cancelation between the two red spikes at $s = 0$, leaving only the integral over the continuum for $s > 4m^2$.

with $\tau(q^2, m^2) = 1 - \frac{4m^2}{q^2}$. The discontinuity of $\Phi_0$ takes the form

$$\text{disc}\,\Phi_0(q^2, m^2) = -\frac{1}{2\pi^2}\left(2\pi i\,\delta(q^2) + 4i\pi m^2\, C_0(q^2 \to 0, m^2)\delta(q^2) - \frac{2m^2}{q^2}\text{disc}\, C_0(q^2, m^2)\right). \tag{9.2.12}$$

There are three contributions to this function: two spikes at $q^2 \equiv s = 0$, one of constant and the other of variable strength, and a cut covering the $q^2 > 4m^2$ region (see Fig. 9.1).
It is straightforward to see that the contributions from the two spikes cancel in the expression above, leaving only the cut for $q^2 > 4m^2$ to contribute to the sum rule. Indeed, introducing the limit

$$C_0(q^2 \to 0, m^2) = -\frac{1}{2m^2} \tag{9.2.13}$$

from (9.2.12), one notices the cancellation of the two localized contributions at $s = 0$. The other contribution can be expressed as

$$\begin{aligned}\text{disc}\, C_0(q^2, m^2) &= \frac{1}{i\pi^2}\int d^4 l \frac{2\pi i\delta_+(l^2 - m^2)2\pi i\delta_+((l-q)^2 - m^2)}{(l-p_1)^2 - m^2 + i\epsilon} \\ &= \frac{2\pi}{iq^2}\log\left(\frac{1+\sqrt{\tau(q^2,m^2)}}{1-\sqrt{\tau(q^2,m^2)}}\right)\theta(q^2 - 4m^2).\end{aligned} \tag{9.2.14}$$

Therefore, one obtain

$$\text{disc}\,\Phi_0(q^2, m^2) = -2i\frac{m^2}{\pi(q^2)^2}\log\frac{1+\sqrt{\tau(q^2,m^2)}}{1-\sqrt{\tau(q^2,m^2)}}\theta(q^2 - 4m^2). \tag{9.2.15}$$

A direct integration of this function gives the sum rule formula

$$\frac{1}{\pi}\int_{4m^2}^{\infty} ds\,\Delta\Phi_0(s, 0, 0, m^2) = \frac{1}{2\pi^2}, \tag{9.2.16}$$

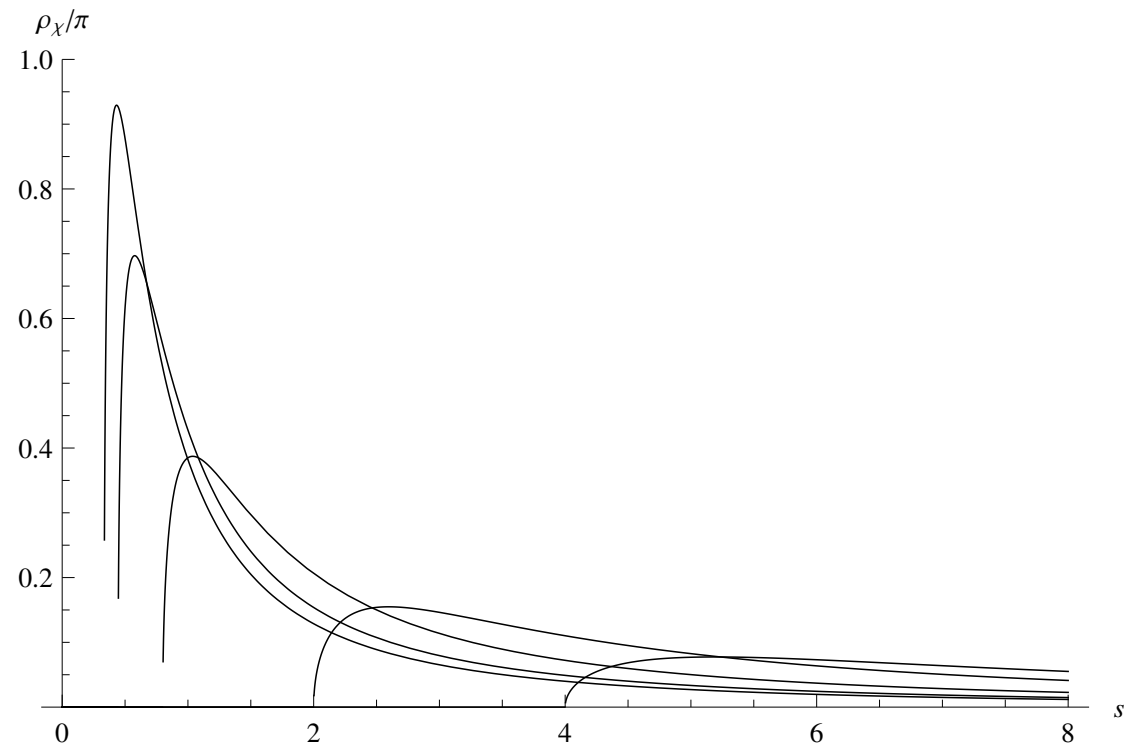

Figure 9.2: Spectral density flow in the $p_1^2 = p_2^2 = 0$ case as a function of the fermion mass $m$. The sum rule is an area law for this spectral flow as one moves towards the conformal point $m \to 0$.

In the on-shell case, being the two contributions at $s = 0$ in $\Delta\Phi_0$ identical and of opposite sign, one can identify the spectral flow of $\Delta\Phi_0$ on the cut.
We now set $m_n^2 \equiv (1/n)m^2$ and reduce the value of the mass parameters for $n \to \infty$.
The sequence $\Delta^{(n)}\Phi_0$ exhibits a remarkable behavior, ultimately converging to a Dirac delta function in the limit $m \to 0$. Specifically, the spectral flow is given by the sequence of functions

$$\lim_{m_n^2 \to 0} \Delta^{(n)}\Phi_0(s, m_n^2) = \lim_{m_n \to 0} \frac{m_n^2}{\pi s^2} \log\left(\frac{1+\sqrt{\tau(s,m_n^2)}}{1-\sqrt{\tau(s,m_n^2)}}\right)\theta(s-4m_n^2) = \frac{1}{2\pi}\delta(s), \tag{9.2.17}$$

where the convergence is understood in the distributional sense. This expression encapsulates the essence of how the spectral densities evolve as the mass parameter $m_n$ approaches zero.
Therefore, the sum rule (9.2.16) shows that if we take the on-shell massive amplitude and perform the $m \to 0$ limit we find a nonzero residue, since the cut degenerates into a pole. Figure 9.2 provides visual insight into this behavior. It displays the sequence of spectral density profiles as the mass diminishes. Notably, the area under each curve remains constant, as dictated by the sum rule governing the spectral flow. This invariance reflects a fundamental property of the anomaly structure: the total contribution from the intermediate states is conserved across the flow.
As $m \to 0$ the density becomes increasingly concentrated near the threshold $s \geq 4m^2$, signifying the replacement of the branch cut by a massless anomaly pole. A notable aspect of (9.2.17) is that it corresponds to the residue of the second term on the right-hand side of eq. (9.1.7), suggesting the possibility of interpreting this contribution as an anomaly pole. While the longitudinal form factor, $\Phi_0$, does not exhibit a pole, the sum rule for its spectral density matches the residue of the second term in Eq. (9.1.7), $\left(1/2\pi^2\right)$. The topological nature of the vertex is intrinsic to this behavior, which appears to contradict the fact that the exchanged state corresponds to a cut rather than a pole.

### 9.2.2 Asymptotic behavior and self-similarity of $\Delta\Phi_0$

While the evolution of the spectral densities $\Delta^{(n)}\Phi_0$ offers a compelling depiction of the anomaly flow, it is also interesting to investigate the large energy behavior of this form factor as $s \to \infty$. This is the limit at which the anomaly pole reappears in the longitudinal form factor. The asymptotic form of $\Phi_0(s)$, as already mentioned, is

$$\Phi_0(s) = \frac{1}{2\pi^2}\frac{1}{s} + O\left(\frac{m^2}{s^2}\ln\frac{s}{m^2}\right) \tag{9.2.18}$$

with the asymptotic behavior of the spectral density given by

$$\text{Im}\Phi_0(s) = O\left(\frac{m^2}{s^2}\ln\frac{s}{m^2}\right), \tag{9.2.19}$$

that is clearly integrable at large $s$. Concerning the behavior of this function under rescaling, the spectral density has the property of being self-similar, a property which is related to the presence of the sum rule, as we are now going to show.
In order to illustrate this point, we proceed as follows. Observe that under a rescaling of $m^2$ and $s$, $\Delta\Phi_0$ transforms as

$$\Delta\Phi_0(\lambda_0\, s, \lambda_0\, m^2) = \frac{1}{\lambda_0}\Delta\Phi_0(s, m^2), \tag{9.2.20}$$

i.e. as a homogeneous function of degree $-1$. Consequently, it satisfies the scaling equation

$$s\frac{\partial\Delta\Phi_0}{\partial s} + m^2\frac{\partial\Delta\Phi_0}{\partial m^2} - \Delta\Phi_0 = 0. \tag{9.2.21}$$

This scaling behavior is both necessary and sufficient for $\Delta\Phi_0$ to satisfy a mass-independent sum rule. To demonstrate this, assume that the on-shell spectral density satisfies the sum rule

$$\int_{4m^2}^{\infty} \Delta\Phi_0(s, m^2)\, ds = a_1, \tag{9.2.22}$$

where $a_1$ is the anomaly. Under a rescaling $m^2 \to \lambda_0\, m^2$, we then have

$$\int_{4\lambda_0\, m^2}^{\infty} \Delta\Phi_0(s, \lambda_0\, m^2)\, ds = a_1. \tag{9.2.23}$$

Next, we perform a change of variables $s = \lambda_0\, s'$, which leads to

$$\int_{4m^2}^{\infty} \Delta\Phi_0(\lambda_0 s', \lambda_0\, m^2)\lambda_0\, ds' = a_1. \tag{9.2.24}$$

Given that $\lambda_0$ is arbitrary, the dependence on $\lambda_0$ disappears if the scaling relation (9.2.20) holds. Therefore, this condition is necessary in order to satisfy the sum rule.

### 9.2.3 The sum rule in the off-shell case

As previously discussed, in the on-shell case the branch cut transitions into a pole in the $m \to 0$ limit. In this scenario, the anomaly pole transforms into a particle pole because, under this specific kinematic condition, the residue of the entire vertex as $q^2 \to 0$ is nonvanishing and equals the anomaly. We now turn to a discussion where the photons are off-shell. As observed in [194], there is no sum rule for the transverse sector. Therefore, we focus once again on the longitudinal anomalous form factor, i.e. $\Phi_0$

$$\text{disc}\,\Phi_0 = \frac{1}{2\pi^2}\left(-4\pi i m^2\, C_0(p_1^2, p_2^2, q^2 \to 0, m^2)\,\delta(q^2) - 2\pi i\,\delta(q^2) + \frac{2m^2}{q^2}\text{disc}\,C_0(p_1^2, p_2^2, q^2, m^2)\right). \tag{9.2.25}$$

The contribution from the continuum is obtained from the s-channel cut by the ordinary cutting rules

$$\mathrm{disc}\, C_0 = \frac{1}{i\pi^2}\int d^4k\, \frac{(-2\pi i)^2\delta_+(k^2-m^2)\delta_+((k-q)^2-m^2)}{(k-p_1)^2-m^2} \tag{9.2.26}$$

that identify the light-cone character of the intermediate state. This integral is evaluated in the rest frame of the invariant $q=(q_0,0,0,0)$. It corresponds to a cut contribution in which the axial-vector line decays into two back-to-back on-shell fermions. This pair can be effectively treated as a pseudoscalar intermediate state, which subsequently decays into two off-shell spin-1 particles. To evaluate the integral, we can use polar coordinates of $\vec{k}$ around the momentum $|p_1|$ together with the relations

$$\delta_+(k^2-m^2) = \frac{\delta(k_0-\sqrt{|k|^2+m^2}}{2\sqrt{|k|^2+m^2}}), \qquad \delta_+((k-q)^2-4m^2)) = \frac{1}{4q_0\sqrt{q^2-4m^2}}\delta(|k|-\frac{1}{2}\sqrt{q^2-4m^2}) \tag{9.2.27}$$

with

$$k_0 = \frac{\sqrt{q^2}}{2} \qquad |k| = \frac{1}{2}\sqrt{q^2-4m^2}. \tag{9.2.28}$$

The integral gets reduced to a simple angular form

$$\int_{-1}^{1}\frac{d\cos\theta}{A+B\cos\theta} = \frac{1}{B}\log\left(\frac{A+B}{A-B}\right) \quad \text{where} \quad A = p_1^2+p_2^2-q^2, \qquad B = 2\lambda^{1/2}\sqrt{q^2-4m^2}. \tag{9.2.29}$$

Collecting all the terms together, the spectral density takes the simple form

$$\frac{\mathrm{disc}\, C_0}{(2i)} = \Delta C_0(p_1^2,p_2^2,q^2,m^2) \equiv \mathrm{Im}\, C_0(p_1^2,p_2^2,q^2,m^2) = -\tfrac{\pi}{4s}\log\tfrac{\chi_+}{\chi_-}\,\theta(q^2-4m^2) \tag{9.2.30}$$

where we have introduced the following quantities

$$\chi_\pm = p_1^2+p_2^2-q^2 \pm 2\lambda^{1/2}\sqrt{q^2-4m^2}. \tag{9.2.31}$$

It is also convenient to introduce the variables

$$z = \frac{1}{2q^2}\left(\sqrt{\lambda}+p_1^2-p_2^2+q^2\right) \qquad \bar{z} = z = \frac{1}{2q^2}\left(-\sqrt{\lambda}+p_1^2-p_2^2+q^2\right) \tag{9.2.32}$$

with

$$w = \frac{1}{2}\left(1+\sqrt{1-\frac{4m^2}{q^2}}\right) \qquad \bar{w} = \frac{1}{2}\left(1-\sqrt{1-\frac{4m^2}{q^2}}\right) \tag{9.2.33}$$

in order to re-express this density in the form

$$\frac{\mathrm{disc}\, C_0}{(2i)} = \Delta C_0(p_1^2,p_2^2,q^2,m^2) = -\frac{\pi\log\left(\frac{m^2-q^2(w-\bar{z})(\bar{w}-z)}{m^2-q^2(w-z)(\bar{w}-\bar{z})}\right)}{q^2(z-\bar{z})}\,\theta(q^2-4m^2). \tag{9.2.34}$$

For the derivation of the explicit form of the discontinuity in eq. (9.2.25), we also need $C_0$ at $q^2=0$, which is given by

$$\lim_{q^2\to 0} C_0(p_1^2,p_2^2,q^2,m^2) = \frac{\log^2\left(\frac{-p_1^2+\sqrt{p_1^2(p_1^2-4m^2)}+2m^2}{2m^2}\right)-\log^2\left(\frac{-p_2^2+\sqrt{p_2^2(p_2^2-4m^2)}+2m^2}{2m^2}\right)}{2(p_1^2-p_2^2)} \tag{9.2.35}$$

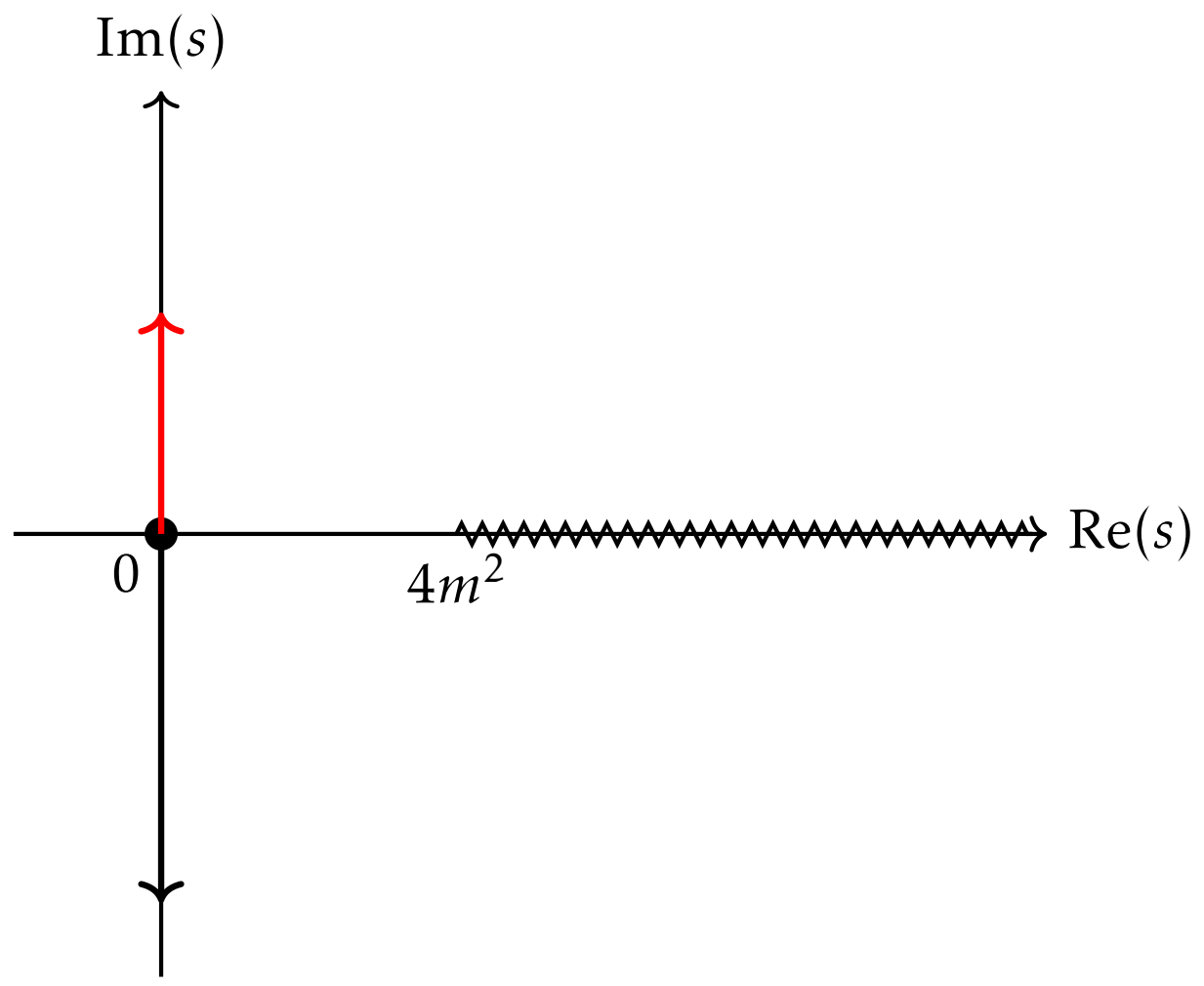

Figure 9.3: Analyticity region of the off-shell anomaly form factor in the complex $s \equiv q^2$-plane, with two bold spikes at $s = 0$. In this case there are cancelations between the cut and the contribution of the black spike, whose strength depends on all the external invariants, after integration over the $q^2 > 0$ region. The sum rule then allows to identify the red spike at $s = 0$ as an anomaly pole.

Taking into account eqs. (9.2.35) and (9.2.34), one can notice an interesting pattern of cancellations once we integrate eq. (9.2.25). Indeed, an explicit computation using the inverse Laplace transform shows that the following identity holds (see the next section)

$$\frac{2im^2}{\pi} C_0(p_1^2, p_2^2, q^2 \to 0, m^2) = \int_{4m^2}^{\infty} ds\, \rho(s, s_1, s_2, m^2) \tag{9.2.36}$$

where

$$\begin{aligned} \rho(s, s_1, s_2, m^2) &\equiv \frac{m^2}{\pi^2 q^2} \text{disc}\, C_0(p_1^2, p_2^2, q^2 m^2) \\ &= -\frac{2im^2 \log\left(\frac{m^2 - s(w-\bar{z})(\bar{w}-z)}{m^2 - s(w-z)(\bar{w}-\bar{z})}\right)}{\pi s^2 (z - \bar{z})} \theta(q^2 - 4m^2), \end{aligned} \tag{9.2.37}$$

Therefore, if we integrate eq. (9.2.25) to compute the sum rule and use eq. (9.2.36), the first and third terms cancel each other, leaving only the constant pole to saturate the sum rule (see Fig. 9.3)

$$\frac{1}{\pi} \int_0^{\infty} ds\, \Delta\Phi_0(q^2, p_1^2, p_2^2, m^2) = \frac{1}{2\pi^2} \int_0^{\infty} ds \delta(s) = \frac{1}{2\pi^2}. \tag{9.2.38}$$

Further insight into this behavior is obtained by analysing (9.2.36) under a global rescaling of all the external invariants and the mass parameter $m^2$. Notice that also in the off-shell correlator the spectral density is self-similar

$$\rho(\lambda_0 s, \lambda_0 s_1, \lambda_0 s_2, \lambda_0 m^2) = \frac{1}{\lambda_0} \rho(s, s_1, s_2, m^2) \tag{9.2.39}$$

and that (9.2.36) is scale invariant if we perform a common rescaling of all the invariants and the mass on both of its sides. It is clear that we are confronted, for this interaction, with a rather special behavior, which can be attributed to the topological properties of the vertex, not shared by other interactions.

### 9.2.4 The use of the Laplace/Borel transforms

As previously stated, in the off-shell case, the cancellation in the spectral function occurs between the integral of $\text{spk}_2$ and the contribution from the continuum, as illustrated in Fig. 9.3. To explore this in greater depth, we introduce a method that derives the result directly from the parametric integral. This approach relies on the use of Borel transforms (see Appendix F), here reformulated in terms of the inverse Laplace transform. The integral of the spectral density of $\Phi_0$ along the cut $\left[4m^2, \infty\right)$ can be evaluated using a technique we will describe. This involves first applying the inverse Laplace transform to the spectral density, followed by taking a specific limit. To proceed, we define, in full generality, the inverse Laplace transform of a generic function $F(s)$ as

$$\mathcal{L}^{-1}\{F(s)\}(t) = \frac{1}{2\pi i}\int_{\gamma-i\infty}^{\gamma+i\infty} F(s)e^{st}\,ds, \tag{9.2.40}$$

For a rational function $F(s)$ with a pole at $s = s_0$, we have

$$\mathcal{L}^{-1}\{\frac{1}{s-s_0}\}(t) = e^{s_0 t} \tag{9.2.41}$$

which provides a nonzero contribution to the dispersion integral as $t \to 0$. The method allows to compute the integral over the continuum of the spectral density quite efficiently. In our case, for the anomaly form factor $\Phi_0$, given a spectral representation of the form

$$\Phi_0(q^2) = \frac{1}{\pi}\int_0^\infty \frac{\Delta\Phi_0(s)}{s-q^2}\,ds \tag{9.2.42}$$

the action of the inverse transform generate the expression

$$\mathcal{L}^{-1}\{\Phi_0\}(t) = \frac{1}{\pi}\mathcal{L}^{-1}\left\{\int_0^\infty \frac{\Delta\Phi_0}{s-q^2}\,ds\right\}(t) = \frac{1}{\pi}\int_0^\infty \Delta\Phi_0(s)e^{st}\,ds \tag{9.2.43}$$

The integral of the spectral density is obtained by the limit

$$\lim_{t\to 0}\mathcal{L}^{-1}\{\Phi_0\}(t) = \frac{1}{\pi}\int_0^\infty \Delta\Phi_0\,ds \tag{9.2.44}$$

This approach allows to derive the expression of the integral of the discontinuity along the cut and verify the sum rules of such form factors.
For this purpose we use the parametric representation

$$\begin{gathered} C_0\left(p_1^2, p_2^2, q^2, m^2, m^2, m^2\right) = \int_0^1 dx \int_0^{1-x} dy \frac{1}{\Delta(x,y)} \\ \Delta(x,y) = m^2 + q^2 y(x+y-1) - p_1^2 xy + p_2^2 x(x+y-1) \end{gathered} \tag{9.2.45}$$

that we insert in the expression of $\Phi_0$ given in (9.1.7). Then, a direct computation gives

$$\mathcal{L}^{-1}\{\Phi_0\}(t) = -\int_0^1 dx \int_0^{1-x} dy\, \frac{g^3\left(m^2 \exp\left(\frac{t\left(-m^2+p_1^2xy-p_2^2x^2-p_2^2xy+p_2^2x\right)}{xy+y^2-y}\right) - 2m^2 + p_1^2xy - p_2^2x^2 - p_2^2xy + p_2^2x\right)}{\pi^2\left(m^2 - p_1^2xy + p_2^2x^2 + p_2^2xy - p_2^2x\right)} \tag{9.2.46}$$

on which we perform the $t \to 0$ limit as described above, in order to derive directly the value of the integral of the discontinuity. In the end, one obtains

$$\mathcal{L}^{-1}\{\Phi_0\}(t \to 0) = \frac{1}{\pi}\int_{4m^2}^{\infty} \Delta\Phi_0(s, p_1^2, p_2^2, m^2)ds = \frac{1}{2\pi^2}. \tag{9.2.47}$$

Such equation holds in the most general case, for off-shell vector lines and a massive fermion. This result is in agreement with the analysis of [64].

## 9.3 The perturbative $\langle TTJ_A \rangle$ in the massive case

In this section, we extend our results from Chapter 5 beyond the conformal limit by introducing a massive fermion. We compute the $\langle TTJ_A \rangle$ correlator perturbatively, taking into account the Feynman diagrams from Section 5.4 while incorporating additional contributions arising from the fermion mass. These mass-dependent effects appear both in the fermion propagator and in the vertices involving fermions and gravitons. Here, we present the final results of these computations.
The $\langle TTJ_A \rangle$ correlator can be decomposed into different sectors. As we will see, this decomposition differs slightly from the massless case presented in Section 5.2. In the massive case, conformal symmetry is broken, introducing additional terms in the decomposition. Indeed, while the energy-momentum tensor remains conserved, the correlator exhibits trace terms due to the presence of a fermion mass

$$p_i \left\langle T^{\mu_1\nu_1} T^{\mu_2\nu_2} J_A^{\mu_3} \right\rangle = 0, \qquad g_{\mu_i\nu_i} \left\langle T^{\mu_1\nu_1} T^{\mu_2\nu_2} J_A^{\mu_3} \right\rangle \neq 0 \qquad i = \{1,2\}. \tag{9.3.1}$$

Therefore, one can write

$$\left\langle T^{\mu_1\nu_1} T^{\mu_2\nu_2} J_A^{\mu_3} \right\rangle = \left\langle t^{\mu_1\nu_1} t^{\mu_2\nu_2} j_A^{\mu_3} \right\rangle + \left\langle t^{\mu_1\nu_1} t^{\mu_2\nu_2} j_{A\,loc}^{\mu_3} \right\rangle + \left\langle t_{loc}^{\mu_1\nu_1} t^{\mu_2\nu_2} j_A^{\mu_3} \right\rangle + \left\langle t^{\mu_1\nu_1} t_{loc}^{\mu_2\nu_2} j_A^{\mu_3} \right\rangle. \tag{9.3.2}$$

$t_{\text{loc}}^{\mu\nu}$ generally contains both a longitudinal and a trace contribution. However, in this case, only the trace contribution remains due to the conservation of the energy-momentum tensor.
Furthermore, note the absence of terms involving multiple occurrences of the "loc" operator in the decomposition. This can be understood from symmetry considerations: one cannot construct tensor structures of odd parity corresponding to such terms. A perturbative expansion further confirms this result

$$\begin{aligned} &g_{\mu_1\nu_1} g_{\mu_2\nu_2} \left\langle T^{\mu_1\nu_1} T^{\mu_2\nu_2} J_A^{\mu_3} \right\rangle = 0, \\ &p_{3_{\mu_3}} g_{\mu_i\nu_i} \left\langle T^{\mu_1\nu_1} T^{\mu_2\nu_2} J_A^{\mu_3} \right\rangle = 0, \qquad i = \{1,2\}. \end{aligned} \tag{9.3.3}$$

proving the consistency of the computation. The axial longitudinal component can be expressed in general as

$$\left\langle t^{\mu_1\nu_1} t^{\mu_2\nu_2} j_{A\,loc}^{\mu_3} \right\rangle = \frac{p_3^{\mu_3}}{p_3^2} \Pi^{\mu_1\nu_1}_{\alpha_1\beta_1}(p_1) \Pi^{\mu_2\nu_2}_{\alpha_2\beta_2}(p_2) \epsilon^{\alpha_1\alpha_2 p_1 p_2} \left( \bar{F}_1 g^{\beta_1\beta_2}(p_1 \cdot p_2) + \bar{F}_2 p_1^{\beta_2} p_2^{\beta_1} \right). \tag{9.3.4}$$

This term is determined by the gravitational anomaly, along with the mass corrections. The explicit expression of the form factors in eq. (9.3.4) is obtained through perturbative computation and is

$$\begin{aligned}
\bar{F}_1 &= \frac{g}{24\pi^2} + \frac{g m^2}{2\pi^2\lambda(s-s_1-s_2)}\Big\{2\Big[\lambda m^2 + s s_1 s_2\Big]C_0\Big(s,s_1,s_2,m^2\Big) + Q(s)\Big[s(s_1+s_2)-(s_1-s_2)^2\Big]\\
&\quad - s_2 Q(s_2)\Big[s+s_1-s_2\Big] - s_1 Q(s_1)\Big[s-s_1+s_2\Big] + \lambda\Big\},\\
\bar{F}_2 &= -\frac{g}{24\pi^2} + \frac{g m^2}{2\pi^2\lambda^2}\Big\{2\Big[\lambda\Big(m^2(-s+s_1+s_2)-2s_1s_2\Big) + 3s_1s_2\Big((s_1-s_2)^2 - s(s_1+s_2)\Big)\Big]C_0\Big(s,s_1,s_2,m^2\Big)\\
&\quad + s_1 Q(s_1)\Big[\lambda + 6s_2(s+s_1-s_2)\Big] + s_2 Q(s_2)\Big[\lambda + 6s_1(s-s_1+s_2)\Big] - Q(s)\Big[12 s s_1 s_2 + \lambda(s_1+s_2)\Big]\\
&\quad + \lambda(-s+s_1+s_2)\Big\}.
\end{aligned} \tag{9.3.5}$$

The trace term of the correlator, on the other hand, can be written as

$$\begin{aligned}
\Big\langle t_{loc}^{\mu_1\nu_1} t^{\mu_2\nu_2} j_A^{\mu_3}\Big\rangle &\equiv \Sigma^{\mu_1\nu_1}_{\alpha_1\beta_1}(p_1)\Pi^{\mu_2\nu_2}_{\alpha_2\beta_2}(p_2)\pi^{\mu_3}_{\alpha_3}(p_3)\Big\langle T^{\mu_1\nu_1}T^{\mu_2\nu_2}J_A^{\mu_3}\Big\rangle\\
&= \frac{\pi^{\mu_1\nu_1}(p_1)}{3}\Pi^{\mu_2\nu_2}_{\alpha_2\beta_2}(p_2)\pi^{\mu_3}_{\alpha_3}(p_3)\Big\langle T^{\mu}_{\mu}\, T^{\mu_2\nu_2}J_A^{\mu_3}\Big\rangle
\end{aligned} \tag{9.3.6}$$

where, in the final step, we used the explicit expression for $\Sigma^{\mu_1\nu_1}_{\alpha_1\beta_1}$ given in (5.2.6) along with the conservation of the energy-momentum tensor. This component is purely a trace term and it is transverse with respect to all the momenta. It can be expressed as

$$\Big\langle t_{loc}^{\mu_1\nu_1} t^{\mu_2\nu_2} j_A^{\mu_3}\Big\rangle = \frac{\pi^{\mu_1\nu_1}(p_1)}{3}\Pi^{\mu_2\nu_2}_{\alpha_2\beta_2}(p_2)\pi^{\mu_3}_{\alpha_3}(p_3)\Big[\bar{F}_3\,\epsilon^{\alpha_2\alpha_3 p_1 p_2}p_1^{\beta_2}\Big] \tag{9.3.7}$$

where

$$\begin{aligned}
\bar{F}_3 &= -\frac{g m^2 s_2}{2\pi^2\lambda^2 s}\Big\{2\lambda s + 2Q(s)\Big[2\lambda s + 3s\Big(s_1^2 + 6s_1s_2 + s_2^2\Big) + 3(s_1+s_2)\Big(\lambda-(s_1-s_2)^2\Big)\Big]\\
&\quad - 2Q(s_1)\Big[s(\lambda+3s_1(s_1+3s_2)) + 3s_1(\lambda-(s_1-s_2)^2)\Big] - 2Q(s_2)\Big[s(\lambda+3s_2(3s_1+s_2)) + 3s_2(\lambda-(s_1-s_2)^2)\Big]\\
&\quad + \Big[\lambda s\Big(4m^2+3(s_1+s_2)\Big) + 24 s s_1 s_2(s_1+s_2) - \Big((s_1-s_2)^2-\lambda\Big)(\lambda+12s_1s_2)\Big]C_0(s,s_1,s_2,m^2)\Big\}.
\end{aligned} \tag{9.3.8}$$

Finally, the transverse-traceless part can be expressed as

$$\begin{aligned}
\langle t^{\mu_1\nu_1}(p_1) t^{\mu_2\nu_2}(p_2) j_A^{\mu_3}(p_3)\rangle &= \Pi^{\mu_1\nu_1}_{\alpha_1\beta_1}(p_1)\Pi^{\mu_2\nu_2}_{\alpha_2\beta_2}(p_2)\pi^{\mu_3}_{\alpha_3}(p_3)\Big[\\
&\quad A_1\varepsilon^{p_1\alpha_1\alpha_2\alpha_3}p_2^{\beta_1}p_3^{\beta_2} - A_1(p_1\leftrightarrow p_2)\varepsilon^{p_2\alpha_1\alpha_2\alpha_3}p_2^{\beta_1}p_3^{\beta_2}\\
&\quad + A_2\varepsilon^{p_1\alpha_1\alpha_2\alpha_3}\delta^{\beta_1\beta_2} - A_2(p_1\leftrightarrow p_2)\varepsilon^{p_2\alpha_1\alpha_2\alpha_3}\delta^{\beta_1\beta_2}\\
&\quad + A_3\varepsilon^{p_1p_2\alpha_1\alpha_2}p_2^{\beta_1}p_3^{\beta_2}p_1^{\alpha_3} + A_4\varepsilon^{p_1p_2\alpha_1\alpha_2}\delta^{\beta_1\beta_2}p_1^{\alpha_3}\Big]
\end{aligned} \tag{9.3.9}$$

where

$$\begin{aligned}
A_1 &= -\frac{g s_2}{24\pi^2\lambda^4}\Big[\bar{A}_{11}Q(s_1) + \bar{A}_{12}Q(s_2) + \bar{A}_{13}Q(s) + \bar{A}_{14}C_0(s,s_1,s_2,m^2) + \bar{A}_{15}\Big],\\
A_2 &= -\frac{g s_2}{48\pi^2\lambda^3}\Big[\bar{A}_{21}Q(s_1) + \bar{A}_{22}Q(s_2) + \bar{A}_{23}Q(s) + \bar{A}_{24}C_0(s,s_1,s_2,m^2) + \bar{A}_{25}\Big],\\
A_3 &= 0,\\
A_4 &= 0.
\end{aligned} \tag{9.3.10}$$

The complete expressions of these contributions can be found in Appendix G.

### 9.3.1 The onhsell case and the pole cancellation

The $\langle TTJ_A\rangle$ correlator takes a form similar to $\langle J_V J_V J_A\rangle$ in the on-shell case. A direct perturbative computation shows that, in this case as well, the only tensor structure present is the longitudinal one

$$\begin{aligned}\langle T^{\mu_1\nu_1}T^{\mu_2\nu_2}J_A^{\mu_3}\rangle &= \frac{g}{96\pi^2}\left[1+\frac{12m^2}{p_3^2}\left(1+2m^2 C_0\left(0,0,p_3^2,m^2\right)\right)\right]\\ &\times\frac{p_3^{\mu_3}}{p_3^2}\left\{\left[\varepsilon^{\nu_1\nu_2 p_1 p_2}\left((p_1\cdot p_2)g^{\mu_1\mu_2}-p_1^{\mu_2}p_2^{\mu_1}\right)+(\mu_1\leftrightarrow\nu_1)\right]+(\mu_2\leftrightarrow\nu_2)\right\}.\end{aligned} \tag{9.3.11}$$

which is proportional to $R\tilde{R}$. This expression is obtained by setting the gravitons on-shell, using Schouten identities, and discarding terms that vanish upon contraction with the polarization tensors. To investigate the presence of a pole in the corresponding form factor, the $C_0(0,0,p_3^2,m^2)$ integral must be expanded around $p_3^2=0$. We can use the expression (9.1.13) or, equivalently, rewrite it as

$$C_0(0,0,p_3^2,m^2)=\frac{\log^2\left(\frac{\sqrt{p_3^2(p_3^2-4m^2)}+2m^2-p_3^2}{2m^2}\right)}{2p_3^2}. \tag{9.3.12}$$

To better understand the behavior of $C_0$ near $p_3^2=0$, we consider the expansion in eq. (9.1.17). Substituting this expansion into (9.3.11) and simplifying the expression reveals the absence of a pole $1/p_3^2$. Specifically, the remaining contributions are finite and free of singularities. Consequently, in the on-shell spectral function, only the cut contribution remains, in complete analogy with the $\langle J_V J_V J_A\rangle$ correlator.

## 9.4 The sum rule of the gravitational chiral anomaly

In this section, we will expand on the previous analysis by demonstrating the presence of a sum rule for the chiral gravitational anomaly in the off-shell case.
By contracting the projectors in eq. (9.3.4), the structure of the longitudinal part of $\langle TTJ_A\rangle$ can be expressed as

$$\begin{aligned}\langle t^{\mu_1\nu_1}t^{\mu_2\nu_2}j^{\mu_3}_{A\,loc}\rangle &= (\bar{F}_1-\bar{F}_2)\frac{p_3^{\mu_3}}{8\,p_3^2}\left\{\left[\varepsilon^{\nu_1\nu_2p_1p_2}\left(p_1\cdot p_2\, g^{\mu_1\mu_2}-p_1^{\mu_2}p_2^{\mu_1}\right)+(\mu_1\leftrightarrow\nu_1)\right]+(\mu_2\leftrightarrow\nu_2)\right\}\\ &+(\bar{F}_1+\bar{F}_2)\,\frac{p_3^{\mu_3}}{8\,p_3^2}\left\{\left[\varepsilon^{\nu_1\nu_2p_1p_2}\left(p_1\cdot p_2\, g^{\mu_1\mu_2}+p_1^{\mu_2}p_2^{\mu_1}\right)+(\mu_1\leftrightarrow\nu_1)\right]+(\mu_2\leftrightarrow\nu_2)\right\}\\ &+(\bar{F}_1+\bar{F}_2)\frac{(p_1\cdot p_2)}{p_1^2p_2^2}\,\frac{p_3^{\mu_3}}{4\,p_3^2}\left\{\left[\varepsilon^{\nu_1\nu_2p_1p_2}\left((p_1\cdot p_2)p_1^{\mu_1}p_2^{\mu_2}-p_2^2p_1^{\mu_1}p_1^{\mu_2}-p_1^2p_2^{\mu_1}p_2^{\mu_2}\right)+(\mu_1\leftrightarrow\nu_1)\right]+(\mu_2\leftrightarrow\nu_2)\right\}\end{aligned} \tag{9.4.1}$$

For simplicity, we now introduce the following quantities

$$\begin{aligned}K_1(s,s_1,s_2)&=s_2(s-s_1)\left(s^2+3ss_1-2s_1^2\right)+s_2^3(3s+2s_1)-s_2^2(s+s_1)(3s+2s_1)+s_1(s-s_1)^3-s_2^4\\ K_2(s,s_1,s_2)&=2m^2\lambda\left(s^2-2s(s_1+s_2)+s_1^2+s_2^2\right)+s_1s_2\left(3s^3-5s^2(s_1+s_2)+s(s_1+s_2)^2+(s_1-s_2)^2(s_1+s_2)\right)\\ K_3(s,s_1,s_2)&=s_2\left(8m^2(s+s_1)+s^2-6ss_1+s_1^2\right)+s_2^2\left(-4m^2+s+s_1\right)-(s-s_1)^2\left(4m^2+s+s_1\right)-s_2^3.\end{aligned} \tag{9.4.2}$$

The first contribution on the right-hand side of eq. (9.4.1) is proportional to $\bar{F}_1 - \bar{F}_2$ and is related to the second functional derivative of the anomaly $R\tilde{R}$ with respect to the metric, defining the anomalous form factor

$$\begin{aligned}\Phi_{TTJ_A} = \frac{1}{g}\frac{\bar{F}_1-\bar{F}_2}{s} = &\frac{m^2 s_1\left(s_2\left(s_1^2-s^2\right)+s_2^2(5s+s_1)+(s_1-s)^3-3s_2^3\right)\mathrm{B}_0\left(s_1,m^2\right)}{\pi^2 s(s-s_1-s_2)\lambda^2}\\ &+\frac{m^2 s_2\left(s_2\left(3s^2+s_1^2\right)+s_2^2(s_1-3s)-(s-s_1)^2(s+3s_1)+s_2^3\right)\mathrm{B}_0\left(s_2,m^2\right)}{\pi^2 s(s-s_1-s_2)\lambda^2}\\ &+\frac{m^2 K_1(s,s_1,s_2)\mathrm{B}_0\left(s,m^2\right)}{\pi^2 s(s-s_1-s_2)\lambda^2}+\frac{m^2 K_2(s,s_1,s_2)\mathrm{C}_0\left(s_1,s_2,s,m^2\right)}{\pi^2 s(s-s_1-s_2)\lambda^2}\\ &\frac{m^2\left(s^2-2s(s_1+s_2)+s_1^2+s_2^2\right)}{\pi^2 s(s-s_1-s_2)\lambda}+\frac{1}{12\pi^2 s}.\end{aligned} \tag{9.4.3}$$

Notice that this form factor behaves as follows at large $s$

$$\Phi_{TTJ_A} \sim \frac{1}{12\pi^2 s}+O\left(\frac{m^2}{s^2}\log\frac{s}{m^2}\right), \tag{9.4.4}$$

similar to the behavior observed in the case of $\langle J_V J_V J_A\rangle$, highlighting the dominance of the anomaly at large $s$.
The second and the third terms in eq. (9.4.1) are proportional to

$$\begin{aligned}\bar{F}_1+\bar{F}_2 = &\frac{2gm^2 s_1 s_2\left(s^2+s(s_1-2s_2)-2s_1^2+s_1 s_2+s_2^2\right)\mathrm{B}_0\left(s_1,m^2\right)}{\pi^2(s-s_1-s_2)\lambda^2}\\ &+\frac{2gm^2 s_1 s_2\left(s_2(s+s_1)+(s-s_1)^2-2s_2^2\right)\mathrm{B}_0\left(s_2,m^2\right)}{\pi^2 s(s-s_1-s_2)\lambda^2}+\frac{2gm^2 s_1 s_2\left(-2s^2+s(s_1+s_2)+(s_1-s_2)^2\right)\mathrm{B}_0\left(s,m^2\right)}{\pi^2(s-s_1-s_2)\lambda^2}\\ &+\frac{gm^2 s_1 s_2 K_3(s,s_1,s_2)\mathrm{C}_0\left(s_1,s_2,s,m^2\right)}{\pi^2(s-s_1-s_2)\lambda^2}-\frac{2g\,m^2 s_1 s_2}{\pi^2(s-s_1-s_2)\lambda}\end{aligned} \tag{9.4.5}$$

Such contributions arise from the breaking of conformal symmetry and vanishes in the massless case or, alternatively, in the physical case of on-shell gravitons ($s_1 = s_2 = 0$).
The combination $\bar{F}_1 - \bar{F}_2$ defines the correct expression of the anomaly form factor $\Phi_{TTJ_A}$, which satisfies a sum rule, as we are going to demonstrate. As mentioned, the tensor structure associated with $\Phi_{TTJ_A}$ corresponds to the second functional derivative of $R\tilde{R}$ with respect to the metric, Fourier-transformed to momentum-space.

### 9.4.1 The spectral function

We start by separating the discontinuity of the anomalous form factor $\Phi_{TTJ_A}$ into pole contributions and the continuum, in the form

$$\mathrm{disc}(\Phi_{TTJ_A}) = \mathrm{disc}(\Phi_{TTJ_A})_{cont}+\mathrm{disc}(\Phi_{TTJ_A})_{pole}, \tag{9.4.6}$$

The discontinuity over the continuum is given by

$$\mathrm{disc}(\Phi_{TTJ_A})_{cont} = \frac{m^2 K_1(s,s_1,s_2)\,\mathrm{disc}\,\mathrm{B}_0\left(s,m^2\right)}{\pi^2 s(s-s_1-s_2)\lambda^2}+\frac{gm^2 K_2(s,s_1,s_2)\,\mathrm{disc}\,\mathrm{C}_0\left(s_1,s_2,s,m^2\right)}{\pi^2 s(s-s_1-s_2)\lambda^2}. \tag{9.4.7}$$

The discontinuity of $C_0$ is reported in eq. (9.2.34), while that of $B_0$ is

$$\text{disc}\, B_0(s,m^2) = 2\pi i \sqrt{1-\frac{4m^2}{s}}\; \theta(s-4m^2). \tag{9.4.8}$$

The pole contributions split at separate locations

$$\text{disc}(\Phi_{TTJ_A})_{pole} = \text{disc}(\Phi_{TTJ_A})_0 + \text{disc}(\Phi_{TTJ_A})_{s_1+s_2} + \text{disc}(\Phi_{TTJ_A})_{s_-} + \text{disc}(\Phi_{TTJ_A})_{s_+}, \tag{9.4.9}$$

where we have indicated the points corresponding to the zeroes of the Källen function $\lambda$ as

$$s_\pm = (\sqrt{s_1} \pm \sqrt{s_2})^2. \tag{9.4.10}$$

The pole at $s=0$ has two spikes

$$\text{disc}(\Phi_{TTJ_A})_0 = 2\pi i \frac{\delta(s)}{12\pi^2} - \frac{2\pi i \delta(s)}{12\pi^2(s_1-s_2)^3(s_1+s_2)}\Psi(s,s_1,s_2,m^2) \tag{9.4.11}$$

with

$$\begin{aligned}\Psi(s,s_1,s_2,m^2) \equiv 12m^2\Big[&\left(s_1^2+s_2^2\right)(s_1-s_2) - (s_1-s_2)\left(s_1^2+s_2^2\right)\text{B}_0\left(s,m^2\right) + s_1\left(s_1^2+2s_1s_2+3s_2^2\right)\text{B}_0\left(s_1,m^2\right) \\ &- s_2\left(3s_1^2+2s_1s_2+s_2^2\right)\text{B}_0\left(s_2,m^2\right) + (s_1-s_2)\left(2m^2\left(s_1^2+s_2^2\right)+s_1s_2(s_1+s_2)\right)\text{C}_0\left(s_1,s_2,s,m^2\right)\Big].\end{aligned} \tag{9.4.12}$$

One can observe that $\Psi$ remains finite across the entire light-cone in the limit $s \to 0$

$$\lim_{s\to 0}\Psi(s,s_1,s_2) < \infty \tag{9.4.13}$$

We have outlined the various contributions to the discontinuity of $\Phi_{TTJ_A}$ in Fig. 9.4. The red spike at $s=0$, which has a constant strength, corresponds to the first term on the right-hand side of (9.4.11). The second contribution, which depends on the external invariants, is represented by the black spike in the same figure.

For $s > 4m^2$, across the cut, we encounter two apparent poles that, as we will demonstrate, do not contribute to the discontinuity. Additionally, there is a pole at $s = s_1+s_2$, which can lie either above or below the cut (i.e. $s > 4m^2$ or $s < 4m^2$), depending on the values of $s_1$ and $s_2$. Such pole contributes to the sum rule through the localized density

$$\begin{aligned}\text{disc}(\Phi_{TTJ_A})_{s_1+s_2} = &\frac{2\pi i\, m^2\delta(s-s_1-s_2)}{4\pi^2(s_1+s_2)}\Big(-2B_0\left(s_1+s_2,m^2\right)+B_0\left(s_1,m^2\right) \\ &+ B_0\left(s_2,m^2\right) + \left(4m^2-s_1-s_2\right)C_0\left(s_1,s_2,s_1+s_2,m^2\right)+2\Big).\end{aligned} \tag{9.4.14}$$

Notice that the UV divergences in the $B_0$'s in all the expressions above cancel automatically, as expected.

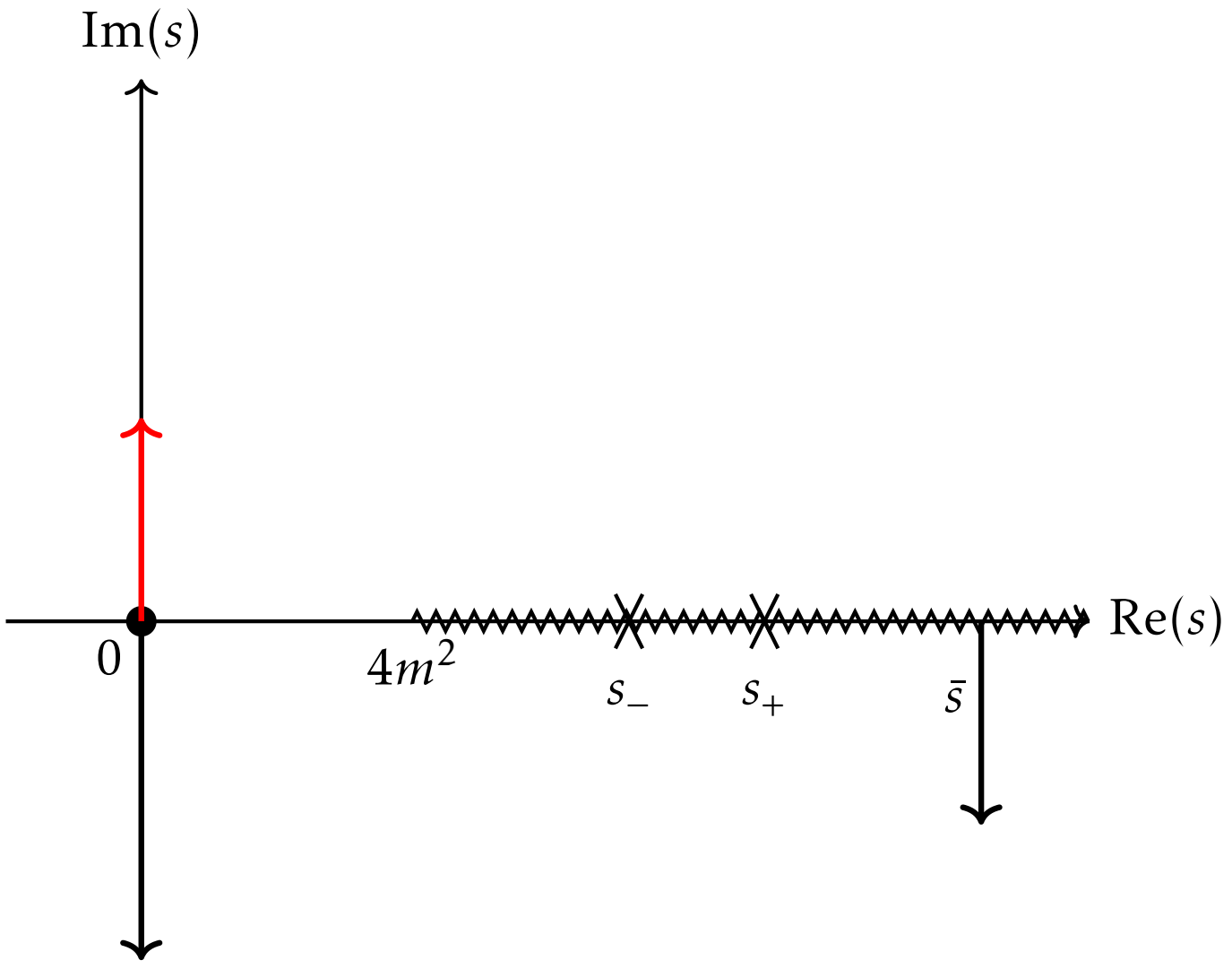

Figure 9.4: The spectral density of the off-shell gravitational anomaly form factor in the complex $s$-plane. Shown are the two spikes at $s = 0$, the anomalous thresholds at $s_\pm = (\sqrt{s_1} \pm \sqrt{s_2})^2$ and the pole at $\bar{s} = s_1 + s_2$. In this case there is a cancelation in the sum rule between the integral of the density over the cut, the black spike at $s = 0$, of variable strength, and the black spike at $\bar{s}$.

### 9.4.2 Cancellation of the anomalous thresholds

The expression for the spectral density can be separated into distinct components, representing contributions from regular kinematical thresholds as well as potential anomalous thresholds

$$\Delta\Phi_{TTJ_A} = \Delta\Phi^{\text{reg}}_{TTJ_A} + \Delta\Phi^{(\lambda)}_{TTJ_A}, \tag{9.4.15}$$

The regular part, $\Delta\Phi^{\text{reg}}_{TTJ_A}$, includes contributions from standard kinematical thresholds located on the cut $s = 4m^2$ and on the poles

$$s = 0, \qquad s = s_1 + s_2 \tag{9.4.16}$$

These thresholds arise from well-understood physical processes. In contrast, the anomalous part, $\Delta\Phi^{\lambda}_{TTJ_A}$, emerges under special conditions when the kinematical factor $\lambda = (s - s_-)(s - s_+) = 0$. The dispersive representation of $\Phi_{TTJ_A}$ may allow terms of the form

$$\begin{aligned}\Delta\Phi_{TTJ_A} =& c_0\delta(s) + c_1(s,s_1,s_2)\theta(s-4m^2)H(s) + c_2(s,s_1,s_2)\delta(s-s_1-s_2)\\ &+ c_3(s,s_1,s_2)\delta'(s-s_-) + c_4(s,s_1,s_2)\delta'(s-s_+) + c_5(s,s_1,s_2)\delta(s-s_-) + c_6(s,s_1,s_2)\delta(s-s_+),\end{aligned} \tag{9.4.17}$$

where the derivative of the $\delta$-function corresponds to higher order poles at $\lambda = 0$. Indeed, the general structure of $\Phi_{TTJ_A}$ is

$$\begin{aligned}\Phi_{TTJ_A} =& \frac{c_0(s,s_1,s_2,m^2)}{s} + \frac{c_1(s,s_1,s_2,m^2)}{s} + (\text{continuum terms}) + \frac{c_2(s,s_1,s_2,m^2)}{s-s_1-s_2} + \frac{c_3(s,s_1,s_2,m^2)}{(s-s_-)^2}\\ &+ \frac{c_4(s,s_1,s_2,m^2)}{(s-s_+)^2} + \frac{c_5(s,s_1,s_2,m^2)}{s-s_-} + \frac{c_6(s,s_1,s_2,m^2)}{s-s_+},\end{aligned} \tag{9.4.18}$$

where $c_3$ and $c_4$ define possible residues at the double poles of $\lambda^2$. These could, in principle, introduce subtractions. These could, in principle, introduce subtractions. However, after a careful evaluation, one can verify that [194]

$$\Phi^{(\lambda)}_{TTJ_A}(s = s_\pm) = 0, \tag{9.4.19}$$

and hence

$$\Delta\Phi^{(\lambda)}_{TTJ_A} = 0. \tag{9.4.20}$$

### 9.4.3 The Laplace/Borel transforms

In order to compute the sum rule for the $\langle TTJ_A\rangle$ correlator, we can use the same technique previously applied to the $\langle J_V J_V J_A\rangle$ correlator. By applying the inverse Laplace transform to both sides of the dispersive representation of the form factor (the Cauchy's relation), we obtain

$$\mathcal{L}^{-1}\{\Phi_{TTJ_A}\}(t) = \mathcal{L}^{-1}\left\{\int_0^\infty \frac{\Delta\Phi_{TTJ_A}}{s-q^2}\,ds\right\}(t) = \frac{1}{\pi}\int_0^\infty \Delta\Phi_{TTJ_A}\, e^{st}\, ds \tag{9.4.21}$$

The integral of the spectral density is obtained by performing the $t \to 0$ limit

$$\lim_{t\to 0}\mathcal{L}^{-1}\{\Phi_{TTJ_A}\}(t) = \frac{1}{\pi}\int_0^\infty \Delta\Phi_{TTJ_A}\, ds. \tag{9.4.22}$$

The technique used to compute the inverse Laplace transform is the same as that described in the case of the $\langle J_V J_V J_A\rangle$ correlator. We employ the parametric form of the two-point and three-point scalar functions, interchanging the order of integration with the inverse Laplace transform. After some lengthy computations, and once the limit on $t$ has been taken, we obtain

$$\frac{1}{\pi}\int_0^\infty \Delta\Phi_{TTJ_A}\, ds = \frac{1}{12\pi^2} \tag{9.4.23}$$

which represents the gravitational anomaly sum rule.

## 9.5 Conclusions

In this chapter, we have explored the perturbative structure of the $\langle J_V J_V J_A\rangle$ and $\langle TTJ_A\rangle$ correlators in the presence of massive fermions. By extending the framework beyond the conformal limit, we aimed to understand how mass deformations affect the fundamental anomaly structures and the consistency of quantum field theories.
A central outcome of our analysis is the confirmation that the anomaly sum rules, both in the chiral gauge and gravitational sectors, remain valid and independent of the mass, placing significant constraints on the spectral densities of the theory. This invariance under mass deformations emphasizes the robustness of these sum rules and underscores their importance as foundational conditions in quantum field theories. These sum rules are deeply connected to the topological and symmetry properties of the theory.

# Chapter 10

# Conclusions and perspectives

Anomalies are an extraordinarily rich subject. In this thesis, we have investigated their role in shaping the structure of parity-odd correlators. Specifically, we have examined chiral and gravitational anomalies, affecting the axial current, and parity-violating trace anomalies, affecting the energy-momentum tensor. Our analysis starts in the context of a CFT in momentum-space. We have shown how the conformal constraints are sufficient to fix parity-odd three-point functions in terms of the anomaly coefficients. Among the interactions studied are the anomalous $\langle J_V J_V J_A\rangle$, $\langle J_A J_A J_A\rangle$, $\langle TTJ_A\rangle$ and $\langle JJT\rangle_{odd}$.

It would be interesting to examine how the conformal constraints affect parity-odd four-point functions. These functions are not completely fixed by conformal invariance, but, as was done for the parity-even case, one could impose additional constraints. Indeed, we know that the additional requirement of dual conformal symmetry is enough to determine the structure of four-point functions [195]. Parity-odd higher-point functions can be particularly interesting since they are related, for example, to the non-Abelian chiral anomaly in the cases of $\langle J_V J_V J_V J_A\rangle$ and $\langle J_V J_V J_V J_V J_A\rangle$. The gravitational anomaly, on the other hand, can be further studied by examining the $\langle TTTJ_A\rangle$ correlator or others involving additional energy-momentum tensors.

The study of correlators in momentum-space is very rich, as one can explore contexts beyond those examined in this thesis. Other examples include non-Abelian field theories, non-conserved currents, gravity with torsion, and dimensions other than four. Additionally, a complete analysis of the $\langle TTT\rangle_{\text{odd}}$ is still lacking. Furthermore, it would be interesting to apply this methodology in different settings, such as the conformal bootstrap and cosmology in general.

In the second part of the thesis, we have worked within the context of thermal field theory. Applications of this study lie in cosmology and condensed matter theory. We have shown how chiral and gravitational anomalies are protected from finite density and temperature corrections. Moreover, we have presented a complete study of the $\langle J_V J_V J_A\rangle$ correlator at finite density, covering both its longitudinal and transverse sectors. A similar study could be conducted for the gravitational anomaly interaction $\langle TTJ_A\rangle$, as we have only examined the longitudinal part of this correlator.

Additionally, one could investigate the conformal anomaly interactions $\langle JJT\rangle$ and $\langle TTT\rangle$, in both their parity-even and parity-odd sectors, at finite temperature and density. We know that the introduction of a scale explicitly breaks conformal symmetry. The consequences of such breaking can also be observed in the conformal anomaly correlators.

Lastly, in the final chapter, we have focused on the structure of the $\langle J_V J_V J_A\rangle$ and $\langle TTJ_A\rangle$ correlators, going beyond the conformal regime by introducing massive fermions. These results have been used to analyze the intricate relationship between anomaly form factors and spectral densities, validating the anomaly sum rules.

After this analysis, one additional question remains. We were able to determine the structure of correlators by imposing conformal invariance conditions. Would it be possible to relax these conditions to account for symmetry breakings, such as finite temperature, density, or massive particles? How would the conformal equations be modified in these cases, and would it be possible to determine their solutions? Would there be a unique solution? We hope to return to these topics in the future.

# Appendix A

# 3K Integrals

The most general solution of the CWIs for our correlators can be written in terms of integrals involving a product of three Bessel functions, namely 3K integrals. In this appendix, we will illustrate such integrals and their properties. For a detailed review on the the topic, see [25, 29, 80].

## A.1 Definition and properties

First, we recall the definition of the general 3K integral

$$I_{\alpha\{\beta_1\beta_2\beta_3\}}(p_1,p_2,p_3) = \int dx x^{\alpha} \prod_{j=1}^{3} p_j^{\beta_j} K_{\beta_j}\left(p_j x\right) \tag{A.1}$$

where $K_\nu$ is a modified Bessel function of the second kind

$$K_\nu(x) = \frac{\pi}{2}\frac{I_{-\nu}(x)-I_\nu(x)}{\sin(\nu\pi)}, \qquad \nu \notin \mathbb{Z} \qquad\qquad I_\nu(x) = \left(\frac{x}{2}\right)^{\nu}\sum_{k=0}^{\infty}\frac{1}{\Gamma(k+1)\Gamma(\nu+1+k)}\left(\frac{x}{2}\right)^{2k} \tag{A.2}$$

with the property

$$K_n(x) = \lim_{\epsilon\to 0} K_{n+\epsilon}(x), \quad n \in \mathbb{Z} \tag{A.3}$$

The triple-K integral depends on four parameters: the power $\alpha$ of the integration variable $x$, and the three Bessel function indices $\beta_j$ . The arguments of the 3K integral are magnitudes of momenta $p_j$ with $j = 1,2,3$. One can notice the integral is invariant under the exchange $(p_j,\beta_j) \leftrightarrow (p_i,\beta_i)$. We will also use the reduced version of the 3K integral defined as

$$J_{N\{k_j\}} = I_{\frac{d}{2}-1+N\{\Delta_j-\frac{d}{2}+k_j\}} \tag{A.4}$$

where we introduced the condensed notation $\{k_j\} = \{k_1,k_2,k_3\}$. The 3K integral satisfies an equation analogous to the dilatation equation with scaling degree

$$\deg\left(J_{N\{k_j\}}\right) = \Delta_t + k_t - 2d - N \tag{A.5}$$

where

$$k_t = k_1 + k_2 + k_3, \qquad\qquad \Delta_t = \Delta_1 + \Delta_2 + \Delta_3 \tag{A.6}$$

From this analysis, it is simple to relate the form factors to the 3K integrals. Indeed, the dilatation WI of each from factor tells us that this needs to be written as a combination of integrals of the following type

$$J_{N+k_t,\{k_1,k_2,k_3\}} \tag{A.7}$$

where $N$ is the number of momenta that the form factor multiplies in the decomposition. Let us now list some useful properties of 3K integrals

$$\begin{aligned}
&\frac{\partial}{\partial p_n} J_{N\{k_j\}} = -p_n J_{N+1\{k_j-\delta_{jn}\}} \\
&J_{N\{k_j+\delta_{jn}\}} = p_n^2 J_{N\{k_j-\delta_{jn}\}} + 2\left(\Delta_n - \frac{d}{2} + k_n\right) J_{N-1\{k_j\}} \\
&\frac{\partial^2}{\partial p_n^2} J_{N\{k_j\}} = J_{N+2\{k_j\}} - 2\left(\Delta_n - \frac{d}{2} + k_n - \frac{1}{2}\right) J_{N+1\{k_j-\delta_{jn}\}}, \\
&K_n J_{N\{k_j\}} \equiv \left(\frac{\partial^2}{\partial p_n^2} + \frac{(d+1-2\Delta_n)}{p_n}\frac{\partial}{\partial p_n}\right) J_{N\{k_j\}} = J_{N+2\{k_j\}} - 2k_n J_{N+1\{k_j-\delta_{jn}\}}, \\
&K_{nm} J_{N\{k_j\}} \equiv (K_n - K_m) J_{N\{k_j\}} = -2k_n J_{N+1\{k_j-\delta_{jn}\}} + 2k_m J_{N+1\{k_j-\delta_{jm}\}}
\end{aligned} \tag{A.8}$$

## A.2 Zero momentum limit

When solving the secondary CWIs, it may be useful to perform a zero momentum limit. In this subsection, we review the behavior of the 3K integrals in the limit $p_3 \to 0$. In this limit, the momentum conservation gives

$$p_1^\mu = -p_2^\mu \qquad \Longrightarrow \qquad p_1 = p_2 \equiv p \tag{A.1}$$

Assuming that $\alpha > \beta_t - 1$ and $\beta_3 > 0$, we can write

$$\lim_{p_3 \to 0} I_{\alpha\{\beta_j\}}(p, p, p_3) = p^{\beta_t - \alpha - 1} \ell_{\alpha\{\beta_j\}} \tag{A.2}$$

where

$$\ell_{\alpha\{\beta_j\}} = \frac{2^{\alpha-3}\Gamma(\beta_3)}{\Gamma(\alpha-\beta_3+1)}\Gamma\left(\frac{\alpha+\beta_t+1}{2}-\beta_3\right)\Gamma\left(\frac{\alpha-\beta_t+1}{2}+\beta_1\right)\Gamma\left(\frac{\alpha-\beta_t+1}{2}+\beta_2\right)\Gamma\left(\frac{\alpha-\beta_t+1}{2}\right) \tag{A.3}$$

We can derive similar formulas for the case $p_1 \to 0$ or $p_2 \to 0$ by considering the fact that 3K integrals are invariant under the exchange $(p_j, \beta_j) \leftrightarrow (p_i, \beta_i)$.

## A.3 Divergences and regularization

The 3K integral defined in (A.1) converges when

$$\alpha > \sum_{i=1}^{3} \left|\beta_i\right| - 1 \quad ; \quad p_1, p_2, p_3 > 0 \tag{A.1}$$

If $\alpha$ does not satisfy this inequality, the integrals must be defined by an analytic continuation. The quantity

$$\delta \equiv \sum_{j=1}^{3} \left|\beta_j\right| - 1 - \alpha \tag{A.2}$$

is the expected degree of divergence. However, when

$$\alpha + 1 \pm \beta_1 \pm \beta_2 \pm \beta_3 = -2k \quad , \quad k = 0, 1, 2, \ldots \tag{A.3}$$

for some non-negative integer $k$ and any choice of the ± sign, the analytic continuation of the 3K integral generally has poles in the regularization parameter. Therefore, if the above condition is satisfied, we need to regularize the integrals. This can be done by shifting the parameters of the 3K integrals as

$$I_{\alpha\{\beta_1,\beta_2,\beta_3\}} \to I_{\tilde{\alpha}\{\tilde{\beta}_1,\tilde{\beta}_2,\tilde{\beta}_3\}} \quad \Longrightarrow \quad J_{N\{k_1,k_2,k_3\}} \to J_{N+u\epsilon\{k_1+v_1\epsilon,k_2+v_2\epsilon,k_3+v_3\epsilon\}} \tag{A.4}$$

where

$$\tilde{\alpha} = \alpha + u\epsilon \quad , \quad \tilde{\beta}_1 = \beta_1 + v_1\epsilon \quad , \quad \tilde{\beta}_2 = \beta_2 + v_2\epsilon \quad , \quad \tilde{\beta}_3 = \beta_3 + v_3\epsilon \tag{A.5}$$

or equivalently by considering

$$d \to d + 2u\epsilon \quad ; \quad \Delta \to \Delta_i + (u + v_i)\,\epsilon \tag{A.6}$$

In general, the regularisation parameters $u$ and $v_i$ are arbitrary. However, in certain cases, there may be some constraints on them. For simplicity, in this paper we consider the same $v_i = v$ for every $i$.

## A.4 3K integrals and Feynman integrals

3K integrals are related to Feynman integrals in momentum-space. The exact relations were first derived in [29, 80]. Here we briefly show the results. Such expressions have been recently used in order to show the connection between the conformal analysis and the perturbative one for the $\langle J_V J_V J_A\rangle$ correlator [34].

Let $K_{d\{\delta_1\delta_2\delta_3\}}$ denote a massless scalar 1-loop 3-point momentum-space integral

$$K_{d\{\delta_1\delta_2\delta_3\}} = \int \frac{\mathrm{d}^d \boldsymbol{k}}{(2\pi)^d} \frac{1}{k^{2\delta_3}\left|\boldsymbol{p}_1 - \boldsymbol{k}\right|^{2\delta_2}\left|\boldsymbol{p}_2 + \boldsymbol{k}\right|^{2\delta_1}} \tag{A.1}$$

Any such integral can be expressed in terms of 3K integrals and vice versa. For scalar integrals the relation reads

$$K_{d\{\delta_1\delta_2\delta_3\}} = \frac{2^{4-\frac{3d}{2}}}{\pi^{\frac{d}{2}}} \times \frac{I_{\frac{d}{2}-1\{\frac{d}{2}+\delta_1-\delta_t,\frac{d}{2}+\delta_2-\delta_t,\frac{d}{2}+\delta_3-\delta_t\}}}{\Gamma(d-\delta_t)\Gamma(\delta_1)\Gamma(\delta_2)\Gamma(\delta_3)} \tag{A.2}$$

where $\delta_t = \delta_1 + \delta_2 + \delta_3$. Its inverse reads

$$\begin{aligned} I_{\alpha\{\beta_1\beta_2\beta_3\}} = 2^{3\alpha-1}\pi^{\alpha+1}\Gamma\left(\frac{\alpha+1+\beta_t}{2}\right)\prod_{j=1}^{3}\Gamma\left(\frac{\alpha+1+2\beta_j-\beta_t}{2}\right) \\ \times K_{2+2\alpha,\left\{\frac{1}{2}(\alpha+1+2\beta_1-\beta_t),\frac{1}{2}(\alpha+1+2\beta_2-\beta_t),\frac{1}{2}(\alpha+1+2\beta_3-\beta_t)\right\}} \end{aligned} \tag{A.3}$$

where $\beta_t = \beta_1 + \beta_2 + \beta_3$. All tensorial massless 1-loop 3-point momentum-space integrals can also be expressed in terms of a number of 3K integrals when their tensorial structure is resolved by standard methods (for the exact expressions in this case see Appendix A.3 of [29]).

## A.5 Appell functions

In the case of scalar primary operators, for example, of scaling dimensions $\Delta_i$, and momenta $p_1, p_2, p_3$, the solutions expressed by 3K integrals can be directly related to the four Appell functions [28] characterized by four pairs of indices $(a_i, b_j)$ $(i, j = 1, 2)$. These are the indices that in the change of variables $(p_1^2, p_2^2, p_3^2) \to (p_1^2, x, y)$, $x = p_2^2/p_1^2$, $y = p_3^2/p_1^2$ reduce the special conformal constraints to Appell hypergeometric equations deprived of $1/x$ or $1/y$ singularities [75,77]. Setting

$$\alpha(a,b) = a + b + \frac{d}{2} - \frac{1}{2}(\Delta_2 + \Delta_3 - \Delta_1) \qquad \beta(a,b) = a + b + d - \frac{1}{2}(\Delta_1 + \Delta_2 + \Delta_3) \tag{A.1}$$

we can introduce the following function

$$F_4(\alpha(a,b), \beta(a,b); \gamma(a), \gamma'(b); x, y) = \sum_{i=0}^{\infty}\sum_{j=0}^{\infty} \frac{(\alpha(a,b), i+j)\,(\beta(a,b), i+j)}{(\gamma(a), i)\,(\gamma'(b), j)} \frac{x^i}{i!} \frac{y^j}{j!} \tag{A.2}$$

where $(\alpha, i) = \Gamma(\alpha + i)/\Gamma(\alpha)$ is the Pochammer symbol. We will refer to $\alpha \ldots \gamma'$ as to the first,..., fourth parameters of $F_4$.

The 4 independent solutions of the Appell system of equations are then all of the form $x^a y^b F_4$, linearly combined in a Bose-symmetric form. Specifically, the solution for the parity-even correlator with three scalar operators takes the general form

$$\Phi(p_1, p_2, p_3) = p_1^{\Delta - 2d} \sum_{a,b} c(a, b, \vec{\Delta})\, x^a y^b\, F_4(\alpha(a,b), \beta(a,b); \gamma(a), \gamma'(b); x, y) \tag{A.3}$$

where the sum runs over the four values $a_i, b_i$ $i = 0, 1$ with arbitrary constants $c(a, b, \vec{\Delta})$, with $\vec{\Delta} = (\Delta_1, \Delta_2, \Delta_3)$. Specifically,

$$\alpha_0 \equiv \alpha(a_0, b_0) = \frac{d}{2} - \frac{\Delta_2 + \Delta_3 - \Delta_1}{2}, \qquad \beta_0 \equiv \beta(b_0) = d - \frac{\Delta_1 + \Delta_2 + \Delta_3}{2}, \tag{A.4}$$

$$\gamma_0 \equiv \gamma(a_0) = \frac{d}{2} + 1 - \Delta_2, \qquad \gamma_0' \equiv \gamma(b_0) = \frac{d}{2} + 1 - \Delta_3 \tag{A.5}$$

for a generic $d$ dimension.

# Appendix B

# Master integrals

In this section we summarize some important relations regarding the master integrals $B_0$ and $C_0$. They are defined by the following expressions in the Euclidean space[1]

$$B_0(p_i^2) \equiv \frac{1}{\pi^2}\int d^d l \frac{1}{l^2\,(l-p_i)^2} = \frac{1}{\varepsilon} + \log\left(-\frac{\mu^2}{\pi p_i^2}\right) - \gamma + 2 \tag{B.1}$$

and

$$\begin{aligned} C_0(p_1^2,p_2^2,p_3^2) \equiv \frac{1}{\pi^2}\int d^d l \frac{1}{l^2(l-p_1)^2(l+p_2)^2} = \\ \frac{1}{\sqrt{\lambda}}\Bigg[\mathrm{Li}_2\left(-\frac{-p_1^2+p_2^2+p_3^2+\sqrt{\lambda}}{p_1^2-p_2^2-p_3^2+\sqrt{\lambda}}\right) + \mathrm{Li}_2\left(-\frac{p_1^2-p_2^2+p_3^2+\sqrt{\lambda}}{-p_1^2+p_2^2-p_3^2+\sqrt{\lambda}}\right) - \mathrm{Li}_2\left(\frac{p_1^2+p_2^2-p_3^2-\sqrt{\lambda}}{p_1^2+p_2^2-p_3^2+\sqrt{\lambda}}\right) \\ + \mathrm{Li}_2\left(-\frac{p_1^2+p_2^2-p_3^2+\sqrt{\lambda}}{-p_1^2-p_2^2+p_3^2+\sqrt{\lambda}}\right) - \mathrm{Li}_2\left(\frac{p_1^2-p_2^2+p_3^2-\sqrt{\lambda}}{p_1^2-p_2^2+p_3^2+\sqrt{\lambda}}\right) - \mathrm{Li}_2\left(\frac{-p_1^2+p_2^2+p_3^2-\sqrt{\lambda}}{-p_1^2+p_2^2+p_3^2+\sqrt{\lambda}}\right)\Bigg] \end{aligned} \tag{B.2}$$

with $\lambda \equiv \lambda(p_1,p_2,p_3)$ defined in (5.4.21). By acting with derivatives on such integrals one finds [75]

$$\frac{\partial}{\partial p_i} B_0\left(p_i^2\right) = \frac{(d-4)}{p_i} B_0\left(p_i^2\right) \tag{B.3}$$

[1] In the Minkowski space, the prefactor of the integrals is $1/(i\pi^2)$.

and

$$\begin{aligned}
\frac{\partial}{\partial p_1} C_0 &= \frac{1}{\lambda p_1}\Bigg\{ 2(d-3)\Big[ \left(p_1^2+p_2^2-p_3^2\right) B_0\left(p_2^2\right) + \left(p_1^2-p_2^2+p_3^2\right) B_0\left(p_3^2\right) - 2p_1^2 B_0(p_1^2)\Big] \\
&\quad + \Big[ (d-4)\left(p_2^2-p_3^2\right)^2 - (d-2)p_1^4 + 2p_1^2\left(p_2^2+p_3^2\right)\Big] C_0\left(p_1^2,p_2^2,p_3^2\right)\Bigg\} \\
\frac{\partial}{\partial p_2} C_0 &= \frac{1}{\lambda p_2}\Bigg\{ 2(d-3)\Big[ \left(p_1^2+p_2^2-p_3^2\right) B_0(p_1^2) + \left(p_2^2+p_3^2-p_1^2\right) B_0\left(p_3^2\right) - 2p_2^2 B_0\left(p_2^2\right)\Big] \\
&\quad + \Big[ 2p_3^2\left(p_2^2-(d-4)p_1^2\right) + \left(p_1^2-p_2^2\right)\left((d-4)p_1^2+(d-2)p_2^2\right) + (d-4)p_3^4\Big] C_0\left(p_1^2,p_2^2,p_3^2\right)\Bigg\} \\
\frac{\partial}{\partial p_3} C_0 &= \frac{1}{\lambda p_3}\Bigg\{ 2(d-3)\Big[ \left(p_1^2-p_2^2+p_3^2\right) B_0(p_1^2) + \left(p_2^2+p_3^2-p_1^2\right) B_0\left(p_2^2\right) - 2p_3^2 B_0\left(p_3^2\right)\Big] \\
&\quad + \Big[ (d-4)\left(p_1^2-p_2^2\right)^2 - (d-2)p_3^4 + 2p_3^2\left(p_1^2+p_2^2\right)\Big] C_0\left(p_1^2,p_2^2,p_3^2\right)\Bigg\}
\end{aligned} \tag{B.4}$$

# Appendix C

# Schouten identities

## C.1 Identities for the $\langle J_V J_V J_A \rangle$

In this section we derive the minimal decomposition to describe $X^{\kappa\mu_1\mu_2\mu_3}$ in eq. (4.2.20). We start by writing all the possible tensor structures

$$
\begin{aligned}
&\varepsilon^{\mu_1\mu_2\mu_3 p_1} p_1^\kappa, \quad \varepsilon^{\mu_1\mu_2\mu_3 p_2} p_1^\kappa, \quad \varepsilon^{\mu_1\mu_2 p_1 p_2} p_1^{\mu_3} p_1^\kappa, \quad \varepsilon^{\mu_1\mu_3 p_1 p_2} p_3^{\mu_2} p_1^\kappa, \quad \varepsilon^{\mu_2\mu_3 p_1 p_2} p_2^{\mu_1} p_1^\kappa,\\
&\varepsilon^{\mu_1\mu_2\mu_3 p_1} p_2^\kappa, \quad \varepsilon^{\mu_1\mu_2\mu_3 p_2} p_2^\kappa, \quad \varepsilon^{\mu_1\mu_2 p_1 p_2} p_1^{\mu_3} p_2^\kappa, \quad \varepsilon^{\mu_1\mu_3 p_1 p_2} p_3^{\mu_2} p_2^\kappa, \quad \varepsilon^{\mu_2\mu_3 p_1 p_2} p_2^{\mu_1} p_2^\kappa,\\
&\varepsilon^{\mu_2\mu_3 p_1 p_2} \delta^{\mu_1\kappa}, \quad \varepsilon^{\mu_1\mu_3 p_1 p_2} \delta^{\mu_2\kappa}, \quad \varepsilon^{\mu_1\mu_2 p_1 p_2} \delta^{\mu_3\kappa},\\
&\varepsilon^{\kappa\mu_1\mu_2\mu_3}, \quad \varepsilon^{\kappa\mu_1\mu_2 p_1} p_1^{\mu_3}, \quad \varepsilon^{\kappa\mu_1\mu_2 p_2} p_1^{\mu_3}, \quad \varepsilon^{\kappa\mu_1\mu_3 p_1} p_3^{\mu_2}, \quad \varepsilon^{\kappa\mu_1\mu_3 p_2} p_3^{\mu_2}, \quad \varepsilon^{\kappa\mu_2\mu_3 p_1} p_2^{\mu_1}, \quad \varepsilon^{\kappa\mu_2\mu_3 p_2} p_2^{\mu_1},\\
&\varepsilon^{\kappa\mu_1 p_1 p_2} \delta^{\mu_2\mu_3}, \quad \varepsilon^{\kappa\mu_1 p_1 p_2} p_3^{\mu_2} p_1^{\mu_3}, \quad \varepsilon^{\kappa\mu_2 p_1 p_2} \delta^{\mu_1\mu_3}, \quad \varepsilon^{\kappa\mu_2 p_1 p_2} p_2^{\mu_1} p_1^{\mu_3}, \quad \varepsilon^{\kappa\mu_3 p_1 p_2} \delta^{\mu_1\mu_2}, \quad \varepsilon^{\kappa\mu_3 p_1 p_2} p_2^{\mu_1} p_3^{\mu_2}. \quad \text{(C.1)}
\end{aligned}
$$

Such tensor structures are not all independent, indeed we are going to show what are the Schouten identites one has to consider in order to find the minimal number of tensor structures. The first two identites are

$$\varepsilon^{[\mu_2\mu_3\kappa p_1}\delta^{\mu_1]\alpha} = 0, \tag{C.2}$$

$$\varepsilon^{[\mu_2\mu_3\kappa p_2}\delta^{\mu_1]\alpha} = 0, \tag{C.3}$$

that can be contracted with $p_{1\alpha}$ and $p_{2\alpha}$ and taking the projectors in front we obtain the four tensor identities

$$
\begin{aligned}
\pi^{\lambda_1}_{\mu_1}\pi^{\lambda_2}_{\mu_2}\pi^{\lambda_3}_{\mu_3}\left(\varepsilon^{p_1\kappa\mu_1\mu_3}p_3^{\mu_2}\right) &= \pi^{\lambda_1}_{\mu_1}\pi^{\lambda_2}_{\mu_2}\pi^{\lambda_3}_{\mu_3}\left(-p_1^2\varepsilon^{\kappa\mu_1\mu_2\mu_3} + \varepsilon^{p_1\mu_1\mu_2\mu_3}p_1^\kappa - \varepsilon^{p_1\kappa\mu_1\mu_2}p_1^{\mu_3}\right)\\
\pi^{\lambda_1}_{\mu_1}\pi^{\lambda_2}_{\mu_2}\pi^{\lambda_3}_{\mu_3}\left(\varepsilon^{p_1\kappa\mu_2\mu_3}p_2^{\mu_1}\right) &= \pi^{\lambda_1}_{\mu_1}\pi^{\lambda_2}_{\mu_2}\pi^{\lambda_3}_{\mu_3}\left(\frac{1}{2}\left(p_1^2+p_2^2-p_3^2\right)\varepsilon^{\kappa\mu_1\mu_2\mu_3} + \varepsilon^{p_1\kappa\mu_1\mu_2}p_1^{\mu_3} + \varepsilon^{p_1\mu_1\mu_2\mu_3}p_2^\kappa\right)\\
\pi^{\lambda_1}_{\mu_1}\pi^{\lambda_2}_{\mu_2}\pi^{\lambda_3}_{\mu_3}\left(\varepsilon^{p_2\kappa\mu_1\mu_3}p_3^{\mu_2}\right) &= \pi^{\lambda_1}_{\mu_1}\pi^{\lambda_2}_{\mu_2}\pi^{\lambda_3}_{\mu_3}\left(\frac{1}{2}\left(p_1^2+p_2^2-p_3^2\right)\varepsilon^{\kappa\mu_1\mu_2\mu_3} + \varepsilon^{p_2\mu_1\mu_2\mu_3}p_1^\kappa - \varepsilon^{p_2\kappa\mu_1\mu_2}p_1^{\mu_3}\right)\\
\pi^{\lambda_1}_{\mu_1}\pi^{\lambda_2}_{\mu_2}\pi^{\lambda_3}_{\mu_3}\left(\varepsilon^{p_2\kappa\mu_2\mu_3}p_2^{\mu_1}\right) &= \pi^{\lambda_1}_{\mu_1}\pi^{\lambda_2}_{\mu_2}\pi^{\lambda_3}_{\mu_3}\left(-p_2^2\varepsilon^{\kappa\mu_1\mu_2\mu_3} + \varepsilon^{p_2\kappa\mu_1\mu_2}p_1^{\mu_3} + \varepsilon^{p_2\mu_1\mu_2\mu_3}p_2^\kappa\right).
\end{aligned} \tag{C.4}
$$

Then, considering the identity

$$\varepsilon^{[\mu_2\mu_3 p_1 p_2}\delta^{\mu_1]\alpha} = 0 \tag{C.5}$$

and contracting with $\delta^\kappa_\alpha$, $p_{1\alpha}$ and $p_{2\alpha}$ we find

$$\begin{aligned}
&\pi^{\lambda_1}_{\mu_1}\pi^{\lambda_2}_{\mu_2}\pi^{\lambda_3}_{\mu_3}\left(\varepsilon^{p_1p_2\mu_1\mu_3}p_3^{\mu_2}\right)=\pi^{\lambda_1}_{\mu_1}\pi^{\lambda_2}_{\mu_2}\pi^{\lambda_3}_{\mu_3}\left(-\frac{p_1^2+p_2^2-p_3^2}{2}\varepsilon^{p_1\mu_1\mu_2\mu_3}-p_1^2\varepsilon^{p_2\mu_1\mu_2\mu_3}-\varepsilon^{p_1p_2\mu_1\mu_2}p_1^{\mu_3}\right)\\
&\pi^{\lambda_1}_{\mu_1}\pi^{\lambda_2}_{\mu_2}\pi^{\lambda_3}_{\mu_3}\left(\varepsilon^{p_1p_2\mu_2\mu_3}p_2^{\mu_1}\right)=\pi^{\lambda_1}_{\mu_1}\pi^{\lambda_2}_{\mu_2}\pi^{\lambda_3}_{\mu_3}\left(\frac{p_1^2+p_2^2-p_3^2}{2}\varepsilon^{p_2\mu_1\mu_2\mu_3}+p_2^2\varepsilon^{p_1\mu_1\mu_2\mu_3}+\varepsilon^{p_1p_2\mu_1\mu_2}p_1^{\mu_3}\right)\\
&\pi^{\lambda_1}_{\mu_1}\pi^{\lambda_2}_{\mu_2}\pi^{\lambda_3}_{\mu_3}\left(\varepsilon^{p_1p_2\mu_1\mu_2}\delta^{\kappa\mu_3}\right)=\pi^{\lambda_1}_{\mu_1}\pi^{\lambda_2}_{\mu_2}\pi^{\lambda_3}_{\mu_3}\left(-\varepsilon^{p_2\mu_1\mu_2\mu_3}p_1^\kappa+\varepsilon^{p_1\mu_1\mu_2\mu_3}p_2^\kappa-\varepsilon^{p_1p_2\mu_2\mu_3}\delta^{\kappa\mu_1}+\varepsilon^{p_1p_2\mu_1\mu_3}\delta^{\kappa\mu_2}\right).
\end{aligned}\tag{C.6}$$

Furthermore, we need to consider the identity

$$\varepsilon^{[\mu_2\kappa p_1p_2}\delta^{\mu_1]\alpha}=0\tag{C.7}$$

that once it is contracted with $p_{1\alpha}$ and $p_{2\alpha}$ we have

$$\begin{aligned}
&\pi^{\lambda_1}_{\mu_1}\pi^{\lambda_2}_{\mu_2}\pi^{\lambda_3}_{\mu_3}\left(\varepsilon^{p_1p_2\kappa\mu_1}p_3^{\mu_2}\right)=\pi^{\lambda_1}_{\mu_1}\pi^{\lambda_2}_{\mu_2}\pi^{\lambda_3}_{\mu_3}\left(\frac{1}{2}\left(p_1^2+p_2^2-p_3^2\right)\varepsilon^{p_1\kappa\mu_1\mu_2}+p_1^2\varepsilon^{p_2\kappa\mu_1\mu_2}+\varepsilon^{p_1p_2\mu_1\mu_2}p_1^\kappa\right),\\
&\pi^{\lambda_1}_{\mu_1}\pi^{\lambda_2}_{\mu_2}\pi^{\lambda_3}_{\mu_3}\left(\varepsilon^{p_1p_2\kappa\mu_2}p_2^{\mu_1}\right)=\pi^{\lambda_1}_{\mu_1}\pi^{\lambda_2}_{\mu_2}\pi^{\lambda_3}_{\mu_3}\left(-\frac{1}{2}\left(p_1^2+p_2^2-p_3^2\right)\varepsilon^{p_2\kappa\mu_1\mu_2}-p_2^2\varepsilon^{p_1\kappa\mu_1\mu_2}+\varepsilon^{p_1p_2\mu_1\mu_2}p_2^\kappa\right),
\end{aligned}\tag{C.8}$$

and it is worth mentioning that the possible contraction with $\delta^{\mu_3}_\alpha$ give again an identity that is not independent taking in consideration the previous tensor identities found.
Finally, we have

$$\varepsilon^{[\kappa\mu_3p_1p_2}\delta^{\mu_1]\mu_2}=0,\tag{C.9}$$

$$\varepsilon^{[\kappa\mu_3p_1p_2}\delta^{\mu_2]\mu_1}=0,\tag{C.10}$$

giving

$$\begin{aligned}
&\pi^{\lambda_1}_{\mu_1}\pi^{\lambda_2}_{\mu_2}\pi^{\lambda_3}_{\mu_3}\left(\varepsilon^{p_1p_2\mu_1\mu_3}\delta^{\mu_2\kappa}\right)=\pi^{\lambda_1}_{\mu_1}\pi^{\lambda_2}_{\mu_2}\pi^{\lambda_3}_{\mu_3}\left(\varepsilon^{p_2\kappa\mu_1\mu_3}p_3^{\mu_2}+\varepsilon^{p_1p_2\kappa\mu_3}\delta^{\mu_1\mu_2}-\varepsilon^{p_1p_2\kappa\mu_1}\delta^{\mu_2\mu_3}\right),\\
&\pi^{\lambda_1}_{\mu_1}\pi^{\lambda_2}_{\mu_2}\pi^{\lambda_3}_{\mu_3}\left(\varepsilon^{p_1p_2\mu_2\mu_3}\delta^{\mu_1\kappa}\right)=\pi^{\lambda_1}_{\mu_1}\pi^{\lambda_2}_{\mu_2}\pi^{\lambda_3}_{\mu_3}\left(\varepsilon^{p_1\kappa\mu_2\mu_3}p_2^{\mu_1}+\varepsilon^{p_1p_2\kappa\mu_3}\delta^{\mu_1\mu_2}-\varepsilon^{p_1p_2\kappa\mu_2}\delta^{\mu_1\mu_3}\right),
\end{aligned}\tag{C.11}$$

and the two identities

$$\varepsilon^{[\kappa\mu_3p_1p_2}p_2^{\mu_1]}=0,\tag{C.12}$$

$$\varepsilon^{[\kappa\mu_3p_1p_2}p_1^{\mu_2]}=0,\tag{C.13}$$

giving the only independent symmetric constraint

$$\begin{aligned}
&\pi^{\lambda_1}_{\mu_1}\pi^{\lambda_2}_{\mu_2}\pi^{\lambda_3}_{\mu_3}\left(\varepsilon^{p_1p_2\kappa\mu_3}p_2^{\mu_1}p_3^{\mu_2}\right)=\\
&=\pi^{\lambda_1}_{\mu_1}\pi^{\lambda_2}_{\mu_2}\pi^{\lambda_3}_{\mu_3}\bigg\{\frac{1}{2}\bigg[-\frac{1}{2}\left(p_1^2+p_2^2-p_3^2\right)\varepsilon^{p_1\kappa\mu_2\mu_3}p_2^{\mu_1}-p_1^2\varepsilon^{p_2\kappa\mu_2\mu_3}p_2^{\mu_1}-\varepsilon^{p_1p_2\mu_2\mu_3}p_1^\kappa p_2^{\mu_1}-\varepsilon^{p_1p_2\kappa\mu_2}p_1^{\mu_3}p_2^{\mu_1}
\end{aligned}$$

$$-\frac{1}{2}\left(p_1^2+p_2^2-p_3^2\right)\varepsilon^{p_2\kappa\mu_1\mu_3}p_3^{\mu_2}-p_2^2\varepsilon^{p_1\kappa\mu_1\mu_3}p_3^{\mu_2}+\varepsilon^{p_1p_2\mu_1\mu_3}p_2^{\kappa}p_3^{\mu_2}-\varepsilon^{p_1p_2\kappa\mu_1}p_1^{\mu_3}p_3^{\mu_2}\Big]\Big]\Big\}. \tag{C.14}$$

In summary, from the analysis above, we conclude that the minimal number of tensor structures to describe $X^{\kappa\mu_1\mu_2\mu_3}$ in (4.2.20) is twelve and in particular we have

$$\begin{aligned}
&\pi^{\lambda_1}_{\mu_1}(p_1)\pi^{\lambda_2}_{\mu_2}(p_2)\pi^{\lambda_3}_{\mu_3}(p_3)\Big(\mathcal{K}^\kappa\langle j^{\mu_1}(p_1)j^{\mu_2}(p_2)J_A^{\mu_3}(p_3)\rangle\Big)=\\
&=\pi^{\lambda_1}_{\mu_1}(p_1)\pi^{\lambda_2}_{\mu_2}(p_2)\pi^{\lambda_3}_{\mu_3}(p_3)\Big[p_1^\kappa\Big(C_{11}\,\varepsilon^{\mu_1\mu_2\mu_3p_1}+C_{12}\,\varepsilon^{\mu_1\mu_2\mu_3p_2}+C_{13}\,\varepsilon^{\mu_1\mu_2p_1p_2}p_1^{\mu_3}\Big)\\
&\quad+p_2^\kappa\Big(C_{21}\,\varepsilon^{\mu_1\mu_2\mu_3p_1}+C_{22}\,\varepsilon^{\mu_1\mu_2\mu_3p_2}+C_{23}\,\varepsilon^{\mu_1\mu_2p_1p_2}p_1^{\mu_3}\Big)+C_{31}\varepsilon^{\kappa\mu_1\mu_2\mu_3}+C_{32}\varepsilon^{\kappa\mu_1\mu_2p_1}p_1^{\mu_3}\\
&\qquad+C_{33}\varepsilon^{\kappa\mu_1\mu_2p_2}p_1^{\mu_3}+C_{34}\varepsilon^{\kappa\mu_1p_1p_2}\delta^{\mu_2\mu_3}+C_{35}\varepsilon^{\kappa\mu_2p_1p_2}\delta^{\mu_1\mu_3}+C_{36}\varepsilon^{\kappa\mu_3p_1p_2}\delta^{\mu_1\mu_2}\Big].
\end{aligned} \tag{C.15}$$

## C.2 Identities for the $\langle TTJ_A\rangle$

In this section we derive the following minimal decomposition used when analyzing the special conformal constraint on the $\langle TTJ_A\rangle$ correlator

$$\begin{aligned}
0=\Pi^{\alpha_1\beta_1}_{\mu_1\nu_1}(p_1)\Pi^{\alpha_2\beta_2}_{\mu_2\nu_2}(p_2)\pi^{\alpha_3}_{\mu_3}(p_3)\Big(\mathcal{K}^\kappa\big\langle T^{\mu_1\nu_1}(p_1)T^{\mu_2\nu_2}(p_2)J_A^{\mu_3}(p_3)\big\rangle\Big)&=\Pi^{\alpha_1\beta_1}_{\mu_1\nu_1}(p_1)\Pi^{\alpha_2\beta_2}_{\mu_2\nu_2}(p_2)\pi^{\alpha_3}_{\mu_3}(p_3)\Big[\\
&p_1^\kappa\Big(C_{11}\varepsilon^{p_1\mu_1\mu_2\mu_3}p_2^{\nu_1}p_3^{\nu_2}+C_{12}\varepsilon^{p_2\mu_1\mu_2\mu_3}p_2^{\nu_1}p_3^{\nu_2}+C_{13}\varepsilon^{p_1\mu_1\mu_2\mu_3}\delta^{\nu_1\nu_2}+C_{14}\varepsilon^{p_2\mu_1\mu_2\mu_3}\delta^{\nu_1\nu_2}\\
&+C_{15}\varepsilon^{p_1p_2\mu_1\mu_2}p_2^{\nu_1}p_3^{\nu_2}p_1^{\mu_3}+C_{16}\varepsilon^{p_1p_2\mu_1\mu_2}\delta^{\nu_1\nu_2}p_1^{\mu_3}\Big)\\
&+p_2^\kappa\Big(C_{21}\varepsilon^{p_1\mu_1\mu_2\mu_3}p_2^{\nu_1}p_3^{\nu_2}+C_{22}\varepsilon^{p_2\mu_1\mu_2\mu_3}p_2^{\nu_1}p_3^{\nu_2}+C_{23}\varepsilon^{p_1\mu_1\mu_2\mu_3}\delta^{\nu_1\nu_2}+C_{24}\varepsilon^{p_2\mu_1\mu_2\mu_3}\delta^{\nu_1\nu_2}\\
&+C_{25}\varepsilon^{p_1p_2\mu_1\mu_2}p_2^{\nu_1}p_3^{\nu_2}p_1^{\mu_3}+C_{26}\varepsilon^{p_1p_2\mu_1\mu_2}\delta^{\nu_1\nu_2}p_1^{\mu_3}\Big)\\
&+\delta^{\kappa\mu_1}\Big(C_{31}\varepsilon^{p_1\mu_2\mu_3\nu_1}p_3^{\nu_2}+C_{32}\varepsilon^{p_2\mu_2\mu_3\nu_1}p_3^{\nu_2}+C_{33}\varepsilon^{p_1p_2\mu_2\nu_1}p_1^{\mu_3}p_3^{\nu_2}+C_{34}\varepsilon^{p_1p_2\mu_2\mu_3}\delta^{\nu_1\nu_2}\Big)\\
&+\delta^{\kappa\mu_2}\Big(C_{41}\varepsilon^{p_1\mu_1\mu_3\nu_2}p_2^{\nu_1}+C_{42}\varepsilon^{p_2\mu_1\mu_3\nu_2}p_2^{\nu_1}+C_{43}\varepsilon^{p_1p_2\mu_1\nu_2}p_1^{\mu_3}p_2^{\nu_1}+C_{44}\varepsilon^{p_1p_2\mu_1\mu_3}\delta^{\nu_1\nu_2}\Big)\\
&+C_{51}\varepsilon^{\kappa\mu_1\mu_2\mu_3}\delta^{\nu_1\nu_2}+C_{52}\varepsilon^{\kappa\mu_1\mu_2\mu_3}p_2^{\nu_1}p_3^{\nu_2}+C_{53}\varepsilon^{p_1\kappa\mu_1\mu_2}p_1^{\mu_3}\delta^{\nu_1\nu_2}+C_{54}\varepsilon^{p_2\kappa\mu_1\mu_2}p_1^{\mu_3}\delta^{\nu_1\nu_2}\Big]
\end{aligned} \tag{C.1}$$

In order to determine such decomposition, first we have to write all the possible tensor structures that can appear in the equation. In particular the tensor related to the primary equations are

$$\begin{aligned}
&\varepsilon^{p_1\mu_1\mu_2\mu_3}p_2^{\nu_1}p_3^{\nu_2}p_1^{\kappa},\quad \varepsilon^{p_1\mu_1\mu_2\mu_3}p_2^{\nu_1}p_3^{\nu_2}p_2^{\kappa},\quad \varepsilon^{p_2\mu_1\mu_2\mu_3}p_2^{\nu_1}p_3^{\nu_2}p_1^{\kappa},\quad \varepsilon^{p_2\mu_1\mu_2\mu_3}p_2^{\nu_1}p_3^{\nu_2}p_2^{\kappa},\\
&\varepsilon^{p_1\mu_1\mu_2\mu_3}\delta^{\nu_1\nu_2}p_1^{\kappa},\quad \varepsilon^{p_1\mu_1\mu_2\mu_3}\delta^{\nu_1\nu_2}p_2^{\kappa},\quad \varepsilon^{p_2\mu_1\mu_2\mu_3}\delta^{\nu_1\nu_2}p_1^{\kappa},\quad \varepsilon^{p_2\mu_1\mu_2\mu_3}\delta^{\nu_1\nu_2}p_2^{\kappa}\\
&\varepsilon^{p_1p_2\mu_1\mu_2}p_2^{\nu_1}p_3^{\nu_2}p_1^{\mu_3}p_1^{\kappa},\quad \varepsilon^{p_1p_2\mu_1\mu_2}p_2^{\nu_1}p_3^{\nu_2}p_1^{\mu_3}p_2^{\kappa},\quad \varepsilon^{p_1p_2\mu_1\mu_2}\delta^{\nu_1\nu_2}p_1^{\mu_3}p_1^{\kappa},\quad \varepsilon^{p_1p_2\mu_1\mu_2}\delta^{\nu_1\nu_2}p_1^{\mu_3}p_2^{\kappa},\\
&\varepsilon^{p_1p_2\mu_1\mu_2}p_2^{\nu_1}\delta^{\nu_2\mu_3}p_1^{\mu_3}p_1^{\kappa},\quad \varepsilon^{p_1p_2\mu_1\mu_2}p_2^{\nu_1}\delta^{\nu_2\mu_3}p_1^{\mu_3}p_2^{\kappa},\quad \varepsilon^{p_1p_2\mu_1\mu_2}p_1^{\nu_2}\delta^{\nu_1\mu_3}p_1^{\mu_3}p_2^{\kappa},
\end{aligned} \tag{C.2}$$

and similar ones for the secondary. However, not all of these tensors are independent. Some of these tensors can be rewritten in terms of each others. We then need to find a set of tensors that form a minimal decomposition. We will illustrate a couple of cases of Schouten identities needed for this purpose. For example we consider the equation

$$0 = \varepsilon^{[p_1 p_2 \mu_1 \mu_2} \delta_{\alpha}^{\mu_3]} \tag{C.3}$$

which can be contracted with $p_1^\alpha$ and $p_2^\alpha$, obtaining

$$\begin{aligned}
&\Pi^{\alpha_1\beta_1}_{\mu_1\nu_1}(p_1)\Pi^{\alpha_2\beta_2}_{\mu_2\nu_2}(p_2)\pi^{\alpha_3}_{\mu_3}(p_3)\left[\varepsilon^{p_1 p_2 \mu_1 \mu_3} p_3^{\mu_2}\right] = \Pi^{\alpha_1\beta_1}_{\mu_1\nu_1}(p_1)\Pi^{\alpha_2\beta_2}_{\mu_2\nu_2}(p_2)\pi^{\alpha_3}_{\mu_3}(p_3)\Bigg[\\
&\qquad \frac{1}{2}\varepsilon^{p_1 \mu_1 \mu_2 \mu_3}\left(-p_1^2 - p_2^2 + p_3^2\right) - \varepsilon^{p_1 p_2 \mu_1 \mu_2} p_1^{\mu_3} - \varepsilon^{p_2 \mu_1 \mu_2 \mu_3} p_1^2\Bigg],\\
&\Pi^{\alpha_1\beta_1}_{\mu_1\nu_1}(p_1)\Pi^{\alpha_2\beta_2}_{\mu_2\nu_2}(p_2)\pi^{\alpha_3}_{\mu_3}(p_3)\left[\varepsilon^{p_1 p_2 \mu_2 \mu_3} p_2^{\mu_1}\right] = \Pi^{\alpha_1\beta_1}_{\mu_1\nu_1}(p_1)\Pi^{\alpha_2\beta_2}_{\mu_2\nu_2}(p_2)\pi^{\alpha_3}_{\mu_3}(p_3)\Bigg[\\
&\qquad \varepsilon^{p_1 \mu_1 \mu_2 \mu_3} p_2^2 + \varepsilon^{p_1 p_2 \mu_1 \mu_2} p_1^{\mu_3} - \frac{1}{2}\varepsilon^{p_2 \mu_1 \mu_2 \mu_3}\left(-p_1^2 - p_2^2 + p_3^2\right)\Bigg].
\end{aligned} \tag{C.4}$$

We then consider the identity

$$0 = \varepsilon^{[p_1 p_2 \mu_1 \mu_2} \delta_{\alpha}^{\kappa]} \tag{C.5}$$

which can be contracted with $p_1^\alpha$ and $p_2^\alpha$, obtaining

$$\begin{aligned}
&\Pi^{\alpha_1\beta_1}_{\mu_1\nu_1}(p_1)\Pi^{\alpha_2\beta_2}_{\mu_2\nu_2}(p_2)\pi^{\alpha_3}_{\mu_3}(p_3)\left[\varepsilon^{p_1 p_2 \kappa \mu_1} p_3^{\mu_2}\right] =\\
&\qquad \Pi^{\alpha_1\beta_1}_{\mu_1\nu_1}(p_1)\Pi^{\alpha_2\beta_2}_{\mu_2\nu_2}(p_2)\pi^{\alpha_3}_{\mu_3}(p_3)\left[-\frac{1}{2}\varepsilon^{p_1 \kappa \mu_1 \mu_2}\left(-p_1^2 - p_2^2 + p_3^2\right) + \varepsilon^{p_1 p_2 \mu_1 \mu_2} p_1^{\kappa} + \varepsilon^{p_2 \kappa \mu_1 \mu_2} p_1^2\right],\\
&\Pi^{\alpha_1\beta_1}_{\mu_1\nu_1}(p_1)\Pi^{\alpha_2\beta_2}_{\mu_2\nu_2}(p_2)\pi^{\alpha_3}_{\mu_3}(p_3)\left[\varepsilon^{p_1 p_2 \kappa \mu_2} p_2^{\mu_1}\right] =\\
&\qquad \Pi^{\alpha_1\beta_1}_{\mu_1\nu_1}(p_1)\Pi^{\alpha_2\beta_2}_{\mu_2\nu_2}(p_2)\pi^{\alpha_3}_{\mu_3}(p_3)\left[-\varepsilon^{p_1 \kappa \mu_1 \mu_2} p_2^2 + \varepsilon^{p_1 p_2 \mu_1 \mu_2} p_2^{\kappa} + \frac{1}{2}\varepsilon^{p_2 \kappa \mu_1 \mu_2}\left(-p_1^2 - p_2^2 + p_3^2\right)\right]
\end{aligned} \tag{C.6}$$

The analysis of the remaining contraints is rather involved, but follows the steps outlined above. In the end, after considering all the possible Schouten identities, one finds a minimal set of independent structures as expressed in eq. (C.1).

## C.3 Identities for the $\langle JJT\rangle_{odd}$

In this section we will derive the following minimal decomposition used when analyzing the special conformal constraints on the $\langle JJT\rangle_{odd}$ correlator

$$\begin{aligned}
0 =&\pi^{\alpha_1}_{\mu_1}(p_1)\pi^{\alpha_2}_{\mu_2}(p_2)\Pi^{\alpha_3\beta_3}_{\mu_3\nu_3}(p_3)\mathcal{K}^k\langle J^{\mu_1}(p_1)J^{\mu_2}(p_2)T^{\mu_3\nu_3}(p_3)\rangle = \pi^{\alpha_1}_{\mu_1}(p_1)\pi^{\alpha_2}_{\mu_2}(p_2)\Pi^{\alpha_3\beta_3}_{\mu_3\nu_3}(p_3)\Bigg[\\
&\Big(C_{11}\varepsilon^{p_1\alpha_1\alpha_2\alpha_3}p_1^{\beta_3}+C_{12}\varepsilon^{p_2\alpha_1\alpha_2\alpha_3}p_1^{\beta_3}+C_{13}\varepsilon^{p_1p_2\alpha_1\alpha_2}p_1^{\alpha_3}p_1^{\beta_3}+C_{14}\varepsilon^{p_1p_2\alpha_2\alpha_3}\delta^{\alpha_1\beta_3}\Big)p_1^{\kappa}+\\
&\Big(C_{21}\varepsilon^{p_1\alpha_1\alpha_2\alpha_3}p_1^{\beta_3}+C_{22}\varepsilon^{p_2\alpha_1\alpha_2\alpha_3}p_1^{\beta_3}+C_{23}\varepsilon^{p_1p_2\alpha_1\alpha_2}p_1^{\alpha_3}p_1^{\beta_3}+C_{24}\varepsilon^{p_1p_2\alpha_2\alpha_3}\delta^{\alpha_1\beta_3}\Big)p_2^{\kappa}+\\
&C_{31}\varepsilon^{\kappa\mu_1\mu_2\mu_3}p_1^{\nu_3}+C_{32}\varepsilon^{p_1\kappa\mu_2\mu_3}\delta^{\mu_1\nu_3}+C_{33}\varepsilon^{p_2\kappa\mu_1\mu_3}\delta^{\mu_2\nu_3}+C_{34}\varepsilon^{p_1p_2\kappa\mu_3}\delta^{\mu_1\mu_2}p_1^{\nu_3}+\\
&C_{41}\delta^{\mu_1\kappa}\varepsilon^{\mu_2\mu_3p_1p_2}p_1^{\nu_3}+C_{51}\delta^{\mu_2\kappa}\varepsilon^{\mu_1\mu_3p_1p_2}p_1^{\nu_3}+C_{61}\delta^{\mu_3\kappa}\varepsilon^{p_1\mu_1\mu_2\nu_3}++C_{62}\delta^{\mu_3\kappa}\varepsilon^{p_2\mu_1\mu_2\nu_3}\Bigg]
\end{aligned} \tag{C.1}$$

Such decomposition is obtained first by writing all the possible tensor structures. In particular the tensor related to the primary equations are

$$\begin{aligned}
&\varepsilon^{p_1\alpha_1\alpha_2\alpha_3}p_1^{\beta_3}p_1^{\kappa}, && \varepsilon^{p_2\alpha_1\alpha_2\alpha_3}p_1^{\beta_3}p_1^{\kappa}, && \varepsilon^{p_1p_2\alpha_1\alpha_2}p_1^{\alpha_3}p_1^{\beta_3}p_1^{\kappa}, && \varepsilon^{p_1p_2\alpha_2\alpha_3}\delta^{\alpha_1\beta_3}p_1^{\kappa}\\
&\cancel{p_2^{\alpha_1}p_1^{\beta_3}\varepsilon^{p_1p_2\alpha_2\alpha_3}p_1^{\kappa}}, && \cancel{p_3^{\alpha_2}p_1^{\beta_3}\varepsilon^{p_1p_2\alpha_1\alpha_3}p_1^{\kappa}}, && \cancel{\delta^{\beta_3\alpha_2}\varepsilon^{p_1p_2\alpha_1\alpha_3}p_1^{\kappa}}\\
&\varepsilon^{p_1\alpha_1\alpha_2\alpha_3}p_1^{\beta_3}p_2^{\kappa}, && \varepsilon^{p_2\alpha_1\alpha_2\alpha_3}p_1^{\beta_3}p_2^{\kappa}, && \varepsilon^{p_1p_2\alpha_1\alpha_2}p_1^{\alpha_3}p_1^{\beta_3}p_2^{\kappa}, && \varepsilon^{p_1p_2\alpha_2\alpha_3}\delta^{\alpha_1\beta_3}p_2^{\kappa}\\
&\cancel{p_2^{\alpha_1}p_1^{\beta_3}\varepsilon^{p_1p_2\alpha_2\alpha_3}p_2^{\kappa}}, && \cancel{p_3^{\alpha_2}p_1^{\beta_3}\varepsilon^{p_1p_2\alpha_1\alpha_3}p_2^{\kappa}}, && \cancel{\delta^{\beta_3\alpha_2}\varepsilon^{p_1p_2\alpha_1\alpha_3}p_2^{\kappa}}
\end{aligned} \tag{C.2}$$

while the tensor related to the secondary ones are

$$\begin{aligned}
&\varepsilon^{\kappa\mu_1\mu_2\mu_3}p_1^{\nu_3}, && \cancel{\varepsilon^{p_1\kappa\mu_2\mu_3}p_2^{\mu_1}p_1^{\nu_3}}, && \cancel{\varepsilon^{p_2\kappa\mu_2\mu_3}p_2^{\mu_1}p_1^{\nu_3}}, && \varepsilon^{p_1\kappa\mu_2\mu_3}\delta^{\mu_1\nu_3}, && \cancel{\varepsilon^{p_2\kappa\mu_2\mu_3}\delta^{\mu_1\nu_3}}\\
&\cancel{\varepsilon^{p_1\kappa\mu_1\mu_3}p_3^{\mu_2}p_1^{\nu_3}}, && \cancel{\varepsilon^{p_2\kappa\mu_1\mu_3}p_3^{\mu_2}p_1^{\nu_3}}, && \cancel{\varepsilon^{p_1\kappa\mu_1\mu_3}\delta^{\mu_2\nu_3}}, && \varepsilon^{p_2\kappa\mu_1\mu_3}\delta^{\mu_2\nu_3}, && \cancel{\varepsilon^{p_1\kappa\mu_1\mu_2}p_1^{\mu_3}p_1^{\nu_3}},\\
&\cancel{\varepsilon^{p_2\kappa\mu_1\mu_2}p_1^{\mu_3}p_1^{\nu_3}}, && \varepsilon^{p_1p_2\kappa\mu_3}\delta^{\mu_1\mu_2}p_1^{\nu_3}, && \cancel{\varepsilon^{p_1p_2\kappa\mu_3}p_2^{\mu_1}p_3^{\mu_2}p_1^{\nu_3}}, && \cancel{\varepsilon^{p_1p_2\kappa\mu_3}\delta^{\nu_3\mu_2}p_2^{\mu_1}}, && \cancel{\varepsilon^{p_1p_2\kappa\mu_3}\delta^{\nu_3\mu_1}p_3^{\mu_2}},\\
&\cancel{\varepsilon^{p_1p_2\kappa\mu_2}\delta^{\mu_1\mu_3}p_1^{\nu_3}}, && \cancel{\varepsilon^{p_1p_2\kappa\mu_2}p_2^{\mu_1}p_1^{\mu_3}p_1^{\nu_3}}, && \cancel{\varepsilon^{p_1p_2\kappa\mu_1}\delta^{\mu_3\mu_2}p_1^{\nu_3}}, && \cancel{\varepsilon^{p_1p_2\kappa\mu_1}p_1^{\nu_3}p_1^{\mu_3}p_3^{\mu_2}},\\
&\cancel{\delta^{\mu_3\kappa}\varepsilon^{\mu_1\mu_2p_1p_2}p_1^{\nu_3}}, && \cancel{\delta^{\mu_3\kappa}\varepsilon^{\nu_3\mu_1p_1p_2}p_3^{\mu_2}}, && \cancel{\delta^{\mu_3\kappa}\varepsilon^{\nu_3\mu_2p_1p_2}p_2^{\mu_1}}, && \delta^{\mu_3\kappa}\varepsilon^{p_1\mu_1\mu_2\nu_3}, && \delta^{\mu_3\kappa}\varepsilon^{p_2\mu_1\mu_2\nu_3},\\
&\delta^{\mu_2\kappa}\varepsilon^{\mu_1\mu_3p_1p_2}p_1^{\nu_3}, && \delta^{\mu_1\kappa}\varepsilon^{\mu_2\mu_3p_1p_2}p_1^{\nu_3}
\end{aligned} \tag{C.3}$$

Not all of these tensors are independent. Indeed all barred tensors can be rewritten in terms of the non-barred ones thanks to the Schouten identities. Ignoring all the barred tensors, we then end up with the minimal decomposition in eq. (C.1). We will now list all the Schouten identities needed in order to eliminate the barred tensor structures. The first one we consider is

$$\varepsilon^{[p_1p_2\mu_1\mu_2}\delta^{\mu_3]}_{\alpha} = 0 \tag{C.4}$$

which can be contracted with $p_{1\alpha}$, $p_{2\alpha}$, $\delta^{\kappa}_{\alpha}$ and $\delta^{\nu_3}_{\alpha}$ obtaining

$$\begin{aligned}
&\pi^{\alpha_1}_{\mu_1}\pi^{\alpha_2}_{\mu_2}\Pi^{\alpha_3\beta_3}_{\mu_3\nu_3}\left(\varepsilon^{p_1p_2\mu_1\mu_3}p_3^{\mu_2}\right)=\pi^{\alpha_1}_{\mu_1}\pi^{\alpha_2}_{\mu_2}\Pi^{\alpha_3\beta_3}_{\mu_3\nu_3}\left(-\frac{p_1^2+p_2^2-p_3^2}{2}\varepsilon^{p_1\mu_1\mu_2\mu_3}-p_1^2\varepsilon^{p_2\mu_1\mu_2\mu_3}-\varepsilon^{p_1p_2\mu_1\mu_2}p_1^{\mu_3}\right)\\
&\pi^{\alpha_1}_{\mu_1}\pi^{\alpha_2}_{\mu_2}\Pi^{\alpha_3\beta_3}_{\mu_3\nu_3}\left(\varepsilon^{p_1p_2\mu_2\mu_3}p_2^{\mu_1}\right)=\pi^{\alpha_1}_{\mu_1}\pi^{\alpha_2}_{\mu_2}\Pi^{\alpha_3\beta_3}_{\mu_3\nu_3}\left(\frac{p_1^2+p_2^2-p_3^2}{2}\varepsilon^{p_2\mu_1\mu_2\mu_3}+p_2^2\varepsilon^{p_1\mu_1\mu_2\mu_3}+\varepsilon^{p_1p_2\mu_1\mu_2}p_1^{\mu_3}\right)\\
&\pi^{\alpha_1}_{\mu_1}\pi^{\alpha_2}_{\mu_2}\Pi^{\alpha_3\beta_3}_{\mu_3\nu_3}\left(\varepsilon^{p_1p_2\mu_1\mu_2}\delta^{\kappa\mu_3}\right)=\pi^{\alpha_1}_{\mu_1}\pi^{\alpha_2}_{\mu_2}\Pi^{\alpha_3\beta_3}_{\mu_3\nu_3}\left(-\varepsilon^{p_2\mu_1\mu_2\mu_3}p_1^{\kappa}+\varepsilon^{p_1\mu_1\mu_2\mu_3}p_2^{\kappa}-\varepsilon^{p_1p_2\mu_2\mu_3}\delta^{\kappa\mu_1}+\varepsilon^{p_1p_2\mu_1\mu_3}\delta^{\kappa\mu_2}\right)\\
&\pi^{\alpha_1}_{\mu_1}\pi^{\alpha_2}_{\mu_2}\Pi^{\alpha_3\beta_3}_{\mu_3\nu_3}\left(\varepsilon^{p_1p_2\mu_1\mu_3}\delta^{\mu_2\nu_3}\right)=\pi^{\alpha_1}_{\mu_1}\pi^{\alpha_2}_{\mu_2}\Pi^{\alpha_3\beta_3}_{\mu_3\nu_3}\left(\varepsilon^{p_1\mu_1\mu_2\mu_3}p_1^{\nu_3}+\varepsilon^{p_2\mu_1\mu_2\mu_3}p_1^{\nu_3}+\varepsilon^{p_1p_2\mu_2\mu_3}\delta^{\mu_1\nu_3}\right)
\end{aligned}\tag{C.5}$$

Then we consider the identities

$$\begin{aligned}
&\varepsilon^{[p_1\mu_1\mu_2\mu_3}\delta^{\kappa]}_{\alpha}=0,\\
&\varepsilon^{[p_2\mu_1\mu_2\mu_3}\delta^{\kappa]}_{\alpha}=0
\end{aligned}\tag{C.6}$$

which contracted with $p_{1\alpha}$, $p_{2\alpha}$ and $\delta^{\nu_3}_{\alpha}$ give

$$\begin{aligned}
&\pi^{\alpha_1}_{\mu_1}\pi^{\alpha_2}_{\mu_2}\Pi^{\alpha_3\beta_3}_{\mu_3\nu_3}\left(\varepsilon^{p_1\kappa\mu_1\mu_3}p_3^{\mu_2}\right)=\pi^{\alpha_1}_{\mu_1}\pi^{\alpha_2}_{\mu_2}\Pi^{\alpha_3\beta_3}_{\mu_3\nu_3}\left(-p_1^2\varepsilon^{\kappa\mu_1\mu_2\mu_3}+\varepsilon^{p_1\mu_1\mu_2\mu_3}p_1^{\kappa}-\varepsilon^{p_1\kappa\mu_1\mu_2}p_1^{\mu_3}\right)\\
&\pi^{\alpha_1}_{\mu_1}\pi^{\alpha_2}_{\mu_2}\Pi^{\alpha_3\beta_3}_{\mu_3\nu_3}\left(\varepsilon^{p_2\kappa\mu_2\mu_3}p_2^{\mu_1}\right)=\pi^{\alpha_1}_{\mu_1}\pi^{\alpha_2}_{\mu_2}\Pi^{\alpha_3\beta_3}_{\mu_3\nu_3}\left(-p_2^2\varepsilon^{\kappa\mu_1\mu_2\mu_3}+\varepsilon^{p_2\kappa\mu_1\mu_2}p_1^{\mu_3}+\varepsilon^{p_2\mu_1\mu_2\mu_3}p_2^{\kappa}\right)\\
&\pi^{\alpha_1}_{\mu_1}\pi^{\alpha_2}_{\mu_2}\Pi^{\alpha_3\beta_3}_{\mu_3\nu_3}\left(\varepsilon^{p_1\kappa\mu_1\mu_2}p_1^{\mu_3}\right)=\pi^{\alpha_1}_{\mu_1}\pi^{\alpha_2}_{\mu_2}\Pi^{\alpha_3\beta_3}_{\mu_3\nu_3}\left(-\frac{p_1^2+p_2^2-p_3^2}{2}\varepsilon^{\kappa\mu_1\mu_2\mu_3}-\varepsilon^{p_1\mu_1\mu_2\mu_3}p_2^{\kappa}+\varepsilon^{p_1\kappa\mu_2\mu_3}p_2^{\mu_1}\right)\\
&\pi^{\alpha_1}_{\mu_1}\pi^{\alpha_2}_{\mu_2}\Pi^{\alpha_3\beta_3}_{\mu_3\nu_3}\left(\varepsilon^{p_2\kappa\mu_1\mu_2}p_1^{\mu_3}\right)=\pi^{\alpha_1}_{\mu_1}\pi^{\alpha_2}_{\mu_2}\Pi^{\alpha_3\beta_3}_{\mu_3\nu_3}\left(\frac{p_1^2+p_2^2-p_3^2}{2}\varepsilon^{\kappa\mu_1\mu_2\mu_3}+\varepsilon^{p_2\mu_1\mu_2\mu_3}p_1^{\kappa}-\varepsilon^{p_2\kappa\mu_1\mu_3}p_3^{\mu_2}\right)
\end{aligned}\tag{C.7}$$

$$\begin{aligned}
&\pi^{\alpha_1}_{\mu_1}\pi^{\alpha_2}_{\mu_2}\Pi^{\alpha_3\beta_3}_{\mu_3\nu_3}\left(\varepsilon^{p_1\kappa\mu_1\mu_3}\delta^{\mu_2\nu_3}\right)=\pi^{\alpha_1}_{\mu_1}\pi^{\alpha_2}_{\mu_2}\Pi^{\alpha_3\beta_3}_{\mu_3\nu_3}\left(\varepsilon^{\kappa\mu_1\mu_2\mu_3}p_1^{\nu_3}-\varepsilon^{p_1\mu_1\mu_2\mu_3}\delta^{\kappa\nu_3}+\varepsilon^{p_1\kappa\mu_2\mu_3}\delta^{\mu_1\nu_3}\right)\\
&\pi^{\alpha_1}_{\mu_1}\pi^{\alpha_2}_{\mu_2}\Pi^{\alpha_3\beta_3}_{\mu_3\nu_3}\left(\varepsilon^{p_2\kappa\mu_2\mu_3}\delta^{\mu_1\nu_3}\right)=\pi^{\alpha_1}_{\mu_1}\pi^{\alpha_2}_{\mu_2}\Pi^{\alpha_3\beta_3}_{\mu_3\nu_3}\left(\varepsilon^{\kappa\mu_1\mu_2\mu_3}p_1^{\nu_3}+\varepsilon^{p_2\mu_1\mu_2\mu_3}\delta^{\kappa\nu_3}+\varepsilon^{p_2\kappa\mu_1\mu_3}\delta^{\mu_2\nu_3}\right)
\end{aligned}\tag{C.8}$$

The identity

$$\varepsilon^{[p_1p_2\mu_1\mu_2}\delta^{\kappa]}_{\alpha}=0\tag{C.9}$$

contracted with $p_{1\alpha}$ and $p_{2\alpha}$ give

$$\begin{aligned}
&\pi^{\alpha_1}_{\mu_1}\pi^{\alpha_2}_{\mu_2}\Pi^{\alpha_3\beta_3}_{\mu_3\nu_3}\left(\varepsilon^{p_1p_2\kappa\mu_1}p_3^{\mu_2}\right)=\pi^{\alpha_1}_{\mu_1}\pi^{\alpha_2}_{\mu_2}\Pi^{\alpha_3\beta_3}_{\mu_3\nu_3}\left(\frac{p_1^2+p_2^2-p_3^2}{2}\varepsilon^{p_1\kappa\mu_1\mu_2}+p_1^2\varepsilon^{p_2\kappa\mu_1\mu_2}+\varepsilon^{p_1p_2\mu_1\mu_2}p_1^{\kappa}\right)\\
&\pi^{\alpha_1}_{\mu_1}\pi^{\alpha_2}_{\mu_2}\Pi^{\alpha_3\beta_3}_{\mu_3\nu_3}\left(\varepsilon^{p_1p_2\kappa\mu_2}p_2^{\mu_1}\right)=\pi^{\alpha_1}_{\mu_1}\pi^{\alpha_2}_{\mu_2}\Pi^{\alpha_3\beta_3}_{\mu_3\nu_3}\left(-\frac{p_1^2+p_2^2-p_3^2}{2}\varepsilon^{p_2\kappa\mu_1\mu_2}-p_2^2\varepsilon^{p_1\kappa\mu_1\mu_2}+\varepsilon^{p_1p_2\mu_1\mu_2}p_2^{\kappa}\right)
\end{aligned}\tag{C.10}$$

It is worth mentioning that one can contract eq. (C.9) with $\delta^{\mu_3}_{\alpha}$ or $\delta^{\nu_3}_{\alpha}$ obtaining other identities that are not independent taking in consideration all the identities of this section. Lastly, we consider the identities

$$\begin{aligned}
&\varepsilon^{[p_1p_2\mu_1\mu_3}\delta^{\kappa]}_{\alpha}=0,\\
&\varepsilon^{[p_1p_2\mu_2\mu_3}\delta^{\kappa]}_{\alpha}=0
\end{aligned}\tag{C.11}$$

Contracting the first identity with $\delta^{\mu_2}_{\alpha}$ and the second one with $\delta^{\mu_1}_{\alpha}$ we obtain

$$\begin{aligned}
\pi^{\alpha_1}_{\mu_1}\pi^{\alpha_2}_{\mu_2}\Pi^{\alpha_3\beta_3}_{\mu_3\nu_3}\Big(\varepsilon^{p_2\kappa\mu_1\mu_3}p_3^{\mu_2}\Big) &= \pi^{\alpha_1}_{\mu_1}\pi^{\alpha_2}_{\mu_2}\Pi^{\alpha_3\beta_3}_{\mu_3\nu_3}\Big(\varepsilon^{p_1p_2\mu_1\mu_3}\delta^{\kappa\mu_2}-\varepsilon^{p_1p_2\kappa\mu_3}\delta^{\mu_1\mu_2}+\varepsilon^{p_1p_2\kappa\mu_1}\delta^{\mu_2\mu_3}\Big)\\
\pi^{\alpha_1}_{\mu_1}\pi^{\alpha_2}_{\mu_2}\Pi^{\alpha_3\beta_3}_{\mu_3\nu_3}\Big(\varepsilon^{p_1\kappa\mu_2\mu_3}p_2^{\mu_1}\Big) &= \pi^{\alpha_1}_{\mu_1}\pi^{\alpha_2}_{\mu_2}\Pi^{\alpha_3\beta_3}_{\mu_3\nu_3}\Big(\varepsilon^{p_1p_2\mu_2\mu_3}\delta^{\kappa\mu_1}-\varepsilon^{p_1p_2\kappa\mu_3}\delta^{\mu_1\mu_2}+\varepsilon^{p_1p_2\kappa\mu_2}\delta^{\mu_1\mu_3}\Big)
\end{aligned}\tag{C.12}$$

while, if we contract the first identity with $p_{2\alpha}$ and the second one with $p_{1\alpha}$, we obtain

$$\begin{aligned}
&\pi^{\alpha_1}_{\mu_1}\pi^{\alpha_2}_{\mu_2}\Pi^{\alpha_3\beta_3}_{\mu_3\nu_3}\Big(\varepsilon^{p_1p_2\kappa\mu_3}p_2^{\mu_1}\Big)p_3^{\mu_2}p_1^{\nu_3} =\\
&\qquad \pi^{\alpha_1}_{\mu_1}\pi^{\alpha_2}_{\mu_2}\Pi^{\alpha_3\beta_3}_{\mu_3\nu_3}\Big(-p_2^2\varepsilon^{p_1\kappa\mu_1\mu_3}-\frac{p_1^2+p_2^2-p_3^2}{2}\varepsilon^{p_2\kappa\mu_1\mu_3}-\varepsilon^{p_1p_2\kappa\mu_1}p_1^{\mu_3}+\varepsilon^{p_1p_2\mu_1\mu_3}p_2^{\kappa}\Big)p_3^{\mu_2}p_1^{\nu_3}\\
&\pi^{\alpha_1}_{\mu_1}\pi^{\alpha_2}_{\mu_2}\Pi^{\alpha_3\beta_3}_{\mu_3\nu_3}\Big(\varepsilon^{p_1p_2\kappa\mu_3}p_2^{\mu_1}\Big)\delta^{\mu_2\nu_3} =\\
&\qquad \pi^{\alpha_1}_{\mu_1}\pi^{\alpha_2}_{\mu_2}\Pi^{\alpha_3\beta_3}_{\mu_3\nu_3}\Big(-p_2^2\varepsilon^{p_1\kappa\mu_1\mu_3}-\frac{p_1^2+p_2^2-p_3^2}{2}\varepsilon^{p_2\kappa\mu_1\mu_3}-\varepsilon^{p_1p_2\kappa\mu_1}p_1^{\mu_3}+\varepsilon^{p_1p_2\mu_1\mu_3}p_2^{\kappa}\Big)\delta^{\mu_2\nu_3}\\
&\pi^{\alpha_1}_{\mu_1}\pi^{\alpha_2}_{\mu_2}\Pi^{\alpha_3\beta_3}_{\mu_3\nu_3}\Big(\varepsilon^{p_1p_2\kappa\mu_3}p_3^{\mu_2}\Big)\delta^{\mu_1\nu_3} =\\
&\qquad \pi^{\alpha_1}_{\mu_1}\pi^{\alpha_2}_{\mu_2}\Pi^{\alpha_3\beta_3}_{\mu_3\nu_3}\Big(-\frac{p_1^2+p_2^2-p_3^2}{2}\varepsilon^{p_1\kappa\mu_2\mu_3}-p_1^2\varepsilon^{p_2\kappa\mu_2\mu_3}-\varepsilon^{p_1p_2\mu_2\mu_3}p_1^{\kappa}-\varepsilon^{p_1p_2\kappa\mu_2}p_1^{\mu_3}\Big)\delta^{\mu_1\nu_3}
\end{aligned}\tag{C.13}$$

and if we contract the first identity with $p_{1\alpha}$ and the second one with $p_{2\alpha}$ we arrive to

$$\begin{aligned}
&\pi^{\alpha_1}_{\mu_1}\pi^{\alpha_2}_{\mu_2}\Pi^{\alpha_3\beta_3}_{\mu_3\nu_3}\Big(\varepsilon^{p_1p_2\kappa\mu_1}p_1^{\mu_3}\delta^{\mu_2\nu_3}\Big) =\\
&\qquad \pi^{\alpha_1}_{\mu_1}\pi^{\alpha_2}_{\mu_2}\Pi^{\alpha_3\beta_3}_{\mu_3\nu_3}\Big(-\frac{p_1^2+p_2^2-p_3^2}{2}\varepsilon^{p_1\kappa\mu_1\mu_3}\delta^{\mu_2\nu_3}-p_1^2\varepsilon^{p_2\kappa\mu_1\mu_3}\delta^{\mu_2\nu_3}-\varepsilon^{p_1p_2\mu_1\mu_3}p_1^{\kappa}\delta^{\mu_2\nu_3}\Big)\\
&\pi^{\alpha_1}_{\mu_1}\pi^{\alpha_2}_{\mu_2}\Pi^{\alpha_3\beta_3}_{\mu_3\nu_3}\Big(\varepsilon^{p_1p_2\kappa\mu_2}p_1^{\mu_3}\delta^{\mu_1\nu_3}\Big) =\\
&\qquad \pi^{\alpha_1}_{\mu_1}\pi^{\alpha_2}_{\mu_2}\Pi^{\alpha_3\beta_3}_{\mu_3\nu_3}\Big(-\frac{p_1^2+p_2^2-p_3^2}{2}\varepsilon^{p_2\kappa\mu_2\mu_3}\delta^{\mu_1\nu_3}-p_2^2\varepsilon^{p_1\kappa\mu_2\mu_3}\delta^{\mu_1\nu_3}+\varepsilon^{p_1p_2\mu_2\mu_3}p_2^{\kappa}\delta^{\mu_1\nu_3}\Big)
\end{aligned}\tag{C.14}$$

If instead we contract both the identities with $\delta^{\nu_3}_{\alpha}$, we don't obtain new independent relations.
In addition to all the identites we have written in this section, we also need to consider all the equations obtained from such relations exchanging $\mu_3 \leftrightarrow \nu_3$.

## C.4 Identities for the $\langle TTJ_A\rangle$ at finite temperature and density

When examining the gravitational anomaly corrections at finite density and temperature, we have introduced the rank-4 tensors $W^{\mu_1\nu_1\mu_2\nu_2}_{i,j}$. Such tensors are parity-odd and depend on the graviton momenta $p_1$ and $p_2$, as well as the momentum of the fermionic loop $l$. There is a set of tensorial relations, known as Schouten identities, which can be used to reduce the expression of $W^{\mu_1\nu_1\mu_2\nu_2}_{i,j}$. These identities arise from the dimensional degeneracies of tensor structures, given that we are working in $d=4$. In particular, in this case we can construct such identities starting from the following format

$$0=\varepsilon^{[l\,p_1p_2\mu_1}\delta^{\mu_2]\alpha}\tag{C.1}$$

Since we are antisymmetrizing over 5 indices and we are working in four dimensions, the result must vanish. The index $\alpha$ can be contracted with a momentum, obtaining

$$\begin{aligned} 0 &= \varepsilon^{[l\,p_1 p_2 \mu_1} p_1^{\mu_2]} \\ 0 &= \varepsilon^{[l\,p_1 p_2 \mu_1} p_2^{\mu_2]} \\ 0 &= \varepsilon^{[l\,p_1 p_2 \mu_1} l^{\mu_2]} \end{aligned} \tag{C.2}$$

or we can pick $\alpha = \{\nu_1, \nu_2\}$

$$\begin{aligned} 0 &= \varepsilon^{[l\,p_1 p_2 \mu_1} \delta^{\mu_2]\nu_1} \\ 0 &= \varepsilon^{[l\,p_1 p_2 \mu_1} \delta^{\mu_2]\nu_2} \end{aligned} \tag{C.3}$$

Note that we do not need to consider Schouten identities where both $\mu_1$ and $\nu_1$ are antisymmetrized since the energy-momentum tensor (and therefore $W_{i,j}^{\mu_1\nu_1\mu_2\nu_2}$) is symmetric under the exchange $\mu_1 \leftrightarrow \nu_1$. The same is true for the indices $\mu_2$ and $\nu_2$.
The identity (C.2) and (C.3) relates tensors with rank less than four. Therefore, we need to complete them with the remaining indices. As an example we can pick the first identity in eq. (C.3), which is a rank-3 equation, and we multiply it with $p_1^{\nu_2}$, $p_2^{\nu_2}$ or $l^{\nu_2}$ ending up with three relations between rank-4 tensors. Proceeding in such way, we end up with 36 identities. Moreover, we can also get new identities from this 36 ones by exchanging $\mu_1 \leftrightarrow \nu_1$ and/or $\mu_2 \leftrightarrow \nu_2$. Therefore, the total number of Schouten identities is $36\times 4 = 144$. All these identities can be used to simplify the structures appearing in the $W$ tensors.

# Appendix D

# The $\langle J_V J_V J_A \rangle$ at finite density

## D.1 Summary on the tensorial decompositions

### D.1.1 General case

In this section, we summarize the steps required to derive the minimal tensorial decomposition for the chiral anomaly interaction.
We start with 60 parity-odd tensor structures and form factors, introduced in eq. (7.4.2). Next, we impose 32 Schouten identities, which reduce the number of form factors to 28, as shown in eq. (7.4.9)

**Schouten Identities:**
**60 → 28.**

The second step involves imposing Bose symmetry on the external photons. We require the vertex $\Gamma^{\lambda\mu\nu}$ to be invariant under the exchange $p_1, \mu \longleftrightarrow p_2, \nu$. This symmetry reduces the expression to 16 form factors, as shown in eq. (7.4.10)

**Bose symmetry:**
**28 → 16.**

In the third step, we impose the vectorial Ward identities which further reduce the number of form factors to 10, as indicated in Eq. (7.4.12):

**Vectorial WIs:**
**16 → 10.**

### D.1.2 Symmetric case

Alternatively one can derive our decomposition of the vertex under the following symmetric assumptions

$$p_1 \cdot \eta = p_2 \cdot \eta = p \cdot \eta, \tag{D.1}$$

$$p_1^2 = p_2^2 = p^2, \tag{D.2}$$

which lead to further reductions in the number of form factors.
The reduction due to Schouten identities remains unchanged

**Schouten Identities:**
**60 → 28.**

However, the symmetric assumptions affect the subsequent steps of the derivation—specifically, the Bose symmetry and vectorial Ward identities. The Bose symmetry conditions now reduce the form factors to 12, as shown in eq. (7.5.3)

**Bose symmetry:**
**28 → 12.**

Finally, the vectorial Ward identities further reduce the number of form factors to 7

**Vectorial WIs:**
**12 → 7.**

In conclusion, under these assumptions, we can express the chiral interaction as shown in eq. (7.5.6).

## D.2 Tensorial structures

In this section, we list the tensor structures that can appear in the expansion of the chiral anomaly vertex at finite temperature and density in eq. (7.4.2), before imposing any symmetry constraint. They are

$$
\begin{array}{lllll}
\varepsilon^{\lambda\mu\nu p_1} & \varepsilon^{\lambda\mu\nu p_2} & \varepsilon^{\lambda\mu\nu\eta} & p_1{}^{\lambda}\varepsilon^{\mu\nu p_1 p_2} & p_1{}^{\lambda}\varepsilon^{\mu\nu p_1\eta} \\
p_1{}^{\lambda}\varepsilon^{\mu\nu p_2\eta} & p_2{}^{\lambda}\varepsilon^{\mu\nu p_1 p_2} & p_2{}^{\lambda}\varepsilon^{\mu\nu p_1\eta} & p_2{}^{\lambda}\varepsilon^{\mu\nu p_2\eta} & \eta^{\lambda}\varepsilon^{\mu\nu p_1 p_2} \\
\eta^{\lambda}\varepsilon^{\mu\nu p_1\eta} & \eta^{\lambda}\varepsilon^{\mu\nu p_2\eta} & p_1{}^{\mu}\varepsilon^{\lambda\nu p_1 p_2} & p_1{}^{\mu}\varepsilon^{\lambda\nu p_1\eta} & p_1{}^{\mu}\varepsilon^{\lambda\nu p_2\eta} \\
p_2{}^{\mu}\varepsilon^{\lambda\nu p_1 p_2} & p_2{}^{\mu}\varepsilon^{\lambda\nu p_1\eta} & p_2{}^{\mu}\varepsilon^{\lambda\nu p_2\eta} & \eta^{\mu}\varepsilon^{\lambda\nu p_1 p_2} & \eta^{\mu}\varepsilon^{\lambda\nu p_1\eta} \\
\eta^{\mu}\varepsilon^{\lambda\nu p_2\eta} & p_1{}^{\nu}\varepsilon^{\lambda\mu p_1 p_2} & p_1{}^{\nu}\varepsilon^{\lambda\mu p_1\eta} & p_1{}^{\nu}\varepsilon^{\lambda\mu p_2\eta} & p_2{}^{\nu}\varepsilon^{\lambda\mu p_1 p_2} \\
p_2{}^{\nu}\varepsilon^{\lambda\mu p_1\eta} & p_2{}^{\nu}\varepsilon^{\lambda\mu p_2\eta} & \eta^{\nu}\varepsilon^{\lambda\mu p_1 p_2} & \eta^{\nu}\varepsilon^{\lambda\mu p_1\eta} & \eta^{\nu}\varepsilon^{\lambda\mu p_2\eta} \\
p_1{}^{\lambda}p_1{}^{\nu}\varepsilon^{\mu p_1 p_2\eta} & p_1{}^{\nu}p_2{}^{\lambda}\varepsilon^{\mu p_1 p_2\eta} & \eta^{\lambda}p_1{}^{\nu}\varepsilon^{\mu p_1 p_2\eta} & p_1{}^{\lambda}p_2{}^{\nu}\varepsilon^{\mu p_1 p_2\eta} & p_2{}^{\lambda}p_2{}^{\nu}\varepsilon^{\mu p_1 p_2\eta} \\
\eta^{\lambda}p_2{}^{\nu}\varepsilon^{\mu p_1 p_2\eta} & \eta^{\nu}p_1{}^{\lambda}\varepsilon^{\mu p_1 p_2\eta} & \eta^{\nu}p_2{}^{\lambda}\varepsilon^{\mu p_1 p_2\eta} & \eta^{\lambda}\eta^{\nu}\varepsilon^{\mu p_1 p_2\eta} & \delta^{\lambda\nu}\varepsilon^{\mu p_1 p_2\eta} \\
p_1{}^{\lambda}p_1{}^{\mu}\varepsilon^{\nu p_1 p_2\eta} & p_1{}^{\mu}p_2{}^{\lambda}\varepsilon^{\nu p_1 p_2\eta} & \eta^{\lambda}p_1{}^{\mu}\varepsilon^{\nu p_1 p_2\eta} & p_1{}^{\lambda}p_2{}^{\mu}\varepsilon^{\nu p_1 p_2\eta} & p_2{}^{\lambda}p_2{}^{\mu}\varepsilon^{\nu p_1 p_2\eta} \\
\eta^{\lambda}p_2{}^{\mu}\varepsilon^{\nu p_1 p_2\eta} & \eta^{\mu}p_1{}^{\lambda}\varepsilon^{\nu p_1 p_2\eta} & \eta^{\mu}p_2{}^{\lambda}\varepsilon^{\nu p_1 p_2\eta} & \eta^{\lambda}\eta^{\mu}\varepsilon^{\nu p_1 p_2\eta} & \delta^{\lambda\mu}\varepsilon^{\nu p_1 p_2\eta} \\
p_1{}^{\mu}p_1{}^{\nu}\varepsilon^{\lambda p_1 p_2\eta} & p_1{}^{\mu}p_2{}^{\nu}\varepsilon^{\lambda p_1 p_2\eta} & \eta^{\nu}p_1{}^{\mu}\varepsilon^{\lambda p_1 p_2\eta} & p_1{}^{\nu}p_2{}^{\mu}\varepsilon^{\lambda p_1 p_2\eta} & p_2{}^{\mu}p_2{}^{\nu}\varepsilon^{\lambda p_1 p_2\eta} \\
\eta^{\nu}p_2{}^{\mu}\varepsilon^{\lambda p_1 p_2\eta} & \eta^{\mu}p_1{}^{\nu}\varepsilon^{\lambda p_1 p_2\eta} & \eta^{\mu}p_2{}^{\nu}\varepsilon^{\lambda p_1 p_2\eta} & \eta^{\mu}\eta^{\nu}\varepsilon^{\lambda p_1 p_2\eta} & \delta^{\mu\nu}\varepsilon^{\lambda p_1 p_2\eta}
\end{array}
\tag{D.1}
$$

## D.3 Bose Symmetry and conservation Ward identities

In this section, we list the constraints we impose on the decomposition (7.4.9). Imposing the Bose symmetry, we find the following set of relations between the form factors in the expansion

$$
\begin{array}{ll}
B_2(p_1,p_2,\eta) = -B_1(p_2,p_1,\eta) & B_3(p_2,p_1,\eta) = -B_3(p_1,p_2,\eta) \\
B_{13}(p_1,p_2,\eta) = -B_7(p_2,p_1,\eta) & B_{14}(p_1,p_2,\eta) = B_9(p_2,p_1,\eta) \\
B_{15}(p_1,p_2,\eta) = B_8(p_2,p_1,\eta) & B_{16}(p_1,p_2,\eta) = -B_4(p_2,p_1,\eta) \\
B_{17}(p_1,p_2,\eta) = B_6(p_2,p_1,\eta) & B_{18}(p_1,p_2,\eta) = B_5(p_2,p_1,\eta) \\
B_{19}(p_1,p_2,\eta) = -B_{10}(p_2,p_1,\eta) & B_{20}(p_1,p_2,\eta) = B_{12}(p_2,p_1,\eta) \\
B_{21}(p_1,p_2,\eta) = B_{11}(p_2,p_1,\eta) & B_{23}(p_2,p_1,\eta) = -B_{23}(p_1,p_2,\eta)
\end{array}
$$

$$B_{25}(p_1,p_2,\eta) = -B_{22}(p_2,p_1,\eta) \qquad B_{26}(p_1,p_2,\eta) = -B_{24}(p_2,p_1,\eta)$$
$$B_{27}(p_2,p_1,\eta) = -B_{27}(p_1,p_2,\eta) \qquad B_{28}(p_2,p_1,\eta) = -B_{28}(p_1,p_2,\eta) \tag{D.1}$$

that reduce their number to 16.
Further simplifications are introduced by requiring that the interaction satisfies the vector Ward identities

$$\begin{aligned}
B_1(p_2,p_1,\eta) &= -{p_1}^2 B_4(p_1,p_2,\eta) - (p_1\cdot p_2) B_7(p_1,p_2,\eta) - (\eta\cdot p_1) B_{10}(p_1,p_2,\eta) \\
B_3(p_1,p_2,\eta) &= -{p_1}^2{p_2}^2 B_{23}(p_2,p_1,\eta) + (p_1\cdot p_2) B_8(p_1,p_2,\eta) + {p_1}^2 (p_1\cdot p_2) B_{22}(p_1,p_2,\eta) + \\
&\quad {p_1}^2(\eta\cdot p_2) B_{24}(p_1,p_2,\eta) - {p_2}^2(\eta\cdot p_1) B_{24}(p_2,p_1,\eta) - (\eta\cdot p_1)(\eta\cdot p_2) B_{27}(p_2,p_1,\eta) \\
B_5(p_2,p_1,\eta) &= -{p_1}^2 B_{23}(p_1,p_2,\eta) + (p_1\cdot p_2) B_{22}(p_2,p_1,\eta) + (\eta\cdot p_1) B_{24}(p_2,p_1,\eta) \\
B_{11}(p_2,p_1,\eta) &= -{p_1}^2 B_{24}(p_1,p_2,\eta) - (\eta\cdot p_1) B_{27}(p_1,p_2,\eta) \\
B_{12}(p_1,p_2,\eta) &= -\frac{{p_1}^2 B_6(p_1,p_2,\eta)}{\eta\cdot p_1} - \frac{(p_1\cdot p_2) B_9(p_1,p_2,\eta)}{\eta\cdot p_1} \\
B_{28}(p_1,p_2,\eta) &= -{p_1}^2 B_{22}(p_1,p_2,\eta) - B_8(p_2,p_1,\eta).
\end{aligned} \tag{D.2}$$

## D.4 Scalar integrals: on-shell case

In this Appendix, we provide some details concerning the analysis of all the integrals that appear in (7.7.7). The procedure is based on the factorization of the radial and angular integrals in dimensional regularization. As explained in the main sections, we indicate by the index "1" the finite density contribution $G^{(1)}$ coming from the insertion of the fermion propagator in a given internal leg of the loop. "0" stands for the ordinary vacuum propagator $G^{(0)}$.
The resulting final integrals will be written using the following basic integrals

$$\Theta[f(t)] = \frac{\pi^{d/2-3/2}}{\Gamma[d/2-1/2]}\int_{-1}^{1} dt\, f(t)(1-t^2)^{d/2-2} \qquad R[f(|\mathbf{k}|)] = \int_0^{\mu} d|\mathbf{k}| f(|\mathbf{k}|) \tag{D.1}$$

We now list the expression of the integrals

$J[p_1\cdot k] = H[p_2\cdot k]$

$$\begin{aligned}
J^{(1,0,0)}[p_1\cdot k] &= -\frac{1}{(2\pi)^{d-1}}\frac{1}{16p_0} R\left[\frac{|\mathbf{k}|^{d-3}}{p_0+k}\right]\Theta[1] \\
J^{(0,1,0)}[p_1\cdot k] &= -\frac{1}{(2\pi)^{d-1}}\frac{1}{8p_0}\,\Theta\left[\frac{1}{1+t}\right] \\
J^{(0,0,1)}[p_1\cdot k] &= \frac{1}{(2\pi)^{d-1}}\frac{1}{16p_0} R\left[\frac{|\mathbf{k}|^{d-3}}{p_0+k}\right]\Theta\left[\frac{1-t}{1+t}\right] + \frac{1}{(2\pi)^{d-1}}\frac{1}{8} R\left[\frac{|\mathbf{k}|^{d-4}}{p_0+k}\right]\Theta\left[\frac{1}{1+t}\right] \\
J^{(1,0,1)}[p_1\cdot k] &= \frac{1}{(2\pi)^{d-2}}\frac{p_0^{d-4}}{16}\Theta[1]\,\theta(\mu-p_0)
\end{aligned} \tag{D.2}$$

$H[p_1\cdot k] = J[p_2\cdot k]$

$$H^{(1,0,0)}[p_1\cdot k] = -\frac{1}{(2\pi)^{d-1}}\frac{1}{16p_0} R\left[\frac{|\mathbf{k}|^{d-3}}{p_0-k}\right]\Theta\left[\frac{1-t}{1+t}\right]$$

$$
\begin{aligned}
H^{(0,1,0)}[p_1\cdot k]) &= -\frac{1}{(2\pi)^{d-1}}\frac{1}{8p_0}R[|\mathbf{k}|^{d-4}]\,\Theta\left[\frac{1}{1+t}\right]-\frac{1}{(2\pi)^{d-1}}\frac{1}{4}R[|\mathbf{k}|^{d-5}]\,\Theta\left[\frac{1}{1-t^2}\right]\\
H^{(0,0,1)}[p_1\cdot k]) &= \frac{1}{(2\pi)^{d-1}}\frac{1}{16p_0}R\left[\frac{|\mathbf{k}|^{d-3}}{p_0+k}\right]\Theta[1]+\frac{1}{(2\pi)^{d-1}}\frac{1}{8}R\left[\frac{|\mathbf{k}|^{d-4}}{p_0+k}\right]\Theta\left[\frac{1}{1-t}\right]\\
H^{(1,0,1)}[p_1\cdot k] &= \frac{1}{(2\pi)^{d-2}}\frac{p_0^{d-4}}{16}\Theta\left[\frac{1-t}{1+t}\right]\theta(\mu-p_0)
\end{aligned}
\tag{D.3}
$$

$J\,[p_1\cdot k\,k^2]=H\,[p_2\cdot k\,k^2]$

$$
\begin{aligned}
J^{(1,0,0)}[(p_1\cdot k)k^2] &= 0\\
J^{(0,1,0)}[(p_1\cdot k)k^2] &= -\frac{1}{(2\pi)^{d-1}}\frac{1}{4}R[|\mathbf{k}|^{d-3}]\,\Theta\left[\frac{1-t}{1+t}\right]\\
J^{(0,0,1)}[(p_1\cdot k)k^2] &= \frac{1}{(2\pi)^{d-1}}\frac{1}{4}R[|\mathbf{k}|^{d-3}]\,\Theta\left[\frac{1-t}{1+t}\right]+\frac{1}{(2\pi)^{d-1}}\frac{p_0}{2}R[|\mathbf{k}|^{d-4}]\,\Theta\left[\frac{1}{1+t}\right]
\end{aligned}
\tag{D.4}
$$

$H\,[p_1\cdot k\,k^2]=J\,[p_2\cdot k\,k^2]=0$
$J\,[(p_1\cdot k)^2]=H\,[(p_2\cdot k)^2]$

$$
\begin{aligned}
J^{(1,0,0)}[(p_1\cdot k)^2] &= -\frac{1}{(2\pi)^{d-1}}\frac{1}{16}R\left[\frac{|\mathbf{k}|^{d-2}}{p_0-k}\right]\Theta[1-t]\\
J^{(0,1,0)}[(p_1\cdot k)^2] &= -\frac{1}{(2\pi)^{d-1}}\frac{1}{8}R[|\mathbf{k}|^{d-3}]\,\Theta\left[\frac{1-t}{1+t}\right]\\
J^{(0,0,1)}[(p_1\cdot k)^2] &= \frac{1}{(2\pi)^{d-1}}\frac{1}{16}R\left[\frac{|\mathbf{k}|^{d-2}}{p_0+k}\right]\Theta\left[\frac{(1-t)^2}{1+t}\right]+\frac{1}{(2\pi)^{d-1}}\frac{p_0}{4}R\left[\frac{|\mathbf{k}|^{d-3}}{p_0+k}\right]\Theta\left[\frac{1-t}{1+t}\right]\\
&\quad+\frac{1}{(2\pi)^{d-1}}\frac{p_0^2}{4}R\left[\frac{|\mathbf{k}|^{d-4}}{p_0+k}\right]\Theta\left[\frac{1}{1+t}\right]\\
J^{(1,0,1)}[(p_1\cdot k)^2] &= \frac{1}{(2\pi)^{d-2}}\frac{p_0^{d-2}}{16}\Theta[1-t]\,\theta(\mu-p_0)
\end{aligned}
\tag{D.5}
$$

$J\,[p_1\cdot k\,p_2\cdot k]=H\,[p_1\cdot k\,p_2\cdot k]$

$$
\begin{aligned}
J^{(1,0,0)}[p_1\cdot k\,p_2\cdot k] &= -\frac{1}{(2\pi)^{d-1}}\frac{1}{16}R\left[\frac{|\mathbf{k}|^{d-2}}{p_0-k}\right]\Theta[1+t]\\
J^{(0,1,0)}[p_1\cdot k\,p_2\cdot k] &= -\frac{1}{(2\pi)^{d-1}}\frac{1}{8}R[|\mathbf{k}|^{d-3}]\,\Theta[1]-\frac{1}{(2\pi)^{d-1}}\frac{p_0}{4}R[|\mathbf{k}|^{d-4}]\,\Theta\left[\frac{1}{1+t}\right]\\
J^{(0,0,1)}[p_1\cdot k\,p_2\cdot k] &= \frac{1}{(2\pi)^{d-1}}\frac{1}{16}R\left[\frac{|\mathbf{k}|^{d-2}}{p_0+k}\right]\Theta[1-t]+\frac{1}{(2\pi)^{d-1}}\frac{p_0}{4}R\left[\frac{|\mathbf{k}|^{d-3}}{p_0+k}\right]\Theta\left[\frac{1}{1+t}\right]\\
&\quad+\frac{1}{(2\pi)^{d-1}}\frac{p_0^2}{4}R\left[\frac{|\mathbf{k}|^{d-4}}{p_0+k}\right]\Theta\left[\frac{1}{1+t}\right]\\
J^{(1,0,1)}[p_1\cdot k\,p_2\cdot k] &= \frac{1}{(2\pi)^{d-2}}\frac{p_0^{d-2}}{16}\Theta[1]\,\theta(\mu-p_0)
\end{aligned}
$$

(D.6)

$$H\,[(p_1\cdot k)^2] = J\,[(p_2\cdot k)^2]$$

$$\begin{aligned}
H^{(1,0,0)}[(p_1\cdot k)^2] &= -\frac{1}{(2\pi)^{d-1}}\frac{1}{16}R\left[\frac{|\mathbf{k}|^{d-2}}{p_0-k}\right]\Theta\left[\frac{(1-t)^2}{1+t}\right]\\
H^{(0,1,0)}[(p_1\cdot k)^2] &= -\frac{1}{(2\pi)^{d-1}}\frac{1}{8}R[|\mathbf{k}|^{d-3}]\Theta\left[\frac{1-t}{1+t}\right]-\frac{1}{(2\pi)^{d-1}}\frac{p_0}{2}R[|\mathbf{k}|^{d-4}]\Theta\left[\frac{1}{1+t}\right]\\
&\quad -\frac{1}{(2\pi)^{d-1}}\frac{p_0^2}{2}R[|\mathbf{k}|^{d-5}]\Theta\left[\frac{1}{1-t^2}\right]\\
H^{(0,0,1)}[(p_1\cdot k)^2] &= \frac{1}{(2\pi)^{d-1}}\frac{1}{16}R\left[\frac{|\mathbf{k}|^{d-2}}{p_0+k}\right]\Theta[1-t]+\frac{1}{(2\pi)^{d-1}}\frac{p_0}{4}R\left[\frac{|\mathbf{k}|^{d-3}}{p_0+k}\right]\Theta[1]\\
&\quad +\frac{1}{(2\pi)^{d-1}}\frac{p_0^2}{4}R\left[\frac{|\mathbf{k}|^{d-4}}{p_0+k}\right]\Theta\left[\frac{1}{1-t}\right]\\
H^{(1,0,1)}[(p_1\cdot k)^2] &= \frac{1}{(2\pi)^{d-2}}\frac{p_0^{d-2}}{16}\Theta\left[\frac{(1-t)^2}{1+t}\right]\theta(\mu-p_0)
\end{aligned}$$

(D.7)

Integrals of the type $(1,1,0)$ vanish, since the two Dirac deltas can be written as

$$\delta(k^2)\,\delta((k-p_1)^2)=\delta(k^2)\,\delta(-2k\cdot p_1)=\delta(k^2)\,\delta\,(-2|\mathbf{k}|p_0(1-\cos\theta))=\delta(k^2)\,\frac{\delta(1-\cos\theta)}{2|\mathbf{k}|p_0} \tag{D.8}$$

The second delta applied on the dimensional regularization factor $(1-t^2)^{d/2-2}$ yields zero, making the whole integral vanish. A similar argument can be made for $\delta(k^2)\,\delta((k-p_2)^2)$, that will be transformed into

$$\delta(k^2)\,\delta((k-p_2)^2)=\delta(k^2)\,\frac{\delta(1+\cos\theta)}{-2|\mathbf{k}|p_0} \tag{D.9}$$

Similarly, the $(0,1,1)$ integrals vanish, since after a translation $k\to k+q$, the deltas can be expressed as

$$\delta((k-q)^2)\,\delta((k-p_1)^2)=\delta((k-q)^2)\,\delta((k+p_1)^2)=\delta(k^2)\,\delta(2k\cdot p_1)=\delta(k^2)\,\frac{\delta(1-\cos\theta)}{-2|\mathbf{k}|p_0}. \tag{D.10}$$

The $(1,1,1)$ integrals are obviously zero as well.

### D.4.1 $d\to 4$ expansions

#### Angular integrals

$$\begin{aligned}
\Theta[1] &= \pi^{d/2-1}\frac{\Gamma(\frac{d}{2}-1)}{\Gamma(\frac{d-1}{2})^2}\to 4\\
\Theta[t] &= 0\\
\Theta[1+t] &= \Theta[1-t]=\Theta[1]\\
\Theta\left[\frac{1}{1+t}\right] &= \Theta\left[\frac{1}{1-t}\right]=\pi^{d/2-1}\frac{\Gamma(\frac{d}{2}-2)}{\Gamma(\frac{d-1}{2})\Gamma(\frac{d-3}{2})}\to\frac{4}{d-4}+2\,(\gamma-2+2\log(16\pi))+O(d-4)
\end{aligned}$$

$$
\begin{aligned}
\Theta\left[\frac{1-t}{1+t}\right] &= \Theta\left[\frac{1+t}{1-t}\right] = (d-2)\,\pi^{d/2-1}\frac{\Gamma(\frac{d}{2}-1)}{2\,\Gamma(\frac{d-1}{2})^2} \to \frac{8}{d-4} + 4(\gamma - 3 + \log(16\pi)) + O(d-4) \\
\Theta\left[\frac{(1-t)^2}{1+t}\right] &= \Theta\left[\frac{(1+t)^2}{1+t}\right] = d\,\pi^{d/2-1}\frac{\Gamma(\frac{d}{2}-1)}{2\,\Gamma(\frac{d-1}{2})^2} \to \frac{16}{d-4} + 4(2\gamma - 7 + 2\log(16\pi)) + O(d-4) \\
\Theta\left[\frac{1}{1-t^2}\right] &= \pi^{d/2-1}\frac{\Gamma(\frac{d}{2}-2)}{\Gamma(\frac{d-1}{2})\Gamma(\frac{d-3}{2})} \to \frac{4}{d-4} + 2(\gamma - 2 + 2\log(16\pi)) + O(d-4)
\end{aligned}
\tag{D.11}
$$

**Radial integrals**

$$
\begin{aligned}
&R\left[\frac{|\mathbf{k}|^{d-2}}{p_0+k}\right] \to \frac{\mu}{2}(\mu - 2p_0) + p_0^2\log\left(1+\frac{\mu}{p_0}\right) && R\left[\frac{|\mathbf{k}|^{d-2}}{p_0-k}\right] \to -\frac{\mu}{2}(\mu + 2p_0) - p_0^2\log\left(1-\frac{\mu}{p_0}\right) \\
&R\left[\frac{|\mathbf{k}|^{d-3}}{p_0+k}\right] \to \mu - p_0\log\left(1+\frac{\mu}{p_0}\right) && R\left[\frac{|\mathbf{k}|^{d-3}}{p_0-k}\right] \to -\mu - p_0\log\left(1-\frac{\mu}{p_0}\right) \\
&R\left[\frac{|\mathbf{k}|^{d-4}}{p_0+k}\right] \to \log\left(1+\frac{\mu}{p_0}\right) && R\left[\frac{|\mathbf{k}|^{d-4}}{p_0-k}\right] \to -\log\left(1-\frac{\mu}{p_0}\right) \\
&R[|\mathbf{k}|^{d-2}] \to \frac{\mu^3}{3} && R[|\mathbf{k}|^{d-3}] \to \frac{\mu^2}{2} \\
&R[|\mathbf{k}|^{d-4}] \to \mu && R[|\mathbf{k}|^{d-5}] \to \frac{1}{d-4} + \log\mu + O(d-4)
\end{aligned}
\tag{D.12}
$$

## D.5 Scalar integrals: off-shell symmetric case

In the following, we analyze the integrals of Appendix D.4 in the off-shell case. Here, we introduce the functions $\chi_\pm$, defined as

$$
\chi_\pm = \sqrt{p_0^2 - M^2}\left(\frac{p_0}{\sqrt{p_0^2-M^2}} \pm \cos\theta\right) = |\mathbf{p}|\left(\frac{p_0}{|\mathbf{p}|} \pm \cos\theta\right). \tag{D.1}
$$

The procedure is based on factorizing radial and angular integrals and then solving only the angular ones.

$H[p_1\cdot k]$

$$
H^{(1,0,0)}[p_1\cdot k] = -\frac{1}{8p_0}\int_0^\mu d|\mathbf{k}|\frac{|\mathbf{k}|}{p_0-|\mathbf{k}|}\int_{-1}^1 d\cos\theta\,\frac{\chi_-}{\chi_+} \tag{D.2}
$$

$$
H^{(0,1,0)}[p_1\cdot k] = \frac{2p_0^2 - M^2}{2}\int_0^\mu d|\mathbf{k}||\mathbf{k}|\int_{-1}^1 d\cos\theta\,\frac{1}{(M^2+2|\mathbf{k}|\chi_+)(M^2-2|\mathbf{k}|\chi_-)} \tag{D.3}
$$

$$
+\frac{1}{2}\int_0^\mu d|\mathbf{k}||\mathbf{k}|^2\int_{-1}^1 d\cos\theta\,\frac{\chi_-}{(M^2+2|\mathbf{k}|\chi_+)(M^2-2|\mathbf{k}|\chi_-)} \tag{D.4}
$$

$$
H^{(0,0,1)}[p_1\cdot k] = \frac{1}{8p_0}\int_0^\mu d|\mathbf{k}|\frac{|\mathbf{k}|^2}{p_0+|\mathbf{k}|}\int_{-1}^1 d\cos\theta\,\frac{\chi_-}{(M^2+2|\mathbf{k}|\chi_-)} \tag{D.5}
$$

$$+\frac{p_0}{4}\int_0^\mu d|\mathbf{k}|\frac{|\mathbf{k}|}{p_0+|\mathbf{k}|}\int_{-1}^1 d\cos\theta\frac{1}{(M^2+2|\mathbf{k}|\chi_-)} \tag{D.6}$$

$J[p_1\cdot k k^2]$

$$J^{(1,0,0)}[p_1\cdot k k^2] = 0 \tag{D.7}$$

$$J^{(0,1,0)}[p_1\cdot k k^2] = \int_0^\mu \frac{d|\mathbf{k}|}{(2\pi)^2}\frac{|\mathbf{k}|}{2}\int_{-1}^1 d\cos\theta\,\frac{M^2+|\mathbf{k}|\chi_-}{M^2-2|\mathbf{k}|\chi_-} \tag{D.8}$$

$$J^{(0,0,1)}[p_1\cdot k k^2] = \int_0^\mu \frac{d|\mathbf{k}|}{(2\pi)^2}\frac{|\mathbf{k}|}{2}\int_{-1}^1 d\cos\theta\,\frac{2p_0^2+|\mathbf{k}|\chi_-}{M^2+2|\mathbf{k}|\chi_+} \tag{D.9}$$

$J[p_2\cdot k k^2]$

$$J^{(1,0,0)}[p_2\cdot k k^2] = 0 \tag{D.10}$$

$$J^{(0,1,0)}[p_2\cdot k k^2] = \int_0^\mu \frac{d|\mathbf{k}|}{(2\pi)^2}\frac{|\mathbf{k}|}{2}\int_{-1}^1 d\cos\theta\,\frac{2p_0^2-M^2+|\mathbf{k}|\chi_-}{M^2-2|\mathbf{k}|\chi_+} \tag{D.11}$$

$$J^{(0,0,1)}[p_2\cdot k k^2] = \int_0^\mu \frac{d|\mathbf{k}|}{(2\pi)^2}\frac{|\mathbf{k}|}{2}\int_{-1}^1 d\cos\theta\,\frac{2p_0^2+|\mathbf{k}|\chi_+}{M^2+2|\mathbf{k}|\chi_+} \tag{D.12}$$

$J[(p_1\cdot k)^2]$

$$J^{(1,0,0)}[(p_1\cdot k)^2] = \frac{1}{8p_0}\int_0^\mu \frac{d|\mathbf{k}|}{(2\pi)^2}\frac{|\mathbf{k}|^3}{p_0-|\mathbf{k}|}\int_{-1}^1 d\cos\theta\,\frac{\chi_-^2}{M^2-2|\mathbf{k}|\chi_-} \tag{D.13}$$

$$J^{(0,1,0)}[(p_1\cdot k)^2] = \int_0^\mu \frac{d|\mathbf{k}|}{(2\pi)^2}\frac{|\mathbf{k}|}{2}\int_{-1}^1 d\cos\theta\,\frac{(M^2+|\mathbf{k}|\chi_-)^2}{(M^2+2|\mathbf{k}|\chi_-)(M^2-2|\mathbf{k}|\chi_+)} \tag{D.14}$$

$$J^{(0,0,1)}[(p_1\cdot k)^2] = \frac{1}{8p_0}\int_0^\mu \frac{d|\mathbf{k}|}{(2\pi)^2}\frac{|\mathbf{k}|}{p_0+|\mathbf{k}|}\int_{-1}^1 d\cos\theta\,\frac{(2p_0^2+|\mathbf{k}|\chi_-)^2}{M^2+2|\mathbf{k}|\chi_+} \tag{D.15}$$

$J[p_1\cdot k p_2\cdot k]$

$$J^{(1,0,0)}[p_1\cdot k p_2\cdot k] = \frac{1}{8p_0}\int_0^\mu \frac{d|\mathbf{k}|}{(2\pi)^2}\frac{|\mathbf{k}|^3}{p_0-|\mathbf{k}|}\int_{-1}^1 d\cos\theta\,\frac{\chi_-\chi_+}{M^2-2|\mathbf{k}|\chi_-} \tag{D.16}$$

$$J^{(0,1,0)}[p_1\cdot k p_2\cdot k] = \int_0^\mu \frac{d|\mathbf{k}|}{(2\pi)^2}\frac{|\mathbf{k}|}{2}\int_{-1}^1 d\cos\theta\,\frac{(M^2+|\mathbf{k}|\chi_-)(2p_0^2-M^2+|\mathbf{k}|\chi_+)}{(M^2+2|\mathbf{k}|\chi_-)(M^2-2|\mathbf{k}|\chi_+)} \tag{D.17}$$

$$J^{(0,0,1)}[p_1\cdot k p_2\cdot k] = \frac{1}{8p_0}\int_0^\mu \frac{d|\mathbf{k}|}{(2\pi)^2}\frac{|\mathbf{k}|}{p_0+|\mathbf{k}|}\int_{-1}^1 d\cos\theta\,\frac{(2p_0^2+|\mathbf{k}|\chi_-)(2p_0^2+|\mathbf{k}|\chi_+)}{M^2+2|\mathbf{k}|\chi_+} \tag{D.18}$$

$J[(p_2\cdot k)^2]$

$$J^{(1,0,0)}[(p_2\cdot k)^2] = \frac{1}{8p_0}\int_0^\mu \frac{d|\mathbf{k}|}{(2\pi)^2}\frac{|\mathbf{k}|^3}{p_0-|\mathbf{k}|}\int_{-1}^1 d\cos\theta\,\frac{\chi_+^2}{M^2-2|\mathbf{k}|\chi_-} \tag{D.19}$$

$$J^{(0,1,0)}[(p_2\cdot k)^2] = \int_0^\mu \frac{d|\mathbf{k}|}{(2\pi)^2}\frac{|\mathbf{k}|}{2}\int_{-1}^1 d\cos\theta\,\frac{(2p_0^2-M^2+|\mathbf{k}|\chi_+)^2}{(M^2+2|\mathbf{k}|\chi_-)(M^2-2|\mathbf{k}|\chi_+)} \tag{D.20}$$

$$J^{(0,0,1)}[(p_2\cdot k)^2] = \frac{1}{8p_0}\int_0^\mu \frac{d|\mathbf{k}|}{(2\pi)^2}\frac{|\mathbf{k}|}{p_0+|\mathbf{k}|}\int_{-1}^1 d\cos\theta\,\frac{(2p_0^2+|\mathbf{k}|\chi_+)^2}{M^2+2|\mathbf{k}|\chi_+} \tag{D.21}$$

### D.5.1 Angular integrals for $p^2 \neq 0$

$H[p_1 \cdot k]$

$$\int_{-1}^{1} d\cos\theta \frac{\chi_-}{\chi_+} = -\frac{2\left(|\mathbf{p}| + \log\left(\frac{p_0 - p}{|\mathbf{p}|+p_0}\right) p_0\right)}{|\mathbf{p}|} \tag{D.22}$$

$$\int_{-1}^{1} d\cos\theta \frac{1}{(M^2+2|\mathbf{k}|\chi_+)(M^2-2|\mathbf{k}|\chi_-)} = \frac{\log\left(\frac{\left(2|\mathbf{k}|(|\mathbf{p}|-p_0)+M^2\right)\left(2|\mathbf{k}|(p_0-|\mathbf{p}|)+M^2\right)}{M^4-4|\mathbf{k}|^2(|\mathbf{p}|+p_0)^2}\right)}{8|\mathbf{k}|^2|\mathbf{p}|p_0} \tag{D.23}$$

$$\begin{aligned}\int_{-1}^{1} d\cos\theta \frac{\chi_-}{(M^2+2|\mathbf{k}|\chi_+)(M^2-2|\mathbf{k}|\chi_-)} &= \frac{|\mathbf{k}|p_0 \log\left(\frac{\left(2|\mathbf{k}||\mathbf{p}|-2|\mathbf{k}|p_0+M^2\right)\left(-2|\mathbf{k}||\mathbf{p}|+2|\mathbf{k}|p_0+M^2\right)}{M^4-4|\mathbf{k}|^2(|\mathbf{p}|+p_0)^2}\right)}{8|\mathbf{k}|^3|\mathbf{p}|p_0} \\ &+ \frac{\left(M^2-2|\mathbf{k}|p_0\right)\tanh^{-1}\left(\frac{2|\mathbf{k}||\mathbf{p}|}{M^2-2|\mathbf{k}|p_0}\right)}{8|\mathbf{k}|^3|\mathbf{p}|p_0} \\ &- \frac{\left(2|\mathbf{k}|p_0+M^2\right)\tanh^{-1}\left(\frac{2|\mathbf{k}||\mathbf{p}|}{2|\mathbf{k}|p_0+M^2}\right)}{8|\mathbf{k}|^3|\mathbf{p}|p_0}\end{aligned} \tag{D.24}$$

$$\int_{-1}^{1} d\cos\theta \frac{\chi_-}{(M^2+2|\mathbf{k}|\chi_-)} = \frac{M^2 \log\frac{\left(-2|\mathbf{k}||\mathbf{p}|+2|\mathbf{k}|p_0+M^2\right)}{\left(2|\mathbf{k}|(|\mathbf{p}|+p_0)+M^2\right)}}{4|\mathbf{k}|^2|\mathbf{p}|} + \frac{1}{|\mathbf{k}|} \tag{D.25}$$

$$\int_{-1}^{1} d\cos\theta \frac{1}{(M^2+2|\mathbf{k}|\chi_-)} = \frac{\log\frac{\left(2|\mathbf{k}|(|\mathbf{p}|+p_0)+M^2\right)}{\left(-2|\mathbf{k}||\mathbf{p}|+2|\mathbf{k}|p_0+M^2\right)}}{2|\mathbf{k}||\mathbf{p}|} \tag{D.26}$$

$J[p_1 \cdot k k^2]$

$$\int_{-1}^{1} d\cos\theta \frac{M^2+|\mathbf{k}|\chi_-}{M^2-2|\mathbf{k}|\chi_-} = \frac{3M^2 \log\frac{\left(2|\mathbf{k}||\mathbf{p}|-2|\mathbf{k}|p_0+M^2\right)}{\left(M^2-2|\mathbf{k}|(|\mathbf{p}|+p_0)\right)}}{4|\mathbf{k}||\mathbf{p}|} - 1 \tag{D.27}$$

$$\int_{-1}^{1} d\cos\theta \frac{2p_0^2+|\mathbf{k}|\chi_-}{M^2+2|\mathbf{k}|\chi_+} = -\frac{\left(4p_0(|\mathbf{k}|+p_0)+M^2\right)\left(\log\frac{\left(-2|\mathbf{k}||\mathbf{p}|+2|\mathbf{k}|p_0+M^2\right)}{\left(2|\mathbf{k}|(|\mathbf{p}|+p_0)+M^2\right)}\right)}{4|\mathbf{k}||\mathbf{p}|} - 1 \tag{D.28}$$

$J[p_2 \cdot k k^2]$

$$\begin{aligned}\int_{-1}^{1} d\cos\theta \frac{2p_0^2-M^2+|\mathbf{k}|\chi_-}{M^2-2|\mathbf{k}|\chi_+} &= \frac{\left(4p_0(|\mathbf{k}|+p_0)-3M^2\right)\left(\log\frac{\left(2|\mathbf{k}||\mathbf{p}|-2|\mathbf{k}|p_0+M^2\right)}{\left(M^2-2|\mathbf{k}|(|\mathbf{p}|+p_0)\right)}\right)}{4|\mathbf{k}||\mathbf{p}|} + 1 \\ \int_{-1}^{1} d\cos\theta \frac{2p_0^2+|\mathbf{k}|\chi_+}{M^2+2|\mathbf{k}|\chi_+} &= \frac{\left(M^2-4p_0^2\right)\left(\log\frac{\left(-2|\mathbf{k}||\mathbf{p}|+2|\mathbf{k}|p_0+M^2\right)}{\left(2|\mathbf{k}|(|\mathbf{p}|+p_0)+M^2\right)}\right)}{4|\mathbf{k}||\mathbf{p}|} + 1\end{aligned} \tag{D.29}$$

$J[(p_1 \cdot k)^2]$

$$\int_{-1}^{1} d\cos\theta \frac{\chi_-^2}{M^2-2|\mathbf{k}|\chi_-} = \frac{M^4\left(\log\frac{\left(2|\mathbf{k}||\mathbf{p}|-2|\mathbf{k}|p_0+M^2\right)}{\left(M^2-2|\mathbf{k}|(|\mathbf{p}|+p_0)\right)} -\right) - 4|\mathbf{k}||\mathbf{p}|\left(2|\mathbf{k}|p_0+M^2\right)}{8|\mathbf{k}|^3|\mathbf{p}|} \tag{D.30}$$

$$\int_{-1}^{1} d\cos\theta \frac{(M^2+|\mathbf{k}|\chi_-)^2}{(M^2+2|\mathbf{k}|\chi_-)(M^2-2|\mathbf{k}|\chi_+)} = \frac{1}{32|\mathbf{k}|^4|\mathbf{p}|p_0}\Big[16|\mathbf{k}|^2|\mathbf{p}|p_0$$

$$+\Big((1-2|\mathbf{k}|)M^2-4|\mathbf{k}|p_0\Big)^2\log\frac{\big(2|\mathbf{k}||\mathbf{p}|-2|\mathbf{k}|p_0+M^2\big)}{(M^2-2|\mathbf{k}|(|\mathbf{p}+p_0))}$$

$$+(1-2|\mathbf{k}|)^2M^4\log\frac{\big(-2|\mathbf{k}||\mathbf{p}|+2|\mathbf{k}|p_0+M^2\big)}{(2|\mathbf{k}|(|\mathbf{p}|+p_0)+M^2)}\Bigg] \qquad \text{(D.31)}$$

$$\int_{-1}^{1}d\cos\theta\,\frac{(2p_0^2+|\mathbf{k}|\chi_-)^2}{M^2+2|\mathbf{k}|\chi_+} = -\frac{1}{8|\mathbf{k}|^3|\mathbf{p}|}\Bigg[4|\mathbf{k}||\mathbf{p}|\Big(2|\mathbf{k}|p_0(4p_0+3)+M^2\Big)$$

$$+\Big(4|\mathbf{k}|p_0(p_0+1)+M^2\Big)^2\left(\log\frac{\big(-2|\mathbf{k}||\mathbf{p}|+2|\mathbf{k}|p_0+M^2\big)}{(2|\mathbf{k}|(|\mathbf{p}|+p_0)+M^2)}\right)\Bigg] \qquad \text{(D.32)}$$

$J[p_1\cdot k\,p_2\cdot k]$

$$\int_{-1}^{1}d\cos\theta\,\frac{\chi_-\chi_+}{M^2-2|\mathbf{k}|\chi_-} = \frac{1}{8|\mathbf{k}|^3|\mathbf{p}|}\Bigg[4|\mathbf{k}||\mathbf{p}|\Big(M^2-2|\mathbf{k}|p_0\Big)$$

$$+M^2\Big(M^2-4|\mathbf{k}|p_0\Big)\left(\log\frac{\big(M^2-2|\mathbf{k}|(|\mathbf{p}|+p_0)\big)}{(2|\mathbf{k}||\mathbf{p}|-2|\mathbf{k}|p_0+M^2)}\right)\Bigg] \qquad \text{(D.33)}$$

$$\int_{-1}^{1}d\cos\theta\,\frac{(M^2+|\mathbf{k}|\chi_-)(2p_0^2-M^2+|\mathbf{k}|\chi_+)}{(M^2+2|\mathbf{k}|\chi_-)(M^2-2|\mathbf{k}|\chi_+)} = \frac{1}{32|\mathbf{k}|^2|\mathbf{p}|p_0}\Bigg[-16|\mathbf{k}|^2|\mathbf{p}|p_0$$

$$-\Big(M^2-4p_0^2\Big)\Big(4|\mathbf{k}|p_0+M^2\Big)\log\frac{\big(2|\mathbf{k}||\mathbf{p}|-2|\mathbf{k}|p_0+M^2\big)}{(M^2-2|\mathbf{k}|(|\mathbf{p}|+p_0))}$$

$$+M^2\Big(4p_0(|\mathbf{k}|+p_0)-M^2\Big)\log\frac{\big(-2|\mathbf{k}||\mathbf{p}|+2|\mathbf{k}|p_0+M^2\big)}{(2|\mathbf{k}|(|\mathbf{p}|+p_0)+M^2)}$$

$$+\Big(4p_0(|\mathbf{k}|+p_0)+M^2\Big)^2\left(\log\frac{\big(-2|\mathbf{k}||\mathbf{p}|+2|\mathbf{k}|p_0+M^2\big)}{(2|\mathbf{k}|(|\mathbf{p}|+p_0)+M^2)}\right)\Bigg] \qquad \text{(D.34)}$$

$J[(p_2\cdot k)^2]$

$$\int_{-1}^{1}d\cos\theta\,\frac{\chi_+^2}{M^2-2|\mathbf{k}|\chi_-} = \frac{1}{8|\mathbf{k}|^3|\mathbf{p}|}\Bigg[4|\mathbf{k}||\mathbf{p}|\Big(6|\mathbf{k}|p_0-M^2\Big)$$

$$+\Big(M^2-4|\mathbf{k}|p_0\Big)^2\left(\log\frac{\big(2|\mathbf{k}||\mathbf{p}|-2|\mathbf{k}|p_0+M^2\big)}{(M^2-2|\mathbf{k}|(|\mathbf{p}|+p_0))}\right)\Bigg]$$

$$\int_{-1}^{1}d\cos\theta\,\frac{(2p_0^2-M^2+|\mathbf{k}|\chi_+)^2}{(M^2+2|\mathbf{k}|\chi_-)(M^2-2|\mathbf{k}|\chi_+)} = \frac{1}{32|\mathbf{k}|^2|\mathbf{p}|p_0}\Bigg[16|\mathbf{k}|^2|\mathbf{p}|p_0+\Big(M^2-4p_0^2\Big)^2\log\frac{\big(2|\mathbf{k}||\mathbf{p}|-2|\mathbf{k}|p_0+M^2\big)}{(M^2-2|\mathbf{k}|(|\mathbf{p}|+p_0))}$$

$$+\Big(M^2-4p_0(|\mathbf{k}|+p_0)\Big)^2\log\frac{\big(-2|\mathbf{k}||\mathbf{p}|+2|\mathbf{k}|p_0+M^2\big)}{(2|\mathbf{k}|(|\mathbf{p}|+p_0)+M^2)}\Bigg]$$

$$\int_{-1}^{1}d\cos\theta\,\frac{(2p_0^2+|\mathbf{k}|\chi_+)^2}{M^2+2|\mathbf{k}|\chi_+} = -\frac{\Big(M^2-4p_0^2\Big)^2\left(\log\frac{\left(-2|\mathbf{k}||\mathbf{p}|+2|\mathbf{k}|p_0+M^2\right)}{(2|\mathbf{k}|(|\mathbf{p}|+p_0)+M^2)}\right)}{8|\mathbf{k}||\mathbf{p}|}+p_0(|\mathbf{k}|+4p_0)-\frac{M^2}{2} \qquad \text{(D.35)}$$

## D.6 Vectorial Ward identities

In this Appendix, we verify the vector Ward identities of the chiral interaction at finite temperature and density through a perturbative computation. We consider the contraction of the $\Gamma$ vertex with the momentum of a photon

$$
\begin{aligned}
&p_{1\mu}\Gamma^{\lambda\mu\nu}(p_1,p_2,\beta) = \\
&p_{1\mu}\int d^4k \mathrm{Tr}[\gamma^\mu \not{k}\gamma^\lambda\gamma^5(\not{k}-\not{q})\gamma^\nu(\not{k}-\not{p}_1)](G_0(k)+G_1(k))(G_0(k-q)+G_1(k-q))(G_0(k-p_1)+G_1(k-p_1)) \\
&+p_{1\mu}\int d^4k \mathrm{Tr}[\gamma^\mu \not{k}\gamma^\lambda\gamma^5(\not{k}-\not{q})\gamma^\nu(\not{k}-\not{p}_2)](G_0(k)+G_1(k))(G_0(k-q)+G_1(k-q))(G_0(k-p_2)+G_1(k-p_2)).
\end{aligned} \tag{D.1}
$$

The algebra we use to perform the full computation is quite straightforward. After passing $p_1$ through the trace, we rewrite in the first integral $\not{p}_1 \to -(\not{k}-\not{p}_1)+\not{k}$, and in the second integral $\not{p}_1 \to -(\not{k}-\not{q})+(\not{k}-\not{p}_2)$. We then have

$$
\begin{aligned}
&p_{1\mu}\Gamma^{\lambda\mu\nu}(p_1,p_2,\beta) = \\
&-\int d^4k \mathrm{Tr}[\not{k}\gamma^\lambda\gamma^5(\not{k}-\not{q})\gamma^\nu](k-p_1)^2(G_0(k)+G_1(k))(G_0(k-q)+G_1(k-q))(G_0(k-p_1)+G_1(k-p_1)) \\
&+\int d^4k \mathrm{Tr}[\gamma^\lambda\gamma^5(\not{k}-\not{q})\gamma^\nu(\not{k}-\not{p}_1)]k^2(G_0(k)+G_1(k))(G_0(k-q)+G_1(k-q))(G_0(k-p_1)+G_1(k-p_1)) \\
&-\int d^4k \mathrm{Tr}[\gamma^\nu\not{k}\gamma^\lambda\gamma^5(\not{k}-\not{p}_2)](k-q)^2(G_0(k)+G_1(k))(G_0(k-q)+G_1(k-q))(G_0(k-p_2)+G_1(k-p_2)) \\
&+\int d^4k \mathrm{Tr}[\gamma^\nu\not{k}\gamma^\lambda\gamma^5(\not{k}-\not{q})](k-p_2)^2(G_0(k)+G_1(k))(G_0(k-q)+G_1(k-q))(G_0(k-p_2)+G_1(k-p_2)).
\end{aligned} \tag{D.2}
$$

We now focus on the terms with only one propagator $G_1$

$$
\begin{aligned}
p_{1\mu}\Gamma^{\lambda\mu\nu}(p_1,p_2,\beta) = &-\int d^4k \mathrm{Tr}[\not{k}\gamma^\lambda\gamma^5(\not{k}-\not{q})\gamma^\nu](G_0(k)G_1(k-q)+G_1(k)G_0(k-q)) \\
&+\int d^4k \mathrm{Tr}[\gamma^\lambda\gamma^5(\not{k}-\not{q})\gamma^\nu(\not{k}-\not{p}_1)](G_0(k-q)G_1(k-p_1)+G_1(k-q)G_0(k-p_1)) \\
&-\int d^4k \mathrm{Tr}[\gamma^\nu\not{k}\gamma^\lambda\gamma^5(\not{k}-\not{p}_2)](G_0(k)G_1(k-p_2)+G_1(k)G_0(k-p_2)) \\
&+\int d^4k \mathrm{Tr}[\gamma^\nu\not{k}\gamma^\lambda\gamma^5(\not{k}-\not{q})](G_0(k-q)G_1(k)+G_1(k-q)G_0(k))+\ldots
\end{aligned} \tag{D.3}
$$

where we denoted with the dots the terms with multiple $G_1$. The first and fourth integrals in the equation above cancel each other. Additionally, by performing a shift in the integration variable, specifically $k \to k-p_1$, we find that the third integral also cancels with the second one. Note that a shift in the integration variable is always allowed in the hot part of the $\Gamma$ diagram since the integrals are not linearly divergent as they are in the cold part. Indeed, we have

$$
\Gamma^{(\mathrm{Dens})\,\lambda\mu\nu} \sim \int_0^\mu dk\, k^\alpha \tag{D.4}
$$

with a power $\alpha > 0$.
Lastly, one can similarly observe that the contributions involving two and three $G_1$ are zero as well in eq. (D.2).

# Appendix E

# The conformal currents and the CWIs

In this section, we retrace the steps of the derivation of the conformal Ward identities, using the $\langle TTJ_A \rangle$ correlator as a working example. A similar procedure can be applied to other correlators. For the derivation of the dilatation and special CWIs, we use the conservation of the conformal currents

$$J_K^\mu = K_\nu T^{\mu\nu} \tag{E.1}$$

where $K$ is a conformal Killing vector (CKV) of the flat metric satisfying the equation

$$\partial_\mu K_\nu + \partial_\nu K_\mu = \frac{2}{d}\eta_{\mu\nu}(\partial \cdot K) \tag{E.2}$$

The conservation of $J_K$ is violated by the presence of a trace anomaly.
The CKVs for the dilatations are

$$K_\mu^{(D)}(x) = x_\mu, \qquad \partial \cdot K^{(D)} = d, \tag{E.3}$$

while for the special conformal transformations, they are given by

$$K^{(S)\kappa}{}_\mu(x) = 2x^\kappa x_\mu - x^2 \delta^\kappa{}_\mu, \qquad \partial \cdot K^{(S)\kappa}(x) = (2d)\, x^\kappa, \qquad \kappa = 1, \dots, d \tag{E.4}$$

We start by assuming that the following surface terms vanish due to the fast fall-off behavior of the correlation function at infinity

$$\begin{aligned} 0 = &\int dx\, \partial_\mu^{(x)} \Big[ K_\nu(x) \langle T^{\mu\nu}(x) T^{\mu_1\nu_1}(x_1) T^{\mu_2\nu_2}(x_2) J_A^{\mu_3}(x_3) \rangle \Big] = \\ &\int dx\, \left(\partial_\mu K_\nu\right) \langle T^{\mu\nu}(x) T^{\mu_1\nu_1}(x_1) T^{\mu_2\nu_2}(x_2) J_A^{\mu_3}(x_3) \rangle + K_\nu \partial_\mu \langle T^{\mu\nu}(x) T^{\mu_1\nu_1}(x_1) T^{\mu_2\nu_2}(x_2) J_A^{\mu_3}(x_3) \rangle. \end{aligned} \tag{E.5}$$

Using the conformal Killing vector equation (E.2), we can then write

$$0 = \int dx \left(\frac{\partial \cdot K}{d}\right) \eta_{\mu\nu} \langle T^{\mu\nu}(x) T^{\mu_1\nu_1}(x_1) T^{\mu_2\nu_2}(x_2) J_A^{\mu_3}(x_3) \rangle + K_\nu \partial_\mu \langle T^{\mu\nu}(x) T^{\mu_1\nu_1}(x_1) T^{\mu_2\nu_2}(x_2) J_A^{\mu_3}(x_3) \rangle. \tag{E.6}$$

On the right-hand side of the last equation, we have the trace and the divergence of a four-point correlator function. We can use the anomalous trace equation and the conservation of the energy-momentum tensor to rewrite these terms. We will demonstrate this in the following.

We first focus on the dilatations. The Killing vectors in this case are given by eq. (E.3). The invariance under diffeomorphisms leads to

$$\nabla^{\mu}\left\langle T_{\mu\nu}\right\rangle - F^{A}_{\mu\nu}\left\langle J^{\mu}_{A}\right\rangle + A_{\nu}\nabla_{\mu}\left\langle J^{\mu}_{A}\right\rangle = 0 \tag{E.7}$$

Applying functional derivatives to this equation and taking the flat limit, we obtain

$$\begin{aligned} 0 = {} & \partial_{\mu}\langle T^{\mu\nu}(x)T^{\mu_1\nu_1}(x_1)T^{\mu_2\nu_2}(x_2)J^{\mu_3}_{A}(x_3)\rangle + \\ & \left[\left(\partial_{\mu}\delta_{xx_1}\right)\delta^{\nu\mu_1}\delta^{\nu_1}_{\lambda} + \left(\partial_{\mu}\delta_{xx_1}\right)\delta^{\nu\nu_1}\delta^{\mu_1}_{\lambda} - \left(\partial^{\nu}\delta_{xx_1}\right)\delta^{\mu_1}_{\mu}\delta^{\nu_1}_{\lambda}\right]\langle T^{\lambda\mu}(x)T^{\mu_2\nu_2}(x_2)J^{\mu_3}_{A}(x_3)\rangle + \\ & \left[\left(\partial_{\mu}\delta_{xx_2}\right)\delta^{\nu\mu_2}\delta^{\nu_2}_{\lambda} + \left(\partial_{\mu}\delta_{xx_2}\right)\delta^{\nu\nu_2}\delta^{\mu_2}_{\lambda} - \left(\partial^{\nu}\delta_{xx_2}\right)\delta^{\mu_2}_{\mu}\delta^{\nu_2}_{\lambda}\right]\langle T^{\lambda\mu}(x)T^{\mu_1\nu_1}(x_1)J^{\mu_3}_{A}(x_3)\rangle + \\ & \left[-\left(\partial_{\nu}\delta_{xx_3}\right)\delta^{\mu_3}_{\mu} + \left(\partial_{\mu}\delta_{xx_3}\right)\delta^{\mu_3}_{\nu} + \delta^{\mu_3}_{\nu}\delta_{xx_3}\partial_{\mu}\right]\langle T^{\mu_1\nu_1}(x_1)T^{\mu_2\nu_2}(x_2)J^{\mu}_{A}(x)\rangle \end{aligned} \tag{E.8}$$

where $\delta_{xy}$ is the dirac delta function $\delta^4(x-y)$ and all derivatives are with respect to the $x$ variable. The anomalous trace equation is instead given by

$$g_{\mu\nu}\langle T^{\mu\nu}\rangle = b_1\,E_4 + b_2\,C^{\mu\nu\rho\sigma}C_{\mu\nu\rho\sigma} + b_3\,\nabla^2 R + b_4\,F^{\mu\nu}F_{\mu\nu} + f_1\,\varepsilon^{\mu\nu\rho\sigma}R_{\alpha\beta\mu\nu}R^{\alpha\beta}_{\ \ \rho\sigma} + f_2\,\varepsilon^{\mu\nu\rho\sigma}F_{\mu\nu}F_{\rho\sigma}, \tag{E.9}$$

Applying functional derivatives and taking the flat limit, we get

$$\delta_{\mu\nu}\langle T^{\mu\nu}(x)T^{\mu_1\nu_1}(x_1)T^{\mu_2\nu_2}(x_2)J^{\mu_3}_{A}(x_3)\rangle = -2\left(\delta_{xx_1} + \delta_{xx_2}\right)\langle T^{\mu_1\nu_1}(x_1)T^{\mu_2\nu_2}(x_2)J^{\mu_3}_{A}(x_3)\rangle \tag{E.10}$$

Inserting equations (E.3), (E.8), and (E.10) into (E.6) and integrating by parts, we obtain the dilatation CWI

$$0 = \left[\sum_i x_i^{\mu}\,\partial^{(x_i)}_{\mu} + 3d - 1\right]\langle T^{\mu_1\nu_1}(x_1)T^{\mu_2\nu_2}(x_2)J^{\mu_3}_{5}(x_3)\rangle \tag{E.11}$$

In momentum-space, we can write it as

$$\left(\sum_{i=1}^{3}\Delta_i - 2d - \sum_{i=1}^{2} p_i^{\mu}\frac{\partial}{\partial p_i^{\mu}}\right)\langle T^{\mu_1\nu_1}(p_1)T^{\mu_2\nu_2}(p_2)J^{\mu_3}_{5}(p_3)\rangle = 0. \tag{E.12}$$

If instead we consider the special conformal transformations, with Killing vectors given in eq. (E.4), proceeding in a similar manner, we arrive at

$$\begin{aligned} 0 = {} & \sum_{i=1}^{3}\left[2x_i^{\kappa}\left(\Delta_i + x_i^{\alpha}\frac{\partial}{\partial x_i^{\alpha}}\right) - x_i^2\delta^{\kappa\alpha}\frac{\partial}{\partial x_i^{\alpha}}\right]\langle T^{\mu_1\nu_1}(x_1)T^{\mu_2\nu_2}(x_2)J^{\mu_3}_{5}(x_3)\rangle \\ & + 2\left[\delta^{\kappa\mu_1}x_{1\alpha} - \delta^{\kappa}_{\alpha}x_1^{\mu_1}\right]\langle T^{\alpha\nu_1}(x_1)T^{\mu_2\nu_2}(x_2)J^{\mu_3}_{5}(x_3)\rangle + 2\left[\delta^{\kappa\nu_1}x_{1\alpha} - \delta^{\kappa}_{\alpha}x_1^{\nu_1}\right]\langle T^{\mu_1\alpha}(x_1)T^{\mu_2\nu_2}(x_2)J^{\mu_3}_{5}(x_3)\rangle \\ & + 2\left[\delta^{\kappa\mu_2}x_{2\alpha} - \delta^{\kappa}_{\alpha}x_2^{\mu_2}\right]\langle T^{\mu_1\nu_1}(x_1)T^{\alpha\nu_2}(x_2)J^{\mu_3}_{5}(x_3)\rangle + 2\left[\delta^{\kappa\nu_2}x_{2\alpha} - \delta^{\kappa}_{\alpha}x_2^{\nu_2}\right]\langle T^{\mu_1\nu_1}(x_1)T^{\mu_2\alpha}(x_2)J^{\mu_3}_{5}(x_3)\rangle \\ & + 2\left[\delta^{\kappa\mu_3}x_{3\alpha} - \delta^{\kappa}_{\alpha}x_3^{\mu_3}\right]\langle T^{\mu_1\nu_1}(x_1)T^{\mu_2\nu_2}(x_2)J^{\alpha}_{5}(x_3)\rangle \end{aligned} \tag{E.13}$$

In momentum-space, we can write it as

$$
\begin{aligned}
0 =& \mathcal{K}^{\kappa} \left\langle T^{\mu_1\nu_1}(p_1)\, T^{\mu_2\nu_2}(p_2)\, J_5^{\mu_3}(p_3) \right\rangle \\
=& \sum_{j=1}^{2} \left( 2\left(\Delta_j - d\right) \frac{\partial}{\partial p_{j\kappa}} - 2 p_j^{\alpha} \frac{\partial}{\partial p_j^{\alpha}} \frac{\partial}{\partial p_{j\kappa}} + \left(p_j\right)^{\kappa} \frac{\partial}{\partial p_j^{\alpha}} \frac{\partial}{\partial p_{j\alpha}} \right) \left\langle T^{\mu_1\nu_1}(p_1)\, T^{\mu_2\nu_2}(p_2)\, J_5^{\mu_3}(p_3) \right\rangle \\
&+ 4 \left( \delta^{\kappa(\mu_1} \frac{\partial}{\partial p_1^{\alpha_1}} - \delta^{\kappa}_{\alpha_1} \delta^{(\mu_1}_{\lambda} \frac{\partial}{\partial p_{1\lambda}} \right) \left\langle T^{\nu_1)\alpha_1}(p_1)\, T^{\mu_2\nu_2}(p_2)\, J_5^{\mu_3}(p_3) \right\rangle \\
&+ 4 \left( \delta^{\kappa(\mu_2} \frac{\partial}{\partial p_2^{\alpha_2}} - \delta^{\kappa}_{\alpha_2} \delta^{(\mu_2}_{\lambda} \frac{\partial}{\partial p_{2\lambda}} \right) \left\langle T^{\nu_2)\alpha_2}(p_2)\, T^{\mu_1\nu_1}(p_1)\, J_5^{\mu_3}(p_3) \right\rangle .
\end{aligned}
\tag{E.14}
$$

# Appendix F

# The Borel transform

The derivation of the spectral density and of the integral can be performed using a limit on the Borel mass after a Borel transform, while the spectral density can be computed applying in sequence the Borel transform. Let's consider the dispersion relation of the function

$$\Pi(q^2) = \frac{1}{\pi}\int_0^\infty \frac{\Delta(s)ds}{(s-q^2)} + \text{subtractions}, \tag{F.1}$$

with a singularity cut starting at $q^2 = 0$. Eq. (F.1) can also be written in the form

$$\Pi(q^2) \;\; = \;\; \frac{1}{\pi}\int_0^\infty ds \int_0^\infty d\alpha\, \Delta(s) e^{-\alpha(s-q^2)}, \quad q^2 < 0, \tag{F.2}$$

where we have used the exponential parametrization of the denominator. The Borel transform in one variable is defined in its differential version by the operator

$$B(Q^2 \to M^2) = \lim_{\substack{Q^2, n\to\infty \\ Q^2/n = M^2}} \frac{1}{(n-1)!}(Q^2)^n \left(-\frac{d}{dQ^2}\right)^n. \tag{F.3}$$

$M^2$ denotes the Borel mass. It satisfies the identity

$$B(Q^2 \to M^2) e^{-\alpha Q^2} = \delta(1-\alpha M^2) \quad \alpha Q^2 > 0. \tag{F.4}$$

Acting on the polarization operator $\Pi(q^2)$ with the Borel transform we get the usual exponential suppression of the higher states

$$M^2 B(-q_1^2 \to M^2)\Pi(q^2) = \frac{1}{\pi}\int_0^\infty ds\, \Delta(s) e^{-s/M^2}. \tag{F.5}$$

The computation of the spectral density can be performed applying once again the Borel transform eq. (F.5), with respect to the inverse Borel mass $1/M^2$, in order to obtain

$$B(1/M^2 \to \nu)(M^2 B(-q^2 \to M^2)\Pi(q^2)) = \frac{1}{\nu}\Delta(1/\nu). \tag{F.6}$$

Therefore, by acting iteratively with Borel transforms on $\Pi(q^2)$, we obtain an expression from which we can easily identify the spectral weight $\Delta(s)$ of eq. (F.1). On the other hand, for the computation of the integral of the spectral density along the discontinuity, we just send the Borel mass $M$ to infinity in (F.5) obtaining

$$\lim_{M\to\infty} M^2 B(-q_1^2 \to M^2)\Pi(q^2) = \frac{1}{\pi}\int_0^\infty ds\, \Delta(s). \tag{F.7}$$

# Appendix G

# The transverse-traceless form factors of the $\langle TTJ_A \rangle$ in the massive case

In this section, we provide the explicit expression for the quantities defined in eq. (9.3.10) in relation to the transverse-traceless sector of the $\langle TTJ_A \rangle$ in the massive case. We have

$$
\begin{aligned}
\bar{A}_{11} =& -2s_1\Big\{2\lambda m^2\Big[21s^3-3s^2(8s_1+7s_2)+s\big(-15s_1^2+100s_1s_2-21s_2^2\big)+3(s_1-s_2)^2(6s_1+7s_2)\Big] \\
&+ s_1\Big[3s^5+s^4(81s_2-12s_1)+2s^3(9s_1^2-40s_1s_2-42s_2^2)-4s^2(3s_1^3+20s_1^2s_2-76s_1s_2^2+21s_2^3)+3s_2(s_1-s_2)^4 \\
&+ s(s_1-s_2)^2(3s_1^2+82s_1s_2+81s_2^2)\Big]\Big\}, \\
\bar{A}_{12} =& -4\lambda m^2\Big[-3s_2^2(s^2+14ss_1-17s_1^2)-3s_2(s-s_1)(s^2-21ss_1-6s_1^2)+s_2^3(7s-24s_1)+2(s-s_1)^3(s+3s_1)-3s_2^4\Big] \\
&+2s_2\Big[s^6-s^5(14s_1+5s_2)+s^4(-11s_1^2+28s_1s_2+10s_2^2)+2s^3(52s_1^3-102s_1^2s_2-5s_2^3)+s^2(-121s_1^4+152s_1^3s_2+ \\
&120s_1^2s_2^2-28s_1s_2^3+5s_2^4)+s(s_1-s_2)^2(38s_1^3+117s_1^2s_2+12s_1s_2^2-s_2^3)+3s_1^2(s_1-s_2)^4\Big], \\
\bar{A}_{13} =& 4\lambda m^2\Big[3s^4+s^3(24s_1-7s_2)+3s^2(-17s_1^2+14s_1s_2+s_2^2)+3s(s_1-s_2)(6s_1^2+21s_1s_2-s_2^2)+2(s_1-s_2)^3(3s_1+s_2)\Big] \\
&+2s\Big[-2s_2^3(5s^3+102ss_1^2-52s_1^3)+s_2^4(10s^2+28ss_1-11s_1^2)-s_2(s-s_1)^2(s^3-12s^2s_1-117ss_1^2-38s_1^3) \\
&+s_2^2(5s^4-28s^3s_1+120s^2s_1^2+152ss_1^3-121s_1^4)+3s_1^2(s-s_1)^4-s_2^5(5s+14s_1)+s_2^6\Big], \\
\bar{A}_{14} =& 24\Big\{\lambda^2m^4\Big[s^2+s(s_1-2s_2)-2s_1^2+s_1s_2+s_2^2\Big]+m^2s_1\Big[2s_2^3(16s^3-15ss_1^2-5s_1^3)-18ss_2^2(s^3+ss_1^2-2s_1^3)-9s_1s_2^5 \\
&+2s_2^6+3s_2^4(-6s^2+3ss_1+5s_1^2)+3s_1s_2(s_1-3s)(s_1-s)^3+(s-s_1)^5(2s+s_1)\Big]+ss_1^2s_2\Big[3s_2^2(-4s^2+4ss_1+s_1^2) \\
&+3s_2(s^3+4s^2s_1-6ss_1^2+s_1^3)+s_2^3(3s-7s_1)+(s-s_1)^3(3s+2s_1)+3s_2^4\Big]\Big\}, \\
\bar{A}_{15} =& \lambda\Big[4\lambda m^2(5s^2+13ss_1-10ss_2-18s_1^2+13s_1s_2+5s_2^2)-2s_2^3(s^2-8ss_1+8s_1^2)-(s-s_1)^3(s^2-5ss_1-2s_1^2) \\
&-2s_2^2(s^3+24s^2s_1-38ss_1^2-5s_1^3)+s_2(s-s_1)(3s^3+19s^2s_1+95ss_1^2-s_1^3)+s_2^4(3s+8s_1)-s_2^5\Big]
\end{aligned}
\tag{G.1}
$$

and

$$
\begin{aligned}
\bar{A}_{21} =& 4s_1\Big[s s_1\Big(s_2^2(9s-19s_1)-2s_2(9s+4s_1)(s-s_1)-(s-s_1)^3+10s_2^3\Big)-\lambda m^2\Big(17s^2-8s(s_1+s_2)-9(s_1-s_2)^2\Big)\Big],\\
\bar{A}_{22} =& 2ss_2\Big[s^4-2s^3(5s_1+2s_2)+6s^2s_2(2s_1+s_2)+s(26s_1^3-60s_1^2s_2+6s_1s_2^2-4s_2^3)-(s_1-s_2)^2(17s_1^2+6s_1s_2-s_2^2)\Big]\\
&-4\lambda m^2\Big[-s_2(s^2-30ss_1+9s_1^2)+2(s-s_1)^2(s+3s_1)-4ss_2^2+3s_2^3\Big],\\
\bar{A}_{23} =& 2s\Big[2s^3(s_1^2+5s_1s_2+2s_2^2)-6s^2(s_1^3-6s_1^2s_2+2s_1s_2^2+s_2^3)+2s(s_1-s_2)\Big(3s_1^3-20s_1^2s_2+s_1s_2^2-2s_2^3\Big)\\
&-(s_1-s_2)^3(2s_1^2+5s_1s_2-s_2^2)-s_2s^4\Big]+4\lambda m^2\Big[3s^3+2s^2(9s_1-2s_2)-2(s_1-s_2)^3-s(s_1-s_2)(19s_1-s_2)\Big],\\
\bar{A}_{24} =& 12\Big(\lambda m^2+s s_1 s_2\Big)\Big[2\lambda m^2(s+s_1-s_2)+s_1\Big(3s^3+s^2(3s_2-5s_1)+s(s_1-s_2)(s_1+5s_2)+(s_1-s_2)^3\Big)\Big],\\
\bar{A}_{25} =& \lambda\Big[20\lambda m^2(s+s_1-s_2)-s^4+2s^3(3s_1+s_2)+4s^2s_1(3s_2-2s_1)+2s(s_1-s_2)(s_1^2+8s_1s_2+s_2^2)+(s_1-s_2)^4\Big].
\end{aligned}
\tag{G.2}
$$

# Appendix H

# The gravitational form factor of hadrons and the conformal anomaly sum rule

In this Appendix, we temporarily pause our discussion of parity-violating interactions to focus on the $\langle TJJ\rangle_{\text{even}}$ correlator. Here, $J$ denotes a non-Abelian vector current in four dimensions ($d = 4$), while $T$ represents the gauge-fixed stress-energy tensor in QCD. For simplicity, from now on we will drop the "even" subscript.
Specifically, we briefly outline the general structure of the correlator decomposition and examine its connection to the gravitational form factor of hadrons. Finally, we conclude by discussing the sum rule of the conformal anomaly interaction. For a more detailed analysis, see [196, 197].

## H.1 The gravitational form factors

The experimental study of the gravitational form factors (GFFs) of the proton and pion provides valuable nonperturbative insights into their interaction with the QCD energy-momentum tensor, offering a window into the internal distributions of energy, spin, pressure, and shear forces. GFFs are formulated through the matrix elements of the energy-momentum tensor between hadronic states, allowing for a detailed characterization of the hadron's internal structure. Quantum corrections induce a breaking of conformal symmetry due to the trace anomaly

$$g_{\mu\nu}\langle T^{\mu\nu}\rangle_{even} \equiv \beta(g)\, F^{a,\mu\nu}F^{a,}{}_{\mu\nu} + (1+\gamma_m)\sum_q m_q\bar{\psi}_q\psi_q\,, \tag{H.1}$$

where $\beta(g)$ is the $\beta$-function of QCD and $\gamma_m$ is the anomalous dimension of the mass operator. The matrix elements of the energy-momentum tensor for a spin 1/2 hadron can be expressed in terms of the GFFs as

$$\langle p',s'|T_{\mu\nu}(0)|p,s\rangle = \bar{u}'\left[A(t)\,\frac{\gamma_{\{\mu}P_{\nu\}}}{2} + B(t)\,\frac{i\,P_{\{\mu}\sigma_{\nu\}\rho}\Delta^\rho}{4M} + D(t)\,\frac{\Delta_\mu\Delta_\nu - g_{\mu\nu}\Delta^2}{4M} + M\sum_{\hat{a}}\bar{c}^{\hat{a}}(t)\,g_{\mu\nu}\right]u \tag{H.2}$$

where $u(p)$ and $\bar{u}(p')$ are proton spinors, $P$ is the average momentum, and $\Delta$ is the momentum transfer. At large momentum transfers, the amplitude decreases rapidly in accordance with quark and gluon counting rules. The general idea behind quark-gluon counting rules is to count the minimum number of constituents (quarks and gluons) participating in the hard scattering process and then determine how the amplitude of the process scales with the momentum transfer, at large values of momentum transfer $t = q^2$, where $q$ is the momentum of the graviton. Nevertheless, it remains accessible within the factorization framework using pion and proton distribution amplitudes, as this approach can be adapted from studies of electromagnetic form factors [198]. One of the major advantages of the EIC is its exceptionally high luminosity, which is expected to reach up to $10^{34} cm^2 s^{-1}$. This high luminosity compensates for the rapid decrease in cross-sections typical of exclusive processes, allowing for a significant number of exclusive events to be detected even at large momentum transfers, where the cross-sections are very small. The large number of collisions ensures that even rare events are observable.

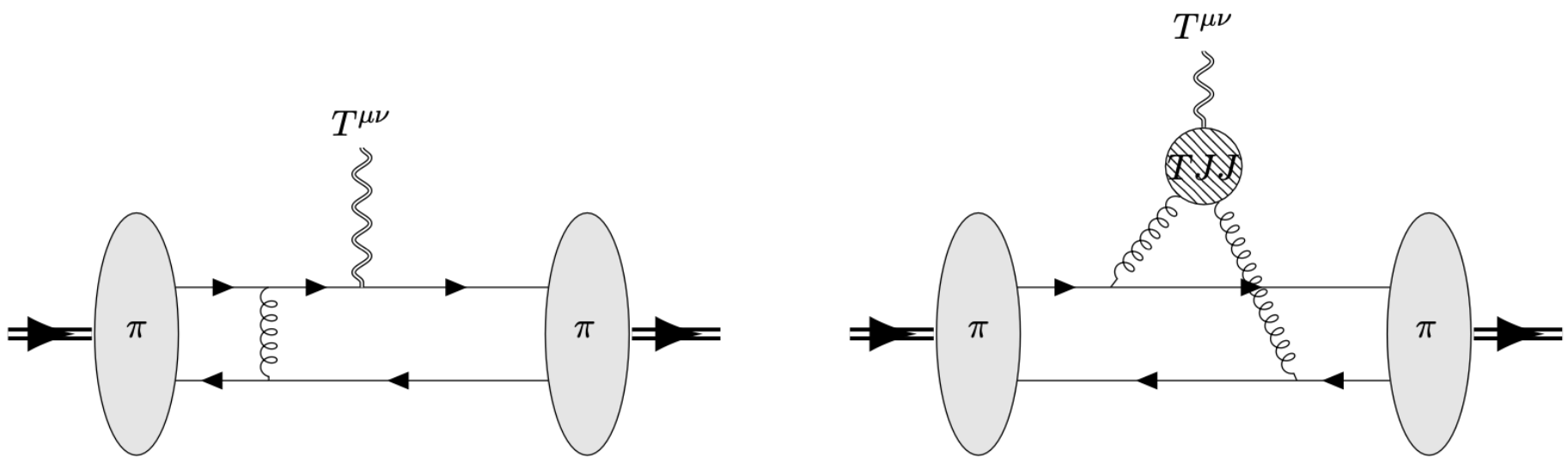

Figure H.1: Factorization picture of the GFF of the pion. Leading order (left). The contribution of the conformal anomaly revealed by the $TJJ$ vertex (right).

We are going to investigate the hard scattering behavior of gravitational form factors (GFFs) at large momentum transfers and higher perturbative orders, where the factorization framework for exclusive processes applies. As shown in Fig. H.1, this method uncovers a conformal anomaly at the $\langle TJJ \rangle$ (graviton/gluon-gluon) vertex at one loop, marked by the emergence of a dilaton interaction mediated by the anomaly. We are going to present an optimal parameterization of the hard scattering amplitude identifying the key form factors from which the dilaton interaction, though subleading in the pertubative expansion, can be singled out at future experiments. Additional details and references of our analysis can be found in [196].

At first sight, a key challenge is the absence of a graviton to directly probe such matrix elements. Gravitational form factors (GFFs) are, however, closely linked to deeply virtual Compton scattering (DVCS) processes, as described in [199]. In DVCS (see Fig. H.2), an electron ($e$) scatters off a nucleon ($N$) in the reaction $eN \to e'N'\gamma$, producing a photon in the final state, controlled by nonforward parton distributions [?, 199]. This process draws significant interest for both pions and protons. In DVCS, a high-energy electron interacts with a hadron, such as a proton or pion, via the exchange of a virtual photon, which then radiates a real photon. By measuring the DVCS invariant amplitudes it is possible to extract informations on the form factors of (H.2). Here, we examine this perturbative correction extending the formalism of conformal field theory in momentum space. We propose a decomposition of the hard scattering amplitude that enables the extraction of the dilaton contribution to these matrix elements. The dilaton can be interpreted as the Nambu Goldstone mode of broken conformal invariance, whose effective interaction can be summarised at phenomenological level in the local form $(\Lambda^{-1}(\varphi FF))$, with $\Lambda$ a typical hadronic scale. Our analysis is built around the exact expression of this interaction in perturbative QCD, which is nonlocal, and can be investigated using the conformal anomaly effective action. We apply methods developed for the analysis of such effective action in gravity [72], specifically adapted to QCD.

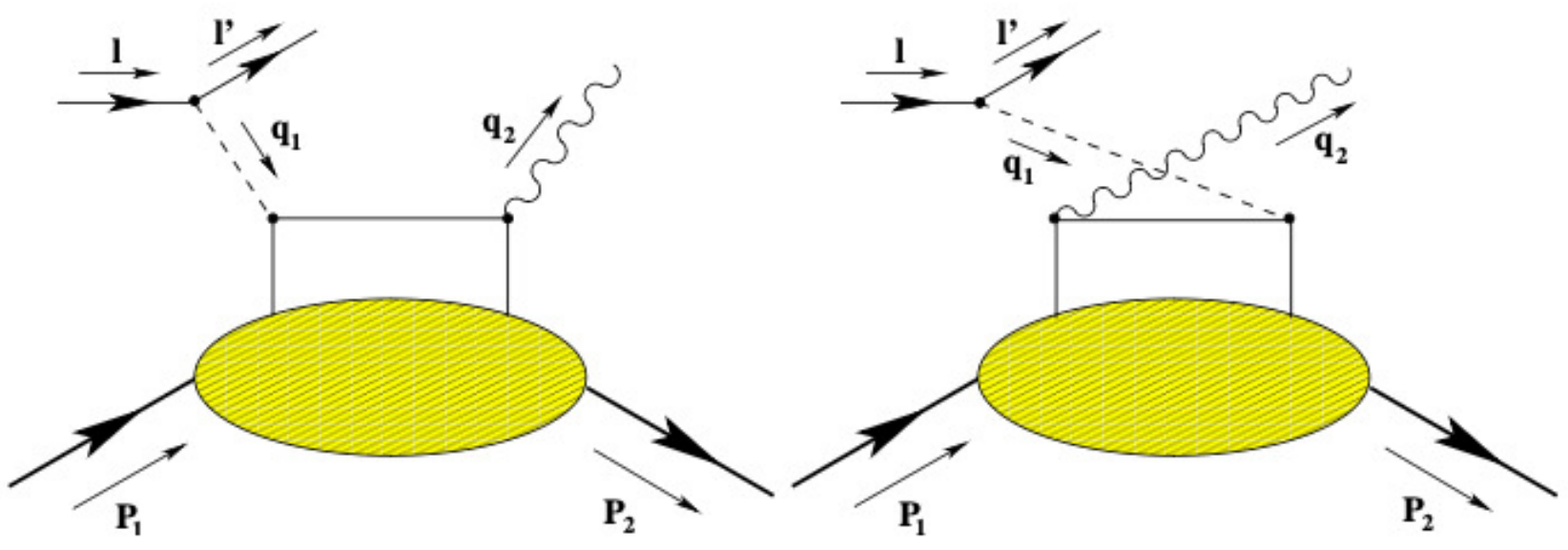

Figure H.2: Factorization in the DVCS process.

## H.2 The $\langle TJJ\rangle$ correlator and the sum rule

Dilaton-like $t$-channel exchanges in the $\langle TJJ\rangle$ correlator were initially studied perturbatively in QED [64, 73] and QCD [74] to explore the coupling of gravity to Standard Model fields [200]. In this section, we focus on the off-shell expansion of the non-Abelian $\langle TJJ\rangle$, previously examined in momentum space through perturbative QCD for on-shell gluons, as detailed in [74]. For a more recent discussion of the trace anomaly in QCD, we refer to [201]. To integrate this interaction into the generalized form factors (GFFs) of hadrons, we extend the on-shell analysis from [74].
We consider the decomposition of the operators $T$ and $J$ in terms of their transverse-traceless, longitudinal and trace components. One can also differentiate between the quark and gluon contributions, that, as we are going to see, behave differently under the application of the conformal constraints. We adopt the following convention for the momenta: the momenta of the two gluons are denoted by $p_1$ and $p_2$, while the momentum of the graviton coupling to the energy-momentum tensor $T$ is denoted by $q$. Additionally, we define the tensor $\hat{\pi}^{\mu\nu} \equiv q^2 g^{\mu\nu} - q^\mu q^\nu$, and introduce the notation $\pi^{\mu\nu} \equiv (1/q^2)\hat{\pi}^{\mu\nu}$. The $\langle TJJ\rangle$ can be expressed in the following general form [196]

$$\begin{aligned}\langle T^{\mu\nu}(q)J^{a\alpha}(p_1)J^{b\beta}(p_2)\rangle &= \langle t^{\mu\nu}(q)j^{a\alpha}(p_1)j^{b\beta}(p_2)\rangle + \langle t^{\mu\nu}(q)j^{a\alpha}_{loc}(p_1)j^{b\beta}(p_2)\rangle_g + \langle t^{\mu\nu}(q)j^{a\alpha}(p_1)j^{b\beta}_{loc}(p_2)\rangle_g \\ &+ 2\mathcal{I}^{\mu\nu\rho}(q)\Big[\delta^{\beta}_{[\rho}p_{2\sigma]}\langle J^{a\alpha}(p_1)J^{b\sigma}(-p_1)\rangle + \delta^{\alpha}_{[\rho}p_{1\sigma]}\langle J^{b\beta}(p_2)J^{a\sigma}(-p_2)\rangle\Big] + \frac{1}{3\,q^2}\hat{\pi}^{\mu\nu}(q)\Big[\mathcal{A}^{\alpha\beta ab} + \mathcal{B}_g^{\alpha\beta ab}\Big],\end{aligned} \tag{H.1}$$

Specifically, we can enumerate all possible tensor that can appear in the transverse-traceless part preserving the symmetry of the correlator

$$\begin{aligned}\langle t^{\mu\nu}(p_1)j^{a\alpha}(p_2)j^{b\beta}(p_3)\rangle = \Pi_1{}^{\mu\nu}_{\mu_1\nu_1}\pi_2{}^{\alpha}_{\alpha_1}\pi_3{}^{\beta}_{\beta_1}\Big(&A_1^{(q)ab}\; p_2^{\mu_1}p_2^{\nu_1}p_3^{\alpha_1}p_1^{\beta_1} + A_2^{(q)ab}\;\delta^{\alpha_1\beta_1}p_2^{\mu_1}p_2^{\nu_1} + A_3^{(q)ab}\;\delta^{\mu_1\alpha_1}p_2^{\nu_1}p_1^{\beta_1} \\ &+A_3^{(q)ab}(p_2\leftrightarrow p_3)\delta^{\mu_1\beta_1}p_2^{\nu_1}p_3^{\alpha_1} + A_4^{(q)ab}\;\delta^{\mu_1\beta_1}\delta^{\alpha_1\nu_1}\Big)\end{aligned} \tag{H.2}$$

We refer to [196] for the explicit expression of the form factors, which are computed perturbatively or using conformal methods. The trace part of the correlator is given by

$$\langle T^{\mu\nu}(q)J^{a\alpha}(p_1)J^{b\beta}(p_2)\rangle_{tr} = \frac{1}{3\,q^2}\hat{\pi}^{\mu\nu}(q)\Big[\mathcal{A}^{\alpha\beta ab} + \mathcal{B}_g^{\alpha\beta ab}\Big]. \tag{H.3}$$

The trace part contains the anomaly contribution $\mathcal{A}^{\alpha\beta ab} = \mathcal{A}^{\alpha\beta}\delta^{ab}$ (where $a$ and $b$ are color indices), and a second term proportional to the equations of motion, $\mathcal{B}_g^{\alpha\beta ab}$. This term should not be considered part of the trace anomaly, although it is part of the trace of the correlator, since the trace operation gives

$$g_{\mu\nu}\langle T^{\mu\nu}(q)J^{a\alpha}(p_1)J^{b\beta}(p_2)\rangle_{tr} = \left[\mathcal{A}^{\alpha\beta ab} + \mathcal{B}_g^{\alpha\beta ab}\right]. \tag{H.4}$$

The first tensor term $\mathcal{A}$ defines the anomaly contribution, as given by the first term on the right-hand-side of (H.1). Its explicit form is

$$\mathcal{A}^{\alpha\beta ab} = \frac{1}{3}\,\frac{g_s^2}{16\pi^2}(11C_A - 2n_f)\delta^{ab}u^{\alpha\beta}(p_1,p_2), \tag{H.5}$$

which is proportional to the QCD $\beta$ function for $n_f$ massless flavours. Here, we have defined the tensor structure

$$u^{\alpha\beta}(p_1,p_2) \equiv (p_1\cdot p_2)g^{\alpha\beta} - p_2^{\alpha}p_1^{\beta}, \tag{H.6}$$

that summarizes the conformal anomaly contribution, being the Fourier transform of the anomaly term at $O(g^2)$

$$u^{\alpha\beta}(p_1,p_2) = -\frac{1}{4}\int d^4x\int d^4y\; e^{-ip_1\cdot x - ip_2\cdot y}\,\frac{\delta^2\{F^a_{\mu\nu}F^{a\mu\nu}(0)\}}{\delta A_\alpha(x)A_\beta(y)}\bigg|_{A=0}. \tag{H.7}$$

The anomaly form factor characterizing the exchange of a dilaton pole in the $\langle TJJ\rangle$ is then distilled from (H.3) in the form

$$\frac{\beta}{q^2}\delta^{ab} \subset \langle T^{\mu\nu}(q)J^{a\alpha}(p_1)J^{b\beta}(p_2)\rangle, \tag{H.8}$$

where the $1/q^2$ anomaly pole, interpreted as the signature of a dilaton exchange in perturbative QCD, has been extracted from the longitudinal projector $\pi^{\mu\nu}$ of the last term on the right-hand side of eq. (H.1).

The analysis of the perturbative amplitude is constrained by ordinary Ward identities, Slavnov-Taylor identities and anomalous conformal Ward identities, as discussed in [196]. Equation (H.1) shows that the structure of the effective vertex corresponding to the $\langle TJJ\rangle$ correlator is modified compared to the ordinary CFT case, as evident from the longitudinal and trace sectors. In the non-Abelian case, the gauge-fixing and ghost sectors of QCD break conformal symmetry. Consequently, the reconstruction of the correlator using momentum-space CFT is valid only for the quark sector. For the full correlator, one must account for additional contributions arising from the exchange of virtual gluons and ghosts within the loops of the $\langle TJJ\rangle$ vertex.

Clearly, when quarks with nonzero mass $m_q$ are introduced, even the quark sector cannot be reconstructed using momentum-space CFT analysis. In this case, for example, the correlator can be computed perturbatively. The decomposition we have presented in this section remains valid. The anomalous tensor $\mathcal{A}^{\alpha\beta ab}$ acquires a more complex structure, encoded in the form factor $\Phi_{TJJ}$

$$\frac{1}{3\,q^2}\mathcal{A}^{\alpha\beta ab} \equiv \Phi_{TJJ}(p_1^2,p_2^2,m_q^2)u^{\alpha\beta}(p_1,p_2)\delta^{ab} \tag{H.9}$$

Proceeding as in Chapter 9, one can verify the existence of a sum rule for the conformal anomaly interaction

$$\frac{1}{\pi}\int_0^\infty \Delta\Phi_{TJJ}\,ds = \lim_{t\to 0}\mathcal{L}^{-1}\left\{\Phi_{TJJ}\right\}(t) = \frac{g^2}{144\pi^2}\left(11C_A - 2n_f\right) \tag{H.10}$$

# Appendix I

# Vertices

In this section we list the explicit expression of all the vertices needed for the perturbative analysis of the $\langle TTJ_A \rangle$ correlator. The momenta of the gravitons and the axial boson are all incoming as well as the momentum indicated with $k_1$. The momentum $k_2$ instead is outgoing. In order to simplify the notation, we introduce the tensor components

$$
\begin{aligned}
A^{\mu\nu\rho\sigma} &\equiv g^{\mu\nu}g^{\rho\sigma} - \frac{1}{2}\left(g^{\mu\rho}g^{\nu\sigma} + g^{\mu\sigma}g^{\nu\rho}\right) \\
B^{\mu\nu\rho\sigma\alpha\beta} &\equiv g^{\alpha\beta}g^{\mu\nu}g^{\rho\sigma} - g^{\alpha\beta}\left(g^{\mu\rho}g^{\nu\sigma} + g^{\mu\sigma}g^{\nu\rho}\right) \\
C^{\mu\nu\rho\sigma\alpha\beta} &\equiv \frac{1}{2}g^{\mu\nu}\left(g^{\alpha\rho}g^{\beta\sigma} + g^{\alpha\sigma}g^{\beta\rho}\right) + \frac{1}{2}g^{\rho\sigma}\left(g^{\alpha\mu}g^{\beta\nu} + g^{\alpha\nu}g^{\beta\mu}\right) \\
D^{\mu\nu\rho\sigma\alpha\beta} &\equiv \frac{1}{2}\left(g^{\alpha\sigma}g^{\beta\mu}g^{\nu\rho} + g^{\alpha\rho}g^{\beta\mu}g^{\nu\sigma} + g^{\alpha\sigma}g^{\beta\nu}g^{\mu\rho} + g^{\alpha\rho}g^{\beta\nu}g^{\mu\sigma}\right) + \\
&\quad \frac{1}{4}\left(g^{\alpha\mu}g^{\beta\sigma}g^{\nu\rho} + g^{\alpha\mu}g^{\beta\rho}g^{\nu\sigma} + g^{\alpha\nu}g^{\beta\sigma}g^{\mu\rho} + g^{\alpha\nu}g^{\beta\rho}g^{\mu\sigma}\right) \\
G^{\alpha\beta\gamma} &\equiv \gamma^\alpha\gamma^\beta\gamma^\gamma - \gamma^\beta\gamma^\alpha\gamma^\gamma + \gamma^\gamma\gamma^\alpha\gamma^\beta - \gamma^\gamma\gamma^\beta\gamma^\alpha
\end{aligned}
\tag{I.1}
$$

The vertices can then be written as

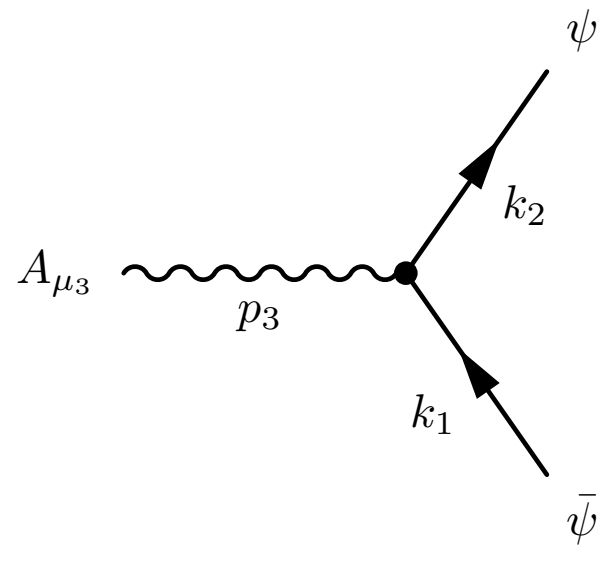

$$
V^{\mu_3}_{A\bar{\psi}\psi} = -ig\gamma^{\mu_3}\gamma^5
$$

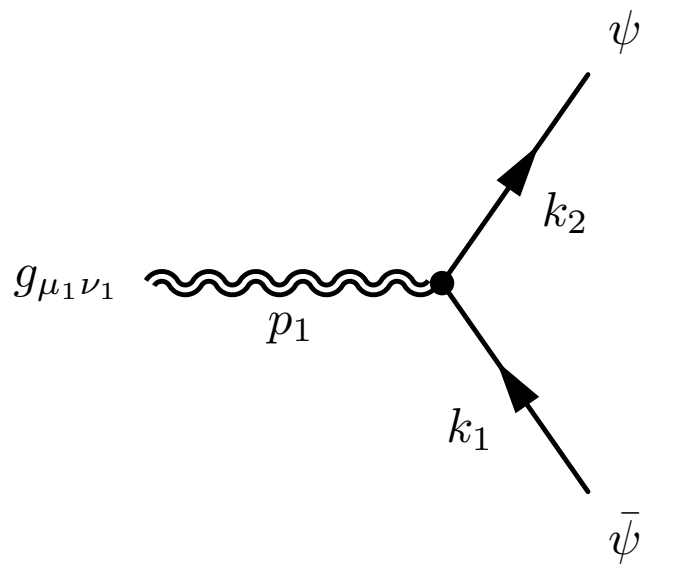

$$V^{\mu_1\nu_1}_{g\bar{\psi}\psi} = \frac{i}{4}A^{\mu_1\nu_1\rho\sigma}(k_1+k_2)_\rho\gamma_\sigma$$

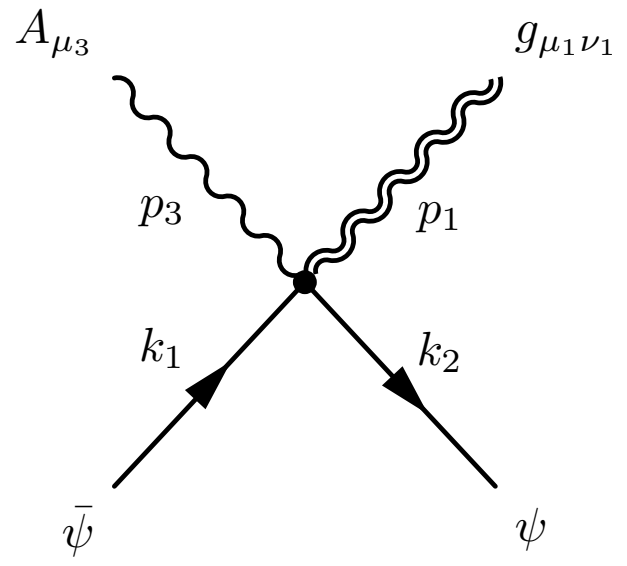

$$V^{\mu_1\nu_1\mu_3}_{gA\bar{\psi}\psi} = -\frac{ig}{2}A^{\mu_1\nu_1\mu_3\rho}\gamma_\rho\gamma^5$$

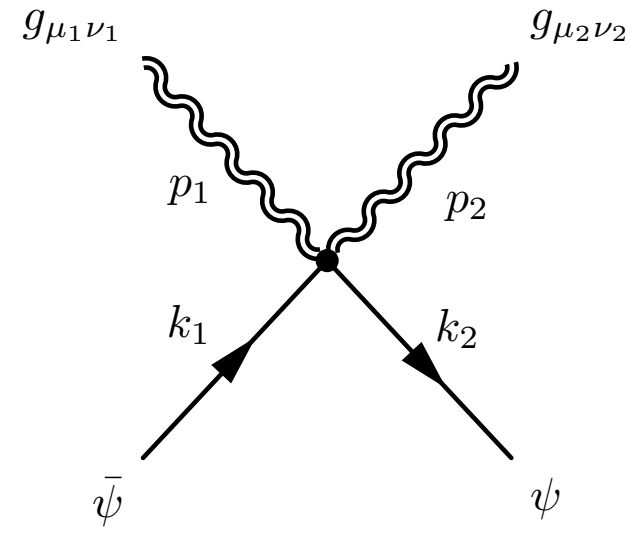

$$\begin{aligned} V^{\mu_1\nu_1\mu_2\nu_2}_{gg\bar{\psi}\psi} = \\ &-\frac{i}{8}\left(B^{\mu_1\nu_1\mu_2\nu_2\alpha\beta} - C^{\mu_1\nu_1\mu_2\nu_2\alpha\beta} + D^{\mu_1\nu_1\mu_2\nu_2\alpha\beta}\right)\gamma_\alpha(k_1+k_2)_\beta \\ &+\frac{i}{128}G^{\alpha\beta\gamma}A^{\mu_1\nu_1\gamma\rho}p_2^\sigma \times \\ &\left(g^{\alpha\mu_2}g^{\beta\sigma}g^{\nu_2\rho} + g^{\alpha\nu_2}g^{\beta\sigma}g^{\mu_2\rho} - g^{\alpha\sigma}g^{\beta\nu_2}g^{\mu_2\rho} - g^{\alpha\sigma}g^{\beta\mu_2}g^{\nu_2\rho}\right) \end{aligned}$$

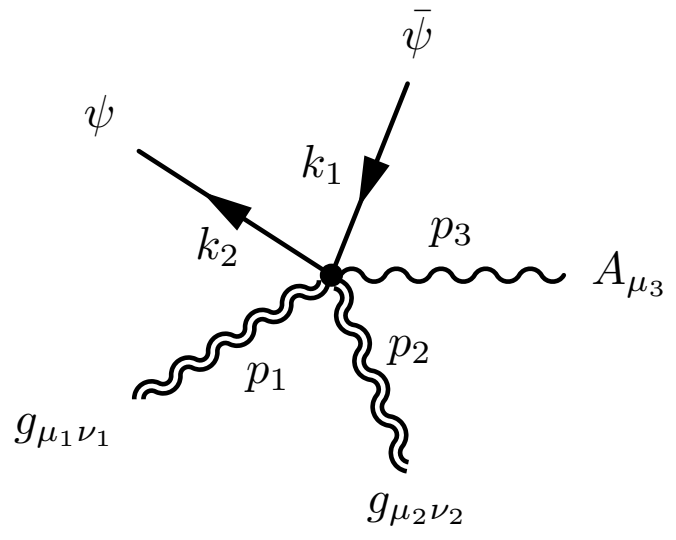

$$\begin{aligned} V^{\mu_1\nu_1\mu_2\nu_2\mu_3}_{ggA\bar{\psi}\psi} = \\ &-\frac{ig}{4}\left(B^{\mu_1\nu_1\mu_2\nu_2\mu_3\lambda} - C^{\mu_1\nu_1\mu_2\nu_2\mu_3\lambda} + D^{\mu_1\nu_1\mu_2\nu_2\mu_3\lambda}\right)\gamma_\lambda\gamma^5 \end{aligned}$$

# Acknowledgements

As this thesis reaches its final page, so too does my PhD journey. It has been a path filled with challenges, growth and invaluable experiences and I am deeply grateful to everyone who has supported me along the way.
First and foremost, I extend my heartfelt gratitude to my supervisor, Claudio Corianò. His dedication, expertise and enthusiasm have been a constant source of inspiration throughout my PhD. Claudio's unwavering commitment to mentoring his students has not only shaped my research but has also been instrumental in my growth as a researcher. Thank you, Claudio, for your invaluable support, encouragement and the countless hours you've devoted to my academic journey.
I am also profoundly grateful to Matteo Maglio, who introduced me to the fascinating world of CFT in momentum-space and ignited the passion that ultimately defined my research focus.
A special thank you goes to my fellow PhD colleagues—Dario, Mario and Riccardo. Sharing this journey with you, through its highs and lows, has been an unforgettable experience. Your camaraderie, insights on life in Salento and friendship made this path both enriching and meaningful.
I am endlessly grateful to my parents for their unwavering support and constant encouragement. Even though they may not fully grasp the content of my work, their belief in me has been a pillar of strength throughout this journey.
To my sister, thank you for (almost) always being a model of initiative and determination. Your drive and ability to make things happen have inspired me to take charge and be more proactive in my own life.
Lastly, I want to thank my friends pursuing careers as judges—Pietro, Guido, Raimonda, Margherita and Iuri—and the soon-to-be physician, Luca. Despite our different academic paths, your dedication and hard work have been a constant source of motivation. I'm grateful for the countless hours we spent studying together in the library, each of us committed to our different respective pursuits (even if you didn't always approve of my study breaks).
To all of you above—thank you for being part of this journey. I am truly grateful for the examples you have set through your actions and viewpoints.